\documentclass[a4paper,11pt]{book}
\usepackage{adjustbox}            %% to align tops of minipages
\usepackage[margin=1in]{geometry} %% a bit more text per line
\usepackage{algorithm}
\usepackage{appendix}
\usepackage{amsmath}
\usepackage{amssymb}
\usepackage{amsfonts}
\usepackage{amsthm}
\usepackage{array}
\usepackage{arydshln}
\usepackage{blindtext}
\usepackage{bm}
\usepackage{csquotes}
\usepackage{dblfloatfix}
\usepackage[english]{babel}
\usepackage{enumitem}
\usepackage{epigraph} 
\usepackage{emptypage}
\usepackage{fixltx2e}
\usepackage{forest}
\usepackage{float}
\usepackage{graphicx}
\usepackage{hyperref}
\usepackage{inconsolata}
\usepackage{lmodern}
\usepackage{listings}
\usepackage{lipsum}
\usepackage{lscape}
\usepackage{multirow}
\usepackage{menukeys}
\usepackage{natbib}
\usepackage[noend]{algpseudocode}
\usepackage{pageno}
\usepackage{picins}
\usepackage{smartdiagram}
\usepackage{setspace}
\usepackage{subfig}
\usepackage[T1]{fontenc}
\usepackage{tabularx} 
\usepackage{tablefootnote}
\usepackage{tikz}
\usepackage{threeparttable}
\usepackage{thmtools}
\usepackage{url}
\usepackage{ulem}
\usepackage[utf8]{inputenc}
\usepackage{wedn}
\usepackage{xcolor}
\usepackage[english]{babel}
\usepackage{blindtext}
\usepackage{graphicx}
\usepackage{wrapfig}
\usetikzlibrary{shadows}
\newcolumntype{C}[1]{>{\centering}p{#1}}
\usetikzlibrary{shadows}
\colorlet{myred}{red!80!black}
\frontmatter
\newcommand*\keystroke[1]{%
  \tikz[baseline=(key.base)]
    \node[%
      draw,
      fill=white,
      drop shadow={shadow xshift=0.25ex,shadow yshift=-0.25ex,fill=black,opacity=0.75},
      rectangle,
      rounded corners=2pt,
      inner sep=1pt,
      line width=0.5pt,
      font=\scriptsize\sffamily
    ](key) {#1\strut}
  ;
}
\newtheorem{dfn}{Definition}[section]
\newtheorem{thm}{Theorem}[section]
\newtheorem{cor}{Corollary}[section]
\newtheorem{prop}{Proposition}[section]

\definecolor{myblue}{RGB}{20,105,176}
\definecolor{constants}{RGB}{127,39,84}
\definecolor{character}{RGB}{100,169,57}
\lstdefinestyle{deltaj}{
        belowcaptionskip=1\baselineskip,
        breaklines=true,
        columns=fullflexible,
        frame=single,
        xleftmargin=\parindent,
        language=R,
        escapeinside={(*}{*)},
        numbers=left,
        stepnumber=1,
        numberblanklines=false,
        basicstyle=\footnotesize\ttfamily,
        keywordstyle=\bfseries\color{green!40!black},
        commentstyle=\itshape\color{purple!40!black},
        identifierstyle=\color{blue},
        stringstyle=\color{orange},
        literate=
         {\{}{{{\color{character}{\{}}}}{1}
         {\}}{{{\color{character}{\}}}}}{1}
         {SPL }{{{\color{constants}{SPL }}}}{1}
         {Features }{{{\color{constants}{Features }}}}{1}
         {Deltas }{{{\color{constants}{Deltas }}}}{1}
         {Constraints }{{{\color{constants}{Constraints }}}}{1}
         {Partitions }{{{\color{constants}{Partitions }}}}{1}
         {Products }{{{\color{constants}{Products }}}}{1}
         {when }{{{\color{constants}{when }}}}{1}
         {\& }{{{\color{constants}{\& }}}}{1}
         {| }{{{\color{constants}{| }}}}{1}
    }
\usetikzlibrary{arrows.meta}

\newcommand{\TikzReturn}{\tikz{\draw[line width=0.1em,{-Triangle[]}] 
(0,0) |- (-1em,-0.4em);}}
  \declaretheoremstyle[
spaceabove=\topsep, spacebelow=\topsep,
headfont=\normalfont\color{myred},
notefont=\mdseries\color{myred}, notebraces={(}{)},
bodyfont=\normalfont,
postheadspace=\newline,
headpunct=,
numberwithin=chapter,
postheadhook=\leavevmode%
  \noindent{\color{myred}\rule{\textwidth}{1pt}}%
qed=\textcolor{myred}{$\blacksquare$}
]{mystyle}
\declaretheorem[style=mystyle]{example}

\newcolumntype{L}{>{\centering\arraybackslash}m{3cm}}
\newenvironment{dedication}
{
   \cleardoublepage
   \thispagestyle{empty}
   \vspace*{\stretch{1}}
   \hfill\begin{minipage}[t]{0.8\textwidth}
   \raggedright
}
{
   \end{minipage}
   \vspace*{\stretch{3}}
   \clearpage
}
\makeatletter
\renewcommand{\@chapapp}{}% Not necessary...
\newenvironment{chapquote}[2][2em]
  {\setlength{\@tempdima}{#1}%
   \def\chapquote@author{#2}%
   \parshape 1 \@tempdima \dimexpr\textwidth-2\@tempdima\relax%
   \itshape}
  {\par\normalfont\hfill--\ \chapquote@author\hspace*{\@tempdima}\par\bigskip}
\makeatother
\title{\Huge \textbf{Digesting Gibbs Sampling Using R}  
}
\author{\textsc{Mahdi Teimouri Yanesari}\thanks{\url{profs.gonbad.ac.ir/fa/teimouri}}}
\begin{document}
\maketitle
\frontmatter
%%%%%%%%%%%%%%%%%%%%%%%%%%%%%%%%%%%%%%%%%%%%%%%%%%%%%%%%%%%%%%%
% Add a dedication paragraph to dedicate your book to someone %
%%%%%%%%%%%%%%%%%%%%%%%%%%%%%%%%%%%%%%%%%%%%%%%%%%%%%%%%%%%%%%%
\begin{dedication}
\begin{center}
{\LARGE{
Dedicated to my parents}}
\end{center}
\vspace*{.15cm}
\end{dedication}
%\begin{center}
%{\LARGE{and}}
%\end{center}
%\vspace*{.15cm}
%\begin{center}
% %   \usefont{T1}{}{bx}{it}
%All rights are reserved. No part of this publication may be reproduced, stored in a retrieval system or transmitted in any form or by any means, electronic, mechanical, photocopying, recording or otherwise, without prior permission of author
%\end{center}
%%%%%%%%%%%%%%%%%%%%%%%%%%%%%%%%%%%%%%%%%%%%%%%%%%%%%%%%%%%%%%%%%%%%%%%%
% Auto-generated table of contents, list of figures and list of tables %
%%%%%%%%%%%%%%%%%%%%%%%%%%%%%%%%%%%%%%%%%%%%%%%%%%%%%%%%%%%%%%%%%%%%%%%%
\tableofcontents
\listoffigures
\listoftables
\mainmatter
%%%%%%%%%%%
% Preface %
%%%%%%%%%%%
%\chapter*{Preface}
%\section*{About the companion website}
%The website\footnote{\url{https://github.com/amberj/latex-book-template}} for this file contains:
%\begin{itemize}
%  \item A link to (freely downlodable) latest version of this document.
%  \item Link to download LaTeX source for this document.
%  \item Miscellaneous material (e.g. suggested readings etc).
%\end{itemize}
%%%%%%%%%%%%%%%%%%%%%%%%%%%%%%%%%%%%
% Give credit where credit is due. %
% Say thanks!                      %
%%%%%%%%%%%%%%%%
% NEW CHAPTER! %
%%%%%%%%%%%%%%%%
\chapter*{Preface}
In general, the statistical simulation approaches are referred to as the {\it{Monte Carlo methods}} as a whole. The broad class of the Monte Carlo methods involves the Markov chain Monte Carlo (MCMC) techniques that attract the attention of researchers from a wide variety of study fields. The main focus of this report is to provide a framework for all users who are interested in implementing the MCMC approaches in their investigations, especially the {\it{Gibbs}} sampling. I have tried, if possible, to eliminate the proofs, but reader is expected to know some topics in elementary calculus (including mathematical function, limit, derivative, partial derivative, simple integral) and statistics (including discrete and continuous random variables, probability distribution, expected value and variance, moment generating function, multivariate distribution, distribution of a functions of random variable, and the central limit theorem). 
\par Almost all of researchers use statistical tools such as {\it{confidence interval}} or {\it{hypothesis testing}} for supporting their study results. Indeed, through both aforementioned tools, investigator can test a statement about the unknown population parameter such as population mean. Herein, the accuracy of decision made about the unknown parameter depends on a concept called {\it{statistic}} that is the sample counterpart of the unknown parameter under testing that is herein, for example, the sample mean or the sample median. The critical question is that which one is better the sample mean or the sample median ? The simulation process enables investigator to answer this question quite well. In fact, investigator will answer this question within a statistical laboratory by simulating sample evidence through the Monte Carlo methods, and then computing the sample mean or median based on simulated evidence. The accuracy of both statistics is accessible easily by repeating the simulation procedure. The Monte Carlo methods are widely used in several fields of study such as biology, engineering, econometrics, medicine, etc. The main concern of this report is placed on showing how user can apply the Monte Carlo methods in practice. I have tried to give the \verb+R+ codes accompanying with examples in order to help user for understanding the theoretical topics in the fields of statistics well. As the last, but not the least, I have used throughout the equivalent phrases ``Gaussian'' and ``normal'' interchangeably. I did my the best to avoid mistakes, specially for the formulas, however, I am sure that still there are several mistakes. Let me know please if you find anyone or have comments through the following email address.
%%%%%%%%%%%%%%%%%%%%%%%%%%%%%%%%%%%%
\section*{Acknowledgements}
\begin{itemize}
\item I would like to express my sincere thanks to my family and my colleagues that indirectly contributed to this work.
\item I would like to express my sincere thanks to Amber Jain (\url{http://amberj.devio.us/}) for a free-to-edit LaTeX template.
%\footnote{\url{https://gonbad.ac.ir}}.
%Professor Don Knuth\footnote{\url{http://www-cs-faculty.stanford.edu/~uno/}} (for \TeX{}) and Leslie Lamport\footnote{\url{http://www.lamport.org/}} (for \LaTeX{}).
%\item I'll also like to thank my colleagues in Gonbad Kavous University \footnote{\url{http://gummi.midnightcoding.org/}} 
%developers and LaTeXila\footnote{\url{http://projects.gnome.org/latexila/}} development team for their awesome \LaTeX{} editors.
%\item I'm deeply indebted my parents and friends for their support and encouragement.
\end{itemize}
%\mbox{}\\
%\mbox{}\\
\noindent Mahdi Teimouri Yanesari, \\
Assistant Professor in Statistics, Gonbad Kavous University, Gonbad Kavous, Iran.\\ 
Email:\url{mahdiba_2001@yahoo.com}\\
%\noindent \url{https://profs.gonbad.ac.ir/fa/teimouri}
%\noindent \url{https://scholar.google.dk/citations?user=jg_oR_4AAAAJ&hl=en}
%\noindent 
%%%%%%%%%%%%%%%%%%%%%%%%%%%%%%%%%%%%%%%%%
\chapter{An introduction to R}
\vspace{-\intextsep}
\vspace{-\baselineskip}
\setlength\intextsep{0pt}
\begin{wrapfigure}[8]{r}{50mm}
%\centering
\includegraphics[width=.95\linewidth]{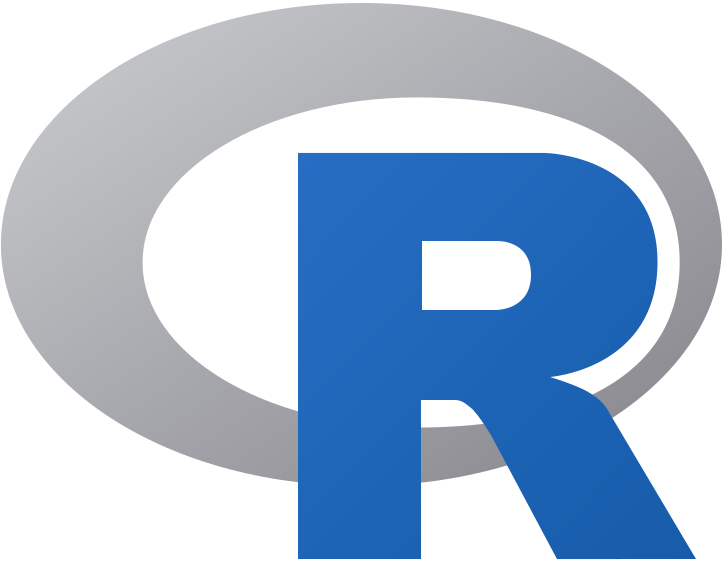}
%%\caption{Basic layout}
\end{wrapfigure}
The \verb+R+ (or \verb+R+ language) is an environment developed mainly for {\it{statistical analysis}}, {\it{statistical computing}}, and {\it{data visualization}}. 
%\vspace{2cm}
%\setlength\intextsep{1pt}
%\blindtext
%\vspace{-\intextsep}
%\begin{wrapfigure}[10]{l}{55mm}
%\begin{wrapfigure}[10]{R}{0.5\textwidth}
%\centering
%\includegraphics[width=0.5\textwidth]{Rlogo}
%\vspace{-10pt}
%%\centering
%%\rule{1pt}{1pt}
%%\vspace{1mm}
%%\vspace{-\baselineskip}
%%\includegraphics[scale=.27]{Rlogo}
%%\vspace{6pt}
%%%\caption{Using Images in \LaTeX}
%\end{wrapfigure}
%\begingroup
%\setlength{\intextsep}{0pt}%
%\setlength{\columnsep}{0pt}%
%\begin{wrapfigure}{r}{i}{o}{0.5\textwidth}
%  \centering\includegraphics[width=\linewidth]{Rlogo}
%  \caption{Basic layout}
%\end{wrapfigure}
%%\lipsum[1]
%\endgroup
%\blindtext
%\pichskip{5pt}% Horizontal gap between picture and text
%\parpic[l][]{%
 % \begin{minipage}{50mm}
%    \includegraphics[width=\linewidth]{Rlogo}%
%%    \captionof{figure}{This is a tiger.}
 % \end{minipage}
%}
%\lipsum[2]
In contrast to {\bf{SPSS}} or {\bf{Minitab}}, \verb+R+ lacks the graphical user interface (GUI) that allows user to accomplish statistical analysis using some graphical components such as button, dialog box, and icon. This means that \verb+R+ is a programming language or {\it{command-line}} interface in which user must type codes within some sessions called \verb+console+ or \verb+script+ for implementing the statistical methods. The first nonformal version on \verb+R+ was created by Ross Ihaka and Robert Gentleman, two statisticians from University of Auckland in August 1993 and then introduced formally in 1996 \citep{ihaka1996r}. The first formal version of \verb+R+, that is version 1.0, was created in February 29, 2000. This language has won a great deal of advocate among data scientists since it  maintains some advantages among them we mention the following:
\begin{enumerate}[label=\roman*.]
\item The \verb+R+ is loaded quickly.% and working on computer with a small hard drive space.
\item The \verb+R+ is user-friendly interface.
\item The \verb+R+ is an open-source software that can be distributed freely. 
\item The \verb+R+ user can call (or interfaces) the \verb+C+ or \verb+Fortran+ for implementing a project faster. 
\item The codes or programs in a new structure called \verb+R+ package that can be exported/imported simply while they are often accompanied by a long-form vignette (documentation) in order to help user for using it.
\end{enumerate}
\par Another computational software written by data analysis researchers at Bell Labs is the \verb+S+ that has its origin from the joint project of Rick Becker and John M. Chambers on May 5 1976 \cite{RJ-2020-028}. The \verb+S+ language was written to provide a framework for research in data analysis and collaborations to apply that research, rather than as a separate project to create a programming language \cite{RJ-2020-028}. The \verb+S+ calls \verb+Fortran+ subroutines and has some extensions such as \verb+S3+ and \verb+S4+. The \verb+R+ and \verb+S3+ are very close together. Herein, we do not deal with the differences between these extensions and their relations with \verb+R+. Some redecorated and absolutely more attractive versions of \verb+R+, called {\bf{R Studio}} or {\bf{Visual R Studio}}, are available for linting, compiling code, highlighting syntax, having a look at data, variables, packages, etc. Herein, we aim to provide a general, but not comprehensive, knowledge for users who are interested in working or being familiar with \verb+R+. Since \verb+R+ is free for distributing or redistributing, the user is able to download it from \url{www.r-project.org}. To do this, the user just needs to enter \url{https://cran.r-project.org/mirrors.html}, and then download \verb+R+, after selecting a mirror from a listing of mirrors that basically are servers located in different locations (countries) which host the same \verb+R+ and corresponding packages. It is suggested to select a close mirror in order to speed downloading. We point out that all notes given here can be found in more details at \url{https://cran.r-project.org/manuals.html}. There is more detailed information on this in \citep{R:Becker+Chambers+Wilks:1988,R:Chambers+Hastie:1992,R:Chambers:1998,R:Nolan+Speed:2000,R:Pinheiro+Bates:2000,R:Therneau+Grambsch:2000,R:Venables+Ripley:2000,R:Harrell:2001,R:fox2002r,R:gareth2013introduction,R:Hothorn+Everitt:2014,R:Bloomfield:2014,R:Daroczi:2015,R:Blangiardo+Cameletti:2015,R:Kohl:2015en,R:Sun:2015,R:Dayal:2015,R:Leemis:2016,
R:SteveMurray:2017,R:hastie2017elements,R:Rahlf:2017,R:kelleyoceanographic2018,R:fox2018r}.
\section{Data and objects in R}
There are, in general, five types of data in \verb+R+ including:
\begin{enumerate}[label=\roman*.]
\item {\it{character}} (or {\it{string}}): the name of people, cities, mountains, or marital status, your gender, etc., that evidently cannot be recorded numerically. 
\item {\it{complex}}: a complex number such as $1+ 2i$ that is represented in \verb+R+ as {\texttt{1+2i}}.
\item {\it{integer}}: a natural or whole number such as 2 that is represented in \verb+R+ when is followed by letter \verb+L+ as \verb+2L+.
\item {\it{logical}}: takes on only two states, that is, \verb+TRUE+ (or \verb+T+) and \verb+FALSE+ (or \verb+F+)
\item {\it{numeric}} (or {\it{double}}): a measurable quantity such as age, height, temperature, etc., that evidently can be recorded numerically, or the chemical, mathematical, or physical constants such as Avogadro's number $\bigl( 6.023\times 10^{23}\bigr)$, the ratio of a circle's circumference to its diameter $(\pi=3.14159265\ldots)$, and the gravity on the Earth $\bigl(g=9.8 m/s^2\bigr)$.
\end{enumerate}
The \verb+R+ user has not permission of access to the stored information directly, but \verb+R+ provides some data structure that is referred to as {\it{object}}. In point of view of some other programming languages {\it{object}} is a {\it{variable}}. Hence, everything in \verb+R+ is an {\it{object}}. All variables, regardless of their type, and all functions are objects. For example, the value such as 2 is a datum of type \verb+"double"+, but if we have \verb+a=2+, then \verb+a+ is an object. Each object in \verb+R+ belongs to a particular {\it{class}} that assigns some {\it{attributes}} or {\it{properties}}. These properties determine how the generic functions operate with the corresponding object. In summary, any object in \verb+R+ has a specific {\it{type}}, is stored in a specific {\it{mode}}, and has a particular class in order to determine the particular {\it{object}} can be used. These features are important when we are working on objects with more  structures such as {\it{vector}}, {\it{data frame}}, {\it{list}}, {\it{matrix}}, etc. 
%\footnote{ The Variables are used to store data values while objects are used to group related data and functions into a single entity}
\subsubsection{R session}
The \verb+R+ has two main sessions including {\it{console}} and {\it{script}} (or code editor). Once we run \verb+R+, then the console session comes up in which a ``>'' prompt appears after which we must write the input command (call). One of the main objects in \verb+R+ is vector that has two types including: {\it{atomic }} and {\it{nonatomic}}. The elements of each atomic vector have the same type while elements of the latter are not of the same type. For example, 
%A general view of \verb+R+ session is given by Figure \ref{}.
suppose we would like to define an atomic vector consisting of numbers one, two, and three. To do this, we may proceed as follows.
{\color{red}{
\begin{verbatim}
R> v0 <- 1:3 # we may write v0 = 1:3
R> v0
\end{verbatim}
}}
and then press the Enter key ``\TikzReturn\ '' to see the output as follows.
{\color{blue}{
\begin{verbatim}
[1] 1 2 3
\end{verbatim}
}}
Throughout, as seen in above, we show the \verb+R+ command and the \verb+R+ output in red- and blue-cored texts, respectively. As it is seen, we can insert an explanatory expression in front of vector \verb+v0+ in above after sharp symbol ``\verb+#+''. We note that another way to define above vector is to run command
{\color{red}{
\begin{verbatim}
R> v1 <- c(1,2,3); v1
\end{verbatim}
}}
Of course, we can write several commands (calls) in one line provided that each command is separated with the next one by {\it{just one}} semicolon ``\verb+;+'' symbol. 
\begin{table}[!h]
\begin{tabular}{ll}%{10mm}{>{\hsize=.10\hsize}X>{\hsize=.95\hsize}X}
&\\
{\texttt{R}} code & output\\
\hline
{\color{red}{\verb+R> mode(v0); typeof(v0); class(v0)+}}&{\color{blue}{\verb+[1] "numeric"+}}\\
&{\color{blue}{\verb+[1] "integer"+}}\\
&{\color{blue}{\verb+[1] "integer"+}}\\
{\color{red}{\verb+R> mode(v1); typeof(v1); class(v1)+}}&{\color{blue}{\verb+[1] "numeric"+}}\\
&{\color{blue}{\verb+[1] "double"+}}\\
&{\color{blue}{\verb+[1] "numeric"+}}\\
\hline
\end{tabular}
\end{table}
\vspace{5mm}
As we see, objects \verb+v0+ and \verb+v1+ have the same \verb+mode+, but different \verb+type+ and \verb+class+. We note that, different objects may have the same class. In practice there is no difference between \verb+v0+ and \verb+v1+ defined in above and can be used interchangeably in many applications. Alternatively, we can produce the output given in above by typing \verb+v1 <-c(1,2,3)+ in the script session and then pressing \keystroke{Ctrl}+ \keystroke{R}. The script is a new window that is appeared by following the track \menu{File>New script}. Once we have created a script session, we follow the track \menu{File>Save as} to save the script under an arbitrary name, for example ``myfile.R'' in which, as we see, suffix  is \verb+.R+. Care should be taken about the location in which the script will be saved. The default path can be changed straightforwardly. The default directory for saving \verb+R+ files is obtained by typing the following command in console or execute it from script session.
{\color{red}{
\begin{verbatim}
R> getwd()
\end{verbatim}
}}
The output is
{\color{blue}{
\begin{verbatim}
[1] "C:/Users/NikPardaz/Documents"
\end{verbatim}
}}
If we no longer interested in directory given above, for changing it, we may put our desired address inside the command \verb+setwd()+. For example, if we would like to change the  path \directory{C/Users/NikPardaz/Documents} to \directory{E/my R work}, then we may use the command given below.
{\color{red}{
\begin{verbatim}
R> setwd("E:/my R work")
\end{verbatim}
}}
\par We note that \verb+R+ is case sensitive that implies, for example, letters ``A'' and ``a'' are different objects. There are different ways in \verb+R+ for getting help with functions and application of phrases. Herein, we suggest the most commonly used ways to get more information about phrase \verb+apply+. The latter is a {\it{built-in}} function that is applied to the margins of an array or matrix. To do this, we may use either command
{\color{red}{
\begin{verbatim}
R> help(apply)
\end{verbatim}
}}
or
{\color{red}{
\begin{verbatim}
R> ?(apply)
\end{verbatim}
}}
The version of \verb+R+ that we are using is very important when installing the external packages. More detailed information on this and related features can be obtained using the command
{\color{red}{
\begin{verbatim}
R> R.version
\end{verbatim}
}}
We may use command 
{\color{red}{
\begin{verbatim}
R> print(v1)
\end{verbatim}
}}
 to observe the elements of vector \verb+v1+. A nonatomic vector such as \verb+v2+ may be defined in \verb+R+ as
{\color{red}{
\begin{verbatim}
R> v2 <- c(1, 2, 'male', 'female')
R> str(v2)
\end{verbatim}
}}
\vspace*{-3mm}
{\color{blue}{
\begin{verbatim}
 chr [1:4] "1" "2" "male" "female"
\end{verbatim}
}}
As it is seen, two last elements of object \verb+v2+ are of type character. If at least one of elements in a given object, such as \verb+v2+, are character, then the given object is considered as a character shown by ``\verb+chr+''. It is worth to note that objects in \verb+R+ whose type is character (nonumeric) must be surrounded by a single or double quotation marks. The elementary mathematical operations including addition, subtraction, and multiplication are meaningless for objects such as \verb+v2+ defined in above. Sometimes, objects \verb+v1+ and \verb+v2+ are called the {\it{numeric}} and {\it{non-numeric}} vectors. Regardless of the its type, the element(s) of each vector can be found easily in \verb+R+. For example, the second element of vector \verb+v1+ is determined by command \verb+v1[2]+. If one needs to determine more elements of a given vector, then the address of those elements must be put inside the square bracket. Let vector \verb+v3+ is defined as
{\color{red}{
\begin{verbatim}
R> v3 <- 1:10 # equivalently we may write v3<-c(1,2,3,4,5,6,7,8,9,10)
\end{verbatim}
}}
For finding the first five elements of \verb+v3+, we may use command
{\color{red}{
\begin{verbatim}
R> v3[1:5] 
\end{verbatim}
}}
Equivalently, we may do this task through the command
{\color{red}{
\begin{verbatim}
R> v3[v3<=5] 
\end{verbatim}
}}
Finally, if we want to save all elements of vector \verb+v3+ that are between 3 and 8 (excluding ending points) in new vector called \verb+v4+, we proceed as follows.
{\color{red}{
\begin{verbatim}
R> v4 <- v3[ v3<=7 & v3>=4 ] 
\end{verbatim}
}}
If the desired elements are note well-addressed, then we have to determine the addresses
through a new vector. For example, for picking up the elements of \verb+v3+ whose addresses are odd (or even), we have
 {\color{red}{
\begin{verbatim}
R> v.odd  <-v3[ c(1,3,5,7, 9) ]# elements of w whose indices are odd number
R> v.even1<-v3[ c(2,4,6,8,10) ]# elements of w whose indices are even number
\end{verbatim}
}}
Sometimes, we are interested eliminating one or mote items from a vector. For example, the last element of vector \verb+v3+ is removed using command \verb+v3[-10]+. Hence, another method for creating vector \verb+v.even+ in above is to use command
 {\color{red}{
\begin{verbatim}
R> v.even2<-v3[-c(1,3,5,7,9)]# elements of v3 whose indices are even numbers
\end{verbatim}
}}
This means that vectors \verb+ v.even1+ and \verb+v.even2+ are the same. The basic mathematical operations such as subtraction (addition) and multiplication (division) are applied to two vectors pointwise provided that the vectors are of the same lengths. For instance, we can compute \verb+v.odd/v_even1+ as follows.
 {\color{red}{
\begin{verbatim}
R> v.div <- v.odd/v.even1  # v.odd divided by v.even1
R> print(v.div) 
\end{verbatim}
}}
\vspace*{-3mm}
  {\color{blue}{
\begin{verbatim}
0.5000000 0.7500000 0.8333333 0.8750000 0.9000000
\end{verbatim}
}}
The name assigned to a vector is quite optional, but we suggest the user to avoid using letters \verb+c+, \verb+D+, \verb+F+, \verb+I+, etc. Typically, an English letter, English letter followed by an integer number, English letter followed by English letter, an English letter followed by dot ``$\cdot$'', or English letter followed by an underscore ``\verb+_+'' is suggested. In above, we assigned the name \verb+v.div+ to vector \verb+v.odd/v.even1+. The format and length of each element in output above can be determined by user. For example, we may use command \verb+options(digits=4)+ to see that 
  {\color{blue}{
\begin{verbatim}
 0.5000 0.7500 0.8333 0.8750 0.9000
\end{verbatim}
}}
As we see, the total digits after the decimal point is 4. If the output is greater than one, then the sum of total digits before and after the decimal point would be 4. If a vector multiplied by a numeric value, then all elements of the given vector is multiplied by that numeric value. This holds true also for subtraction, division, and addition operations. We have
 {\color{red}{
\begin{verbatim}
R> v4 <- 1:3
R> v5 <- 3:5
R> v6 <- v4/v5 
\end{verbatim}
}}
\vspace*{-3mm}
  {\color{blue}{
\begin{verbatim}
0.3333 0.5000 0.6000
\end{verbatim}
}}
Above three commands can be written in shorter form in order to save the space. For this purpose, all commands must be written sequentially, but each is separated from the others by symbol ``\verb+;+''. We have
 {\color{red}{
\begin{verbatim}
R> v4 <- 1:3; v5 <- 3:5; v6 <- v4/v5  
\end{verbatim}
}}
Either both commands \verb+ls()+ or \verb+objects()+ can be used to show a listing of variables and objects created when working with \verb+R+. Moreover, the objects created during an \verb+R+ session are stored in machine's memory and are ready when  \verb+R+ is started at later time. We suggest to remove all variables at the end of each \verb+R+ session. For example, if we would like to remove either object \verb+x+ or both of objects \verb+x+ and \verb+y+ from the set of objects, we can use command
{\color{red}{
\begin{verbatim}
R> rm('x')
\end{verbatim}
}}
and 
 {\color{red}{
\begin{verbatim}
R> rm(list=c('x', 'y'))
\end{verbatim}
}}
respectively. In general command 
 {\color{red}{
\begin{verbatim}
R> rm(list=ls())
\end{verbatim}
}}
can be used to remove all objects from the current session. We note that, in general, \verb+R+ does not discriminate between commands \verb+rm('x')+ and \verb+rm("x")+. It is worth to note that the \verb+R+ user if free to put blank or empty space between object(s) and  
round brackets \verb+( )+, square brackets \verb+[ ]+, and curly bracket \verb+{ }+. For example, all commands \verb+rm("x"     )+, \verb+rm(      "x")+, \verb+rm      ("x")+, and  
\verb+rm(list=c('x',        'y'))+ are correct and do the same task. Furthermore, the \verb+R+ is case sensitive language. That is, the {\it{capital}} and {\it{small}} letters have different meaning and do different tasks. 
For instance, objects \verb+x+ and \verb+X+ are two different objects and command \verb+Rm("x")+ is meaningless. When using the latter, the output would be an error message as follows.
 {\color{red}{
\begin{verbatim}
R> Rm("x")
\end{verbatim}
}}
\vspace*{-5mm}
 {\color{blue}{
\begin{verbatim}
Error in Rm("x") : could not find function "Rm"
\end{verbatim}
}}
The reminder of this subsection is devoted to introduce more complicated objects (or data structures) in \verb+R+ including sequence, data frame, matrix, array, and list. 
\subsubsection{Sequence}
Herein, we introduce two important methods for creating vectors that have specified structures that typically are helpful in statistical analysis. For repeating some specified numeric values under some given pattern, we may use \verb+rep(.)+. For example, in order to create sequence \verb+1,1,2,2,3,3,4,4+, we may proceed as follows.
 {\color{red}{
\begin{verbatim}
R> r1 <- rep(1:4, each = 2)
\end{verbatim}
}}
Furthermore, for creating sequence \verb+1,2,3,4,1,2,3,4+, we have
{\color{red}{
\begin{verbatim}
R> r2 <- rep(1:4, 2)
\end{verbatim}
}}
and finally, we write
{\color{red}{
\begin{verbatim}
R> r3 <- rep(1:4, 1:4)
\end{verbatim}
}}
to produce sequence \verb+1,2,2,3,3,3,4,4,4,4+.
\par For producing numeric values within an interval, $[a,b]$ say, with a given increment \verb+delta+ (or length \verb+n+), we use command \verb+seq(a,b,delta)+ (or \verb+seq(a,b,length=n)+). For example, in order to create vector \verb+0,0.1,0.2,...,0.9,1+, we can use the following command.
 {\color{red}{
\begin{verbatim}
R> s1 <- seq(0, 1, 0.1)
\end{verbatim}
}}
The output is
 {\color{blue}{
\begin{verbatim}
[1] 0.0 0.1 0.2 0.3 0.4 0.5 0.6 0.7 0.8 0.9 1.0
\end{verbatim}
}}
Alternatively, above output can be created using command
{\color{red}{
\begin{verbatim}
R> s2 <- seq(0, 1, length=11)
\end{verbatim}
}}
\par When working with vectors, we must take care about the observations that are recorded as missing or Not Available (\verb+NA+), Not Available as Number (\verb+NAN+), and infinity (\verb+Inf+). In \verb+R+, all observations that have been not recorded during the sampling stage are denoted by \verb+NA+. The symbol \verb+NAN+ accounts for the undetermined (or undefined) cases such as is 0/0 (or $\sqrt{-1}$), and hence is called {\it{Not Available as Number}}. Evidently, when a nonzero value is divided by zero, the result would be \verb+Inf+. The mathematical operations cannot be applied on \verb+NA+, \verb+NAN+, and \verb+Inf+ cases. Hence, to avoid problems arising from these cases, we should have our diagnostics. One simple way, is the use of command
\verb+summary+. We have
{\color{red}{
\begin{verbatim}
R> z3 <- c(NA, 1:5)
R> summary(z3)
\end{verbatim}
}}
\vspace*{-5mm}
{\color{blue}{
\begin{verbatim}
   Min. 1st Qu.  Median    Mean 3rd Qu.    Max.    NA's 
   1.00    3.25    5.50    5.50    7.75   10.00       1 
\end{verbatim}
}}
Based on above output, there is one missing value in vector \verb+z3+. Typically, we need to apply the mathematical/statistical functions on elements of a vector. Table \ref{R-function-statistical-vector} shows a listing of statistical functions in \verb+R+ with a widespread use in applications.
\begin{table}
\centering
\caption{Some commands for numeric vector algebra in {\texttt{R}}.}
\vspace*{-3mm}
%\resizebox{\textwidth}{!}{%
\begin{tabular}{ll}%{10mm}{>{\hsize=.10\hsize}X>{\hsize=.95\hsize}X}
\hline
Function& output/task\\
\hline
\multicolumn{1}{m{3cm}}{{\texttt{all(x, na.rm=FALSE)}}}&\multicolumn{1}{m{10cm}}{{\texttt{logical}} is {\texttt{TRUE}} if all elements of vector {\texttt{x}} holds true  for the given property, otherwise is {\texttt{FALSE}}.}\\
\multicolumn{1}{m{3cm}}{{\texttt{any(x, na.rm=FALSE)}}}&\multicolumn{1}{m{10cm}}{{\texttt{logical}} is {\texttt{TRUE}} if at least one elements of vector {\texttt{x}} holds true for some given property, otherwise is {\texttt{FALSE}}.}\\
\verb+as.vector(x)+&considering sequence \verb+x+ as a vector.\\
\verb+cbind(x1,x2,...)+&creating a matrix whose columns are vectors \verb+x1,x2,...+.\\
\multicolumn{1}{m{3cm}}{{\texttt{ceiling(x)}}}&\multicolumn{1}{m{10cm}}{
rounding elements of {\texttt{x}} to the smallest integer greater than {\texttt{x}}.}\\
\verb+cumsum(x)+&cumulative sum of elements of \verb+x+.\\
\verb+diff(x,lag=k)+&difference of \verb+x+ values with lag \verb+k+.\\
\multicolumn{1}{m{2cm}}{{\texttt{floor(x)}}}&\multicolumn{1}{m{12cm}}{
rounding elements of {\texttt{x}} to the greatest integer less than {\texttt{x}}.}\\
\verb+intersect(x,y)+&intersect of \verb+x+ and \verb+y+.\\
\verb+is.numeric(x)+&logical: if \verb+FALSE+, then \verb+x+ is not a numeric vector.\\ 
&and \verb+TRUE+ otherwise.\\
\verb+length(x)+& length of \verb+x+.\\
\verb+max(x,na.rm=FALSE)+&maximum of \verb+x+, by default \verb+NA+ is not removed.\\
\verb+min(x,na.rm=FALSE)+&minimum of \verb+x+, by default \verb+NA+ is not removed.\\
\verb+mean(x)+&average of \verb+x+.\\
\verb+median(x,na.rm=FALSE)+&median of \verb+x+.\\
\verb+numeric(n)+&a numeric vector of length \verb+n+.\\
\verb+NULL+&a numeric vector regardless of its length.\\
\verb+prodsum(x)+&cumulative product of elements of \verb+x+.\\
\verb+quantile(x,q)+&\verb+q+th quantile of \verb+x+.\\
\verb+range(x,na.rm=FALSE)+&range of \verb+x+, by default \verb+NA+ is not removed.\\
\verb+rbind(x1,x2,...)+&creating a matrix whose rows are vectors \verb+x1,x2,...+.\\
\verb+rev(x)+&representing elements of \verb+x+ in reversed representation of \verb+x+.\\
\verb+round(x)+&rounding elements of \verb+x+ to the nearest integer.\\
\verb+sd(x,na.rm=FALSE)+&standard deviation of \verb+x+, by default \verb+NA+ is not removed.\\
\verb+sum(x,na.rm=FALSE)+&sum of \verb+x+, by default \verb+NA+ is not removed.\\
\verb+summary(x)+&6-point summary of \verb+x+.\\
\verb+union(x,y)+&union of \verb+x+ and \verb+y+.\\
\verb+var(x,na.rm=FALSE)+&variance of \verb+x+, by default \verb+NA+ is not removed.\\
\verb+which.min(x)+&index in which \verb+min(x)+ occurs.\\
\verb+which.max(x,na.rm=FALSE)+&index in which \verb+max(x)+ occurs.\\
\hline
\end{tabular}
%}
\label{R-function-statistical-vector}
\end{table}
%\begin{table}
%\centering
%\caption{Some mathematical functions for vector algebra.}
%\begin{tabular}{llll}%{10mm}{>{\hsize=.10\hsize}X>{\hsize=.95\hsize}X}
%\hline
%Function& output\\
%\hline
%\verb+abs(x)+ & absolute value of the variable/vector x&\verb+abs(x)+&\\
%\hline
%\end{tabular}
%\label{R-function-mathematical-vector}
%\end{table}
\subsubsection{Data frame}
A rectangular structure of data with different types in \verb+R+ can be constructed using a concept called {\it{data frame}}. The data frame can be used to represent all types of data including numeric and non-numeric, but in a rectangular (balanced) form. The general command for data frame is 
\begin{verbatim}
data.frame(name1=object1,name2=object2,...,row.names=NULL)
\end{verbatim}
For example of data frame may be
 {\color{red}{
\begin{verbatim}
R> name <- c('row1', 'row2', 'row3')
R> df1 <- data.frame(col1=1, col2=v4, col3=v5, row.names=name)
R> df1
\end{verbatim}
}}
The output becomes
 {\color{blue}{
\begin{verbatim}
     col1 col2 col3
row1    1    1    3
row2    1    2    4
row3    1    3    5
\end{verbatim}
}}
As another example, let we want to construct a {\it{design of experiment}} in which there are two factors (each of three levels) with one repetition. To do this, we may construct a data frame as follows.
\begin{table}[!h]
\begin{tabular}{ll}%{10mm}{>{\hsize=.10\hsize}X>{\hsize=.95\hsize}X}
&\\
{\texttt{R}} code & output\\
\hline
{\color{red}{\verb+R> Level<-c('low','medium','high')+}}&
{\color{blue}{\verb+   Factor1 Factor2 response+}}\\
{\color{red}{\verb+R> n.L<-length(Level); Rep<-1+}}&
{\color{blue}{\verb+1      low     low        1+}}\\
{\color{red}{\verb+R> factor1<-rep(Level,times=Rep,+}}&
{\color{blue}{\verb+2      low     low        2+}}\\
{\color{red}{{\texttt{+ each=n.L*n.L)}}}}&
{\color{blue}{\verb+3      low     low        3+}}\\
{\color{red}{\verb+R> factor2<-rep(Level,times=n.L,+}}&
{\color{blue}{\verb+4      low     medium     4+}}\\
{\color{red}{{\texttt{+ each=n.L)}}}}&
{\color{blue}{\verb+5      low     medium     5+}}\\
{\color{red}{{\texttt{R> df1<-data.frame(Factor1=}}}}&
{\color{blue}{\verb+6      low     medium     6+}}\\
{\color{red}{{\texttt{+ factor1, Factor2=factor2,}}}}&
{\color{blue}{\verb+7      low     high       7+}}\\
{\color{red}{{\texttt{+ response=1:27)}}}}&
{\color{blue}{\verb+8      low     high       8+}}\\
&{\color{blue}{\verb+...    ...     ...      ...+}}\\
&{\color{blue}{\verb+26     high    high      26+}}\\
&{\color{blue}{\verb+27     high    high      27+}}\\
\hline
\end{tabular}
\end{table}
\vspace{5mm}
\par Sometimes we need to manipulate or organize some part of a data frame (or matrix) in terms of its other parts. A useful function for this purpose is \verb+subset(df,subset,select)+ in which, \verb+df+ denotes the data frame involving whole data, \verb+subset+ is the constraint we are willing to apply, and \verb+select+ is the name of column(s) to be select from a data frame. For example, in \verb+airquality+ data, suppose we are willing to selecting all levels of variables \verb+Temp+ and \verb+Wind+ for which variables \verb+Ozone+ and \verb+Solar.R+ are not missing. To do this is, we may use command \verb+subset+ as follows.
 {\color{red}{
\begin{verbatim}
dd<-subset(airquality, !(Ozone=="NA") & !(Solar.R=="NA"), select=c(Wind,Temp))
\end{verbatim}
}}
When data frame is large, then we would prefer to manipulate the data frame in terms of its column's name rather than the column's order or number. In such a situation, one may use the \verb+R+ package \verb+dplyr+ available at R Comprehensive archive \url{https://cran.r-project.org/web/packages/dplyr/index.html}.
\subsubsection{List}
A useful and informative structure of data can be constructed in \verb+R+ using a concept called
{\it{list}}. The list can be used to represent all types of data including matrix, numeric data, non-numeric data, and logical. To create a list (called here \verb+L1+) consisting of vector  \verb+v2+ and data frame \verb+df1+ defined earlier, we can write
 {\color{red}{
\begin{verbatim}
R> L1 <- list(x1 = v2, x2 = df1); L1 
\end{verbatim}
}}  
\vspace{-3mm}
 {\color{blue}{
\begin{verbatim}
$x1
[1] "1"      "2"      "male"   "female"

$x2
  x y z
1 1 1 4
2 1 2 5
3 1 3 6
\end{verbatim}
}}  
To access to each part of data frame, for example the first part, we may write \verb+L$x1+ and also for the second row of the second part we may write \verb+L$x2[2,]+, and finally elements consisting of digit 3 in last part of list \verb+L+ is accessible through command \verb+L$x3[4:6]+. Of course, alternatively, the latter is accessible using \verb+L[[2]][2,]+. We note that if we use command \verb+list(v2,df1)+, then names of two parts in the list \verb+L+ will be disappeared.
\subsubsection{Matrix}
One of the most commonly used structures in \verb+R+ is matrix. Recall that elements in each column of data frame are of the same type while different columns may have different types. The matrix is a rectangular form of data with the same type for which the numeric one is commonly used. In general, for constructing a matrix called \verb+A+, we use the command given by the following.
{\color{red}{
\begin{verbatim}
R> A <- matrix(data=NA, nrow=1, ncol=1, byrow=FALSE, dimnames=NULL) 
\end{verbatim}
}}
Herein \verb+data+, \verb+nrow+, \verb+ncol+, \verb+byrow+, and \verb+dimnames+ are arguments of function matrix. Table \ref{R-function-matrix-1} below gives more explanations about the duty of each argument.
\vspace{5mm}
\begin{table}[!h]
\centering
\caption{Arguments of {\texttt{matrix}} in {\texttt{R}}.}
\vspace*{-3mm}
\begin{tabular}{ll}%{10mm}{>{\hsize=.10\hsize}X>{\hsize=.95\hsize}X}
\hline
Argument& output/task\\
\hline
\verb+data+ & a numeric vector that can include even \verb+NA+, \verb+NAN+, and \verb+Inf+.\\
\verb+nrow+& number of rows.\\
\verb+ncol+ & number of columns.\\
\multicolumn{1}{m{2cm}}{{\texttt{byrow}}}&\multicolumn{1}{m{12cm}}{
{\texttt{logical}}: if {\texttt{FALSE}} (by default), then the matrix is filled by columns, otherwise the matrix is filled by rows.}\\
\multicolumn{1}{m{2cm}}{{\texttt{dimnames}}}&\multicolumn{1}{m{12cm}}{a list of length 2 that gives the row and column names, respectively.}\\
\hline
\end{tabular}
\label{R-function-matrix-1}
\end{table}
\vspace{5mm}
It is worthwhile to note that \verb+R+ automatically fills out the elements of matrix {\it{column-by-column}}. 
If user forgets the order of arguments in \verb+matrix+ function, it suffices to run command \verb+args(matrix)+ as  
{\color{red}{
\begin{verbatim}
R> args(matrix)
\end{verbatim}
}}
to see the arguments of function \verb+matrix+ as follows. 
{\color{blue}{
\begin{verbatim}
function (data=NA, nrow=1, ncol=1, byrow=FALSE, dimnames=NULL) 
NULL
\end{verbatim}
}}
Hence, if constructing a $2\times2$ matrix $A$ represented as
\begin{align*}
A=\left[\begin{matrix} 
1&3\\
2&4\\
\end{matrix}\right].
\end{align*}
is of interest, then we may use two commands to define matrix A as follows. 
\begin{table}[!h]
\begin{tabular}{ll}%{10mm}{>{\hsize=.10\hsize}X>{\hsize=.95\hsize}X}
&\\
{\texttt{R}} code & output\\
\hline
{\color{red}{\verb+R>A<-matrix(c(1,2,3,4),nrow=2,ncol=2)# method 1+}}&{\color{blue}{\verb+     [,1] [,2]+}}\\
{\color{red}{\verb+R>A<-matrix(1:4,nrow=2,ncol=2)# method 2+}}&{\color{blue}{\verb+ [1,]   1    3+}}\\
{\color{red}{\verb+R>A+}}&{\color{blue}{\verb+ [2,]   2    4+}}\\
\hline
\end{tabular}
\end{table}
\vspace{5mm}
If we want to assign names to the rows and columns, and further, fill out the matrix $A$ by rows, we proceed as follows.  
%{\color{red}{
%\begin{verbatim}
%R> L2<- list(c("row1", "row2"), c("col1", "col2"))
%R> A <- matrix(1:4, nrow=2, ncol=2, byrow=TRUE, dimnames=L2); A
% \end{verbatim}
%}}       
%The output is
%{\color{blue}{
%\begin{verbatim}
%     col1 col2
%row1    1    2
%row2    3    4
%\end{verbatim}
%}}
\begin{table}[!h]
\begin{tabular}{ll}%{10mm}{>{\hsize=.10\hsize}X>{\hsize=.95\hsize}X}
&\\
{\texttt{R}} code & output\\
\hline
{\color{red}{\verb+R>L2<-list(c("row1","row2"),c("col1","col2"))+}}&{\color{blue}{\verb+     col1 col2+}}\\
{\color{red}{\verb+R>A<-matrix(1:4,nrow=2,ncol=2,byrow=TRUE,+}}&{\color{blue}{\verb+row1    1    2+}}\\
{\color{red}{{\texttt{+ dimnames=L2)}}}}&{\color{blue}{\verb+row2    3    4+}}\\
\hline
\end{tabular}
\end{table}
\vspace{5mm}
To access $(i,j)$th element of matrix \verb+A+, we must write \verb+A[i,j]+. Moreover, the elements of $i$th row of matrix \verb+A+ are obtained by command \verb+A[i, ]+, and accordingly, the elements of $j$th column of matrix \verb+A+ are obtained by command \verb+A[ ,j]+. 
For instance, the output of commands \verb+A[1,1]+, \verb+A[1, ]+, and \verb+A[, 2]+ are given as follows.
\vspace{5mm}
\begin{table}[!h]
\begin{tabular}{lll}%{10mm}{>{\hsize=.10\hsize}X>{\hsize=.95\hsize}X}
{\color{red}{\verb+ R> A[1,1] +}}&{\color{red}{\verb+ R> A[1,] +}}&{\color{red}{\verb+ R> A[,2] +}}\\
{\color{blue}{\verb+ [1] 1 +}}&{\color{blue}{\verb+ [1] 1  2 +}}&{\color{blue}{\verb+ [1] 2  4 +}}\\
\end{tabular}
\end{table}
\par The matrix algebra typically is used in statistical methods. Table \ref{R-function-matrix-2} lists the most commonly used functions (commands) associated with the matrix algebra.
\vspace{0.5cm}
\begin{table}[!h]
\centering
\caption{Some commands for matrix algebra in {\texttt{R}}.}
\vspace*{-3mm}
\begin{tabular}{ll}%{\textwidth}{>{\hsize=.10\hsize}X>{\hsize=.95\hsize}X}
\hline
Function& output/task\\
\hline
\verb+A*B+ & pointwise product of the given matrices \verb+A+ and \verb+B+.\\
\verb+A%*%B+ & product of the given matrices \verb+A+ and \verb+A+.\\
\multicolumn{1}{m{3cm}}{{\texttt{as.matrix(df)}}}&\multicolumn{1}{m{11cm}}{considers a data frame {\texttt{df}} as a matrix.}\\
\multicolumn{1}{m{3cm}}{{\texttt{asplit(df,MARGIN)}}}&\multicolumn{1}{m{11cm}}{splitting an array, data frame, list, or matrix by its margins ({\texttt{1}} for rows, {\texttt{2}} for columns, and {\texttt{c(1,2)}} for rows and columns).}\\
\verb+crossprod(A,B)+ &computing \verb+t(A)%*%B+. \\
\verb+colMeans(A)+ &computing average of columns of \verb+A+.\\
\verb+colnames(A)+ &assigning names to columns of \verb+A+.\\
\verb+colSums(A)+ &computing sum of columns of \verb+A+.\\
\verb+det(A)+ &computing determinant of matrix \verb+A+.\\
\verb+dim(A)+ &dimensions of matrix \verb+A+.\\
\verb+diag(A)+ &diagonal of matrix \verb+A+.\\
\multicolumn{1}{m{3cm}}{{\texttt{diag(a)}}}&\multicolumn{1}{m{11cm}}{constructing identity matrix whose diagonal elements are elements of vector {\texttt{a}}.}\\
\verb+eigen(A)+ &eigen vectors and values of matrix {\texttt{A}}.\\
\verb+identical(A,B)+ &logical: \verb+TRUE+ if matrices \verb+A+ and \verb+B+ are identical; and\\ & \verb+FALSE+ otherwise.\\
\multicolumn{1}{m{3cm}}{{\texttt{is.matrix(df)}}}&\multicolumn{1}{m{11cm}}{
{\texttt{logical}}: {\texttt{TRUE}} is the given data frame {\texttt{df}} is a matrix, and {\texttt{FALSE}} otherwise.}\\
\verb+isSymmetric(A)+ &logical: \verb+TRUE+ if matrix \verb+A+ is symmetric, and \verb+FALSE+ otherwise.\\
\verb+mahalanobis(x,mu,S)+& Mahalanobis distance, that is $({\text{x-mu}})^{\top} {\text{S}}^{-1}({\text{x-mu}})$.\\
\verb+ncol(A)+& the number of columns of matrix (data frame) \verb+A+.\\
\verb+qr(A)+ &QR decomposition of matrix \verb+A+.\\
\verb+rowMeans(A)+ & computing average of rows of \verb+A+.\\
\verb+rownames(A)+ & assigning names to rows of \verb+A+.\\
\verb+rowSums(A)+ & computing sum of rows of \verb+A+.\\
\verb+scale(A,center=+ &matrix whose columns are standardized columns of matrix \verb+A+.\\
{\texttt{TRUE,scale=TRUE)}} &\\
\verb+solve(A)+ &inverse of matrix \verb+A+.\\
\verb+t(A)+ &transpose of matrix \verb+A+.\\
\hline
\end{tabular}
\label{R-function-matrix-2}
\end{table}
\subsubsection{Array}
The array can be a one-, two-, or even three-dimensional object. The general command for array is \verb+array(data=NA, dim=c(nr,nc,nz), dimnames=NULL)+ in which \verb+data+ is the vector or sequence of array's elements, \verb+nr+ is the number of rows, \verb+nc+ is the number of columns, \verb+nz+ is the number of third coordinate, and \verb+dimnames+ is a list of names for three coordinates. There is also one- or two-dimensional array. The array can be reduced to construct a matrix. A simple example of three-, two-, and one-dimensional array can be constructed using vector \verb+1:8+,  \verb+1:4+, and  \verb+1:2+ as follows.
{\color{red}{
\begin{verbatim}
R> array(1:8,dim=c(2,2,2))  R> array(1:4,dim=c(2,2))   R> array(1:2,dim=2)
\end{verbatim}
}}
\vspace{-3mm}
{\color{blue}{
\begin{verbatim}
, , 1                            [,1] [,2]             [1] 1 2
     [,1] [,2]              [1,]    1    3
[1,]    1    3              [2,]    2    4
[2,]    2    4
, , 2
     [,1] [,2]
[1,]    5    7
[2,]    6    8
\end{verbatim}
}}

%\begin{table}[!h]
%\begin{tabular}{lll}%{10mm}{>{\hsize=.10\hsize}X>{\hsize=.95\hsize}X}
%&\\
%{\color{red}{\verb+R>array(1:8,dim=c(2,2,2))+}}&{\color{red}{\verb+R>array(1:4,dim=c(2,2))+}}&{\color{red}{\verb+R>array(1:2,dim=2)+}}\\
%{\color{blue}{\verb+, , 1                   +}}&{\color{blue}{\verb+     [,1] [,2]+}}&{\color{blue}{\verb+[1] 1 2+}}\\
%{\color{blue}{\verb+     [,1] [,2]+}}&{\color{blue}{\verb+[1,]    1    3+}}&\\
%{\color{blue}{\verb+[1,]    1    3+}}&{\color{blue}{\verb+[2,]    2    4+}}&\\
%{\color{blue}{\verb+[2,]    2    4+}}&&\\
%{\color{blue}{\verb+, , 2+}}&&\\
%{\color{blue}{\verb+     [,1] [,2]+}}&&\\
%{\color{blue}{\verb+[1,]    5    7+}}&&\\
%{\color{blue}{\verb+[2,]    6    8+}}&&\\
%\end{tabular}
%\end{table}
\par A useful summary about a data frame may be obtained though call \verb+str(...)+. This some extension of command \verb+type+ defined formerly for vector object. The  \verb+str(...)+ displays the internal structure of an complicated object such as array, data frame, list, or matrix. For example, recall data frame \verb+df1+, we have
{\color{red}{
\begin{verbatim}
R> str(df1)
\end{verbatim}
}}
\vspace{-6mm}
{\color{blue}{
\begin{verbatim}
'data.frame':  27 obs. of 3 variables:
 $Factor1:Factor w/ 3 levels "high","low","medium":2 2 2 2 2 2 2 2 2 3 ...
 $Factor2:Factor w/ 3 levels "high","low","medium":2 2 2 3 3 3 1 1 1 2 ...
 $response:int 1 2 3 4 5 6 7 8 9 10 ...
\end{verbatim}
}}
The phrase \verb+'data.frame'+ in output indicates that the type of object is data frame that consists of 27 observations on three variables. Then, a summary is given on each of three variables \verb+Factor1+, \verb+Factor2+, and \verb+response+. 
\section{Condition and repeat}
In what follows, we deal with briefly some tools that are useful for creating desired data structure, data refinement, comparisons, etc. Some tools are needed to create a user-defined function  well. These tools have been defined for \verb+R+ and include \verb+if(...)+, \verb+ifelse(...)+, \verb+while(...)+, an \verb+for(...)+. In what follows, we briefly discuss aforementioned tools.
\subsection{Conditional decision using if(...)}
For accomplishing some decision under some predetermined {\it{condition}} (or nature state), we use this command. In general, let nature space \verb+Omega+ consist of mutually disjoint states \verb+State 1+, \verb+State 2+, and \verb+State 3+ so that 
$$
\verb+State 1+  \cup  \verb+State 2+  \cup  \verb+State 3+ =\verb+Omega+.
$$
Furthermore, suppose we are willing to make \verb+decision 1+ (if \verb+State 1+ holds true), \verb+decision 2+ (if \verb+State 2+ holds true), and \verb+decision 3+ (if \verb+State 3+ holds true). For example, we can use the command \verb+if(...)+ for evaluating the sign of a real given number \verb+x+. To this end, the pertinent pseudo and \verb+R+ codes are given by the following.
\vspace{0.5cm}
\begin{table}[!h]
\begin{tabular}{ll}%{10mm}{>{\hsize=.10\hsize}X>{\hsize=.95\hsize}X}
Pseudo code for  command & {\texttt{R}} code for {\texttt{if(...)}} command\\
\hline
\verb+if(State 1 holds true)+~~~&\verb+ if(x==0) +\\
\verb+{decision 1 +&\verb+ {print(0)  # decision 1+\\
\verb+ }else if(State 2 holds true){ +~~~&\verb+  }else if(x>0){+\\
\verb+ decision 2 +&\verb+  print(1)  # decision 2+\\
\verb+ }else{ +&\verb+  }else{ +\\
\verb+ decision 3}+&\verb+  print(-1) # decision 3+\\
\verb+}+&\verb+ } +\\
\hline
\end{tabular}
\end{table}
\vspace{0.5cm}
\par If nature states consist of only two states, then we may use a simpler form of \verb+if(...)+ command called \verb+ifelse(test,yes,no)+. Of course, the \verb+if(...)+ command still works well. Suppose, we are interested in turning nominal variable \verb+x+ that takes on either \verb+'male'+ or \verb+'female'+ into a numeric variable that takes on either \verb+1+ or \verb+0+, respectively. The pertinent \verb+if(...)+ and \verb+ifelse(test,yes,no)+ commands are as follows.
{\color{red}{
\begin{verbatim}
R> if(x=='male')    R> x<-ifelse(x=='male',1,0) 
+  {
+  x<-1 
+  } else { 
+  x<-0
+  }
\end{verbatim}
}}
While using \verb+if(...)+, we may use some operators. Table \ref{R-nonprimitive-built-in-function-if-command} lists some commonly used operators. 
\vspace{0.5cm}
\begin{table}[!h]
\centering
\caption{Some commonly used operators in {\texttt{R}}.}
\vspace*{-3mm}
\begin{tabular}{ll}%{\textwidth}{>{\hsize=.10\hsize}X>{\hsize=.95\hsize}X}
\hline
Operator& output/task\\
\hline
\verb+<=+& less than or equal.\\
\verb+>=+& greater than or equal.\\
\verb+!=+&not equals.\\
\verb+a%%b+ & remainder of division a/b.\\
\verb+a%/%b+ & quotient of division a/b.\\
\multicolumn{1}{m{2cm}}{{\texttt{a==round(a)}}}&\multicolumn{1}{m{12cm}}{{\texttt{logical}}: returns {\texttt{TRUE}} if {\texttt{a}} is an integer value, otherwise returns {\texttt{FALSE}}.}\\
\multicolumn{1}{m{2cm}}{{\texttt{a|b}}}&\multicolumn{1}{m{12cm}}{
{\texttt{logical}}: it applies element-wise to vectors {\texttt{a}} and {\texttt{b}}, and returns a vector of logical values.}\\
 \multicolumn{1}{m{2cm}}{{\texttt{a||b}}}&\multicolumn{1}{m{12cm}}{
{\texttt{logical}}: it applies to numeric values {\texttt{a}} and {\texttt{b}}, and returns a single logical value. The output is {\texttt{TRUE}} if either {\texttt{a}} or {\texttt{b}} holds true.}\\
 \multicolumn{1}{m{2cm}}{{\texttt{a\symbol{38}b}}}&\multicolumn{1}{m{12cm}}{
{\texttt{logical}}: it applies element-wise to vectors {\texttt{a}} and {\texttt{b}}, and returns a vector of logical values. Each element of output is {\texttt{TRUE}} if either {\texttt{a}} or {\texttt{b}} holds true.}\\
\multicolumn{1}{m{2cm}}{{\texttt{a\symbol{38}\symbol{38}b}}}&\multicolumn{1}{m{12cm}}{
{\texttt{logical}}: it applies to numeric values {\texttt{a}} and {\texttt{b}}, and returns a single logical value. The output is {\texttt{TRUE}} if both of {\texttt{a}} and {\texttt{b}} hold true.}\\
\hline
\end{tabular}
\label{R-nonprimitive-built-in-function-if-command}
\end{table}
It is worth to note that operator \verb+a||b+ evaluates \verb+b+ only if \verb+a+ is \verb+FALSE+ \citep{R:fox2018r}. Let \verb+c1<- c(1,1,2,2,1,2,1)+ and \verb+c2<- c(1,2,2,2,1,1,1)+ denote the clustering results of two populations each consisting of two clusters \verb+1+ and \verb+2+. This fact that two clusters \verb+c1+ and \verb+c2+ are not fully matched can be shown by the following commands.
\vspace{5mm}
\begin{table}[!h]
\begin{tabular}{lll}%{10mm}{>{\hsize=.10\hsize}X>{\hsize=.95\hsize}X}
{\color{red}{\verb+R> a1<-ifelse( +}}&{\color{red}{\verb+R> a2<-ifelse(+}}&{\color{red}{\verb+R> if(all(c1-c2==0)+}}\\
{\color{red}{{\texttt{+all(c1-c2 == 0),}}}}&{\color{red}{{\texttt{+any(c1/c2 != 1),}}}}&{\color{red}{{\texttt{+|| all(c1/c2==1)}}}}\\
{\color{red}{{\texttt{+"full", "not full")}}}}&{\color{red}{{\texttt{+"not full", "full")}}}}&{\color{red}{{\texttt{+a3<-"full" else a3<-"not full")}}}}\\
{\color{red}{\verb+R> a1+}}&{\color{red}{\verb+R> a2+}}&{\color{red}{\verb+R> a3+}}\\
{\color{blue}{\verb+[1] "not full"+}}&{\color{blue}{\verb+[1] "not full"+}}&{\color{blue}{\verb+[1] "not full"+}}\\
\end{tabular}
\end{table}
\subsection{Repeat using while(...)}
For repeating some task until some statement holds true, we may use command \verb+while(...)+. This command essentially needs some initialization in relation to the given statement. In general, we should consider some counter 
\begin{verbatim}
initialization
while(statement holds true)
{
body of while
}
\end{verbatim} 
Suppose we would like to print sum of nonnegative integer numbers provided that sum of these numbers does not exceed 10. Below, we give the command \verb+while(...)+ for this purpose.
%\begin{table}[!h]
%\begin{tabular}{ll}%{10mm}{>{\hsize=.10\hsize}X>{\hsize=.95\hsize}X}
%command & output\\
%\hline
%{\color{red}\texttt{R> Sum <- 0; i <- 0}}}&{\color{blue}{\verb+ [1] 0 +}}\\
%{\color{red}\texttt{R> while(Sum < 10)}}}&{\color{blue}{\verb+ [1] 1 +}}\\
%{\color{red}{ \texttt{ +{ } }}&{\color{blue}{\verb+ [1] 3 +}}\\
%{\color{red}{\texttt{+ Sum <- Sum + i }}}&{\color{blue}{\verb+ [1] 6 +}}\\
%{\color{red}{\texttt{+ print(Sum)}}}&{\color{blue}{\verb+ [1] 10 +}}\\
%{\color{red}{\texttt{+  i <- i + 1}}}&\\
%{\color{red}{\texttt{+}  }  }}&\\
%\end{tabular}
%\end{table}
%\begin{lstlisting}[style=deltaj]
%R> Sum <- 0; i <- 0  #initialization
%R> while(Sum < 10) {   # statement  = Sum < 10
%+ Sum <- Sum + i
%+ print( Sum )
%+ i <- i + 1
%+}
%\end{lstlisting}
\begin{table}[!h]
\begin{tabular}{lc}%{10mm}{>{\hsize=.10\hsize}X>{\hsize=.95\hsize}X}
&\\
{\texttt{R}} code & output\\
\hline
{\color{red}{\verb+R> Sum <- 0; i <- 0 #initialization+}}&{\color{blue}{\verb+[1] 0+}}\\
{\color{red}{\verb+R> while(Sum < 10){ #while(statement)+}}&{\color{blue}{\verb+[1] 1+}}\\
{\color{red}{\texttt{+ Sum <- Sum + i}}}&{\color{blue}{\verb+[1] 3+}}\\
{\color{red}{\texttt{+ print( Sum )}}}&{\color{blue}{\verb+[1] 6+}}\\
{\color{red}{\texttt{+ i <- i + 1} }}&{\color{blue}{\verb+ [1] 10+}}\\
 {\color{red}{\texttt{+ }\verb+}+}}&{\color{blue}{\verb+ +}}\\
 \hline
\end{tabular}
\end{table}
\subsection{Repeat using for(...)}
For doing some task repeatedly for a given number, we may use command \verb+for(...)+. Note that in contrast to  \verb+while(...)+, the number of iterations in \verb+for(...)+ is fixed and known. Let the non-numeric vector \verb+sex+ is defined as \verb+sex=c('male','female','male','female','female','male')+ that represents the sex of six persons. If we are willing to replace \verb+'male'+ and \verb+'female'+, respectively, with \verb+1+ and \verb+0+, then we can command \verb+for(...)+ as follows.
\begin{lstlisting}[style=deltaj]
R> sex <- c('male','female','male','female','female','male')
R> n <- length(sex)
R> x <- rep(0, n) # creating a numeric vector called x of size n
R> for(i in 1:n)
+ {
+	 	if( sex[i] == 'female') 
+ 	{
+ 		x[i] <- 0
+ 	}else{
+ 		x[i] <- 1
+   	 }
+ }
\end{lstlisting}
We note that in order to constructing vector \verb+x+ in above, we may obtain \verb+x+ simply using command \verb+x<-ifelse(sex=='female',0,1)+. The next example uses  command \verb+for(...)+ to create a vector involving the names of rows (or columns) for a $10\times 10$ matrix. The pertinent \verb+R+ code and output are as follows.
{\color{red}{
\begin{verbatim} 
R> N1<-rep(NA,10)                  R> N2<-rep(NA,10)
R> for(i in 1:length(N1)){         R> for(i in 1:length(N2)){
+ N1[i]<-paste('a',                + N2[i]<-paste(letters[i],
+ sep="-", i)}                     + sep="", i) }
\end{verbatim} 
}}
\vspace{-5mm}
{\color{blue}{
\begin{verbatim}
[1] "a-1"  "a-2"  "a-3"  "a-4"     [1] "a1"  "b2"  "c3"  "d4"
[5] "a-5"  "a-6"  "a-7"  "a-8"     [5] "e5"  "f6"  "g7"  "h8"
[9] "a-9"  "a-10"                  [9] "i9"  "j10"
\end{verbatim} 
}}
%%%%%%%%%%%%%%%%%%%%%%%%%%%%%%%%%%%%%%%%
\section{Data importing and exporting}
Herein, we deal with reading {\it{built-in}} data sets existing in \verb+R+and further discussing   its nice property of exporting (writing) and importing (reading) data from other applications. 
\subsection{Reading built-in data}
There are some applied data sets available in \verb+R+ that are known as {\it{built-in}} data. The list of data is ready for using by running command \verb+R>data()+. There appears a list of several data sets. For calling some special data set, we need just to put the name of data inside the former command. Herein, we are willing to call two famous data sets known as \verb+airquality+ \cite{chambers2018graphical}, \verb+iris+ \cite{fisher1936} data. The first set is a data frame with 153 observations on 6 variables representing the daily air quality measurements in New York fro May to September 1973. The second set consists of 50 measurements of features sepal length, sepal width, petal length, and petal width sampled from species Setosa, Versicolor, and Virginica of iris flower. For example, we may run the command 
{\color{red}{
\begin{verbatim} 
R> data(airquality)
R> head(airquality,6) #we may use tail(airquality,10) for last 10 lines
\end{verbatim}
}}
to see the first six rows of \verb+airquality+ data as follows.
{{
{\color{blue}{
\begin{verbatim}
  Ozone Solar.R Wind Temp Month Day									
1    41     190  7.4   67     5   1
2    36     118  8.0   72     5   2
3    12     149 12.6   74     5   3
4    18     313 11.5   62     5   4
5    NA      NA 14.3   56     5   5
6    28      NA 14.9   66     5   6
\end{verbatim}
}}
}}
Alternatively, one may use command \verb+View(airquality)+ to have a look at data in a new frame.  It is worth to note that if you have access to some built-in data called \verb+x+ that other do not, one way to move it is as follows. First run command \verb+dput(x)+ and then copy the blue-colored output that starts with {\color{blue}{\texttt{structure(list(...}}} and then insert it at a new variable called \verb+y+ or email the blue-colored output to any destination.  
\subsection{Importing data from R package}
We can read data in \verb+R+ fro other applications. If data are appended to an \verb+R+ package, then we need to load the package that involves the data set. This is done through command \verb+library(package name)+ or \verb+require(package name)+. For example, to access the \verb+DBH+ data existing in \verb+R+ package \verb+ForestFit+, we proceed as follows.
{\color{red}{
\begin{verbatim} 
R> library(ForestFit) # loading package
R> data(DBH)          # loading the pertinent data called DBH
R> head(DBH,2)        # representing first 2 lines of data
\end{verbatim}
}}
%%%%%%%%%%%%%%%%%%%%%%%%%%%%%%
%%%%%%%%%%%%%%%%%%%%%%%%%%%%%%
\subsection{Importing data from other application}
The prepared data for reading should be in tabular structure such as data frame or matrix and the acceptable formats are variants of \verb+txt+ including comma-, pipe-, or tab-delimited. Herein, the phrase delimiter refers to an object that separates the entries of two columns (fields). The comma-delimited format is also known as the comma separated values (CSV). Typically, the raw data are available form an external resource and reading data using command \verb+read.table(...)+ saves the user time and avoids some potential error arising from manual data entry. The main difference between all above mentioned types of \verb+txt+ data is the type of delimiter between columns. In general, command \verb+read.table(...)+ is a function that reads a tabular structure of data and converts it into a data frame. The arguments and the pertinent details are given in Table \ref{Read-table-table-1}.
\vspace{0.5cm}
\begin{table}[!h]
\centering
\caption{Details about main arguments of {\texttt{read.table(...)}} in {\texttt{R}}.}
\vspace*{-3mm}
\begin{tabular}{ll}%{\textwidth}{>{\hsize=.10\hsize}X>{\hsize=.95\hsize}X}
\hline
Argument& output/task\\
\hline
\multicolumn{1}{m{1.5cm}}{{\texttt{file}}}&\multicolumn{1}{m{12.5cm}}{
the full address of file that is ready for reading. Each row of the tabular data would be a line of the output file.}\\
\multicolumn{1}{m{1.5cm}}{{\texttt{header}}}&\multicolumn{1}{m{12.5cm}}{
logical: if {\texttt{TRUE}}, then the file contains the names of the variables as its first line; otherwise, a vector of  letter {\texttt{"V''}} followed by column number will be assigned to columns as the name.}\\
\verb+row.names+& a vector of row names.\\
\multicolumn{1}{m{1.5cm}}{{\texttt{col.names}}}&\multicolumn{1}{m{12.5cm}}{
a vector of names for the variables. The default is a set consist of letters {\texttt{"V''}} followed by the column number.}\\
\multicolumn{1}{m{1.5cm}}{{\texttt{sep}}}&\multicolumn{1}{m{12.5cm}}{
this parameter allows to set different delimiters including comma, pipe, semicolon
 tab, and one space for separating columns. It is assumed that the entries at
each line of the imported file are separated by the character in front of {\texttt{sep}} parameter. If columns are comma-, pipe-, or tab-delimited, then {\texttt{sep=","}}, {\texttt{sep=''|''}}, or {\texttt{{''\textbackslash{}t''}}}, is recommended, respectively.}\\
\multicolumn{1}{m{1.5cm}}{{\texttt{na.strings}}}&\multicolumn{1}{m{12.5cm}}{
a character or string that is recognized as missing value of input data. By default {\texttt{na.strings="NA"}}. If not, put the missing symbol instead of {\texttt{NA}} inside the quotation marks.}\\
\verb+dec+& a character assumed for decimal points.\\
\hline
\end{tabular}
\label{Read-table-table-1}
\end{table}
\vspace{5mm}
Although \verb+read.table(...)+ can be used for all formats, but we may use commands \verb+read.csv(...)+, \verb+read.csv2(...)+, and \verb+read.delim(...)+ for reading tabular data with different delimiters as shown by Table \ref{Read-table-csv-1}.
\vspace{5mm}
\begin{table}[!h]
\centering
\caption{Functions for importing data in {\texttt{R}}.}
\vspace*{-3mm}
\begin{tabular}{ll}%{\textwidth}{>{\hsize=.10\hsize}X>{\hsize=.95\hsize}X}
\hline
Function& input data specifications\\
\hline
\verb+read.csv(...)+&comma delimited (separated), that is \verb+sep=','+, and \verb+dec='.'+\\
\verb+read.csv2(...)+&semi colon separates, that is \verb+sep=';'+, and \verb+dec=','+\\
\verb+read.delim(...)+&tab separated, that is \verb+'\+{\texttt{t'}}, and \verb+dec='.'+\\
\hline
\end{tabular}
\label{Read-table-csv-1}
\end{table}
\vspace{5mm}
\par For example, suppose we have a tabular data whose name is \verb+A.csv+ as
\begin{verbatim}
"" ,"a","b","c","d","e"
"a",  1,  2,  3,  4,  .
"b",  6,  7,  8,  9, 10
"c", 11, 12, 13, 14, 15
"d", 16, 17, 18, 19,  *
\end{verbatim}
existing at address \directory{E/my R work/A.csv}. As it may be seen, two characters ``\verb+.+''  and ``\verb+*+'' are denoting missing data and further letters \verb+a,b,c,d,e+ are assigned as names to the rows and columns. Using command \verb+read.csv(...)+, then  the following output is obtained.
\vspace{5mm}
\begin{table}[!h]
{{\begin{tabular}{ll}%{10mm}{>{\hsize=.10\hsize}X>{\hsize=.95\hsize}X}
Command& output\\
\hline
{\color{red}{\verb+R> out <- read.csv("E:/my R code/A.csv",+}}&{\color{blue}{\verb+   X   a  b  c  d  e+}}\\
{\color{red}{{\texttt{+}} \verb+header = T, na.strings=c('*', '.'))+}}&~~~~{\color{blue}{\verb+1 a  1  2  3  4 NA+}}\\
{\color{red}{\verb+R> out+}}&~~~~{\color{blue}{\verb+2 b  6  7  8  9 10+}}\\
&~~~~{\color{blue}{\verb+3 c 11 12 13 14 15+}}\\
&~~~~{\color{blue}{\verb+4 d 16 17 18 19 NA+}}\\
\hline
\end{tabular}}}
\end{table}
\subsection{Importing xls and xlsx data}
Sometimes the original data are needed to read from Microsoft Excel application. One way to read such data in \verb+R+ is to install the package \verb+readxl+ \footnote{\url{https://cran.r-project.org/package=readxl}.} developed for \verb+R (>= 3.6)+. For instance, suppose the well known built-in data \verb+mtcars+ is available in \verb+xls+ format at path \directory{E/my R work/Mtcars.xls} and we are willing to import it using package \verb+readxl+. To do this, we proceed as follows.
{\color{red}{
\begin{verbatim}
R> library(readxl)
R> datxl1 <- read_excel("E:/my R work/Mtcars.xls", sheet=1, range="A1:L33")
R> datxl2 <- as.data.frame(datxl1)
 \end{verbatim}
}}
All \verb+Mtcars.xls+ data are collected in just one sheet. If there are other sheets, then the argument \verb+sheet+ allows user to import another existing data sheet. If user interested in importing a predetermined range of data sheet, then the argument \verb+range+ can be changed appropriately. The range of \verb+Mtcars.xls+ data is {\texttt{''A1:L33''}}. For example, some other types of importing a predetermined part of data sheet is given as follows. 
{\color{red}{
\begin{verbatim}
read_excel("Mtcars.xls", range="mtcars!B1:D5")
read_excel("Mtcars.xls", range=cell_rows(1:4))
read_excel("Mtcars.xls", range=cell_cols("B:D"))
\end{verbatim}
}}
Alternatively, for \verb+xls+ format as above, the commands \verb+read_xls+ also works. Unless stated otherwise, any blank cell in the original Excel data sheet are considered as a missing value. Any change is permitted using argument {\texttt{na='' ''}}. More detailed information on other arguments is accessible running \verb+args(read_excel)+. Moreover, for data of format \verb+xlsx+, the command \verb+read_xlsx(...)+
\subsection{Data exporting}
Likewise, data can be exported fro \verb+R+ for other applications as an input. Table \ref{Write-table-1} shows the arguments of command \verb+write.table(...)+. If the format of output  data is \verb+csv+, then the command \verb+write.csv(...)+ is preferable. 
\vspace{5mm}
\begin{table}[!h]
\centering
\caption{Details about main arguments of {\texttt{write.table(...)}} in {\texttt{R}}.}
\vspace*{-3mm}
\begin{tabular}{ll}%{\textwidth}{>{\hsize=.10\hsize}X>{\hsize=.95\hsize}X}
\hline
Argument& output/task\\
\hline
\verb+x+& the name of file to be exported.\\
\texttt{file=''''}& the full address of the output file.\\
\verb+append+ &logical: By default is \verb+FALSE+ that means the old output is replaced with\\
& the new one, otherwise the new output will be added at the end of the old one. \\
\texttt{na=''NA''} & the missing values in the output file is denoted by ``\verb+NA+'' by default, otherwise the\\
&character must be determined by user.\\ 
\hline
\end{tabular}
\label{Write-table-1}
\end{table}
\vspace{5mm}
For example, we may apply command \verb+write.table(...)+ to the first 6 lines of the \verb+airquality+ data to produce a \verb+txt+ file, namely \verb+air.txt+, for which columns are tab-separated, decimal points are represented as slash``\verb+/+', missing data are denoted by ``\verb+*+'' rather than ``\verb+NA+'', and finally row names are removed.
\vspace{5mm}
\begin{table}[!h]
{\small{\begin{tabular}{ll}%{10mm}{>{\hsize=.10\hsize}X>{\hsize=.95\hsize}X}
Command& output\\
\hline
{\color{red}{\verb+R>write.table( head(airquality,6), +}}&{\color{blue}{\verb+"Ozone"	"Solar.R" "Wind" "Temp"	"Month"	"Day"+}}\\
{\color{red}{{\texttt{+"E:/my R code/air.txt", sep=}}\verb+'\t'+,}}&
~~~~~{\color{blue}{\verb+41+~~~~~~~\verb+190+~~~~~~~~\verb+7/4+~~~~~~~\verb+67+~~~~~~~~\verb+5+~~~~~~~~~~\verb+1+}}\\
{\color{red}{{\texttt{+}}\verb+ dec='/', na='*')+}}&
~~~~~{\color{blue}{\verb+36+~~~~~~~\verb+118+~~~~~~~~\verb+8+~~~~~~~~~~~\verb+72+~~~~~~~\verb+5+~~~~~~~~~~\verb+2+}}\\
&~~~~~{\color{blue}{\verb+12+~~~~~~~\verb+149+~~~~~~~~\verb+12/6+~~~~~~\verb+74+~~~~~~~\verb+5+~~~~~~~~~~\verb+3+}}\\
&~~~~~{\color{blue}{\verb+18+~~~~~~~\verb+313+~~~~~~~~\verb+11/5+~~~~~~\verb+62+~~~~~~~\verb+5+~~~~~~~~~~\verb+4+}}\\
&~~~~~{\color{blue}{\verb+*+~~~~~~~~~\verb+*+~~~~~~~~~~\verb+14/3+~~~~~~~\verb+56+~~~~~~~\verb+5+~~~~~~~~~~\verb+5+}}\\
&~~~~~{\color{blue}{\verb+28+~~~~~~~\verb+*+~~~~~~~~~~\verb+14/99+~~~~~~\verb+66+~~~~~~~\verb+5+~~~~~~~~~~\verb+6+}}\\
\hline
\end{tabular}}}
\end{table}

%%%%%%%%%%%%%%%%%%%%%%%%%%%%%%%%%%%%%%%%
\subsection{Apply family}
A wide range of functions can be applied to the members of array, data frame, matrix, and vector through \verb+apply+ family of functions. Herein, we will briefly discuss these functions.
\begin{enumerate}
\item \verb+apply(A, MARGIN, FUN,...)+: The data structure \verb+A+ can be matrix, array, or data frame (with numeric values). If \verb+MARGIN=1 (or 2)+, then function \verb+FUN+ is applied to the rows (or columns) of \verb+A+. The output of \verb+apply+ would be a vector. It it worth to note that ``\verb+...+'' accounts for some additional arguments that are passed to the function \verb+apply+. For example, we may add argument \verb+na.rm=T+ that means the missing vales are removed when applying function \verb+FUN+ to the coordinates of data structure \verb+A+. For instance, the sum of rows, mean of columns, and variance of rows of matrix \verb+A+ denoted as above is obtained as follows. 
{\color{red}{
\begin{verbatim}
R> apply(A, 1, sum)     R> apply(A, 2, mean)     R> apply(A, 1, var)
\end{verbatim}
}}
\vspace*{-5mm}
{\color{blue}{
\begin{verbatim}
[1] 4   6               [1] 1.5   3.5            [1] 2   2
\end{verbatim}
}}
\item \verb+lapply(A, FUN,...)+: The data structure \verb+A+ can be data frame, list, matrix, or vector.
The command \verb+lapply+ applies function \verb+FUN+ to each element of object \verb+A+. The output of \verb+lapply+ would be a list. For example, let function \verb+f1+ is defined as
{\color{red}{
\begin{verbatim}
R> f1<- function(x) 10*x
 \end{verbatim}
}} 
\vspace{-5mm}
then the command \verb+lapply(df2, f1)+ applies function \verb+f1+ to the columns (members) of data frame \verb+df2+ defines as \verb+df2 <- data.frame(x=1:2, y=c(-2,2))+. The output is
{\color{blue}{
\begin{verbatim}
$x
[1] 10 20
$y
[1] -20  20
 \end{verbatim}
}} 
\vspace{-5mm}
\item \verb+sapply(A, INDEX, FUN,...)+: Both functions \verb+lapply+ and \verb+sapply+ do the same task, but the input of \verb+sapply+ is vector, data frame, or list and moreover, the output of the \verb+lapply+ is a list while the results of \verb+sapply+ is vector. Let we are interested in standardizing the columns of data frame \verb+df1+ defined earlier. To do this, we have 
\vspace{5mm}
\begin{table}[!h]
\centering
{{\begin{tabular}{ll}%{10mm}{>{\hsize=.10\hsize}X>{\hsize=.95\hsize}X}
\hline
Command& output\\
\hline
{\color{red}{\verb+R> sapply(df1, function(x)+}}&{\color{blue}{\verb+     col1 col2 col3+}}\\
{\color{red}{{\texttt{+  c(x-mean(x))/sd(x) )}}}}&{\color{blue}{\verb+[1,]  NaN   -1   -1+}}\\
&{\color{blue}{\verb+[2,]  NaN    0    0+}}\\
&{\color{blue}{\verb+[3,]  NaN    1    1+}}\\
\hline
\end{tabular}}}
\end{table}
\par Not surprisingly, the first column of output in above is \verb+NAN+ since we have undetermined case $0/0$. 
\item \verb+tapply(A, INDEX, FUN,...)+: The function \verb+tapply+ in \verb+R+ can be used when we want to apply some function to a vector that is grouped by some other vector. The input of \verb+tapply+ can be array, data frame, matrix, or list. In what follows, suppose we would like to compute the variance of variable \verb+Sepal.Length+ within \verb+iris+ data in terms of variable \verb+Species+. We have
%\begin{table}[!h]
%\centering
%\begin{tabular}{ll}%{10mm}{>{\hsize=.10\hsize}X>{\hsize=.95\hsize}X}
%{\color{red}{\verb+R>data(iris)+}}&{\color{red}{\verb+R>data(iris)+}}\\
%{\color{red}{\verb+R>tapply(iris$Petal.Length,+}}&{\color{red}{\verb+R>tapply(iris$Petal.Length,+}}\\
%{\color{red}{{\texttt{+}}\verb+ iris$Petal.Width<1,var)+}}&{\color{red}{{\texttt{+}}\verb+ iris$Species=="setosa",var)+}}\\
%{\color{blue}{\verb+  FALSE   TRUE+}}&{\color{blue}{\verb+FALSE   TRUE+}}\\
%{\color{blue}{\verb+  0.68157980 0.03015918+}}&{\color{blue}{\verb+0.68157980 0.03015918+}}\\
%\end{tabular}
%\end{table}
{\color{red}{
\begin{verbatim} 
R>data(iris)                      R>data(iris)
R>tapply(iris$Petal.Length,       R>tapply(iris$Petal.Length,
+ iris$Petal.Width<1,var)         + iris$Species=="setosa",var)
 \end{verbatim}
}} 
\vspace*{-14mm}
{\color{blue}{
\begin{verbatim} 
FALSE        TRUE        			        FALSE        TRUE
0.68157980   0.03015918  	        0.68157980   0.03015918
 \end{verbatim}
}} 
\par As it may be seen, not surprisingly, both parts in output above are the same due to the fact that \verb+iris$Petal.Width<1+ and \verb+iris$Species=="setosa"+ are equal sets. We conclude that the variance of \verb+Petal.Length+ when    variable \verb+Petal.Width+ is less (more) than one is 0.03015918 (0.68157980). The function \verb+tapply+ can be employed for applying \verb+FUN+ on a vector that is grouped by two or more variables. For example, suppose we are willing to compute the average of \verb+response+ variable involved in data frame \verb+df1+ introduced earlier while two constraints including \verb+Factor1='low'+ and \verb+Factor2='high'+ are applied. The pertinent command and output are given by the following.
%\begin{table}[!h]
%\centering
%\begin{tabular}{ll}%{10mm}{>{\hsize=.10\hsize}X>{\hsize=.95\hsize}X}
%{\color{red}{\verb+R>tapply(df1$response, list(df1$+}}&
%{\color{red}{\verb+R>tapply(df1$response, list(Factor1=+}}\\
%{\color{red}{{\texttt{+}}\verb+ Factor1=='low', df1$Factor2==+}}& {\color{red}{{\texttt{+}}\verb+ df1$Factor1, Factor2=df1$Factor2),+}}\\
%{\color{red}{{\texttt{+}}\verb+ 'high'), mean)+}}& {\color{red}{{\texttt{+}}\verb+ mean)+}}\\
%{\color{blue}{\verb+        FALSE   TRUE+}}&{\color{blue}{\verb+                     Factor2+}}\\
%{\color{blue}{\verb+ FALSE  17.0   21.5+}}&{\color{blue}{\verb+    Factor1     high low medium+}}\\
%{\color{blue}{\verb+ TRUE    3.5    8.0+}}&{\color{blue}{\verb+        high     26  20     23+}}\\
%                                                                 &{\color{blue}{\verb+         low      8   2      5+}}\\
%                                                                 &{\color{blue}{\verb+      medium     17  11     14+}}\\
%\end{tabular}
%\end{table}
{\color{red}{
\begin{verbatim} 
R>tapply(df1$response, list(      R>tapply(df1$response, list(
+ df1$Factor1=='low', df1$        + Factor1=df1$Factor1, Factor2=
+ Factor2=='high'), mean)         + df1$Factor2), mean)
\end{verbatim} 
}}
\vspace{-7mm}
{\color{blue}{
\begin{verbatim} 
       FALSE  TRUE                          Factor2
FALSE   17.0  21.5                Factor1    high low medium
 TRUE    3.5   8.0                  high       26  20     23
                                    low         8   2      5
                                    medium     17  11     14
\end{verbatim} 
}}
\end{enumerate}
%%%%%%%%%%%%%%%%%%%%%%%%%%%%%
\section{Graphical summary}
Data visualization is one of the most crucial elements in data analysis. A main part of statistical evaluations and comparisons is based on graphical summary. We may use different visualization facilities in \verb+R+, depending on the type of data set. In general, graphical summaries are either high-level or low-level. High-level graphical facilities always create a new plot window and erase the current plot if necessary; rather the low-level ones, in general, do not create a new plot window. Among the high-level summaries, for example, we may mention the scatterplot for representing two sets of numeric data running \verb+plot(...)+. If two data sets are of non-numeric or different types, then a contingency table may be suggested for graphical summary representation using \verb+table(...)+.
%Figure \ref{visualization facilities} displays a list of important high- and low-level graphical summaries. 
In what follows, we list some important high- and low-level graphical facilities. 
%%%%%%%%%%%%%%%%%%%%%%%%%%%%%
%\vspace{5mm}
%\resizebox{.75\totalheight}{!}{
%%\newcolumntype{C}[1]{>{}p{#1}}
%\begin{forest}
%for tree={
%  if level=0{align=center}{% allow multi-line text and set alignment
%    align={@{}C{45mm}@{}},
%  },
%  grow=east,
%  draw,
%  font=\sffamily\bfseries,
%  edge path={
%    \noexpand\path [draw, \forestoption{edge}] (!u.parent anchor) -- +(5mm,0) |- (.child anchor)\forestoption{edge label};
%  },
%  parent anchor=east,
%  child anchor=west,
%  l sep=8mm,
%  tier/.wrap pgfmath arg={tier #1}{level()},
%  edge={ultra thick, rounded corners=2pt},
%  fill=white,
%  rounded corners=2pt,
%  drop shadow,
%}
%[Graphical summary
%  [Low-level
%    [abline(...)]
%    [axis(...)]
%    [curve(...)]
%    [expression(...)]
%    [legend(...)]
%    [lines(...)]
%    [par(...)]
%    [segments(...)]
%    [text(...)]
%  ]
%  [High-level
%    [plot(...)]
%    [pers(...)]
%    [pairs(...)]
%    [image(...)]
%    [hist(...)]
%    [dotchart(...)]
%    [contour(...)]
%    [coplot(...)]
%    [boxplot(...)]
%    [barplot(...)]
%  ]
%]
%\end{forest}
%}
%\label{visualization facilities}
%\caption{Some important visualization facilities within {\texttt{R}}.}
%  \end{center}
%  \vspace{-20pt}
%  \vspace{1pt}
%\end{wrapfigure}
\begin{enumerate}[label=\roman*.]
\item{{\bf{High-level}}}: \verb+plot(...)+, \verb+pers(...)+, \verb+pairs(...)+, \verb+image(...)+, \verb+hist(...)+, \verb+dotchart(...)+, \verb+contour(...)+, \verb+coplot(...)+, \verb+boxplot(x~y,...)+, and \verb+barplot(...)+.
\item{{\bf{Low-level}}}: \verb+abline(...)+, \verb+axis(...)+, \verb+curve(...)+, \verb+expression(...)+, \verb+legend(...)+, \verb+lines(...)+, \verb+par(...)+, \verb+segments(...)+, and \verb+text(...)+.
\end{enumerate}
%\begin{forest} [CP [DP,draw,circle,name=N] [\vdots [,phantom] [VP [DP] [V’ [V,name=N0] [DP,name=object](N0)]]]] \draw[->,dotted] (object) to[out=south west,in=south] (N); 
%\end{forest}
%%%%%%%%%%%%%%%%%%%%%%%%%%%%%
%%%%%%%%%%%%%%%%%%%%%%%%%%%%%
In what follows we deal with the \verb+plot(...)+ among the other high-level graphical facilities. 
\subsection{Plot(...)}
\begin{figure}[!h]
\center
\includegraphics[width=55mm,height=55mm]{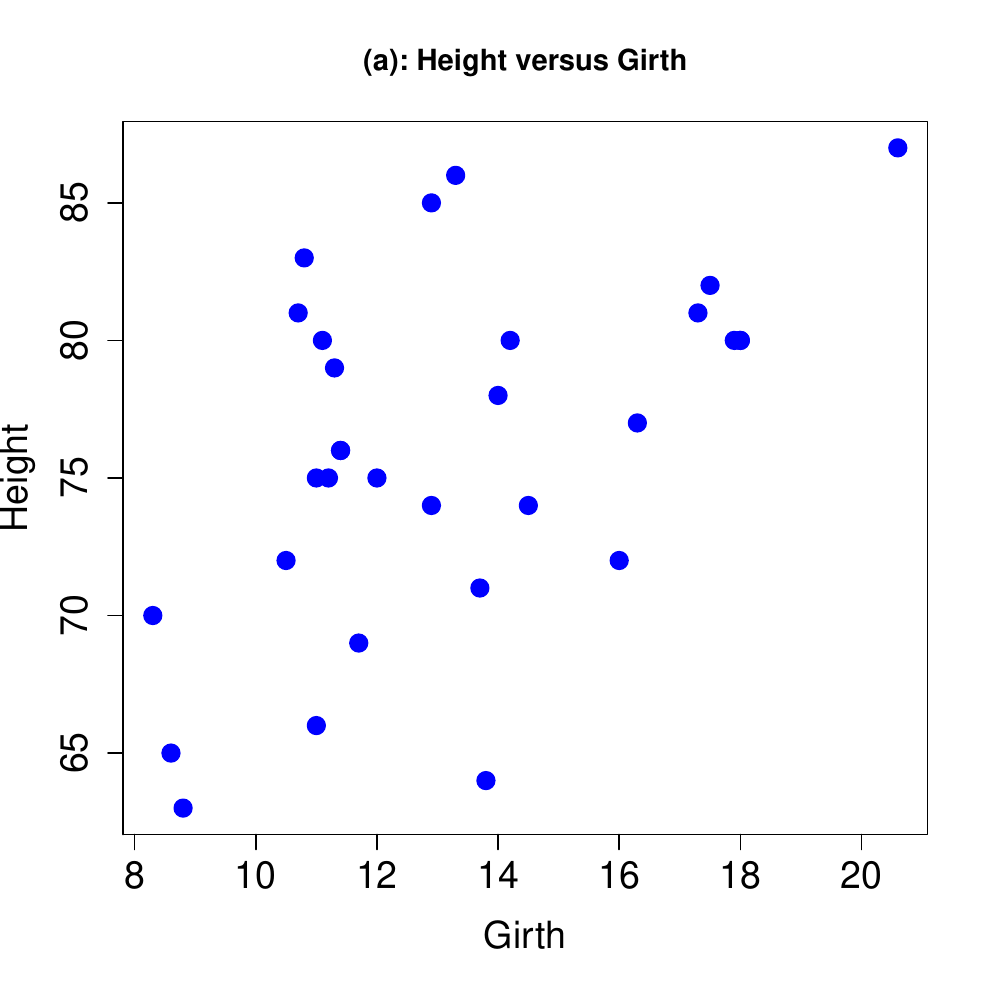}
\includegraphics[width=55mm,height=55mm]{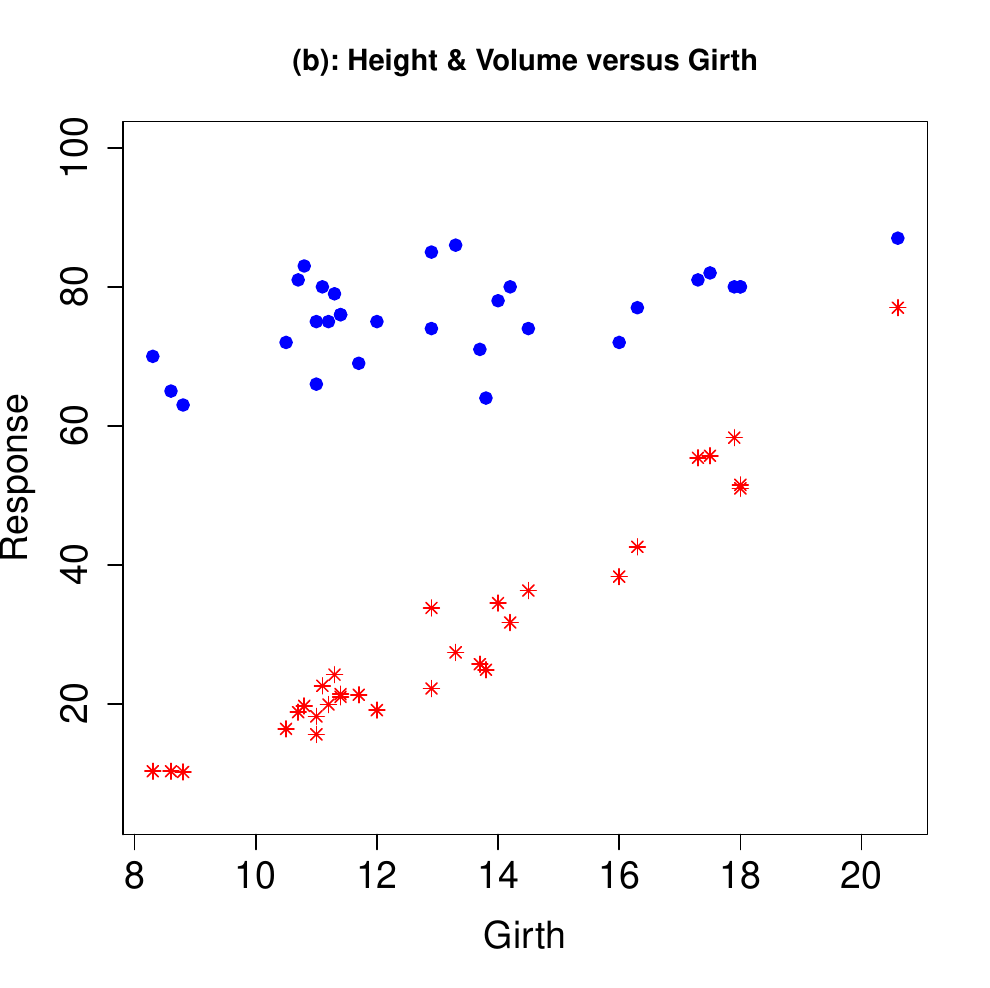}
\caption{Scatterplot of (a): height vs. girth and (b): height and volume vs. girth.}
\label{plot-scatterplot-R-introduction-trees-data}
\end{figure}

We may call \verb+plot(x,y,...)+ to display a two-dimensional visual summary of horizontal (\verb+x+) and vertical (\verb+y+) coordinates of numerical data points. The three dots \verb+...+ represent further arguments to be passed to object \verb+plot(x,y,...)+ that are optional. Table \ref{R-primitive-built-in-function-plot-1} lists some descriptions for some important arguments of this object.
\begin{table}[!h]
\centering
\caption{Details about main arguments of {\texttt{plot}} in {\texttt{R}}.}
\vspace*{-3mm}
\begin{tabular}{ll}%{\textwidth}{>{\hsize=.10\hsize}X>{\hsize=.95\hsize}X}
\hline
Argument& output/task\\
\hline
\verb+cex+& the size of symbol suggested by \verb+pch+.\\
\verb+cex.axis+& the size of axes labels.\\
\verb+cex.lab+& the size of axes title.\\
\verb+cex.main+& the size of main title.\\
\verb+col+& name of color defined for plot points inside the quotation mark.\\
\multicolumn{1}{m{2.5cm}}{{\texttt{font}}}&\multicolumn{1}{m{11cm}}{
the font type and font style of axes labels that is determined using object {\texttt{Hershey}}+ for font type and its style.}\\
\multicolumn{1}{m{2.5cm}}{{\texttt{lty}}}&\multicolumn{1}{m{11cm}}{
an integer number representing the line type: 0=blank, 1=solid (default), 2=dashed, 3=dotted, 4=dotdash, 5=longdash, 6=twodash.}\\           
\verb+lwd+& the thickness (width) of lines and symbols.\\
\verb+mgp=c(a,b,c)+& it adjusts distance between axes lines and axes title (through \verb+a+),\\
& axes labels (through \verb+b+), and axes ticks (through \verb+c+)\\
\multicolumn{1}{m{2.5cm}}{{\texttt{pch}}}&\multicolumn{1}{m{11cm}}{
an integer number from 0 to 20 denoting the symbol of data points shown as Figure \ref{plot-scatterplot-R-pch}.}\\
\multicolumn{1}{m{2.5cm}}{{\texttt{type}}}&\multicolumn{1}{m{11cm}}{
it is {\texttt{''b''}} (for both points and lines), {\texttt{''c''}} (for empty points joined by lines), {\texttt{''h''}} (for histogram-like vertical lines), {\texttt{''l''}} (for lines),  
{\texttt{''n''}} (for any points or lines), {\texttt{''o''}} (for overplotted points and lines), {\texttt{''p''}} (for points), {\texttt{''s''}} (for a stair steps), and etc.}\\
\verb+xlab+& label for horizontal axis inside the quotation mark.\\
\verb+ylab+& label for vertical axis inside the quotation mark.\\
\verb+xlim+& \verb+c(a,b)+ that restricts the range of horizontal axis from \verb+a+ to \verb+b+.\\
\verb+ylim+& \verb+c(a,b)+ that restricts the range of vertical axis from \verb+a+ to \verb+b+.\\
\verb+yaxt="n"+& removes the labels of vertical axis.\\
\verb+xaxt="n"+& removes the labels of horizontal axis.\\
\hline
\end{tabular}
\label{R-primitive-built-in-function-plot-1}
\end{table}
Herein, we call \verb+trees+ built-in data using command \verb+data(trees)+ for our graphical illustration. This data set consists of 31 measurements of felled black cherry trees on variables: girth (diameter), height, and volume\footnote{Diameter is recorded at around 1.3 meters (or 4 to 6 feet) above the ground}. Figure \ref{plot-scatterplot-R-introduction-trees-data}(a) shows the scatterplot of height versus girth and Figure \ref{plot-scatterplot-R-introduction-trees-data}(b) displays the scatterplot of height and  volume versus girth simultaneously. The command for producing these figures are given as follows.
%%%%%%%%%%%%%%%%%%%%%%%%%%%%%%
 \begin{figure}[!htb]
    \begin{minipage}{0.3\textwidth}
      \includegraphics[width=\linewidth]{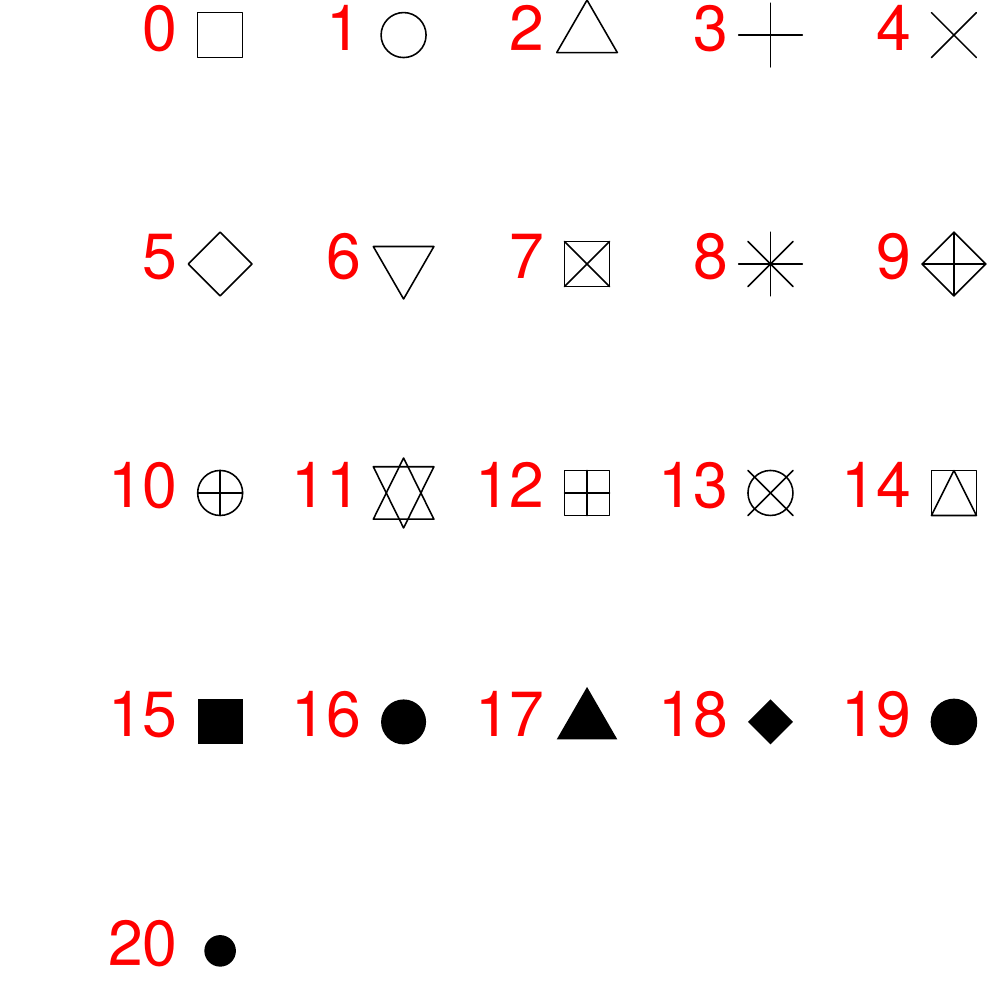}
    \end{minipage}
  ~$\underset{\xrightarrow{\hspace*{2cm}}}{{\texttt{R code}}}$~
      \begin{minipage}{0.3\textwidth}
{\color{red}{
\begin{verbatim} 
R> par(mar=c(0,0,0,0))
R> w=c(rep('black',21),
+  rep('white',4))
R> y=rev(rep(1:5,each=5))
R> x=c(rep(1:5,5))
R> plot(x,y,pch=0:20,cex=4,
+  ylim=c(1,5),xlim=c(0,5), 
+  axes=F,xlab=' ',ylab=' ',col=w)
R> text(x[1:21]-0.15,y[1:21],0:20,
+  pos=2,adj=1,col='red',cex=2.5)
\end{verbatim} 
}}
\end{minipage}
\caption{Commonly used symbols for {\texttt{pch}} in {\texttt{R}} and the pertinent code.}\label{plot-scatterplot-R-pch}
\end{figure}
{\color{red}{
\begin{verbatim} 
R> # commands for producing Subfigure(a)
R>data(trees); x<-trees$Girth; y<-trees$Height; z<-trees$Volume; m<-1.5
R> plot(x, y, main="Height versus Girth", xlab="Girth", ylab="Height",
+ type="p", col="blue", pch=19, cex=m, lwd=m, cex.lab=m, cex.axis=m)
R> # commands for producing Subfigure(b)
R> plot(x, y, main="Height & Volume versus Girth", xlab="Girth", ylab=
+ "Response", type="p", col="blue", pch=19, ylim=c(5,100), lwd=m, 
+ cex.lab=m, cex.axis=m)
R> lines(x,z,type="p",col="red",pch=8)
\end{verbatim} 
}}
After producing figure, user can save the produced figure freely in such formats as \verb+bmp+, \verb+jpeg+, \verb+pdf+, \verb+png+, and \verb+tiff+. Herein, we use command \verb+pdf(...)+ for this end. Details about arguments of these calls are given using, for example, \verb+?tiff+. Suppose the name of created figure is \verb+FigA+ and we would like to save it with portable document format (pdf) in track \directory{E/my R work}. To do this, we have
{\color{red}{
\begin{verbatim} 
R> setwd("E:/my R work")
R> pdf(file="FigA.pdf",width=4,height=4) # width and height are in inches
R> # put here command plot(x,y,...)
R> # put here command line(x,y,...)
R> dev.off()
\end{verbatim} 
}}
More manipulations can be made on produced plot using low-level graphical facilities. Table \ref{R-primitive-built-in-function-plot-low-level-facilities-1}.
\vspace{5mm}
\begin{table}[!h]
\centering
\caption{Details about arguments of low-level graphical facilities.}
\vspace*{-3mm}
{\small{
\begin{tabular}{ll}%{\textwidth}{>{\hsize=.10\hsize}X>{\hsize=.95\hsize}X}
\hline
Function/argument& output/task\\
\hline
\hline
\multicolumn{2}{m{14cm}}{{\texttt{abline(a=NULL,b=NULL,h=NULL,v=NULL,...)}}~~~
it adds one or more straight lines to the plot. Parameters {\texttt{a}} and {\texttt{b}} are, accordingly, the intercept and slope of the fitted regression line. Parameters {\texttt{h}} and {\texttt{v}} are, respectively, the position of drawn horizontal and vertical lines to the plot.}\\
\hline
\hline
\multicolumn{2}{l}{{\texttt{axis(side,at=NULL,labels=TRUE,...)}}}\\
\multicolumn{1}{m{1.5cm}}{{\texttt{side}}}&\multicolumn{1}{m{10.5cm}}{an integer number indicating the side of the plot we are willing to work on. That includes {\texttt{1}} (for {\texttt{below}}), {\texttt{2}} (for {\texttt{left}}), {\texttt{3}} (for {\texttt{above}}), and {\texttt{4}} (for {\texttt{right}}).}\\
%\verb+side+&an integer number indicating the side of the plot we are \\
%&willing to work on. That includes \verb+1+, \verb+2+, \verb+3+, and \verb+4+ for below,\\
%&left, above and right, respectively.\\
\multicolumn{1}{m{1.5cm}}{{\texttt{at}}}&\multicolumn{1}{m{10.5cm}}{
a vector of numerical values indicating the location at which
the tick marks are shown.}\\
\multicolumn{1}{m{1.5cm}}{{\texttt{labels}}}&\multicolumn{1}{m{10.5cm}}{
a vector of numeric or non-numeric data to be inserted at points determined in {\texttt{at}} argument.}\\
\multicolumn{1}{m{1.5cm}}{{\texttt{las}}}&\multicolumn{1}{m{10.5cm}}{
an integer ({\texttt{0=}}always parallel to the axis [default] and {\texttt{1=}}always perpendicular to the axis) value.}\\
\multicolumn{1}{m{1.5cm}}{{\texttt{hadj}}}&\multicolumn{1}{m{10.5cm}}{
a numeric number for moving the position of all axis labels simultaneously parallel to the reading direction.}\\
\multicolumn{1}{m{1.5cm}}{{\texttt{padj}}}&\multicolumn{1}{m{10.5cm}}{
a numeric number for moving the position of all axis labels simultaneously perpendicular to the reading direction.}\\
\hline
\hline
\multicolumn{2}{m{14cm}}{{\texttt{curve(expr,from=NULL,to=NULL,n=101,add=FALSE)}}~~~
it works similar to call {\texttt{plot(...)}} for depicting a curve with formula {\texttt{expr}} over the given range starting at {\texttt{from}} and ending at {\texttt{to}}. If we want to add the output of call {\texttt{curve(...)}} to the output of plot, we must set {\texttt{add=TRUE}}.}\\
\hline
\hline
\multicolumn{2}{m{14cm}}{{\texttt{expression(paste(...))}}~~~ this command is useful when we want to add some annotations that typically involves mathematical formula. A full list of to a full list of mathematical symbols and formula run {\texttt{demo(plotmath)}}.}\\ 
\hline
\hline
\multicolumn{2}{l}{{\texttt{legend('position',legend,lwd,col,horiz=FALSE,...)}}}\\
\multicolumn{1}{m{1.5cm}}{{\texttt{position}}}&\multicolumn{1}{m{10.5cm}}{one of {\texttt{top}}, {\texttt{bottom}}, {\texttt{topleft}}, {\texttt{topright}}, {\texttt{bottomleft}}, or {\texttt{bottomright}}. If some position else is desired, the {\texttt{x}} and {\texttt{y}} coordinates must be given instead.}\\
%\verb+side+&an integer number indicating the side of the plot we are \\
%&willing to work on. That includes \verb+1+, \verb+2+, \verb+3+, and \verb+4+ for below,\\
%&left, above and right, respectively.\\
\multicolumn{1}{m{1.5cm}}{{\texttt{legend}}}&\multicolumn{1}{m{10.5cm}}{a vector of explanations for each line existing in the plot.}\\
\multicolumn{1}{m{1.5cm}}{{\texttt{lwd}}}&\multicolumn{1}{m{10.5cm}}{width of lines (or symbols) existing in argument {\text{legend}}.}\\
\multicolumn{1}{m{1.5cm}}{{\texttt{col}}}&\multicolumn{1}{m{10.5cm}}{color of lines existing in legend.}\\
\multicolumn{1}{m{1.5cm}}{{\texttt{horiz=FALSE}}}&\multicolumn{1}{m{10.5cm}}{logical: if {\texttt{TRUE}}, then the expressions are shown horizontally.}\\  
\hline
\hline
\multicolumn{2}{m{14cm}}{
{\texttt{lines(x,y,...)}}~~~it adds a new curve to object  {\texttt{plot(...)}} and its arguments are the same as {\texttt{plot(...)}}.}\\
\hline
\hline
\multicolumn{2}{m{14cm}}{ {\texttt{par(...)}}~~~it includes 72 graphical parameters or queries. The full list of these parameters is accessible by running {\texttt{par(...)}}. It is worth to note that if user changes the setting of object {\texttt{par}}, then these change will be applied to all graphs whenever the session is working. To restore the default settings, for other plots, calling {\texttt{on.exit(...)}} is suggested.}\\
\hline
\hline
\multicolumn{2}{m{14cm}}{{\texttt{segments(x0,y0,x1=x0,y1=y0,...)}}~~~it draws a line from starting point whose {\texttt{x}} and {\texttt{y}} coordinates are, accordingly, {\texttt{x0}} and {\texttt{y0}} to end point whose {\texttt{x}} and {\texttt{y}} coordinates are, accordingly, {\texttt{x1}} and {\texttt{y1}}.}\\
\hline
\hline
\multicolumn{2}{m{14cm}}{{\texttt{text(a,b,'note',font,...)}}~~~it puts a {\texttt{note}} at a position whose {\texttt{x}} and {\texttt{y}} coordinates are {\texttt{a}} and {\texttt{b}}, respectively. Parameter {\texttt{font}} determines the {\texttt{note}}, for which a {\texttt{list}} of fonts and styles can be found in object {\texttt{Hershey}}. Finally, three dots {\texttt{...}} represents further arguments that must be passed, for example {\texttt{col}} (for color), {\texttt{cex}} (for size), etc.}\\
\hline
\end{tabular}}}
\label{R-primitive-built-in-function-plot-low-level-facilities-1}
\end{table}
Figure \ref{plot-scatterplot-R-introduction-data-supplementary} shows two sample plots produced based on some supplementary arguments given in Table \ref{R-primitive-built-in-function-plot-low-level-facilities-1}.
\begin{figure}[!h]
\center
\includegraphics[width=55mm,height=55mm]{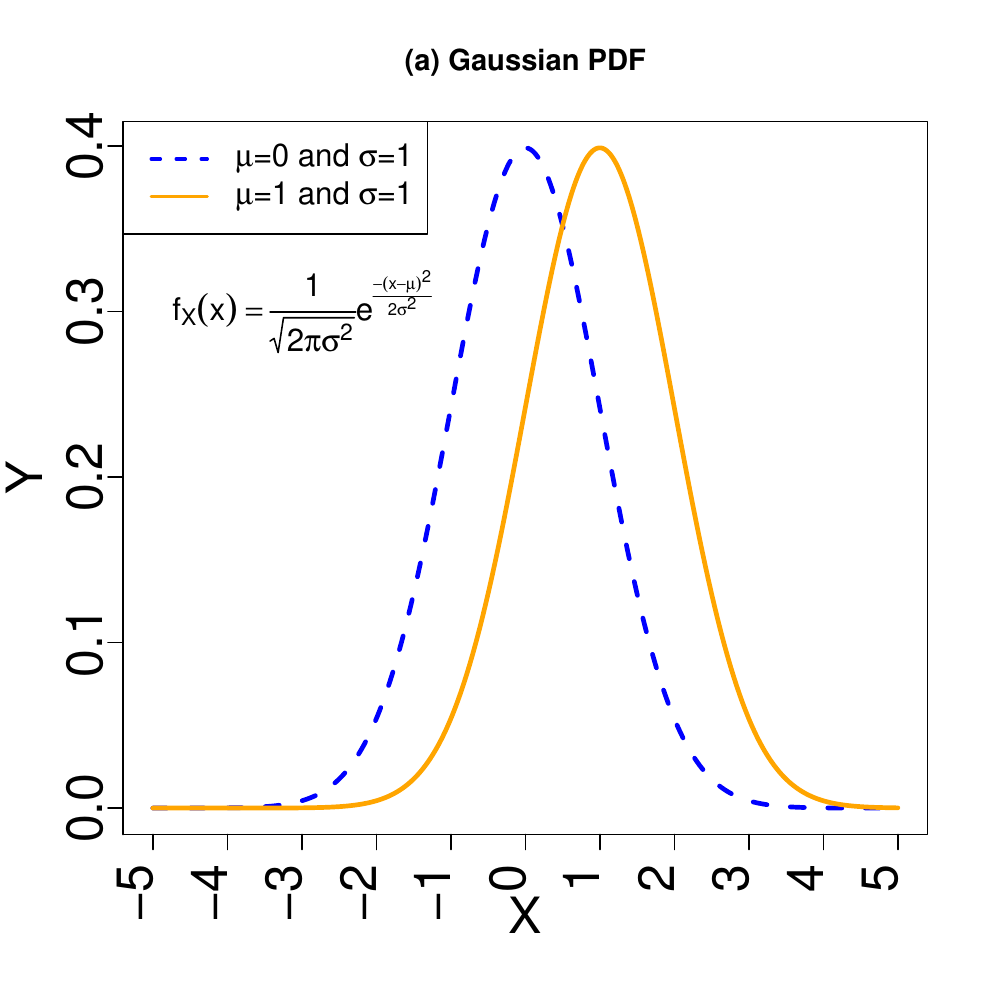}
\includegraphics[width=55mm,height=55mm]{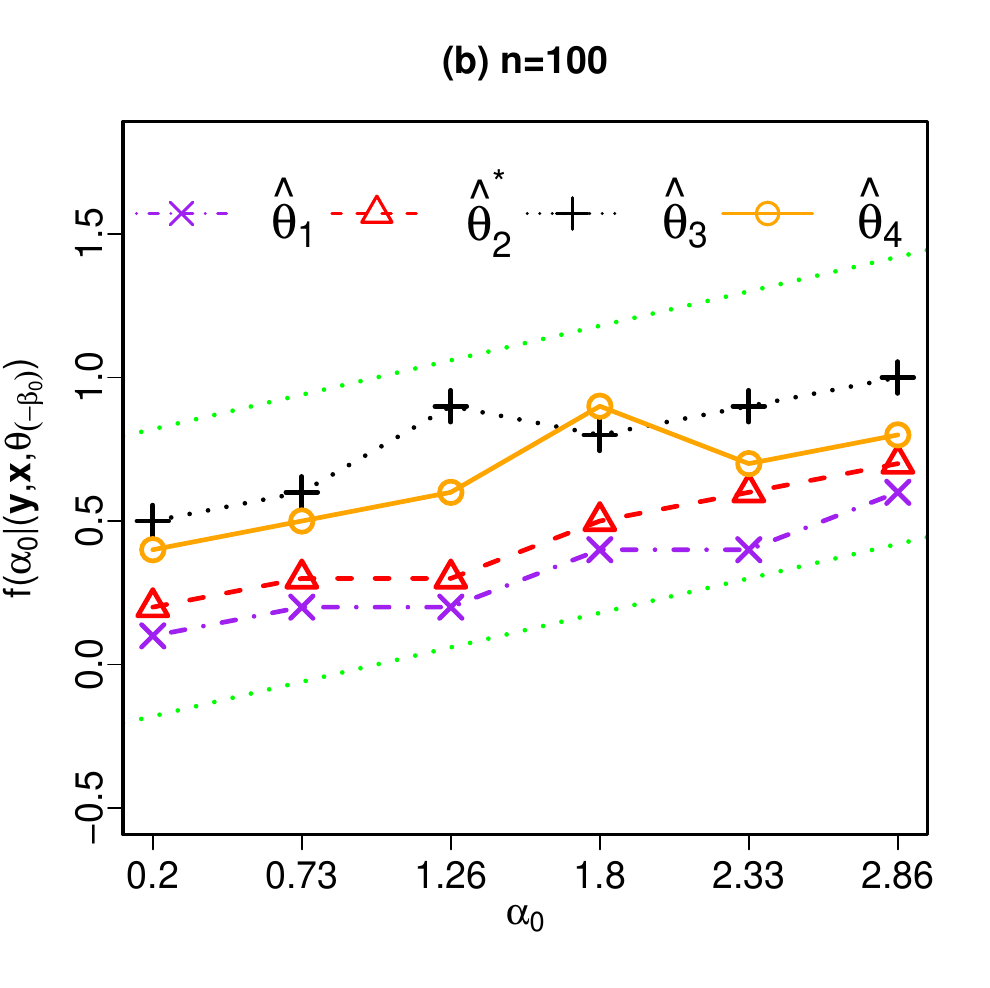}
\caption{Two sample plots produced based on some supplementary calls of Table \ref{plot-scatterplot-R-introduction-data-supplementary}.}
\label{plot-scatterplot-R-introduction-data-supplementary}
\end{figure}
The corresponding \verb+R+ codes are given as follows.
{\small{
{\color{red}{
\begin{verbatim}
R># R code for producing Figure 1.2(a)
R>x<-seq(-5,5,.01); posi<-seq(-5,5,1)
R>plot(x,dnorm(x),col="blue",type='l',xlab='X',main="
+(a) Gaussian PDF",mgp=c(2.5,.5,0), ylab='Y',cex=1,
+lty=2,lwd=3,xaxt='n',cex.lab=2,cex.axis=2,pch=8)
R>lines(x,dnorm(x,1,1),col='orange',lwd=3, pch=19)
R>Leg1<-expression( paste(mu,"=0 and ",sigma,"=1"))
R>Leg2<-expression( paste(mu,"=1 and ",sigma,"=1"))
R>mytext<-expression(paste(f[X](x)==frac(1,sqrt(2*
+pi*sigma^2))*e^{frac(-(x-mu)^2,2*sigma^2)} ))
R>text(-3,0.3,mytext,cex=1.25)
R>axis(1,at=posi,lab=posi,hadj=1,padj=0,las=2,cex.axis=2)
R>legend('topleft',legend=c(Leg1,Leg2),lty=c(2,1),col=
+c('blue','orange'),cex=1.25,pch=c(8,19),lwd=2)
R># R code for producing Figure 1.2(b)
R>x1<-c(0.1,0.2,0.2,0.4,0.4,0.6);x2<-c(0.2,0.3,0.3,0.5,0.6,0.7)
R>x3<-c(0.5,0.6,0.9,0.8,0.9,  1);x4<-c(0.4,0.5,0.6,0.9,0.7,0.8)
R>vmle  <-expression(paste( hat(theta)[1]   ))
R>vstar <-expression(paste(hat(theta)[2]^'*'))
R>vhat1 <-expression(paste( hat(theta)[3]   ))
R>vhat2 <-expression(paste( hat(theta)[4]   ))
R>valpha<-expression(paste( alpha[0] ))
R>vylab <- expression(paste("f(",alpha[0],"|(", bold(y),",",
+bold(x),",", bold(theta)[(-beta[0])],")"))
R>x0 <- seq(0.2, 5, .5); v1<-c(vmle, vstar, vhat1, vhat2); m<-1.5;
R>plot(x1,cex.lab=m,cex.main=m,cex=2,main="(b) n=100",cex.axis=m,
+col="purple",lty=4,lwd=3,mgp=c(2.45,0.5,0),pch=4,type="o",
+xaxt='n',xlab=valpha,ylab=vylab,ylim=c(-0.5,1.8))
R>lines(x2, cex=2, col="red"   ,lty=2, lwd=3, pch=2 ,type="o")
R>lines(x3, cex=2, col="black" ,lty=3, lwd=3, pch=3 ,type="o")
R>lines(x4, cex=2, col="orange",lty=1, lwd=3, pch=1 ,type="o")
R>axis(1, at=seq(1,length(x0)), cex.axis=m,+labels=c("0.2","0.73",
+"1.26","1.8","2.33","2.86","3.4","3.93","4.46","5"))
R>cols<-c("purple","red","black","orange")
R>legend("topleft",horiz=TRUE, legend=v1,bty="n",lwd=2,cex=2,
+col=cols,pch=c(4,2,3,1),lty=c(4,2,3,1))
R>abline(a=0.7,b=0.12,lty=3,lwd=3,col='green')
R>abline(a=-0.3,b=0.12,lty=3,lwd=3,col='green')
R>box(lwd=2)
\end{verbatim}
}}}}
\subsection{Pairwise visualization}
\begin{figure}[!htb]
    \begin{minipage}{0.5\textwidth}
\includegraphics[width=85mm,height=85mm]{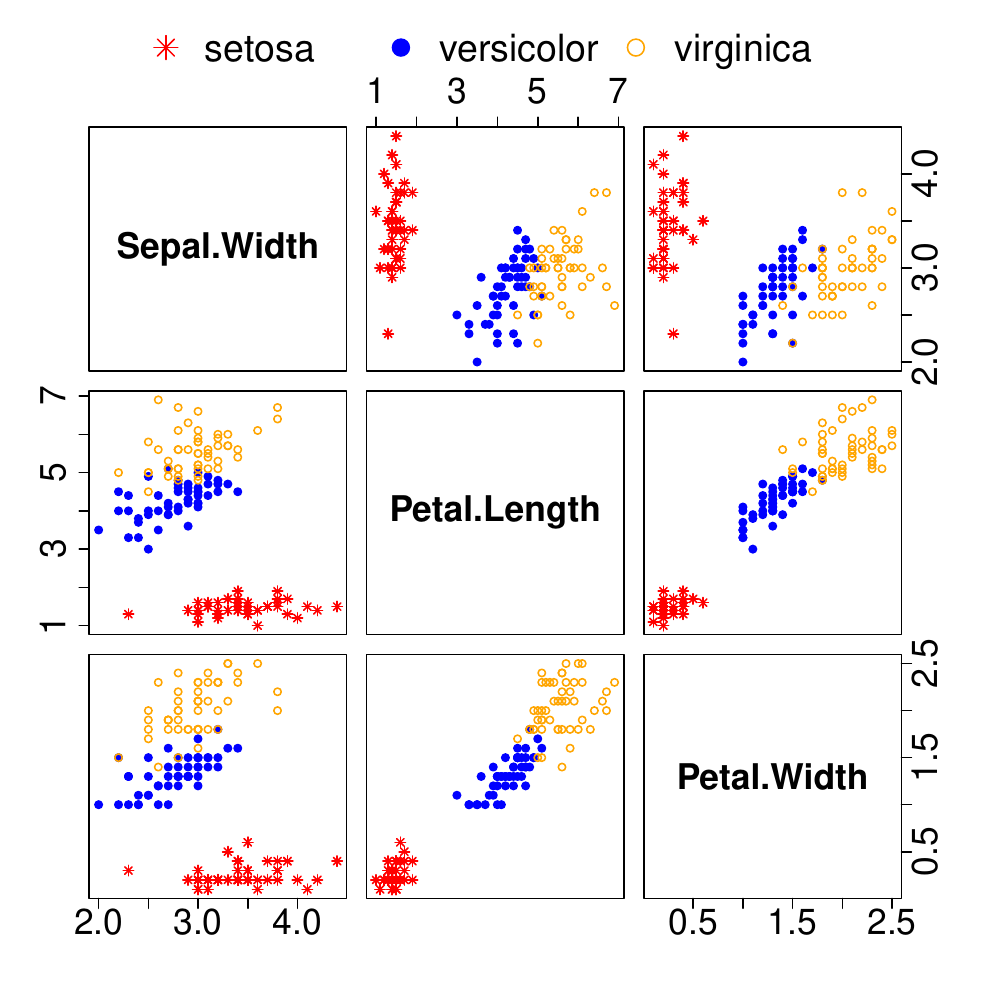}
    \end{minipage}
  ~$\underset{\xrightarrow{\hspace*{.5cm}}}{{\texttt{R code}}}$~
      \begin{minipage}{0.7\textwidth}
      {\small{
{\color{red}{
\begin{verbatim} 
R> c0<-unclass(iris$Species)
R> par(xpd=TRUE)
R> Col<-c("red","blue","orange")
R> pairs(iris[,2:4],main="",pch= 
R> c(8,19,21)[c0],col=Col[c0],
+  labels=c("Sepal.Width",
+  "Petal.Length","Petal.Width"),
+  cex.labels=2.2, font.labels=2,
+  cex.axis=2.2)
R> L <-c("setosa","versicolor",
+  "virginica")
R> legend(0.1,1.05,bty="n",cex=
+  1.5,col=Col,pch=c(8,19,21),
+  legend=L,horiz=T)
R> dev.off()
\end{verbatim}
}}}}
\end{minipage}
\caption{Pairwise scatterplots of {\texttt{iris}} data.}
\label{plot-scatterplot-R-introduction-data-pairs-1}
\end{figure}

\begin{figure}[!htb]
    \begin{minipage}{0.5\textwidth}
\includegraphics[width=85mm,height=85mm]{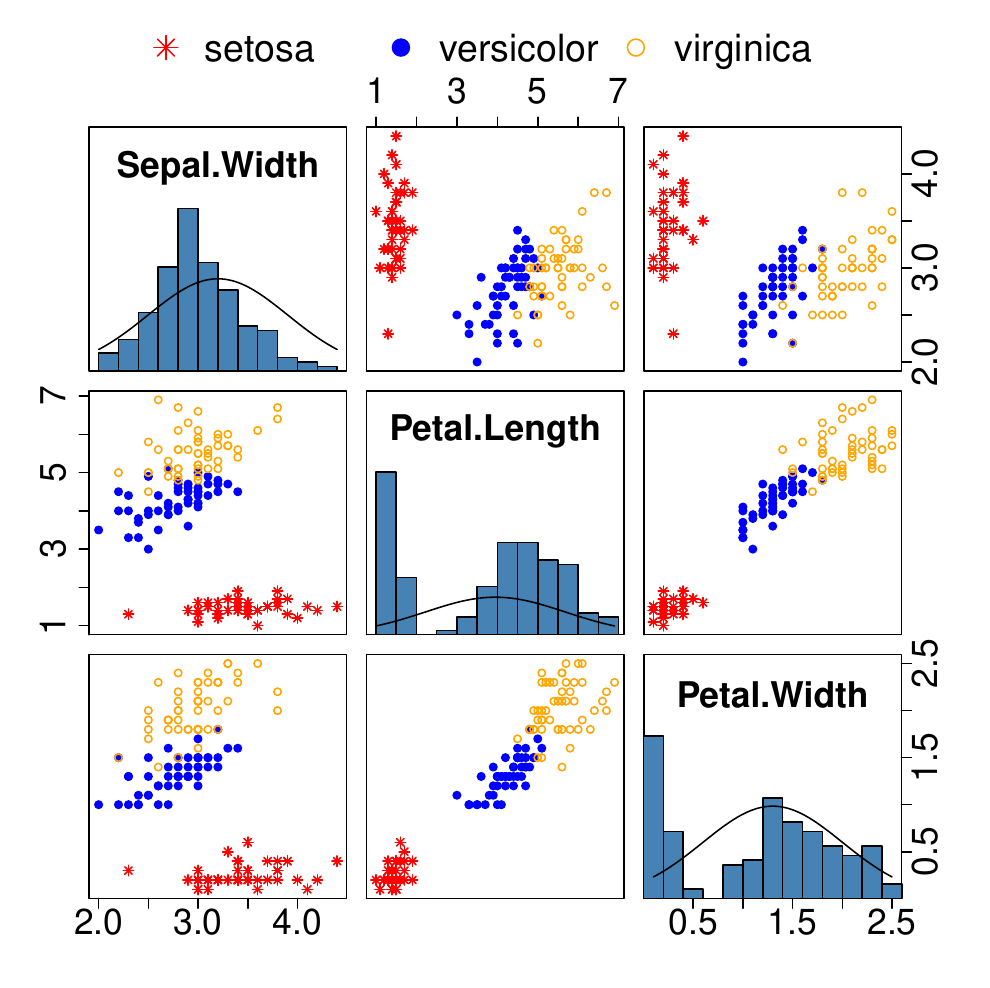}
    \end{minipage}
  ~$\underset{\xrightarrow{\hspace*{.5cm}}}{{\texttt{R code}}}$~
      \begin{minipage}{0.7\textwidth}
      {\small{
{\color{red}{
\begin{verbatim} 
R> panel.hist<-function(x,col,...){
+ usr<-par("usr");on.exit(par(usr))
+ par(usr=c(usr[1:2],0,1.5))
+ h<-hist(x, plot=FALSE)
+ breaks<-h$breaks
+ nB<-length(breaks)
+ y<-h$counts; y<-y/max(y)
+ rect(breaks[-nB],0,breaks[-1],
+ y,col="steelblue",...)
+ curve(dnorm(x,mean=mean(x),
+ sd=sd(x)),add=T)
+}
R> c0<-unclass(iris$Species)
R> par(xpd=TRUE)
R> Col<-c("red","blue","orange")
R> pairs(iris[,2:4],main="",pch= 
R> c(8,19,21)[c0],col=Col[c0],
+  diag.panel =panel.hist,
+  labels=c("Sepal.Width",
+  "Petal.Length","Petal.Width"),
+  cex.labels=2.2, font.labels=2,
+  cex.axis=2.2)
R> L <-c("setosa","versicolor",
+  "virginica")
R> legend(0.1,1.05,bty="n",cex=
+  1.5,col=Col,pch=c(8,19,21),
+  legend=L,horiz=T)
R> dev.off()
\end{verbatim}
}}}}
\end{minipage}
\caption{Pairwise scatterplots of {\texttt{iris}} data with fitted Gaussian curve to the marginals.}
\label{plot-scatterplot-R-introduction-data-pairs-2}
\end{figure}

When there are more than two sets of numeric data, then a more useful graphical summary about the variables involved in study and associated relationships can be found using object \verb+pairs(...)+. There is a long list of arguments for this object , among them in order to save space, we mention the \verb+lower.panel+, \verb+upper.panel+, and \verb+diag.panel+ that are more important than others. Suppose we are interested in relationship between last three variables, excluding \verb+Species+ \verb+iris+ data. A graphical summary may be obtained such as Figure \ref{plot-scatterplot-R-introduction-data-pairs-1} that displays the pairwise scatterplots of these three variables. More changes on produced pairwise scatterplot is possible by changing the arguments \verb+lower.panel+, \verb+upper.panel+, and \verb+diag.panel+ appropriately. Herein, we use our user-defined function by setting \verb+diag.panel=panel.hist+ for presenting the histogram of each marginal within the diagonal cells of pairwise scatterplots of Figure \ref{plot-scatterplot-R-introduction-data-pairs-2} while the superimposed appropriate Gaussian curve helps in discovering departure from normality.

%%%%%%%%%%%%%%%%%%%%%%%%%%%%%
%%%%%%%%%%%%%%%%%%%%%%%%%%%%%%%%%
\subsection{Contingency table}
One useful graphical summary associated with both numeric and non-numeric data is contingency table. For constructing a contingency table we may use object \verb+cut(...)+ that splits numeric data into a number of mutually disjoint intervals (groups). Details about arguments of this object is given by Table \ref{R-primitive-built-in-function-cut-1}.
\vspace{5mm}
\begin{table}[!h]
\centering
\caption{Details about arguments of {\texttt{cut(...)}}.}
\vspace*{-3mm}
{\small{
\begin{tabular}{ll}%{\textwidth}{>{\hsize=.10\hsize}X>{\hsize=.95\hsize}X}
\hline
Argument& output/task\\
\hline
{\texttt{x}}&a given numeric vector for grouping.\\
\multicolumn{1}{m{2.5cm}}{{\texttt{breaks (br)}}}&\multicolumn{1}{m{11cm}}{
either a numeric vector representing cut points or a single value>2 representing the number of groups.}\\
\multicolumn{1}{m{2.5cm}}{{\texttt{labels=NULL}}}&\multicolumn{1}{m{11cm}}{optional labels for the created groups.}\\
\multicolumn{1}{m{2.5cm}}{{\texttt{include.lowest}}}&\multicolumn{1}{m{11cm}}{{\texttt{logical}}: if {\texttt{TRUE}} and {\texttt{x[i]}} coincides with the lower bound of group, then group includes {\texttt{x[i]}}, otherwise {\texttt{x[i]}} belongs to another group.}\\
\multicolumn{1}{m{2.5cm}}{{\texttt{right}}}&\multicolumn{1}{m{11cm}}{
{\texttt{logical}}: if {\texttt{TRUE}}, then each group takes the form {\texttt{(a,b]}}, otherwise it would be of the form {\texttt{[a,b)}}.}\\
\multicolumn{1}{m{2.5cm}}{{\texttt{...}}}&\multicolumn{1}{m{11cm}}{a number of arguments to be passed.}\\
\hline
\end{tabular}}}
\label{R-primitive-built-in-function-cut-1}
\end{table}
\vspace{5mm}
Recalling \verb+women+ data, we call \verb+cut(...)+ for grouping this set of data as follows.
\vspace{5mm}
\begin{table}[!h]
\centering
\vspace*{-3mm}
{\small{
\begin{tabular}{ll}%{\textwidth}{>{\hsize=.10\hsize}X>{\hsize=.95\hsize}X}
\hline
Command& output\\
\hline
{\color{red}{\verb+R>br.h<-c(50,60,70,80)+}}&{\color{blue}{\verb+         W                +}}\\
{\color{red}{{\texttt{R>br.w<-c(110,130,150,170)}}}}&{\color{blue}{\verb+H         (110,130] (130,150] (150,170]+}}\\
{\color{red}{\verb+R>H<-cut(women$height,br=br.h)+}}&{\color{blue}{\verb+  (50,60]         3         0         0           +}}\\
{\color{red}{\verb+R>W<-cut(women$weight,br=br.w)+}}&{\color{blue}{\verb+  (60,70]         3         6         1+}}\\
{\color{red}{\verb+R>table(H,W)+}}&{\color{blue}{\verb+  (70,80]         0         0         2+}}\\
\hline
\end{tabular}}}
\end{table}
\vspace{5mm}
\par
\begin{figure}[!htb]
    \begin{minipage}{0.4\textwidth}
      \includegraphics[width=65mm,height=75mm]{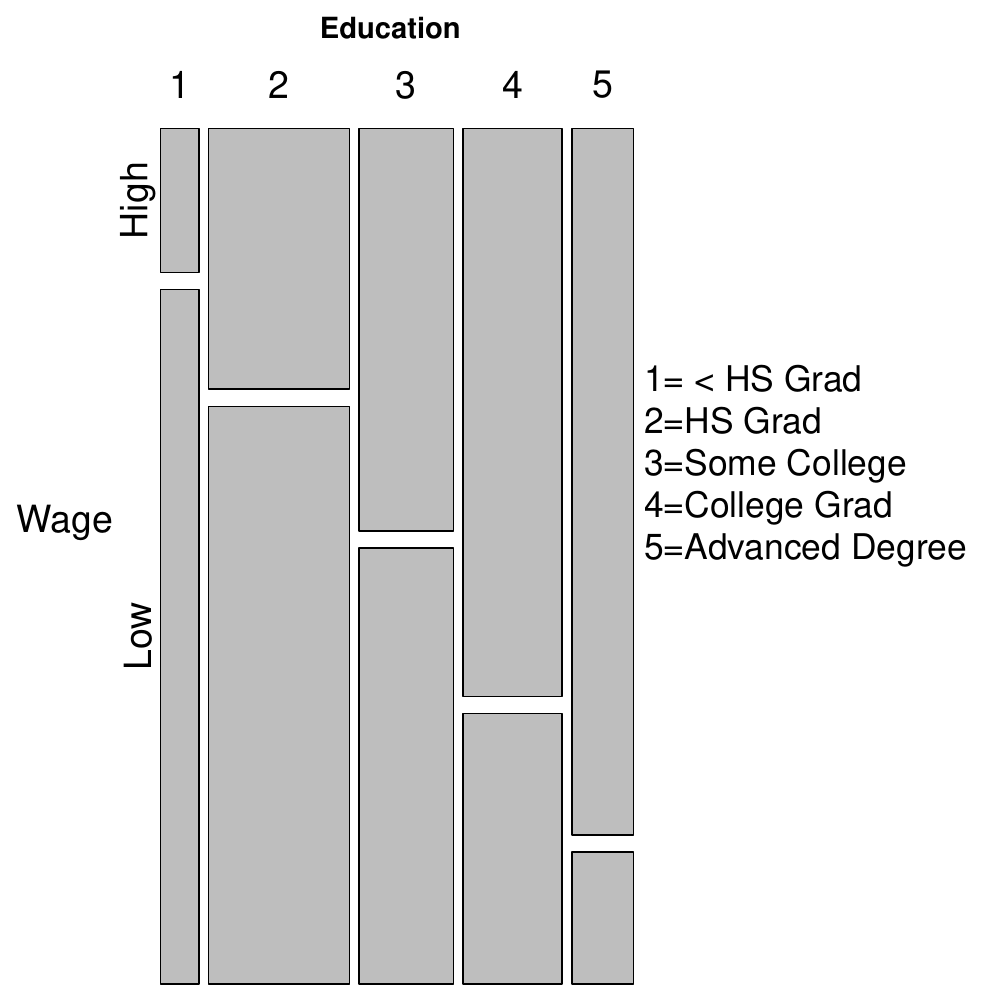}
    \end{minipage}
  ~$\underset{\xrightarrow{\hspace*{.5cm}}}{{\texttt{R code}}}$~
      \begin{minipage}{0.7\textwidth}
{\color{red}{
\begin{verbatim} 
R> library(ISLR); data(Wage)
R> W.Level <-(ifelse(Wage$wage>
+  median(Wage$wage),"High","Low"))
R> n<- length(Wage$education)
R> E.Level <- unclass(Wage$education)
R> Table <- table(E.Level, W.Level)
R> par(mar=c(2, 4, 2, 11), xpd=TRUE)
R> mosaicplot(Table, cex.axis=1.5, 
+  xlab="", ylab="",main="Education")
R> text(-.15,0.50,"Wage",cex=1.5,xpd=NA)
R> L <-c('1= < HS Grad','2=HS Grad','3=
+  Some College','4=College Grad','5=
+  Advanced Degree')
R> legend('topright',inset=c(-.6,0.3),L,
+  horiz=F, cex=1.4, bty="n")
R> dev.off()
\end{verbatim}
}}
\end{minipage}
\caption{Mosaicplot for {\texttt{Wage}} data.}
\label{plot-scatterplot-R-introduction-data-contingency-table-1}
\end{figure}

 The focus of the second example is in \verb+wage+ (wage and other information of 3000 male workers in the Mid-Atlantic region) data available at package \verb+ISLR+. Suppose we are interested in relationship between marital status and wage variables. The former is non-numeric while the latter is numeric. One way to deal with relation between these two variables, first we may divide wage variable into two levels \verb+High+ (greater than median) and \verb+Low+ (less than median) and then investigating the chi-squared independence test.
%{\color{red}{
%\begin{verbatim}
%R> library(ISLR)
%R> data(Wage)
%R> W.L <-(ifelse(Wage$wage>median(Wage$wage),"High","Low"))
%R> n<-length(Wage$education) 
%R> E.L<-z<-rep(NA, n)
%R> for(i in 1:5)
%+{
%+z <- grep(paste0(i,'.'),Wage$education); E.L[z]<-i
%+}
%R> TG <- table(E.L,W.L)
%R> par(mar=c(0, 4, 2, 11), xpd=TRUE)
%R> mosaicplot(TG,cex.axis = 1.5, xlab="", ylab = "", 
%+ main="Education")
%R> text(-.15,.50, "Wage", cex=1.5, xpd=NA)
%R> L <- c('1= < HS Grad','2=HS Grad','3=Some College',
%+'4=College Grad','5=Advanced Degree')
%R> legend('topright',inset=c(-.6,.3),L, horiz=F, cex=
%+1.4, bty = "n")\end{verbatim}
%R> dev.off()
%}}
The output of chi-square independence test, after running \verb+chisq.test(Table)+, is given by the following. 
{\color{blue}{
\begin{verbatim}
        Pearson's Chi-squared test
data:  Table
X-squared = 541.37, df = 4, p-value < 2.2e-16
\end{verbatim}
}}
Obviously, as it may be seen from Figure \ref{plot-scatterplot-R-introduction-data-contingency-table-1}, we decide to reject the null hypothesis of $H_{0}$: ``Wage and education variables are independent'' against $H_{1}$: ``Wage and education variables are not independent''. 
%%%%%%%%%%%%%%%%%%%%%%%%%%%%%%
% \begin{figure}[!h]
%\center
%\includegraphics[width=55mm,height=55mm]{plot-scatterplot-R-introduction-data-contingency-table-1.pdf}
%\caption{Mosaicplot for {\texttt{Wage}} data.}
%\label{plot-scatterplot-R-introduction-data-contingency-table-1}
%\end{figure} 
\section{Function in R}
When we would like to accomplish some task(s) repeatedly, it is recommended to use some structure called {\it{function}} in order to saving space, reducing the computational costs, debugging errors and warnings easily, etc., within an \verb+R+ project. In general, an \verb+R+ function is either {\it{built-in}} or {\it{user-defined}}. Tables \ref{R-function-statistical-vector}, \ref{R-function-matrix-2}, and \ref{R-primitive-built-in-function-1} involve some built-in functions. The user-defined function are those created by user for some special purpose. Each function has four parts including {\it{name}}, {\it{input}}, {\it{body}}, and {\it{output}}. Two key points when creating a user-defined function are: i) to avoid constructing a function whose built-in counterpart does exist and ii) the name of the user-defined and existing built-in functions must be different; otherwise the user-defined function will {\it{mask}} the built-in function. The general structure of user-defined function is given as follows.
\begin{verbatim}
name <- function(input)
{
body of function
return(output) # we may write output instead
}
\end{verbatim}
The input may be array, data frame, list, matrix, single value, vector, or a combination of these. The body of function is a set of \verb+R+ commands in order to achieve the goal (output). If the output has just one part, then user is free to write \verb+output+ rather than \verb+return(output)+ for returning the output of function. It sounds worthwhile to note that if the body of function consists of only one sentence then the curly brackets (braces) and \verb+return(output)+ can be removed in order to relax the created function.
Table \ref{R-primitive-built-in-function-1} lists with some popular primitive built-in functions. 
\vspace{0.5cm}
\begin{table}[!h]
\centering
\caption{Some primitive built-in functions in {\texttt{R}}.}
\vspace*{-3mm}
\begin{tabular}{ll}%{\textwidth}{>{\hsize=.10\hsize}X>{\hsize=.95\hsize}X}
\hline
Function& output/task\\
\hline
\verb+abs(x)+ & absolute value of \verb+x+.\\
\verb+atan(x)+& inverse of trigonometric function \verb+tan(x)+.\\
\verb+exp(x)+ & natural exponential of \verb+x+.\\
\verb+Im(x)+ & imaginary part of \verb+x+.\\
\verb+log(x)+ & logarithm of \verb+x+ based on number \verb+y=exp(1)+ by default.\\
\verb+lgamma(x)+ & logarithm of $\Gamma($\verb+x+$).$\\
\verb+Re(x)+ & real part of complex number \verb+x+.\\
\verb+sign(x)+ & takes on values -1, 0, or 1, if \verb+x<0+, \verb+x=0+, or, \verb+x>0+, respectively.\\
\verb+sqrt(x)+ & square root of \verb+x+.\\
\verb+x^y+ & \verb+x+ to power \verb+y+ (often symbol \verb+^+ is created by \keystroke{Shift}+ \keystroke{6}).\\
\hline
\end{tabular}
\label{R-primitive-built-in-function-1}
\end{table}
%\vspace{0.5cm}
%\begin{table}[!h]
%\centering
%\caption{Some primitive built-in functions in {\texttt{R}}.}
%\vspace*{-3mm}
%\begin{tabular}{ll}%{\textwidth}{>{\hsize=.10\hsize}X>{\hsize=.95\hsize}X}
%\hline
%Function& output/task\\
%\hline
%\verb+abs(x)+ & absolute value of \verb+x+\\
%\verb+atan(x)+& inverse of trigonometric function \verb+tan(.)+\\
%\verb+exp(x)+ & natural exponential of \verb+x+\\
%\verb+Im(x)+ & imaginary part of \verb+x+\\
%\verb+log(x)+ & logarithm of \verb+x+ based on number \verb+y=exp(1)+ by default \\
%\verb+lgamma(x)+ & logarithm of $\Gamma($\verb+x+$)$\\
%&otherwise one must change \verb+y+ to another positive value\\
%\verb+Re(x)+ & real part of \verb+x+\\
%\verb+sign(x)+ & takes on values -1, 0, or 1, if \verb+x<0+, \verb+x=0+, or, \verb+x>0+, respectively\\
%\verb+sqrt(x)+ & square root of \verb+x+\\
%\verb+x^y+ & \verb+x+ to power \verb+y+ (often symbol \verb+^+ is created by \keystroke{Shift}+ \keystroke{6})\\
%\hline
%\end{tabular}
%\label{R-nonprimitive-built-in-function-1}
%\end{table}
%\vspace{0.5cm}
\par For example, if we want repeatedly standardize a numeric vector, then we may create a function called \verb+my.scale+ for this purpose as follows.
{\color{red}{
\begin{verbatim}
R> my.scale <- function(v) (v - mean(v) )/sd(v)
\end{verbatim}
}}
Of course, we do not suggest to use the user-defined function \verb+my.scale+ for practical purposes since it built-in counterpart exists. Hence, the user-defined are mainly developed to perform those tasks for which the built in function is not applicable. As another example, suppose we would like to create a function that computes the sign of a given real value. Recall that we have just computed the sign of vector \verb+sex+. A little more work required to create function called \verb+my.sign+ for computing sign of given vector \verb+x+. The function \verb+my.sign+ is given as follows.
\begin{lstlisting}[style=deltaj]
R> my.sign <- function(x)
+{
+ n <- length(x)
+ y <- rep(0, n) # a numeric vector called y of size n as output
+ for(i in 1:n)
+ {
+	 	if( x[i] == 0) 
+ 	{
+ 		y[i] <- 0
+ 	}else if (x[i] > 0){
+ 		y[i] <- 1
+   	 }else{
+      		y[i] <- -1
+ 		}
+	}
+ return(y)
+}
\end{lstlisting} 
It is worthwhile to note that each user-defined function must be checked to see does it work well or not, and next priorities are speed, memory usage, and etc.  
\subsection{Built-in function for character}
Working with character or string is also important. The built-in functions have been developed effectively to deal with such type of variables in \verb+R+. Herein, among them, we briefly introduce \verb+grep+, \verb+grepl+, \verb+gregexpr+, \verb+letters+, \verb+regexec+, \verb+regexpr+, \verb+sub+, \verb+gsub+, \verb+substr+, \verb+strsplit+, \verb+toupper+, \verb+tolower+. These functions, in general, have been developed for finding some given pattern within a non-numeric vector. Among aforementioned commands, we are willing to discuss briefly the commands \verb+grep+ and \verb+substr+. Details about arguments of \verb+grep+ are given in Table \ref{R-function-character-grep-1}.
\vspace{5mm}
\begin{table}[!h]
\centering
\caption{Details about main arguments of {\texttt{grep}} in {\texttt{R}}.}
\vspace*{-3mm}
\begin{tabular}{ll}%{10mm}{>{\hsize=.10\hsize}X>{\hsize=.95\hsize}X}
\hline
Argument& output/task\\
\hline
\multicolumn{1}{m{2cm}}{{\texttt{pattern}}}&\multicolumn{1}{m{12cm}}{
a sequence of character(s) that we want to find it within some given non-numeric vector.}\\
\verb+x+& given non-numeric vector.\\
\multicolumn{1}{m{2cm}}{{\texttt{ignore.case}}}&\multicolumn{1}{m{12cm}}{
{\texttt{logical}}: if {\texttt{TRUE}}, the matched sequence will be found regardless of being upper or lowercase.}\\
\multicolumn{1}{m{2cm}}{{\texttt{value}}}&\multicolumn{1}{m{12cm}}{
{\texttt{logical}}: if {\texttt{TRUE}}, then it returns the vector of matched sequence; otherwise, the indices vector is returned.}\\
\hline
\end{tabular}
\label{R-function-character-grep-1}
\end{table}
\vspace{5mm}
Suppose we would like to find pattern \verb+'Bd'+ inside the non-numeric atomic vector \verb+x <- c('abcd','Bdcd','abcdabcd','bd2')+. Hence, we define
{\color{red}{
\begin{verbatim}
R> x <- c('abcd','Bdcd','abcdabcd','bd2'); pattern <- 'Bd'
\end{verbatim}
}}
We run command {\texttt{grep(...)}} for four different combinations of its arguments as follows. 
{\color{red}{
\begin{verbatim}
R>grep(pattern,x,ignore.case=T,value=T) R>grep(pattern,x,ignore.case=T,value=F)
\end{verbatim}
}}
\vspace{-4mm}
{\color{blue}{
\begin{verbatim}
[1] "Bdcd" "bd2"                        [1] 2 4
\end{verbatim}
}}
\vspace{-4mm}
{\color{red}{
\begin{verbatim}
R>grep(pattern,x,ignore.case=F,value=T) R>grep(pattern,x,ignore.case=F,value=F)
\end{verbatim}
}}
\vspace{-4mm}
{\color{blue}{
\begin{verbatim}
[1] "Bdcd"                              [1] 2
\end{verbatim}
}}
Details about arguments of function \verb+substr(...)+ is given in Table \ref{R-function-character-substr-1}. This command is mainly used to extract or replace substrings in a given non-numeric vector.
\vspace{5mm}
\begin{table}[!h]
\centering
\caption{Arguments of {\texttt{substr(...)}} in {\texttt{R}}.}
\vspace*{-3mm}
\begin{tabular}{ll}%{10mm}{>{\hsize=.10\hsize}X>{\hsize=.95\hsize}X}
\hline
Argument&output/task\\
\hline
\verb+x+& a non-numeric vector.\\
\verb+start+ & first element to be extracted (or replaced).\\
\verb+stop+& last element to be extracted (or replaced).\\
\verb+value+& a string to be replaced.\\
\hline
\end{tabular}
\label{R-function-character-substr-1}
\end{table}
For example, if we want to substitute the string \verb+'sm'+ into the second and third places of elements of vector \verb+x+ defined earlier, or further, extract the second and third elements of vector \verb+x+. The corresponding commands and outputs are given as follows.
\begin{table}[!h]
\begin{tabular}{ll}%{10mm}{>{\hsize=.10\hsize}X>{\hsize=.95\hsize}X}
&\\
{\texttt{R}} code & output\\
\hline
{\color{red}{{\texttt{R>x<-c('abcd','Bdcd','abcdabcd','bd2')}}}}&\\
{\color{red}{{\texttt{R>value<-'sm'; substr(x,2,3)<-value}}}}&\\
{\color{red}{{\texttt{R>x}}}}&{\color{blue}{{\texttt{[1] "asmd" "Bsmd" "asmdabcd" "bsm"}}}}\\
&\\
{\color{red}{{\texttt{R>x<-c('abcd','Bdcd','abcdabcd','bd2')}}}}&\\
{\color{red}{\texttt{R>substr(x,2,3)}}}&{\color{blue}{{\texttt{[1] "bc" "dc" "bc" "d2"}}}}\\
\hline
\end{tabular}
\end{table}
In a further example, e are willing to separate digits of a given number. To this end, we may use \verb+gsub+ command to separate digits of 2024 as follows.
\begin{table}[!h]
\begin{tabular}{ll}%{10mm}{>{\hsize=.10\hsize}X>{\hsize=.95\hsize}X}
&\\
{\texttt{R}} code & output\\
\hline
{\color{red}{\verb+R> val1 <- as.character(2024)+}}&{\color{blue}{\verb+[1]  2   0   2   4+}}\\
{\color{red}{\verb+R> val2 <- gsub("(.)", "\\1 ", val1)+}}&\\
{\color{red}{\verb+R> val3 <- strsplit(val2, " ")[[1]]+}}&\\
{\color{red}{\verb+R> val4 <- as.numeric(val3); val4+}}&\\
\hline
\end{tabular}
\end{table}
%%%%%%%%%%%%%%%%%%%%%%%%%%%%%
%%%%%%%%%%%%%%%%%%%%%%%%%%%%%
\section{External R package}
\subsection{External R package initialization}
As mentioned before, attaching \verb+R+ packages is privileged to users for implementing the advanced or more technical statistical methods. Herein, we would like to deal with utilizing an external \verb+R+ package. To this end, the first step is prepare the package's documentation (or Reference manual) that is given in pdf available at \url{https://cran.r-project.org/web/packages/pakage name}. The documentation gives necessary information on \verb+R+ package. There are several features about each \verb+R+ package that, for example, may be seen at \url{https://cran.r-project.org/doc/manuals/r-devel/R-exts.html#The-DESCRIPTION-file}. These features must be investigated whenever user wants to install and then use a package, among them we mention the following.  
\begin{enumerate}[label=\roman*.]
\item {\texttt{Depends}}: This field gives a numbers of comma-separated package names that the desired package depends on. The prerequisite packages will be installed while the current package is installed. This field also may specify a dependence on a certain version of \verb+R+. For instance, if the desired package works only with version 4 or later, we may see {\texttt{R (>= 4.0)}} in the ``Depends'' field.
\item {\texttt{Suggests}}: This field has the same characteristics as field ``Depends'' does. But, the list of packages in front of this field are not necessary to install. The prerequisite package(s) are just used for implementing examples, tests existing in the desired package. 
\item {\texttt{Package source}}: This fields contains the compressed source code of the desired package. For example, the compressed source code of package \verb+nortest+ is given by \verb+nortest_1.0-4.tar.gz+ in this field.
\end{enumerate}
\subsection{External R package installation}
In this part, we are willing to deal with installing an external \verb+R+package. To this end, we select the \verb+nortest+ available at \url{https://cran.r-project.org/package=nortest}. The first step for installing an external \verb+R+package is study the documentation file of \verb+nortest+ as mentioned before. If all features hold true, then we use the steps given by the following for using this package for all practical purposes.
{\color{red}{
\begin{verbatim}
R > address = "E:/my R code"
R > install.packages("nortest", lib=address,
+   rpos="https://cran.r-project.org/package=nortest")
\end{verbatim}
}}
Running above two lines there appears a box entitles ``Secure CRAN mirrors'' that lists all available mirrors for installing the desired package. We suggest to select the nearest country, if you country is not listed there, and then approving to continue the installing process. Depending on the desired package's dependency on other packages, the installation process may take several seconds or minutes. Finally, a blue-colored message appears in console session indicating that the installation process is accomplished successfully. After installing, we must call the desired package through 
{\color{red}{
\begin{verbatim}
R > library(nortest) # or require(nortest)
\end{verbatim}
}}
It is worth to mention that the argument \verb+address+ is quite optional and determined by user. The argument \verb+rpos+ is obtained from the last line in the package's page at CRAN. The above process must be done for using all external packages. Sometimes, due to some technical problems, user is not able to install the desired package. One way to avoid such a issue is to install the desired package manually. To this end, user just needs to download the package's compressed source code available at {\texttt{Package source}} part. Then, we determine the address of the package source through
{\color{red}{
\begin{verbatim}
install.packages(file.choose( ), repos = NULL, type="source")
\end{verbatim}
}}
to complete the installation process.
\subsection{Normality test}
Testing the normality of a data set is a vital statistical hypothesis in many parametric and non-parametric statistical methods
\citep{thode2002testing,stephens2017tests}. Though implementing the goodness-of-fit tests does not need 
an in-depth \verb+R+ programming skill, but we would prefer to use some external package for this purpose rather to create our user-defined function. After installing package \verb+nortest+, having a look at topics, we can see that function \verb+ad.test+ has been developed for accomplishing the Anderson-Darling test of normality. There is more detailed information on this function in the pertinent page. Also, the package \verb+nortest+ allows user for running more normality test such as Cramer-von Mises, Lilliefors (or Kolmogorov-Smirnov), Pearson chi-square, and Shapiro-Wilk. In what follows, we give the pertinent commands and outputs when these goodness-of-fit tests are applied to a data set of size $n=100$ randomly generated from Gaussian distribution with mean 5 and standard deviation 3. Each test consists of the {\it{test statistic}} and corresponding $p$-value. Since all p-values are greater than significance level, $\alpha=0.05$ say, then {\it{based on the sample evidence}}, we decide to accept the null hypothesis of $H_{0}$: ``data follow Gaussian distribution'' against $H_{1}$: ``data do not follow Gaussian distribution'' at this level. 
\vspace{5mm}
\begin{table}[!h]
\begin{tabular}{ll}%{10mm}{>{\hsize=.10\hsize}X>{\hsize=.95\hsize}X}
{\color{red}{\verb+R> set.seed(20241204)+}}&{\color{red}{\verb+R> ad.test(x)+}}\\
{\color{red}{\verb+R> x<-rnorm(100,mean=5,sd=3)+}}&{\small{{\color{blue}{\verb+Anderson-Darling normality test+}}}}\\
{\color{red}{\verb+R> #we use rnorm() to generates+}}&{\small{{\color{blue}{\verb+data:  x+}}}}\\
{\color{red}{\verb+R> #from Gaussian distribution+}}&{\small{{\color{blue}{\verb+A = 0.3784, p-value = 0.4007+}}}}\\
{\color{red}{\verb++}}&{\color{blue}{\verb++}}\\
{\color{red}{\verb+R> cvm.test(x)+}}&{\color{red}{\verb+R> lillie.test(x)+}}\\
{\small{{\color{blue}{\verb+Cramer-von Mises normality test+}}}}&{\small{{\color{blue}{\verb+Lilliefors (Kolmogorov-Smirnov) normality test+}}}}\\
{\small{{\color{blue}{\verb+data:  x+}}}}&{\small{{\color{blue}{\verb+data:  x+}}}}\\
{\small{{\color{blue}{\verb+W = 0.055022, p-value = 0.4364+}}}}&{\small{{\color{blue}{\verb+D = 0.063051, p-value = 0.4239+}}}}\\
{\small{{\color{red}{\verb++}}}}&{\small{{\color{blue}{\verb++}}}}\\
{\small{{\color{red}{\verb+R> pearson.test(x)+}}}}&{\color{red}{\verb+R> sf.test(x)+}}\\
{\small{{\color{blue}{\verb+Pearson chi-square normality test+}}}}&{\small{{\color{blue}{\verb+Shapiro-Francia normality test+}}}}\\
{\small{{\color{blue}{\verb+data:  x+}}}}&{\small{{\color{blue}{\verb+data:  x+}}}}\\
{\small{{\color{blue}{\verb+P = 9.72, p-value = 0.4654+}}}}&{\small{{\color{blue}{\verb+W = 0.98913, p-value = 0.5087+}}}}\\
\end{tabular}
\end{table}
\vspace{5mm}
\par
\begin{figure}[!h]
\center
\includegraphics[width=55mm,height=55mm]{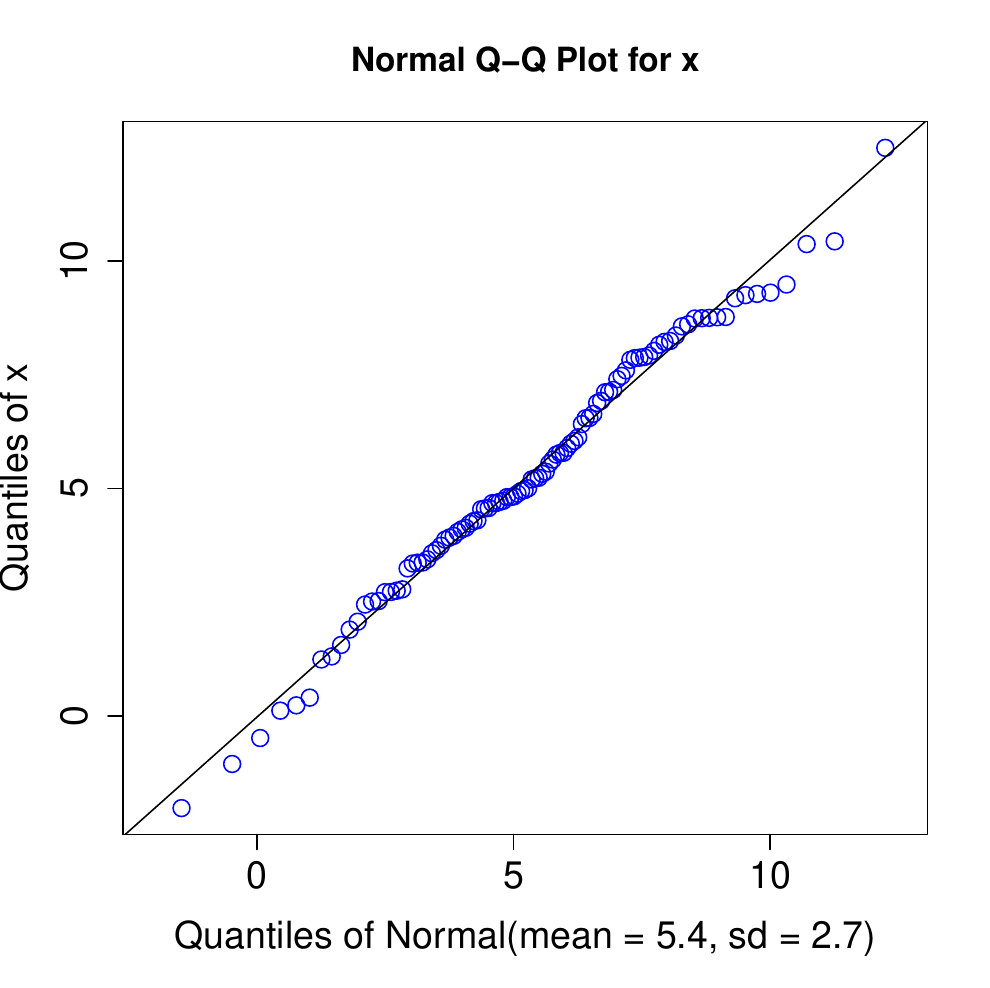}
\includegraphics[width=55mm,height=55mm]{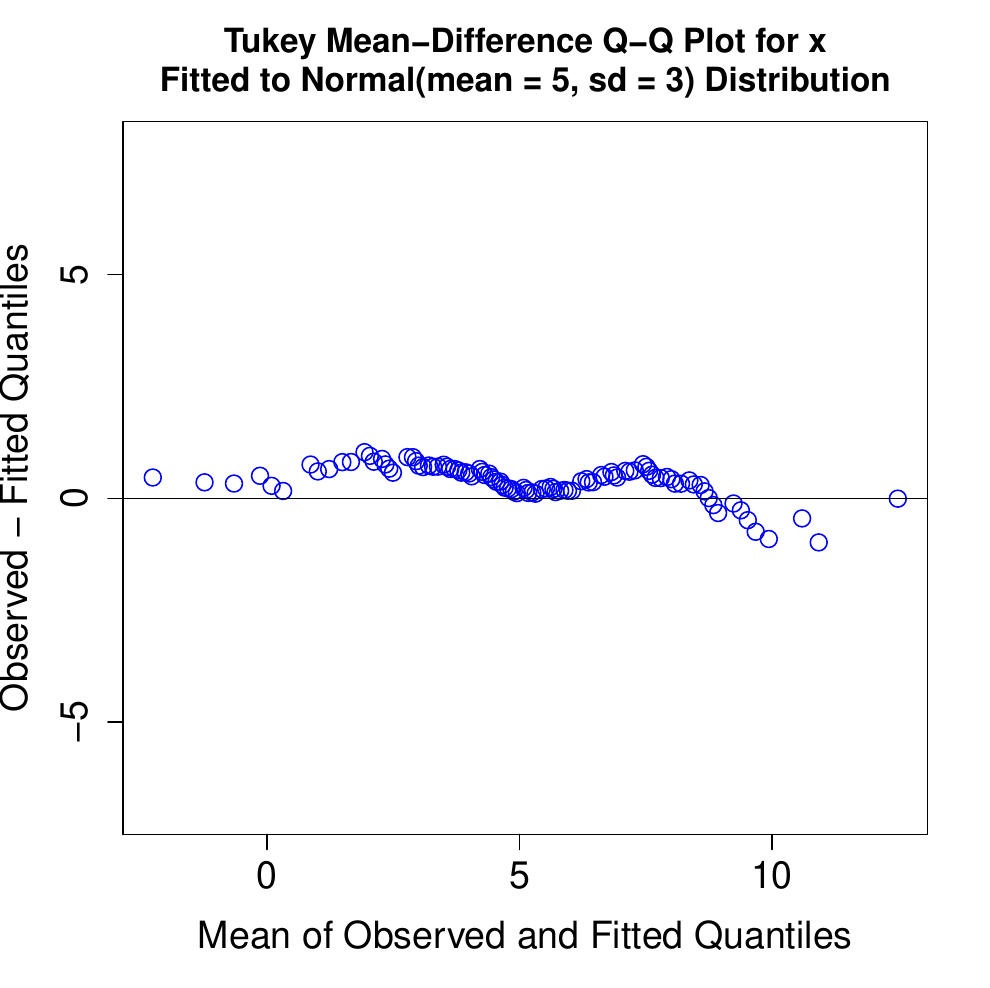}
\caption{(left subfigure): q-q plot and (right subfigure): Tukey mean-difference q-q plot.}
\label{plot-qqplot-R-introduction}
\end{figure}

 More diagnostics for assessing the normality are discussed through \verb+R+ package \verb+EnvStats+ available at \url{https://cran.r-project.org/package=EnvStats}. It is worthwhile to note that this package needs an \verb+R(>= 3.5.0)+ and imports packages \verb+MASS+, \verb+ggplot2+, \verb+nortest+. Hence, installing \verb+EnvStats+, we should take care about the imported packages and their dependencies. The package \verb+EnvStats+ is enable to draw quantile-quantile plot (q-q plot). Suppose we would like to display the q-q plot of the data already generated from Gaussian distribution. To this end, we run the commands as
\vspace{2mm}
{\color{red}{
\begin{verbatim}
R> library(EnvStats); set.seed(20241204); x<-rnorm(100,mean=5,sd=3)
R> qqPlot(x, dist = "norm", estimate.params = T, digits = 2, 
+ points.col = "blue",  add.line = TRUE)
R> qqPlot(x, dist = "norm", param.list=list(mean=5, sd=3), 
+ estimate.params = F, digits = 2, points.col = "blue",  add.line = 
+ TRUE, plot.type="Tukey Mean-Difference Q-Q")
\end{verbatim}
}}
to display graphical summaries depicted in Figure \ref{plot-qqplot-R-introduction}.
It is worthwhile to say that for producing the normal q-q plot (left subfigure), we set argument \verb+estimate.params=T+ and hence the package automatically sets \verb+param.list=list(mean=mu, sd=sigma)+ in which \verb+mu=5.4+ and \verb+sigma=2.7+ are the average and standard deviation of sample evidence \verb+x+. For producing the Tukey mean-difference q-q plot (right subfigure), we set \verb+estimate.params=F+ and hence we need to put the true mean and standard deviation using argument \verb+param.list=list(mean=5, sd=3)+. 
\section{Regression analysis in R}
The simple linear, multiple linear, and non-linear regression analysis are among applied topics is statistics with widespread use in other fields of study. Herein, we are willing to discuss how user can implement the regression analysis in \verb+R+. 
\subsection{Linear regression}
Suppose the amount of some variable that is known in the literature as {\it{response}} or {\it{dependent}} variable depends {\it{linearly}} on one (simple case) or more (multiple case) variable(s) namely {\it{independent}} variable(s) or {\it{covariate}}. The linear regression enables us to estimate the value of dependent variable $y$ when a level of independent $x$ is given. The set of dependent and independent variables should be provided in a data frame structure. For practical purposes, we will
focus on {\it{built-in}} \verb+women+ data that consists of two variables average heights and weights for American women. Hence, for detecting how the dependent (weight) and independent (height) variables are linearly correlated, we proceed as follows.
 {\color{red}{
\begin{verbatim}
R> data(women)
R> dat <- data.frame(y=women$weight, x=women$height)
R> model <- lm(y~x, data=dat) # lm accounts for linear model 
R> out <- summary(model)
\end{verbatim}
}} 
The summary of regression analysis such as estimated coefficients, estimated residuals, etc., are used for further analysis. Table \ref{Simple-Regression-model-1} lists some commands to access more detailed summary related to the simple linear regression analysis. 
\vspace{5mm}
\begin{table}[!h]
\centering
\caption{Details about output of {\texttt{lm(...)}} in {\texttt{R}}.}
\vspace*{-3mm}
\begin{tabular}{ll}%{\textwidth}{>{\hsize=.10\hsize}X>{\hsize=.95\hsize}X}
\hline
Command& output\\
\hline
 \verb+out$coefficients[1,1]+ & estimated intercept.\\
 \verb+out$coefficients[2,1]+ & estimated slope.\\
 \verb+out$coefficients[1,2]+ & standard error of estimated intercept.\\
 \verb+out$coefficients[2,2]+ & standard error of estimated slope.\\
{\texttt{out\symbol{36}coefficients[1,4]}}&p-value corresponds to null. hypothesis of intercept is zero.\\
{\texttt{out\symbol{36}coefficients[2,4]}}&p-value corresponds to null hypothesis of slope is zero.\\
 \verb+out$sigma+ & estimated response variable standard error.\\
 \verb+out$r.squared+ & coefficient of determination $R^{2}$.\\
 \verb+confint(model,level=0.95)+& 0.95\% confidence interval for regression coefficients.\\
 \verb+out$residuals+& residuals of fitted model.\\
 \verb+anova(model)+& analysis of variance table.\\
 \multicolumn{1}{m{3cm}}{{\texttt{predict(object,new,...)}}}&\multicolumn{1}{m{10cm}}{
object is {\texttt{lm(y x)}} and {\texttt{new}} is a single or vector of values for which a confidence ({\texttt{"c"}}) or prediction ({\texttt{"p"}}) interval at
{\texttt{level=0.95}} is constructed.}\\
\hline
\end{tabular}
\label{Simple-Regression-model-1}
\end{table}
\vspace{5mm}
Figure \ref{plot-qqplot-R-introduction} displays our user-defined function created for depicting the dependent and independent scatterplot. The pertinent \verb+R+ function called \verb+reg_plot(x,y,...)+ is given as follows.
\begin{lstlisting}[style=deltaj]
R> reg_plot <- function(x,y,PI=FALSE, interval=c("c","p"),...)
+ { # ''c'' for confidence and ''p'' for prediction interval
+	model <- lm(y~x)
+	out <- summary(model)
+	b0 <- round(out$coefficien[1,1],4)
+	b1 <- round(out$coefficien[2,1],4)
+	note1 <- paste("Y=", b0, "+", b1, "X")
+	note2 <- paste("R.squared=",round(out$r.squared,4)) 
+  n <- length(x)
+	new <- data.frame(x=seq(min(x), max(x), length =n))
+	if(PI == FALSE)
+	{
+		plot(x,y,...)
+		curve(b0 + b1*x, add=T)
+		lines(new$x, b0 + b1*new$x)
+		text(mean(x), max(y), note1)
+		text(mean(x), quantile(y, 0.90), note2)
+	}else if(PI == TRUE){
+		pred <- predict(model, new, interval = interval)
+		plot(x,y,...)
+		x0<-c(new$x[1:n], new$x[n:1])
+		y0<-c(pred[,2], pred[n:1,3])
+		polygon(x0,y0,col="steelblue", border="red")
+		text(mean(x), max(y), note1)
+		text(mean(x), quantile(y, 0.90), note2)
+		points(x,y,...)
+		curve(b0 + b1*x, add=T)
+	}
+}
\end{lstlisting}
\begin{figure}[!h]
\center
\includegraphics[width=55mm,height=55mm]{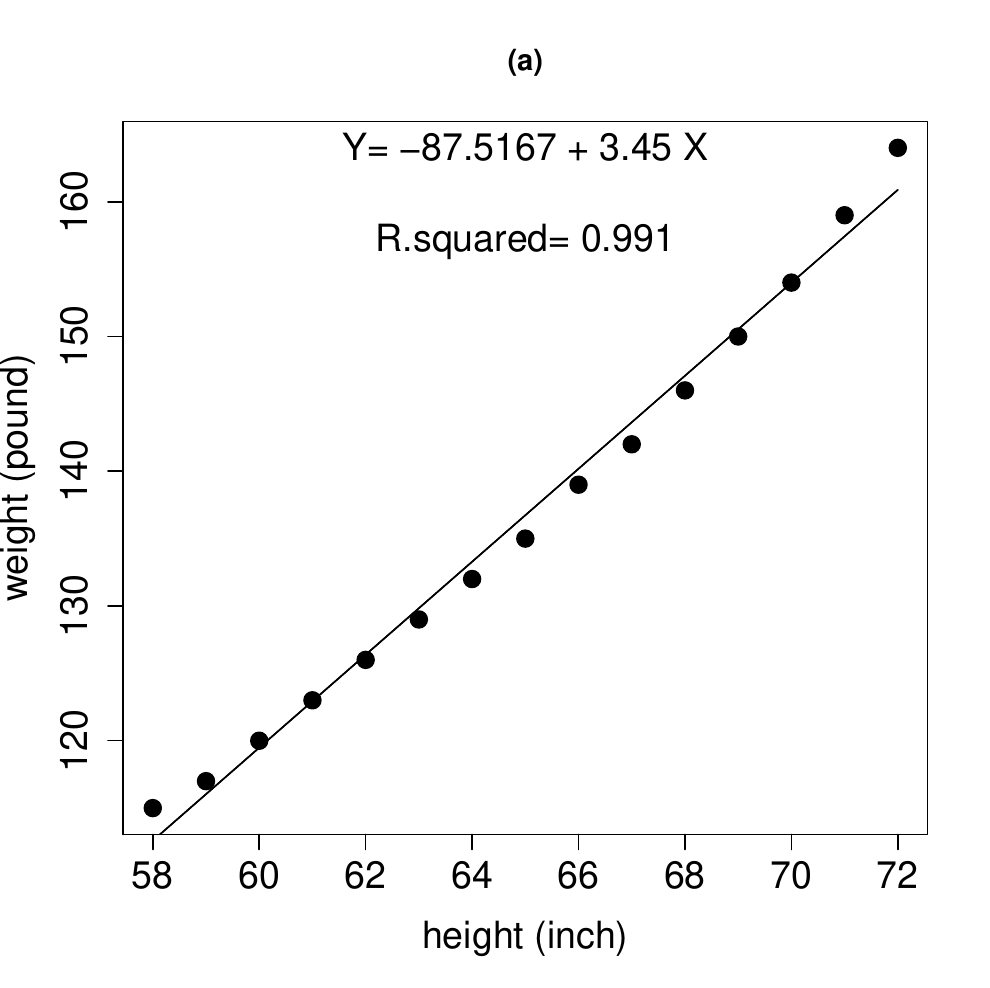}\includegraphics[width=55mm,height=55mm]{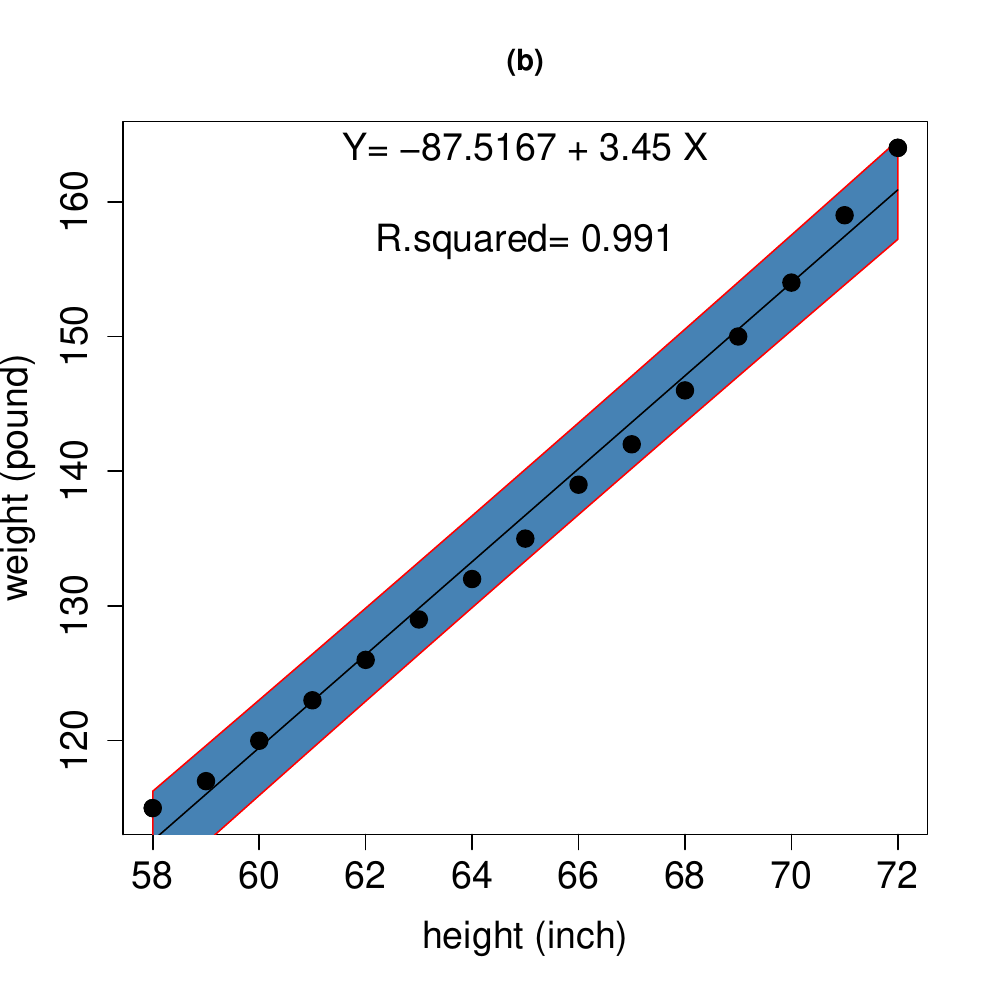}
\caption{(a): Scatteplot of {\texttt{women}} data: (a) fitted regression line and (b): prediction interval.
}
\label{plot-simple-regression-R-introduction}
\end{figure}
We run the \verb+R+ commands as follows for producing Figure \ref{plot-simple-regression-R-introduction}.
{\color{red}{
\begin{verbatim}
R> data(women)
R> x <- women$height; y <- women$weight
R> reg_plot(x,y,PI=F,interval="p",pch=19,cex=1.5, 
+ cex.lab=1.5,cex.axis=1.5,main="(a)",
+ xlab="height (inch)",ylab="weight (pound)")# for subfigure (a)
R> reg_plot(x,y,PI=T,interval="p",pch=19,cex=1.5, 
+ cex.lab=1.5,cex.axis=1.5,main="(b)",
+ xlab="height (inch)",ylab="weight (pound)")# for subfigure (b)
\end{verbatim}
}}
\subsection{Non-linear regression}
In some applications relation between dependent and independent variables is not linear. There are infinite number of non-linear mathematical models for fitting to data. Herein, we would like to deal with the growth model. An application of such models can be found in forestry for describing the relation between tree's height (h in m/foot) and its diameter at breast height (dbh in cm/inch) \cite{teimouri2020forestfit}. Among the most commonly used growth models, we focus on Weibull model whose relation is given by \cite{yang1978potential}:
\begin{align}\label{Weibull-growth-model-1}
h=1.3+b1\bigl[1-e^{-b2 d^{b3}}\bigr],
\end{align}
where $b1>0$, $b2>0$ and $b3>0$ are the model's parameters \footnote{ The constant 1.3 is model's intercept since the experimenter's height at breast level is on the average 1.3 m.}. There are different ways for estimating the model's parameters based on the nonlinear least squares method, among them herein, we suggest the use of function \verb+nls(formula,data,start,...)+. Detail on main arguments of the former function is shown by Table \ref{Nonlinear-Regression-model-1}.
\vspace{5mm}
\begin{table}[!h]
\centering
\caption{Details about main arguments of {\texttt{nls(...)}} in {\texttt{R}}.}
\vspace*{-3mm}
\begin{tabular}{ll}%{10mm}{>{\hsize=.10\hsize}X>{\hsize=.95\hsize}X}
\hline
Argument& output/task\\
\hline
 \verb+formula+& mathematical expression such as formula (\ref{Weibull-growth-model-1}).\\
 \verb+data+& data frame consisting of dependent and independent data.\\
 \verb+start+& vector of starting (initial) values of the model's parameters.\\
\hline
\end{tabular}
\label{Nonlinear-Regression-model-1}
\end{table}
\vspace{5mm}
\par The \verb+nls(...)+ uses some iterative algorithms to estimate the parameters. But, the performance (being convergent or not) of these algorithms in general is sensitive on the initial values. Not surprisingly, if starting values are close to the global maximum, then algorithm converges soon. Otherwise, the algorithm will not converge and \verb+nls(...)+ fails to estimate model's parameters. One way to deal this practical issue is the use of function \verb+tryCatch(...)+ that has mainly four parts described as Table \ref{TryCatch-Regression-model-1}.
\vspace{5mm}
\begin{table}[!h]
\centering
\caption{Details about main arguments of {\texttt{tryCatch(...)}} in {\texttt{R}}.}
\vspace*{-3mm}
\begin{tabular}{ll}%{10mm}{>{\hsize=.10\hsize}X>{\hsize=.95\hsize}X}
\hline
Argument& output/task\\
\hline
\verb+expr+& the task to be accomplished.\\
\multicolumn{1}{m{1.5cm}}{{\texttt{error}}}&\multicolumn{1}{m{12.5cm}}{
the task to be done if the given task in part {\texttt{expr}} if an error is caught.}\\
\multicolumn{1}{m{1.5cm}}{{\texttt{warning}}}&\multicolumn{1}{m{12.5cm}}{
the task to be done if accomplishing the given task in part {\texttt{expr}} does
not fail, but a warning is caught.}\\
\multicolumn{1}{m{1.5cm}}{{\texttt{finally}}}&\multicolumn{1}{m{12.5cm}}{
the task to be done regardless of accomplishing the given task in part {\texttt{expr}} is successful or not successful.}\\
\hline
\end{tabular}
\label{TryCatch-Regression-model-1}
\end{table}
\vspace{5mm}
In what follows, we run function \verb+nls(...)+ for \verb+DBH+ data involved in package \verb+ForestFit+ under two scenarios including: i) true parameters are assumed to be known and function \verb+tryCatch(...)+ is not employed and ii) true parameters are assumed to be unknown and function \verb+tryCatch(...)+ is employed. We deal with these two scenarios as follows.
\begin{enumerate}[label=\roman*.]
\item {\bf{First scenario}}: We assume that true parameters are $b1=25.011$, $ b2=0.0167$, and $b3=1.1561$. Using this setting, we use the command
{\color{red}{
\begin{verbatim}
R> library(ForestFit); data(DBH)
R> d <- DBH[DBH[,1]==55,10]; h <- DBH[DBH[,1]==55,11]
R> start <- c(25.011,0.016, 1.156)
R> relation <- as.formula(h~1.3+b1*(1-exp(-b2*d^b3)))
R> out <- summary( nls( relation, data=data.frame(h=h, d=d), 
+ start = list( b1 = start[1], b2 = start[2], b3 = start[3] )) )
R> out
\end{verbatim}
}} 
to see the output as follows.
{\color{blue}{
\begin{verbatim}
Formula: h ~ 1.3 + b1 * (1 - exp(-b2 * d^b3))
Parameters:
    Estimate Std. Error t value Pr(>|t|)    
b1 25.011868   2.400674  10.419 1.26e-14 ***
b2  0.016704   0.004924   3.392  0.00129 ** 
b3  1.156198   0.124108   9.316 6.63e-13 ***
---
Signif. codes:  0 ‘***’ 0.001 ‘**’ 0.01 ‘*’ 0.05 ‘.’ 0.1 ‘ ’ 1
Residual standard error: 1.558 on 55 degrees of freedom
Number of iterations to convergence: 4 
Achieved convergence tolerance: 8.725e-06
\end{verbatim}
}} 
\item {\bf{Second scenario}}: We let the model's parameter are not precisely known. Hence, we use the function \verb+tryCatch(...)+ to implement the \verb+nls(...)+ function. The pertinent \verb+R+ code and output are given by the following.
{\color{red}{
\begin{verbatim}
R> library(ForestFit); data(DBH)
R> d <- DBH[DBH[,1]==55,10]
R> h <- DBH[DBH[,1]==55,11]
R> start <- c(1, 1, 1)
R> relation<-as.formula(h~1.3+b1*(1-exp(-b2*d^b3)))
R> f<-function(x, par)
+  {
+    b1<-par[1]; b2<-par[2]; b3<-par[3]; 1.3+b1*(1-exp(-b2*x^b3))
+  }
R> it <- 1; i <- 0
R> while(it <= 1)
+ {
+ r1<-runif(1,max(0,start[1]-10),start[1]+10) # b1 is selected randomly  
+ r2<-runif(1,max(0,start[2]-10),start[2]+10) # b2 is selected randomly 
+ r3<-runif(1,max(0,start[3]-10),start[3]+10) # b3 is selected randomly 
+ out <- tryCatch(summary( nls( relation, data=data.frame(h=h, d=d), 
+ start = list( b1 = r1, b2 = r2, b3 = r3 ) ) ),
+	error=function(e)( "fail" )  )
+   if( out[1] == "fail" )
+   {
+      it <- 1
+   }else{
+      it <- 2
+   }
+ i <- i + 1
+}
R> out
\end{verbatim}
}} 
As it may be seen, we set the starting value as \verb+start=c(1,1,1)+ in second scenario and function \verb+tryCatch(...)+ allows to test different starting values during each test within \verb+while(...)+ function. As expected, the estimated regression coefficients under both both scenario are the same. We further applied the function  \verb+tryCatch(...)+ to estimate three parameters of the logistic growth curve. It suffices to consider as follows.
{\color{red}{
\begin{verbatim}
R> relation<-as.formula(h~1.3+b1/(1+b2*exp(-b3*d)))
R> f<-function(x, par)
+ {
+    b1<-par[1]; b2<-par[2]; b3<-par[3]; 1.3+b1/(1+b2*exp(-b3*x))
+ }
\end{verbatim}
}} 
Figure \ref{plot-R-introduction-Weibull-logistic-growth-model-trycatch} displays the scatterplot of height and diameter variables while the estimated Weibull (left subfigure) and logistic (right subfigure) growth curves are added to the plot. It is worth to note that the output corresponds to fitting logistic growth curve is as follows.
{\color{blue}{
\begin{verbatim}
Formula: h ~ 1.3 + b1/(1 + b2 * exp(-b3 * d))
Parameters:
    Estimate Std. Error t value Pr(>|t|)    
b1 21.688329   0.814105  26.641  < 2e-16 ***
b2  6.878022   1.212616   5.672 5.42e-07 ***
b3  0.086529   0.009079   9.530 3.04e-13 ***
---
Signif. codes:  0 ‘***’ 0.001 ‘**’ 0.01 ‘*’ 0.05 ‘.’ 0.1 ‘ ’ 1
Residual standard error: 1.802 on 55 degrees of freedom
Number of iterations to convergence: 9 
Achieved convergence tolerance: 6.395e-06
\end{verbatim}
}}
\begin{figure}[!h]
\center
\includegraphics[width=55mm,height=55mm]{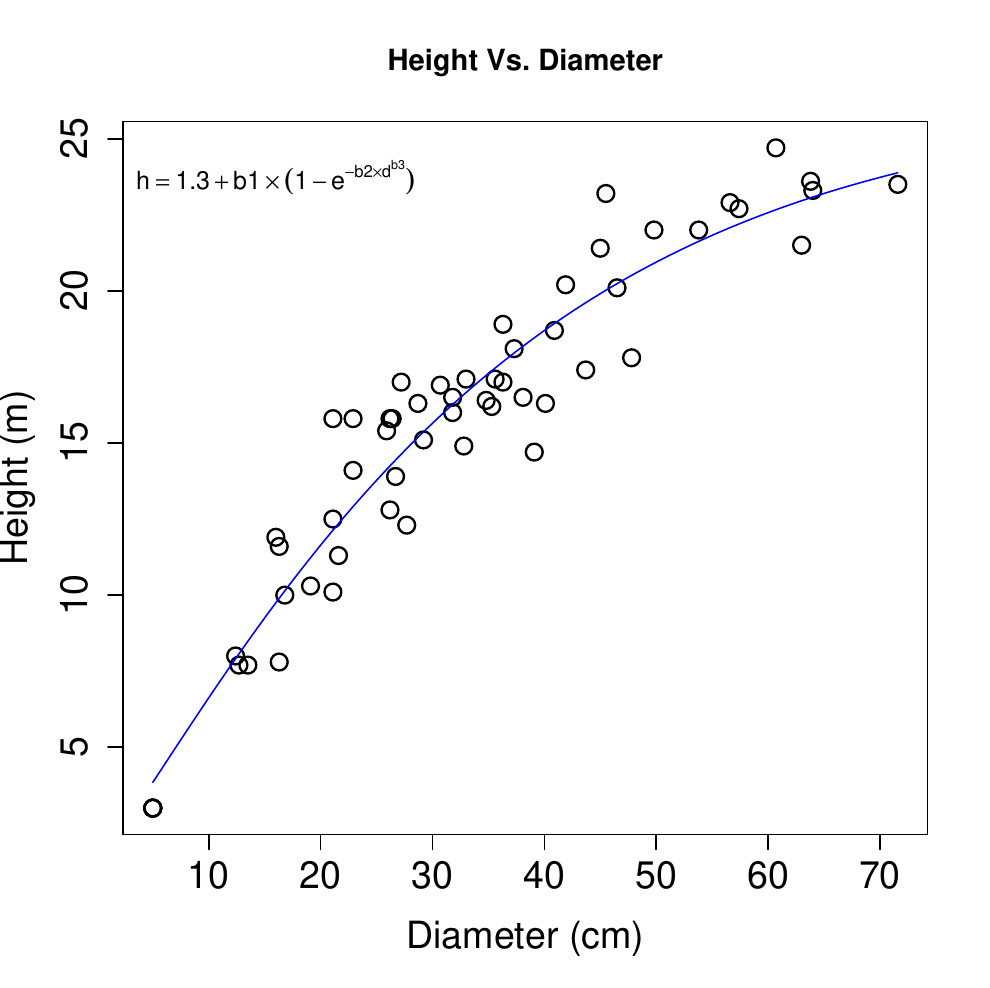}
\includegraphics[width=55mm,height=55mm]{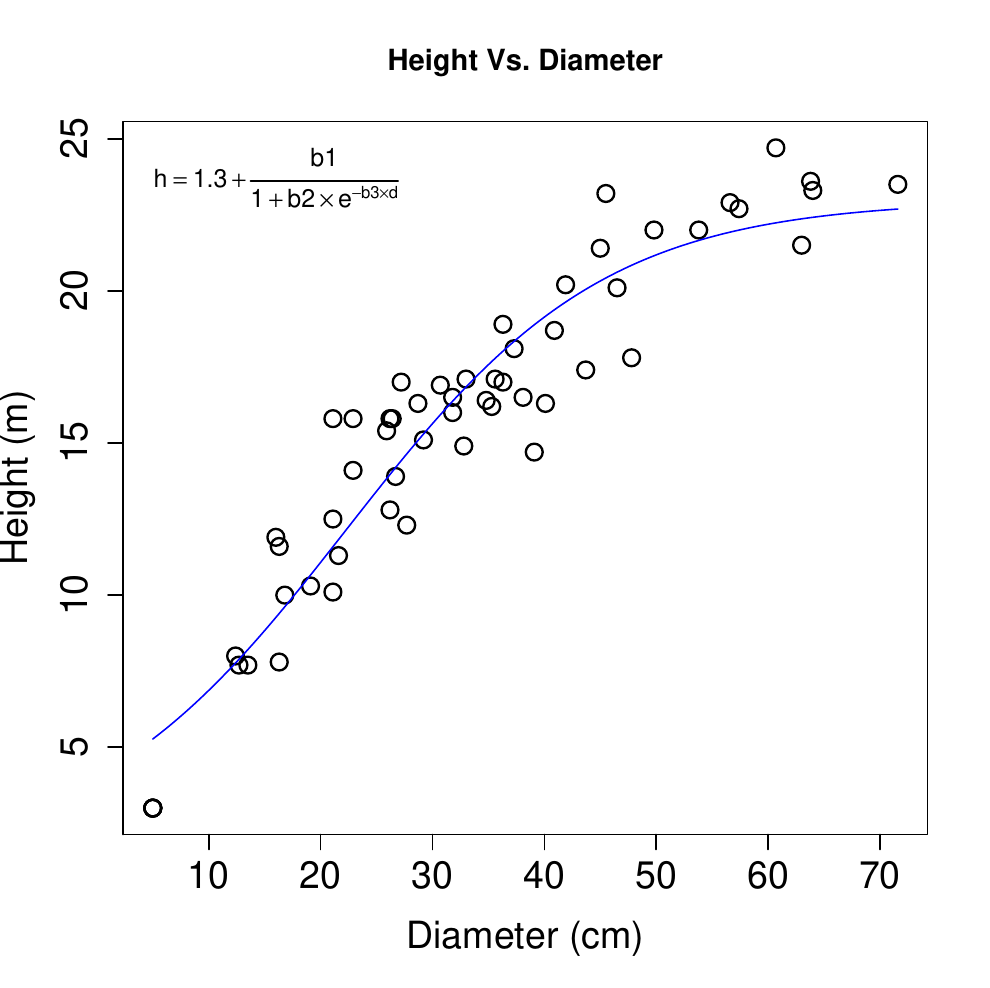}
\caption{Scatterplot of height and diameter data with fitted Weibull (left subfigure) and logistic (right subfigure) growth curves.}
\label{plot-R-introduction-Weibull-logistic-growth-model-trycatch}
\end{figure}
\end{enumerate}
%%%%%%%%%%%%%%%%%%%%%%%%%%%%%%%%%%%%%%%%%%%%%%%%%%%%%%%%%
\chapter{Common simulation techniques}
This section focuses on some important simulation approaches including: {\it{Metropolis-Hastings algorithm}}, {\it{Inverse transform method}}, {\it{Distributional identity}}, {\it{Rejection sampling}}, {\it{Adaptive rejection sampling}}, and {\it{Ratio of uniforms method}}, and {\it{Generalized ratio of uniforms method}}. We further deal with the methods for simulating from multivariate Gaussian distributions. Throughout, we assume that generation from uniform distribution on (0,1), that is ${\cal{U}}(0,1)$, is accomplished by the computer machine.
%%%%%%%%%%%
\section{Metropolis-Hastings algorithm}\label{MHsection0}
%\begin{chapquote}{John W. Tukey, June 16, 1915 - July 26, 2000}
%% \textit{\cite{fernholz2000conversation}}}
%``Be approximately right rather than exactly wrong.''
%\end{chapquote}
%\subsection{Markov chain}\label{MHsection}
In this part, we describe the Metropolis-Hastings algorithm. To this end, first, we review some preliminaries.
% p a general description of the Metropolis-Hastings (MH) algorithm followed by one illustration implemented in \verb+R+ environment. First, we recall three basic definitions as follows.
\begin{dfn}\label{def-stochasticprocess}
A collection of sample evidence (random variables) indexed by time set $t \in \{0,1,2,\ldots\}$, represented as $\bigl\{X_t\big\}_{t}$, is called a stochastic or random process.
\end{dfn}
\begin{dfn}\label{def-Markov}
A stochastic process $\bigl\{X_t\big\}_{t\geq0}$ is said to be Markov process if 
\begin{align}
p\bigl(X_{n} =x_{n} \big \vert X_{0}=x_{0},\cdots,X_{n-1}=x_{n-1} \bigr) =p\bigl(X_{n} =x_{n} \big \vert X_{n-1}=x_{n-1} \bigr), 
\end{align}
that means the position (or state) of a Markov process (chain) at time $n$ just depends on its position at previous state $n-1$ for $n\geq 1$. 
\end{dfn}
\begin{dfn}\label{def-irreducible}
For a given pair $(i,j)$, if there is a positive probability that chain will ever visit state $j$ while it starts with state $i$, then this chain is called {\it{irreducible}}.
\end{dfn}
%Alternatively, Definition \label{def-irreducible} states that if a chain is irreducible, then State there exists some $n \geq 1$ such that $P^{n}_{ij}=P\bigl(X_{n}=j \big \vert X_{0}=i\bigr)>0$, that is, by starting with state $i$, after $n$ steps, there is a positive probability that the chain visits state $j$. 
Let $\pi_j$ denote the long-run proportion of times that  an irreducible Markov chain visits state $j$. It can be shown that $\pi_j$ exists, independent of the chain's initial state $i$, and is given as the solution of the equations
\begin{align}\label{stationary-probabilities}
\pi_j=\sum_{i=1}^{N} \pi_i P_{ij},~{\text{for}}~j=1,\cdots,N.
\end{align}
The system of equations in (\ref{stationary-probabilities}) can be represented in matrix form as 
\begin{align}%\label{stationary-probabilities-matrix} 
\boldsymbol{\pi}^{\top}=
\left[\begin{matrix}
\pi_{1}\\
\pi_{2}\\
\vdots\\
\pi_{j}\\
\vdots\\
\pi_{N}
\end{matrix}\right]^{\top}=
\left[\begin{matrix}
\pi_{1}\\
\pi_{2}\\
\vdots\\
\pi_{j}\\
\vdots\\
\pi_{N}
\end{matrix}\right]^{\top}\times
\left[\begin{matrix}
 P_{11}& P_{12}&\cdots&P_{1N}\\
 P_{21}& P_{22}&\cdots&P_{2N}\\
 \vdots&\vdots&\vdots&\vdots\\
  P_{j1}& P_{j2}&\cdots&P_{jN}\\
 \vdots&\vdots&\vdots&\vdots\\
 P_{N1}& P_{N2}&\cdots&P_{NN}\\
\end{matrix}\right].
\end{align}
The vector $\boldsymbol{\pi}$ is known as the {\it{limiting distribution}} or the {\it{stationary distribution}} and ${\pi}_j$ is called the {\it{limiting probability}} that the Markov chain visits state $j$. 
\begin{dfn}\label{def-aperiodic}
An irreducible Markov chain is said to be {\it{aperiodic}} if we have
\begin{align}
P\bigl(X_{n}=j \big \vert X_{0}=j\bigr)>0,~{\text{for}}~n\geq0. 
\end{align}
\end{dfn}
\begin{dfn}
A Markov chain with limiting probability $\pi_j$ and transition probabilities $P_{ij}$ is said to be {\it{time-reversible}} if we have $\pi_iP_{ij}=\pi_jP_{ji}$ for all $i\neq j$.
\end{dfn}
Suppose we are willing to simulate from random variable for which we have $P(X=j)=\pi_j$ for $j=1,\cdots,m$ ($m$ is large). One way is the use of MH algorithm that constructs a time-reversible Markov chain $\{X_{n}\}_{n\geq 0}$ whose limiting probability is $\pi_j$. To this end, we consider an irreducible Markov chain with integer states $\{1,\cdots,m\}$ whose transition probability matrix is $\boldsymbol{Q}$, see \cite{ross2022simulation}. Let the $(i,j)$-element of $\boldsymbol{Q}$ is denoted as $q(i,j)$. If $X_n=i$, then generation $X$ is drawn from distribution with probability mass function $P(X=j)=q(i,j)$ for $j=1,\cdots,m$. If $X=j$, then $X_{n+1}$ would be $i$ and $j$, accordingly, with probabilities $1-\alpha(i,j)$ and $\alpha(i,j)$. The transition probabilities of the constructed Markov chain becomes
\begin{align*}
P_{i,j}=&q(i,j)\alpha(i,j),~ {\text{for}}~ j\neq i, \\
P_{i,i}=&q(i,i)+\sum_{k\neq i}^{}q(i,k)\bigl[1-\alpha(i,k)\bigr]. 
\end{align*}
The time-reversibility of this Markov chain with limiting distribution $\pi_j$ holds true if
\begin{align*}%\label{time-reversibility-1}
\pi_iP_{i,j}=\pi_jP_{j,i}~{\text{for}}~j\neq i,
\end{align*}
that implies
 \begin{align}\label{time-reversibility-2}
\pi_{i}q(i,j)\alpha(i,j)=\pi_{j}q(j,i)\alpha(j,i).
\end{align}
It is not hard to see that the identity in (\ref{time-reversibility-2}) is satisfied if we have
\begin{align}\label{acceptance-1}
\alpha(i,j)=\min \left\{1,\frac{\pi_{j}q(j,i)}{\pi_{i}q(i,j)}\right\}.
\end{align}
The limiting probability $\pi_{j}$ may be represented as $\pi_{j}=b_j/B$ in which $B=\sum_{j=1}^{m}b_j$ for $b_j>0$ and sufficiently large $m$. Evidently, factor $B$ plays the role of normalizing constant, and hence the right-hand side (RHS) of (\ref{acceptance-1}) can be represented as 
\begin{align}\label{acceptance-2}
\alpha(i,j)=\min \left\{1,\frac{b_{j}q(j,i)}{b_{i}q(i,j)}\right\}.
\end{align}
The steps of the MH algorithm is given by Algorithm \ref{MH-algorithm-1}.
\begin{algorithm}
\caption{MH-algorithm for generation one sample from target PDF $\pi_{j}=b_{j}/B$}
 \label{MH-algorithm-1}
\begin{algorithmic}[1]
%\Procedure{Generation a sample from from p}{}     
\State Construct an irreducible Markov chain with transition probabilities $q(i,j)$ (for $i,j=1,\cdots,m$) and suggest an integer number $c$ such that $1\leq c \leq m$;
\State Set $n=0$ and $x_{0}=c$;
\State Generate $x_{new}$ form probability mass function $P(X_{new}=j)=q\bigl(x_{n},j\bigr)$;
\State Generate a random variable $u\sim{\cal{U}}(0,1)$;
      \If{$u<\bigl[b_xq\bigl(x_{new},x_{n}\bigr)\bigr]/\bigl[b_{x_{n}}q\bigl(x_{n},x_{new}\bigr)\bigr]$}
        \State $x_{new} \leftarrow x$
    %\Comment{\small{$x^{k+1}$ is accepted}}
    % that means chain moves from state $x^{k}$ to the state $x^{k+1}$}} 
 \Else   
 \State $x_{new} \leftarrow x_{n}$;
     %   \Comment{reject candidate that means chain remains at state $x^{k}$}
     \EndIf
     \State {\bf{end}}
\State $n \leftarrow n +1$ and $x_{n} \leftarrow x_{new}$;
\State go to step 3.
% \EndProcedure
\end{algorithmic}
\end{algorithm}
In practice, the MH algorithm is use frequently to simulate from continuous distribution with probability density function (PDF) $f(\cdot\vert \theta)$ that coincides with the limiting distribution $\pi_{j}$, of irreducible Markov chain. But, as it is seen from Algorithm \ref{MH-algorithm-1}, the MH algorithm produces dependent generations sine the next generation is created based on the current one. Hence, all estimators constructed based on the generated samples are biased and inefficient. To avoid this problem, the initial generated samples generated sample are removed from the whole \citep{robert2004metropolis}. The length $L$, of the removed initial samples is known as {\it{burn-in}} or {\it{warm-up}} for which some criteria has been determined \citep{hobert2004mixture}. It can be shown that after sufficiently large number $M$ of generations, the generated samples are statistically uncorrelated and converge to the chain's limiting distribution that coincides with distribution of PDF $f(\cdot\vert\theta)$. 
We follow the steps of Algorithm \ref{MH-algorithm-2} for implementing the MH algorithm to simulate one generation from PDF $f(\cdot\vert \theta)$. 
\par
\begin{figure}[!h]
\center
\includegraphics[width=55mm,height=55mm]{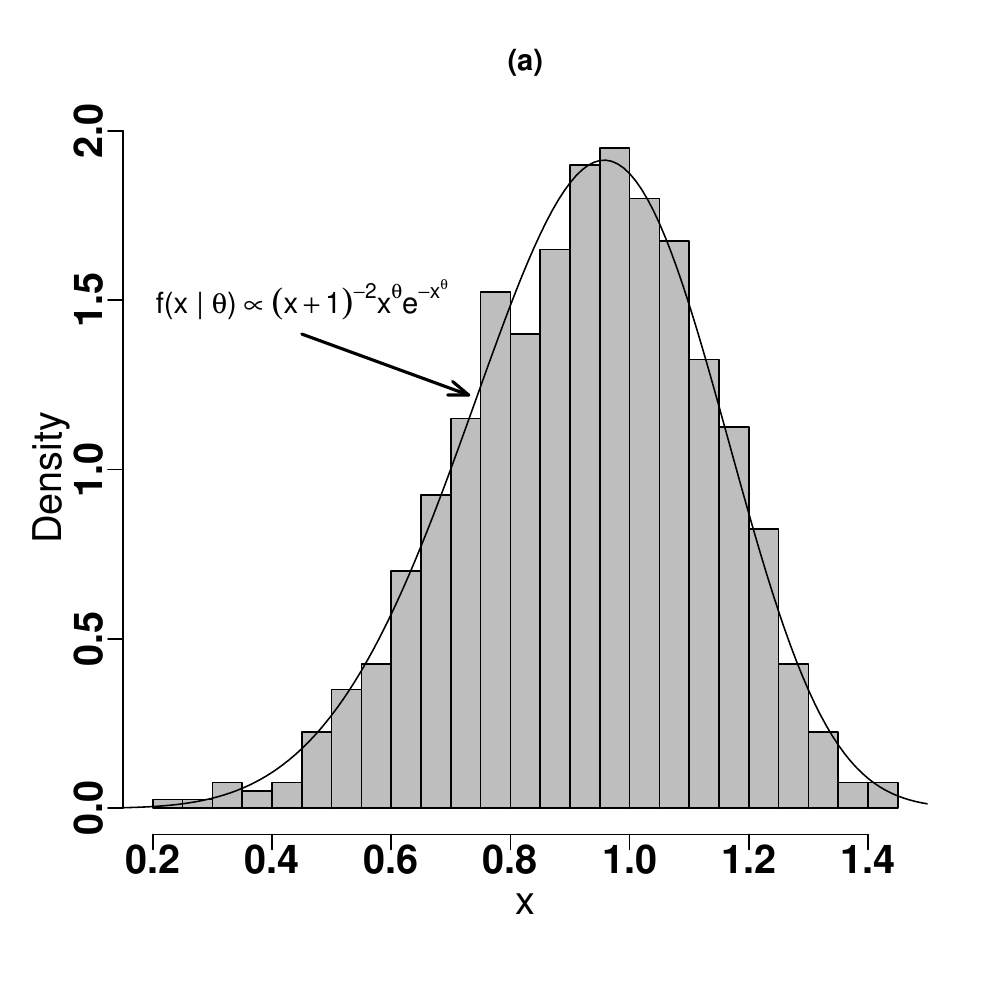}
\includegraphics[width=55mm,height=55mm]{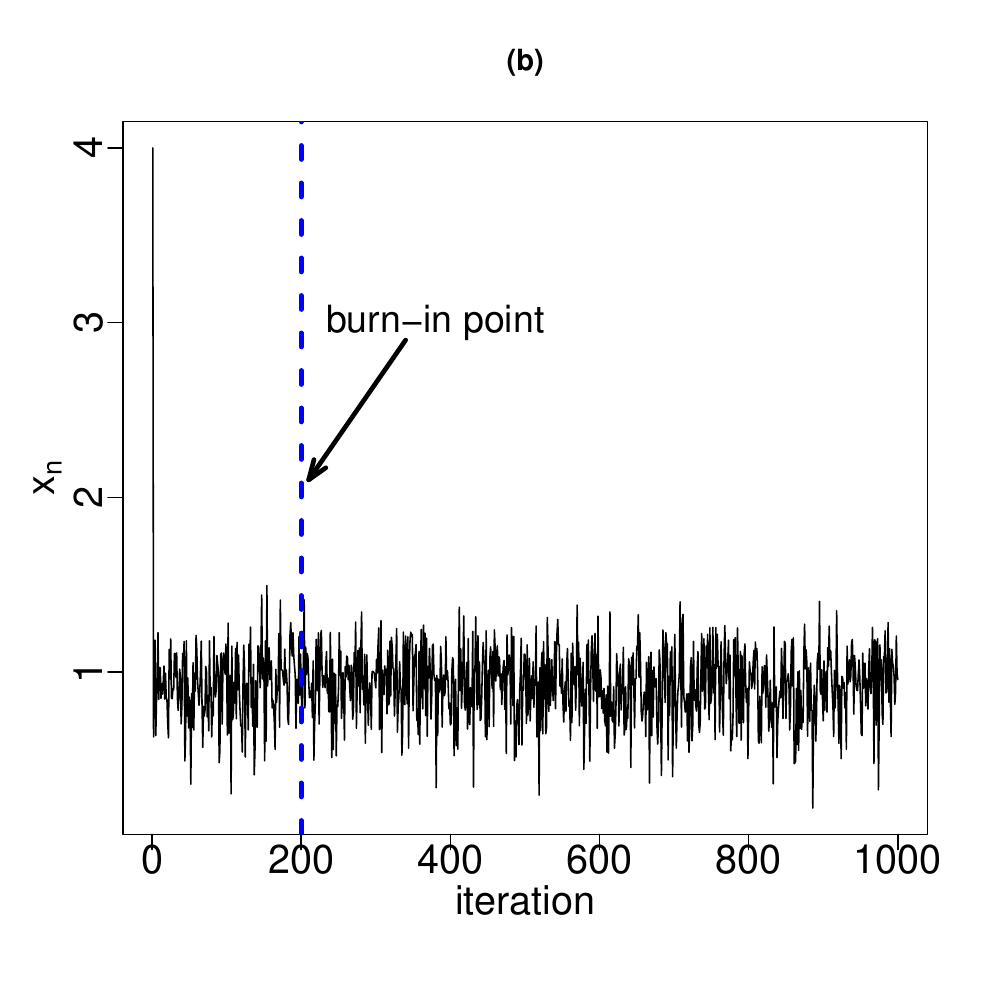}
\caption{(a): Histogram (constructed based on chain's output between 201$th$ up to 1000$th$ iterations); superimposed is $f(x\vert \theta=5)$ and (b): Chain's motion across iterations for sampling from $f(x\vert \theta=5)$ with starting value $x_0=4$.}
\label{fig1}
\end{figure}

 It is worthwhile to mention that $q(\cdot\vert \cdot)$ is known as the {\it{proposal}} or {\it{candidate}} distribution and $f(\cdot\vert \theta)$ is also referred to as {\it{target}} distribution in the literature. The proposal must be  easy to sampling from and moreover we have ${\cal{S}}_{q}={\cal{S}}_{f}$ \footnote{${\cal{S}}_{f}=\bigl\{ x \in {\mathbb{R}} \big \vert 
f(x\vert \theta) \geq 0, \int_{{\cal{S}}_{f}} f(x\vert \theta)dx=1\bigr\}$.}. If the next state is obtained by adding some random variable to the current state then we have a {\it{random walk}} proposal and the algorithm is called sometimes random walk Metropolis algorithm. Furthermore, if the random variable follows a symmetric distribution, then the acceptance rate or transition probability takes a simpler form. In such a case, we have $q\bigl(x_{n+1}\vert x_{n}\bigr)=q\bigl(x_{n}\vert x_{n+1}\bigr)$. The original Metropolis \citep{metropolis1953equation} algorithm was constructed based on a symmetric random walk proposal.
%random variable The have th The o that is generated  the  mean (or median) of the proposal is considered as the current state can be itsand so $q\bigl(x_{n+1} \big \vert x_{n} \bigr)$ can be interpreted as the chance of moving to the next state $x_{n+1}$ ($x_{new}$) given that  the current state on the average is $x_{n}$. 
For example, suppose that chain's current state is $X_{n}$ and we use a standard Gaussian distribution, as the proposal, for determining the next state $X_{n+1}$. It follows that
\begin{align*}
X_{n+1} =X_{n} +Y,
\end{align*}
where $Y\sim{\cal{N}}\bigl(\mu= 0, \sigma=1\bigr)$. It is easy to check that PDF of the next state becomes ${\cal{N}}\bigl( x_{n+1} \big\vert \mu= x_{n}, \sigma=1\bigr)$ or equivalently
\begin{align*}%\label{proposal-1}
q\bigl(x_{n+1} \big \vert x_{n}\bigr)={\cal{N}}\bigl( x_{n+1} \big\vert \mu= x_{n}, \sigma=1\bigr).
\end{align*}
Obviously, since ${\cal{N}}\bigl( x_{n+1} \big\vert \mu= x_{n}, \sigma=1\bigr)={\cal{N}}\bigl( x_{n} \big\vert \mu= x_{n+1}, \sigma=1\bigr)$,  we have 
\begin{align}\label{proposal-2}
q\bigl(x_{n+1} \big \vert x_{n}\bigr)=q\bigl(x_{n} \big \vert x_{n+1}\bigr).
\end{align}
The above argument holds true for any $\sigma>0$. Therefore the acceptance rate given in Algorithm has the simpler form
 \begin{align*}
A=\min \left\{1,\frac{f\bigl( x_{n+1}\big\vert \theta\bigr)}{f\bigl(x_{n}\big\vert\theta\bigr)}\right\}.
\end{align*}
%As mentioned above, within a Metropolis algorithm the proposal has a symmetric PDF that is easy to sampling from. 
Another example for a random walk Metropolis algorithm is obtained by considering the uniform distribution on $(a-3b, a+3 b)$ as the proposal in which $a \in \mathbb{R}$ and $b>0$. In this case, we have $q\bigl(x_{n+1}\vert x_{n}\bigr)=q\bigl(x_{n}\vert x_{n+1}\bigr)=1/(6\sigma)$. The algorithm proposed by \cite{metropolis1953equation} was then developed by \cite{hastings1970monte} for general case including both symmetric and asymmetric [$q(x_{n+1}\vert x_{n})\neq q(x_{n+1}\vert x_{n})$] cases. Today the latter case is known as the Metropolis-Hasting algorithm. A privilege of the MH algorithm is its capability to apply for the cases that the target distribution needs only to be specified up to proportionality constant. The care must taking into account that target PDF must never be zero at the initial state $x_0$ and the candidate $q(\cdot\vert \cdot)$ must have a broad enough support to visit any state of ${\cal{S}}_{f}$ with a positive probability while gets as simple as possible form. We give more details on finding a suitable proposal through the following example. 
\vspace{5mm}
\begin{algorithm}
\caption{MH-algorithm for generation one sample from target PDF $f(\cdot\vert\theta)$}
\label{MH-algorithm-2}
\begin{algorithmic}[1]
%\Procedure{Generation a sample from from p}{}     
\State Set $n=0$, read $L$, and choose the current state (initial generation) $x_{0}$;
    \While{$n \leq L$}  %\Comment{put some comments here}
\State Sample $x_{new}$ from the proposal distribution with PDF $q\bigl(\cdot \big \vert x_{n}\bigr)$;
\State Compute 
 \begin{align*}
A=\min \left\{1,\frac{f\bigl( x_{new}\big\vert \theta\bigr)q\bigl(x_{n} \big \vert x_{new} \bigr)}{f\bigl(x_{n}\big\vert\theta\bigr)q\bigl(x_{new} \big \vert x_{n}\bigr)}\right\};
\end{align*}
\State Generate $u\sim{\cal{U}}(0, 1)$;
      \If{$u<A$}
        \State Set $n \leftarrow n +1$ and $x_{n} \leftarrow x_{new}$; %, $n \leftarrow n +1$, and $x_{n} \leftarrow y$;
        %\Comment{\small{$x^{k+1}$ is accepted}}
        % that means chain moves from state $x^{k}$ to the state $x^{k+1}$}} 
 \Else   
 \State $y \leftarrow x_{n}$, $n \leftarrow n +1$, and $x_{n} \leftarrow y$;% and $n \leftarrow n +1$;
     %   \Comment{reject candidate that means chain remains at state $x^{k}$}
     \EndIf
          \State {\bf{end}}
     \EndWhile  %\label{roy's loop}
     \State {\bf{end}}
     \State Accept $x_{L}$ as a generation form target distribution $f$.
% \EndProcedure
\end{algorithmic}
\end{algorithm}
%The MH works by generating sequence $\{x_{1},x_{2},\cdots,x_{n}\}$ from limiting distribution $\pi(j)$ that theoretically follows  for $n \rightarrow \infty$. More precisely, suppose the current state of Markov chain is $x_{n}$, then within a Metropolis algorithm framework, for moving form the current state (generation) $x_{n}$ to the next state (generation) $x_{n+1}$, the transition probability\footnote{The transition probability is called also as acceptance rate represented as $A_{\theta}$.} $p\bigl(x_{n}, x_{n+1}\bigr)$, is
%\begin{align}\label{ratio}
%p\bigl(x_{n} \rightarrow x_{n+1}\bigr)=\min \left\{1,\frac{f\bigl( x_{n+1}\big\vert \theta\bigr)q\bigl(x_{n} \big \vert x_{n+1} \bigr)}{f\bigl( x_{n}\big\vert\theta\bigr)q\bigl(x_{n+1} \big \vert x_{n}\bigr)}\right\}.
%\end{align}
%Since moving to next state $x_{n+1}$ depends on the current state $x_n$, hence the sequence of the generated samples constitutes a correlated stochastic process. But, despite existence such dependency, the simulated generations are approximately distributed as $f(\cdot\vert \theta)$. This dependency has its origin from the Markovian nature of the simulation so that the next generation is highly depends on the current state (generation).
%%%%%%%%%%%%%%%%%%%%%%%%%%%%
%\item The transition probability $p\bigl(x_{n}, x_{n+1}\bigr)$ should not be small, letting chain to visit any state of space.
\begin{example}\label{exam-weibull}%\lipsum[]
Let $f(x\vert \theta) \propto(x+1)^{-2} x^{\theta}\exp\{-x^\theta\}$ for $x>0$ denote the PDF of target distribution in which $\theta>0$ is the family parameter. Considering the Weibull distribution with shape parameter $\theta$ and scale unity as the proposal distribution whose PDF is denoted as ${\cal{W}}(\cdot \vert \theta, 1)$, we represent the acceptance rate (transition probability) as
\begin{align}\label{ratioexample}
A=&\min \left\{1,\frac{(x_{n+1}+1)^{-2} x_{n+1}^{\theta}\exp\{-x_{n+1}^\theta\}{\cal{W}}(x_{n}\vert \theta, 1)}{(x_n+1)^{-2} x_{n}^{\theta}\exp\{-x_{n}^\theta\}{\cal{W}}(x_{n+1}\vert \theta, 1)}\right\}.\nonumber\\
=&\min \left\{1,\frac{(x_{n}+1)^2 x_{n+1}}{(x_{n+1}+1)^2 x_{n}}\right\}.
\end{align}
Fortunately, though using an asymmetric proposal, the acceptance rate has a simple form. Setting $\theta=5$, we follow the steps given by the following for simulating $800$ realizations from $f(x\vert \theta=5)$. It is worth to highlight that we run Algorithm \ref{MH-algorithm-2} for $Ls=1000$ times of which first $Lb=200$ (burn-in point) generations are removed.
\begin{itemize}
\item Set $n=0$, $Lb=200$, $Ls=1000$, and the initial state (generation) $x_{0}=1$;
\begin{enumerate}
\item Generate $x_{new}\sim {\cal{W}}(\theta=5,1)$
\item Compute $A=\min \left\{1,\frac{(x_{n}+1)^2 x_{new}}{(x_{new}+1)^2 x_{n}}\right\}$;
\item Generate $u\sim {\cal{U}}(0, 1)$;
\item If $u<A$, then $n \leftarrow n+1$ and $x_{n} \leftarrow x_{new}$;
%(chain moves from state $x_{n}$ to state $x_{new}$) and 
\item Repeat algorithm from step 1;
\end{enumerate}
\item Stop algorithm if $n=Lb$ and then accept $x_{Lb}$ as a generation from $f(x\vert \theta=5)$. 
\end{itemize}
\end{example}
The \verb+R+ code, for generating from PDF $f(x\vert \theta=5)$ given in Example \ref{exam-weibull} is given as follows.
%\begin{verbatim}
%\end{verbatim}
\begin{lstlisting}[style=deltaj]
R> set.seed(20240412)
R> theta <- 5
R> Ls <- 1000  # total number of iterations
R> Lb <- 200		# burn-in point
R> x <- rep(NA, Ls)
R> x0 <- 1; x[1]<- x0 # initial value
R> for(n in 1:Ls)
+ {
+ 	x.new <- rweibull(1, shape = theta, scale = 1)
+ 	A <- 2*log(x[n] + 1) + log(x.new) - 
+							2*log(x.new + 1) - log(x[n])
+ 	if( runif(1) < exp( A ) )
+ 	{
+ 		x[n+1] <- x.new  # x.new is accepted
+ 	}else{
+ 		x[n+1] <- x[n]		# x.new is rejected
+ 	}
+ }
R > hist(x[(Lb+1):Ls], prob=T, br=20)
\end{lstlisting}
As it is seen in example above, the Weibull choice of candidate yields a simple form of transition probability. Figure \ref{fig1}(b) shows the chain's motion across $Ls=1000$ iterations while the initial state is $x_{0}=4$. As it is seen, the convergence occurs very soon. There have been introduced some extensions of the MH algorithm in the literature. We refer reader to for more details about extensions such as \citep{robert2004metropolis,roberts2009examples,robert2010introducing}.
\vspace{5mm}
There have been introduced some extensions of the MH algorithm in the literature. We refer reader to for more details about extensions such as \citep{robert2004metropolis,roberts2009examples,robert2010introducing}.
\section{Inverse transform method}\label{inverse-transform-section}
Throughout this chapter let random variable $X$ from which drawing sample is of interest has closed form cumulative distribution function (CDF), denoted herein by $F(\cdot)$, or is computed using accurate computer algorithms such as Gaussian distribution. In what follows $X\sim F(\cdot)$ or $X\sim f(\cdot)$ will denote the random variable $X$ follows CDF $F(\cdot)$ or probability density function (PDF) $f(\cdot)$. The following proposition plays main role for simulating from $X$ through the Inverse transform method. 
\begin{prop}\label{prop-inverse-transform} Let $X\sim F(\cdot)$ and $U\sim{\cal{U}}(0,1)$. Then,
\begin{align*}
X \mathop=\limits^d F^{-1}(U),
\end{align*}
where ``$\mathop=\limits^d$'' denotes the identity in distribution and $F^{-1}(\cdot)$ is  inverse of CDF or the quantile function.
\end{prop}
Based on Proposition (\ref{prop-inverse-transform}) in above, for simulating from random variable with CDF $F(\cdot)$ (or PDF $f(\cdot)$), first, one can simulate a generation such as $u$ from ${\cal{U}}(0,1)$ and then a sample such as $x$ is produced by computing $F^{-1}(u)$. In what follows, we give an example that accomplishes this method simply for drawing sample from exponential distribution.
\begin{example}\label{exam-weibullsimulation-rejection}%\lipsum*[]
Let $X\sim {\cal{EXP}}(\lambda)$ with CDF given by
\begin{align}\label{cdf-exponential}
F(x)= 1- \exp\{-\lambda x\},
\end{align}
where $\lambda>0$ is the family rate parameter. It easy to see that
\begin{align}
F^{-1}(u)= -\frac{1}{\lambda} \log(1-u),  0<u<1.
\end{align}
Hence, a generation from exponential family with CDF in (\ref{cdf-exponential}) becomes $x=-\log(1-u)/\lambda$ in which $u$ is drawn from ${\cal{U}}(0,1)$. The \verb+R+ function \verb+rexp0+ given by
\begin{lstlisting}[style=deltaj]
R> rexp0 <- function(n, lambda = 1)
+{
+x <- -1/lambda * log( 1- runif(n) )
+return(x)
+}
\end{lstlisting}
can be used for generating $n$ independent realizations form ${\cal{EXP}}(\lambda)$. But, it should be noted that \verb+R+ has its own function \verb+rexp(n, rate=lambda)+ for generating from this family. So, we would prefer to use the latter in simulation setting rather than the former \verb+R+ code.   
\end{example}
\par
\begin{figure}[!h]
\center
\includegraphics[width=55mm,height=55mm]{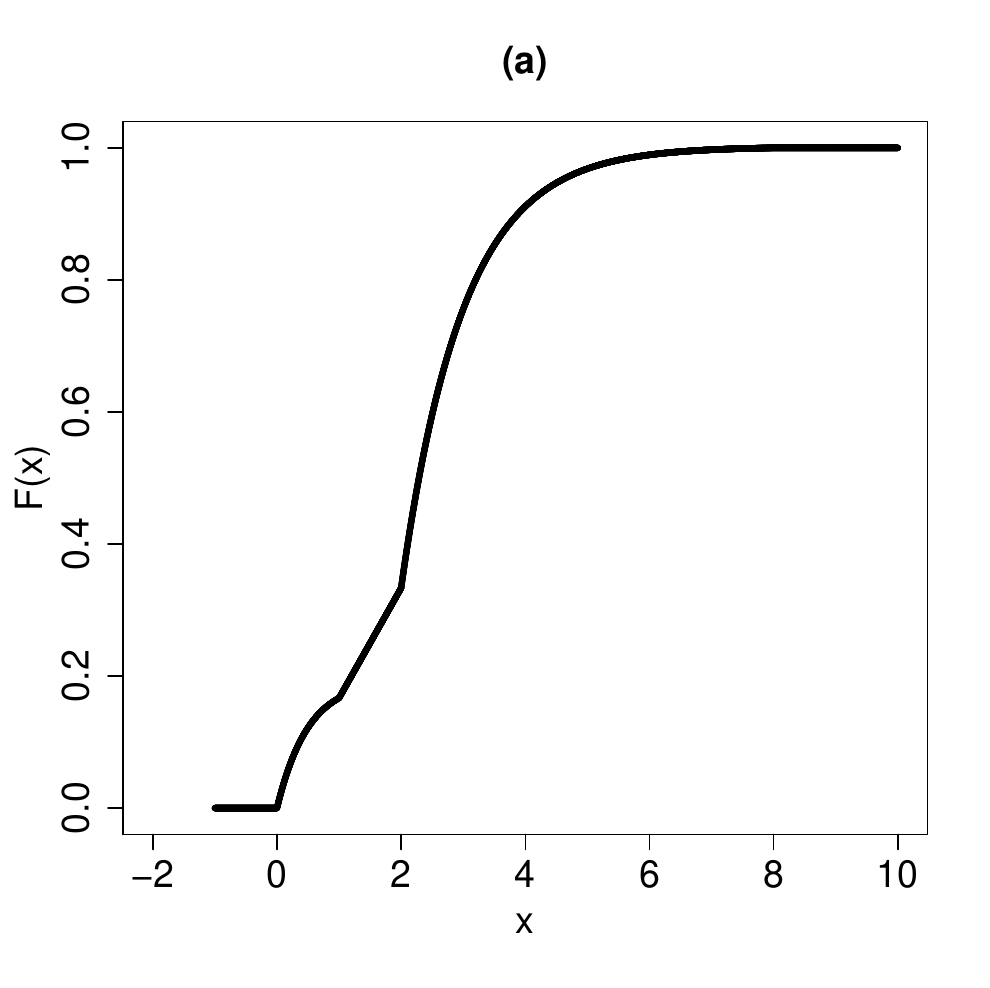}
\includegraphics[width=55mm,height=55mm]{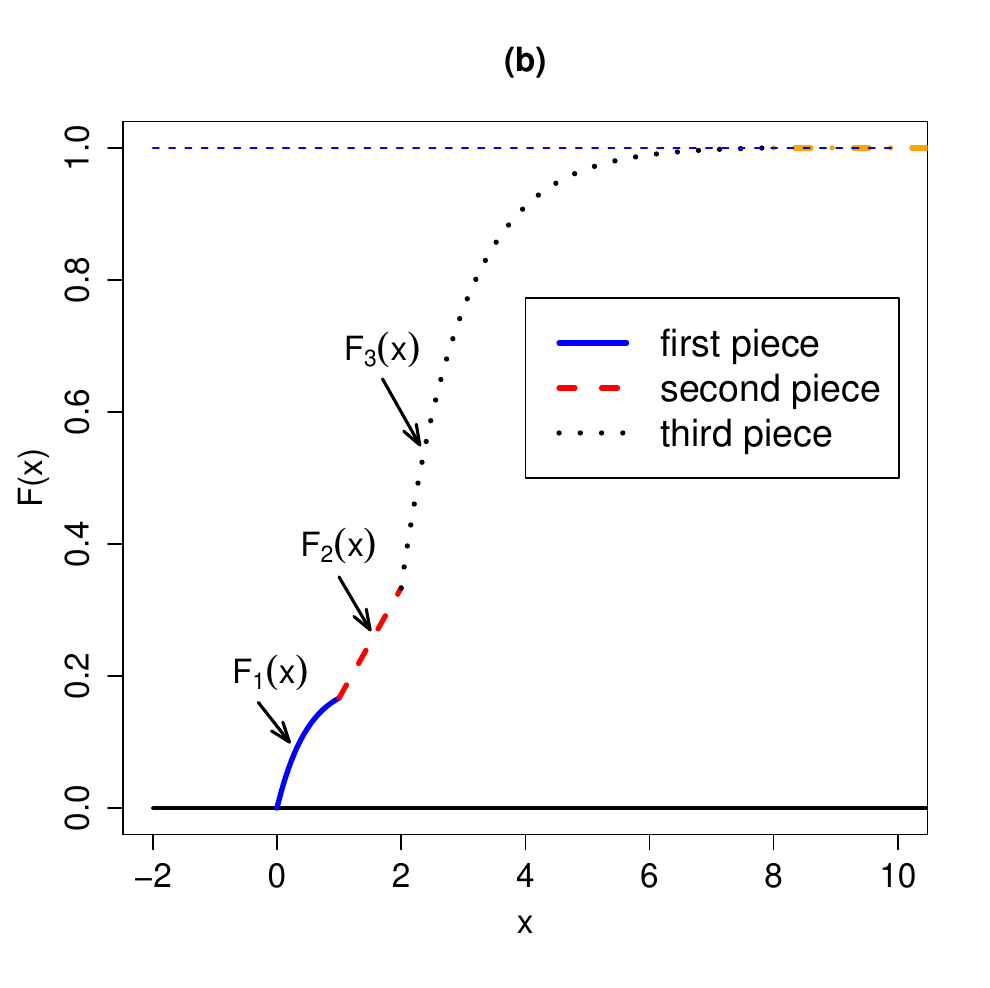}
\caption{(a): a schematic of three-piece CDF and (b): the corresponding pieces.}
\label{plot-piecewise-CDF-example-1}
\end{figure}

 In order to simulate from some well-known statistical families of distributions in \verb+R+, we may run the commands given in Table \ref{Simulation-R-command}. When the CDF is a piecewise function, then the quantile function can be exploited by inverting the CDF on each piece for which $0<F(\cdot)<1$. Assuming that random variable $X$ has a $m$-piece CDF represented by
\begin{align}\label{piecewise-CDF}
F(x)=    \begin{cases}
~~~~~~0,~ & {\text{if}}\ ~~~~~~~~~~~ x<a_{0},\\
F_{1}(x),~ & {\text{if}}\ ~~~~a_{0}\leq x< a_{1},\\
F_{2}(x),~ & {\text{if}}\ ~~~~a_{1}\leq x< a_{2},\\
~~~\vdots&~~~~~~~~~~~~~~\vdots\\
%F_{i}(x),~ & {\text{if}}\ ~a_{i-1}< x< a_{i},\\
%~~~\vdots&~~~~~~~~~~~~~~\vdots\\
F_{m}(x),~ & {\text{if}}\ a_{m-1}\leq x< a_{m},\\
~~~~~~1,~ & {\text{if}}\ ~~~a_{m}\leq x.\\
    \end{cases}
\end{align}
For example, schematic of a three-piece CDF 
\begin{align}\label{piecewise-CDF}
F(x)=    \begin{cases}
0, & {\text{if}}\ ~~~~~~~x<0,\\
\frac{1}{6} \times\frac{1-\exp\{-2x\} }{ 1 - \exp\{-2\} }, & {\text{if}}\ ~~0\leq x< 1,\\
\frac{1}{6}+\frac{x-1}{6},~ & {\text{if}}\ ~~1\leq x< 2,\\
\frac{1}{3}+\frac{2}{3} \times \frac{\exp\{-x\} - \exp\{-2\} }{ \exp\{-8\} - \exp\{-2\} }, & {\text{if}}\ ~~2\leq x< 8,\\
1, & {\text{if}}\ ~~8\leq x,\\
    \end{cases}
\end{align}
is displayed in Figure \ref{plot-piecewise-CDF-example-1}, We emphasize that  $F_{i}(x)$ (for $i=1,\cdots,m$) are not CDF. Algorithm \ref{Simulation-piecewise-CDF-1} describes how to simulate from a random variable that possesses a piecewise CDF.
%, but $F_{i}(x)=\sum_{j=1}^{i}F_{j}(x)$ (for $i=1,\cdots,m$).

\begin{figure}[!h]
\center
\includegraphics[width=55mm,height=55mm]{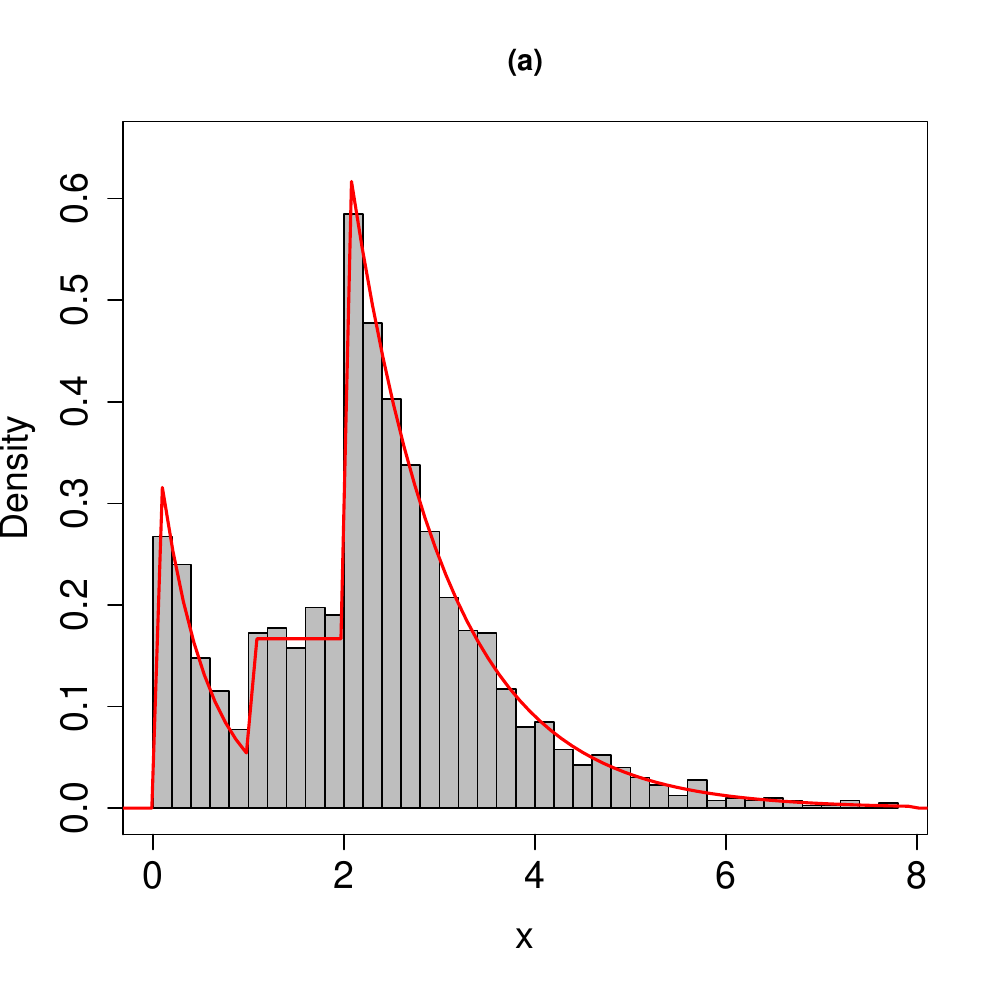}\includegraphics[width=55mm,height=55mm]{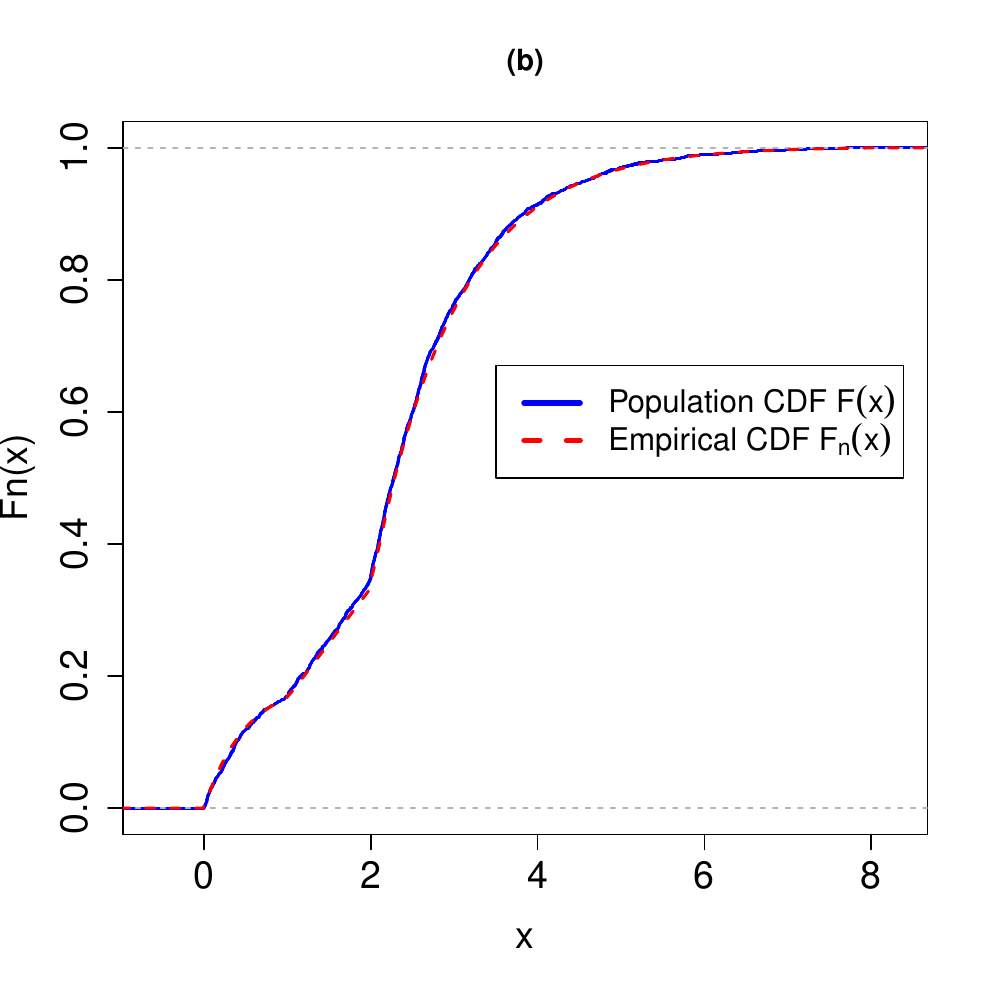}
\caption{(a): Histogram of $n=2000$ simulated data from family with CDF (\ref{piecewise-CDF}). The red-colored superimposed curve is the corresponding PDF. (b): The population CDF and empirical CDF.}
\label{plot-piecewise-histogram-example-2}
\end{figure}

\begin{algorithm}
\caption{Simulation from random variable with piecewise CDF}
\label{Simulation-piecewise-CDF-1}
\begin{algorithmic}[1]
%\Procedure{}{}     %  \Comment{This is a test}
 %   \State System Initialization
 \State Read $n$ and sequence $\bigl\{F(a_{0}),F(a_{1}),\cdots,F(a_{m})\bigr\}$;
 \State Generate $u$ from ${\cal{U}}(0, 1)$;
        \If{$F(a_{0}) < u < F(a_{1})$}
        \State $x = F_{1}^{-1}(u)$;
        \EndIf
        \If{$F(a_{1}) < u < F(a_{2})$}
        \State $x = F_{2}^{-1}(u)$;
            \EndIf
           \State ~~~~~~~~~~~~~~\vdots
        \If{$F(a_{m-2}) < u < F(a_{m-1})$}
        \State $x = F_{m-1}^{-1}(u)$;
            \EndIf
        \If{$F(a_{m-1}) < u < F(a_{m})$}
        \State $x = F_{m}^{-1}(u)$;
       % \EndIf
    \EndIf
\State return $x$ as a generation from CDF $F(\cdot)$.
%\EndProcedure
\end{algorithmic}
\end{algorithm}
\section{Distributional identity}\label{distributional-identity-section}
Herein, we present some examples in which generating from the distribution of interest is accomplished through its relation with other distribution(s) from which sampling is an easy or easier task. One of the most widely used distribution, that is the Gaussian, can be simulated through this approach. Although the quantile function of the Gaussian distribution can be computed numerically in \verb+R+ an then the inverse transform method can be applied to sample from, alternatively an efficient method known as the Box-Muller transform proposed by \cite{box1958note} is suggested for this purpose as follows. It needs to generate independent random variables $U_{1},U_{2}\sim{\cal{U}}(0,1)$ and then it can be proved that
\begin{align*}%\label{Gaussian-simulation}
Z_{1}=\sqrt{-2 \log U_{1}} \cos(2\pi U_{2}),\\
Z_{2}=\sqrt{-2 \log U_{1}} \sin(2\pi U_{2}),
\end{align*}
independently follow a standard Gaussian distribution. Consequently, either $\mu +\sigma Z_{1}$ or $\mu +\sigma Z_{2}$ follows ${\cal{N}}(\mu,\sigma^2)$. If evaluating the sine or cosine functions wastes the computational time, one can use the method proposed by  \cite{marsaglia1964convenient} that is described below by Algorithm \ref{Simulation-Marsaglia-Bray}.
\begin{algorithm}
\caption{Marsaglia-Bray method for simulating Gaussian random variable}
\label{Simulation-Marsaglia-Bray}
\begin{algorithmic}[1]
%\Procedure{}{}     %  \Comment{This is a test}
 %   \State System Initialization
 \State Generate $U_{1}$ and $U_{2}$ independently from ${\cal{U}}(-1, 1)$;
 \State Compute $W=U^{2}_{1}+U^{2}_{2}$;
        \If{$W<1$}
        \State $Y =W^{-1/2} \times \sqrt{-2 \log W}$;
        \State $Z_{1}=Y U_{1}$ and $Z_{2}=Y U_{2}$;
        \Else
        \State Come back to step 2 and repeat algorithm;
        \EndIf
        \State {\bf{end}}
\State Return $\mu +\sigma Z_{2}$ and $\mu + \sigma Z_{2}$ as two independent generations from ${\cal{N}}(\mu,\sigma^2)$. 
%\EndProcedure
\end{algorithmic}
\end{algorithm}
The pertinent \verb+R+ function \verb+rnorm0+ given below can be used for generating Gaussian realizations based on Marsaglia-Bray method.
\begin{lstlisting}[style=deltaj]
R> rnorm0 <- function(n, mu, sigma)
+{ # n is supposed to be an even number
+i <- 1
+Z <- rep(NA, n) 
+	while (i <= n/2)
+	{
+	U1 <- 2*(runif(1) - 0.5); U2 <- 2*(runif(1) - 0.5)
+	W <- U1^2 + U2^2
+		if(W < 1)
+		{
+		Y <- sqrt( -2*log(W)/W )
+		Z1 <- Y*U1; Z2 <- Y*U2; 
+		Z[ (2*i - 1):(2*i) ] <- c(Z1, Z2)
+		i <- i + 1
+		}
+	}
+ return( mu + sigma*Z )
+}
\end{lstlisting}

\subsection{Multinomial distribution}
%we could use multinomial distribution for simulating a random variable with piecewise CDF. To this end, we represent the multinomial distribution. 
If random vector $\boldsymbol{X}=(X_1,\cdots,X_{\text{K}})^{\top}$ follows a multinomial distribution, then its PDF is given by
\begin{align}\label{Multinomial-PDF}
{\cal{MUL}}\bigl(\boldsymbol{x}\big \vert N, \boldsymbol{\omega}\bigr)=\frac{\Gamma(N+1)}{\prod_{k=1}^{{\text{K}}}\Gamma\bigl(N_{k}+1\bigr)}\omega_{1}^{x_{1}}\omega_{2}^{x_{2}}\cdots \omega_{{\text{K}}}^{x_{{\text{K}}}},
\end{align}
where $N=\sum_{k=1}^{{\text{K}}}x_{k}$ in which $x_{k} \geq 0$. The elements of family parameter $\boldsymbol{\omega}=\bigl(\omega_1,\cdots,\omega_{\text{K}}\bigr)^{\top}$ are constrained to be nonnegative and sum up to one, that is $\sum_{k=1}^{{\text{K}}}\omega_{k}=1$. We write $\boldsymbol{X} \sim {\cal{MUL}}(n,\boldsymbol{\omega})$ to indicate that random vector $\boldsymbol{X}$ follows a multinomial distribution with PDF given by (\ref{Multinomial-PDF}). In order to simulate from ${\cal{MUL}}\bigl(\boldsymbol{x}\big \vert n, \boldsymbol{\omega}\bigr)$ in \verb+R+, we can use command \verb+rmultinom(n, size, prob)+ in which \verb+n+ is the sample size, \verb+size+ $=N$, and \verb+prob+ $=\boldsymbol{\omega}$. If \verb+size+ $=N=1$, and ${{\text{K}}}=2$ (or \verb+prob+=$(\omega_{1},\omega_{2})^{\top}$), then multinomial distribution turns into a Bernoulli distribution with success probability $\omega_1$. 
\subsection{Finite mixture model}
The PDF of a $\text{K}$-component finite mixture model is given by
\begin{align}\label{mixture-PDFk}
g({y} \vert \boldsymbol{\Psi})=\sum_{k=1}^{\text{K}} \omega_{k} f\bigl({y}\bigl \vert \boldsymbol{\theta}_{k}\bigr),
\end{align}
where $\boldsymbol{\Psi}=(\boldsymbol{\omega}^{\top},\boldsymbol{\theta}_1,\cdots, \boldsymbol{\theta}_\text{K})$ is the parameter space in which $\boldsymbol{\omega}=(\omega_1,\dots,\omega_\text{K})^{\top}$ is vector of mixing parameters such that $\sum_{k=1}^{\text{K}}\omega_{k}=1$. In (\ref{mixture-PDFk}), $f({y} \vert\boldsymbol{\theta}_{k})$ is the PDF of $k$-th component and hence the PDF $g({y} \vert \boldsymbol{\Psi})$ is proper since 
%for $\boldsymbol{\Theta}_{k}=\bigl(\boldsymbol{\theta}_k, \boldsymbol{\mu}_k, {\Sigma}_k, {\Lambda}_k\bigr)$
\begin{align*}%\label{Gibbs-mixture-PDF2}
\int_{\mathbb{R}^{p}}g(\boldsymbol{y} \vert \boldsymbol{\Psi})d\boldsymbol{y}=\sum_{k=1}^{\text{K}} \omega_{k} \int_{\mathbb{R}^{p}}f\bigl(\boldsymbol{y}\bigl \vert \boldsymbol{\theta}_{k}\bigr)d\boldsymbol{y}=\sum_{k=1}^{\text{K}} \omega_{k}=1.
\end{align*}
The weight parameter $\boldsymbol{\omega}$ provides the chance of observing sample from components. Sometimes, representation (\ref{mixture-PDFk}) is called the {\it{finite mixture model}} since the quantity ${\text{K}}$ or number of clusters is assumed to be finite. Herein, we we suggest two approaches for generating from finite mixture model. Under the first method, for generating $n$ realizations from a ${\text{K}}$-component mixture model, one can the multinomial distribution. In what follows, Algorithm \ref{Simulating-mixture-Gaussian-model-1} gives a pseudo code describing how to draw a sample of size $n$ from a ${\text{K}}$-component Gaussian mixture model under the first method. 
\vspace{5mm}
\begin{algorithm}[!h]
\caption{Simulating from ${\text{K}}$-component Gaussian mixture model: first method}
    \label{Simulating-mixture-Gaussian-model-1}
\begin{algorithmic}[1]
\State Read $n$, ${\text{K}}$, and $\boldsymbol{\Psi}=(\boldsymbol{\omega},\boldsymbol{\theta}^{\top}_1,\cdots, \boldsymbol{\theta}^{\top}_\text{K})^{\top}$;
\State Set ${Y}$ a vector of length $n$; 
    \State Set $i=1$;
    \While{$i \leq n$}  %\Comment{put some comments here}
\State Sample $\boldsymbol{u}\sim {\cal{MUL}}(1,\boldsymbol{\omega})$;
\State Set $k_{max} =\bigl\{k \big \vert \boldsymbol{u}[k]=1\bigr\}$ where $\boldsymbol{u}[k]$ is the $k$th element of vector $\boldsymbol{u}$ for 
\State $k=1,\cdots,{\text{K}}$;
\State Sample ${y}$ from PDF $f\bigl(\cdot\bigl \vert \boldsymbol{\theta}_{k_{max}}\bigr)$
\State Set ${Y}[i] \leftarrow {y}$;
    \State Set $i \leftarrow i+1$;
             \EndWhile  %\label{roy's loop}
             \State {\bf{end}}
        \State Accept ${Y}$ as sample of size $n$ from ${\text{K}}$-component finite mixture model. 
\end{algorithmic}
\end{algorithm}
\vspace{5mm}
Under the second method, we assume that the number of simulated realization under each component, that is $n_{k}$, is known or is determined as $n_{k}=\lceil n \times \omega_{k} \rceil$ (for $k=1,\cdots, {\text{K}}-1$) provided that $n_{\text{K}}=n-\sum_{k=1}^{{\text{K}}-1} n_{k}$. Herein, the generic symbol $\lfloor n \times \omega_{k}\rfloor$ denotes the greatest integer value equal or less than $n\times \omega_{k}$. A sample of size $n$ form ${\text{K}}$-component mixture model is obtained by merging all ${\text{K}}$ simulated vectors each of size $n_{\text{K}}$ from PDF $f\bigl({y}\big \vert \boldsymbol{\theta}_{k}\bigr)$. Algorithm \ref{Simulating-mixture-Gaussian-model-2} gives a pseudo code describing how to accomplish this task.
\vspace{5mm}
\begin{algorithm}
\caption{Simulating from ${\text{K}}$-component Gaussian mixture model: second method}
    \label{Simulating-mixture-Gaussian-model-2}
\begin{algorithmic}[1]
\State Read ${\text{K}}$, $n_{1},\cdots,n_{{\text{K}}}$, and $\boldsymbol{\Psi}=(\boldsymbol{\omega},\boldsymbol{\theta}^{\top}_1,\cdots, \boldsymbol{\theta}^{\top}_\text{K})^{\top}$;
\State Set ${Y}$ as a vector of length $n$; 
    \State Set $k=1$;
    \While{$k \leq {{\text{K}}}$}  %\Comment{put some comments here}
\State Draw sample $\boldsymbol{y}_{1}$ of size $n_{k}$ from PDF $f\bigl(\cdot\bigl \vert \boldsymbol{\theta}_{k}\bigr)$;
    \State Set $k \leftarrow k+1$;
             \EndWhile  %\label{roy's loop}
             \State {\bf{end}}
        \State Accept vector $(\boldsymbol{y}_{1},\cdots,\boldsymbol{y}_{{\text{K}}})^{\top}$ as sample of size $n$ from ${\text{K}}$-component finite mixture model. 
\end{algorithmic}
\end{algorithm}
\begin{example}\label{exam-simulation-two-component-Gaussian-1}%\lipsum*[]
Suppose we are interested in generating a sample of $n=300$ observations from 
two-component Gaussian mixture model with PDF given by
\begin{align}\label{exam-simulation-two-component-Gaussian-2}
g({y} \vert \boldsymbol{\Psi})= \omega_1\times{\cal{N}}\bigl(y\big \vert {\mu}_{1},\sigma^{2}_{1}\bigr)+\omega_2\times{\cal{N}}\bigl(y\big \vert{\mu}_{2},\sigma^{2}_{2}\bigr)
\end{align}
where $\boldsymbol{\Psi}=\bigl(\omega_1,\omega_2,\boldsymbol{\theta}^{\top}_1,\boldsymbol{\theta}^{\top}_2\bigr)^{\top}$ is whole parameter vector whose elements are
\begin{align}\label{exam-simulation-two-component-Gaussian-3}
&\omega_1=\frac{6}{10}, \boldsymbol{\theta}_{1}=\bigl({\mu}_1, \sigma_{1}\bigr)^{\top}=\bigl(-3, \sqrt{2}\bigr)^{\top}, \nonumber\\
&\omega_2=\frac{4}{10}, \boldsymbol{\theta}_{2}=\bigl({\mu}_2, \sigma_{2}\bigr)^{\top}=\bigl(0,1\bigr)^{\top}.
\end{align}
The \verb+R+ functions \verb+rmixnorm1+ and \verb+rmixnorm2+ have been developed based on Algorithm \ref{Simulating-mixture-Gaussian-model-1} and Algorithm \ref{Simulating-mixture-Gaussian-model-2}, respectively, to generate from a two-component Gaussian mixture model (\ref{exam-simulation-two-component-Gaussian-2}) with whole parameter given in (\ref{exam-simulation-two-component-Gaussian-3}). Replacing one of the commands given in Table \ref{Simulation-R-command} with \verb+rnorm(n=1, mu=mu[[k]], sigma=sigma[[k]])+ in line 17 of function \verb+rminorm1+ to simulate from other finite mixture models. 
\par For example, the command \verb+rgamma(n=1, shape=shape[[k]], rate=rate[[k]])+ should be replaced with \verb+rnorm(n=1, mu=mu[[k]], sigma=sigma[[k]])+ for simulating from a two-component gamma mixture model with PDF given by 
\begin{align}\label{exam-simulation-two-component-gamma-4}
g({y} \vert \boldsymbol{\Psi})= \frac{6}{10}\times{\cal{G}}\bigl(y\big \vert {a}_{1},b_{1}\bigr)+\frac{4}{10}\times{\cal{G}}\bigl(y\big \vert a_{2}, b_{2}\bigr),
\end{align}
if demanded. The corresponding whole parameter vector becomes
\begin{align}\label{exam-simulation-two-component-gamma-5}
&\omega_1=\frac{6}{10}, \boldsymbol{\theta}_{1}=\bigl(a_1, b_{1}\bigr)^{\top}=(6, 0.6)^{\top}, \nonumber\\
&\omega_2=\frac{4}{10}, \boldsymbol{\theta}_{2}=\bigl(a_2, b_{2}\bigr)^{\top}=(2,1)^{\top}.
\end{align}
Figure \ref{plot-mixture-simulation-Gaussian-univariate} displays the histogram of $n=5000$ simulated data from two-component Gaussian (\ref{exam-simulation-two-component-Gaussian-2}) and gamma (\ref{exam-simulation-two-component-gamma-4}) mixture models. For producing Figure \ref{plot-mixture-simulation-Gaussian-univariate}(a), we used function \verb+rmixnorm1+.
\begin{figure}[!h]
\center
\includegraphics[width=55mm,height=55mm]{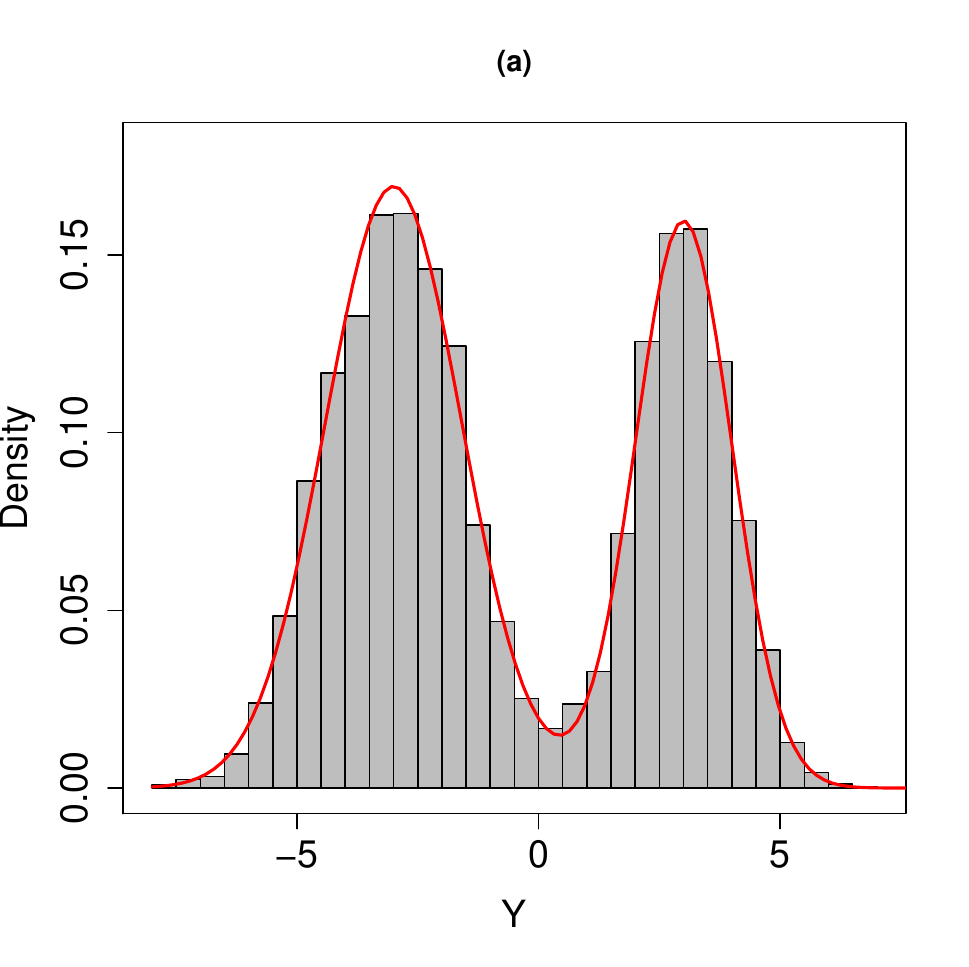}
\includegraphics[width=55mm,height=55mm]{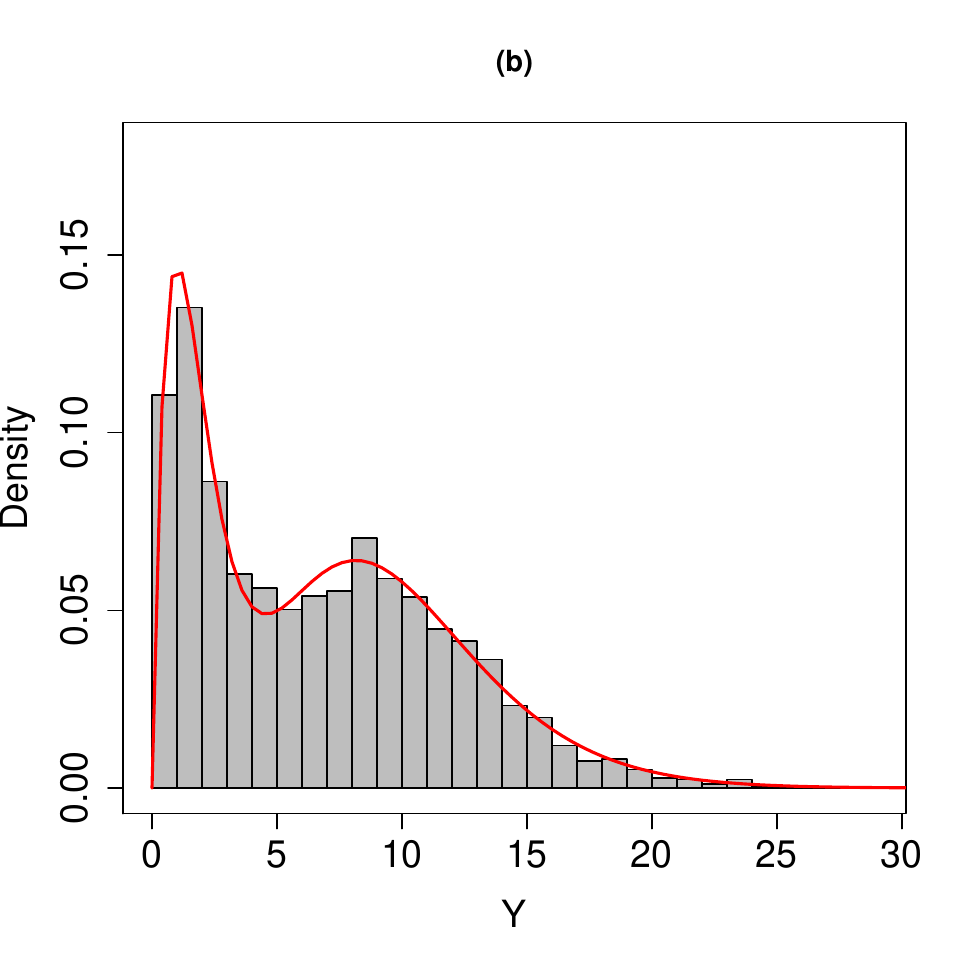}
\caption{(a): histogram of $n=5000$ simulated data from: (a) two-component Gaussian mixture model (\ref{exam-simulation-two-component-Gaussian-2}) and (b): two-component gamma mixture model (\ref{exam-simulation-two-component-gamma-4}). The red-colored superimposed curve is the corresponding PDFs given in (\ref{exam-simulation-two-component-Gaussian-2}) and (\ref{exam-simulation-two-component-gamma-5}), respectively.}
\label{plot-mixture-simulation-Gaussian-univariate}
\end{figure}
\end{example}
\begin{lstlisting}[style=deltaj]
R> set.seed(20240713)
R> n <- 5000; K <- 2; omega <- c(0.6, 0.4);  mu1 <- -3; mu2 <- 0; sigma1 <- 1; sigma2 <- 2 
R> mu <- list(mu1, mu2); sigma <- list(sigma1, sigma2)
R> rmixnorm1 <- function(n, mu, sigma, omega)
+	{
+	p <- length( mu[[1]] )
+	Y <- label <- rep(0, n)
+		for(i in 1:n)
+		{
+		r_MUL <- rmultinom(n = 1, size = 1, omega)
+		k <- apply(r_MUL, 2, which.max)
+		label[i] <- k
+		Y[i] <- rnorm(n = 1, mean = mu[[k]], sd = sigma[[k]])
+		}
+	ls <- list( "sample" = Y, "label" = label )
+	return(ls)
+	}
R> Y <- rmixnorm1(n, mu, sigma, omega)
R> plot( Y$sample, col = Y$label )
\end{lstlisting}
\begin{lstlisting}[style=deltaj]
R> set.seed(20240713)
R> n <- 5000; K <- 2; omega <- c(0.6, 0.4); mu1 <- -3; mu2 <- 3; sigma1 <- sqrt(2); sigma2 <- 1 
R> mu <- list(mu1, mu2); sigma <- list(sigma1, sigma2)
R> rmixnorm2 <- function(n, mu, sigma, omega)
+ {
+ 	Y <- rep(0, n)
+ 	n_k <- rep(0, K)
+ 	n_k <- floor(n*omega)
+ 	cn_k <- cumsum(n_k)
+ 	n_k[K] <- n - sum(n_k[1:(K-1)])
+ 	Y[ 1: cn_k[1] ] <- rnorm(n = n_k[1], mean = mu[[1]], sd = sigma[[1]])
+ 		for(k in 2:K) Y[ ( (k - 1)*cn_k[k-1] + 1): cn_k[k] ] <- 
+ 				rnorm(n = n_k[k], mean = mu[[k]], sd = sigma[[k]])
+ 	ls <- list( "sample" = Y, "label" = rep(1:K, n_k) )
+ return(ls)
+ }
\end{lstlisting}
%two multivariate statistical distributions of the same family whose PDFs are $f\bigl(\boldsymbol{y}\big \vert \boldsymbol{\theta}_{1}\bigr)$ and $f\bigl(\boldsymbol{y}\big \vert \boldsymbol{\theta}_{2}\bigr)$, respectively. The corresponding weights are 0.4 and 0.6 and hence the distribution of ten observations is a mixture of two distributions fitted to two clusters whose PDF is
%\begin{align}\label{mixture-PDF2}
%g(\boldsymbol{y} \vert \boldsymbol{\Psi})=0.4 f\bigl(\boldsymbol{y}\big \vert \boldsymbol{\theta}_{1}\bigr) + 0.6  f\bigl(\boldsymbol{y}\big \vert \boldsymbol{\theta}_{2}\bigr).
%\end{align}
%In (\ref{mixture-PDF2}), the vector $\boldsymbol{\omega}=(0.40,0.60)^{\top}$, is known as the weight vector that means 40\% of observations in Figure \ref{plot-mixture-two-component-schematic}(d) are coming from the first component and the remaining 60\% are coming from the second component.
\subsection{Dirichlet distribution} 
If random vector $\boldsymbol{X}=(X_1,\cdots,X_{\text{K}})^{\top}$ follows a Dirichlet distribution with parameter vector $\boldsymbol{\rho}$, say $\boldsymbol{X}\sim {\cal{DIR}}(\boldsymbol{\rho})$, then the PDF of $\boldsymbol{X}$ is
\begin{align}\label{Dirichlet-PDF-31}
{\cal{DIR}}(\boldsymbol{x}\vert \boldsymbol{\rho})=\frac{\Gamma\bigl(\rho_1 +\cdots+\rho_{{\text{K}}}\bigr)}{\prod_{k=1}^{{\text{K}}}\Gamma\bigl(\rho_{k}\bigr)}x_{1}^{\rho_{1}-1}\times x_{2}^{\rho_{2}-1}\times \cdots \times x_{{\text{K}}}^{\rho_{{\text{K}}}-1},
\end{align}
where $0< x_{k}<1$ and ${\rho}_{k}>0$ (for $k=1,\cdots,{\text{K}}$), provided that $\sum_{k=1}^{{\text{K}}}x_{k}=1$. If ${{\text{K}}}=2$, then Dirichlet distribution turns into Beta distribution with shape parameters $\rho_{1}>0$ and $\rho_{2}>0$, denoted by ${\cal{BET}}(\rho_{1},\rho_{2})$, whose PDF is
\begin{align}\label{Beta-PDF}
{\cal{BET}}(x\big \vert\rho_{1},\rho_{2})=\frac{\Gamma\bigl(\rho_{1} +\rho_{2}\bigr)}{\Gamma\bigl(\rho_{1})\Gamma\bigl(\rho_{2})}x^{\rho_{1}-1}(1-x)^{\rho_{2}-1},
\end{align}
where $0<x<1$. Furthermore, it is worth to note that the $i$th marginal of random vector ${\cal{DIR}}(\boldsymbol{x}\vert \boldsymbol{\rho})$ follows ${\cal{BET}}(\rho_{i},\sum_{k=1}^{{{\text{K}}}}\rho_{k}-\rho_{i})$. 
\par For generating form ${\cal{BET}}(\rho_{1},\rho_{2})$ in \verb+R+, we can use command \verb+rbeta(n, shape1, shape2, ncp = 0)+ in which \verb+n+ is sample size, \verb+shape1+ $=\rho_{1}$, and \verb+shape2+ $=\rho_{2}$. Let $X_{k}\sim {\cal{G}}(\rho_{k},\beta)$ (for $k=1,\cdots,{{\text{K}}}$), then random vector $(X_{1}/Y,\cdots,X_{{\text{K}}}/Y)^{\top}$ follows a Dirichlet distribution with parameter vector $\boldsymbol{\rho}=(\rho_{1}\cdots,\rho_{{\text{K}}})^{\top}$ in which $Y=\sum_{k=1}^{{{\text{K}}}}X_{k}$.
\begin{lstlisting}[style=deltaj]
R> rdirichlet <- function(n, omega)
+{
+	K <- length(omega); beta <- 1
+	X <- matrix(0, nrow = n, ncol = K)
+	for(k in 1:K) X[, k] <- rgamma(n,  shape = omega[k], rate = beta)
+return( X/rowSums(X) )
+}
\end{lstlisting}
%\begin{algorithm}
%\caption{Simulation from random variable with piecewise CDF}
%\label{Simulation-piecewise-CDF-2}
%\begin{algorithmic}[1]
%%\Procedure{}{}     %  \Comment{This is a test}
% %   \State System Initialization
% \State Read $n$ and sequence $\bigl\{F(a_{0}),F(a_{1}),\cdots,F(a_{m})\bigr\}$;
% \State Set $\boldsymbol{s}=(s_{1},\cdots,s_{m})^{\top}$ in which $s_{i}=F(a_{i})-F(a_{i-1})$ for $i=1,\cdots,m$;
% \State Generate $u$ from ${\cal{MUL}}(1, m, \boldsymbol{s})$;
%        \If{$u=$}
%        \State $x = F_{1}^{-1}(u)$;
%        \EndIf
%        \If{$F(a_{1}) < u < F(a_{2})$}
%        \State $x = F_{2}^{-1}(u)$;
%            \EndIf
%           \State ~~~~~~~~~~~~~~\vdots
%        \If{$F(a_{m-2}) < u < F(a_{m-1})$}
%        \State $x = F_{m-1}^{-1}(u)$;
%            \EndIf
%        \If{$F(a_{m-1}) < u < F(a_{m})$}
%        \State $x = F_{m}^{-1}(u)$;
%    \EndIf
%\State return $x$ as a generation from CDF $F(\cdot)$.
%\end{algorithmic}
%\end{algorithm}
%%%%%%%%%%%%%%%%%%%%%%%%%%%%%
\subsection{Chi-squared distribution}
The random variable $X$ is said to have chi-squared distribution with $\nu$ degrees of freedom if its PDF is given by
\begin{align}\label{pdf-chi-square-1}
f(x\vert {\nu}) =\frac{2^{-\frac{\nu}{2}}}{\Gamma\bigl(\nu/2\bigr)}x^{\frac{\nu}{2}-1}\exp\Bigl\{-\frac {x}{2}\Bigr\},
\end{align}
where $\nu \in \mathbb{N}$ is the family parameter. The chi-squared distribution is a special case of gamma distribution. This means that if $X\sim {\cal{G}}(k/2, 1/2)$, then $X\sim {{\chi}}(k)$. For Simulating random variable $X$ following a chi-squared distribution in \verb+R+, we may use command \verb+ rchisq(n, df, ncp=0)+ in which \verb+n+ is the sample size, \verb+df+ is degrees of freedom (here is parameter $\nu$), and, by default, the non-centrality parameter \verb+bncp+ is set to be zero. In fact, $X$ can be represented as the sum of $\nu$ squared standard Gaussian random variables as $X=\sum_{i=1}^{\nu}Z_{i}^{2}$. Hence, as an elementary way, one can simulate a chi-squared random variable $X$ through adding up $\nu$ number(s) independent standard Gaussian random variables. Of course, herein, we suggest to use either \verb+rgamma(n, shape = nu/2, rate = 1/2)+ or \verb+rchisq(n, df = nu, ncp = 0)+.  When the non-centrality parameter is nonzero, then the PDF of a chi-squared distribution can be written as
\begin{align}\label{pdf-chi-square-2}
f(x\vert {\nu}, \lambda) =\frac{1}{2}\Bigl(\frac{x}{2}\Bigr)^{\frac{\nu}{4}-\frac{1}{2}}
\exp\Bigl\{-\frac {x+\lambda}{2}\Bigr\} I_{\nu/2-1 }(\sqrt{\lambda x}),
\end{align}
where $I_{u}(\cdot)$ is the modified Bessel function of the first kind that is given by
\begin{align}
 I_{i}(y)=\Bigl(\frac{y}{2}\Bigr)^{i}\sum _{j=0}^{\infty }\Bigl(\frac{y^{2}}{4}\Bigr)^{j}\frac {1}{\Gamma (j+1)\Gamma (i +j+1)}.
\end{align}
Alternatively, the PDF of each chi-squared distribution can be represented as
\begin{align}\label{pdf-chi-square-3}
f(x\vert {\nu}, \lambda) = \sum_{i=0}^{\infty}\frac{\exp\bigl\{-\lambda/2\bigr\}}{\Gamma (i +1)}\Bigl(\frac{\lambda}{2}\Bigr)^{i}f(x\vert {\nu+2 i}).
\end{align}
As it is seen from (\ref{pdf-chi-square-3}), when $\lambda\neq 0$, then PDF of a chi-squared can be expressed in terms a mixture model whose weights are equal to a Poisson mass function. Therefore, for simulating from a chi-squared distribution with nonzero non-centrality parameter, we adopt the following \verb+R+ code. 
\begin{lstlisting}[style=deltaj]
R> rchisq0 <- function(n, nu = 1, lambda = 0)
+ {
+ X <- rep(NA, n)
+ P <- rpois(n, lambda = lambda/2)
+ for(i in 1:n) X[i] <- rchisq(1, df = nu + 2*P[i])
+ return(X)
+ }
\end{lstlisting}
\subsection{Inverse Gaussian distribution}
For simulating from Inverse Gaussian distribution, denoted as ${\cal{IG}}(\lambda, \mu)$, we follow the method proposed by \cite{michael1976generating}. The positive random variable $X$ is said to follow ${\cal{IG}}(\lambda, \mu)$ if its PDF is given by 
\begin{align}\label{pdf-gi}
f(x\vert\boldsymbol{\theta}) =\sqrt{\frac{\lambda}{2\pi x^3}}\exp\Bigl\{-\frac { \lambda(x-\mu)^{2}}{2\mu^2 x}\Bigr\},
\end{align}
where $\boldsymbol{\theta}=(\lambda>0,\mu>0)^{\top}$. Comparing by the RHS of (\ref{pdf-gig}) and (\ref{pdf-gi}), it is evident that generalized Inverse Gaussian distribution specializes to Inverse Gaussian distribution when $a=-1/2$ and $b=0$. In words, ${\cal{IG}}(\lambda, \mu)={\cal{GIG}}(a=-1/2, b=\lambda/\mu^2, c=\lambda)$. For simulating from ${\cal{IG}}(\lambda, \mu)$, first we note that \cite{shuster1968inverse}:
\begin{align}\label{pdf-gi-1}
Y=g(X)=\frac{\lambda(X-\mu)^2}{\mu^2 X} \sim \chi_{(1)},
\end{align}
where $\chi_{(n)}$ accounts for a chi-square distribution with $n$ degrees of freedom. So, we may think that generating from $X$ is accomplished straightforwardly by drawing a sample from $\chi_{(1)}$ and then $X$ is recovered by a simple inverse transformation. But, care must be taken into the fact that transformation in (\ref{pdf-gi-1}) may have two roots and the problem is here to choose among these two distinct roots $x_{1}$ and $x_{2}$. Overall, based on method of \cite{michael1976generating}, for a given $y$ in (\ref{pdf-gi-1}) there are $k$ distinct roots $\{x_{1},\cdots,x_{k}\}$ (here, we have $k=2$). The problem turns into determining the multinomial probabilities associated with each of $x_{i}s$ (for $i=1,\cdots,k$). Here, we investigate the case in which $X$ is a continuous random variable. Let $f(\cdot)$ and $F(\cdot)$, respectively, denote the PDF and CDF of random variable $X$ that we are interested in sampling from and $p_{i}(y)$ is the conditional (or multinomial) probability with which the $i$th root must be chosen. It is shown that \cite{michael1976generating}: 
\begin{align}\label{pdf-gi-1}
p_{i}(y)=\biggl[1+\sum_{i=1, i\neq j}^{k} \Big \vert \frac{g^{\prime}(x_{i})}{g^{\prime}(x_{j})}\Big \vert \times \frac{f(x_{j})}{f(x_{i})}\biggr]^{-1},
\end{align}
where $g^{\prime}(x)$ denotes the first derivative of transformation $g(x)$ with respect to $x$. Computing $p_{i}(y)$ in (\ref{pdf-gi-1}) for $i=1,\cdots,k$, the $i$th root is chosen as a generation from $X$ with probability $p_{i}(y)$. In what follows, we describe this method when $X\sim {\cal{IG}}(\lambda, \mu)$. Evidently, $g(x)$ given by (\ref{pdf-gi-1}) has two roots, namely $x_{1}$ and $x_{2}$. Then more algebra shows $f(x_{2})/f(x_{1})=(x_{1}/\mu)^{3}$ and $g^{\prime}(x_{i})/g^{\prime}(x_{j})=-(\mu/x_{1})^{2}$. Hence, for a given $y\sim \chi_{(1)}$, the roots $x_{1}$ and $x_{2}$ are
\begin{align*}%\label{pdf-gi-1}
x_{1}=\mu+\frac{\mu^2 y}{2\lambda}-\frac{\mu}{2\lambda}\sqrt{4\mu \lambda y +\mu^2 y^2 },
\end{align*}
and $x_{2}=\mu^2 /x_{1}$ with probabilities
\begin{align*}%\label{pdf-gi-1}
p_{1}(y)=\biggl[ 1+\Big \vert \frac{g^{\prime}(x_{1})}{g^{\prime}(x_{2})}\Big \vert \times \frac{f(x_{2})}{f(x_{1})}\biggr]^{-1}=\frac{\mu}{\mu+x_{1}},
\end{align*}
 and $p_{2}(y)=1-p_{1}(y)=x_{1}/(\mu+x_{1})$, respectively. We adopt the Algorithm \ref{Simulating from Inverse Gaussian distribution} for simulating from $X\sim {\cal{IG}}(\lambda, \mu)$ as follows. 
 \vspace{5mm}
\begin{algorithm}
\caption{Simulating from Inverse Gaussian distribution}
\label{Simulating from Inverse Gaussian distribution}
\begin{algorithmic}[1]
%\Procedure{Generation a sample from from p}{} 
\State Read parameters $\lambda$ and $\mu$;   
\State Generate $u$ from ${\cal{U}}(0, 1)$;
\State Generate $y$ from $\chi_{(1)}$;
\State Set $x_{1}=\mu+\frac{\mu^2 y}{2\lambda}-\frac{\mu}{2\lambda}\sqrt{4\mu \lambda y +\mu^2 y^2 }$;
      \If{$u<\mu/(\mu+x_{1})$}
        \State $x=x_{1}$;
 \Else   
 \State $x=\mu^2/x_{1}$;
     \EndIf
          \State {\bf{end}}
     \State Accept $x$ as a generation form ${\cal{IG}}(\lambda, \mu)$.
% \EndProcedure
\end{algorithmic}
\end{algorithm}
\vspace{5mm}
The corresponding \verb+R+ function \verb+rig(n, lambda, mu)+, based on Algorithm \ref{Simulating from Inverse Gaussian distribution}, for generating $n$ realizations from ${\cal{IG}}(\lambda, \mu)$ is given by the following.
\begin{lstlisting}[style=deltaj]
R> rig <- function(n, lambda, mu)
+   {
+     y <- rchisq(n, df =1)
+     u <- runif(n)
+     x_1 <- mu + mu^2*y/(2*lambda) - mu/(2*lambda)*sqrt(4*mu*lambda*y + mu^2*y^2)
+     x_2 <- mu^2/x_1
+     x <- ifelse( u < mu/(mu + x_1), x_1, x_2 )
+    return(x)
+ }
\end{lstlisting}
\subsection{$\alpha$-Stable distribution}
Herein, we briefly describe the family of $\alpha$-stable distributions. This family of distributions
was introduced by Paul Pierre L\'{e}vy, a French mathematician, in his work on sums of independent identically distributed random variables around 1924. Almost all of members of this family either have not closed form for PDF or are only expressed in terms of the special functions \citep{zolotarev1986one,taqqu1994}. So, this family of distributions are represented in terms of their characteristic function. The characteristic function of $\alpha$-stable distribution has different parameterizations ( forms). The most commonly used parameterizations for is given as follows \citep{nolan1998parameterizations}.
\begin{eqnarray}\label{stable-chf}
\displaystyle
E\bigl(\exp\{j t X\}\bigr)=\left\{\begin{array}{c}
\displaystyle
\exp\left\{-\lvert \sigma t \rvert^\alpha
\left[1-j\beta
\
\displaystyle
\mathrm{sign}(t)\tan\bigl(\frac {\pi \alpha}{2}\bigr)\right]+j t\mu \right\},
\mathrm{{if}}
\
\alpha \ne 1,
\\
\displaystyle
\exp\left\{-\lvert \sigma t \rvert \Bigl[1+j\beta
\
\displaystyle
\mathrm{sign}(t)\frac {2}{\pi}\log \lvert t \rvert \Bigr]+j t\mu \right\},~~~~
\mathrm{if}
\
\alpha= 1,
\end{array} \right.
\end{eqnarray}
where $j^{2}=-1$ and $\text{sign(u)}$ is the sign function that takes values -1 and +1 for $u<0$ and $u\geq 0$, respectively. Other forms for characteristic function of $\alpha$-stable distributions are {A}, {M}, {B}, {C}, and {E} \citep{zolotarev1986one}. The characteristic function given in (\ref{stable-chf}) is known in the literature as $S_{1}$ parameterization that is a slightly different version of {A} parameterization. We write $X\sim S(\alpha,\beta, \sigma, \mu)$ to denote that random variable $X$ follows an $\alpha$-{stable} distribution with parameters $\alpha$, $\beta$, $\sigma$, and $\mu$ in $S_{1}$ parameterization. Here, parameter $\alpha \in (0,2]$ is the index of stability, which determines the tail thickness of the density function. When $\alpha=2$, the tail thickness is at its thinnest while smaller values of $\alpha$ would yield thicker tails for the  distribution. Similarly, the parameter $\beta \in [-1, 1]$ is the degree of skewness (-1 for totally skewed to the left, zero for symmetrical, and +1 for totally skewed to the right), $\sigma \in \mathbb{R}^{+}$, is the dispersion (scale) around the location parameter $\mu$ and $\mu \in \mathbb{R}$, is the horizontal shift of the density function. It is important to note that the parameters $\sigma$ and $\mu$, in general, are not synonymous with standard deviation and location parameter, as they are in the special case of Gaussian distribution. The only members of this class with closed form PDF are L\'{e}vy ($\alpha=0.5$ and $\beta=1$), Cauchy ($\alpha=1$ and $\beta=0$) and Gaussian ($\alpha=2$). If $\beta=0$, and $\mu=0$, then the class of zero-location symmetric $\alpha$-stable (S$\alpha$S) distribution can be identified with the following characteristic function
\begin{align} \label{chf1}
\phi(t)=\exp\bigl\{- \sigma^{\alpha} \vert t \vert^{\alpha}\bigr\}.
\end{align}
The known members of S$\alpha$S distribution with closed from density function are Cauchy ($\alpha=1$) and Gaussian ($\alpha=2$). Here, it is noteworthy to mention that the $k$th moment of each $\alpha$-{stable} distribution is finite if $k<\alpha$. This means that variance of non-Gaussian $\alpha$-stable distribution is not finite and its first moment is finite if $\alpha>1$. For $\alpha>1$ then $\mu$ becomes the expected value of distribution. For $\alpha<1$, the support of class $S(\alpha, 1, \sigma, 0)$ only covers the positive semi-axis. For this reason, if $\alpha<1$, $\beta=1$, and $\mu=0$, then the $\alpha$-{{stable}} distribution family is also called positive $\alpha$-{stable} distribution. An important member of the positive $\alpha$-{stable} family is $S(\alpha/2,1, [\cos(\pi \alpha/4)]^{2/\alpha}, 0)$ whose distribution function is given by \citep{ibragimov1959}:
\begin{equation} \label{positivestable-cdf-1}
F_{P}(p \vert \alpha)=\frac{1}{\pi}\int_{0}^{\pi}\exp\Bigl\{-p^{-\frac{\alpha}{2-\alpha}}A(u)^{}\Bigr\}du,
\end{equation} 
where $[A(u)]^{2/(2-\alpha)}$ in the right-hand side of (\ref{positivestable-cdf-1}) is called the Zolotarev's function \cite{devroye2009random}, defined for $0<\alpha < 2$ as
\begin{equation} \label{A}
A(u)=
%\left[
\frac{\Bigl\{\sin\bigl[ \bigl(\frac{\alpha}{2}\bigr)u\bigr]\Bigr\}^{\frac{\alpha}{2}}\Bigl\{\sin \Bigl[\bigl (1-\frac{\alpha}{2}\bigr) u\Bigr]\Bigr\}^{1-\frac{\alpha}{2}}}{\sin (u)}
%\right]^{\frac{2}{2-\alpha}}.
\end{equation} 
Let $W\sim {\cal{EXP}}(1)$ and are $U\sim {\cal{U}}(0,\pi)$ independent. It follows from (\ref{positivestable-cdf-1}) that 
\begin{equation} \label{positivestable-cdf-2}
F_{P}(p \vert \alpha)=P\Bigl\{W>p^{-\frac{\alpha}{2-\alpha}}A(U)\Bigr\}.
\end{equation} 
A straightforward method was suggested by \cite{kanter1975} for simulating $P\sim S(\alpha/2,1, [\cos(\pi \alpha/4)]^{2/\alpha}, 0)$ as follows.
\begin{equation} \label{positivestable-cdf-2}
P\mathop=\limits^d \Bigl(\frac{A(U)}{W}\Bigr)^{\frac{2-\alpha}{\alpha}}.
\end{equation} 
Let $B(\alpha)=1-\vert 1- \alpha \vert$ and
\begin{align} \label{positivestable-cdf-3}
a^{\alpha}=\sin\Bigl(\frac{\pi B(\alpha)(1+\beta)}{2}\Bigr)/\sin\bigl(\pi B(\alpha)\bigr),\nonumber\\
b^{\alpha}=\sin\Bigl(\frac{\pi B(\alpha)(1-\beta)}{2}\Bigr)/\sin\bigl(\pi B(\alpha)\bigr).\nonumber\\
\end{align}  
Furthermore, suppose $Z\sim S(\alpha,\beta,1, 0)$ and $P_{1},P_{2}\sim S(\alpha/2,1, [\cos(\pi \alpha/4)]^{2/\alpha}, 0)$ are independent. A method for simulating $Z$ in standard case ($\sigma=1$ and $\mu=0$) was proposed by \cite{chambers1976method} in $B$ parameterization based on the fact that $Z \mathop=\limits^d a P_{1}+b P_{2}$. This method was elaborated to generate from $Z$ in commonly used parameterization (\ref{stable-chf}) with more detailed information by \cite{weron1996on}. We refer reader to \cite{nolan2003stable} for a corrected version of the method for simulating $Z\sim S(\alpha,\beta,1, 0)$. Let $\theta_0=\arctan(\beta \tan(\pi\alpha/2))$. We have
\begin{eqnarray}\label{stable-simulation-1}
\displaystyle
Z\mathop=\limits^d \left\{\begin{array}{c}
\displaystyle
\frac{2}{\pi}\biggl[\bigl(\frac{\pi}{2}+\beta U_1\bigr)\tan\bigl(U_1\bigr)-\beta\log\biggl(\frac{\frac{\pi}{2} W\cos(U_1)}{\frac{\pi}{2}+\beta U_1}\biggr)\biggr]
\displaystyle,~~~~~~~~~~
\mathrm{if}
\
\alpha= 1,\\
\displaystyle
\frac{
\sin\bigl[\alpha \bigl(\theta_{0}+U_1\bigr)\bigr]}{\bigl[\cos\bigl(\alpha\theta_{0}\bigr) \cos(U_1)\bigr]^{1/\alpha}}\biggl[\frac{\cos\bigl(\alpha\theta_{0} +(1-\alpha)U_1\bigr)}{W}\biggr]^{(1-\alpha)/\alpha}
\
\displaystyle,~
\mathrm{{if}}
\
\alpha \ne 1,
\end{array} \right.
\end{eqnarray}
where $W\sim {\cal{EXP}}(1)$ is independent of $U_{1}\sim {\cal{U}}(-\pi/2,\pi/2)$. For general case ($\sigma\neq 1$ and $\mu \neq0$), we use the following transformation for generating from $X\sim S(\alpha,\beta,\sigma, \mu)$.
\begin{eqnarray}\label{stable-simulation-2}
\displaystyle
X\mathop=\limits^d \left\{\begin{array}{c}
\displaystyle
\sigma Z+ \mu + \beta \frac{2}{\pi} \sigma \log \sigma,~\mathrm{if}~~\alpha=1,\\
\sigma Z+ \mu,~~~~~~~~~~~~~~~~~ \mathrm{{if}}~~\alpha \ne 1,
\end{array} \right.
\end{eqnarray}
where $Z\sim S(\alpha,\beta,1, 0)$ is defined as (\ref{stable-simulation-1}). The class of $\alpha$-stable distributions possesses several nice properties. For instance, suppose $X_{1},X_{2},X_{3}\sim S(\alpha,\beta,\sigma, \mu)$ are independent. Then, for any given positive constants such as $a$ and $b$, we have
\begin{align*}
aX_{1}+bX_{2}\mathop=\limits^d \bigl(a^{\alpha}+b^{\alpha}\bigr)^{1/\alpha} X_{3}.
\end{align*}  
Furthermore, if $X_{1}\sim S(\alpha,\beta,\sigma, \mu)$ and $X_{2}\sim S(\alpha,-\beta,\sigma, \mu)$, then $X_{1}\mathop=\limits^d -X_{2}$. The latter is called reflection property \citep{nolan2003stable}. The \verb+R+ function \verb+stable(n, alpha, beta, sigma, mu)+, given below, can be used for simulating $n$ realizations of $X\sim S(\alpha,\beta,\sigma, \mu)$ in $S_1$ parameterization.
\begin{lstlisting}[style=deltaj]
R<- rstable <- function(n, alpha, beta, sigma, mu)
+{
+	u1 <- pi*( runif(n) - 1/2 )
+	theta0 <- atan( beta*tan(pi*alpha/2) )/alpha
+	w <- -log( runif(n) )
+		if (alpha == 1)
+		{
+		z <- 2/pi*( (pi/2 + beta*u1)*tan(u1) - beta*log( ( pi/2*w*cos(u1) )/(pi/2 + +beta*u1) ) )
+		x <- sigma*z + 2/pi*beta*sigma*log(sigma) + mu
+		}else{
+		z <- sin(alpha*(theta0 + u1))/(cos(alpha*theta0)*cos(u1))^(1/+alpha)*(cos(alpha*theta0 + (alpha - 1)*u1)/w)^( (1 - alpha)/alpha )
+		x <- sigma*z + mu
+		}
+return(x)
+}
\end{lstlisting}
\subsection{Birnbaum-Saunders distribution}\label{distribution-BS}
The Birnbaum-Saunders (BS) distribution has attracted much attention in life-testing and survival analysis. One interesting feature of the BS distribution is the role of the central limit theorem for constructing this distribution. Suppose a sequence of periodic loading are applied to a segment of metal, thereby a sequence of cracks with non-zero length in metal is produced. The total time $T$, until failure is called BS random variable. The CDF of the BS random variable $T$, is given by
\begin{align}\label{pdf-BS-1}
F(t\vert \boldsymbol{\theta})= \Phi\biggl(\frac{1}{\alpha}\bigg[\sqrt{\frac{t}{\beta}} - \sqrt{\frac{\beta}{t}}\bigg]\biggr),
\end{align}
where $t>0$, $\Phi(\cdot)$ is the standard Gaussian CDF, and $\boldsymbol{\theta}=(\alpha,\beta)^{\top}$. Here, $\alpha>0$ and $\beta>0$ are the shape and scale parameters of the BS distribution, respectively. The BS distribution has received much attention in a wide range of fields. A review of its applications can be found in \cite{ng2003modified,teimouri2023fast} and references therein. Differentiating the CDF given in (\ref{pdf-BS-1}), then the PDF of the BS distribution is given by
\begin{align}\label{pdf-BS-2}
f(t\vert\boldsymbol{\theta})=\frac{\sqrt{\frac{t}{\beta}}+ \sqrt{\frac{\beta}{t}}}{2\sqrt{2\pi}\alpha t} \exp\biggl\{-\frac{1}{2\alpha^2}\biggl(\sqrt{\frac{t}{\beta}} - \sqrt{\frac{\beta}{t}}\biggr)^2\biggr\}.
\end{align}
We write $T\sim {\cal{BS}}(\alpha,\beta)$ to indicate that random Variable $T$ follows a BS distribution with CDF (\ref{pdf-BS-1}). If random variable $T$ follows the BS distribution with PDF given by (\ref{pdf-BS-2}) can be represented as a mixture of two GIG distributions as
\begin{align}\label{pdf-BS-3}
f(t \vert \boldsymbol{\theta})= \frac{1}{2} g\bigl(t\big \vert \boldsymbol{\theta}_{1}\bigr) + \frac{1}{2}g\bigl(t\big \vert \boldsymbol{\theta}_{2}\bigr),
\end{align}
where $g(\cdot\vert\boldsymbol{\theta}_{i})$ (for $i=1,2$) is the PDF of GIG distribution\citep{hormann2014generating}, $\boldsymbol{\theta}_{1}=\bigl(1/2,1/(\alpha^2\beta),\beta/\alpha^2 \bigr)^{\top}$, and $\boldsymbol{\theta}_{2}=\bigl(-1/2,1/(\alpha^2\beta),\beta/\alpha^2 \bigr)^{\top}$. This means that each BS random variable can be simulated through a two-component GIG mixture model with weight (or mixing) vector $\boldsymbol{\omega}=(0.5,0.5)^{\top}$. It is not hard to check that
\begin{align}
 X\mathop=\limits^d{\frac {1}{2}}\biggl(\sqrt{\frac{T}{\beta }}-\sqrt{\frac{\beta } {T}}\biggr),
\end{align}
where $X\sim {\cal{N}}(0,\alpha^2/4)$ or equivalently $T=\beta \bigl[1+2X^{2}+2X(1+X^{2})^{1/2}\bigr]$. Among distributional properties of BS distribution, we mention that if $T\sim {\cal{BS}}(\alpha,\beta)$, then $1/T\sim {\cal{BS}}(\alpha,1/\beta)$. The \verb+R+ function \verb+rbs(n, alpha, beta)+, given by the following, can be used for simulating $n$ realizations of $T\sim {\cal{BS}}(\alpha,\beta)$.
\begin{lstlisting}[style=deltaj]
R> rbs <- function(n, alpha, beta)
+{
+	X <- alpha/2*rnorm(n)
+	T <- beta*( 1 + 2*X^2 + 2*X*sqrt( 1 + X^2 ) )
+ return(T)
+}
\end{lstlisting}

%%%%%%%%%%%%%%%%%%%%%%%%%%%%%%%%%%%%%%%%%%%%%%%%%%%%%%%%%
%\chapter{Rejection sampling}
%\begin{chapquote}{Bradley Efron, Stanford University}
%``Statistics did not come
%naturally to me. Dad’s
%keeping score for the
%baseball league helped a lot'' \footnote{\url{https://rss.onlinelibrary.wiley.com/doi/pdf/10.1111/j.1740-9713.2010.00460.x}}
%\end{chapquote}
\section{Rejection sampling}
\begin{figure}[h]
\center
\includegraphics[width=55mm,height=55mm]{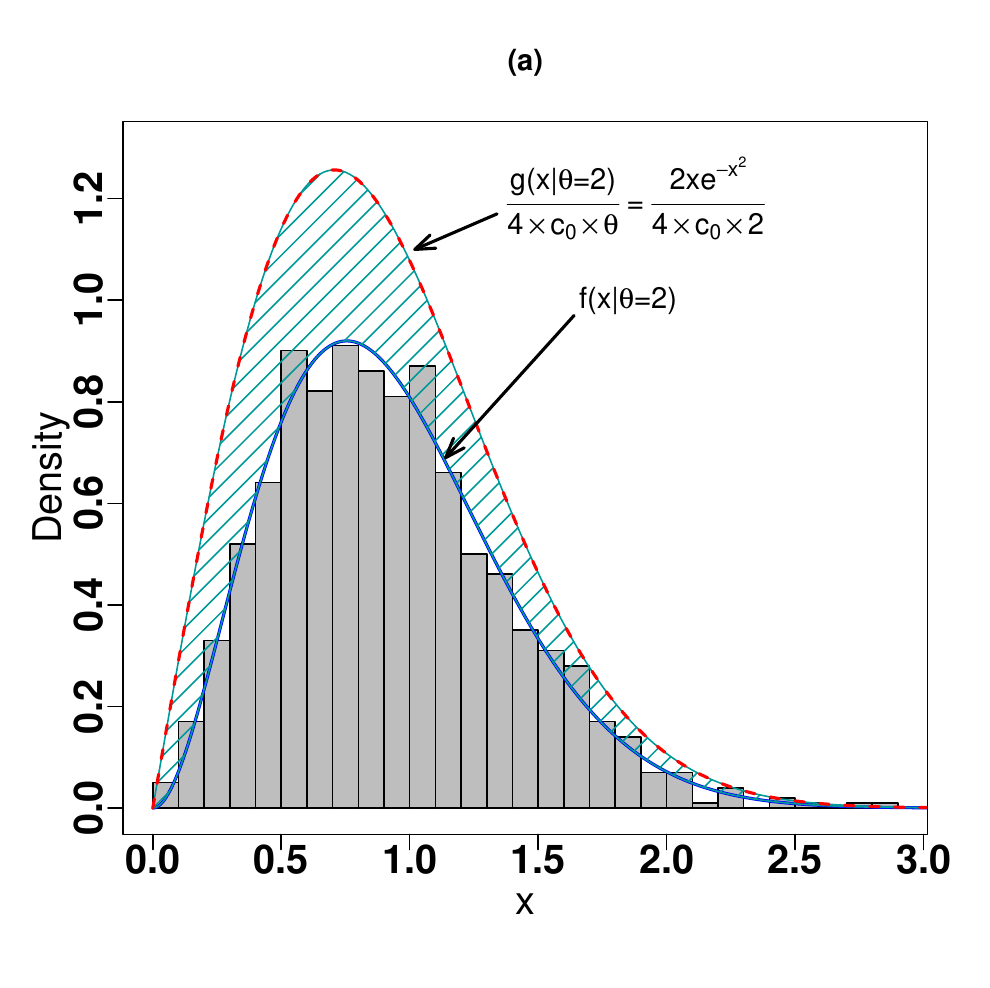}
\includegraphics[width=55mm,height=55mm]{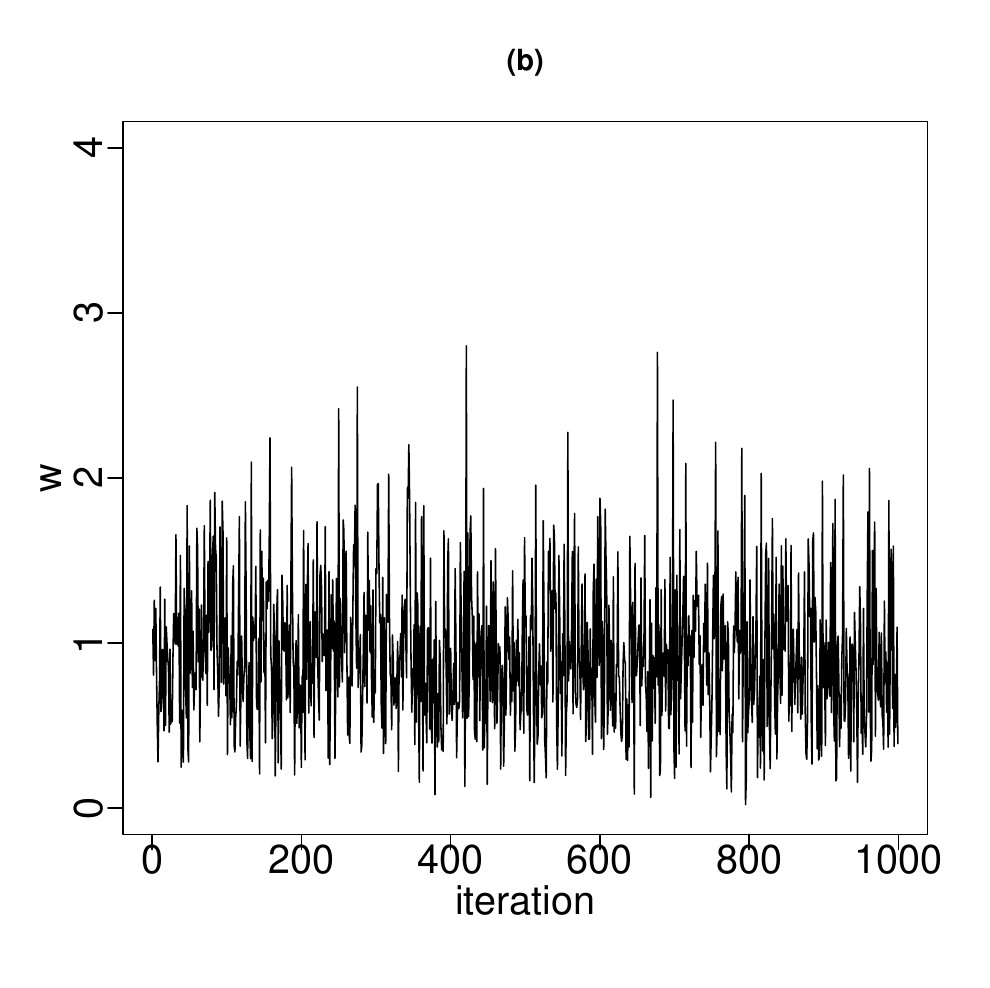}
\caption{(a): Histogram constructed based on $1000$ samples generated from ${\cal{W}}(\theta=2, 1)$. Superimposed are $M g(x\vert \theta)$ (red line) and ${\cal{W}}(x\vert \theta=2, 1)$ (blue line). (b): Generated samples across iterations for sampling from $f(x\vert \theta=2)={\cal{W}}(x\vert \theta=2, 1)$.}
\label{fig2}
\end{figure}

Suppose sampling form PDF $f(\cdot\vert \theta)$ is not easy but it is dominated by another PDF such as $g(\cdot\vert \theta)$ so that
\begin{align*}
\underset{x \in {\cal{S}}_{f}}{\operatorname{sup}}\frac{f(x\vert\theta)}{g(x\vert\theta)} \leq M,
\end{align*}
where $M\geq 1$ may depend on $\theta$. A proposal with PDF $g(x\vert\theta)$ which is as possible as close to $f(x\vert\theta)$ and easy-to-sample from would be the key objective in each rejection sampling scheme. The following four-step algorithm shows how a rejection sampling scheme works \citep{devroye1986}. It is proved that average numbers of trials for generating a sample in each rejection sampling scheme is $M^{-1}$, see \cite{ross2022simulation}. Figure \ref{fig2}(b) shows the generated samples across $1000$ iterations.
\vspace{5mm}
\begin{algorithm}
\caption{Rejection sampling technique}
\label{Rejection sampling technique}
\begin{algorithmic}[1]
%\Procedure{}{}     %  \Comment{This is a test}
 %   \State System Initialization
 %   \State Read the value 
%    \If{$condition = True$}
%        \State Do this
%        \If{$Condition \geq 1$}
%        \State Do that
%        \ElsIf{$Condition \neq 5$}
%        \State Do another
%        \State Do that as well
%        \Else
%        \State Do otherwise
%        \EndIf
%    \EndIf
\State Choose suitable proposal with PDF $g(\cdot \vert \theta)$ and determine $M$;
\State Generate $y\sim g(\cdot \vert \theta)$;
\State Generate $u\sim {\cal{U}}(0, 1)$;
\State Squeezing test: If $ u<\frac{f(y\vert\theta)}{M \times g(y\vert\theta)}$, then accept $y$ as a generation from $f(\cdot\vert\theta)$,
  otherwise return to step 2 and repeat algorithm.
%\EndProcedure
\end{algorithmic}
\end{algorithm}
\vspace{5mm}
\begin{example}\label{exam-weibullsimulation-rejection}%\lipsum*[]
Suppose $f(x\vert \theta)$ is defined as in Example \ref{exam-weibull}. Herein, we can rewrite $f(x\vert \theta)= (x+1)^{-2} x^{\theta}\exp\{-x^\theta\}/c_{0}$ where $c_{0}$ is normalizing constant guarantees $\int_{{\cal{S}}_{f}}f(x\vert \theta) dx=1$. Considering $g(x\vert\theta)={\cal{W}}(x\vert \theta, 1)$ as the proposal, it is easy to check that
\begin{align}\label{ratioexample2}
%\underset{x \in {\cal{S}}_{f}}{\operatorname{sup}}
\frac{f(x\vert\theta)}{g(x\vert\theta)}=
%\underset{x \in (0,\infty)}{\operatorname{sup}} 
\frac{x}{c_{0}\theta(x+1)^2}.
%=\frac{\theta}{4}
%=c(\theta).
\end{align}
Differentiating the RHS of (\ref{ratioexample2}) and setting the resultant to 0 reveals that the maximal value of RHS of (\ref{ratioexample2}) is $1/(4c_{0}\theta)$ that is obtained for $x=1$. That is
\begin{align*}
\frac{f(x\vert\theta)}{g(x\vert\theta)} \leq
\frac{1}{4c_{0}\theta}=M.
\end{align*}
%Herein, if we set $\theta=5$, then $c_0=0.0490788$. 
It follows that
\begin{align*}
\frac{f(x\vert\theta)}{M\times g(x\vert\theta)} =\frac{4x}{(x+1)^2}.
\end{align*}
Based on Algorithm \ref{Rejection sampling technique}, the rejection sampling scheme for generating realization from $f(\cdot\vert\theta=2)$ is described as follows.
\begin{enumerate}
\item Generate $y\sim {\cal{W}}(\theta=2, 1)$;
\item Generate $u\sim {\cal{U}}(0, 1)$;
\item If $u<4y/(y+1)^2$, then accept $y$ as a generation from $f(\cdot\vert\theta=2)$; otherwise go to step 1 and repeat algorithm.
\end{enumerate}
\end{example}
In Example \ref{exam-weibullsimulation-rejection} above, the usage time study for generating $n=1000$ realizations is around 0.01 second. The pertaining \verb+R+ code, for producing Figure \ref{fig2}(b), is given as follows.
\begin{lstlisting}[style=deltaj]
R > set.seed(20240519)
R > n <- 10000  # size of generations
R > theta <- 2
R > w <- rep(0, n) # vector of generated realizations
R > j <- 1
R > while(j <= n)
+	{
+		y <- rweibull(1, shape = theta, scale = 1)
+		u <- runif(1) 
+		if( u < 4*y/(y+1)^2 )
+		{
+			w[j] <- y        # y is accepted
+			j <- j + 1
+		}
+	}
R > plot(w)
\end{lstlisting}
\begin{example}\label{exam-skewnormalsimulation-rejection}%\lipsum*[2]
Let random variable $X$ follow a distribution with PDF given by \cite{azzalini1985class}:
\begin{align}\label{PDF-sg}
f(x\vert\mu, \sigma,\lambda)=\frac{2}{\sigma}\phi\Bigl(\frac{x-\mu}{\sigma}\Bigr)\Phi\Bigl(\lambda\frac{x-\mu}{\sigma}\Bigr),
\end{align}
where $\phi(\cdot)$ and $\Phi(\cdot)$ represent accordingly the PDF and CDF of the standard Gaussian distribution. Herein, $\lambda \in {\mathbb{R}}$ is the skewness parameter (the cases $\lambda \rightarrow-\infty$, $\lambda=0$, and $\lambda \rightarrow+\infty$ yield totally skewed to the left, symmetric (ordinary Gaussian), and totally skewed to the right distributions, respectively), $\mu \in {\mathbb{R}}$ is the location parameter, and $\sigma \in {\mathbb{R}}^{+}$ denotes the scale parameter. Hereafter, we write $X\sim{\cal{SG}}(\mu, \sigma,\lambda)$ to denote that $X$ follows the distribution with PDF given by (\ref{PDF-sg}). Without loss of generality, let $\mu=0$ and $\sigma=1$. A direct method proposed by \cite{henze1986probabilistic} for simulating from ${\cal{SG}}(0, 1,\lambda)$ as follows. Let $Z_1$ and $Z_2$ denote two independent copies of the standard Gaussian random variate. Then, for random variable $X\sim{\cal{SG}}(0, 1,\lambda)$ we can write
\begin{align}\label{Rep-sg}
X\mathop=\limits^d \frac{\lambda \big \vert Z_1\big \vert + Z_2}{\sqrt{1+\lambda^2}}.
\end{align}
For simulating from  ${\cal{SG}}(0, 1,\lambda)$ using the rejection method, let $f(x\vert 0, 1,\lambda=4)$ to be the PDF of ${\cal{SG}}(x\vert 0, 1,\lambda=4)$. We proceed by considering the fact that
\begin{align}\label{rejection-sg-bound}
f(x\vert 0, 1,\lambda)\leq 2 g(x\vert \theta)=2 \phi(x).
\end{align}
Hence, simulating from ${\cal{SG}}(0, 1,\lambda)$, through the rejection sampling, is possible by considering the proposal PDF to be $\phi(x)$ and $M=2$. Figure \ref{fig-rejection-skewgaussian}(a) shows histogram of generated realizations as well as the superimposed target and proposal PDFs for $\lambda=4$. The sampler's motion across 500 iterations are displayed in Figure \ref{fig2}(b).
\begin{figure}[h]
\center
\includegraphics[width=55mm,height=55mm]{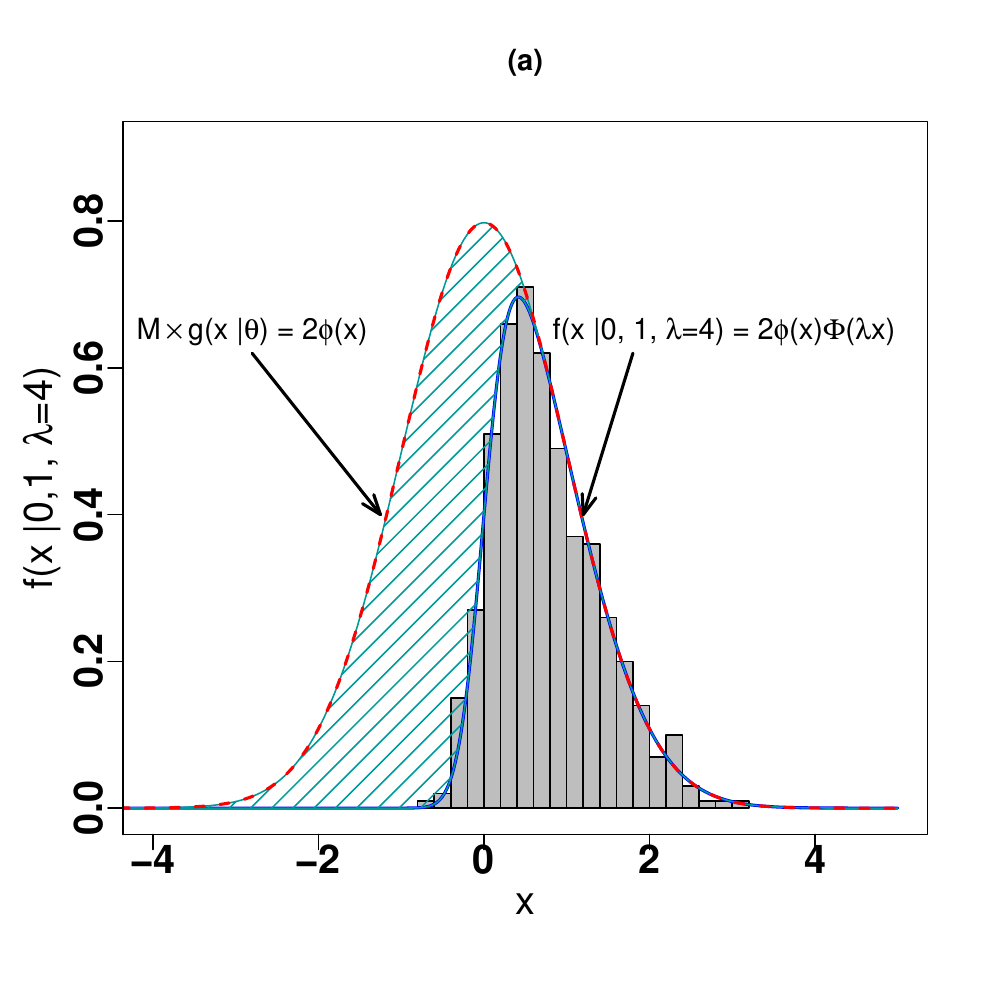}
\includegraphics[width=55mm,height=55mm]{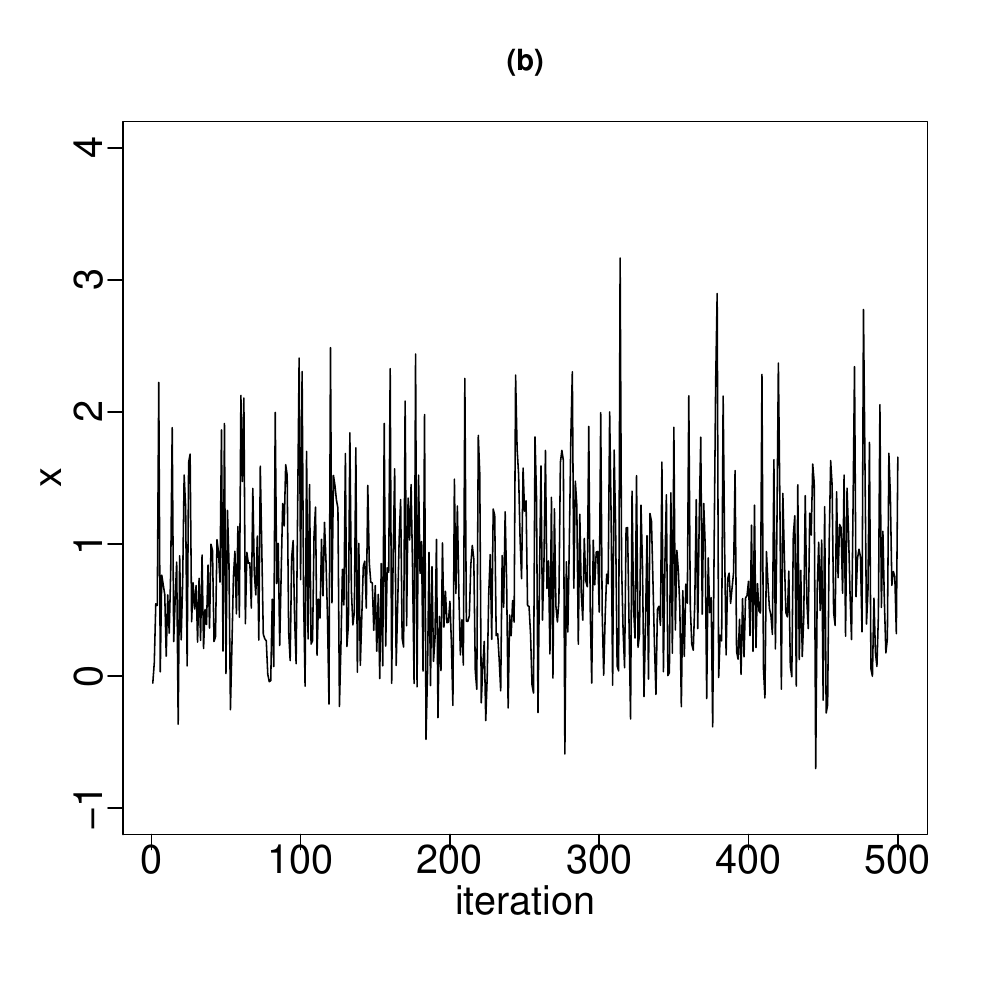}
\caption{(a): Histogram constructed based on 500 samples generated from ${\cal{SG}}(0, 1,\lambda=4)$. Superimposed are $M g(x\vert \theta)=2\phi(x)$ (red line) and $f(x\vert 0, 1,\lambda=4)$ (blue line). (b): Generations across iterations for sampling from $f(x\vert 0, 1,\lambda=4)\sim {\cal{SG}}(0, 1,\lambda=4)$.}
\label{fig-rejection-skewgaussian}
\end{figure}
\end{example}
The pertaining \verb+R+ code, for producing Figure \ref{fig-rejection-skewgaussian}(b) is given as follows.
\begin{lstlisting}[style=deltaj]
R> set.seed(20240225)
R> n <- 500
R> i <- 1; x <- rep(0, n); lambda <- 4
R> while(i <= n)
+ {
+ z <- rnorm(1);
+		 if( runif(1) < pnorm(lambda*z) )
+		   {
+		     x[i] <- z; i <- i + 1
+		   }
+  }
R > plot(x)
\end{lstlisting}
%%%%%%%%%%%%%%%%%%%%%%%%%%%%%%%%
\section{Adaptive rejection sampling}
Suppose $f(x\vert\theta)$ is a log-concave PDF over ${\cal{S}}_{f}$ that equivalently means ${\cal{H}}(x)=\log f(x\vert\theta)$ is a concave function of $x$. A method was proposed by \cite{luc1986} for generating realization form log-concave distribution  provided that the mode (a point at which derivative of $f(x\vert\theta)$ with respect to $x$ is zero) of $f(x\vert\theta)$ to be known. More precisely, the proposed method is a rejection sampling scheme using a three-piece exponential proposal in which the center piece touches the target PDF $f(x\vert\theta)$ at its mode. 
%\cite{hormann2004automatic}
%The rejection sampling approach does not work well if finding optimum $c$ can be a difficult task. 
While this method is not adaptive, the adaptive rejection sampling (ARS) scheme \citep{gilks1992adaptive} develops this method by improving the proposal distribution in each iteration such that the updated proposal proposes generation that passes the {\it{squeezing}} test with higher probability than the previous iteration. 
%%To this end, ARS algorithm assumes that $f(x\vert\theta)$ is called log-concave over ${\cal{S}}_{f}$ or ${\cal{H}}(x)=\log f(x\vert\theta)$ is a concave function of $x$. 
In each ARS scheme, %%for $(x, a) \in {\cal{S}}_{f}$, function 
${\cal{H}}(x)$ can be bounded by piecewise linear upper hull in which each tangent line is defined as 
\begin{align}\label{uk}
U_{k}(x)=(x-a){\cal{H}}^{\prime}(a)+{\cal{H}}(a),
\end{align}
for $x \in [z_{i-1},z_{i}]$ and $i=1,\cdots,k$ where
\begin{align*}
z_{i}=\frac{{\cal{H}}\bigl(x_{i+1}\bigr)-{\cal{H}}\bigl(x_{i}\bigr)-x_{i+1} {\cal{H}}^{\prime}\bigl(x_{i+1}\bigr) +x_{i}{\cal{H}}^{\prime}\bigl(x_{i}\bigr)}{{\cal{H}}^{\prime}\bigl(x_{i}\bigr)-{\cal{H}}^{\prime}\bigl(x_{i+1}\bigr)}.
\end{align*}
We note that the tangent lines are constructed at abscissa (points) $T_{k}=\{x_1 \leq x_2 \leq \cdots \leq x_k  \in {\cal{S}}_{f}\}$ and intersect at $z_i$ where $z_0$ and $z_k$ are the lower ($-\infty$ if ${\cal{S}}_{f}$ is not bounded from below) and the upper ($\infty$ if ${\cal{S}}_{f}$ is not bounded from above) bounds of ${\cal{S}}_{f}$, accordingly. Likewise, we can define a piecewise linear lower hull constructed based on chords between adjacent abscissae in $T_{k}$ as 
\begin{align}\label{lk}
L_{k}(x)=\frac{{\cal{H}}\bigl(x_{i+1}-x\bigr){\cal{H}}\bigl(x_{i}\bigr)-
                  {\cal{H}}\bigl(x-x_{i}\bigr){\cal{H}}\bigl(x_{i+1}\bigr)}{
                  x_{i+1}x_{i}},
\end{align}
where $x \in [x_{i},x_{i+1}]$ for $i=1,\cdots,k-1$. If $x<x_1$(or $x>x_k$), then we set $L_{k}(x)=-\infty$. Notice that log-concavity of $f(x\vert\theta)$ ensures that $L_{k}(x) \leq {\cal{H}}(x) \leq U_{k}(x)$ for $x \in {\cal{S}}_{f}$. Each ARS algorithm proceeds three {\it{initialization}}, {\it{sampling}}, and {\it{updating}} steps for generating $n$ realizations from PDF $f(x\vert\theta)$ given by the following.   
\begin{itemize}
\item {\it{initialization step}}: Construct the abscissae in $T_{k}$. If ${\cal{S}}_{f}$ is left-unbounded, then set $x_{1}$ such that ${\cal{H}}^{\prime}\bigl(x_{1}\bigr)<0$ and if ${\cal{S}}_{f}$ is right-unbounded, then set $x_{k}$ such that ${\cal{H}}^{\prime}\bigl(x_{k}\bigr)<0$. Furthermore, compute $L_{k}(x)$, $U_{k}(x)$, and $S_{k}(x)$ where
\begin{align*}
s_{k}(x)=\frac{\exp\{U_{k}(x)\}}{\int_{}\exp\{U_{k}(y)\} dy},
\end{align*}
and functions $L_{k}(x)$ and $U_{k}(x)$ as defined in (\ref{lk}) and (\ref{uk}), accordingly.
\item {\it{sampling step}}: Generate $x^{*}$ from $s_{k}(x)$ and $u\sim \text{Unif}(0, 1)$ independently; and then perform squeezing test that accepts $x^{*}$ if $w\leq \exp\{L_{k}(x^{*})-U_{k}(x^{*})\}$; otherwise compute the rejection test that accepts $x^{*}$ if $w\leq \exp\{{\cal{H}}\bigl(x^{*}\bigr)-U_{k}(x^{*})\}$; otherwise reject $x^{*}$.
\item {\it{updating step}}: If evaluating ${\cal{H}}\bigl(x^{*}\bigr)$ and ${\cal{H}}^{\prime}\bigl(x^{*}\bigr)$ is necessary, then add $x^{*}$ to
$T_{k}$ in order to form updated $T_{k}$ as $T^{*}_{k}=\bigl\{ \{x_1 \leq x_2 \leq \cdots \leq x_k\} \cup x^{*}  \in {\cal{S}}_{f}\bigr \}$. Return to the {\it{initialization}} step while the set of $k+1$ abscissae in $T^{*}_{k}$ is  relabeled in ascending order and repeat these three steps if $n$ realizations have been not yet obtained. We note that for initialization, a set of two starting abscissae $(k \geq 2)$ is necessary.  
\end{itemize}
We note that a given PDF $f(.\vert\theta)$ is unimodal if and only if $f(.\vert\theta)$ is log-concave. So, for simulating from a distribution with unimodal PDF $f(.\vert\theta)$, the ARS algorithm would be a suitable candidate provided that ${\cal{H}}(.)$ has closed form.
%
%to theby tangent line $T_{a}(x)$, that is, ${\cal{H}}(x)\leq T_{a}(x)$ where Herein, ${\cal{H}}^{\cdot}(a)$ denotes the first derivative of ${\cal{H}}^{\cdot}(a)$ with respect to $x$ computed at abscissa (point) $a$. The ARS scheme employs the proposal $l(\vert \theta)$ such that
%\begin{align*}
%f(x\vert \theta)\leq \exp\{l(x\vert \theta)\}= \underset{i=1,\ldots,k}{\operatorname{min}} T_{a_{i}}(x)
%\end{align*}
%where $x_{1}\leq x_{2} \leq \cdots \leq x_{k}$ are the abscissae of the tangent lines. Indeed the set of tangents between $x_{i}$ and $x_{i+1}$, for $i=1,\cdots, k-1$, constitute a piecewise linear upper hull to the ${\cal{H}}(x)$ that Hence, for $x \in [z_{i-1},z_{i}]$ and $i=1,\cdots,k$, we have
%$T_{k}(x)=\bigl(x-x_{i}\bigr){\cal{H}}^{\cdot}\bigl(x_{i}\bigr)+{\cal{H}}\bigl(x_{i}\bigr)$. Let the upper squeezing function to be
%Sometimes finding sharp upper bound $c$ is not possible in a rejection sampling scheme. In such a case, if $f(x\vert\theta)$ is a log-concave function of $x$, then generation from $f(\cdot\vert \theta)$ can be drawn efficiently using a method called in the literature as the adaptive rejection sampling \citep{gilks1992adaptive}. We note that pdf $f(x\vert\theta)$ is called log-concave if 
%\begin{align*}
%\frac{\partial^2}{\partial x^2}  \log f(x\vert\theta)\leq 0.
%\end{align*}
%Fortunately, similar to the MH algorithm, the rejection sampling scheme can be applied for the cases that the target distribution is only known up to proportionality constant.
\begin{example}\label{exam-gigsimulation-adaptiverejection}%\lipsum*[]
Let $f(x\vert \boldsymbol{\theta})$ represent the PDF of generalized inverse Gaussian (denoted as ${\cal{GIG}}(a,b,c)$) distribution defined as 
\begin{align}\label{pdf-gig}
f(x\vert\boldsymbol{\theta}) =\Bigl(\frac{b}{c}\Bigr)^{\frac{a}{2}}\frac {x^{a-1}}{2{\cal{K}}_{a}(\sqrt{bc})}\exp\Bigl\{-\frac { b x}{2}-\frac{c}{2x}\Bigr\},
\end{align}
where $\boldsymbol{\theta}=(a,b,c)^{\top}$. It is easy to check that $f(x\vert\boldsymbol{\theta})$ is log-concave for $a \geq 1$. Implementing the ARS algorithm is executable through the computer code. The package \texttt{ars} developed for this purpose that is available at \texttt{https://CRAN.R-project.org/package=ars}.   
%in Example 2. It ollows that
%\begin{align*}
%\frac{\partial^2}{\partial x^2}  \log f(x\vert\theta)=&
%\frac{\partial^2}{\partial x^2}  \Bigl[-2 \log (x+1)+ \theta \log x -x^\theta - \log c_{0}\Bigr]\nonumber\\
%=&\frac{2}{(x+1)^2}-\frac{\theta}{x^2} -\theta (\theta-1)x^{\theta-2}.
%\end{align*}
%Graphical display shows that the pdf $f(x\vert \theta)$  is log-concave for $\theta\geq 1.1$ and hence the adaptive rejection method can be used for simulating from this pdf for $\theta\geq 1.1$.
\end{example}
%\section{Slice sampling}
%Let $f(x\vert{\theta})$, that my be known only up to a proportionality constant $z$, is an unimodal pdf of random variable $X \in \mathbb{R}$ from which sampling is of interest. By considering an extra auxiliary variable $Y$, we can sample from the joint distribution of $(X,Y)$. To this end, assuming $U = \{(X, Y) \vert 0<Y<f(X\vert{\theta})\}$, the joint density for $(X,Y)$ is
%\begin{eqnarray*}
%\displaystyle
%p(x,y\vert \theta)=\left\{\begin{array}{c}
%\displaystyle
%\frac{1}{z},~~~~~~\mathrm{{if}}~~~ 0 < y < f(x\vert{\theta}),\\
%\displaystyle
%0,~~~~~~\mathrm{otherwise},~~~~~~~~~~~
%\end{array} \right.
%\end{eqnarray*}
%where $z=\int_{{\cal{S}}_{f}} f(x\vert{\theta}) dx$. Based on the joint pdf $p(x,y\vert \theta)$ defined ass above, the marginal pdf of $X$ is
%\begin{align*}
%p(x\vert \theta)=\int_{0}^{ f(x\vert{\theta})} \frac{1}{z} dy=\frac{f(x\vert{\theta})}{z}.
%\end{align*}
%Obviously, from generating $x$, one can proceed by sampling form joint distribution with pdf $p(x,y\vert \theta)$ defined as above and then remove $y$.  In practice, this is carried out by sampling alternately from the
%conditional distribution of $Y$ given the current $X$, that is, $Y\vert X \sim \text{Unif}\bigl(0, f(X\vert{\theta})\bigr)$, and then from the conditional distribution of $X$ given the current $Y$ that is uniform on region (or {\it{slice}}) $S$ defined as $S=\{X \vert Y <f(X\vert{\theta})\}$.
\subsection{Two-dimensional single rejection sampling}
Let $f(x\vert{\theta})$, that my be known only up to a proportionality constant $z$, is an unimodal PDF of random variable $X \in \mathbb{R}$ from which sampling is of interest. If it is possible to consider an extra auxiliary variable $Y$, thereby  the marginal PDF o $X$ can be marginalized by integrating out $Y$ as
\begin{align*}
f(x\vert{\theta})=\int f(x,y\vert{\theta}) dy,
\end{align*}
then one can use the {\it{two-dimensional rejection sampling}} scheme in order to generate sample $x$ with PDF $f(x\vert{\theta})$. For this purpose, let $l(x,y\vert\theta)$ denote the joint PDF of the proposal (candidate) such that 
\begin{align*}
\underset{(x,y) \in {\cal{S}}_{f(x,y\vert\theta)}}{\operatorname{sup}}\frac{f(x,y\vert\theta)}{l(x,y\vert\theta)} = c.
\end{align*}
We use the following algorithm for simulating from $f(x\vert{\theta})$ through the {\it{two-dimensional single rejection}} sampling scheme.
% otherwise we refer to another Monte Carlo simulation technique, called adaptive rejection sampling (ARS) algorithm. The ARS algorithm is used to simulate realization when posterior pdf is log-concave (i. e., the second derivative of $\pi(\theta\vert\boldsymbol{x})$ with respect to $\theta$ is negative). In such a case, the ARS algorithm developed by \cite{gilks1992adaptive} is highly efficient.
\begin{enumerate}
\item Generate pair $(X,Y)$ or the joint PDF $l(x,y\theta)$;
\item Generate $u\sim \text{Unif}(0, 1)$;
\item If $u<f(x,y\vert\theta)/\bigl[c \times l(x,y\vert\theta)\bigr]$, then accept $X$ as a generation from $f(x\vert\theta)$; otherwise return to step 1 and repeat the algorithm.
\end{enumerate}
The efficiency of a {\it{two-dimensional single rejection}} sampling scheme depends on quantity $c$ and the complexity of sampling from two-dimensional candidate $l(x,y\vert\theta)$. If $l(x,y\vert\theta)$ is easy to sampling from and $c$ is as small as possible, then this algorithm works efficiently.
\begin{example}\label{exam-doublerejection-1}%\lipsum*[]
Let $X$ denotes a random variable following an exponentially tilted $\alpha$-stable distribution with tilting parameter $\lambda>0$ if it follows a distribution whose PDF is given by
\begin{align}\label{pdfet}
f(x\vert \boldsymbol{\theta})=\exp\bigl\{-x \lambda +\lambda^{\frac{\alpha}{2}}\bigr\} d(x\vert \alpha) ,
\end{align}
where $\boldsymbol{\theta}=(\alpha, \lambda)^{\top}$ for $0< \alpha \leq 2$ is the family parameter vector and $d(x\vert \alpha)$ is the PDF of a positive $\alpha$-stable distribution, see \cite{devroye2009random}. The PDF $d(x \vert \alpha)$ has not closed form, but can be represented as 
\begin{align} \label{pdfp}
d(x \vert \alpha)=\frac{1}{\pi}\int_{0}^{\pi}\exp\Bigl\{-x^{-\frac{\alpha}{2-\alpha}}A(y)\Bigr\}dy,
\end{align}
where 
\begin{align*}
A(y)=\frac{\Bigl\{\sin\bigl[ \bigl(\frac{\alpha}{2}\bigr)y\bigr]\Bigr\}^{\frac{\alpha}{2}}\Bigl\{\sin \Bigl[\bigl (1-\frac{\alpha}{2}\bigr) y\Bigr]\Bigr\}^{1-\frac{\alpha}{2}}}{\sin (y)}.
\end{align*} 
We note that $A(y)^{2/(2-\alpha)}$ is called the Zolotarev's function \citep{devroye2009random}. From (\ref{pdfet}) and (\ref{pdfp}), it can be easily seen that $f(x\vert \boldsymbol{\theta})$ is the marginal of a joint PDF. Herein, we do not aim to discuss this example in details and just mention that for simulating from $f(x ,y\vert \boldsymbol{\theta})$, a bivariate distribution with PDF
\begin{align}\label{lbi}
l(x ,y\vert \boldsymbol{\theta})=\frac{1}{\pi}\times\frac{x^{m-1}}{\Gamma(m)}\exp\{-x\},
\end{align}
is chosen as candidate. Depending on values of $\alpha$ and $\lambda$, the gamma PDF with shape parameter $m=\alpha \lambda^{\alpha}$ in RHS of (\ref{lbi}) can be replaced with a gamma PDF with shape parameter $m=(1-\alpha) \lambda^{\alpha}+1$ or truncated Gaussian PDF. The reader is referred to \citep{hofert2011sampling,qu2021random} for a thorough treatment of this example. For simulating from PDF (\ref{pdfet}), there is an open source code called \texttt{copula} has been developed for \verb+R+ language that is available at \texttt{https://CRAN.R-project.org/package=copula}.
\end{example}
\begin{example}\label{exam-doublerejection-2}%\lipsum*[]
Suppose $p$-dimensional random vector $\boldsymbol{X}=(X_{1},\cdots,X_{p})^{\top}$ follows a truncated Gaussian distribution on region with positive Lebesgue measure $\boldsymbol{R}$ denoted as  
\begin{align}\label{exam-doublerejection-truncated-Gaussian-1}
{\cal{TN}}_{\boldsymbol{R}}(\boldsymbol{\mu},\Sigma),~~\boldsymbol{R}=\bigl\{ \boldsymbol{x}\in \mathbb{R}^{p} \big \vert \boldsymbol{a} \leq A \boldsymbol{x} \leq \boldsymbol{b}\bigr\},
\end{align}
where $\boldsymbol{a}=(a_{1},\cdots,a_{p})^{\top}$ and $\boldsymbol{b}=(b_{1},\cdots,b_{p})^{\top}$ are vectors of finite or infinite constants. Moreover, matrix $A$ is a $p\times p$ full rank matrix that constructs a set of $p$ independent linear constraints. As pointed out by \citep{geweke1991efficient} the marginals of $\boldsymbol{X}$ with distribution given by (\ref{exam-doublerejection-truncated-Gaussian-1}) are not truncated Gaussian, but the full conditionals, that is distribution of $X_{i}$ (for $i=1,\cdots,p$) {\it{conditional}} on all of other members of $\boldsymbol{X}$ are truncated Gaussian. To verify this claim, we shall confine ourselves here to suppose $p=2$ and $A=\boldsymbol{I}_{p}$ in which $\boldsymbol{I}_{2}$ accounts for a $2 \times 2$ identity matrix. Therefore (\ref{exam-doublerejection-truncated-Gaussian-1}) can be rewritten as
\begin{align}\label{exam-doublerejection-truncated-Gaussian-2}
{\cal{TN}}_{\boldsymbol{R}}(\boldsymbol{\mu},\Sigma),~~\boldsymbol{R}=\bigl\{ \boldsymbol{x}=(x_{1},x_{2})^{\top} \in \mathbb{R}^{2} \big \vert a_{1}\leq x_{1} \leq b_{1}, a_{2}\leq x_{2} \leq b_{2}\bigr\},
\end{align}
where $\boldsymbol{\mu}=(\mu_{1},\mu_{2})^{\top}$ and $\Sigma=\bigl[(\sigma_{11},\sigma_{12})^{\top},(\sigma_{21},\sigma_{22})^{\top}\bigr]$. In univariate case, we may write ${\cal{TN}}_{R_{i}}\bigl(\mu_{i}^{*},\sigma_{i}^{*}\bigr)$ denote the family of truncated univariate Gaussian distributions with mean $\mu_{i}^{*}$ and variance $\sigma_{i}^{*}$ on interval $R_{i}=\bigl\{x_{i} \in \mathbb{R} \big \vert a_i\leq x_i \leq b_i\bigr\}$, for $i=1,\cdots,p$. For computing the marginal PDF, e.g. PDF of $X_{1}$, corresponds to the joint PDF (\ref{exam-doublerejection-truncated-Gaussian-2}), we have
\begin{align}\label{exam-doublerejection-truncated-Gaussian-3}
f_{X_{1}}(x_{1}\vert \boldsymbol{\mu}, \Sigma)=&\frac{1}{{\cal{P}}_{G}}\int_{a_{2}}^{b_{2}}
\boldsymbol{\phi}_{2}\bigl(\boldsymbol{x}\vert\boldsymbol{\mu},{\Sigma}\bigr)d x_{2},\nonumber\\
=&\frac{1}{{\cal{P}}_{G}}{\phi}\bigl(x_{1}\big \vert \mu_{1},\sigma_{11}\bigr)
\int_{a_{2}}^{b_{2}}{\phi}\bigl(x_{2}\big \vert {\mu}_{2.1}, \sigma_{2.1}\bigr)d x_{2},\nonumber\\
=&\frac{1}{{\cal{P}}_{G}}{\phi}\bigl(x_{1}\big \vert \mu_{1},\sigma_{11}\bigr)\Bigl[\Phi\bigl(b_{2}\big \vert {\mu}_{2.1}, \sigma_{2.1}\bigr)-\Phi\bigl(a_{2}\big \vert {\mu}_{2.1}, \sigma_{2.1}\bigr)\Bigr],
\end{align}
where ${\cal{P}}_{G}=\boldsymbol{\Phi}_{2}\bigl(\boldsymbol{b}\big\vert\boldsymbol{\mu}, {\Sigma}\bigr)-\boldsymbol{\Phi}_{2}\bigl(\boldsymbol{a}\big\vert\boldsymbol{\mu}, {\Sigma}\bigr)$, $x_{1}\in R_{1}$, and 
\begin{align}
\mu_{2.1}=&\mu_{2}+\sigma_{21}\sigma_{11}^{-1}\bigl(x_{1}- {\mu}_{1}\bigr),
\label{exam-doublerejection-truncated-Gaussian-mu1.2}\\
\sigma_{2.1}=&\sigma_{22}-\sigma_{21}\sigma_{11}^{-1}\sigma_{12}.\label{exam-doublerejection-truncated-Gaussian-sigma1.2}
\end{align} 
%in which $R_{i}=\bigl\{x\in \mathbb{R} \big \vert x \in [a_{i}, b_{i}]\bigr \}$ (for $i=1,2$) 
Obviously, since both terms within the square bracket in the RHS of (\ref{exam-doublerejection-truncated-Gaussian-3}) depend on $x_1$ through ${\mu}_{2.1}$ given by (\ref{exam-doublerejection-truncated-Gaussian-mu1.2}), hence $f_{X_{1}}(x_{1}\vert \boldsymbol{\mu}, \Sigma)$ is no longer the PDF of a truncated Gaussian distribution unless ${\mu}_{2.1}$ is independent of $x_{1}$ that occurs only when $\sigma_{12}=\sigma_{21}=0$. This is while both of the full conditionals $X_{1}\vert X_{2}$ and $X_{2}\vert X_{1}$ correspond to the joint PDF in (\ref{exam-doublerejection-truncated-Gaussian-2}) are Gaussian. For example $X_{2}\big \vert X_{1}\sim {\cal{TN}}_{R_{2}}\bigl({\mu}_{2.1}, \sigma_{2.1}\bigr)$ for $\boldsymbol{x} \in \boldsymbol{R}$. Herein, we are willing to apply the two-dimensional single rejection scheme for sampling from distribution (\ref{exam-doublerejection-truncated-Gaussian-2}) whose PDF may be rewritten as
\begin{align*}%\label{exam-doublerejection-truncated-Gaussian-4}
f_{X_{1},X_{2}}\bigl(x_{1},x_{2}\big\vert \boldsymbol{\mu}, \Sigma\bigr)=&\frac{1}{{\cal{P}}_{G}}\phi\Bigl(\frac{x_{1}-\mu_{1}}{\sqrt{\sigma_{11}}}\Bigr)\phi\Bigl(\frac{x_{2}-\mu_{2.1}}{\sqrt{\sigma_{2.1}}}\Bigr)\times{\mathbb{I}}_{\boldsymbol{R}}\bigl(x_{1},x_{2}\bigr)\nonumber\\
=&c_{1}\phi\Bigl(\frac{x_{1}-\mu_{1}}{\sqrt{\sigma_{11}}}\Bigr)\phi\Bigl(\frac{x_{2}-\mu_{2.1}}{\sqrt{\sigma_{2.1}}}\Bigr)\times{\mathbb{I}}_{\boldsymbol{R}}\bigl(x_{1},x_{2}\bigr),
\end{align*}
where $c_{1}$ is a constant independent of $(x_{1},x_{2})^{
\top}$ and ${\mu}_{2.1}$ and $\sigma_{2.1}$ are defined in (\ref{exam-doublerejection-truncated-Gaussian-mu1.2}) and (\ref{exam-doublerejection-truncated-Gaussian-sigma1.2}), respectively. Furthermore, ${\mathbb{I}}_{\boldsymbol{R}}(x,y)$ is the well-known indicator function defined for the set $\boldsymbol{R}$ defined as
\begin{eqnarray}\label{exam-doublerejection-truncated-Gaussian-5}
\displaystyle
\mathbb{I}_{\boldsymbol{R}}\bigl(x_{1},x_{2}\bigr)=\left\{\begin{array}{c}
\displaystyle
1,~\mathrm{{if}}\ x_{1}~\mathrm{{and}}~x_{2} \in \boldsymbol{R},\\
\displaystyle
0,~\mathrm{if}\ x_{1}~\mathrm{{or}}~~~x_{2} \notin \boldsymbol{R}.
\end{array} \right.
\end{eqnarray}
 We may consider the distribution of proposal as the product of ${\cal{TN}}_{R_{1}}\bigl(\mu_{1},\sigma_{11}\bigr)$ and ${\cal{U}}\bigl(a_2, b_2\bigr)$ with PDF
\begin{align*}%\label{exam-doublerejection-truncated-Gaussian-6}
l_{X_{1},X_{2}}\bigl(x_{1},x_{2}\big\vert \boldsymbol{\mu}, \Sigma\bigr)=&\frac{\phi\bigl(\frac{x_{1}-\mu_{1}}{\sqrt{\sigma_{11}}}\bigr)\times{\mathbb{I}}_{{R}_{1}}(x_{1})}{\Phi\bigl(b_{1}\big\vert \mu_{1},\sqrt{\sigma_{11}}\bigr)-\Phi\bigl(a_{1}\big\vert \mu_{1},\sqrt{\sigma_{11}})}\times\frac{{\mathbb{I}}_{{R}_{2}}(x_{2})}{b_{2}-a_{2}},\nonumber\\
=&c_{2}\times \phi\Bigl(\frac{x_{1}-\mu_{1}}{\sqrt{\sigma_{11}}}\Bigr)\times{\mathbb{I}}_{{R}_{1}}\bigl(x_{1}\bigr)\times{\mathbb{I}}_{{R}_{2}}\bigl(x_{2}\bigr),
\end{align*}
where $c_{2}$ is independent of $(x_{1},x_{2})^{\top}$. It follows that
\begin{align*}%\label{exam-doublerejection-truncated-Gaussian-7}
f_{X_{1},X_{2}}\bigl(x_{1},x_{2}\big\vert \boldsymbol{\mu}, \Sigma\bigr)&\leq \frac{c_{1}}{\sqrt{2\pi \sigma_{2.1}}} \phi\Bigl(\frac{x_{1}-\mu_{1}}{\sqrt{\sigma_{11}}}\Bigr)\times{\mathbb{I}}_{\boldsymbol{R}}\bigl(x_{1},x_{2}\bigr)\nonumber\\
&\leq c_{1}\frac{b_{2}-a_{2}}{\sqrt{2\pi \times \sigma_{2.1}}}
 l_{X_{1},X_{2}}\bigl(x_{1},x_{2}\big\vert \boldsymbol{\mu}, \Sigma\bigr)
 \nonumber\\
&= c\times l_{X_{1},X_{2}}\bigl(x_{1},x_{2}\big\vert \boldsymbol{\mu}, \Sigma\bigr).
\end{align*}
Hence,
\begin{align*}%\label{exam-doublerejection-truncated-Gaussian-8}
\frac{f_{X_{1},X_{2}}\bigl(x_{1},x_{2}\big\vert \boldsymbol{\mu}, \Sigma\bigr)}{ c\times l_{X_{1},X_{2}}\bigl(x_{1},x_{2}\big\vert \boldsymbol{\mu}, \Sigma\bigr)}=\exp\Bigl\{\frac{-\bigl(x_{2}-\mu_{2.1}\bigr)^2}{2\sigma_{2.1}}\Bigr\}.
\end{align*}
In what follows Algorithm \ref{exam-doublerejection-truncated-Gaussian-algorithm} describes how to sample from from truncated bivariate Gaussian distribution. 
\vspace{5mm}
\begin{algorithm}
\caption{Two-dimensional single rejection scheme for sampling from truncated bivariate Gaussian distribution}
\label{exam-doublerejection-truncated-Gaussian-algorithm}
\begin{algorithmic}[1]
\State Read $\boldsymbol{a}$, $\boldsymbol{b}$, $\boldsymbol{\mu}$, and $\Sigma$ given by (\ref{exam-doublerejection-truncated-Gaussian-2});
%\State Set $i=1$;
%\While{$i < n$}  %\Comment{put some comments here}
\State Generate $u\sim {\cal{U}}(0, 1)$;
\State Generate $x_{1}\sim {\cal{TN}}_{R_{1}}\bigl({\mu}_{1}, \sigma_{11}\bigr)$;
\State Generate $x_{2}\sim {\cal{U}}\bigl(a_{2}, b_{2}\bigr)$;
\State Squeezing test: If $ -2\sigma_{2.1}\log u> (x_{2}-\mu_{2.1})^2$ where ${\mu}_{2.1}$ and ${\sigma}_{2.1}$ are given by (\ref{exam-doublerejection-truncated-Gaussian-mu1.2}) and (\ref{exam-doublerejection-truncated-Gaussian-sigma1.2}), respectively, then accept $(x_1,x_2)^{\top}$ as a generation from PDF given by (\ref{exam-doublerejection-truncated-Gaussian-2}), otherwise return to step 2 and repeat algorithm;
%   
%   \State $\boldsymbol{Z}^{(t+1)} \leftarrow \boldsymbol{Z}^{(t)}$;
%   \State $\boldsymbol{X}^{(t+1)}=D^{-1}\boldsymbol{Z}^{(t+1)} $
%   \State {\bf{end}}
\end{algorithmic}
\end{algorithm}
\vspace{5mm}
In general where $A$ is not necessarily an identical full rank matrix, we suggest the use of method proposed by \cite{geweke1991efficient} for simulating from truncated bivariate Gaussian distribution that will be discussed in Section \ref{Simulating from truncated Gaussian distribution}. Here we give a brief discussion of this method. 
%\begin{align*}
%\boldsymbol{\alpha}=&\boldsymbol{a}-A\boldsymbol{\mu},\\%\label{a-star},\\
%\boldsymbol{\beta}=&\boldsymbol{b}-A\boldsymbol{\mu}.%\label{b-star}.
%\end{align*} 
For simulating form PDF given by (\ref{exam-doublerejection-truncated-Gaussian-1}), first one needs to simulate from $\boldsymbol{Z}\sim {\cal{TN}}_{\boldsymbol{T}}(\boldsymbol{0}, A\Sigma A^{\top})$ where $\boldsymbol{T}=\bigl\{ \boldsymbol{z}\in \mathbb{R}^{p}\big \vert \boldsymbol{a}-A\boldsymbol{\mu}\leq \boldsymbol{z} \leq \boldsymbol{b}-A\boldsymbol{\mu} \bigr\}$, and then accept $\boldsymbol{X}=\boldsymbol{\mu}+A^{-1}\boldsymbol{Z}$ as a generation from PDF given by (\ref{exam-doublerejection-truncated-Gaussian-1}). The \verb+R+ function below is provided for generating $n$ sample from truncated bivariate Gaussian distribution through Two-dimensional single rejection scheme.
\begin{lstlisting}[style=deltaj]
R> rtbinorm_rejection <- function(n, Mu, Sigma, a, b, A)
+ {
+ V <- A%*%Sigma%*%t(A) 
+ T1 <- matrix( cbind(a - A%*%Mu, b - A%*%Mu), nrow = 2, ncol = 2)
+ X <- Z <- matrix(0, nrow = n, ncol = 2)
+ sigma2.1 <- V[2, 2] - V[1, 2]^2/V[1, 1]
+ sigma1 <- sqrt( V[1, 1])
+ j <- 1
+ 	while(j < n)
+ 	{
+ 		u <- runif(3)
+ 		cdf.a <- pnorm(T1[1, 1], 0, sigma1)
+ 		cdf.b <- pnorm(T1[1, 2], 0, sigma1)
+ 		x <- qnorm( u[1]*(cdf.b - cdf.a) + cdf.a,
+ 					 mean = 0, sd = sigma1 )
+ 		mu2.1 <- V[1, 2]/V[1, 1]*x
+ 		y <- (T1[2, 2] - T1[2, 1])*u[2] + T1[2, 1]
+ 			if( -2*log(u[3]) > (y - mu2.1)^2/sigma2.1 )
+ 			{
+ 				Z[j, ] <- c(x, y)
+ 				j <- j + 1
+ 			}
+ 	}
+ X <- matrix(Mu, nrow = n, ncol = 2, byrow = T) + Z%*%solve(A)
+ return(X)
+ }
\end{lstlisting}

\end{example}
\section{Ratio of uniforms method}
The Ratio of uniforms method is in fact a rejection method that introduced by \cite{kinderman1977computer}. Suppose we are interested in generating random variable $X$ with PDF is $f(x)$ that may be known only up to a proportionality constant $k>0$, that is $f(x)=k \times h(x)$ \footnote{The function $h(x)$  sometimes is called quasi PDF.}. If random vector $(U,V)^{\top}$ is distributed uniformly on
${\text{C}}_{h}=\bigl\{(u,v) \in \mathbb{R}^{2}~\big \vert ~0\leq u \leq \sqrt{h(v/u)}\bigr\}$ with PDF $g(u,v)$ given by
\begin{eqnarray}\label{ratio-uniforms-method-1}
\displaystyle
g(u,v)= \left\{\begin{array}{c}
\displaystyle
2k,~~\mathrm{if}\ 0\leq u \leq \sqrt{h(v/u)},\\
~0,~~~otherwise.~~~~~~~~~~~~~
\end{array} \right.
\end{eqnarray}
Consider two transformations $X=V/U$ and $Y=U$, it can be shown that the distribution of random $X$ follows a distribution with PDF $f(x)$. To show this claim, we notice that Jacobian of inverted transformations, that is $U=Y$ and $V=XY$, is $-Y$. Therefore
\begin{align*}
f(x,y)=\vert -y \vert \times g(v/u,u)=y\times 2k, ~~ 0<y<\sqrt{h(x)}.
\end{align*} 
An expression for the marginal PDF of $X$ is obtained by integrating out $y$ as
\begin{align*}
\int_{0}^{\sqrt{h(x)}}f(x,y) dy=\int_{0}^{\sqrt{h(x)}}2ky dy=kh(x)=f(x).
\end{align*} 
For generating $X$, first we generate two realizations from ${\cal{U}}(0,a)$ and ${\cal{U}}(b,c)$ in which
\begin{align*}
a=\sup_{x} \sqrt{h(x)},~~b=-\sup_{x} x\sqrt{h(x)}=\inf_{x} x\sqrt{h(x)},~~c=\sup_{x} x\sqrt{h(x)}.
\end{align*}
The support of random vector $(U,V)^{\top}$, that is $\bigl\{ (u,v) \big \vert u \in (0,a) ~{\text{and}}~v \in (b,c)\bigr \}$, is a bounding rectangle that encloses the region ${\text{C}}_{h}$. If simulated vector $(U,V)^{\top}$ lies inside region ${\text{C}}_{h}$, then we accept $V/U$ as a generation of $X$. Hence, if $h(x)$ and $x^2h(x)$ are finite, then ratio of uniforms method proceeds as follows for simulating $X$ with PDF $f(x)$.
\begin{enumerate}[label=\roman*.]
\item Generate $U\sim{\cal{U}}(0,a)$ and $V\sim {\cal{U}}(b,c)$ independently;
\item If $U^2<h(V/U)$, then accept $X=V/U$ from PDF $f(x)$, otherwise go back step (i).
\end{enumerate}
Evidently, the ratio of uniforms method is a rejection method. The acceptance rate is
\begin{align*}
\int_{}^{}h(x)dx \times [2a(c-b)]^{-1}.
\end{align*}
As pointed out by \cite[Theorem 2]{wakefield1991efficient}, for all location families of distributions that are unimodal, the acceptance probability for the ratio of uniforms method maximizes when the location parameter is zero. Therefore, for example, to generate $X\sim{\cal{N}}(\mu,\sigma^2)$, first we simulate from $Y={\cal{N}}(0,\sigma^2)$ and then we have $Y+\mu \sim{\cal{N}}(\mu,\sigma^2)$.
\begin{example}\label{exam-ratio-of-uniform-method}%\lipsum*[]
Suppose we are interested in simulating from $X\sim{\cal{N}}(0,1)$ through the ratio of uniforms method. Let $h(x)=\exp\{-x^2/2\}$, we can see that
\begin{align*}
{\text{C}}_{h}&=
\Bigl\{ (u,v) \in \mathbb{R}^{2}~\Big \vert~    0\leq u \leq  \exp\bigl\{-1/2\times v^2/u^2\bigr\} \Bigr\}\nonumber\\
&=\Bigl\{ (u,v) \in \mathbb{R}^{2}~\Big  \vert~ v^2 \leq -4 u^2 \log u \Bigr\},
\end{align*}
and
\begin{align*}
a=\sup_{x} \sqrt{h(x)}=1,~~b=-\sup_{x} x\sqrt{h(x)}=-\sqrt{2e},~~c=\sup_{x} x\sqrt{h(x)}=\sqrt{2e},
\end{align*}
where $e=\exp\{1\}=2.718281\ldots$. We follow a two-step procedure given below for simulating $X$.
\begin{enumerate}[label=\roman*.]
\item Generate $U\sim{\cal{U}}(0,1)$ and $V\sim {\cal{U}}(-\sqrt{2/e},\sqrt{2/e})$ independently;
\item If $U^2<\exp\bigl\{-1/2\times V^2/U^2\bigr\}(V/U)$, then accept $X=V/U$ from PDF standard Gaussian distribution, otherwise go back to step (i).
\end{enumerate}
The acceptance rate for this method is 
\begin{align*}%\label{pdfet}
\int_{}^{}h(x)dx \times [2a(c-b)]^{-1}=\frac{\sqrt{2\pi}}{4\sqrt{2\exp\{-1\}}}\approx 0.731.
\end{align*}
The pertaining \verb+R+ function called \verb+rnorm_ru+, for generating $n$ realizations from $X\sim{\cal{N}}(0,1)$ based on the ratio of uniforms method is given as follows.
\begin{lstlisting}[style=deltaj]
R> rnorm_ru <- function(n)
+{
+	X <- rep(NA, n)
+	j <- 1
+	while( j <= n )
+	{
+		U <- runif(1)
+		V <- (2*runif(1) - 1)*sqrt( 2*exp(-1) )
+		if( V^2 < -4*U^2*log( U ) )
+		{
+			X[j] <- V/U
+			j <- j + 1
+		}
+	}
+return(X)
+}
\end{lstlisting}
\end{example}
\begin{example}\label{exam-doublerejection}%\lipsum*[]
Let  $X\sim{\cal{G}}(\alpha,\beta)$ with $\alpha>1$. We know that $X \mathop=\limits^d G/\beta$ in which $G\sim{\cal{G}}(\alpha,1)$, hence without loss of generality, for simulating $X$ through the ratio of uniforms method, we assume that $\beta=1$. Once we have simulated $G$, the quantity $X$ is simply produced as $X=G/\beta$. Setting $h(x)=x^{\alpha-1}\exp\{-x\}$, we have
\begin{align*}
{\text{C}}_{h}&=
\Bigl\{ (u,v) \in \mathbb{R}^{2}~\Big \vert~    u<\sqrt{(v/u)^{\alpha-1}\exp\bigl\{-v/u\bigr\} } \Bigr\}\nonumber\\
&=
\Bigl\{ (u,v) \in \mathbb{R}^{2}~\Big  \vert~ 2\log u < (\alpha - 1) \log(v/u) - v/u \Bigr\},
\end{align*}
and
\begin{align*}
a=\sup_{x} \sqrt{h(x)}=\Bigl(\frac{\alpha-1}{e}\Bigr)^{(\alpha-1)/2},~~b=-\sup_{x} x\sqrt{h(x)}=0,~~c=\sup_{x} x\sqrt{h(x)}=\Bigl(\frac{\alpha+1}{e}\Bigr)^{(\alpha+1)/2},
\end{align*}
where $e=\exp\{1\}=2.718281\ldots$. We follow a three-step procedure given below for simulating $X$.
\begin{enumerate}[label=\roman*.]
\item Generate $U\sim{\cal{U}}(0,1)$ and $V\sim {\cal{U}}(-\sqrt{2/e},\sqrt{2/e})$ independently;
\item If $U^2<(V/U)^{\alpha-1}\exp\{-V/U\bigr\}$, then go next step, otherwise go back to step (i).
\item Accept $X=G/\beta$ as a generation from ${\cal{G}}(\alpha,\beta)$.
\end{enumerate}
\begin{lstlisting}[style=deltaj]
R> rgamma_ru <- function(n, alpha, beta = 1)
+{
+	if (alpha <= 1) stop( "shape parameter must be greater than one." )
+	a <- ( (alpha - 1)*exp(-1) )^( (alpha - 1)/2 )
+	b <- 0
+      c0 <- ( (alpha + 1)*exp(-1) )^( (alpha + 1)/2 )
+	X <- rep(NA, n)
+	j <- 1
+	while( j <= n )
+	{
+		U <- runif(1, 0, a); V <- runif(1, 0, c0)
+		upper <- 1/2*( (alpha - 1)*log(V/U) - V/U )
+		if( log(U) < upper )
+		{
+			X[j] <- 1/beta*V/U
+			j <- j + 1
+		}
+	}
+return(X)
+}
\end{lstlisting}
When $\alpha<1$, then $h(x)$ appears to be unbounded and hence the ratio of uniforms method cannot be applied for simulating from gamma distribution.
\end{example}
\section{Generalized ratio of uniforms method}
The ratio of uniforms method discussed earlier has been extended by \cite{wakefield1991efficient}. Let $g(\cdot): \mathbb{R}^{+} \rightarrow \mathbb{R}^{+}$ denote a strictly increasing function such that $g(0)=0$. If random vector $(U,V)^{\top}$ is distributed uniformly on region 
\begin{align}\label{region-generalized-ratio-to-uniform-1}
{\text{C}}_{h,g}=
\Bigl\{ (u,v) \in \mathbb{R}^{2}~\Big \vert~ 0< u \leq  g^{-1}\Bigl[ c
h\Bigl(\frac{v}{g^{\prime}(u)}\Bigr) 
\Bigr]\Bigr\},
\end{align}
where $c>0$ is a constant, then $X=V/g^{\prime}(U)$ follows a distribution with PDF $f(x)= h(x)/{\int h(x)}$ in which $h(x)$ is the quasi PDF. We note that $g^{\prime}(\cdot)$ denotes the first derivative of $g(\cdot)$. The bounding rectangle $(0, a_{g})\times (b_{g}, c_{g})$ corresponds to the region ${\text{C}}_{h,g}$ is constructed by determining
\begin{align}\label{region-generalized-ratio-to-uniform-2}
a_{g}=&\sup_{x} g^{-1}\bigl[ c h(x)\bigr]\nonumber\\
b_{g}=&\inf_{x\leq 0} x g^{\prime}\bigl[g^{-1}\bigl( c h(x)\bigr) \bigr]\leq 0\nonumber\\
c_{g}=&\sup_{x\geq 0} x g^{\prime}\bigl[g^{-1}\bigl( c h(x)\bigr) \bigr]\geq 0.
\end{align}
A simple choice for $g(\cdot)$ is then the power function for which $g^{\prime}(\cdot)$ is also strictly increasing is $g(x)=u^{r+1}/(r+1)$ (for $r \geq 0$) and hence $c=(1+r)^{-1}$. The original method is recovered for $r=1$ for which the region ${\text{C}}_{h,g}$ given in (\ref{region-generalized-ratio-to-uniform-1}) turns into
\begin{align}\label{region-generalized-ratio-to-uniform-3}
{\text{C}}_{h}(r)=
\Bigl\{ (u,v) \in \mathbb{R}^{2}~\Big \vert~ 0< u \leq  \Bigl[ 
h\Bigl(\frac{v}{u^{r}}\Bigr) 
\Bigr]^{1/(r+1)}\Bigr\}.
\end{align}
The area of ${\text{C}}_{h}(r)$ given in (\ref{region-generalized-ratio-to-uniform-3}) becomes 
$(1+r)^{-1}\int h(x) dx$. If $h(x)$ and $x^{r+1}[h(x)]^{r}$ are bounded, then the bounds of rectangle (\ref{region-generalized-ratio-to-uniform-3}) are
\begin{align}\label{region-generalized-ratio-to-uniform-4}
a(r)=&\sup_{x} \bigl[  h(x)\bigr]^{1/(r+1)}\nonumber\\
b(r)=&\inf_{x\leq 0} x \bigl[ h(x) \bigr]^{r/(r+1)}\leq 0\nonumber\\
c(r)=&\sup_{x\geq 0} x \bigl[ h(x) \bigr]^{r/(r+1)}\geq 0.
\end{align}
For a pair $(u,v)^{\top}$ generated from rectangle (\ref{region-generalized-ratio-to-uniform-3}), the acceptance probability is
 \begin{align}\label{region-generalized-ratio-to-uniform-5}
p(r)=\frac{\int h(x)}{(r+1) \times a(r) \times [c(r)- b(r)]}.
\end{align}
\begin{example}\label{exam-generalized-ratio-of-uniform-method}%\lipsum*[]
Following the Example \ref{exam-ratio-of-uniform-method}, we would like to simulate from $X\sim{\cal{N}}(0,1)$ through the generalized ratio of uniforms method. Let $h(x)=\exp\{-x^2/2\}$ for which $\int h(x)dx=\sqrt{2\pi}$. Furthermore, we can see $a(r)=1$ and $b(r)=-c(r)=-\sqrt{1+1/r}\exp\{-1/2\}$. Substituting these quantities into (\ref{region-generalized-ratio-to-uniform-5}) yields
\begin{align*}
p(r)=\frac{\sqrt{2\pi r \exp\{1\}}}{ 2(1+r)^{3/2}}.
\end{align*}
The maximum value of $p(r)$ is obtained when $r=0.5$, that is $p(0.5)=0.795$. Comparing with the acceptance rate of the original method discussed in Example \ref{exam-ratio-of-uniform-method} that is $p(1)=0.731$, the generalized ratio of uniforms method is more efficient.
\end{example}
Table \ref{Simulation-R-command} lists some \verb+R+ commands for simulating from widely used univariate distributions. For other distributions, user may use the approaches discussed earlier in this chapter or those discussed in \cite{luc1986,ross2022simulation}.
\vspace{5mm}
\begin{table}[!h]
\centering
\caption{Commands for simulating from some statistical families in {\texttt{R}}.}
\vspace*{-3mm}
\begin{tabular}{lll}%{\textwidth}{>{\hsize=.10\hsize}X>{\hsize=.95\hsize}X}
\hline
Family&command&PDF
\\
\hline
${\cal{U}}(a,b)$& \verb+runif(n,min=a,max=b)+ &$\frac{1}{b-a}$\\
${\cal{BET}}(a,b)$& \verb+rbeta(n,shape1=a,shape2=b,ncp=0)+ &$\frac{1}{B(a,b)}x^{a-1}(1-x)^{b-1}$\\
${\cal{G}}(a,b)$& \verb+rgamma(n,shape=a,rate=b)+ &
$\frac{b^a}{\Gamma(a)} x^{a-1}\exp\bigl\{- b x\bigr\}$\\
$\chi_{(\nu)}$& \verb+rchisq(n,df=nu,ncp=0)+ &$\frac{2^{-\frac{\nu}{2}}}{\Gamma(\nu/2)}x^{\frac{\nu}{2}-1}\exp\bigl\{-\frac {x}{2}\bigr\}$\\
${\cal{N}}(\mu,\sigma^2)$& \verb+rnorm(n,mean=mu,sd=sigma)+ &${\frac {1}{ {\sqrt {2\pi }\sigma} }} \exp\bigl\{-{\frac{(x-\mu )^{2}}{2\sigma^{2}}}\bigr\}$\\
${\cal{W}}(a,b)$& \verb+rweibull(n,shape=a,scale=b)+ &$\frac{a}{b} \bigl(\frac{x}{b}\bigr)^{a-1} \exp\bigl\{-\bigl(\frac{x}{b}\bigr)^{a}\bigr\}$ \\
${\cal{LN}}(\mu, \sigma)$& \verb+rlognorm(n,meanlog=mu,sdlog=sigma)+ &${\frac {1}{ x\sigma {\sqrt {2\pi }} }} \exp\bigl\{-{\frac{(\log x-\mu )^{2}}{2\sigma^{2}}}\bigr\}$\\
${\cal{T}}(\nu)$ & \verb+rt(n,df=nu,ncp=0)+ & $\frac{ \Gamma \bigl(\frac{\nu+1}{2}\bigr)}{\sqrt{\nu \pi}\Gamma(\nu/2)} \bigl[ 1 +\frac{x^2}{\nu}\bigr]^{-\frac{\nu+1}{2}}$\\
\hline
\end{tabular}
\label{Simulation-R-command}
\end{table}
\section{Simulating multivariate distributions}\label{sec31}
This part dealing with simulating from multivariate distributions with widespread use in applications. These include Gaussian, Wishart, and Students $t$. 
\subsection{Gaussian distribution}
Let random vector $\boldsymbol{X}=\bigl(X_1,\cdots,X_p\bigr)^{\top}$ follows ${\cal{N}}_{p}(\boldsymbol{\mu},\Sigma)$. Each Gaussian distribution is characterized by its mean (location) vector $\boldsymbol{\mu}=(\mu_1,\cdots,\mu_p)^{\top}$ and covariance (scale) matrix $\Sigma$. The PDF of $\boldsymbol{X}$ is shown by ${\cal{N}}_{p}(\cdot\big \vert \boldsymbol{\mu},\Sigma)$ that is
\begin{align}\label{PDF-Gaussian}
{\cal{N}}_{p}(\boldsymbol{x}\big \vert \boldsymbol{\mu},\Sigma)=\frac{1}{{\text{C}}_{G}}\exp\Bigl\{-\frac{\delta(\boldsymbol{x}-\boldsymbol{\mu},\Sigma)}{2}\Bigr\},
\end{align}
where 
\begin{align}
{\text{C}}_{G}&=(2\pi)^{\frac{p}{2}}\big \vert \Sigma \big \vert^{\frac{1}{2}},\label{cg}\\
\delta(\boldsymbol{x}-\boldsymbol{\mu},\Sigma)=&(\boldsymbol{x}-\boldsymbol{\mu})^{\top}\Sigma^{-1}(\boldsymbol{x}-\boldsymbol{\mu}).\label{delta}
\end{align}
The quantity ${\text{C}}_{G}$ is the normalizing constant and expression (\ref{delta}) is  the well-known {\it{Mahalanobis}} distance. We write ${\cal{N}}_{p}(\boldsymbol{\mu},\Sigma)$ to show that random vector $\boldsymbol{X}$ follows a $p$-dimensional Gaussian distribution with location vector $\boldsymbol{\mu}$ and covariance matrix $\Sigma$ whose PDF is given by (\ref{PDF-Gaussian}). The structure of $\Sigma$ may be shown as 
\begin{equation}
{\small{\Sigma=\left(
\begin{matrix}
\Sigma_{1,1}&\Sigma_{1,2}&\cdots&\Sigma_{1,p}\\
\Sigma_{2,1}&\Sigma_{2,2}&\cdots&\Sigma_{2,p}\\
\vdots & \vdots &\ddots&\vdots\\
\Sigma_{p,1}&\Sigma_{p,2}&\cdots&\Sigma_{p,p}
\end{matrix}\right)
=\left(
\begin{matrix}
{\text{var}}(X_{1})&{\text{cov}}(X_{1},X_{2})&\cdots&{\text{cov}}(X_{1},X_{p})\\
{\text{cov}}(X_{2},X_{1})&{\text{var}}(X_{2})&\cdots&{\text{cov}}(X_{2},X_{p})\\
\vdots & \vdots &\ddots&\vdots\\
{\text{cov}}(X_{p},X_{1})&{\text{cov}}(X_{p},X_{2})&\cdots&{\text{var}}(X_{p})
\end{matrix}\right)}}.
\end{equation}
The matrix $\Sigma$ is a symmetric and positive definite matrix and sometimes is called the variance-covariance matrix. This is because the diagonal elements of $\Sigma$ are variances, that is ${\text{var}}(X_{i})=\Sigma_{i,i}$, and its $(i,j)$th off-diagonal element represents the linear association between marginal variables $X_{i}$ and $X_{j}$, that is 
${\text{cov}}(X_{i},X_{j})=\Sigma_{i,j}$. 
\subsection{Properties of Gaussian distribution}\label{property-Gaussian}
Let $\boldsymbol{X}\sim{\cal{N}}_{p}(\boldsymbol{\mu},\Sigma)$, herein, we mention three interesting properties of Gaussian distribution.
\begin{enumerate}[label=\roman*.]
\item Both of expected value and variance of $\boldsymbol{X}$ are finite and given by $E(\boldsymbol{X})=\boldsymbol{\mu}$ and ${\text{var}}(\boldsymbol{X})=\Sigma$. The ML estimator of $\boldsymbol{\mu}$ and $\Sigma$, based on sample $\bigl\{\boldsymbol{x}_1,\cdots,\boldsymbol{x}_n\bigr\}$ is obtained readily as
\begin{align}
\hat{\boldsymbol{\mu}}_{ML}&=\bar{\boldsymbol{X}}=\frac{1}{n}\sum_{i=1}^{n}\boldsymbol{x}_{i},\nonumber\\
\hat{\Sigma}_{ML}&=S=\frac{1}{n}\sum_{i=1}^{n}\bigl(\boldsymbol{x}_{i}-\bar{\boldsymbol{x}}\bigr)\bigl(\boldsymbol{x}_{i}-\bar{\boldsymbol{x}}\bigr)^{\top},
\end{align}
respectively.
\item For a given $d \times p$ constant matrix such as $A$ and constant vector $\boldsymbol{c}$ of length $d$, we have 
\begin{align}\label{property-simulation-Gaussian-1}
\boldsymbol{c} + A \boldsymbol{X} \sim {\cal{N}}_{d}\bigl(\boldsymbol{c} + A\boldsymbol{\mu}, A\Sigma A^{\top}\bigr).
\end{align}
\item Let for given constant $1\leq q <p$, the random vector $\boldsymbol{X}$ is represented as
\begin{equation}
\boldsymbol{X}=\left(
\begin{matrix}
\boldsymbol{X}_1\\
\boldsymbol{X}_2
\end{matrix}
\right),
\end{equation}
in which $\boldsymbol{X}_1=\bigl({X}_1,\cdots,{X}_q\bigr)^{\top}$ and $\boldsymbol{X}_2=\bigl({X}_{q+1},\cdots,{X}_p\bigr)^{\top}$. The mean and scale parameters correspondingly can be represented as
\begin{align}
\boldsymbol{\mu}=\left(
\begin{matrix}
\boldsymbol{\mu}_1\\
\boldsymbol{\mu}_2
\end{matrix}\right)=
\left(
\begin{matrix}
E\bigl(\boldsymbol{X}_1\bigr)\\
E\bigl(\boldsymbol{X}_2\bigr)
\end{matrix}\right),
\end{align}
and
\begin{align}
\left(
\begin{array}{c;{2pt/2pt}c}
\boldsymbol{\Sigma}_{11} & \boldsymbol{\Sigma}_{12} \\
\hdashline[2pt/2pt]
\boldsymbol{\Sigma}_{21} & \boldsymbol{\Sigma}_{22}
\end{array}
\right)={\tiny{\left(
\begin{array}{ccc;{3pt/3pt}ccc}
\Sigma_{1,1}&\cdots&\Sigma_{1,q}&\Sigma_{1,q+1}&\cdots&\Sigma_{1,p}\\
\Sigma_{2,1}&\cdots&\Sigma_{2,q}&\Sigma_{2,q+1}&\cdots&\Sigma_{2,p}\\
\vdots & \vdots &\vdots&\vdots\\
\Sigma_{q,1}&\cdots&\Sigma_{q,q}&\Sigma_{q,q+1}&\cdots&\Sigma_{q,p}\\
\hdashline[3pt/3pt]
\Sigma_{q+1,1}&\cdots&\Sigma_{q+1,q}&\Sigma_{q+1,q+1}&\cdots&\Sigma_{q+1,p}\\
\vdots & \vdots &\vdots&\vdots\\
\Sigma_{p,1}&\cdots&\Sigma_{p,q}&\Sigma_{p,q+1}&\cdots&\Sigma_{p,p}\\
\end{array}\right)
}},
\end{align}
where $\boldsymbol{\Sigma}_{11}={\text{var}}(\boldsymbol{X}_1)$, $\boldsymbol{\Sigma}_{12}=\boldsymbol{\Sigma}_{21}={\text{cov}}(\boldsymbol{X}_1,\boldsymbol{X}_2)$ and $\boldsymbol{\Sigma}_{22}={\text{var}}(\boldsymbol{X}_2)$.
% Furthermore, symmetric sub-matrices are of size $q\times q$, $q\times (p-q)$, $(p-q)\times q$, and $(p-q)\times (p-q)$, respectively. 
It can be shown that the conditional distribution $\boldsymbol{X}_1$ given $\boldsymbol{X}_2=\boldsymbol{x}_{2}$ is ${\cal{N}}_{q}\bigl(\boldsymbol{\mu}_{\boldsymbol{X}_1\vert \boldsymbol{X}_{2}}, \Sigma_{\boldsymbol{X}_1\vert \boldsymbol{X}_{2}}\bigr)$ where
\begin{align}
\boldsymbol{\mu}_{\boldsymbol{X}_1\vert \boldsymbol{X}_{2}}&=\boldsymbol{\mu}_1 + \boldsymbol{\Sigma}_{12}\bigl(\boldsymbol{\Sigma}_{22}\bigr)^{-1}\bigl(\boldsymbol{x}_{2}-\boldsymbol{\mu}_1\bigr),\nonumber\\
\Sigma_{\boldsymbol{X}_1\vert \boldsymbol{X}_{2}}&=\boldsymbol{\Sigma}_{11}-\boldsymbol{\Sigma}_{12}\bigl(\boldsymbol{\Sigma}_{22}\bigr)^{-1}\boldsymbol{\Sigma}^{\top}_{12}.
\end{align}
\item Sub-vector $\boldsymbol{X}_1=\bigl({X}_1,\cdots,{X}_q\bigr)^{\top}$, for $q=1,\cdots,p$, follows ${\cal{N}}_{q}\bigl(\boldsymbol{\mu}_1,\Sigma_{11}\bigr)$.
\item If ${\text{cov}}\bigl({X}_i,{X}_j\bigr)=0$, for $i \neq j=1,\cdots,p$, then ${X}_i$ and $X_j$ are independent.
\end{enumerate}
\subsection{Generation from Gaussian distribution}\label{spectral-representation}
Several methods have been introduced for sampling from Gaussian distribution in the literature. Herein, we represent the most three commonly used approaches that are based on decomposition (factorization) of scale matrix $\Sigma$.  
\begin{enumerate}[label=\roman*.]
\item {\bf{Cholesky decomposition}}: The first approach is based on the Cholesky decomposition of scale matrix $\Sigma$. If lower triangular matrix $L$ denotes the Cholesky decomposition of $\Sigma$, then we can write $\Sigma = L L^{\top}$. Furthermore, suppose $\boldsymbol{Z}=\bigl(Z_{1},\cdots, Z_{p}\bigr)^{\top}$ is a vector $p$ independent univariate standard Gaussian variates. Using property  (\ref{property-simulation-Gaussian-1}) while setting $\boldsymbol{c}=\boldsymbol{\mu}$, $A=L$, and considering the fact that $\boldsymbol{Z}\sim{\cal{N}}_{p}\bigl(\boldsymbol{0}, \boldsymbol{I}_{p} \bigr)$ in which $\boldsymbol{I}_{p}$ denotes a $p \times p$ identity matrix, we can easily see that 
\begin{align}
\boldsymbol{\mu} + L \boldsymbol{Z} \sim {\cal{N}}_{p}\bigl(\boldsymbol{\mu}, \Sigma \bigr).
\end{align}
\item {\bf{spectral decomposition}}: The second way relies on the spectral or eigenvalue decomposition of $\Sigma$. Let $\bigl\{\lambda_1,\cdots,\lambda_p\bigr\}$ is sequence of eigenvalues of $\Sigma$ and columns of $p \times p$ matrix $V$ are associated eigenvectors. It is well known that $\Sigma$ can be represented as
\begin{align}\label{spectral-decomposition}
\Sigma=&V
\left(
\begin{matrix}
\lambda_{1}       &0                  &0    	 				  &\cdots  &0\\
                 0       &\lambda_{2} & 0   	 				  & \cdots &0\\
                 0       & 0                 & \lambda_{3}  	       &           &0\\
      \vdots           &\vdots           &         				  & \ddots &\\
              0          &   0               & 0   	 				  &           &\lambda_{p}
\end{matrix}
\right) V^{-1}\nonumber\\
=&V D V^{\top},
\end{align}
where $D$ is a $p \times p$ diagonal matrix whose positive diagonal elements are eigenvalues of $\Sigma$. Furthermore, it should be noted that 
\begin{enumerate}
\item Since $V$ is orthonormal then we have $V^{-1}=V^{\top}$.
\item Usually the software represents the eigenvalues in ascending order meaning that $\lambda_1>\lambda_2>\cdots>\lambda_p>0$.
\item $\Sigma ^{n}=V D^{n} V^{\top}$, for $n \in \mathbb{N}$.
\item $\Sigma ^{-1}=V D^{-1} V^{\top}$.
\item 
\begin{align*}
D^{-1}=
\left(
\begin{matrix}
\frac{1}{\lambda_{1}}   &0                  &0    	 				  &\cdots  &0\\
                 0       &\frac{1}{\lambda_{2}}  & 0   	 				  & \cdots &0\\
                 0       & 0                 & \frac{1}{\lambda_{3}}    	  &           &0\\
      \vdots           &\vdots           &         				  & \ddots &\\
              0          &   0               & 0   	 				  &           &\frac{1}{\lambda_{p}} 
\end{matrix}
\right).
\end{align*}
\end{enumerate}
Using first property given above that states $V^{-1}=V^{\top}$, it is easy to check that $\Sigma = U U^{\top}$ where $U=VD^{1/2}$. So recalling property (ii) given in Sub-section \ref{property-simulation-Gaussian-1} while choosing $\boldsymbol{c}=\boldsymbol{\mu}$, $A=U$, and $\boldsymbol{Z}\sim{\cal{N}}_{p}\bigl(\boldsymbol{0}, \boldsymbol{I}_{p} \bigr)$, it turns out that 
\begin{align}
\boldsymbol{\mu} + VD^{\frac{1}{2}} \boldsymbol{Z} \sim {\cal{N}}_{p}\bigl(\boldsymbol{\mu}, \Sigma \bigr).
\end{align}
\end{enumerate}
In what follows,a simple the \verb+R+ code is given for simulating from $p$-dimensional Gaussian distribution through the Cholesky decomposition.
\begin{lstlisting}[style=deltaj]
R> rmvnorm <- function(n, mean = Mu, scale = Sigma)
+{
+  p <- length(Mu)
+  L <- t( chol(Sigma) )
+  Z <- matrix( rnorm(n*p), nrow = p, ncol = n )
+  L <- t( chol(Sigma) )
+  X <- matrix(Mu, nrow = p, ncol = 1)%*%rep(1, n) + L %*%Z 
+  t(X) 
+}
\end{lstlisting}
There are several \verb+R+ packages that can be used for simulating from multivariate Gaussian distribution. 
%\section{Wishart distribution}
\subsection{Skew Gaussian distribution}\label{skew-Gaussian}
Here, 
%we consider to draw Bayesian inference on parameters of a model which contains two latent variable. To this end, first, 
we introduce the class of skew Gaussian distributions that is known in the literature as the canonical fundamental unrestricted skew Gaussian distribution, see \cite{arellano2006unification}.
We write $\boldsymbol{X}\sim \text{SG}_{p,q}(\boldsymbol{\mu},\Sigma,{\Lambda})$ to denote that $p$-dimensional random vector $\boldsymbol{X}$ follows a canonical fundamental unrestricted skew Gaussian distribution with PDF given by  
\begin{align}\label{ursn}
f_{}(\boldsymbol{x}\big \vert \boldsymbol{\mu},{\Sigma},{{\Lambda}}) =2^q \boldsymbol{\phi}_{p}\bigl(\boldsymbol{x}\vert\boldsymbol{\mu},{\Omega}\bigr)
\boldsymbol{\Phi}_{q}\bigl(\boldsymbol{m}\vert\boldsymbol{0}_{q},{\Delta}\bigr),
\end{align}
where ${\Omega}={\Sigma}+{{\Lambda}}{{\Lambda}}^{\top}$, ${\Delta}=\boldsymbol{I}_q-{{\Lambda}}^{\top}{\Omega}^{-1}{{\Lambda}}$, $\boldsymbol{m}={\Lambda}^{\top}{{\Omega}}^{-1}(\boldsymbol{y}-\boldsymbol{\mu})$, and $\boldsymbol{I}_{q}$ denotes the
$q \times q$ identity matrix. Further, $\boldsymbol{\phi}_{p}\bigl(.\big \vert \boldsymbol{\mu},{\Omega}\bigr)$ denotes the PDF of a $p$-dimensional Gaussian distribution with location vector $\boldsymbol{\mu}$ and dispersion matrix $\Sigma$, and $\boldsymbol{\Phi}_{q}\bigl(.\big\vert\boldsymbol{0}_{q}, {\Delta}\bigr)$ is the CDF of a $q$-dimensional Gaussian distribution with location vector $\boldsymbol{0}_{q}$ (a vector of zeros of length $q$) and dispersion matrix ${\Delta}$. The random vector $\boldsymbol{X}$ admits stochastic representation as follows, see \citep{arellano2005fundamental,arellano2006unification,arellano2007bayesian}.
\begin{align}\label{ursnrep0}
\boldsymbol{X} \mathop=\limits^d\boldsymbol{\mu}+{\Lambda}\big \vert \boldsymbol{Z}_0 \big \vert+ {\Sigma}^{\frac{1}{2}}\boldsymbol{Z}_1,
\end{align}
where $ \mathop=\limits^d$ means ``distributed as" and
\begin{align}\label{ursnrep}
\biggl[\begin{matrix}
\boldsymbol{Z}_0\\ \boldsymbol{Z}_1
\end{matrix}\biggr]
\sim {\cal{N}}_{q+p}\biggl(\biggl[\begin{matrix}
\boldsymbol{0}_{q}\\ \boldsymbol{0}_{p}
\end{matrix}\biggr],
\biggl[\begin{matrix}
\boldsymbol{I}_{q}&\boldsymbol{0}_{q\times p}\\
\boldsymbol{0}_{p\times q}& {\Sigma}\\
\end{matrix}\biggr]
\biggr).
\end{align}
The Bayesian paradigm for $\text{SG}_{p,q}(\boldsymbol{\mu},\Sigma,{\Lambda})$ distribution with representation (\ref{ursnrep0}) has been developed by \cite{liseo2013bayesian}. It can be shown that \citep{maleki2019robust,morales2022moments}:
\begin{align}\label{ursnrep}
\biggl[\begin{matrix}
\boldsymbol{Z}_0\\ \boldsymbol{X}
\end{matrix}\biggr]
\sim {\cal{N}}_{q+p}\biggl(\biggl[\begin{matrix}
\boldsymbol{0}_{q}\\ \boldsymbol{0}_{p}
\end{matrix}\biggr],
\biggl[\begin{matrix}
\boldsymbol{I}_{q}&\boldsymbol{\Lambda}^{\top}\\
\boldsymbol{\Lambda}& {\Sigma}+\boldsymbol{\Lambda}\boldsymbol{\Lambda}^{\top}\\
\end{matrix}\biggr]
\biggr).
\end{align}
%%%%%%%%%%%%%%%%%%%%%%%
The following \verb+R+ code can be used for simulating $n$ realizations from skew Gaussian distribution.
\begin{lstlisting}[style=deltaj]
R> rsn <- function(n, mean = Mu, scale = Sigma, shape = Lambda)
+	{
+		p <- length(Mu)
+		q <- length(Lambda[1, ])
+		Y <- matrix(NA, nrow = p, ncol = n)
+		Z1 <- rmvnorm( n, mean = rep(0, p), scale = Sigma )
+		Z0 <- abs( rmvnorm(n, mean = rep(0, q), scale = diag(q)) )
+		Y <- matrix(Mu, nrow = p, ncol = 1)%*%rep(1, n) + Lambda%*%t(Z0) + t(Z1) 
+		 t(Y)
+  }
\end{lstlisting}
%%%%%%%%%%%%%%%%%%%%%%%%%%%%%%%%%%%%%%%%%%%%%%%%%
%%%%%%%%%%%%%%%%%%%%%%%%%%%%%%%%%%%%%%%%%%%%%%%%%
%%%%%%%%%%%%%%%%%%%%%%%%%%%%%%%%%%%%%%%%%%%%%%%%%
%%%%%%%%%%%%%%%%%%%%%%%%%%%%%%%%%%%%%%%%%%%%%%%%%
%%%%%%%%%%%%%%%%%%%%%%%%%%%%%%%%%%%%%%%%%%%%%%%%%
\chapter{Importance Sampling}\label{sec33}
The Monte Carlo techniques essentially can be seen as a tool for computing single or multiple integrals. This feature of the Monte Carlo techniques that focuses on computing integrals is called {\it{importance}} sampling. It is worth noting that the {\it{importance}} sampling aims to compute the expected value of an estimator that is typically computing an integral. This concern of this chapter is placed on computing the single and multiple integrals that are widely used in statistical analyses.
\section{Importance sampling}
Let $h(x)$ to be a continuous real function and $f(x)$ is a given PDF. Suppose we would like to compute the expectation
\begin{align}\label{importance-sampling-1}
E_{f}\bigl(h(X)\bigr)=\int_{\mathbb{R}} h(x)f(x)dx,
\end{align}
where the generic symbol $E_{f}(h(X))$ indicates that the expectation of random variable $h(X)$ is computed when $X$ follows the PDF $f$. Usually, computing (\ref{importance-sampling-1}) directly is not a simple task and mus be evaluated numerically. The importance sampling is a tool to get round this difficulty. To this end,  
%The importance sampling enables us to compute above integral by considering a suitable PDF as the candidate.To this end,
notice that the integral (\ref{importance-sampling-1}) can be rewritten as
 \begin{align}\label{importance-sampling-2}
E_{f}\bigl(h(X)\bigr)=\int_{\mathbb{R}}h(y)\frac{f(y)}{g(y)}g(y)dy=E_{g}\Bigl(h(X)\frac{g(X)}{f(X)}\Bigr)=E_{g}\bigl(h(X) w(X)\bigr),
\end{align}
where $w(u)=f(u)/g(u)$. Herein, the PDF  $g(\cdot)$ is called the importance or instrumental PDF and its support dominates that of the PDF $f(x)$, that is ${\cal{S}}_{f} \subset {\cal{S}}_{g}$. By the law of large numbers, the RHS of (\ref{importance-sampling-2}) is estimated as
\begin{align}\label{importance-sampling-3}
\hat{\mu}=\widehat{E_{f}\bigl(h(X)\bigr)}= \frac{1}{n}\sum_{i=1}^{n}w(x_{i})h(x_{i}), 
\end{align}
where $x_1,\cdots,x_n$ come independently from PDF $g(\cdot)$. Hence, by generating a large number of realizations from PDF $g(\cdot)$ the preceding expectation is approximated well. for a wide ranges of applications the target PDF $f(\cdot)$ is just known up to a normalization constant. In such cases, estimator of $E_{f}\bigl(h(X)\bigr)$ is given by
\begin{align}\label{importance-sampling-4}
\tilde{\mu}=\widetilde{E_{f}\bigl(h(X)\bigr)}= \frac{\sum_{i=1}^{n}w(x_{i})h(x_{i})}{\sum_{i=1}^{n}w(x_{i})},
\end{align}
that is self-normalized. The bias and variance of both estimators $\hat{\mu}$ and $\tilde{\mu}$ can be evaluated \citep{liu2001monte}. We have
\begin{align}
E_{g}\bigl(\hat{\mu}\bigr)&=\mu\label{importance-sampling-e1}\\
E_{g}\bigl(\tilde{\mu}\bigr)&\approx\mu+\frac{\mu \times \text{var}_{g}(w(X))-\text{cov}_{g}\bigl(w(X),w(X)h(X)\bigr)}{n}\label{importance-sampling-e2}
\end{align}
and
\begin{align}
\text{var}_{g}\bigl(\hat{\mu}\bigr)&=\frac{\text{var}_{g}\bigl(w(X)h(X)\bigr)}{n}\label{importance-sampling-var1}\\
\text{var}_{g}\bigl(\tilde{\mu}\bigr)&\approx\frac{\text{var}_{g}\bigl(w(X)h(X)\bigr)
-2\mu\times\text{cov}_{g}\bigl(w(X),w(X)h(X)\bigr)+\mu^2\times \text{var}_{g}\bigl(w(X)\bigr)}{n}\label{importance-sampling-var2}
\end{align}
It is clear from (\ref{importance-sampling-e1}) and (\ref{importance-sampling-e2}) that $\hat{\mu}$ is unbiased estimator for $\mu$ while the estimator $\tilde{\mu}$ is biased. Of course, in some situations the variance (standard error) of the self-normalized estimator $\tilde{\mu}$, as it is seen from RHS of (\ref{importance-sampling-var1}), can be smaller than that of $\hat{\mu}$ that is given by (\ref{importance-sampling-var2}). The main feature in each importance sampling problem is to determine the instrumental PDF $g(\cdot)$. Although, theoretically, there are many choices of $g(\cdot)$, but we have to find the best choice that yields as small as possible standard error for $\hat{\mu}$ (or $\tilde{\mu}$). Fortunately, there is a specific rule available for this purpose. It is proved that the optimal choice $g_{o}(x)$ becomes \citep{johansen2010monte}:
\begin{align}\label{importance-sampling-gchoice}
g_{o}(x)=\frac{\vert h(x)\vert f(x)}{\int_{\mathbb{R}}\vert h(u)\vert f(u)du}.
\end{align}
\begin{example}\label{exam-importance-sampling-1}%\lipsum*[]
%Recalling from Example \label{exam-empirical-gamma}, suppose $x_1,x_2,\cdots,x_n$ follow independently ${\cal{G}}(\alpha,\beta)$ with unknown $\alpha$ and known $\beta$. We can write
%\begin{align*}
%\pi\bigl(\alpha \big \vert \boldsymbol{x},\boldsymbol{\theta}_{0}\bigr) =\frac{f(\boldsymbol{x} \vert \alpha)\pi\bigl(\alpha\big \vert \boldsymbol{\theta}_{0}\bigr)}{m\bigl(\boldsymbol{x}\vert \boldsymbol{\theta}_{0}\bigr)},
%\end{align*} 
%where
%\begin{align*}
%m\bigl(\boldsymbol{x}\vert \boldsymbol{\theta}_{0}\bigr)&=\int_{0}^{\infty}\frac{\beta_{0}^{\alpha_{0}}\alpha_{0}^{\beta_{0}-1}\beta^{n\alpha}\alpha^{\alpha_{0}-1}}{\Gamma(\alpha_0)\bigl[\Gamma(\alpha)\bigr]^{n}}\Bigl[\prod_{i=1}^{n}x_{i}\Bigr]^{\alpha-1}\exp\bigl\{-\beta \sum_{i=1}^{n}x_{i}-\beta_{0}\alpha\bigr\}d\alpha.
%\end{align*} 
Suppose we are interested in computing ${{\Omega}}=E\bigl(\log(1+X)\bigr)$ where $X \sim {\cal{G}}(\alpha,\beta)$. We may choose three instrumentals as: 1- exponential distribution with rate parameter $\beta$, say ${\cal{E}}(\beta)$, 2-  gamma distribution, say ${\cal{G}}(\alpha,\beta)$, and the optimal one with PDF $g_{0}(x)=h(x){\cal{G}}(x\vert \alpha,\beta)$.
Under the first choice, we have
\begin{align}
\hat{\Omega}=\frac{\beta^{\alpha-1}}{\Gamma(\alpha)} \frac{1}{N}\sum_{i=1}^{N}x_{i}^{\alpha - 1}\log( x_{i} + 1 ) 
\end{align}
where $x_{i}$s come independently from ${\cal{E}}(\beta)$. In the case of second choice of instrumental, we have
\begin{align}
\hat{\Omega}=\frac{1}{N}\sum_{i=1}^{N}x_{i},
\end{align}
where $x_{i}$s come independently ${\cal{G}}(\alpha,\beta)$. In the case of $g_{o}(x)$ and ${\cal{SG}}(0, 1,\lambda)$, respectively.
and second choices of instrumental, we have
\begin{align}
\hat{\Omega}=N \biggl[\sum_{i=1}^{N}\frac{1}{\log \bigl( x_{i}+1\bigr)}\biggr]^{-1},
\end{align}
where $x_{i}$s come independently from PDF given by (\ref{exam-importance-sampling-gamma-go}). 
\begin{align}\label{exam-importance-sampling-gamma-go}
g_{o}(x)= \frac{ \log (x+1) {\cal{G}}(x\vert \alpha,\beta) }{\int_{0}^{\infty} \log (u+1) {\cal{G}}(u\vert \alpha,\beta) du},
\end{align}
For sampling from $g_{o}(x)$ given above, we designate a rejection scheme given by the following steps.
\begin{enumerate}
\item Sample realization $y$ from ${\cal{G}}(\alpha+1,\beta)$ as the proposal distribution;
\item Sample independent realization $u$ from ${\cal{U}}(0, 1)$;
\item if $y<\log(y+1)/y$, then accept $y$ as a generation from PDF $g_{o}(x)$ given by (\ref{exam-importance-sampling-gamma-go}); otherwise return to step (1) an repeat this procedure until having acceptance. 
\end{enumerate}
The \verb+R+ function \verb+rlgamma+, given below, generates sample from $g_{o}(x)$ based on above rejection scheme.
\begin{lstlisting}[style=deltaj]
R> rlgamma <- function(N, alpha, beta)
+   {
+   g <- rep(0, N) # vector of generated realizations
+   j <- 1
+   while(j <= N)
+	{
+		y <- rgamma(1, shape = alpha + 1, rate = beta)
+		u <- runif(1) 
+  	  A <- log(y+1)/(y)
+			if( u < A )
+			{
+			g[j] <- y        # y is accepted
+			j <- j + 1
+			}
+   }
+ return(g)
+ }
\end{lstlisting}
Table \ref{importance-sampling-gamma} represents the results of simulation study based on $M=5000$ runs of samples each of size $N=5000$. It follows from this table that  the second instrumental, that is ${\cal{SG}}(0, 1,\lambda)$, gives the best summary. Not surprisingly, the optimal instrumental does not work as well as or better than others since ${\text{var}}_{g_{o}}\bigl(W(X)\bigr)$ is inflated by the small values in the generated data from this instrumental. It is worth noting that the $g_{o}(\cdot)$ works well if $W(x)=f(x)/g_{o}(x)$ is bounded. For our case, this condition may fail since $W(x)=f(x)/g_{o}(x)=1/\vert x\vert$ becomes large in presence of even a few number of small realizations. 
\vspace{5mm}
\begin{table}[h!]
\center
\caption{Summary statistics for computing $\Omega=E\bigl(\log(1+X)\bigr)$ where $X \sim {\cal{G}}(\alpha,\beta)$ through three instrumentals.} 
\vspace*{-3mm}
\begin{tabular}{llccccc} 
\cline{1-7} 
& & &\multicolumn{4}{c}{summary}\\ \cline{4-7} 
$(\alpha,\beta)$&$\Omega$&instrumental&bias&SE&minimum&maximum\\ \cline{1-7} 
\multicolumn{1}{l}{\multirow{3}{*}{(0.5,0.5)}}&\multicolumn{1}{c}{\multirow{3}{*}{0.6695}}&
\multicolumn{1}{l}{${\cal{E}}(\beta)$}								&1.1e-04  &5.4e-05    & 0.5047 & 0.5611 \\  
&&\multicolumn{1}{l}{${\cal{G}}(\alpha,\beta)$} 					&-4.9e-05 &3.4e-06    & 0.5271 & 0.5404 \\  
&&\multicolumn{1}{l}{$g_{o}$}											&6.2e-03  &1.9e-03    & 0.0649 & 0.6249\\ \cline{1-7}                                                            			                  
\multicolumn{1}{l}{\multirow{3}{*}{(0.5,5)}}& \multicolumn{1}{c}{\multirow{3}{*}{0.0884}} &  \multicolumn{1}{l}{${\cal{E}}(\beta)$}								&-4.8e-05 &-4.9-05   & 0.0830    & 0.0948\\ 
&&\multicolumn{1}{l}{${\cal{G}}(\alpha,\beta)$} 					&-5.9e-07 & -5.9e-07  & 0.0864 & 0.0902\\
&&\multicolumn{1}{l}{$g_{o}$}											&1.4e-03  & 1.5e-03   & 0.0066 & 0.1076\\ \cline{1-7}                                                           			                  
\multicolumn{1}{l}{\multirow{3}{*}{(5,0.5)}}& \multicolumn{1}{c}{\multirow{3}{*}{2.3152}} &                                  			                  
\multicolumn{1}{l}{${\cal{E}}(\beta)$}								& 1.2e-04 &3.4e-05    & 2.2940 & 2.3356\\   
&&\multicolumn{1}{l}{${\cal{G}}(\alpha,\beta)$} 					&-4.1e-03 &1.2e-01    & 1.4422 & 6.0980\\ 
&&\multicolumn{1}{l}{$g_{o}$}											&-2.9e-05 &-4.1e-05   & 2.2926 & 2.3363\\ \cline{1-7}                                                           			                  
\multicolumn{1}{l}{\multirow{3}{*}{(5,5)}}& \multicolumn{1}{c}{\multirow{3}{*}{0.6695}}   &                                     			                  
\multicolumn{1}{l}{${\cal{E}}(\beta)$}								&-1.8e-05 & 3.4e-05   & 0.6586 & 0.6814\\   
&&\multicolumn{1}{l}{${\cal{G}}(\alpha,\beta)$} 					&1.1e-03  & 1.1e-01   & 0.4109 & 1.5190\\ 
&&\multicolumn{1}{l}{$g_{o}$}											&1.2e-05  & 4.1e-05   & 0.6562 & 0.6848\\ \cline{1-7}
\end{tabular} 
%\tabnote{Hint: CI is short form for confidence interval.}                                                                                                           
\label{importance-sampling-gamma}
\end{table}
%where $f$ is the PDF of ${\cal{G}}\bigl(\alpha_{0},\beta_{0}\bigr)$ and 
% \begin{align}\label{exam-importance-sampling-h}
%h(\alpha)=\bigl[\Gamma(\alpha)\bigr]^{-n}\beta^{n\alpha}\Bigl[\prod_{i=1}^{n}x_{i}\Bigr]^{\alpha-1}.
%\end{align} 
%So, the optimal choice of instrumental PDF becomes
%  \begin{align}\label{exam-importance-sampling-goptimal}
%g_{o}(\alpha)=&\frac{\bigl[\Gamma(\alpha)\bigr]^{-n}\beta^{n\alpha}\Bigl[\prod_{i=1}^{n}x_{i}\Bigr]^{\alpha-1} {\cal{G}}\bigl(\alpha \big \vert \alpha_{0},\beta_{0}\bigr)}{\int_{0}^{\infty}
%\bigl[\Gamma(u)\bigr]^{n}\beta^{nu}\Bigl[\prod_{i=1}^{n}x_{i}\Bigr]^{u-1} {\cal{G}}\bigl(u \big \vert \alpha_{0},\beta_{0}\bigr)du}\nonumber\\
%=&\frac{\bigl[\Gamma(\alpha)\bigr]^{-n}\beta^{n\alpha}}{m_{0}}\Bigl[\prod_{i=1}^{n}x_{i}\Bigr]^{\alpha-1} {\cal{G}}\bigl(\alpha \big \vert \alpha_{0},\beta_{0}\bigr),
%\end{align}
%where $m_{0}$ is normalizing constant independent of $\alpha$. The importance sampling technique suggests the estimator of $\mu$ in (\ref{exam-importance-sampling-mu}) as
%
%\begin{align}\label{exam-importance-sampling-mu}
%\tilde{\mu}= \frac{\sum_{i=1}^{N}w(\alpha_{i})h(\alpha_{i})}{\sum_{i=1}^{N}w(\alpha_{i})},
%\end{align}
%where $\alpha_{1},\cdots,\alpha_{N}$ come from PDF $g_{o}(\alpha)$ in (\ref{exam-importance-sampling-goptimal}), $h(\alpha)$ is given by (\ref{exam-importance-sampling-h}), and 
%\begin{align}\label{exam-importance-sampling-w}
%w(\alpha)=\frac{{\cal{G}}\bigl(\alpha \big \vert \alpha_{0},\beta_{0}\bigr)}{g_{o}(\alpha)}=\frac{m_{0}}{h(\alpha)}.
%\end{align}
\end{example}
Next example shows that if above conditions are not met, then the importance sampling using the optimal choice $g_{o}(\cdot)$ may fail to show the best performance.
\begin{example}\label{exam-importance-sampling-sg1}%\lipsum*[]
Suppose we would like to compute $\Omega=E(\vert X\vert)$ where $X\sim {\cal{SG}}(0,1,\lambda)$ and expression of ${\cal{SG}}(x\vert 0,1,\lambda)$ is given by (\ref{PDF-sg}). To this end, we consider three instrumentals including:  ${\cal{N}}(0, 1)$, ${\cal{SG}}(0, 1,\lambda)$, and $g_{o}(x)$. The PDF of the latter is given by
\begin{align}\label{exam-importance-sampling-go}
g_{o}(x)= \frac{ 2\vert x \vert \phi(x)\Phi(\lambda x)}{\int_{-\infty}^{\infty}2\vert u \vert \phi(u)\Phi(\lambda u) du},
\end{align}
where quantity $c$ plays the role of normalizing constant. But, first, we need to simulate from $g_{o}(x)$ given by (\ref{exam-importance-sampling-go}). In what follows we give two methods for this purpose without proof.
\begin{enumerate}
\item Sample independent realizations $z_1$ and $z_2$ from ${\cal{N}}(0, 1)$;
\item Sample independent realization $u\sim {\cal{U}}(0, 1)$;
\item if $z_1<\lambda z_2$ and $4u<\big \vert z_2 \big \vert$, then accept $z_2$ as a generation from PDF (\ref{exam-importance-sampling-go}); otherwise return to step (1) an repeat this procedure until having acceptance. 
\end{enumerate}
The second method is as follows.
\begin{enumerate}
\item Sample realization $w\sim {\cal{W}}(2, 1/\sqrt{2})$;
\item Sample independent realization $u\sim{\cal{U}}(0, 1)$;
\item If $u<\Phi(\lambda w)$, then accept $w$ as a generation from PDF (\ref{exam-importance-sampling-go}); otherwise accept $-w$ as a generation from PDF (\ref{exam-importance-sampling-go}). 
\end{enumerate}
Evidently, the second algorithm is more efficient than the first one in terms of the time of implementing. For computing $\hat{\Omega}$, we consider under the first choice of instrumental, we have
\begin{align}
\hat{\Omega}=\frac{2}{N}\sum_{i=1}^{N}\big \vert x_{i}\big \vert \Phi\bigl(\lambda x_{i}\bigr),
\end{align}
where $x_{i}$s come independently from ${\cal{N}}(0, 1)$. In the case of second choice of instrumental, we have
\begin{align}
\hat{\Omega}=\frac{1}{N}\sum_{i=1}^{N}x_{i},
\end{align}
where $x_{i}$s come independently ${\cal{SG}}(0, 1,\lambda)$, respectively.
In the case of $g_{o}(x)$   and ${\cal{SG}}(0, 1,\lambda)$, respectively.
and second choices of instrumental, we have
\begin{align}
\hat{\Omega}=N\Bigl [\sum_{i=1}^{N}\frac{1}{\big \vert x_{i}\big \vert}\Bigr]^{-1},
\end{align}
where $x_{i}$s come independently from PDF given by (\ref{exam-importance-sampling-go}). Table represents the results of simulation study based on 5000 runs of samples each of size $N=5000$. As it may be seen from Table \ref{importance-sampling-univariate-sg}, overall, the second instrumental, that is ${\cal{SG}}(0, 1,\lambda)$, gives the best summary. Not surprisingly, the optimal instrumental does not work as well as or better than others since ${\text{var}}_{g_{o}}\bigl(W(X)\bigr)$ is inflated by the small values in the generated data from this instrumental. It is worth noting that the $g_{o}(\cdot)$ works well if $W(x)=f(x)/g_{o}(x)$ is bounded. For our case, this condition may fail since $W(x)=f(x)/g_{o}(x)=1/\vert x\vert$ becomes large in presence of even a few number of small realizations. 
\begin{table}[ht]
\center
\caption{Summary statistics for computing $\Omega=E(\vert X\vert)$ where $X\sim {\cal{SG}}(0,1,\lambda)$ through three instrumentals.}  
\begin{tabular}{cccccc} 
\cline{1-6} 
& & \multicolumn{4}{c}{summary}\\ \cline{3-6} 
$\lambda$&instrumental&bias&SE&minimum&maximum\\ \cline{1-6}
\multicolumn{1}{c}{\multirow{3}{*}{$\lambda=0$}}&
\multicolumn{1}{c}{{{${\cal{N}}(0, 1)$}}}&5.1e-05&0.0083&  0.7688&  0.8250\\  
&\multicolumn{1}{c}{{{${\cal{SG}}(0, 1,\lambda=0)$}}} &3.8e-05&0.0084&  0.7663&  0.8259\\  
&\multicolumn{1}{c}{{{$g_{o}(\cdot)$}}}&1.0e-03&0.0218&  0.4925&  0.8574\\ 
\cline{1-6}
\multicolumn{1}{c}{\multirow{3}{*}{$\lambda=5$}}& 
\multicolumn{1}{c}{{{${\cal{N}}(0, 1)$}}}&4.3e-04&0.0168&  0.7267&  0.8560\\ 
&\multicolumn{1}{c}{{{${\cal{SG}}(0, 1,\lambda=5)$}}} &3.5e-05&0.0084&  0.7693&  0.8290\\
&\multicolumn{1}{c}{{{$g_{o}(\cdot)$}}}&1.1e-03&0.0214&  0.5341&  0.8573\\
\cline{1-6}
\multicolumn{1}{c}{\multirow{3}{*}{$\lambda=10$}}& 
\multicolumn{1}{c}{{{${\cal{N}}(0, 1)$}}}&7.2e-05&0.0165&  0.7367&  0.8553\\   
&\multicolumn{1}{c}{{{${\cal{SG}}(0, 1,\lambda=10)$}}} &2.9e-05&0.0084&  0.7683&  0.8311\\ 
&\multicolumn{1}{c}{{{$g_{o}(\cdot)$}}}&6.9e-04&0.0221&  0.2998&  0.8533\\ 
\cline{1-6}
\\
\end{tabular} 
%\tabnote{Hint: CI is short form for confidence interval.}                                                                                                           
\label{importance-sampling-univariate-sg}
\end{table}
\end{example}
\section{Variance reduction}
Suppose $\hat{\theta}$ is an unbiased candidate (estimator) for estimating unknown parameter $\theta$. This means that $E(\hat{\theta})={\theta}$. For a better unbiased estimator $\hat{\theta}^{*}$, if there exists, we have $var\bigl(\hat{\theta}^{*}\bigr)<var\bigr(\hat{\theta}\bigr)$. Here, we are interested in finding $\hat{\theta}^{*}$. To this end, we recall the well-known Rao-Blackwellization approach that states the standard error of an given biased (or unbiased) estimator $\hat{\theta}$ can be reduced by conditioning on a {\it{sufficient}} statistics (estimator). Overall, we can write
\begin{align}
var\bigr(\hat{\theta}\bigr)= E\bigl[var\bigr(\hat{\theta}\big \vert T\bigr)\bigr] + var\bigl[E\bigr(\hat{\theta}\big \vert T\bigr)\bigr].
\end{align}
This implies
\begin{align}
var\bigl[E\bigr(\hat{\theta}\big \vert T\bigr)\bigr] < var\bigr(\hat{\theta}\bigr).
\end{align}
Hence, $\hat{\theta}^{*}=E\bigr(\hat{\theta}\big \vert T\bigr)$ is an unbiased estimator whose standard error is smaller than that of the $\hat{\theta}$. 
\begin{example}\label{exam-importance-sampling-1}%\lipsum*[]
Suppose $X_{1},X_{2}\sim {\cal{N}}(\mu,1)$ are independent random variables. It is known that $T=X_{1}+X_{2} \sim {\cal{N}}(2\mu, 2)$ is a sufficient statistic for unknown parameter $\mu$. Moreover, we know that $\hat{\theta}=X^{2}_{1}-1$ is an unbiased estimator for $\mu^{2}$ with variance 1. But, an estimator of $\mu^{2}$ with smaller standard error, called here $\hat{\theta}^{*}$, can be found using Rao-Blackwellization procedure as follows. First, we can see that the PDF of $X_{1}$ given $T=X_{1}+X_{2}$ is 
\begin{align*}%\label{pdf-Gaussian-Rao-blackwell}
f_{X_{1}\vert T}(x_{1}\vert T=t)&=\frac{f_{X_{1},X_{2}}(x_{1}, t-x_{1})}{f_{T}(t)}\nonumber\\
&=\frac{f_{X_{1}}(x_{1})f_{X_{2}}(x_{1}, t-x_{1})}{f_{T}(t)}\nonumber\\
&=\frac{\sqrt{2}}{\sqrt{2\pi}}\exp\bigl\{-\bigl(x_{1}-t/2\bigr)^{2}\bigr\}.
\end{align*}
So, $X_{1} \big \vert X_{1}+X_{2}=t \sim {\cal{N}}\bigl(t/2, 1/2\bigr)$. It follows that
\begin{align*}
\hat{\theta}^{*}=&E\bigl[ \hat{\theta} \big \vert T=t \bigr]\\
=&E\bigl[ X^{2}_{1} \big \vert X_{1}+X_{2}=t \bigr]-1\\
=&var\bigl[X_{1}\big \vert X_{1}+X_{2}=t \bigr] +\bigl\{E\bigl[X_{1}\big \vert X_{1}+X_{2}=t\bigr]\bigr\}^{2}-1\\
=&\frac{1}{2}+\frac{t^{2}}{4}-1\\
=&\bar{X}^{2}-\frac{1}{2}.
\end{align*}
It is evident that both estimators $\hat{\theta}$ and $\hat{\theta}^{*}$ are unbiased for $\mu^2$, that is $E\bigl(\hat{\theta}\bigr)=E\bigl(\hat{\theta}^{*}\bigr)=\mu^2$, but  $var\bigl(\hat{\theta}^{*}\bigr)<var\bigl(\hat{\theta}\bigr)$.
\end{example}
\section{Estimators with negative covariance}\label{Estimators-negative-covariance}
Suppose both of estimators $\hat{\theta}_{1}$ and $\hat{\theta}_{2}$ are unbiased for unknown parameter ${\theta}$. If $cov(\hat{\theta}_{1},\hat{\theta}_{2})<0$, then $\hat{\theta}^{*}=\bigl[\hat{\theta}_{1}+\hat{\theta}_{2}\bigr]/2$ is also unbiased estimator for ${\theta}$ with smaller standard error than either $\hat{\theta}_{1}$ or $\hat{\theta}_{2}$. This because
\begin{align*}
var\bigl(\hat{\theta}^{*}\bigr)=&\frac{var\bigl(\hat{\theta}_{1}\bigr)+var\bigl(\hat{\theta}_{2}\bigr)}{4}+\frac{cov(\hat{\theta}_{1},\hat{\theta}_{2})}{2}
\\
\leq &\frac{var\bigl(\hat{\theta}_{1}\bigr)+var\bigl(\hat{\theta}_{2}\bigr)}{4}\\
<&\frac{1}{2} \max \bigl\{var\bigl(\hat{\theta}_{1}\bigr),var\bigl(\hat{\theta}_{2}\bigr)\bigr\}.
\end{align*}
\begin{example}\label{exam-importance-sampling-BS}%\lipsum*[]
Let random variable $X$ follows a BS distribution with PDF given by (\ref{pdf-BS-1}) in which parameter $\beta$ is known. It can be shown that \cite{ng2003modified}:
\begin{align*}%\label{moment-BS-1}
E(X)=\beta\Bigl(1+\frac{\alpha^2}{2}\Bigr).
\end{align*}
Since $1/X\sim{\cal{BS}}(\alpha, 1/\beta)$, we can write
\begin{align*}%\label{moment-BS-2}
E\Bigl(\frac{1}{X}\Bigr)=\frac{1}{\beta}\Bigl(1+\frac{\alpha^2}{2}\Bigr).
\end{align*}
Based on a random sample of size $n$ denoted by $x_1,\cdots,x_n$ drawn from ${\cal{BS}}(\alpha, 1)$, one can see that $\hat{\theta}_{1}=2(S-1)$ and $\hat{\theta}_{2}=2(R-1)$ are unbiased estimators for $\alpha^2$ in which $S=1/n\sum_{i=1}^{n}x_{i}$ and $R=1/n\sum_{i=1}^{n}1/x_{i}$. Intuitively, the statistics $S$ and $R$ have negative correlation, and hence, we can construct a new estimator represented as
\begin{align*}%\label{moment-BS-3}
\hat{\theta}^{*}=\frac{\hat{\theta}_{1}+\hat{\theta}_{2}}{2},
\end{align*}
whose standard error is less that that of $\hat{\theta}_{1}$ and $\hat{\theta}_{2}$. We performed a small comparison study in which four estimators including $\hat{\theta}_{MLE}$ (ML estimator of $\alpha^2$), $\hat{\theta}^{*}$, $\hat{\theta}_{1}$, and $\hat{\theta}_{2}$ take part for estimating $\alpha^2$ when $\beta$ is assumed to be known. To this end, aforementioned four competitors have been computed for based on random sample of size $n=100$ for $N=1000$ runs. It is worth to note that for simulating for BS distribution we used the \verb+R+ function \verb+rbs+ given in \ref{distribution-BS} and the ML estimator of parameter $\alpha^2$ is computed using package \verb+bibs+ available at \verb+https://cran.r-project.org/web/packages/bibs/index.html+, then the $\hat{\theta}_{MLE}$ is easily obtained as the squared ML estimator of $\alpha$ by the general nice invariance property of the ML estimator. Figure \ref{plot-importance-BS-1} shows the average and standard error of these competitors. As it is seen, in terms of average and  standard error, both estimators $\hat{\theta}_{MLE}$ and $\hat{\theta}^{*}$ shows the same performance and simultaneously outperform $\hat{\theta}_{1}$ and $\hat{\theta}_{2}$. The corresponding \verb+R+ code is given by the following.
\begin{lstlisting}[style=deltaj]
R> library("bibs")
R> N <- 1000; n <- 100; alpha <- seq(0.2, 5, length = 10); beta <- 1
R> r <- s <- alpha2_mle <- alpha2_star <- alpha2_hat1 <- alpha2_hat2 <- rep(0, N) 
R> bias <- rmse <- matrix(0, nrow = length(alpha), ncol = 4 )
R> for(j in 1:length(alpha) ){ 
+ for(i in 1:N){
+ x <- rbs(n, alpha[j], beta); s[i] <- mean(x); r[i] <- mean(1/x)
+ alpha2_mle[i] <- mlebs(x, c(alpha[j], beta), method = "Nelder-Mead", CI = 0.95)$Estimates[[1]]
+ alpha2_hat1[i] <- ( 2*(r[i]*beta - 1) ); alpha2_hat2[i] <- ( 2*(s[i]/beta - 1) )
+ alpha2_star[i] <- ( alpha2_hat1[i] + alpha2_hat2[i] )/2
+ }
+ hat <- as.matrix( cbind(alpha2_mle^2, alpha2_star, alpha2_hat1, alpha2_hat2), nrow = N, ncol = 4 )
+ shifted <- sweep(hat, 2, alpha[j]^2, FUN = '-')
+ bias[j, ] <- apply( shifted, 2, mean )
+ rmse[j, ] <- sqrt( apply( shifted, 2, var ) )
+ }
\end{lstlisting}
 \begin{figure}[!h]
\center
\includegraphics[width=55mm,height=55mm]{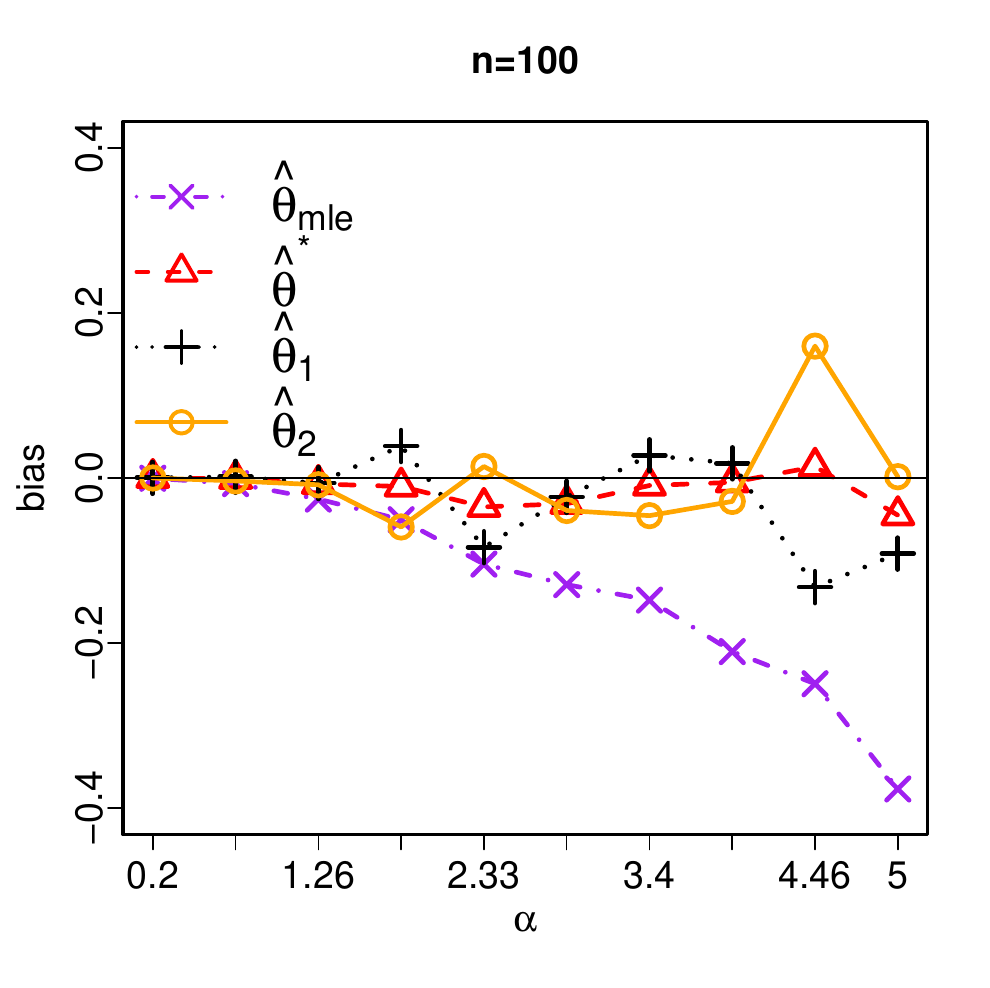}
\includegraphics[width=55mm,height=55mm]{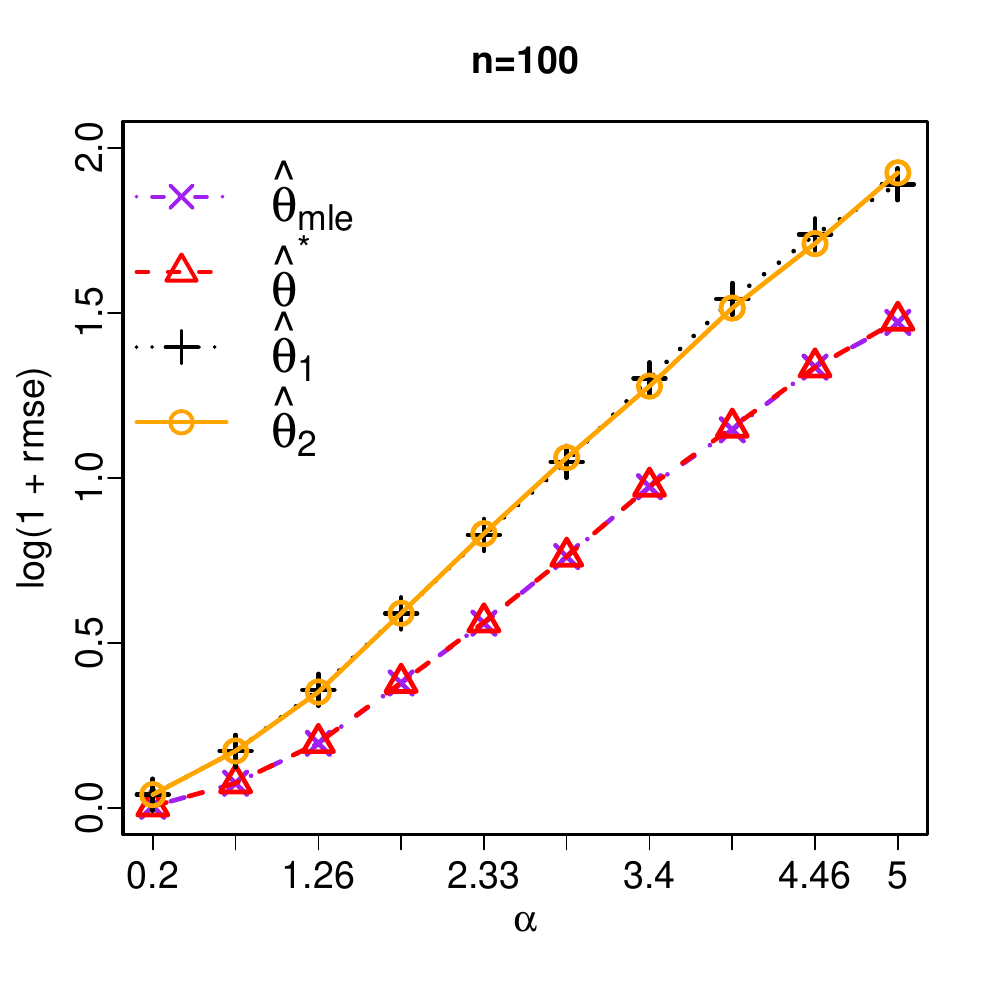}\\
\includegraphics[width=55mm,height=55mm]{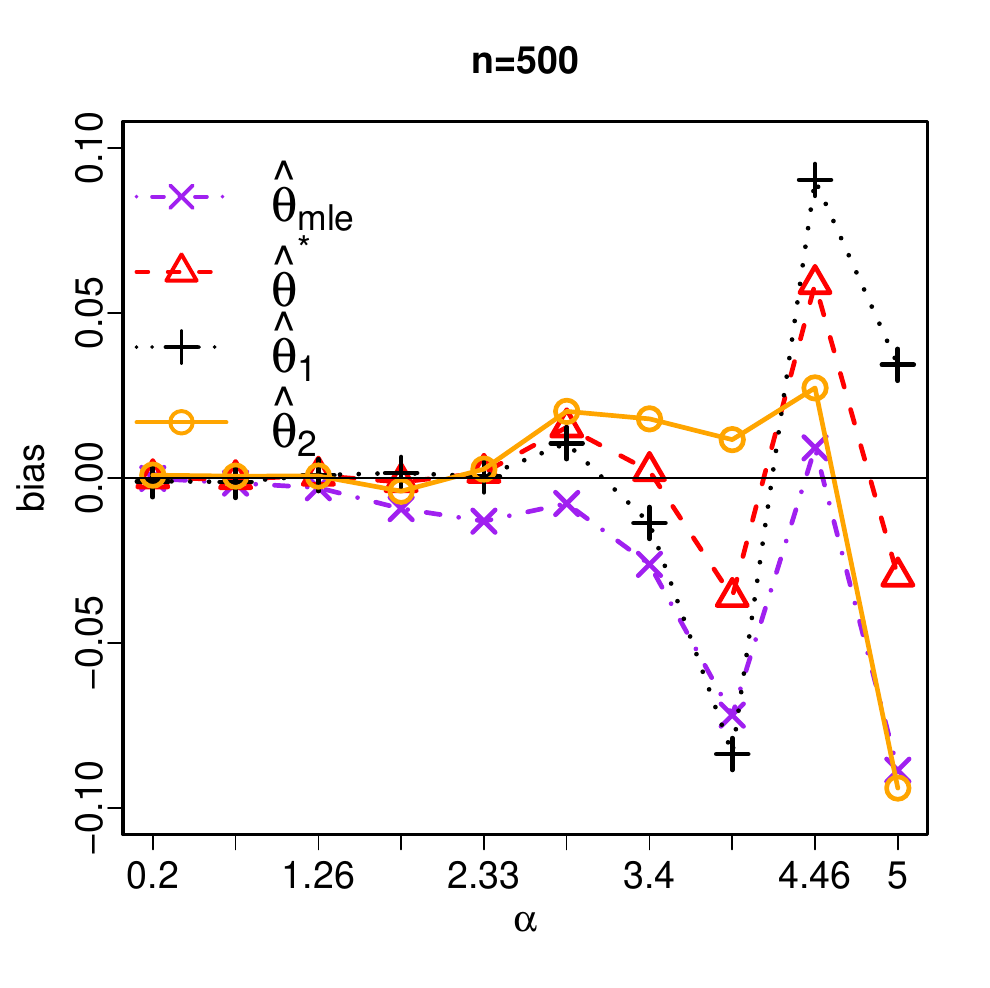}
\includegraphics[width=55mm,height=55mm]{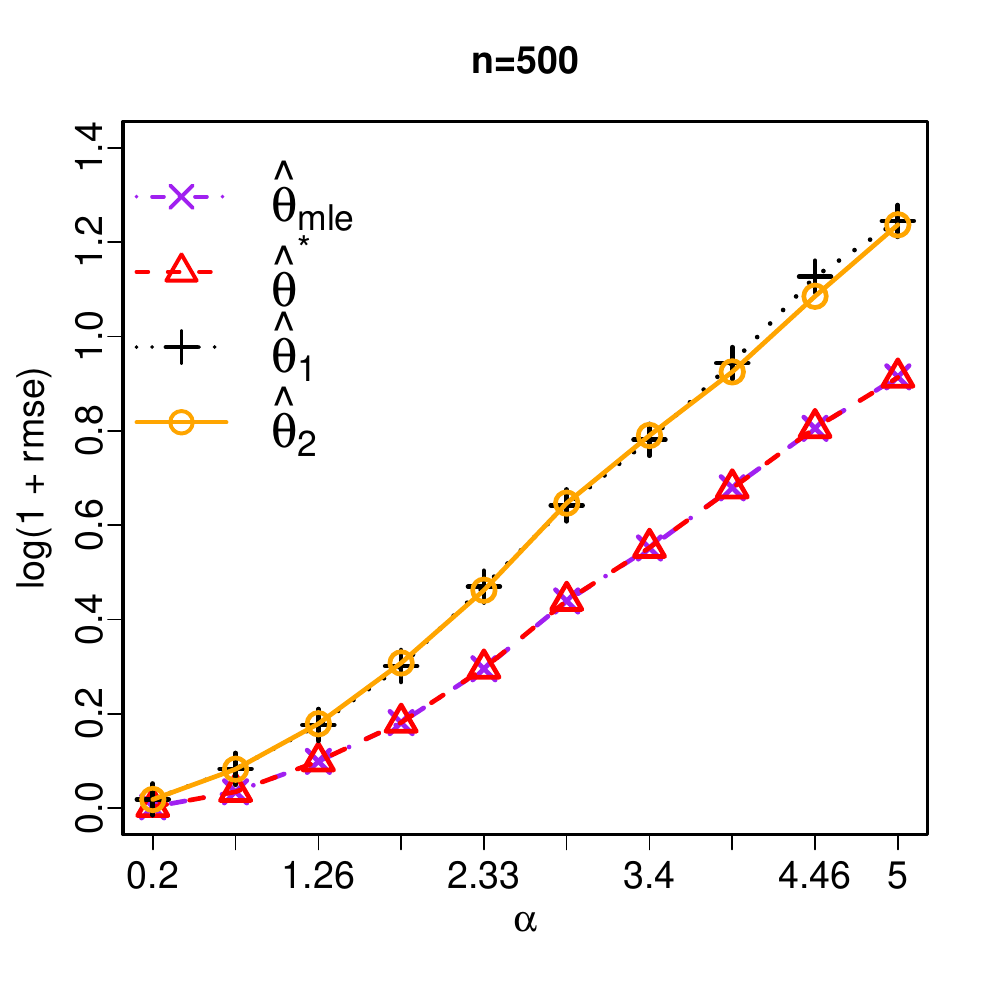}
\caption{The top row shows the bias and {\text{log(1+rmse)}} for $\hat{\theta}_{MLE}$, $\hat{\theta}^{*}$, $\hat{\theta}_{1}$, and $\hat{\theta}_{2}$ computed based on $N=1000$ independent sample each of size $n=100$ generated from ${\cal{BS}}(\alpha, \beta=1)$. The bottom row displays the same graph with the same parameters except $n=500$.}
\label{plot-importance-BS-1}
\end{figure}
\end{example}
\begin{example}\label{exam-importance-sampling-BS}%\lipsum*[]
Let $U,V\sim{\cal{U}}(-1,1)$ are independent random variables distributed jointly on $(-1,1) \times (-1,1)$ (shown by a square in Figure \ref{plot-circle-square-pi-approximation}). 
 \begin{figure}[!h]
\center
\includegraphics[width=65mm,height=65mm]{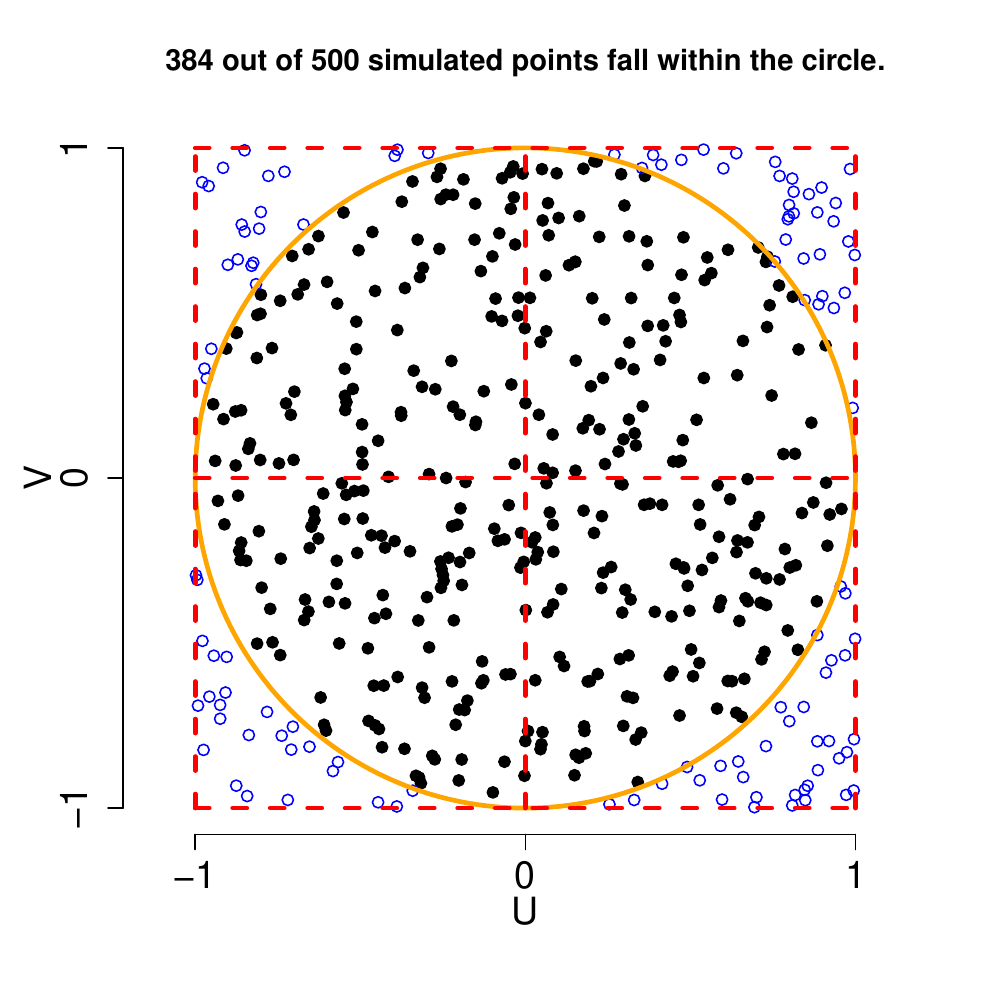}
\caption{Components of random vector $(U,V)^{\top}$ come independently form ${\cal{U}}(-1,1)$.}
\label{plot-circle-square-pi-approximation}
\end{figure}
It may bee seen that the probability of falling a point inside the circle in Figure \ref{plot-circle-square-pi-approximation} is $\pi/4$. Hence, the constant $\pi$ is approximately as
\begin{align}\label{approximation-pi}
  \hat{\pi} = 4 \times \frac{{\text{number of pairs lie inside the circle}}}{{\text{number of pairs do not lie inside the circle}}}.
\end{align} 
Therefore, the constant $\pi$ can be estimated by simulating a sufficiently large ($n$, say) numbers of point distributed uniformly on $(-1,1) \times (-1,1)$ and then computing the  ratio \ref{approximation-pi}. To this end, we define
\begin{eqnarray}\label{indicator-function-circle}
\displaystyle
\mathbb{I}_{{\cal{C}}}(u_{i},v_{i})=\left\{\begin{array}{c}
\displaystyle
1,~\mathrm{{if}}\ i {\text{th pair}}~(u_{i},v_{i})^{\top}~{\text{lies inside the circle}},~~~\\
\displaystyle
0,~\mathrm{if}\  i {\text{th pair}}~(u_{i},v_{i})^{\top}~{\text{lies outside the circle}},
\end{array} \right.
\end{eqnarray}
and then, based on the law of large numbers, we have
\begin{align}
 \hat{\pi} = 4 \times \frac{\sum_{i=1}^{n}\mathbb{I}_{{\cal{C}}}(u_{i},v_{i})}{n}.
\end{align} 
It is easy to check that $E(\hat{\pi})=\pi$ and $var( \hat{\pi})=4\pi(1-\pi/4)/n=2.69676/n$. Here, we interested in improving estimator $\hat{\pi}$ by conditioning on marginal variable $U$ as follows.
\begin{align*}
\hat{\pi}_{1}=& \frac{4}{n}\sum_{i=1}^{n} E\bigl[ \mathbb{I}_{{\cal{C}}}(U_{i},V_{i}) \big \vert U_{i}=u_{i}\bigr]\nonumber\\
=&\frac{4}{n}\sum_{i=1}^{n} P\bigl[ U^{2}_{i}+V^{2}_{i}\leq 1 \big \vert U_{i}=u_{i}\bigr]\nonumber\\
=&\frac{4}{n}\sum_{i=1}^{n}P\bigl[ V^{2}_{i}\leq 1 - u^{2}_{i}\big \vert U_{i}=u_{i}\bigr]\nonumber\\
=&\frac{4}{n}\sum_{i=1}^{n}P\bigl[ V^{2}_{i}\leq 1 - u^{2}_{i}\bigr]\nonumber\\
=&\frac{4}{n}\sum_{i=1}^{n}\sqrt{1-u^{2}_{i}},
\end{align*} 
where $U_{i}\sim {\cal{U}}(-1,1)$. This yields
\begin{align*}
\hat{\pi}_{1}=&\frac{4}{n}\sum_{i=1}^{n}\sqrt{1-U^{2}_{i}}.
\end{align*} 
 Therefore, $\hat{\pi}_{1}$ is obtained, first by generating $u_{1},\cdots,u_{n}\sim {\cal{U}}(-1,1)$, and then computing $\hat{\pi}_{1}=4/n\sum_{i=1}^{n}\sqrt{1-u^{2}_{i}}$. It is easy to check that
\begin{equation}
  \begin{split}
E\bigl(\hat{\pi}_{1}\bigr)&=\frac{4}{n}\sum_{i=1}^{n}E\Bigl(\sqrt{1-U^{2}_{i}}\Bigr)\nonumber\\
&=\frac{4}{n}\sum_{i=1}^{n}\int_{-1}^{1}\sqrt{1-y^2}\frac{1}{2}dy\nonumber\\
&=\frac{4}{n}\sum_{i=1}^{n} E\Bigl(\sqrt{1-Z^{2}_{i}}\Bigr)\nonumber\\
&=\pi \label{importance-sampling-pi-star-1},
  \end{split}
%\quad\leftrightarrow\quad
  \begin{split}
var\bigl(\hat{\pi}_{1}\bigr)&=\frac{16}{n^2}\sum_{i=1}^{n}\Bigl[E\bigl(1-Z_{i}^{2}\bigr)-\Bigl(\frac{\pi}{4}\Bigr)^2\Bigr],\nonumber\\
&=\frac{16}{n}\Bigl[\frac{2}{3}-\Bigl(\frac{\pi}{4}\Bigr)^2\Bigr]\nonumber\\
&=\frac{0.79706}{n}.
\\\\\\
  \end{split}
\end{equation}
where $Z_{1},\cdots,Z_{n}\sim{\cal{U}}(0,1)$. As it is seen, $\hat{\pi}_{1}$ yields around 70.44\% reduction in variance for estimating $\pi$ compared to $\hat{\pi}$. Yet the performance of estimator $\hat{\pi}_{1}$ can be improved using the method introduced in Section \ref{Estimators-negative-covariance}. It follows from above that quantity
\begin{align*} 
\frac{4}{n}\sum_{i=1}^{n} \sqrt{1-Z^{2}_{i}}
\end{align*} 
with variance ${0.79706/n}$. We know that $Z\sim1-Z \sim{\cal{U}}(0,1)$ and so the quantity 
\begin{align*} 
\frac{4}{n}\sum_{i=1}^{n} \sqrt{1-(1-Z)^{2}_{i}}
\end{align*}
is also unbiased estimator for $\pi$ with the same variance. But, evidently $Z$ and $1-Z$ are correlated negatively and hence     
\begin{align*} 
\hat{\pi}_{2}=\frac{2}{n}\sum_{i=1}^{n} \sqrt{1-Z^{2}_{i}} +\frac{2}{n}\sum_{i=1}^{n} \sqrt{1-(1-Z_{i})^{2}},
\end{align*}
is an unbiased estimator for $\pi$ whose variance is smaller than that of $\hat{\pi}_{1}$.
%(\ref{importance-sampling-pi-star-1})
 \begin{figure}[!h]
\center
\includegraphics[width=55mm,height=55mm]{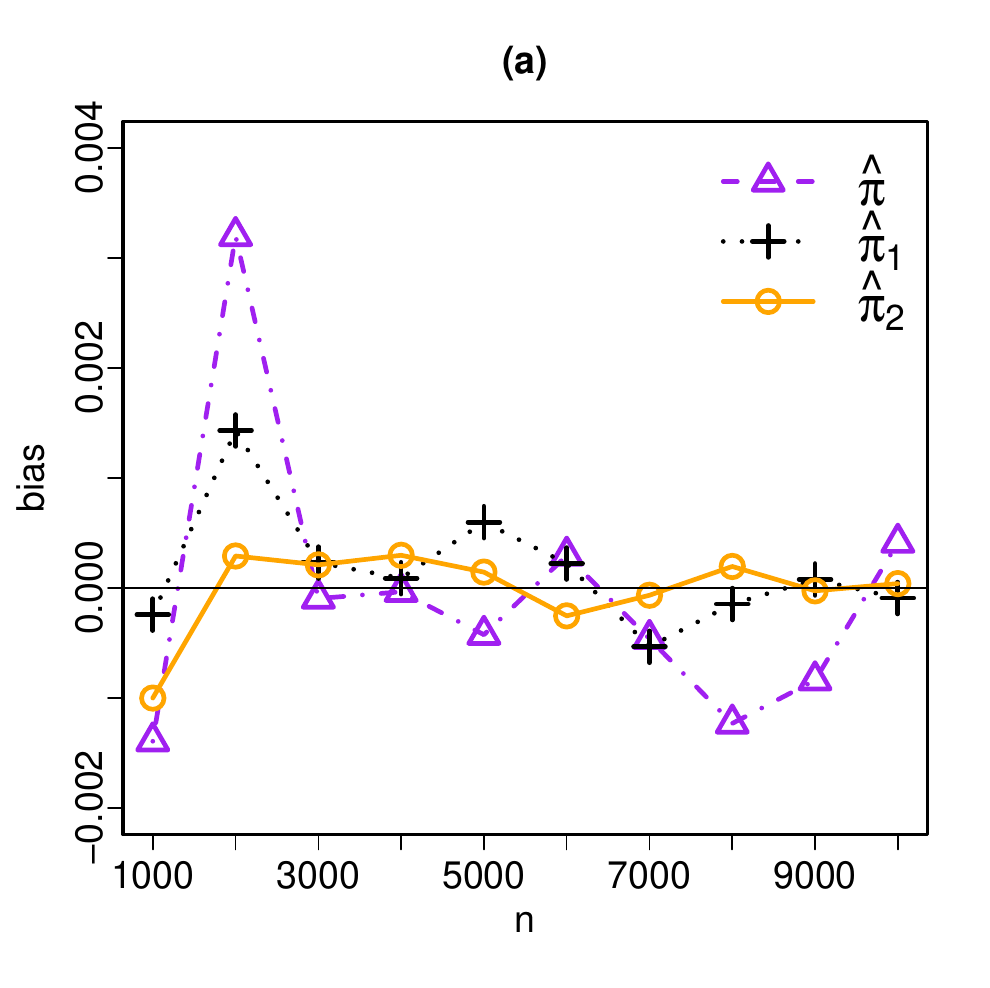}
\includegraphics[width=55mm,height=55mm]{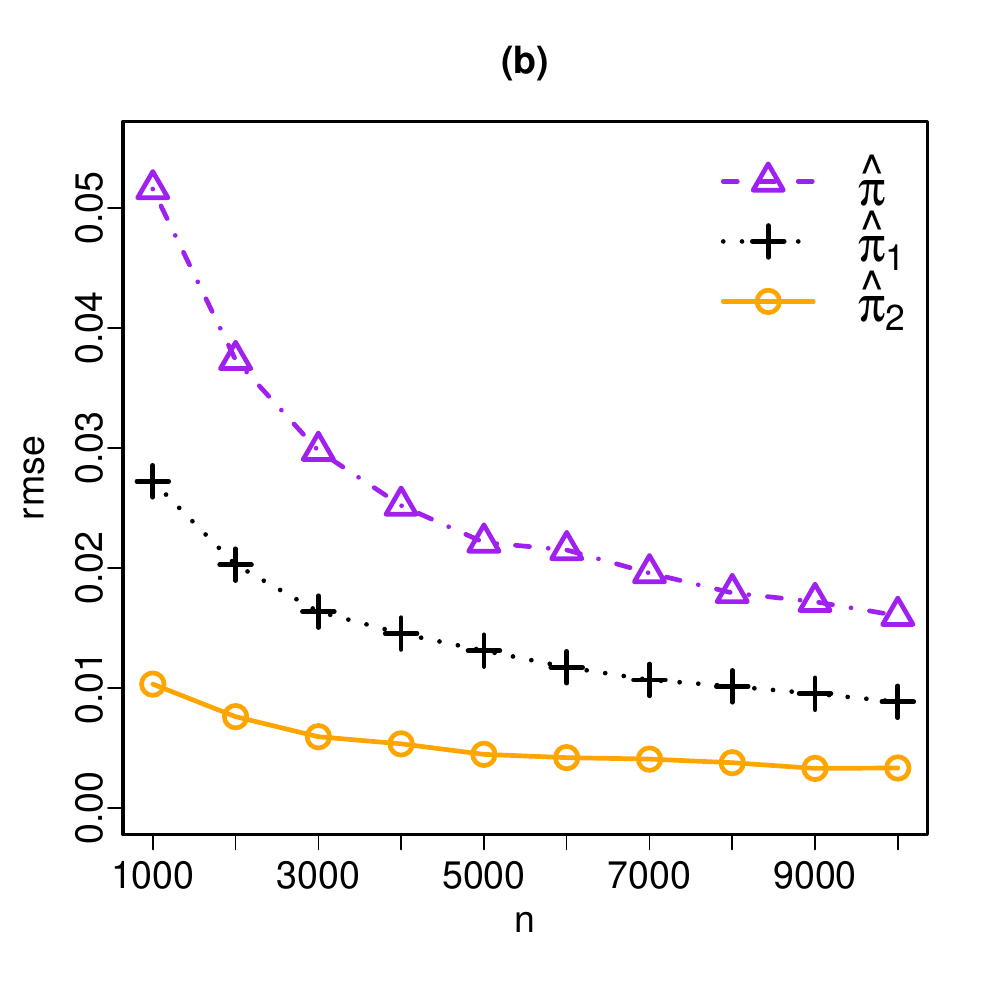}\\
\caption{The bias (a) and rmse (b) of $\hat{\pi}$, $\hat{\pi}_{1}$, and $\hat{\pi}_{2}$ computed based on $N=1000$ independent samples of size $n$.}
\label{plot-importance-pi-1}
\end{figure}
As it may seen from Figure \ref{plot-importance-pi-1}, the $\hat{\pi}_{2}$ outperforms $\hat{\pi}_{1}$ and $\hat{\pi}$, in terms of both bias and rmse criteria. The pertinent \verb+R+ code for computing the bias and rmse of $\hat{\pi}$, $\hat{\pi}_{1}$, and $\hat{\pi}_{2}$ is given by the following.
\begin{lstlisting}[style=deltaj]
R> set.seed(20241012)
R> N <- 1000; n <- 100; sample<- seq(1000, 10000, 1000)
R> pi_hat <- pi_hat_star1 <- pi_hat_star2 <- rep(0, N) 
R> bias <- rmse <- matrix(0, nrow = length(sample), ncol = 3 )
R> for(j in 1:length(sample) )
+  { 
+ 	for(i in 1:N)
+	{
+ 		u <- runif(sample[j], -1, 1); v <- runif(sample[j], -1, 1); Z <- runif(sample[j], 0, 1)
+ 		pi_hat[i] <- 4*mean( ifelse(u^2 + v^2 <= 1, 1, 0) )
+ 		pi_hat_star1[i] <- 4*mean( sqrt(1 - u^2) )
+ 		pi_hat_star2[i] <- (   4*mean( sqrt(1 - Z^2) ) + 4*mean( sqrt(1 - (1 - Z)^2) )   )/2
+ 	}
+ 		hat <- as.matrix( cbind(pi_hat, pi_hat_star1, pi_hat_star2), nrow = N, ncol = 3 )
+ 		colnames(hat) <- NULL
+ 		shifted <- sweep(hat, 2, pi, FUN = '-')
+ 		bias[j, ] <- apply( shifted, 2, mean )
+ 		rmse[j, ] <- sqrt( apply( shifted, 2, var ) )
+ }
\end{lstlisting}
\end{example}

\section{Computing the CDF of Gaussian distribution}
In this section, we proceed to provide the compute the Monte Carlo approximation for the CDF of the Gaussian distribution.
\subsection{Computing the CDF of univariate Gaussian distribution}
Indeed computing the CDF of a Gaussian distribution is an importance sampling problem. Following example gives more detailed information. 
\begin{example}\label{exam-importance-sampling-sg2}%\lipsum*[]
Let $Z\sim {\cal{N}}(0,1)$ and we are interested in computing $P(Z<2)$. Evidently the given probability can be represented as
\begin{align}
P(Z<2)&=\frac{1}{\sqrt{2\pi}}\int_{-\infty}^{2}\exp\Bigl\{-\frac{z^{2}}{2}\Bigr\}dz\nonumber\\
&=\frac{1}{\sqrt{2\pi}}\int_{-\infty}^{\infty}\mathbb{I}_{(-\infty,2)}(z)\times \exp\Bigl\{-\frac{z^{2}}{2}\Bigr\}dz\nonumber\\
&=E\Bigl(\mathbb{I}_{(-\infty,2)}(Z)\Bigr),
\end{align} 
where $\mathbb{I}_{{\cal{S}}}(x)$ is the well-known indicator function defined as
\begin{eqnarray}\label{indicator-function}
\displaystyle
\mathbb{I}_{{\cal{S}}}(x)=\left\{\begin{array}{c}
\displaystyle
1,~~\mathrm{{if}}\ x \in  {\cal{S}},\\
\displaystyle
0,~~\mathrm{if}\ x \notin  {\cal{S}}.
\end{array} \right.
\end{eqnarray}
Hence, the quantity can be readily approximated by considering the standard univariate Gaussian  as the instrumental distribution. It follows, based on the law of large numbers, that
\begin{align}\label{approximation-CDF-univariate-Gaussian}
P(Z\leq 2)&\approx \frac{1}{N}\sum_{i=1}^{N} \mathbb{I}_{(-\infty,2)}\bigl(z_{i}\bigr)=\frac{{\text{number of $z_{i}$s less or equal 2 in a sample of size $N$}}}{N},
\end{align} 
where $z_{1},\cdots,z_{N}$ independently follow ${\cal{N}}(0,1)$ and $N$ is a sufficiently large integer value. This means that the Monte Carlo approximation of $P(Z<2)$ given in (\ref{approximation-CDF-univariate-Gaussian}) is the average of samples that are less than two. The Central limit theorem guarantees that estimator (\ref{approximation-CDF-univariate-Gaussian}) is unbiased for $P(Z<2)$ with variance 
\begin{align*}%\label{variance-CDF-estimator-univariate-Gaussian}
\frac{\Phi(2\vert 0, 1)\bigl[1-\Phi(2\vert 0, 1)\bigr]}{N}.
\end{align*} 
\end{example}
Of course, the command \verb+integrate(f,lower,upper,...)+ (when \verb+f+ is the PDF of Gaussian distribution) or \verb+pnorm(q,mean=0,sd=1,...)+ can be used in \verb+R+ environment for computing the CDF of Gaussian distribution. However, one could use the well known {\it{Gaussian quadrature}} rules discussed in next Chapter.
\section{Computing the CDF of multivariate Gaussian distribution}\label{Computing-CDF-multivariate-Gaussian}
Due to significant importance of the multivariate Gaussian distribution in a wide range of study fields, computing its CDF has a long history. Several efforts have been made for computing the CDF of multivariate Gaussian distribution. For more detailed information, we refer the reader to \cite{botev2017normal} and references therein. One of the pioneers in this context was \cite{genz1992numerical} who suggested the separation of variables (SOV) method. Let $\boldsymbol{X}\sim {\cal{N}}_{p}(\boldsymbol{\mu},\Sigma)$ and $L$ denote the Cholesky decomposition of $\Sigma$, that is $\Sigma=LL^{\top}$ in which $L$ is a $p \times p$ lower triangular matrix. For computing ${\cal{P}}=P(\boldsymbol{a}<\boldsymbol{X}<\boldsymbol{b})$, in which $\boldsymbol{a}$ and $\boldsymbol{b}$ are vectors of real constants, the SOV method can be described as follows. Consider transformation $\boldsymbol{x}=L\boldsymbol{y}$, we have 
\begin{align*}%\label{multivariate-Gaussian-CDF-computation-1}
\boldsymbol{x}^{\top}\Sigma^{-1}\boldsymbol{x}=\boldsymbol{y}^{\top}L^{\top}L^{-\top}L^{-1}L\boldsymbol{y}=\boldsymbol{y}^{\top}\boldsymbol{y}=\sum_{i=1}^{p}y^{2}_{i},
\end{align*}
and $d\boldsymbol{x}=\vert L\vert d\boldsymbol{y}=\sqrt{\vert \Sigma\vert} d\boldsymbol{y}$. Hence, $\boldsymbol{a}<L\boldsymbol{y}<\boldsymbol{b}$ or equivalently
\begin{eqnarray*}%\label{multivariate-Gaussian-CDF-computation-2}
\frac{a_{1}}{L_{1,1}}\leq&{y}_{1}&\leq\frac{b_{1}}{L_{1,1}}\nonumber\\
\frac{a_{2}-L_{2,1}y_{1}}{L_{2,2}}\leq&{y}_{2}&\leq
\frac{b_{2}-L_{2,1}y_{1}}{L_{2,2}}\nonumber\\
%\frac{a_{3}-L_{3,1}y_{1}-L_{3,2}y_{2}}{L_{3,3}}\leq&{y}_{3}&\leq
%\frac{b_{3}-L_{3,1}y_{1}-L_{3,2}y_{2}}{L_{3,3}}\nonumber\\
&\vdots&\nonumber\\
\frac{a_{p}-\sum_{i=1}^{p-1}L_{p,i}y_{i}}{L_{p,p}}\leq &{y}_{p}& \leq 
\frac{b_{p}-\sum_{i=1}^{p-1}L_{p,i}y_{i}}{L_{p,p}}\nonumber.
\end{eqnarray*}
This type of change of variable reduces the problem of computing ${\cal{P}}$ to $p$ univariate integration as follows.
\begin{align}\label{multivariate-Gaussian-CDF-computation-3}
{\cal{P}}=(2\pi)^{-\frac{p}{2}}\int_{a_{1}^{\prime}}^{b_{1}^{\prime}}\exp\Bigl\{-\frac{y_{1}^{2}}{2}\Bigr\}\int_{a_{2}^{\prime}}^{b_{2}^{\prime}}\exp\Bigl\{-\frac{y_{2}^{2}}{2}\Bigr\}\cdots\int_{a_{p}^{\prime}}^{b_{p}^{\prime}}\exp\Bigl\{-\frac{y_{m}^{2}}{2}\Bigr\}d\boldsymbol{y},
\end{align} 
where $a^{\prime}_{i}=\bigl(a_{i}-\sum_{j=1}^{i-1}L_{i,j}y_{j}\bigr)/L_{i,i}$ and $b^{\prime}_{i}=\bigl(b_{i}-\sum_{j=1}^{i-1}L_{i,j}y_{j}\bigr)/L_{i,i}$. By applying another transformation of the form $z_i=\Phi\bigl(y_{i}\bigr)$ (for $i=1,\cdots,p$), we obtain
\begin{align}\label{multivariate-Gaussian-CDF-computation-4}
{\cal{P}}=\int_{d_{1}}^{e_{1}} \int_{d_{2}}^{e_{2}}\cdots\int_{d_{p}}^{e_{p}}d\boldsymbol{z},
\end{align} 
where
\begin{align*} 
d_{i}=&\Phi\biggl(\frac{a_{i}-\sum_{j=1}^{i-1}L_{i,j}\Phi^{-1}(z_{j})}{L_{i,i}}\biggr),\\
e_{i}=&\Phi\biggl(\frac{b_{i}-\sum_{j=1}^{i-1}L_{i,j}\Phi^{-1}(z_{j})}{L_{i,i}}\biggr).
\end{align*} 
 Although the integrand in (\ref{multivariate-Gaussian-CDF-computation-4}) has a simpler form than (\ref{multivariate-Gaussian-CDF-computation-3}), but its integral region is complicated. As a solution for this issue, we apply the transformation $w_i=(z_i-d_i)/(e_i-d_i)$ to see that
\begin{align}\label{multivariate-Gaussian-CDF-computation-5}
{\cal{P}}
%\int_{0}^{1}\bigl(e^{\prime}_{1}-d^{\prime}_{1}\bigr) \int_{0}^{1} \bigl(e^{\prime}_{2}-d^{\prime}_{2}\bigr)\cdots \int_{0}^{1}\bigl(e^{\prime}_{p}-d^{\prime}_{p}\bigr)\int_{0}^{1}d\boldsymbol{w},\nonumber\\
&=\int_{0}^{1}\bigl(e^{\prime}_{1}-d^{\prime}_{1}\bigr)d{w_{1}}\int_{0}^{1} \bigl(e^{\prime}_{2}-d^{\prime}_{2}\bigr)d{w_{2}}\cdots \int_{0}^{1}\bigl(e^{\prime}_{p-1}-d^{\prime}_{p-1}\bigr)d{w_{p-1}}\int_{0}^{1}\bigl(e^{\prime}_{p}-d^{\prime}_{p}\bigr)d{w_{p}},\nonumber\\
&=\bigl(e^{\prime}_{1}-d^{\prime}_{1}\bigr)\int_{0}^{1}\bigl(e^{\prime}_{2}-d^{\prime}_{2}\bigr)d{w_{1}}\int_{0}^{1} \bigl(e^{\prime}_{3}-d^{\prime}_{3}\bigr)d{w_{2}}\cdots\int_{0}^{1}\bigl(e^{\prime}_{p}-d^{\prime}_{p}\bigr)d{w_{p-1}}\int_{0}^{1}d{w_{p}},\nonumber\\
&=\bigl(e^{\prime}_{1}-d^{\prime}_{1}\bigr)\int_{0}^{1}\bigl(e^{\prime}_{2}-d^{\prime}_{2}\bigr)d{w_{1}}\int_{0}^{1} \bigl(e^{\prime}_{3}-d^{\prime}_{3}\bigr)d{w_{2}}\cdots\int_{0}^{1}\bigl(e^{\prime}_{p}-d^{\prime}_{p}\bigr)d{w_{p-1}},
\end{align}
where 
\begin{align*} 
d^{\prime}_{i}=&\Phi\biggl(\frac{a_{i}-\sum_{j=1}^{i-1}L_{i,j}\Phi^{-1}\bigl(d_{j}+w_{j}(e_{j}-d_{j})\bigr)}{L_{i,i}}\biggr),\\
e^{\prime}_{i}=&\Phi\biggl(\frac{b_{i}-\sum_{j=1}^{i-1}L_{i,j}\Phi^{-1}\bigl(d_{j}+w_{j}(e_{j}-d_{j})\bigr)}{L_{i,i}}\biggr).
\end{align*} 
It is worth to note that in the RHS of (\ref{multivariate-Gaussian-CDF-computation-5}) the integrand within $i$th integral do not depend on $w_i$ and so moves back as an integrand to $(i-1)$th integral. Henceforth, the number of integrals is $p-1$. Furthermore, the quantity $e^{\prime}_{1}-d^{\prime}_{1}$ is independent of $w$. 
\par 
For computing ${\cal{P}}$ in (\ref{multivariate-Gaussian-CDF-computation-5}), the method proposed by \cite{genz1992numerical} approximates ${\cal{P}}$ through the Monte Carlo method in which $w_i$ (for $i=1,\cdots,p-1$) is drawn from ${\cal{U}}(0,1)$ as the instrumental distribution. This method as originally developed for zero-mean Gaussian distribution. In general, when $\boldsymbol{\mu}$ of, the constants $\boldsymbol{a}$ and $\boldsymbol{b}$ are replaced with $\boldsymbol{a}-\boldsymbol{\mu}$ and $\boldsymbol{b}-\boldsymbol{\mu}$, respectively. The following algorithm describes the method of \cite{genz1992numerical} in more details.
\vspace{5mm}
\begin{algorithm}
\caption{Computing CDF of the multivariate Gaussian distribution}
\label{Computing-CDF-multivariate-Gaussian-distribution}
\begin{algorithmic}[1]
\State Read $\boldsymbol{a}$, $\boldsymbol{b}$, $\boldsymbol{\mu}$, $\Sigma$, $\epsilon$, and ${\text{N}}_{max}$;
\State Compute the Cholesky decomposition $L$ of $\Sigma$;
\State Set ${\text{Sum}}_{int}=0$, ${\text{Var}}_{int}=0$, and $\alpha=2.5$; 
\State Set $\boldsymbol{b}= \boldsymbol{b} -\boldsymbol{\mu}$ and $\boldsymbol{a}= \boldsymbol{a} -\boldsymbol{\mu}$;
\State Set $e_{1}=\Phi\bigl(b_{1}/L_{1,1}\bigr)$, $d_{1}=\Phi\bigl(a_{1}/L_{1,1}\bigr)$, and $f_{1}=e_{1}-d_{1}$;
\State Set $N=0$ and ${\text{Error}}=10$;
\While{ ${\text{Error}} \geq \epsilon$ and $N < {\text{N}}_{max}$ }
  %\Comment{put some comments here}
\State Generate realizations $w_{1},w_{2},\cdots,w_{p-1}$ from ${\cal{U}}(0,1)$;
 \State Set $i= 2$;
\While{ $i \leq p$}
 \State Set $j = i - 1$;
\State  $y_{j} = \Phi^{-1}\Bigl( d_{j} + w_{j} \bigl(e_{j} - d_{j}\bigr) \Bigr)$;
\State  $L_y = \sum_{k=1}^{j} L_{i, k}\times y_{k} $;
\State  $d_{i}= \Phi\Bigl( \frac{ a_{i} - L_{y} }{L_{i, i}} \Bigr)$;
 \State $e_{i}= \Phi\Bigl( \frac{ b_{i} - L_{y} }{L_{i, i}} \Bigr)$;
 \State $f_{i}= \bigl(e_{i}-d_{i} \bigr) f_{j}$;
 \State Set $i= i +1$;
     \EndWhile  %\label{roy's loop}
     \State {\bf{end}}
     \State $N = N + 1$;
\State $\delta= \frac{f_{p}- {\text{Sum}}_{int}}{N}$;
\State ${\text{Sum}}_{int} ={\text{Sum}}_{int} + \delta$;
\State ${\text{Var}}_{int} = (N-2)  \frac{{\text{Var}}_{int}}{N} + \delta^2$;
\State ${\text{Error}} = \alpha \sqrt{{\text{Var}}_{int}}$;
        %\Comment{another comment}
        %\State $var3 \leftarrow var4$
     \EndWhile  %\label{roy's loop}
     \State {\bf{end}}
     \State Return ${\text{Sum}}_{int}$ as an estimation of CDF of multivariate Gaussian distribution.
%\EndProcedure
\end{algorithmic}
\end{algorithm}
In the following, the \verb+R+ function \verb+pmvGauss+ is written for implementing the \cite{genz1992numerical} algorithm.
\begin{lstlisting}[style=deltaj]
R> pmvGauss <- function(a, b, Mu, Sigma, epsilon = 10e-5, N_max = 200)
+ {
+   p <- length(Mu)
+	 L <- t( chol(Sigma) )
+	 alpha <- 2.5
+	 Error <- 100
+	 d <- e <- f <- w <- y <- rep(0, p)
+	 a <- a - Mu
+	 b <- b - Mu
+	 Int_sum <- N <- Var_sum <- 0
+	 d[1] <- pnorm( a[1]/L[1, 1] )
+	 e[1] <- pnorm( b[1]/L[1, 1] )
+	 f[1] <- e[1] - d[1] 
+	 	while( Error >= epsilon & N < N_max )
+	 	{
+	 	w <- runif(p - 1)
+	 	i<-2
+	 		while( i <= p ){
+	 		j <- i - 1
+	 		y[j] <- qnorm( d[j] + w[j]*(e[j] - d[j]) )
+	 		L_y  <- sum( L[i,(1:j)]*y[1:j] )
+	 		d[i] <- pnorm( (a[i] - L_y )/L[i, i] )
+	 		e[i] <- pnorm( (b[i] - L_y )/L[i, i] )
+	 		f[i] <- (e[i] - d[i])*f[j]
+	 		i<-i+1
+	 	}
+	 	N <- N + 1
+	 	delta <- (f[p] - Int_sum)/N
+	 	Int_sum <- Int_sum  + delta
+	 	Var_sum <- (N - 2)*Var_sum/N + delta^2
+	 	Error <- alpha*sqrt(Var_sum)
+	 	}
+	 ls <- list("CDF" = Int_sum, "error" = Error)
+	 return(ls)
+ }
\end{lstlisting}  
\section{Univariate Gaussian quadrature}
The {\it{quadrature rules}} involve methods for approximating an integral 
\begin{align}\label{quadrature-Gaussian-broad-sense-1}
I= \int_{a}^{b} f(x) dx \approx {\cal{Q}}(f)=\sum_{i=1}^{k}\omega_i f\bigl(x_i\bigr)
\end{align}
where quantities $\omega_i$ and $x_i$ are the $i$th {\it{weight}} and {\it{node}} of the quadrature rule, respectively. The idea behind a quadrature rule is to approximate the integrand $f(x)$ through a simple function. Reasonably, an appropriate choice is a polynomial $P(x)$, of degree not greater than $k-1$ that interpolate $f(x)$ at $x_i$ (for $i=1,\cdot,k$). Suppose, for a given function $f(x)$, we are interested in approximating the integral of the form $f(x)$ over interval $(a,b)$ using a two-point ($k=2$) quadrature rule. We have
\begin{align}\label{quadrature-Gaussian-broad-sense-2}
\int_{a}^{b} f(x) dx \approx \omega_1 f\bigl(x_1\bigr) + \omega_2 f\bigl(x_2\bigr),
\end{align}
where the value of above integral is known when all four unknown constants including $\omega_1$ (first weight), $\omega_2$ (second weight), $x_1$ (first node), and $x_2$ (second node) are determined. To this end, we assume $f(x)$ is a polynomial of degree three represented as $f(x) = a_{0} + a_{1}x + a_{2}x^{2} + a_{3}x^{3}$. Although we can let for larger values of $m$, but we set $m=2$ for simplicity. So, substituting the third-order polynomial $f(x)= a_{0} + a_{1}x + a_{2}x^{2} + a_{3}x^{3}$ into the LHS of (\ref{quadrature-Gaussian-broad-sense-2}), it follows that
\begin{align}\label{quadrature-Gaussian-broad-sense-3} 
\int_{a}^{b} f(x) \ dx &= \int_{a}^{b} a_0+a_1x+a_2x^2+a_3x^3 dx\nonumber
\\ &= \begin{bmatrix}
 a_0x+a_1\frac{x^2}{2}+a_2\frac{x^3}{3}+a_3\frac{x^4}{4} \end{bmatrix}_{a}^{b}\nonumber\\ 
 &= a_{0}( b - a) + a_{1}\Bigl( \frac{b^{2} - a^{2}}{2} \Bigr) + a_{2}\Bigl( \frac{b^{3} - a^{3}}{3} \Bigr) + a_{3}\Bigl( \frac{b^{4} - a^{4}}{4} \Bigr).
 \end{align}
Representing the RHS of (\ref{quadrature-Gaussian-broad-sense-2}) in terms of third-order polynomial yields
\begin{align}\label{quadrature-Gaussian-broad-sense-4}
\omega_{1}f(x_{1}) + \omega_{2}f(x_{2})= \omega_{1}\bigl( a_{0} + a_{1}x_{1} + a_{2}x_{1}^{2} + a_{3}x_{1}^{3} \bigr) + \omega_{2}\bigl( a_{0} + a_{1}x_{2} + a_{2}x_{2}^{2} + a_{3}x_{2}^{3} \bigr).
\end{align}
Equating the RHSs of (\ref{quadrature-Gaussian-broad-sense-3}) and (\ref{quadrature-Gaussian-broad-sense-4}) yields a set of four equations as follows.
\begin{align} \label{quadrature-Gaussian-broad-sense-5}
\left\{\begin{array}{l}
 b - a = \omega_{1} + \omega_{2}, \\ 
 \frac{b^2 - a^2}{2} = \omega_{1}x_{1} + \omega_{2}x_{2}, \\ 
 \frac{b^3 - a^3}{3} = \omega_{1}{x_{1}}^{2} + \omega_{2}{x_{2}}^{2}, \\ 
 \frac{b^4 - a^4}{4} = \omega_{1} {x_{1}}^{3} + \omega_{2} {x_{2}}^{3}.
\end{array} \right.
\end{align}
Solving the set of four equations given in (\ref{quadrature-Gaussian-broad-sense-5}), we have
\begin{align} \label{quadrature-Gaussian-broad-sense-6}
\omega_{1} &= \frac{b - a}{2},\nonumber\\ 
\omega_{2} &= \frac{b - a}{2}, \nonumber\\ 
x_{1} &= - \frac{1}{\sqrt{3}}\Bigl( \frac{b - a}{2} \Bigr) + \frac{b + a}{2}, \nonumber\\ 
x_{2} &= \frac{1}{\sqrt{3}}\Bigl( \frac{b - a}{2} \Bigr) + \frac{b + a}{2}.
 \end{align}
Substitute the RHS of (\ref{quadrature-Gaussian-broad-sense-6}) into the RHS of (\ref{quadrature-Gaussian-broad-sense-2}) to see that
\begin{align} 
\int_{a}^{b} f(x) dx &\approx \omega_{1}f\bigl( x_{1} \bigr) + \omega_{2}f\left( x_{2} \right)\nonumber\\ &= \frac{b - a}{2}f\Bigl( - \frac{1}{\sqrt{3}}\frac{b - a}{2} + \frac{b + a}{2} \Bigr)+ \frac{b - a}{2}f\Bigl( \frac{1}{\sqrt{3}}\frac{b - a}{2} + \frac{b + a}{2} \Bigr).
 \end{align}
Following example shows how much a quadrature can be reliable.
\begin{example}\label{Gaussian-quadrature-broad-sense-exam1}%\lipsum*[]
Let $f(x)=x\exp\{-x\}$ and we would like to approximate 
\begin{align}\label{Gaussian-quadrature-broad-sense-1}
I=\int_{0}^{3} x\exp\{-x\} dx, 
 \end{align}
based on Gaussian quadrature of $m=2$ points. We compute two weights and nodes through  (\ref{quadrature-Gaussian-broad-sense-5}). It is easy to check that
\begin{align} 
I= \approx 1.5\times f(0.63397)+ 1.5\times f(2.36602)=0.83755, 
 \end{align}
 where the exact value of above integral is $0.80085$. Figure \ref{plot-Gaussian-quadrature-exam1} displays the integrand and the corresponding nodes.
 \begin{figure}[!h]
\center
\includegraphics[width=55mm,height=55mm]{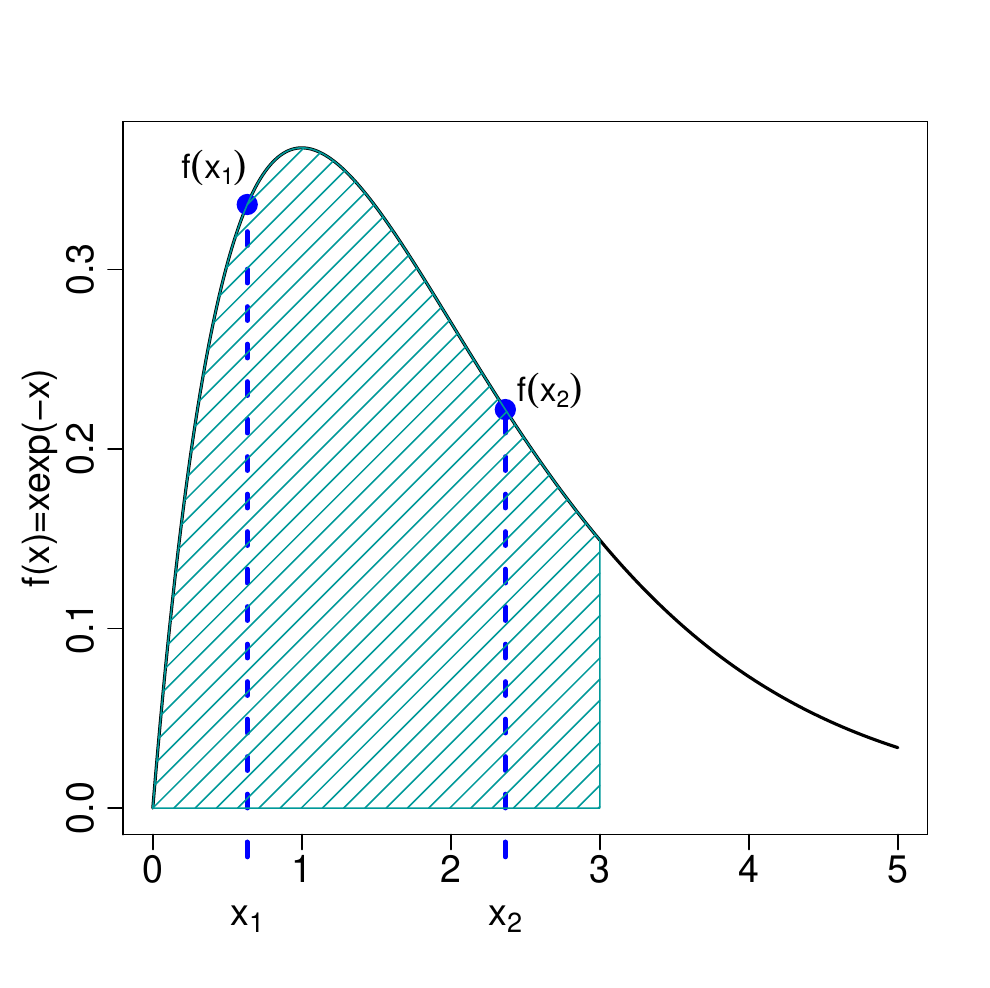}
\caption{Plot of integrand $f(x)=x\exp\{-x\}$ and nodes $x_1$ and $x_2$. The area under the integrand is shown by shaded area.}
\label{plot-Gaussian-quadrature-exam1}
\end{figure}
\end{example}
The above reasoning can be extended for $k>2$ that yields more accurate approximation of integral (\ref{quadrature-Gaussian-broad-sense-2}). In general, using the quadrature as described above, one can approximate the given definite integral using a quadrature rule with an arbitrary $k$ numbers of nodes and weights. In general, we can write
\begin{align*}%\label{quadrature-Gaussian-broad-sense-1}
I= \int_{a}^{b} f(x) dx= \int_{a}^{b} P(x) dx=\int_{a}^{b} \sum_{i=1}^{k}L_i(x) f\bigl(x_i\bigr) dx=\sum_{i=1}^{k}\omega_i f\bigl(x_i\bigr),
\end{align*}
where 
\begin{align*}%\label{quadrature-Gaussian-broad-sense-1}
\omega_i= \int_{a}^{b} L_{i}(x) dx,
\end{align*}
in which $L_{i}(x)$ is the well-known Lagrange interpolating polynomial of order $k-1$.  defined as
\begin{align*}%\label{quadrature-Gaussian-broad-sense-1}
L_{i}(x)=\prod_{j=1, j \neq i}^{n}\frac{x-x_i}{x_j-x_i}.
\end{align*}
In what follows, among several quadrature rules introduced in the literature, we briefly describe the {\it{Gaussian quadrature}} rules for computing a definite integral.
\subsection{Gaussian quadrature rules}
The quadrature rule presented in the previous sub-section works based on a general polynomial. Herein, we describe the well-known {\it{Gaussian quadrature}} that works based on a special class of polynomials known as the {\it{orthogonal polynomials}} and can be applied for approximating accurately a wide range of definite integrals \citep{dahlquist2003numerical}. For presentation and theoretical purposes, the following definitions are useful.
% in discussing Gaussian quadrature rules based on orthogonal polynomials.
\begin{dfn} \citep{foupouagnigni2018orthogonal} The function $w(\cdot):(a,b) \rightarrow \mathbb{R}^{+}$ that satisfies the following three conditions is said to be a weight function.
\begin{enumerate}[label=\roman*.]
\item $w(x)$ is measurable.
\item The $k$th moment of $w(x)$ must be finite, that is $\int_{a}^{b}\vert x \vert^{k}w(x)dx<\infty$ for $k=0,1,2,\cdots$.
\item For a given polynomial such as $g(x)$ if $\int_{a}^{b}g(x)w(x)dx=0$ then we must have $g(x)=0$.
\end{enumerate}
\end{dfn}
\begin{dfn}\citep{foupouagnigni2018orthogonal} The sequence of polynomials $\{f_{0}(x)$,$f_{1}(x)$,$f_{2}(x)$,$\cdots\}$ constitutes a sequence of orthogonal polynomials with respect to weight function $w(\cdot):(a,b) \rightarrow \mathbb{R}^{+}$ if we have
\begin{align*} 
\langle f_{n}, f_{m}\rangle=\int_{a}^{b} f_{n}(x) f_{m}(x) w(x)dx={\text{C}}_{n}\times {\cal{D}}_{m,n},
\end{align*} 
where $\langle ., .\rangle$ represents the inner product operator, ${\text{C}}_{n}$ is a constant, and ${\cal{D}}_{m,n}$ is the Kronecker delta defined as
\begin{eqnarray*}%\label{delta-Kronecker}
\displaystyle
{\cal{D}}_{m,n}=\left\{\begin{array}{c}
\displaystyle
1,~~\mathrm{{if}}\ m=n,\\
\displaystyle
0,~~\mathrm{if}\ m \neq n.
\end{array} \right.
\end{eqnarray*}
\end{dfn}
There are several types of the orthogonal polynomials such as {\it{Chebyshev of first kind}} ($T_{1,n}$), {\it{Chebyshev of second kind}} ($T_{2,n}$), {\it{generalized Laguerre}} (${\cal{GL}}_{n}^{\alpha}$), {\it{Hermite}} ($H_{n}$), {\it{Jacobi}} ($J_{n}^{\alpha,\beta}$), {\it{Laguerre}} (${\cal{L}}_{n}$), and {\it{Legendre}} ($L_{n}$), to name a few. In general, the orthogonal polynomials are solution of differential equation and obtained through a recurrence formula. In what follows, we describe briefly the {\it{Hermite}} and {\it{Laguerre}} classes of orthogonal polynomials with widespread applications in statistics.
\begin{itemize}
\item {\bf{Hermite polynomial}}: the Hermite polynomial of degree $n$, shown by $H_{n}(x)$, 
is orthogonal with respect to weight function $w(x)=\exp\{-x^2\}$. This means that
\begin{align}\label{Laguerre-Hermite-1} 
\langle H_{n}, H_{m}\rangle=\int_{-\infty}^{\infty} H_{n}(x) H_{m}(x) \exp\bigl\{-{x^2}{}\bigr\}dx=2^{n}\sqrt{\pi}\Gamma(n+1)\times {\cal{D}}_{m,n}.
%\langle H_{n}, H_{m}\rangle=\int_{-\infty}^{\infty} H_{n}(x) H_{m}(x) \exp\Bigl\{-{x^2}{2}\Bigr\}dx=\sqrt{2\pi}\Gamma(n+1)\times {\cal{D}}_{m,n},
\end{align}
The expansion representation for $H_{n}(x)$ is
\begin{align}\label{polynomial-Hermite-3} 
H_{n}(x)= \Gamma(n+1) \sum_{i=0}^{\lfloor \frac{n}{2}\rfloor}\frac{(-1)^{i}(2x)^{n-2i} }{
\Gamma(i+1)\Gamma(n-2i+1)},
\end{align} 
where $\lfloor n/2\rfloor$ denotes the greatest integer less or equal $n/2$ (for $n=0,1,2,\cdots$). This class of orthogonal polynomials satisfies with the recurrence relations
%\begin{eqnarray}\label{polynomial-Hermite-2}
%\displaystyle
%\left\{\begin{array}{c}
%\displaystyle
%\frac{d H_{n}(x)}{dx}- n H_{n-1}(x)=0,~~~~~~~~~~~~~~~~\mathrm{if}\ n \geq 1,\\
%\displaystyle
%\frac{d^2 H_{n}(x)}{dx^2}- x \frac{d H_{n}(x)}{dx}+ n H_{n}(x)=0,~~\mathrm{if}\ n \geq 0,
%\end{array} \right.
%\end{eqnarray}
\begin{align*}%\label{polynomial-Hermite-2}
\frac{d H_{n}(x)}{dx}- 2x  H_{n}(x) + H_{n+1}(x)=0,
\end{align*} 
%that is
%\begin{align}\label{polynomial-Hermite-3} 
%H_{n}(x)= \Gamma(n+1) \sum_{i=0}^{\lfloor \frac{n}{2}\rfloor}\frac{(-1)^{i}(2x)^{n-2i} }{
%\Gamma(i+1)\Gamma(n-2i+1)},
%%H_{n}(x)= \Gamma(n+1) \sum_{i=0}^{\lfloor \frac{n}{2}\rfloor}\frac{(-1)^{i}x^{n-2i} }{2^{i}\Gamma(i+1)\Gamma(n-2i+1)},
%\end{align} 
 %The differential equation (\ref{polynomial-Hermite-2}) satisfies the recurrence formula
for $n \geq 0$ and
\begin{align}\label{polynomial-Hermite-4} 
H_{n+1}(x) - 2x H_{n}(x) + 2n H_{n-1}(x) =0,
%H_{n+1}(x) - x H_{n}(x) + n H_{n-1}(x) =0,
\end{align} 
for $n \geq 1$. 
Using expression %(\ref{polynomial-Hermite-3}) and 
(\ref{polynomial-Hermite-4}),
%% and (\ref{polynomial-Hermite-5})
 the first six members of Hermite polynomials are as follows \citep{arfken2011mathematical}.
\begin{align*}
H_{0}(x)&=1,\nonumber\\
H_{1}(x)&=2x,\nonumber\\
H_{2}(x)&=4x^{2}-2,\nonumber\\
H_{3}(x)&=8x^{3}-12x,\nonumber\\
H_{4}(x)&=16x^{4}-48x^{2}+12,\nonumber\\
H_{5}(x)&=32x^{5}-160x^{3}+120x. \nonumber
\end{align*}
%\begin{align}
%{H} _{0}(x)&=1,\\
%{H} _{1}(x)&=x,\\
%{H} _{2}(x)&=x^{2}-1,\\
%{H} _{3}(x)&=x^{3}-3x,\\
%{H} _{4}(x)&=x^{4}-6x^{2}+3,\\
%{H} _{5}(x)&=x^{5}-10x^{3}+15x.
%{H} _{6}(x)&=x^{6}-15x^{4}+45x^{2}-15,\\
%{H} _{7}(x)&=x^{7}-21x^{5}+105x^{3}-105x,\\
%{H} _{8}(x)&=x^{8}-28x^{6}+210x^{4}-420x^{2}+105,\\
%{H} _{9}(x)&=x^{9}-36x^{7}+378x^{5}-1260x^{3}+945x,\\
%{H} _{10}(x)&=x^{10}-45x^{8}+630x^{6}-3150x^{4}+4725x^{2}-945.
 \begin{figure}[h]
\center
\includegraphics[width=55mm,height=55mm]{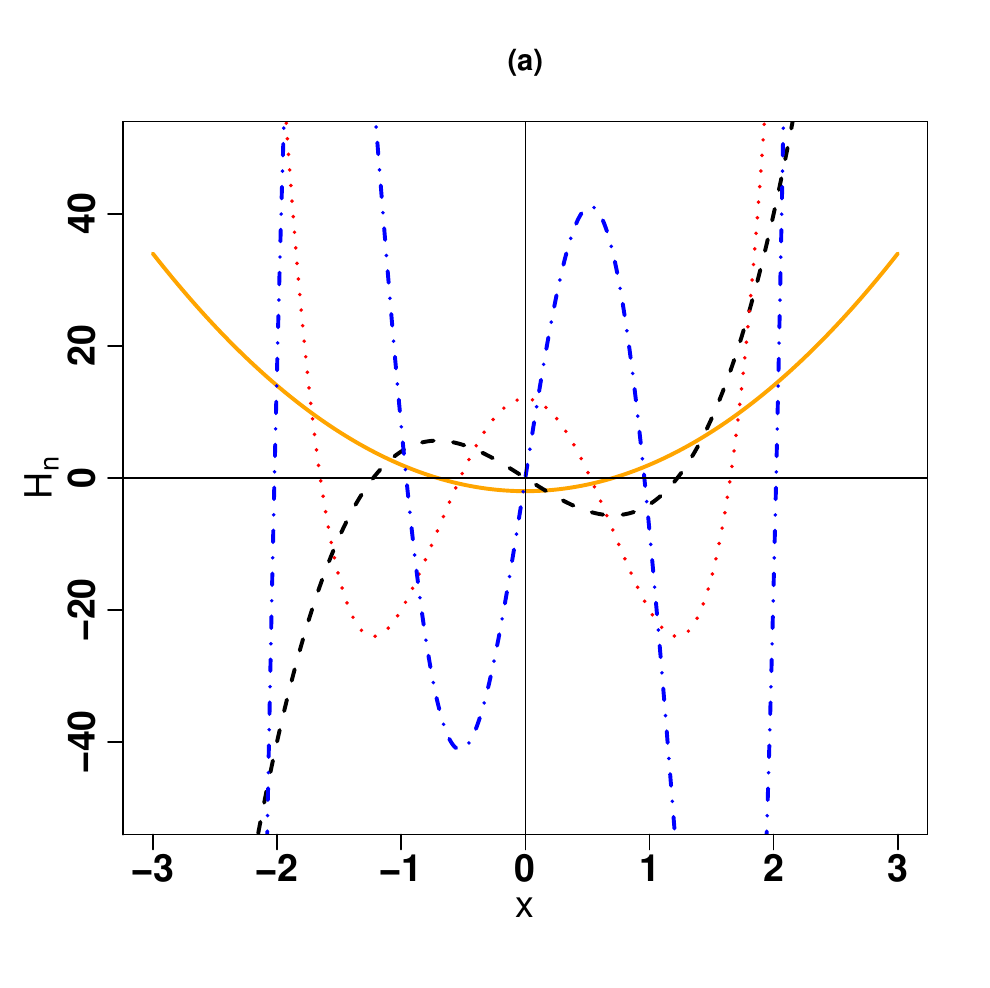}
\includegraphics[width=55mm,height=55mm]{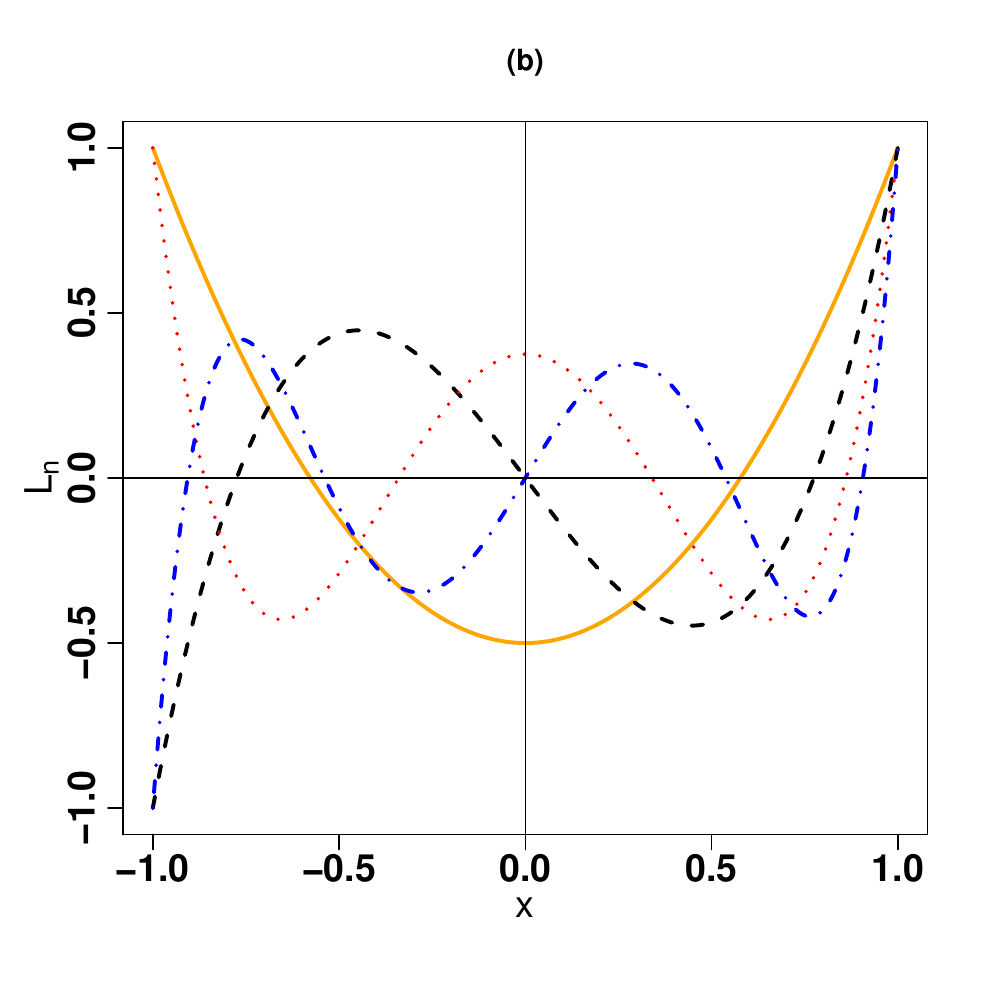}
\caption{Graph of (a): Hermite and (b): Legendre polynomials of orders $n=2$ (solid orange line), $n=3$ (dashed black line), $n=4$ (dotted red line), and $n=5$ (dotted-dashed  blue line). }
\label{Gaussian-quadrature-Hermite-Legendre-function}
\end{figure}
Figure \ref{Gaussian-quadrature-Hermite-Legendre-function} displays the graph of Hermite and  Legendre polynomials of orders $n=2,3,4,5$.
\item {\bf{Laguerre polynomial:}} The Laguerre polynomial of degree $n$, shown by ${\cal{L}}_{n}(x)$, is orthogonal with respect to weight function $w(x)=\exp\{-x\}$. This means that
\begin{align}\label{Laguerre-Laguerre-1} 
\langle {\cal{L}}_{n}, {\cal{L}}_{m}\rangle=\int_{0}^{\infty} {\cal{L}}_{n}(x) {\cal{L}}_{m}(x) \exp\{-x\}dx={\cal{D}}_{m,n}.
\end{align}
This class of orthogonal polynomials is solution of the differential equation
\begin{align}\label{Laguerre-Laguerre-2} 
x \frac{d^2 {\cal{L}}_{n}(x)}{dx^2}+ (1-x) \frac{d {\cal{L}}_{n}(x)}{dx}+ n {\cal{L}}_{n}(x)=0,
\end{align}
whose solution becomes
\begin{align}\label{Laguerre-Laguerre-3}
{\cal{L}}_{n}(x)= \Gamma(n+1) \sum_{i=0}^{n}\frac{(-1)^{i}x^{i}}{\Gamma(n-i+1)\bigl[\Gamma(i+1)]^{2}}.
\end{align} 
Using expression (\ref{Laguerre-Laguerre-3}), the first six members of Laguerre polynomials are as follows.
\begin{align*}
{\cal{L}} _{0}(x)&=1,\\
{\cal{L}} _{1}(x)&=1-x,\\
{\cal{L}} _{2}(x)&=\frac{x^{2}-4x+2}{2},\\
{\cal{L}} _{3}(x)&=-\frac{x^{3}-9x^{2}+18x-6}{6},\\
{\cal{L}} _{4}(x)&=\frac{x^{4}-16x^{3}+72x^{2}-96x+24}{24},\\
{\cal{L}} _{5}(x)&=\frac{-x^{5}+25x^{4}-200x^{3}+600x^{2}-600x+120}{120}.
\end{align*}
\end{itemize}
Table \ref{orthogonal-polynomials-1} shows a list of most commonly used orthogonal polynomials in statistics.
\vspace{5mm}
\begin{table}[!h]
%{\small{
\center
\caption{Most commonly used orthogonal polynomials.}  
\begin{tabular}{ccccc} \cline{1-5} 
$P_{n}(\cdot)$&symbol&${\text{C}_{n}}$&$[a,b]$&$w(x)$\\ \cline{1-5}
$\begin{matrix} {\text{Chebyshev}} \\
                       {\text{(first kind)}}  \end{matrix}$&
$T_{1,n}$&$\begin{cases} \pi & \mathrm{if}\ n =0, \\
                                                \frac{\pi}{2} &  \mathrm{if}\ n \neq 0.  \end{cases}$
&$(-1,1)$&$\bigl(1-x^2\bigr)^{-\frac{1}{2}}$\\\\
$\begin{matrix} {\text{Chebyshev}} \\
                       {\text{(second kind)}}  \end{matrix}$&
$T_{2,n}$&$\frac{\pi}{2}$&$[-1,1]$&$\bigl(1-x^2\bigr)^{\frac{1}{2}}$\\\\
$\begin{matrix}{\text{Generalized}} \\ {\text{Laguerre}}\end{matrix}$&
${\cal{L}}_{n}^{\alpha}$&$\frac{\Gamma(n+\alpha+1)}{\Gamma(n+1)}$&
$[0,\infty)$&$\begin{matrix} x^{\alpha}\exp\bigl\{-x\bigr\}, \\
                        \alpha>-1  \end{matrix}$\\\\
Hermite&$H_{n}$&$2^{n}\sqrt{\pi}\Gamma(n+1)$&$(-\infty,\infty)$&$\exp\{-{x^2}{}\}$\\\\
Jacobi&$J_{n}^{\alpha,\beta}$&
$\begin{matrix} \frac{2^{\alpha +\beta +1}}{2n+\alpha +\beta +1}\times \\
                        \frac {\Gamma (n+\alpha +1)\Gamma (n+\beta +1)}{\Gamma (n+\alpha +\beta +1)\Gamma (n+1)}  \end{matrix}$                        &$(-1,1)$&
%$\frac{2^{\alpha +\beta +1}}{2n+\alpha +\beta +1}\frac {\Gamma (n+\alpha +1)\Gamma (n+\beta +1)}{\Gamma (n+\alpha +\beta +1)\Gamma (n+1)}$&$(-1,1)$&
$\begin{matrix}
(1-x)^{\alpha}(1+x)^{\beta}\mathrm{,}\\ \alpha,\beta>-1\end{matrix}$\\\\
Laguerre&${\cal{L}}_{n}$&1&$[0,\infty)$&$\exp\bigl\{-x\bigr\}$\\\\
Legendre&$L_{n}$&$\frac{2}{2n+1}$&$[-1,1]$&1\\
\cline{1-5}
\end{tabular} 
%}}
\label{orthogonal-polynomials-1}
\end{table}
\vspace{5mm}
There are different types of the Gaussian quadrature rules have been developed in the literature for approximating the integral (\ref{quadrature-Gaussian}). Particular Gauss quadrature rule arises from particular choice of the interval and the form of the weight function $w(x)$. For example, the Gauss-Legendre quadrature rule is applied when 
$[a, b] = [-1,1]$ and $w(x) = 1$, and so is employed for approximating integral of the form 
\begin{align*}
\int_{-1}^{1}f(x)dx.
\end{align*}
Moreover, for approximating integral of the form 
\begin{align*}
\int_{0}^{\infty}\exp\{-x\}f(x)dx,
\end{align*}
we may use the Gauss-Laguerre quadrature rule since as it is seen the range of integral is $[a,b] = [0,\infty)$ and $w(x)=\exp\{-x\}$, and finally, a $k$-point Gauss-Chebyshev rule is useful when $[a,b]=[-1,1]$. The bounds of integral (\ref{quadrature-Gaussian}) can be changed from $(a,b)$ to $[-1,1]$ through a simple transformation when $a\neq-1$ and $b\neq1$. For integrand of the form $w(x)f(x)$, in which $w(x)$ is given in Table \ref{orthogonal-polynomials-1}, an orthogonal polynomial up to degree $k+1$ is fitted to integrand $w(x)f(x)$ that intercepts integrand $w(x)f(x)$ at $k$ nodes $\{x_{n}\}_{n=1}^{k}$. Once we have found $k$ pairs of nodes and associated weights, shown by $\{\bigl(x_{n},\omega_{n}\bigr)\}_{n=1}^{k}$, the integral can be approximated based on a $k$-point Gaussian quadrature rule as
\begin{align}\label{quadrature-Gaussian} 
\int_{a}^{b} w(x)f(x) dx \approx \sum_{n=1}^{k} \omega_{n}f\bigl(x_{n}\bigr).
\end{align} 
%%%%%%%%%%%%%%%%%%%%%%%%%%%%%%%%%%%%%%%%%
The nodes and weighs for Gauss-Chebyshev quadrature are determined simply. The class ${{T}}_{1,n}(\cdot)$ and ${{T}}_{2,n}(\cdot)$ of orthogonal polynomials are given by
\begin{align}\label{polynomial-Chebyshev-1}
    \begin{cases}
{{T}}_{1,n}(x)=\cos\bigl(n \theta\bigr),&{\text{for first kind,}}\\
{{T}}_{2,n}(x)=\frac{\sin\bigl((n+1)\theta\bigr)}{\sin(\theta)},&{\text{for second kind,}}
 \end{cases}
 \end{align} 
 %\begin{align}
%\left\{
%\begin{array}{ll}
%{{T}}_{1,n}(x)=\cos\bigl(n \theta\bigr),~&\mathrm{for~first~kind},\\
%{{T}}_{2,n}(x)=\frac{\sin\bigl((n+1)\theta\bigr)}{\sin(\theta)},~&\mathrm{for~second~kind},
%  \end{array}
%\end{align}
where $x=\cos(\theta)$ for $\theta \in [0,\pi]$ \citep{rivlin2020chebyshev}. These two classes of orthogonal polynomials admit the recurrence relation 
\begin{align}\label{polynomial-Chebyshev-2}
{{T}}_{1,n+1}(x)=2x{{T}}_{1,n}(x) -{{T}}_{1,n-1}(x),~\mathrm{for}~n \geq 1,
\end{align}
where ${T}_{1,0}(x)=1$ and ${T}_{1,1}(x)=x$. The nodes are given by
\begin{align}\label{polynomial-Chebyshev-3}
    \begin{cases}
x_n=\cos\Bigl(\frac{2n-1}{2k}\pi\Bigr),&{\text{for first kind,}}\\
x_n=\cos\Bigl(\frac{n}{k+1}\pi\Bigr),&{\text{for second kind,}}
 \end{cases}
 \end{align}
%\begin{align}\label{polynomial-Chebyshev-2}
%\left\{
%\begin{array}{ll}
%x_n=\cos\Bigl(\frac{2n-1}{2k}\pi\Bigr),~&\mathrm{for~first~kind},\nonumber\\
%x_n=\cos\Bigl(\frac{n}{k+1}\pi\Bigr),~&\mathrm{for~second~kind},
%    \end{array}
%\end{align}
where $n=1,\cdots,k$. The pertaining weights for the first and second kinds are the same and distributed uniformly, that is $\omega_n=\pi/k$, for $n=1,\cdots,k$.
%%%%%%%%%%%%%%%%%%%%%%%%%%%%%%%%%%%%%%%%%
The Gaussian quadrature rules based on the other orthogonal polynomials 
% nodes and weights, that is for where $\{\bigl(x_{n},\omega_{n}\bigr)\}_{n=1}^{k}$ 
are determined through the Golub-Welsch algorithm \citep{golub1969calculation} as follows. 
\subsection{Golub-Welsch algorithm}
The fundamental Gaussian quadrature theorem states that the nodes $x_i$s are the roots of the orthogonal polynomial. It can be shown that each orthogonal polynomial of degree $k$ has distinct real roots on interval $[a,b]$ \cite[pp. 53]{shen2011spectral}. There are two simple methods for computing the roots of orthogonal polynomial \cite[pp. 55]{shen2011spectral}. Except for the Chebyshev case, the roots of other orthogonal polynomials may be obtained trough the {\it{eigenvalue}} or {\it{iterative}} method. In what follows, we describe briefly the {\it{eigenvalue method}} or {\it{Golub-Welsch algorithm}}, see \citep[pp. 55]{shen2011spectral}. 
\begin{itemize}
\item {\bf{nodes}}: Let $P_{n}(x)$ denote the orthogonal polynomial. It can be seen that the orthogonal polynomials given on Table \ref{orthogonal-polynomials-1} admits the recurrence formula as \cite[Theorem 3.1]{shen2011spectral}:
\begin{align}\label{quadrature-Gaussian-recurrence} 
%xP{i}(x)=\alpha_{i+1}P_{i+1}(x) +\beta_{i}P_{i}(x) +\alpha_{i}P_{i-1}(x).
xP_{n}(x)=a_{n}P_{n+1}(x) +b_{n}P_{n}(x) +c_{n}P_{n-1}(x),
\end{align} 
for $c_{0}P_{-1}(x)=0$. The zeros of the orthogonal polynomial $P_{n}(x)$ are eigenvalues of symmetric threediagonal matrix $A$ given by \citep{gil2007numerical}:
\begin{align}\label{quadrature-Gaussian-recurrence-A}
A=
\left(
\begin{matrix}
b_{0}                  &\sqrt{a_0c_1}  &       	   									 &      &\\
\sqrt{a_0c_1}      &b_{1}              & \sqrt{a_1c_2}    	 				      &       &\\
   				        &\sqrt{a_1c_2}  & b_{2}    &             					 &\\
              	        &                       &             &  \ddots    					 & \sqrt{a_{k-2}c_{k-1}}\\
                          &                       &      	   & \sqrt{a_{k-2}c_{k-1}}\ &b_{k-1} 
\end{matrix}
\right),
\end{align}
where 
\begin{align*}
\alpha_{0}P_{-1}(x)=0, for i\geq 0.
%;~~b_{i}=\beta_{i}
%,~{\text{for}}~
%;~~\alpha_{i}=\sqrt{c_{i}a_{i-1}},~{\text{for}}~ i\geq 1.  
\end{align*}
The eigenvalues ${e}_{1},\cdots,{e}_{k}$ of matrix $A$ are also nodes of the Gaussian quadrature rule, that is $e_{n}=x_{n}$, for $n=1,\cdots,k$.
\item {\bf{weight}}: Once we have computed the eigenvalues of $A$ in (\ref{quadrature-Gaussian-recurrence-A}), the $n$th eight $\omega_{n}$ is obtained as 
\begin{align}\label{orthogonal-polynomials-weights}
\omega_{n}= \mu_{0}\frac{\bigl(\boldsymbol{v}_{n}[1]\bigr)^{2}}{\lVert \boldsymbol{v}_{n} \rVert},
\end{align}
in which
\begin{align*}
\mu_0=\int_{a}^{b}w(x)dx,
\end{align*}
$\lVert \cdot \rVert$ denotes the Euclidean norm, $\{\boldsymbol{v}_{1},\cdots,\boldsymbol{v}_{k}\}$ are the eigenvectors correspond to $\{{e}_{1},\cdots,{e}_{k}\}$, and $\boldsymbol{v}_{n}[1]$ is the first element of eigenvector $\boldsymbol{v}_{n}$. 
\end{itemize}
In what follows, we compute the Gaussian quadrature rule for orthogonal polynomials presented in Table \ref{orthogonal-polynomials-1}.
\begin{itemize} 
%%%%%%%%%%%%%%%%%%%%%%%%%%%%%%%%%%%%%%%%%
\item {\bf{nodes and weighs for generalized Laguerre polynomial}}: The class ${\cal{L}}^{\alpha}_{n}(\cdot)$ of orthogonal polynomials admits the recurrence relation 
\begin{align}\label{polynomial-Laguerre-generalized-1}
(n+1) {\cal{L}}^{\alpha}_{n+1}(x)=(2n+\alpha+1-x){\cal{L}}^{\alpha}_{n}(x) -(n+\alpha){\cal{L}}^{\alpha}_{n-1}(x),
\end{align}
where ${\cal{L}}^{\alpha}_{0}(x)=1$. Comparing the recurrence relations given in (\ref{quadrature-Gaussian-recurrence}) and (\ref{polynomial-Laguerre-generalized-1}), we obtain $b_n=0$, for $n=0,1,2,\cdots$, and
\begin{align}\label{polynomial-Laguerre-generalized-2}
    \begin{cases}
a_n=-(n+1) ~{\text{and}}~ b_n=(2n+\alpha+1), & {\text{if}}~n \geq 0,\\
c_n=-(n+\alpha),& {\text{if}}~n \geq 1.
 \end{cases}
 \end{align}
 %\begin{align}\label{polynomial-Laguerre-generalized-2}
%%\left.
%\left\{
% \begin{array}{ll}
%a_n=-(n+1)~\mathrm{and}~b_n=(2n+\alpha+1),~~ & \mathrm{if}\ n \geq 0,\\
%c_n=-(n+\alpha),~~ & \mathrm{if}\ n \geq 1.
%    \end{array}
% % \right \}  \Longrightarrow \alpha_{i}=\frac{i}{\sqrt{4i^2-1}},~~ i \geq 1.
%\end{align}
In this case, matrix $A$ in (\ref{quadrature-Gaussian-recurrence-A}) can be reconstructed by using the constants $a_n$, $b_n$, and $c_n$ provided in (\ref{polynomial-Laguerre-generalized-2}) for computing eigenvalues of the reconstructed matrix $A$ as the nodes. For computing weights, it follows from Table \ref{orthogonal-polynomials-1} that
\begin{align*}
\mu_0=\int_{a}^{b} w(x)dx= \int_{0}^{\infty} x^{\alpha} \exp\{-x\}dx=\Gamma(\alpha+1).
\end{align*}
Hence the weights $\omega_n$ (for $n=1,\cdots,k$) are obtained by considering $\mu_0=\Gamma(\alpha+1)$ in the RHS of (\ref{orthogonal-polynomials-weights}) in which $\boldsymbol{v}_n$  (for $n=1,\cdots,k$) is the eigenvector corresponds to $i$th eigenvalue of matrix $A$ constructed based on information given by (\ref{polynomial-Laguerre-generalized-2}).
%%%%%%%%%%%%%%%%%%%%%%%%%%%%%%%%%%%%%%%%%
\item {\bf{nodes and weighs for Hermite polynomial}}: The class ${{H}}^{\alpha}_{n}(\cdot)$ of orthogonal polynomials admits the recurrence relation given by (\ref{polynomial-Hermite-4}) for $n\geq 1$.
%\begin{align}\label{polynomial-Hermite-1}
%H_{n+1}(x)=2xH_{n}(x) -2nH_{n-1}(x),
%\end{align}
 Comparing the recurrence relations given in (\ref{quadrature-Gaussian-recurrence}) and (\ref{polynomial-Hermite-4}), we obtain
\begin{align}\label{polynomial-Hermite-2}
    \begin{cases}
a_n=\frac{1}{2}~{\text{and}}~b_n=0,~~ & {\text{if}}\ n \geq 0,\\
c_n=n,~~ & {\text{if}}\ n \geq 1.
    \end{cases}
\end{align}
In this case, matrix $A$ in (\ref{quadrature-Gaussian-recurrence-A}) can be reconstructed by using the constants $a_n$, $b_n$, and $c_n$ provided in (\ref{polynomial-Hermite-2}) for computing eigenvalues of the reconstructed matrix $A$ as the nodes. For computing weights, it follows from Table \ref{orthogonal-polynomials-1} that
\begin{align*}
\mu_0=\int_{a}^{b} w(x)dx= \int_{-\infty}^{\infty} \exp\{-{x^2}{}\}dx=\sqrt{2\pi}.
\end{align*}
Hence the weights $\omega_n$ (for $n=1,\cdots,k$) are obtained by considering $\mu_0=\sqrt{\pi}$ in the RHS of (\ref{orthogonal-polynomials-weights}) in which $\boldsymbol{v}_n$  (for $n=1,\cdots,k$) is the eigenvector corresponds to $i$th eigenvalue of matrix $A$ constructed based on information given by (\ref{polynomial-Hermite-2}).
%%%%%%%%%%%%%%%%%%%%%%%%%%%%%%%%%%%%
\item {\bf{nodes and weighs for Jacobi polynomial}}: The class $J^{\alpha,\beta}_{i}(\cdot)$ of orthogonal polynomials admits the recurrence relation 
\begin{align}\label{polynomial-Jacobi-1}
J^{\alpha,\beta}_{n+1}(x)=\bigl(x a^{\alpha,\beta}_{n}-b^{\alpha,\beta}_{n}\bigr)J^{\alpha,\beta}_{n}(x) -c^{\alpha,\beta}_{n}J^{\alpha,\beta}_{n-1}(x),
\end{align}
with $J^{\alpha,\beta}_{0}(x)=1$ and $J^{\alpha,\beta}_{1}(x)=x(\alpha+\beta+2)/2+(\alpha-\beta)/2$, for $n=1,2,\cdots$. Furthermore,
\begin{align}\label{polynomial-Jacobi-2}
a^{\alpha,\beta}_{n}&=\frac{(2n+\alpha+\beta+1)(2n+\alpha+\beta+2)}{2(n+1)(n+\alpha+\beta+1)},\nonumber\\
b^{\alpha,\beta}_{n}&=\frac{(\beta^2-\alpha^2)(2n+\alpha+\beta+1)}{2(n+1)(n+\alpha+\beta+1)(2n+\alpha+\beta)},\nonumber\\
c^{\alpha,\beta}_{n}&=\frac{(n+\alpha)(n+\beta)(2n+\alpha+\beta+2)}{(n+1)(n+\alpha+\beta+1)(2n+\alpha+\beta)}.
\end{align}
Rearranging the RHS of (\ref{polynomial-Jacobi-1}) and comparing it with the recurrence relation given in (\ref{quadrature-Gaussian-recurrence}), we obtain
\begin{align}\label{polynomial-Jacobi-3}
\begin{cases}
a_{0}=\frac{2}{\alpha+\beta+2}~{\text{and}}~b_{0}=\frac{\beta-\alpha}{\alpha+\beta+2}, & {\text{if}}~ \ n = 0,\\
a_{n}=\frac{2(n+1)(n+\alpha+\beta+1)}{(2n+\alpha+\beta+1)(2n+\alpha+\beta+2)}, 
& {\text{if}}~ \ n \geq 1,\\
b_{n}=\frac{\beta^2-\alpha^2}{(2n+\alpha+\beta)(2n+\alpha+\beta+2)},
& {\text{if}}~ \ n \geq 1,\\
c_{n}=\frac{2(n+\alpha)(n+\beta)}{(2n+\alpha+\beta)(2n+\alpha+\beta+1)},
& {\text{if}}~ \ n \geq 1.
\end{cases}
\end{align}
In this case, matrix $A$ in (\ref{quadrature-Gaussian-recurrence-A}) can be reconstructed by using the constants $a_{n}$, $b_{n}$, and $c_{n}$ provided in (\ref{polynomial-Jacobi-3}). As before, the nodes are eigenvalues of the reconstructed matrix $A$ and for computing weights, it follows from Table \ref{orthogonal-polynomials-1} that
\begin{align*}
\mu_0=\int_{a}^{b} w(x)dx= \int_{-1}^{1} (1-x)^{\alpha}(1+x)^{\beta} dx=
\frac{2^{\alpha +\beta+1}\Gamma (\alpha +1)\Gamma (\beta +1)}{\Gamma (\alpha +\beta +1)}.
\end{align*}
Hence the weights $\omega_{n}$ (for $n=1,\cdots,k$) are obtained by considering $\mu_{0}=2$ in the RHS of (\ref{orthogonal-polynomials-weights}) in which $\boldsymbol{v}_{n}$  (for $n=1,\cdots,k$) is the eigenvector corresponds to $n$th eigenvalue of matrix $A$ constructed based on information given by (\ref{polynomial-Jacobi-3}).
%%%%%%%%%%%%%%%%%%%%%%%%%%%%%%%%%%%%%%%%
\item {\bf{nodes and weighs for Legendre polynomial}}: The class $L_{n}(\cdot)$ of orthogonal polynomials admits the recurrence relation 
\begin{align}\label{polynomial-Legendre-1}
x L_{n}(x)=\frac{n+1}{2n+1}L_{n+1}(x) +\frac{n}{2n+1}L_{n-1}(x),
\end{align}
with $L_{0}(x)=1$ and $L_{1}(x)=x$, for $n=1,2,\cdots$. Comparing the recurrence relations given in (\ref{quadrature-Gaussian-recurrence}) and (\ref{polynomial-Legendre-1}), we obtain $b_n=0$, for $n=0,1,2,\cdots$, and
%\begin{align}\label{polynomial-Legendre-2}
%%\left.
%\left\{
%    \begin{array}{ll}
%      a_{n}=\frac{n+1}{2n+1},~~ & \mathrm{if}\ n \geq 0,\\
%      c_{n}=\frac{n}{2n+1},~~& \mathrm{if}\ n \geq 1.
%    \end{array}
% % \right \}  \Longrightarrow \alpha_{i}=\frac{i}{\sqrt{4i^2-1}},~~ i \geq 1.
%\end{align}
\begin{align}\label{polynomial-Legendre-2}
\begin{cases}
a_{n}=\frac{n+1}{2n+1},~& {\text{if}}\ n \geq 0,\\
c_{n}=\frac{n}{2n+1},~& {\text{if}}\ n \geq 1.
\end{cases}
\end{align}
In this case, matrix $A$ in (\ref{quadrature-Gaussian-recurrence-A}) can be reconstructed by setting $\sqrt{a_{n-1}c_{n}}=n/\sqrt{4n^2-1}$ (for $n=1,\cdots,k$) in the RHS of (\ref{quadrature-Gaussian-recurrence-A}). The nodes are eigenvalues of the reconstructed matrix $A$. For computing weights, it follows from Table \ref{orthogonal-polynomials-1} that
\begin{align*}
\mu_0=\int_{a}^{b} w(x)dx= \int_{-1}^{1} dx=2.
\end{align*}
Hence the weights $\omega_{n}$ (for $n=1,\cdots,k$) are obtained by considering $\mu_{0}=2$ in the RHS of (\ref{orthogonal-polynomials-weights}) in which $\boldsymbol{v}_{n}$ (for $n=1,\cdots,k$) is the eigenvector corresponds to $n$th eigenvalue of matrix $A$ constructed based on information given by (\ref{polynomial-Legendre-2}).
\end{itemize}
It should be noted that the accuracy of a Gaussian quadrature depends on the structure of the orthogonal polynomial. In general, if $f(x)$ in (\ref{quadrature-Gaussian}) has continuous derivatives of order $2k$, then the error in approximating integral (\ref{quadrature-Gaussian}) becomes \citep{kahaner1989numerical}:
\begin{align*}
{\text{Error}}=\int_{a}^{b} w(x)f(x)dx-\sum_{i=1}^{k}\omega_i f(x_i)=\frac{f^{(2k)}(\eta)}{\Gamma(2k+1)} \int_{a}^{b} w(x) P^{2}_{k}(x)dx,
\end{align*}
where $f^{(2k)}(\eta)$ denotes the $(2k)$th derivative of $f(x)$ for some $a<\eta<b$. In special case, when $w(x)=1$, we have
 \begin{align*}
{\text{Error}}=\frac{(b-a)^{2k+1}\bigl[\Gamma(k+1)\bigr]^4}{
(2k+1)\bigl[\Gamma(2k+1)\bigr]^3} f^{(2k)}(\eta).
\end{align*}
As a useful criterion, the goodness of approximation $\hat{I}$ for integral $I$ is reflected by absolute relative error (ARE) criteria defined as
 \begin{align}\label{ARE}
{\text{ARE}}=\Bigl \vert \frac{I-\hat{I}}{I}\Bigr \vert.
\end{align}
The following \verb+R+ function \verb+quad_rule+ is provided to compute the nodes and weights for implementing the Gaussian quadrature rules discussed earlier. We note that the argument \verb+type+ includes: \verb+CH1+ (for Chebyshev of first kind), \verb+CH2+ (for Chebyshev of second kind), \verb+HE+ (for Hermite), \verb+GL+ (for generalized Laguerre), \verb+JA+ (for Jacobi), and \verb+LE+ (for Legendre). Furthermore, arguments \verb+alpha+ and \verb+beta+ are parameters of Gaussian quadrature based on generalized Laguerre or Jacobi orthogonal polynomials. 
\begin{lstlisting}[style=deltaj]
R> quad_rule <- function(n, type = "HE", alpha = 0, beta = 0)
+{
+	if (any(c(alpha,beta) <= -1))stop(" alpha and beta must be greater than -1.")
+	A <- diag(0, n)
+	if ( type == "CH1" )
+	{
+		ia <- 1:n
+		node <- cos( (2*ia - 1)/(2*n)*pi )
+		weight <- pi/n
+	}
+	if ( type == "CH2" )
+	{
+		ia <- 1:n
+		node <- cos( ia/(n + 1)*pi )
+		weight <- pi/n
+	}
+	if ( type == "HE" )
+	{
+		a_nc_n <- 1:(n - 1)/2 
+		A[row(A) - col(A) == 1] <- A[row(A) - col(A) == -1] <- sqrt( a_nc_n ) 
+		A_eigen <- eigen(A)
+		node <- A_eigen$values
+		weight <- sqrt(pi)*( A_eigen$vector[1, ] )^2
+	}
+	if ( type == "GL" )
+	{
+		ib <- seq(0, n - 1)
+		b_n <- 2*ib + alpha + 1
+		ia <- seq(0, n - 2)
+		diag(A) <- b_n
+		A[row(A) - col(A) == 1] <- A[row(A) - col(A) == -1] <- 
+		sqrt( (ia + 1)*(ia + 1 + alpha) )
+		A_eigen <- eigen(A)
+		node <- A_eigen$values
+		weight <- gamma(alpha + 1)*( A_eigen$vector[1, ] )^2
+	}
+	if ( type == "JA" )
+	{
+		ib <- seq(1, n - 1)
+		b_n <- (beta^2 - alpha^2)/( (2*ib + alpha + beta)*
+									(2*ib + alpha + beta + 2) )
+		ia <- seq(1, n - 2)
+		a_n <- 2*(ia + 1)*(ia + alpha + beta + 1)/( (2*ia + alpha + beta + 1)*
+									(2*ia + alpha + beta + 2) )
+		ic <- ia + 1
+		c_n <- 2*(ic + alpha)*(ic + beta )/( (2*ic + alpha + beta)*
+									 (2*ic + alpha + beta + 1) ) 
+		diag(A) <- c(0, b_n )
+		A[row(A) - col(A) == 1] <- A[row(A) - col(A) == -1] <-  c(0, sqrt( a_n*c_n ) )
+		A[1, 1] <- (beta - alpha)/( alpha + beta + 2 )
+		A[1, 2] <- sqrt( 4*(1 + alpha)*(1 + beta )/
+											  ( (2 + alpha + beta)^2*(3 + alpha + beta) ) )
+		A[2, 1] <- A[1, 2] 
+		A_eigen <- eigen(A)
+		node <- A_eigen$values
+		weight <- 2^(alpha + beta + 1)*(alpha + beta + 1)*
+											beta(alpha + 1, beta + 1)*( A_eigen$vector[1, ] )^2
+	}
+	if ( type == "LE" )
+	{
+		ia <- seq(0, n - 2)
+		a_n <- (ia + 1)/(2*ia + 1)
+		ic <- ia + 1
+		c_n <- ic/(2*ic + 1)
+		A[row(A) - col(A) == 1] <- A[row(A) - col(A) == -1] <- sqrt( a_n*c_n )
+		A_eigen <- eigen(A)
+		node <- A_eigen$values
+		weight <- 2*( A_eigen$vector[1, ] )^2
+	}
+	ls <- list( "node" = node , "weight" = weight)
+	return(ls)
+}
\end{lstlisting}
%%%%%%%%%%%%%%%%%%%%%%%%%%%%%%%%%%%%%%
\begin{example}\label{quadrature-exam2}%\lipsum*[]
Let's to approximate integral (\ref{Gaussian-quadrature-broad-sense-1}) of Example \ref{Gaussian-quadrature-broad-sense-exam1} through the Gaussian-Legendre rule. To this end, first, we need to convert the range of integration from $[0,3]$ to $[-1,1]$. Using a change of variable $z=2u/3-1$ it turns out that
\begin{align}\label{Gaussian-quadrature-exam2-1}
I=\int_{0}^{3} x\exp\{-x\} dx=&\frac{9}{4}\exp\Bigl\{-\frac{3}{2}\Bigr\}\int_{-1}^{1} (z+1)\exp\Bigl\{-\frac{3}{2}z\Bigr\} dz\nonumber\\
=&\frac{9}{4}\exp\Bigl\{-\frac{3}{2}\Bigr\}\int_{-1}^{1} w(z)f(z)dz,  
\end{align}
where $w(z)=1$ and $f(z)=(z+1)\exp\bigl\{-3z/2\bigr\}$. Hence, the sequence of nodes and weights can be computed using function \verb+quad_rule+, when \verb+type=+``\verb+LE+''. Computing the sequence $\{x_i,\omega_i\}_{i=1}^{k}$, we can write
\begin{align}\label{Gaussian-quadrature-exam2-2}
I\approx \frac{9}{4}\exp\Bigl\{-\frac{3}{2}\Bigr\}\sum_{i=1}^{k} \omega_{i} \times  \bigl(x_i+1\bigr)\exp\Bigl\{-\frac{3}{2}x_i\Bigr\}.
\end{align}
The pertaining AREs for $k=2,3,5$ and 10 are 4.58e-02, 1.50e-03 1.75e-07, and 3.18e-15, respectively. We remind that the general quadrature rule applied in Example \ref{Gaussian-quadrature-broad-sense-exam1} yields 0.83755 with ${\text{ARE}}=0.045$. Evidently, he Gaussian quadrature shows superior performance with ${\text{ARE}}=0.045$ just by adding one more node.
\end{example}
%%%%%%%%%%%%%%%%%%%%%%%%%%%%%%%%%%%%%%
\begin{example}\label{Gaussian-quadrature-exam3}%\lipsum*[]
Suppose we are interested in approximating 
\begin{align}\label{Gaussian-quadrature-exam3-1}
I(p)=\int_{0}^{\infty}  x ^{p} \log  x \exp\bigl\{-x\bigr\} dx, 
\end{align}
where $p>0$. Since the range of integrand $I$ is $(0,\infty)$ and involves the term $\exp\{-x\}$, using class ${\cal{L}}^{p}_{k}$ of orthogonal polynomials, a Gaussian quadrature rule is applied for approximating $I(p)$ for different settings of $k$ and $p$. To this end, we compute the sequence of nodes and weights using function \verb+quad_rule+, when \verb+type=+``\verb+GL+'' and \verb+alpha=+$p$. Computing sequence $\{x_i,\omega_i\}_{i=1}^{k}$, we approximate integral $I(p)$ as follows.
\begin{align}\label{Gaussian-quadrature-exam3-2}
I(p)\approx \sum_{i=1}^{k} \omega_{i} \times \bigl(x_i\bigr)^{p} \log x_i.
\end{align} 
Table \ref{Gaussian-quadrature-exam3-tab} shows the ARE results for computing integral $I(p)$ in (\ref{Gaussian-quadrature-exam3-1}). It turns out from Table \ref{Gaussian-quadrature-exam3-tab} that for larger $p$, smaller $k$ provides the accuracy at desired level.
\begin{table}[h!]
\centering
\caption{ARE for approximating integral in (\ref{Gaussian-quadrature-exam3-1}).}
%using Gauss-generalized Laguerre rule.}
\begin{tabular}{clcccc} 
\cline{1-6} 
&& \multicolumn{4}{c}{ARE}\\ \cline{3-6} 
p &Exact value of $I(p)$ & $k=5$&$k=10$&$k=15$&$k=20$\\ \cline{1-6} 
0 &-0.5772156    &0.214375& 0.108274& 7.24e-02& 5.44e-02\\
1 & 0.42278433   &4.50e-01& 0.012404& 5.70e-03& 3.26e-03\\
2 & 1.84556867   &5.69e-03& 9.12e-04& 2.96e-04& 1.30e-04\\
3 & 7.53670601  &1.53e-03& 1.51e-04& 3.54e-05& 1.22e-05\\
4 &36.1468240   &5.64e-04& 3.55e-05& 6.14e-06& 1.68e-06\\
\cline{1-6} 
\end{tabular} 
\label{Gaussian-quadrature-exam3-tab}
\end{table}
%, but we consider the Gauss-Legendre rule for this purpose. As Table \ref{orthogonal-polynomials-1} shows, this method has been developed to approximate an integral on the interval $[-1,1]$ when $w(x)=1$. Hence, for computing above integral using the Gauss-Legendre rule, care must be taken as the range of integration must be changed from $(0,\infty)$ to $[-1,1]$. Fortunately, this integrals can be truncated on a $[-1,1]$ because of fast decay of the integrand due to exponential term with negative quadratic exponent $-x^2/2$. To this end, considering a predetermined precision $\epsilon$, the upper limit of integration becomes $\alpha=\sqrt{-2 \log \epsilon}$ for $\epsilon=2e-9$. So, applying a change of variable of the form $t=2x/\alpha-1$, it turns out that
%\begin{align}\label{exam-quadrature-Gaussian-4} 
%I\approx&\frac{\alpha^{p+1}}{2^{p-1}}\int_{-1}^{1} (t+1)^{p} \log \Bigl(\frac{\alpha}{2}(t+1)\Bigr) \exp\Bigl\{- \frac{\alpha^2(t+1)^2}{8}\Bigr\} dt\nonumber\\
%\approx&
%\frac{\alpha^{p+1}}{2^{p-1}}  \sum_{i=1}^{m} \omega_{i} \times \bigl(x_i+1\bigr)^{p} \log \Bigl(\frac{\alpha}{2}\bigl(x_i+1\bigr)\Bigr) \exp\Bigl\{- \frac{\alpha^2\bigl(x_i+1\bigr)^2}{8}\Bigr\},
%\end{align} 
% where the vectors of nodes and weights, i.e., $\{(x_i, \omega_i)\}_{j=1}^{k}$, are given in Table \ref{Gaussian-quadrature-exam2-tab} for $k=30$. 
%Using information of Table \ref{Gaussian-quadrature-exam2-tab} for the RHS of (\ref{exam-quadrature-Gaussian-4}), an approximation of integral (\ref{Gaussian-quadrature-exam2-1}) is 0.4572323 with relative error 5.551115e-17. 
 \begin{figure}[!h]
\center
\includegraphics[width=55mm,height=55mm]{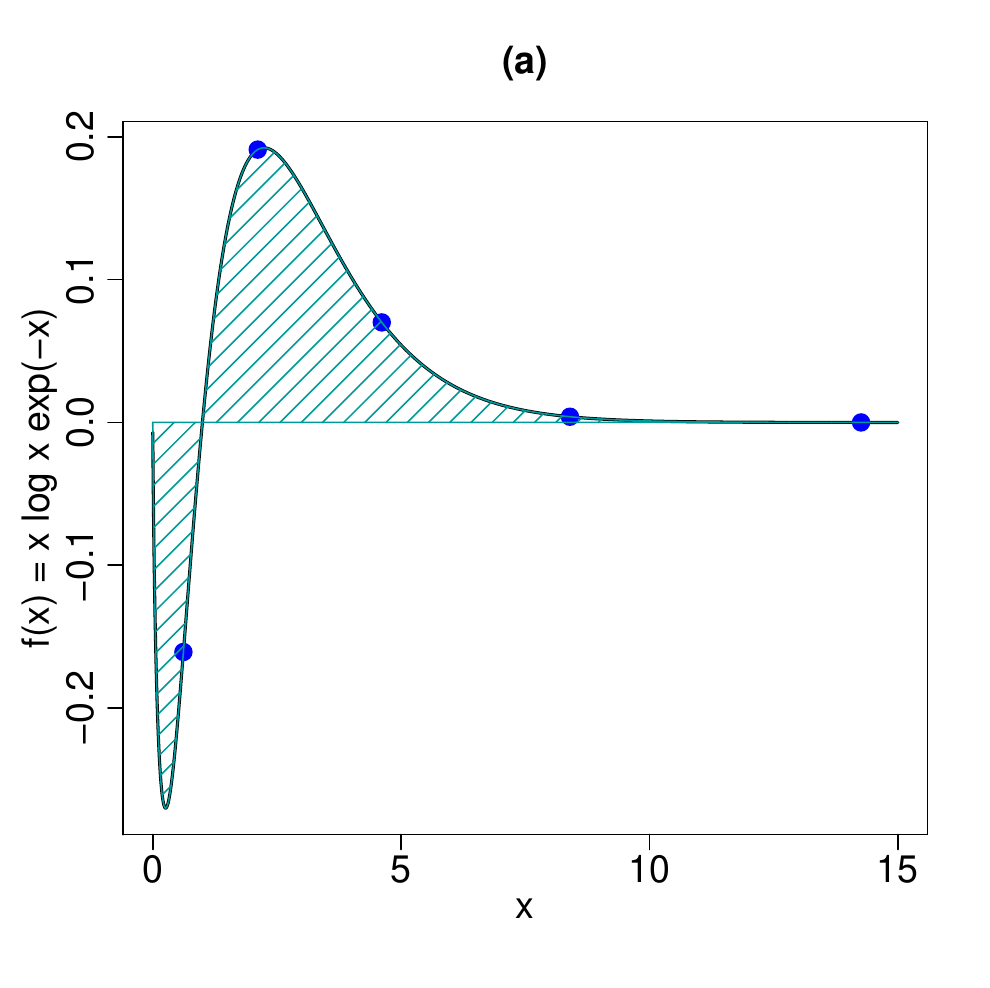}
\includegraphics[width=55mm,height=55mm]{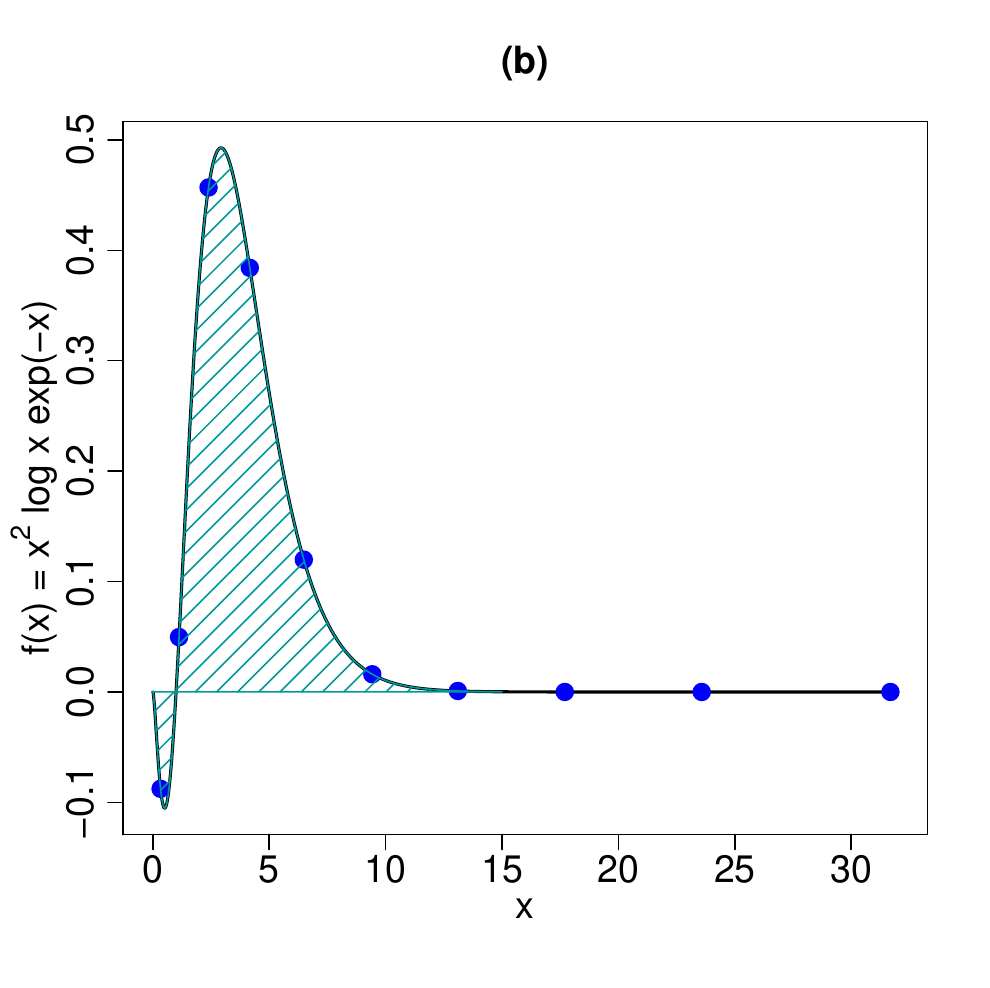}
\caption{Graph of integrand (\ref{Gaussian-quadrature-exam3-1}). (a): $p=1$ and $k=5$, (b): $p=2$ and $k=10$.}
\label{Gaussian-quadrature-exam3-fig}
\end{figure}
Figure \ref{Gaussian-quadrature-exam3-fig} shows the integrand of integral $I(p)$ in (\ref{Gaussian-quadrature-exam3-1}) on which x-coordinates of bullets are the position of the nodes. 
\end{example}
%%%%%%%%%%%%%%%%%%%%%%%%%%%%%%%
\begin{example}\label{Gaussian-quadrature-exam4}%\lipsum*[]
Suppose we are interested in approximating 
\begin{align}\label{Gaussian-quadrature-exam4-1}
I(z,\alpha)=\frac{1}{2\pi}\int_{-\infty}^{\infty} \exp\bigl\{-u^{\alpha}\bigr\} \cos(zu)du, 
 \end{align}
where $0<\alpha\leq 2$. In fact, integral $I(z,\alpha)$ is the PDF of symmetric standard $\alpha$-stable distribution at point $z$ \citep{nolan2020univariate}. For $\alpha=2$, each standard symmetric $\alpha$-stable distribution turns into a zero-mean Gaussian distribution with scale $1/\sqrt{2}$. Hence, we have $I(z, 2)={\cal{N}}(z\vert 0, 1/2)$. Herein, using a Gauss-Hermite rule, we proceed to approximate integral $I(z,\alpha)$. Table \ref{Gaussian-quadrature-exam4-tab} shows the ARE for different settings of $k$ and $z$. As it is seen, the ARE for large $z$ decreases and a larger $k$ needs to provide the accuracy of desired level. 
%the RHS of (\ref{exam-quadrature-Gaussian-4}), an approximation of integral (\ref{exam-quadrature-Gaussian-3}) is 0.4572323 with relative error 5.551115e-17. 
%Herein, we consider the Gauss-Hermite rule for this purpose. As Table \ref{exam2-quadrature-Gaussian-1} shows, this method has been developed to approximate an integral on the interval $[-1,1]$ for $w(x)=1$. Fortunately, the range of integral can be truncated on a $[-1,1]$ by considering a predetermined precision $\epsilon$, the upper limit of integration becomes $\alpha=\sqrt{-2 \log \epsilon}$ for $\epsilon=2e-9$. So, applying a change of variable of the form $t=2x/\alpha-1$, it turns out that
%\begin{align}\label{exam-quadrature-Gaussian-4} 
%I\approx&\frac{\alpha^{p+1}}{2^{p-1}}\int_{-1}^{1} (t+1)^{p} \log \Bigl(\frac{\alpha}{2}(t+1)\Bigr) \exp\Bigl\{- \frac{\alpha^2(t+1)^2}{8}\Bigr\} dt\nonumber\\
%\approx&
%\frac{\alpha^{p+1}}{2^{p-1}}  \sum_{i=1}^{m} \omega_{i} \times \bigl(x_i+1\bigr)^{p} \log \Bigl(\frac{\alpha}{2}\bigl(x_i+1\bigr)\Bigr) \exp\Bigl\{- \frac{\alpha^2\bigl(x_i+1\bigr)^2}{8}\Bigr\},
%\end{align} 
% where the vectors of nodes and weights, i.e., $\{(x_i, \omega_i)\}_{j=1}^{k}$, are given in Table \ref{orthogonal-polynomials-2} for $k=30$. 
\begin{table}[!h]
\centering
\caption{ARE for approximating integral in (\ref{Gaussian-quadrature-exam4-1}) for $\alpha=2$ using Gauss-Hermite rule.}
\begin{tabular}{cccccc} 
\cline{1-6} 
&& \multicolumn{4}{c}{ARE}\\ \cline{3-6} 
z &Exact value & $k=5$&$k=10$&$k=15$&$k=20$\\ \cline{1-6} 
0 &0.28209479 &1.96e-16& 1.96e-16& 1.96e-16& 5.90e-16\\
%1774
1 &0.21969564 &1.18e-06& 2.65e-15& 8.84e-16& 1.51e-15\\
%4734
2 &0.10377687&1.80e-03& 2.56e-09& 3.61e-15& 5.08e-15\\
%4355 
3 &0.02973257&2.02e-01& 1.63e-05& 9.84e-11& 1.08e-14\\
%2306 
4 &0.00516674&9.18174& 1.27e-02& 1.35e-06& 2.64e-11\\
%6339 
5 &0.00054457&284.672& 3.568332& 3.47e-03& 6.29e-07\\
\cline{1-6} 
%1057 
\end{tabular} 
\label{Gaussian-quadrature-exam4-tab}
\end{table}
Figure \ref{Gaussian-quadrature-exam4-fig} displays the pertaining integrand on which x-coordinates of bullets are the position of the nodes that are symmetrically distributed around origin. It it worth to note that integrand in (\ref{Gaussian-quadrature-exam4-1}) is an oscillatory function and 
%As it is see from Figure \ref{Gaussian-quadrature-exam3-fig}, 
when $z$ is large the number of oscillations is large \citep[pp. 68]{nolan2020univariate}.
 \begin{figure}[!h]
\center
\includegraphics[width=55mm,height=55mm]{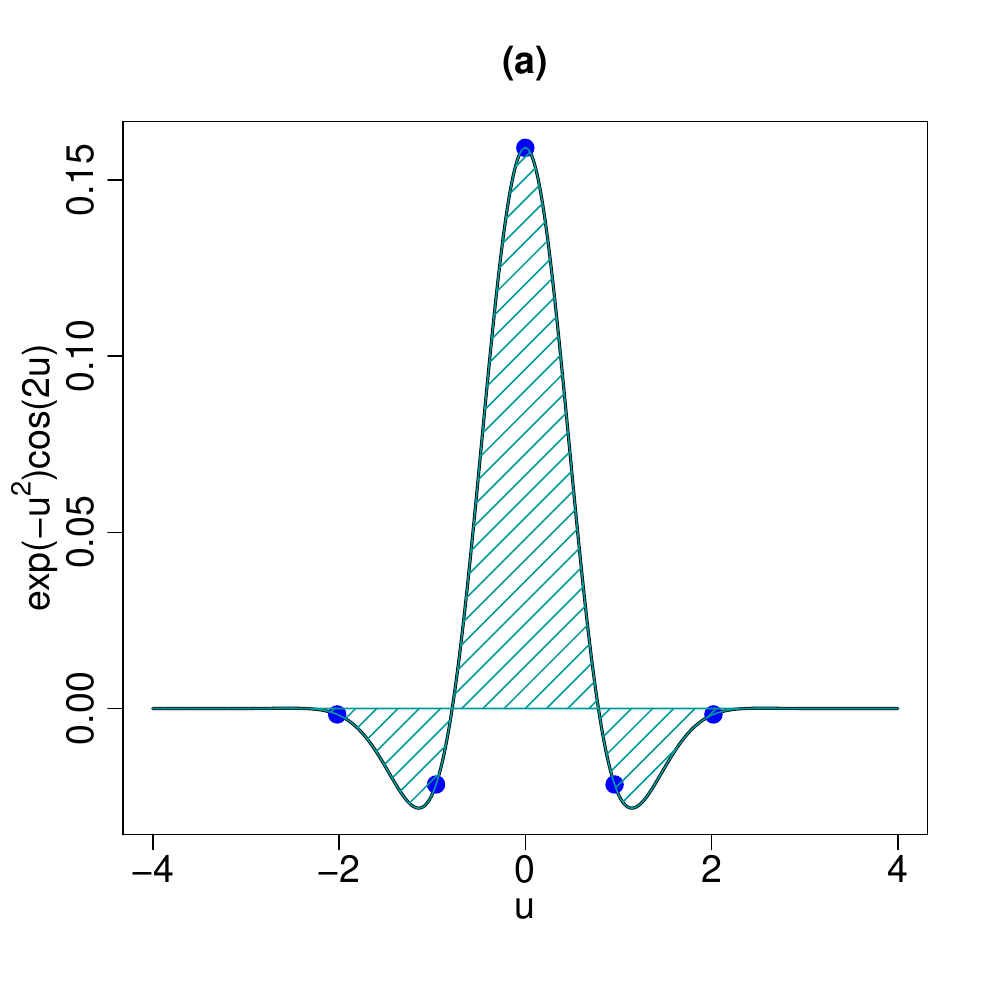}
\includegraphics[width=55mm,height=55mm]{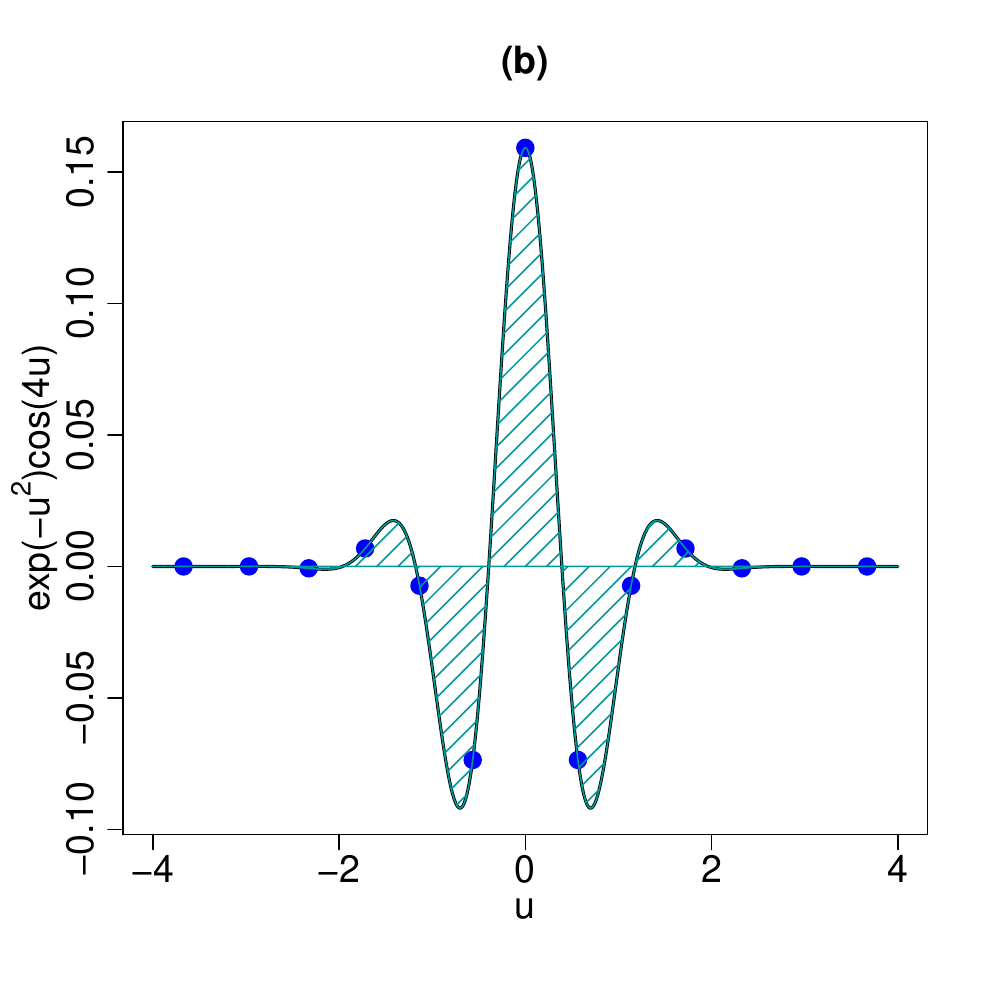}
\caption{Graph of integrand (\ref{Gaussian-quadrature-exam4-1}) for (a): $z=2$ and $k=5$, (b):$z=4$ and $k=15$. }
%The x-coordinate of each blue-colored bullet is the position of the corresponding node.}
\label{Gaussian-quadrature-exam4-fig}
\end{figure}
\end{example}
\section{Multivariate Gaussian quadrature}\label{multivariate-Gaussian-quadrature}
Each univariate $k$-point Gaussian quadrature rule represented as (\ref{quadrature-Gaussian}) can be straightforwardly extended to the $p$-dimensional case with $k^p$ points. Let $f(\boldsymbol{x})$ denote real function with domain $\boldsymbol{x} \in \mathbb{R}^{p}$ and we are willing to approximate the multiple integral of the form 
%\begin{align}
%\int_{\boldsymbol{a}}^{\boldsymbol{b}} w(\boldsymbol{x})f(\boldsymbol{x})d\boldsymbol{x}=\int_{a_{1}}^{b_{1}}\cdots\int_{a_{p}}^{b_{p}}
%w\bigl(x_{1}\bigr) \cdots w\bigl(x_{p}\bigr)f\bigl(x_{1},\cdots,x_{p}\bigr) dx_{1}\cdots dx_{p}
%-\sum_{i=1}^{k}\omega_i f(x_i)=\frac{f^{(2k)}(\eta)}{\Gamma(2k+1)} \int_{a}^{b} w(x) P^{2}_{k}(x)dx,
%\end{align}
\begin{align}\label{Gaussian-quadrature-multivariate-gamma-1}
\int_{\boldsymbol{a}}^{\boldsymbol{b}} w(\boldsymbol{x})f(\boldsymbol{x})d\boldsymbol{x}=\int_{a_{1}}^{b_{1}}\cdots\int_{a_{p}}^{b_{p}}
w\bigl(x_{1},\cdots,x_{p}\bigr)f\bigl(x_{1},\cdots,x_{p}\bigr) dx_{1}\cdots dx_{p},
\end{align}
in which elements of either $\boldsymbol{a}=(a_1,\cdots,a_{p})^{\top}$ or $\boldsymbol{b}=(b_1,\cdots,b_{p})^{\top}$ may be finite or infinite. It is assumed that the weight function is decomposed as the product of $p$ marginal weight functions as $w(\boldsymbol{x})=w_{1}(x_{1})w_{2}(x_{2})\cdots w_{p}(x_{p})$. Hence, the integral (\ref{Gaussian-quadrature-multivariate-gamma-1}) can be represented as
\begin{align*}%\label{Gaussian-quadrature-multivariate-2}
\int_{\boldsymbol{a}}^{\boldsymbol{b}} w(\boldsymbol{x})f(\boldsymbol{x})d\boldsymbol{x}=\int_{a_{1}}^{b_{1}}\cdots\int_{a_{p}}^{b_{p}}
w_{1}\bigl(x_{1}\bigr)w_{2}\bigl(x_{2}\bigr)\cdots w_{p}\bigl(x_{p}\bigr)f\bigl(x_{1},\cdots,x_{p}\bigr) dx_{1}\cdots dx_{p}.
\end{align*}
 The Gaussian quadrature in multivariate case allows to approximate the integral (\ref{Gaussian-quadrature-multivariate-gamma-1}) as
\begin{align}\label{Gaussian-quadrature-multivariate-gamma-2}
\int_{\boldsymbol{a}}^{\boldsymbol{b}} w(\boldsymbol{x})f(\boldsymbol{x})d\boldsymbol{x}\approx \sum_{i_{1}=1}^{k_{1}}\sum_{i_{2}=1}^{k_{2}}\cdots \sum_{i_{p}=1}^{k_{p}}
\omega_{1i_{1}}\omega_{2i_{2}}\cdots\omega_{p i_{p}}f\bigl(x_{1   i_{1}},x_{2 i_{2}},\cdots,x_{p i_{p}}\bigr),
\end{align}
where $\{\bigl({x}_{i j},{\omega}_{i j}\bigr)\}_{j=1}^{k_{i}}$ is the set of $k_{i}$ nodes and weights associated  with the $i$th coordinate under one of seven Gauss quadrature rules given by Table \ref{orthogonal-polynomials-1}, for $i=1,\cdots,p$. It is worthwhile to note that the Gauss quadrature rule applied to two distinct coordinates may be different. For example, for approximating a double integral, we may compute $k_{1}$ pairs of nodes and weights for the first coordinate using the Gauss-Laguerre rule while the set of $k_{2}$ nodes and weights for the second coordinate are obtained using the Gauss-Legendre rule.
\begin{example}\label{Gaussian-quadrature-multivariate-gamma-exam-1}%\lipsum*[]
Let the PDF of random vector $\boldsymbol{X}=\bigl(X_{1},X_{2}\bigr)^{\top}$  is given by
\begin{align}\label{Gaussian-quadrature-multivariate-gamma-4}
g(x_{1}, x_{2}\vert \boldsymbol{\theta})=\frac{\beta^{\alpha}x_{1}^{\alpha+\lambda-1}x_{2}^{\lambda-1}}{\Gamma(\lambda)\Gamma{(\alpha)}}\exp\bigl\{-\beta x_{1}-x_{1}x_{2}\bigr\},
 \end{align}
where $x_{1}>0$, $x_{2}>0$, and $\boldsymbol{\theta}=\bigl(\alpha>0,\beta>0,\lambda>0\bigr)^{\top}$. For computing the associated CDF that can be expressed in terms of integral representation as
\begin{align}\label{Gaussian-quadrature-multivariate-gamma-5}
G(q_1, q_2\vert \boldsymbol{\theta})=&\frac{\beta^{\alpha}}{\Gamma(\lambda)\Gamma(\alpha)}\int_{0}^{q_1}\int_{0}^{q_2} x_{1}^{\alpha+\lambda-1}x_{2}^{\lambda-1}\exp\bigl\{-\beta x_{1}-x_{1}x_{2}\bigr\} dx_{2}dx_{1},
\end{align}
where $q_1 \in \mathbb{R}^{+}$ and $q_2 \in \mathbb{R}^{+}$, we would like to use the Gaussian quadrature rule. To this end, we  apply the Gauss-Legendre rule to both coordinates of integral in the RHS of (\ref{Gaussian-quadrature-multivariate-gamma-5}). Hence, using a simple change of variable, we have
\begin{align}\label{Gaussian-quadrature-multivariate-gamma-6}
G(q_1, q_2\vert \boldsymbol{\theta})=&\int_{-1}^{1}\int_{-1}^{1} \frac{\beta^{\alpha}}{\Gamma(\lambda)\Gamma(\alpha)}\Bigl[\frac{x_{1}+1}{2}q_1\Bigr]^{\alpha+\lambda-1}\Bigl[\frac{x_{2}+1}{2}q_2\Bigr]^{\lambda-1}\nonumber\\
&\times \exp\Bigl\{-\frac{x_{1}+1}{2}q_1\Bigl[\beta+\frac{x_{2}+1}{2}q_2\Bigr]\Bigr\} dx_{2}dx_{1}\nonumber\\
=&\frac{q_1q_2}{4}\int_{-1}^{1}\int_{-1}^{1}w(\boldsymbol{x})f(x_{1},x_{2})dx_{2}dx_{1},
\end{align}
in which 
\begin{align}
w(\boldsymbol{x})=&w(x_{1})w(x_{2})=1\times 1=1,%\label{Gaussian-quadrature-multivariate-7}
\\
 f(u,v)=&\frac{\beta^{\alpha}}{\Gamma(\lambda)\Gamma(\alpha)}\Bigl[\frac{u+1}{2}q_1\Bigr]^{\alpha+\lambda-1}\Bigl[\frac{v+1}{2}q_2\Bigr]^{\lambda-1} \exp\Bigl\{-\frac{u+1}{2}q_1\Bigl[\beta+\frac{v+1}{2}q_2\Bigr]\Bigr\} 
\label{Gaussian-quadrature-multivariate-gamma-7}.
 \end{align}
It follows form (\ref{Gaussian-quadrature-multivariate-gamma-2}) that 
\begin{align}\label{Gaussian-quadrature-multivariate-gamma-8}
G(q_1, q_2\vert \boldsymbol{\theta})
\approx&\frac{q_1q_2}{4} \sum_{i=1}^{k_{1}}\sum_{j=1}^{k_{2}} 
\omega_{1i}\omega_{2j} f(x_{1i},x_{2j}),
\end{align}
where $\omega_{pi}$ and $x_{pi}$ denote, accordingly, the $i$th weight and node associated with the $p$th coordinate, for $p=1,2$, and $f(\cdot,\cdot)$ is given by (\ref{Gaussian-quadrature-multivariate-gamma-8}). Furthermore, constants $k_1$ and $k_2$ are the number of nodes (or weights) corresponds to quadrature rule considered for the first and second coordinates, respectively. Evidently, herein, we should use the Gauss-Legendre rule. In what follows, we give \verb+R+ code for approximating $G(q_1, q_2\vert \boldsymbol{\theta})$.
\vspace{5mm}
\begin{lstlisting}[style=deltaj]
R > G_quad <- function(q1, q2, k, theta)
+ {
+ alpha <- theta[1]; beta <- theta[2]; lambda <- theta[3]
+ out <- quad_rule(k, type = "LE", alpha = 0, beta = 0)
+ weight_grid <- x_grid <- matrix(0, nrow = k^2, ncol = 2)
+ weight_grid <- apply( expand.grid(out$weight, out$weight), 1, prod)
+ x_grid <- expand.grid(out$node, out$node)
+ x1 <- (x_grid[, 1] + 1)*q1/2 
+ x2 <- (x_grid[, 2] + 1)*q2/2
+ f <- x1^(alpha + lambda - 1)*x2^(lambda - 1)*exp(-beta*x1 - x1*x2)
+ I1 <- sum( weight_grid*f ) 
+ q1*q2/4*beta^alpha/(gamma(lambda)*gamma(alpha))*I1
+ }
\end{lstlisting}
It should be noted that the function \verb+G_quad(...)+ depends on \verb+quad_rule(...)+ that is given after (\ref{ARE}). We end this example with a brief investigation on performance of \verb+G_quad(...)+ for approximating CDF given by (\ref{Gaussian-quadrature-multivariate-gamma-5}). Table \ref{table-Gaussian-quadrature-multivariate-gamma-1} shows the pertinent results.
%\vspace{1cm}
%\begin{table}[htp]
%\begin{center}
%\begin{tabular}{|l|l|l|}\hline
%semimajor axis $a$       & 300000  & 262413    \\ \hline
%true anamoly $f$         & 163.76   & 176.08   \\\hline
%$r_p$                    & 6678 & 1689
% \tablefootnote{spacecraft will hit earth on way back since $r_p<r_{earth}$} \\\hline
%\end{tabular}
%\caption{Summary table for non-tangential per and post flyby the moon}
%\label{tab:part_3_1_summary}
%\end{center}
%\end{table}
\vspace{5mm}
\begin{table}[h]
\begin{threeparttable}
\caption{ARE for approximating CDF (\ref{Gaussian-quadrature-multivariate-gamma-5}) for $\beta=1$ using Gauss-Legendre rule.}
\label{table-Gaussian-quadrature-multivariate-gamma-1}
\begin{tabular}{lllccccc} 
\cline{1-8} 
$q_1=0.5$     &$q_2=0.5$       &\multicolumn{5}{c}{ARE}                   \\ \cline{1-8}
    &   &Exact value\tnote{1}                 & $k=10$      &$k=40$       & $k=80$     & $k=160$    \\ \cline{1-8}
$\alpha=0.5$  &$\lambda=0.5$  & 0.17066603 & 0.04304235  & 0.01115619  & 0.00561266 & 0.00281506 \\
$\alpha=0.5$  &$\lambda=  5$  & 3.2578e-07 & 4.5364e-07  & 4.53441e-07  & 4.5341e-07 & 4.5341e-07 \\
$\alpha=  5$  &$\lambda=0.5$  & 8.1943e-05 & 0.04438234  & 0.01150114  & 0.00578534 & 0.00290082 \\
$\alpha=  5$  &$\lambda=  5$  & 5.5782e-10 & 5.1988e-10  & 5.1988e-10  & 5.1988e-10 & 5.1988e-10 \\ \cline{1-8}
$q_1=0.5$     &$q_2=10$       &\multicolumn{5}{c}{}                   \\ \cline{1-8}
$\alpha=0.5$  &$\lambda=0.5$ & 0.48776719 &0.06751127 &  0.01745956 &  0.00878285 &  0.00440495 \\ 
$\alpha=0.5$  &$\lambda=  5$ & 0.05560379 &7.0189e-09 &  5.0508e-10 &  5.0508e-10 &  5.0508e-10 \\
$\alpha=  5$  &$\lambda=0.5$ & 1.7098e-04 &0.09553933  &  0.02465960 &  0.01240306 &  0.00622006 \\
$\alpha=  5$  &$\lambda=  5$ & 6.8194e-05 &9.2085e-10  &  9.2843e-10 &  9.2843e-10 &  9.2843e-10 \\ \cline{1-8}
$q_1=10$     &$q_2=0.5$       &\multicolumn{5}{c}{}                   \\ \cline{1-8}
$\alpha=0.5$  &$\lambda=0.5$ & 0.39181881 &0.04766594  &  0.01234977 &  0.00621302 &  0.00311616\\ 
$\alpha=0.5$  &$\lambda=  5$ & 0.00118855 &5.0360e-06 &  1.9364e-11 &  1.9364e-11 &  1.9364e-11\\
$\alpha=  5$  &$\lambda=0.5$ & 0.92144044 &0.07497320  &  0.01938347 &  0.00975048 &  0.00489024\\
$\alpha=  5$  &$\lambda=  5$ & 0.12532847 &3.4583e-07 &  2.5685e-10 &  2.5685e-10 &  2.5685e-10\\ \cline{1-8}
$q_1=10$     &$q_2=10$       &\multicolumn{5}{c}{}                   \\ \cline{1-8}
$\alpha=0.5$  &$\lambda=0.5$ & 0.80501002 &0.10896581  &  0.02690028 &  0.01352624 &  0.00678325\\ 
$\alpha=0.5$  &$\lambda=  5$ & 0.34088538 &1.7070e-02 &  5.3893e-09&  2.5295e-10 &  2.5120e-10\\
$\alpha=  5$  &$\lambda=0.5$ & 0.97074572 &0.34194629  &  0.08258106 &  0.04142828 &  0.02076378\\
$\alpha=  5$  &$\lambda=  5$ & 0.97017575 &1.3694e-02 &  2.0986e-11 &  2.0992e-11 &  2.0983e-11\\
\cline{1-8}    
%\multicolumn{4}{l}{\small *THIS IS A NICE FOOTNOTE.} \\
 %\tablefootnote{spacecraft will hit earth on way back since } \\
\end{tabular} 
\begin{tablenotes}
\item[1] The exact value has been computed using \verb+Maple+ software up to eight decimal places.
\end{tablenotes}
%\footnote{Hint: The exact value has been computed using \verb+Maple+ software up to eight decimal places.}                                                                                                           
\end{threeparttable}
\end{table}
\begin{figure}[!h]
\center
\includegraphics[width=55mm,height=55mm]{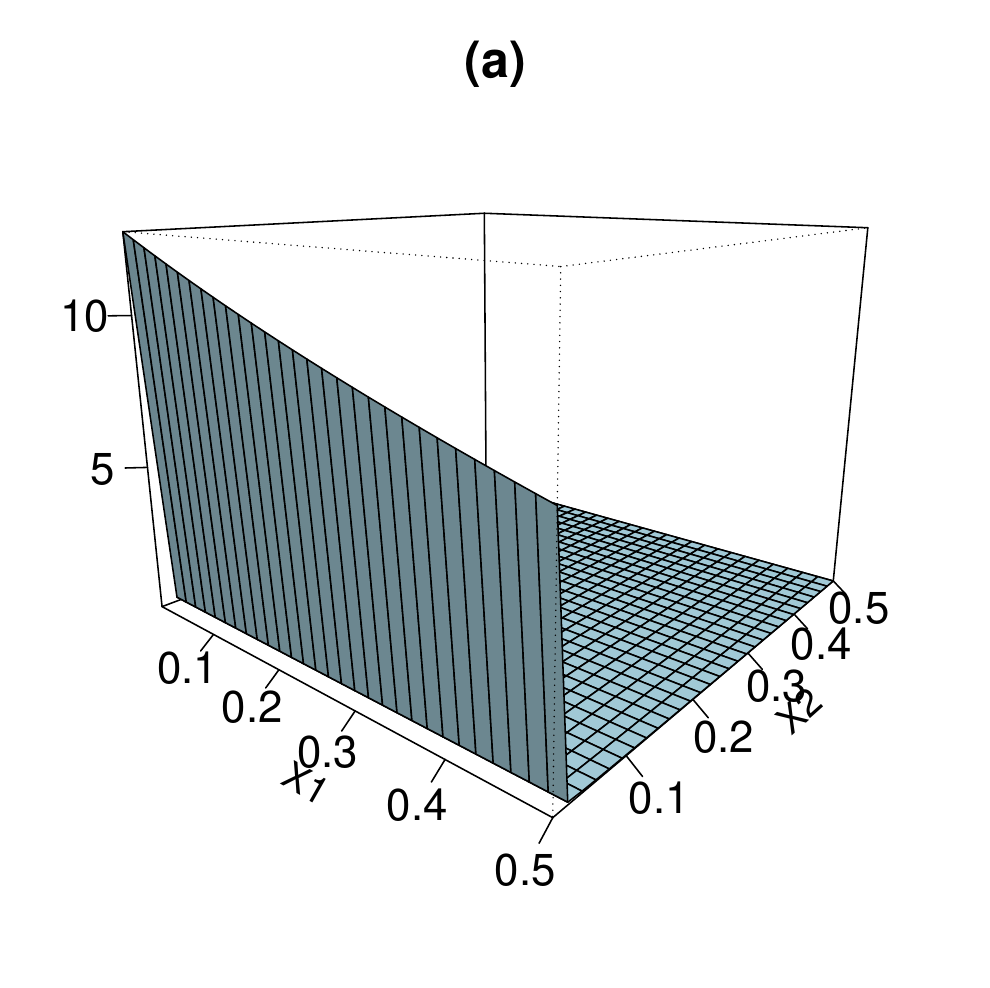}
\includegraphics[width=55mm,height=55mm]{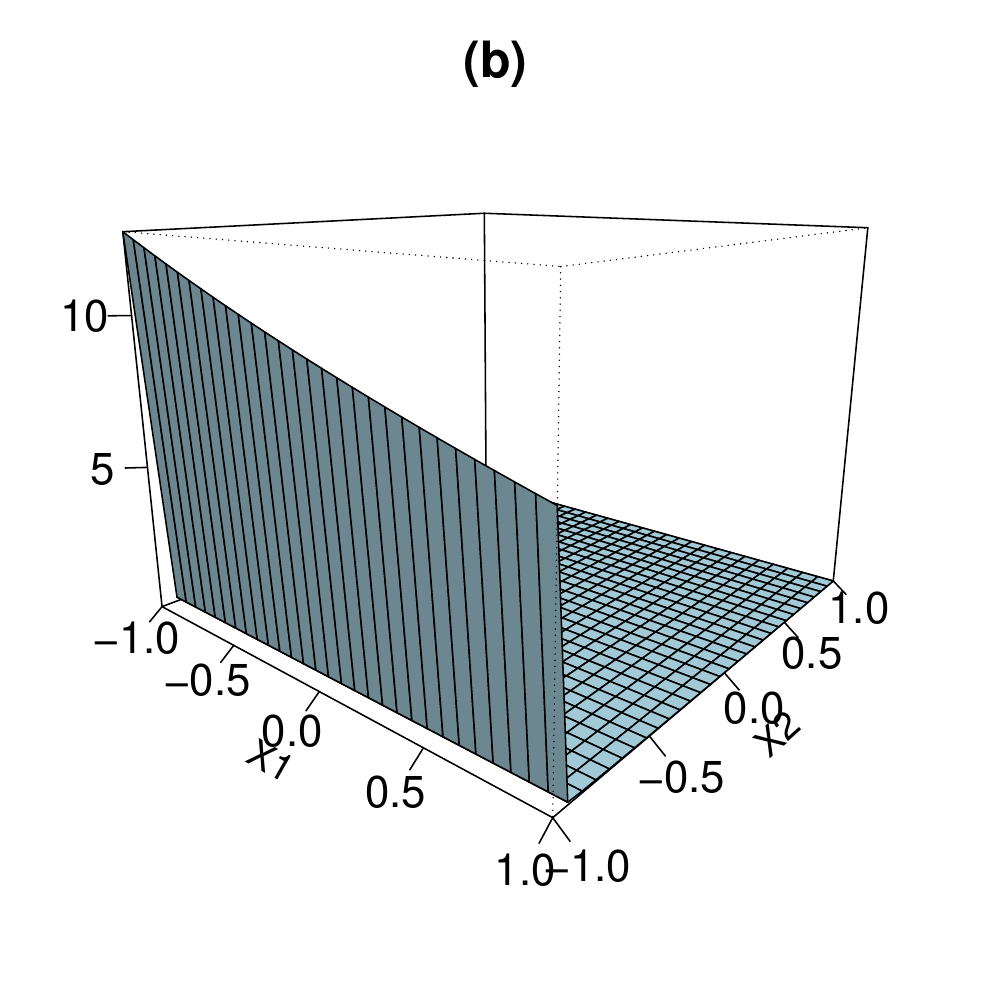}\\
\includegraphics[width=55mm,height=55mm]{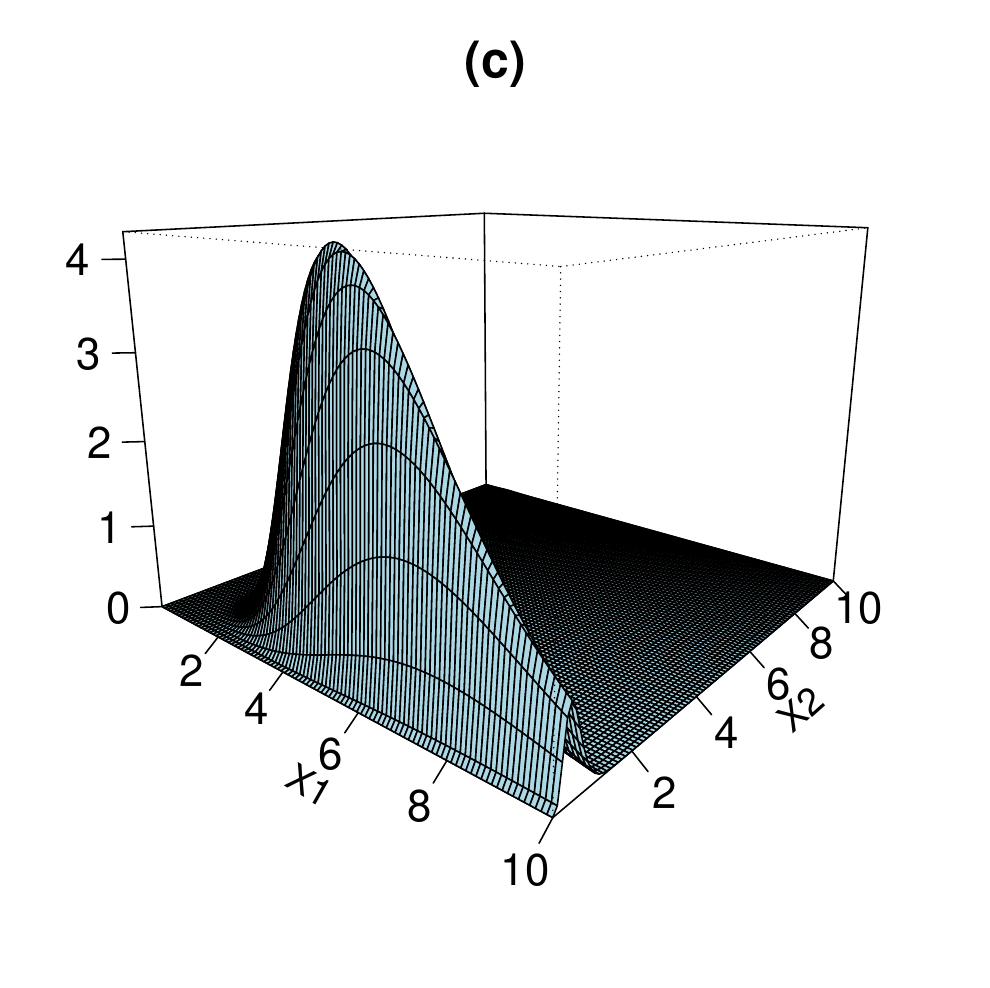}
\includegraphics[width=55mm,height=55mm]{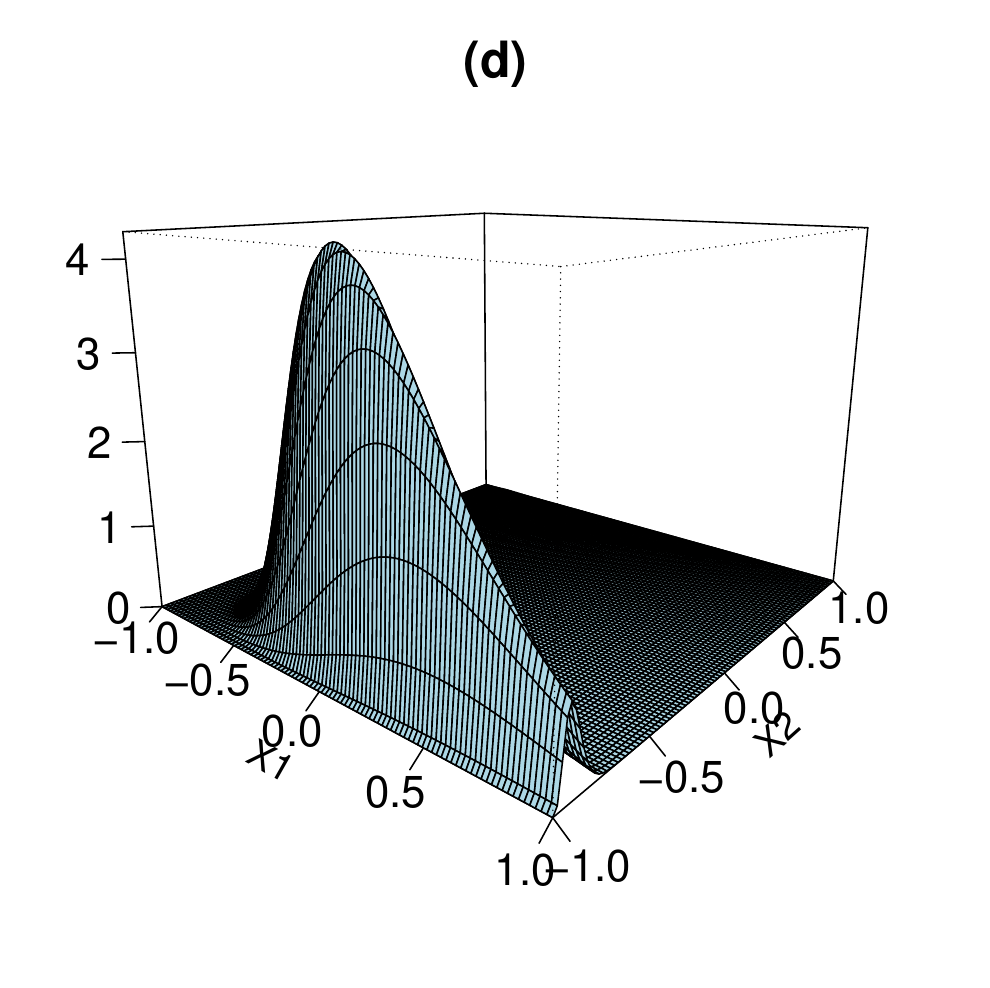}\\
\includegraphics[width=55mm,height=55mm]{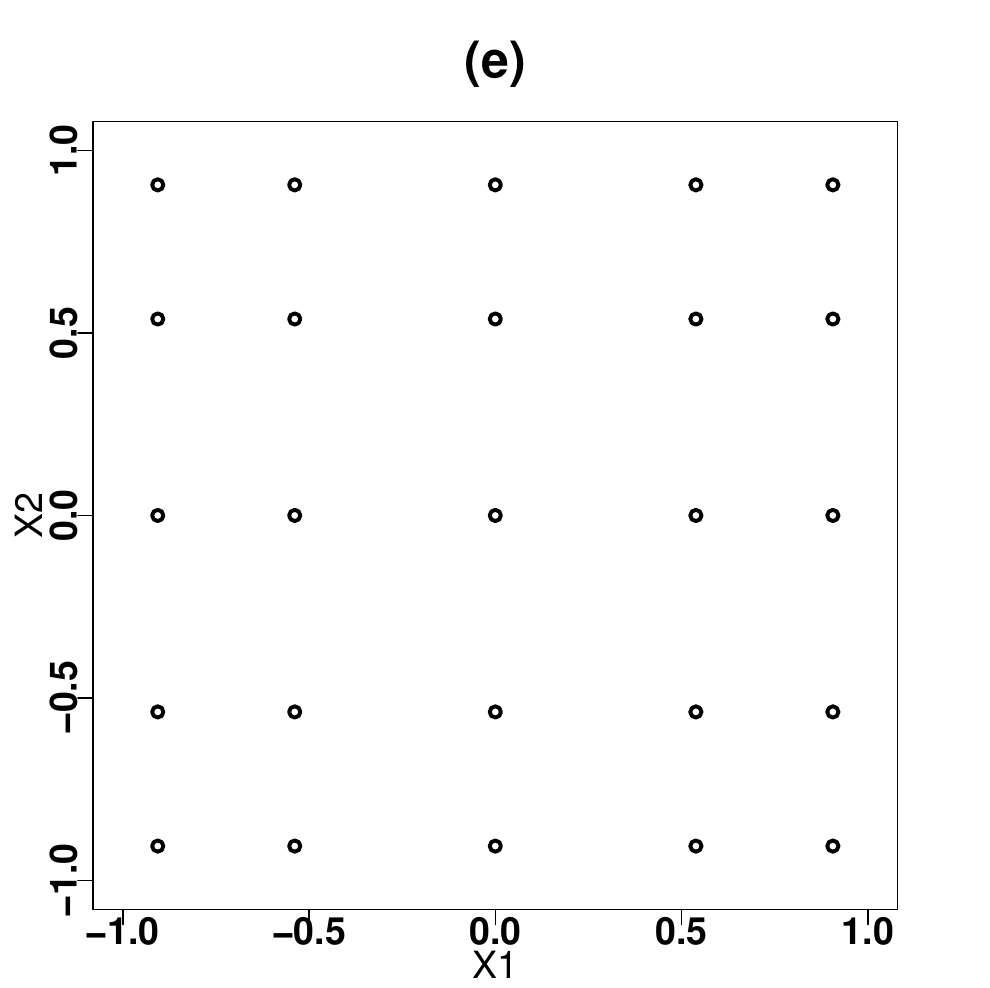}
\includegraphics[width=55mm,height=55mm]{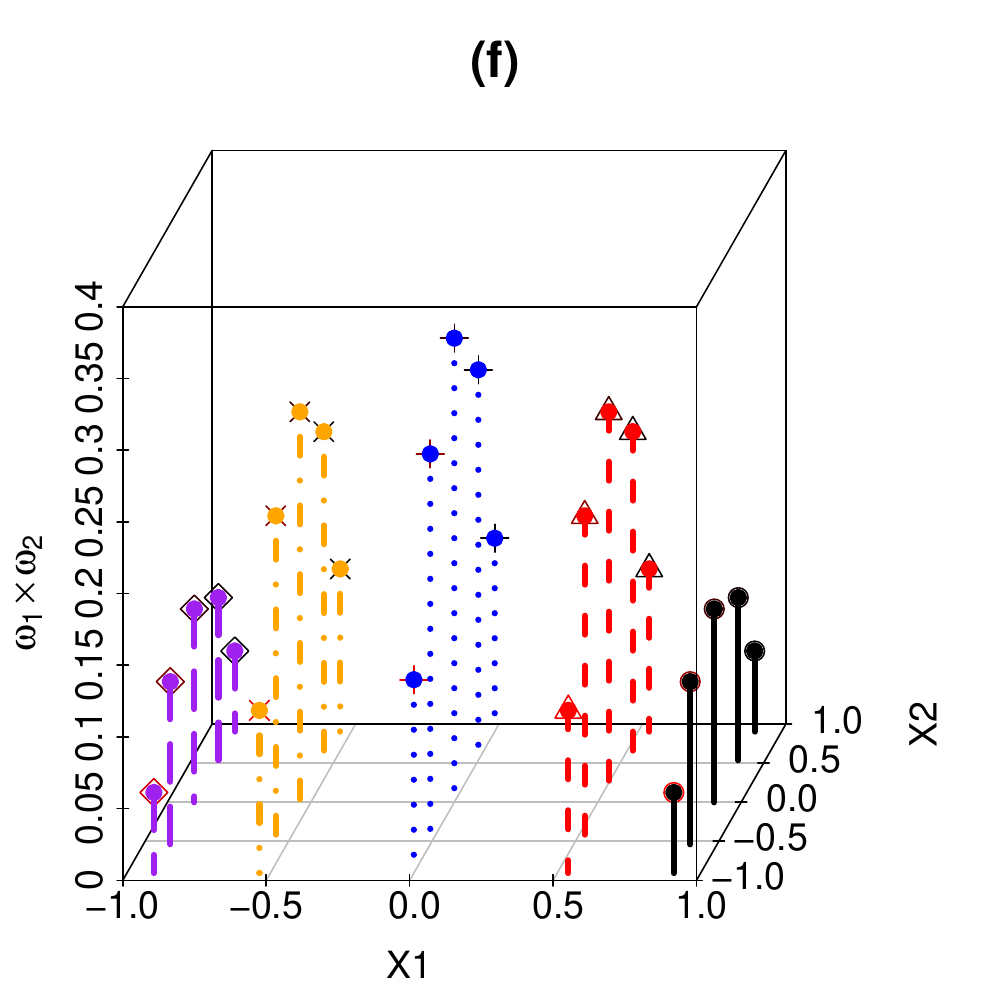}
\caption{(a): Graph of $g\bigl(x_{1}, x_{2}\vert \boldsymbol{\theta}=(0.5,1,0.5)^{\top}\bigr)$ on region $[0,0.5]\times[0,0.5]$, (b): Graph of $g\bigl(x_{1}, x_{2}\vert \boldsymbol{\theta}=(0.5,1,0.5)^{\top}\bigr)$ transformed to region $[-1,1]\times[-1,1]$, (c): Graph of $g\bigl(x_{1}, x_{2}\vert \boldsymbol{\theta}=(5,1,5)^{\top}\bigr)$ on region $[0,10]\times[0,10]$, (b): Graph of $g\bigl(x_{1}, x_{2}\vert \boldsymbol{\theta}=(5,1,5)^{\top}\bigr)$ transformed to region $[-1,1]\times[-1,1]$, (e): Coordinates of nodes computed using Gauss-Legendre rule with $k_1=k_2=5$ points, and (f): Product of weights computed using Gauss-Legendre rule with $k_1=k_2=5$ points.}
\label{fig-Gaussian-quadrature-multivariate-gamma-1}
\end{figure}

\vspace{5mm}
As it is seen from Table \ref{table-Gaussian-quadrature-multivariate-gamma-1}, when $\lambda$ is small or both elements of pair $(\alpha,\lambda)$ are small, then the performance of Gaussian quadrature significantly lessened for $k=10$ compared to other settings of $k$. Figure \ref{fig-Gaussian-quadrature-multivariate-gamma-1} (a)-(d)  display the PDF (\ref{Gaussian-quadrature-multivariate-gamma-4}) for two settings of parameter vector $\boldsymbol{\theta}$. Specifications of the Gauss-Legendre rule with $k_1=k_2=5$ points for approximating $G\bigl(0.5, 0.5\vert (0.5,1,0.5)^{\top}\bigr)$ and $G\bigl(10,10\vert (5,1,5)^{\top}\bigr)$ are shown by Figure \ref{fig-Gaussian-quadrature-multivariate-gamma-1} (e)-(f).
\end{example}
\begin{example}\label{Gaussian-quadrature-multivariate-BS-exam-1}%\lipsum*[]
Recall the univariate BS distribution whose CDF is given by (\ref{pdf-BS-1}). 
We write $\boldsymbol{T}\sim {\cal{BS}}(\boldsymbol{\theta})$ to denote random vector $\boldsymbol{T}=\bigl(T_{1},T_{2}\bigr)^{\top}$ follows a bivariate BS distribution with PDF given by \citep{kundu2010bivariate}:
 \begin{align}\label{Gaussian-quadrature-multivariate-BS-1}
f\bigl(t_{1},t_{2} \big \vert\boldsymbol{\theta}\bigr)=\boldsymbol{\phi}_{2}\bigl(\boldsymbol{z}\big \vert \boldsymbol{0}, \Sigma\bigr)
 \prod_{i=1}^{2}\frac{1}{2\alpha_{i}\beta_{i}}
\Big[\Bigl(\frac{\beta_{i}}{t_{i}}\Bigr)^{\frac{1}{2}}+
\Bigl(\frac{\beta_{i}}{t_{i}}\Bigr)^{\frac{3}{2}} \Bigr],
\end{align}
where $\boldsymbol{\theta}=\bigl(\alpha_{1},\beta_{1},\alpha_{2},\beta_{2},\rho\bigr)^{\top}$ is the family parameter vector, $\Sigma=[(1,\rho)^{\top},(\rho,1)^{\top}]$, and $\boldsymbol{z}=\bigl(z_{1},z_{2}\bigr)^{\top}$ with
 \begin{align}\label{Gaussian-quadrature-multivariate-BS-2}
z_{i}=\frac{1}{\alpha_{i}}
\bigg[\sqrt{\frac{t_{i}}{\beta_{i}}}-\sqrt{\frac{\beta_{i}}{t_{i}}} \biggr],~i=1,2.
\end{align}
We further note that
 \begin{align}\label{Gaussian-quadrature-multivariate-BS-3}
P\bigl(T_{1}\leq t_{1},T_{2}\leq t_{2} \vert \boldsymbol{\theta}\bigr)=\boldsymbol{\Phi}_{2}\bigl(\boldsymbol{z}\big \vert \boldsymbol{0}, \Sigma\bigr).
\end{align}
where $z_{i}$ is given by (\ref{Gaussian-quadrature-multivariate-BS-2}). 
%By definition 
% \begin{align}\label{Gaussian-quadrature-multivariate-BS-4}
%E\bigl(T^{m}_{1}T^{n}_{2}\bigr)=\int_{0}^{\infty}\int_{0}^{\infty}
%t^{m}_{1}t^{n}_{2}f\bigl(t_{1},t_{2} \big \vert\boldsymbol{\theta}\bigr)dt_{1}
%dt_{2}.
%\end{align}
Let $\xi_{i}=\xi_{i}\bigl(\alpha_{i},\rho\bigr)=2(1-\rho^2)\alpha_{i}^{2}$ for $i=1,2$. Then more algebra shows
  \begin{align}\label{Gaussian-quadrature-multivariate-BS-5}
E\bigl(T^{m}_{1}T^{n}_{2}\bigr)=&\int_{0}^{\infty}\int_{0}^{\infty}
t^{m}_{1}t^{n}_{2} 
f\bigl(t_{1},t_{2} \big \vert\boldsymbol{\theta}\bigr)dt_{1}
dt_{2}\nonumber\\
%\frac{\beta^{m}_{1}\xi_{1}^{m+1}\beta^{n}_{2} \xi_{2}^{n+1}}{8\alpha_{1}\alpha_{2}\pi (1-\rho^2)}\int_{0}^{\infty}\int_{0}^{\infty}\exp\bigl\{-t_{1}-t_{2}\bigr\}t_{1}^{m} t_{2}^{n}
%\nonumber\\
%&\times\exp\biggl\{\frac{\rho}{(1-\rho^2)\alpha_{1}\alpha_{2}}\biggl[\sqrt{\xi_{1}\xi_{2}t_{1}t_{2}}-\frac{\alpha_{1}}{\alpha_{2}}\sqrt{\frac{t_{1}}{t_{2}}}-\frac{\alpha_{2}}{\alpha_{1}}\sqrt{\frac{t_{2}}{t_{1}}}+\frac{1}{\sqrt{\xi_{1}\xi_{2}t_{1}t_{2}}}\biggr]\biggr\}\nonumber\\
%&
\approx & \int_{L_{1}}^{U_{1}}\int_{L_{2}}^{U_{2}}
t^{m}_{1}t^{n}_{2} \boldsymbol{\phi}_{2}\bigl(\boldsymbol{z}\big \vert \boldsymbol{0}, \Sigma\bigr)
 \prod_{i=1}^{2}\frac{1}{2\alpha_{i}\beta_{i}}
\Big[\Bigl(\frac{\beta_{i}}{t_{i}}\Bigr)^{\frac{1}{2}}+
\Bigl(\frac{\beta_{i}}{t_{i}}\Bigr)^{\frac{3}{2}} \Bigr]dt_{i},
%\times \prod_{i=1}^{2}
%\exp\Bigl\{\frac{2}{\xi_{i}}-\frac{1}{\xi^{2}_{i}t_{i}}\Bigr\}\frac{1}{\sqrt{\xi_{i}t_{i}}}\Bigl[1+\frac{1}{\xi_{i}t_{i}}\Bigr]dt_{i}.
\end{align}
where the quantities $L_{i}$ and $U_{i}$ are chosen in such a way as to constitute on a region beyond which integrand in (\ref{Gaussian-quadrature-multivariate-BS-5}) is sufficiently small. To this end, we set
\begin{align}
L_{i}=&\bigg\{t_{i} \in \mathbb{R}^{+} \bigg \vert \frac{1}{\alpha_{i}}
\bigg[\sqrt{\frac{t_{i}}{\beta_{i}}}-\sqrt{\frac{\beta_{i}}{t_{i}}} \biggr]=-\eta\bigg\},\label{Gaussian-quadrature-multivariate-BS-6}\\
U_{i}=&\bigg\{t_{i} \in \mathbb{R}^{+} \bigg \vert \frac{1}{\alpha_{i}}
\bigg[\sqrt{\frac{t_{i}}{\beta_{i}}}-\sqrt{\frac{\beta_{i}}{t_{i}}} \biggr]=\eta\bigg\},\label{Gaussian-quadrature-multivariate-BS-7}
\end{align}
where $\eta$ is a positive constant. Since $z_{i}$ is an increasing function of $t_{i}$, then by solving non-linear equations in (\ref{Gaussian-quadrature-multivariate-BS-6})-(\ref{Gaussian-quadrature-multivariate-BS-7}), we obtain
\begin{align}
L_{i}=&- \eta\alpha_{i}\beta_{i} \Bigl[ \frac{\eta}{2}\alpha_{i} + \frac{1}{2} \sqrt{\eta^2 \alpha_{i}^2+4}\Bigr]+\beta_{i}\bigl(\eta^2 \alpha_{i}^2+1\bigr),\\
U_{i}=&- \eta\alpha_{i}\beta_{i} \Bigl[ \frac{\eta}{2}\alpha_{i} - \frac{1}{2} \sqrt{\eta^2 \alpha_{i}^2+4}\Bigr]+\beta_{i}\bigl(\eta^2 \alpha_{i}^2+1\bigr).
\end{align}
Herein, we set $\eta=6$ that guarantees the joint Gaussian PDF, and hence integrand in the RHS of (\ref{Gaussian-quadrature-multivariate-BS-5}), is sufficiently small. Using an appropriate transformation of the form $y_{i}=2(t_{i}-L_{i})/(U_{i}-L_{i})-1$, for $i=1,2$, then the RHS of (\ref{Gaussian-quadrature-multivariate-BS-5}) becomes
  \begin{align}\label{Gaussian-quadrature-multivariate-BS-8}
E\bigl(T^{m}_{1}T^{n}_{2}\bigr)\approx&\prod_{i=1}^{2}\frac{U_{i}-L_{i}}{4\alpha_{i}\beta_{i}}\times\int_{-1}^{1}\int_{-1}^{1}
\eta^{m}_{1}\eta^{n}_{2} \boldsymbol{\phi}_{2}\bigl(\boldsymbol{y}\big \vert \boldsymbol{0}, \Sigma\bigr)
\Big[\Bigl(\frac{\beta_{i}}{y_{i}}\Bigr)^{\frac{1}{2}}+
\Bigl(\frac{\beta_{i}}{y_{i}}\Bigr)^{\frac{3}{2}} \Bigr]dy_{1}dy_{2},
\end{align}
where $\eta_{i}=(y_{i}+1)(U_{i}-L_{i})/2+L_{i}$ for $i=1,2$. The RHS of (\ref{Gaussian-quadrature-multivariate-BS-8}) can be approximated through a $k$-point Gauss- Legendre rule. Therefore
\begin{align}\label{Gaussian-quadrature-multivariate-BS-9}
E\bigl(T^{m}_{1}T^{n}_{2}\bigr)\approx&\prod_{r=1}^{2}\frac{U_{r}-L_{r}}{4\alpha_{r}\beta_{r}}\times \sum_{i=1}^{k}\sum_{j=1}^{k} \omega_{1i}\omega_{2j} \Bigl[\frac{1}{2}(y_{1i}+1)(U_{1}-L_{1})+L_{1}\Bigr]^m\nonumber\\
&\times \Bigl[\frac{1}{2}(y_{2j}+1)(U_{2}-L_{2})+L_{2}\Bigr]^n
\boldsymbol{\phi}_{2}\Bigl((y_{1i},y_{2j})^{\top}\Big \vert \boldsymbol{0}, \Sigma\Bigr)\nonumber\\
&\times \Big[\Bigl(\frac{\beta_{1}}{y_{1i}}\Bigr)^{\frac{1}{2}}+\Bigl(\frac{\beta_{1}}{y_{1i}}\Bigr)^{\frac{3}{2}} \Bigr] \Big[\Bigl(\frac{\beta_{2}}{y_{2j}}\Bigr)^{\frac{1}{2}}+
\Bigl(\frac{\beta_{2}}{y_{2j}}\Bigr)^{\frac{3}{2}} \Bigr],
\end{align}
where $\omega_{rs}$ and $y_{rs}$ are, accordingly, the $s$th (for $s=1,\cdots,k$) weight and node for the $r$th coordinate (for $r=1,2$) computed based on a $k$-point Gauss-Legendre rule. A simulation study have been carried out to check the performance of approximation (\ref{Gaussian-quadrature-multivariate-BS-9}). The results are shown in Table \ref{table-Gaussian-quadrature-multivariate-BS-1}. We notice that when $\rho \neq 0$, then the Gauss-Legendre rule works well only when $k$ is large.
\vspace{5mm}
\begin{table}[h]
\begin{threeparttable}
\caption{Results for approximating $E\bigl(T^{2}_{1}T^{2}_{2}\bigr)$ using $k$-point Gauss-Legendre rule.}
\label{table-Gaussian-quadrature-multivariate-BS-1}
\begin{tabular}{lllccc} 
\cline{1-6} 
$\rho=0.99$               &                          &                        &\multicolumn{3}{c}{Approximation}                       \\ \cline{1-6}
                            &                          &Exact value\tnote{1}    & $k=30$         &$k=60$         & $k=120$       \\ \cline{1-6}
$\alpha_1=0.2$              & $\alpha_2=0.2$           &1.36981507              &     2.19135117 & 1.39966739    &1.36981553    \\
$\alpha_1=2.0$              & $\alpha_2=0.2$           &74.3634327              &    74.53853374 & 74.35577417   &74.36313027    \\
$\alpha_1=2.0$              & $\alpha_2=2.0$           &17422.2127              & 17724.84078171 & 17421.54059427&17421.52544734 \\
\cline{1-6}
$\rho=0.50$                &                          &                        &\multicolumn{3}{c}{}                       \\ \cline{1-6}
                            &                          &    &          &         &      \\ \cline{1-6}
$\alpha_1=0.2$              & $\alpha_2=0.2$           &1.26793909              &    1.26793903  & 1.26793903    &1.26793903   \\
$\alpha_1=2.0$              & $\alpha_2=0.2$           &52.2184563              &   52.21790783  & 52.21827096   &52.21835271   \\
$\alpha_1=2.0$              & $\alpha_2=2.0$           &5412.76146              & 5412.62822259  & 5412.65315429 &5412.65234609 \\
\cline{1-6}
$\rho=0.00$                 &                          &                        &\multicolumn{3}{c}{}                          \\ \cline{1-6}
                            &                          &   &          &         &  \\ \cline{1-6}
$\alpha_1=0.2$              & $\alpha_2=0.2$           &1.17158976              &    1.17158974  & 1.17158974    &1.17158974    \\
$\alpha_1=2.0$              & $\alpha_2=0.2$           &35.7191999              &   35.71896967  & 35.71904935   &35.71917926    \\
$\alpha_1=2.0$              & $\alpha_2=2.0$           &1088.99999              & 1088.98597622  & 1088.99083451 &1088.99875603  \\
\cline{1-6}                
\end{tabular}
\begin{tablenotes}
\item[1] The exact value has been computed using \verb+Maple+ software.
\end{tablenotes}
\end{threeparttable}
\end{table}
\end{example}
\begin{example}\label{Gaussian-quadrature-multivariate-double-1}%\lipsum*[]
Suppose we are interested in computing double integral
\begin{align}\label{Gaussian-quadrature-multivariate-double-2}
I\bigl(\boldsymbol{\theta}\bigr)=\int_{a}^{b}\int_{c(x_{1})}^{d(x_{1})}
f\bigl(x_{1},x_{2} \big \vert\boldsymbol{\theta}\bigr)dx_{2}dx_{1}.
\end{align}
For approximating $I\bigl(\boldsymbol{\theta}\bigr)$, we may proceed based on a  Gauss-Legendre rule as follows. First we apply a $k_1$-point Gauss-Legendre rule for inner integral as 
\begin{align}\label{Gaussian-quadrature-multivariate-double-3}
\int_{c(x_{1})}^{d(x_{1})}f\bigl(x_{1},x_{2} \big \vert\boldsymbol{\theta}\bigr)dx_{2}=&\int_{-1}^{1}f\Bigl[x_{1}, \frac{\bigl(x_{2}+1\bigr)\bigl(d(x_{1})-c(x_{1})\bigr)}{2}+c(x_{1}) \Big \vert\boldsymbol{\theta}\Bigr]dx_{2}\nonumber\\
\approx&\frac{d(x_{1})-c(x_{1})}{2}\sum_{j=1}^{k_2}\omega_{2j}f\Bigl[x_{1}, \frac{\bigl(x_{2j}+1\bigr)\bigl(d(x_{1})-c(x_{1})\bigr)}{2}+c(x_{1})\Big \vert\boldsymbol{\theta}\Bigr],
\end{align}
where $\omega_{2j}$ and $x_{2j}$, for $j=1,\cdots,k_{2}$ are, accordingly, the weights and nodes constructed based on a $k_2$-point Gauss-Legendre rule. Then, we complete computing $I\bigl(\boldsymbol{\theta}\bigr)$ by approximating the outer integral as follows.
\begin{align}\label{Gaussian-quadrature-multivariate-double-4}
I\bigl(\boldsymbol{\theta}\bigr)\approx&\frac{b-a}{4}\sum_{i=1}^{k_1}\sum_{j=1}^{k_2}\omega_{1i}\omega_{2j}\bigl[d(x_{1i})-c(x_{1i})\bigr]\nonumber\\
&\times f\Bigl[\frac{\bigl(x_{1i}+1\bigr)\bigl(b-a\bigr)}{2}+a, \frac{\bigl(x_{2j}+1\bigr)\bigl(d(x_{1i})-c(x_{1i})\bigr)}{2}+c(x_{1i})\Big \vert\boldsymbol{\theta}\Bigr],
\end{align}
where $\omega_{1i}$ and $x_{1i}$, for $i=1,\cdots,k_{1}$ are, accordingly, the  weights and nodes constructed based on a $k_1$-point Gauss-Legendre rule. 
\end{example}
\begin{example}\label{Gaussian-quadrature-multivariate-Cholesky-exam-1}%\lipsum*[]
Indeed, computing the probability of a bivariate Gaussian distribution is simply expressed in terms of $I\bigl(\boldsymbol{\theta}\bigr)$ given in (\ref{Gaussian-quadrature-multivariate-double-2}). Let $\boldsymbol{X}=\bigl(X_{1},X_{2}\bigr)^{\top}$ follow a Gaussian distribution with mean $\boldsymbol{\mu}=\bigl(\mu_{1},\mu_{2}\bigr)^{\top}$ and covariance matrix $\Sigma=\bigl[(\Sigma_{1,1},\Sigma_{1,2})^{\top},(\Sigma_{2,1},\Sigma_{2,2})^{\top}\bigr]$. Recall from Subsection \ref{Computing-CDF-multivariate-Gaussian} that discusses computing ${\cal{P}}=P(\boldsymbol{a}\leq \boldsymbol{X}\leq \boldsymbol{b})$ using a transformation $\boldsymbol{y}=L(\boldsymbol{x}-\boldsymbol{\mu})$ in which $L$ denotes the Cholesky decomposition of $\Sigma$. We have 
\begin{align}\label{Gaussian-quadrature-multivariate-Cholesky-1}
{\cal{P}}=\frac{1}{2\pi}
\int_{
\frac{a_{1}-\mu_{1}}{L_{1,1}}
}^{
\frac{b_{1}-\mu_{1}}{L_{1,1}}
}\int_{c(y_{1})}^{d(y_{1})}
\exp\Bigl\{-\frac{y^{2}_{1}+y^{2}_{2}}{2}\Bigr\}dy_{2}dy_{1},
\end{align}
where $\boldsymbol{a}=\bigl(a_{1},a_{2}\bigr)^{\top}$, $\boldsymbol{b}=\bigl(b_{1},b_{2}\bigr)^{\top}$, and 
\begin{align*}%\label{Gaussian-quadrature-multivariate-Cholesky-2}
c(y_{1})=\frac{a_{2}-\mu_{2}-L_{2,1}y_{1}}{L_{2,2}},\\
d(y_{1})=\frac{b_{2}-\mu_{2}-L_{2,1}y_{1}}{L_{2,2}}.
\end{align*}
Following (\ref{Gaussian-quadrature-multivariate-double-1}) for approximating the RHS of (\ref{Gaussian-quadrature-multivariate-Cholesky-1}), it turns out that
\begin{align}\label{Gaussian-quadrature-multivariate-Cholesky-3}
{\cal{P}}\approx&\frac{(b_1-a_1)(b_2-a_2)}{4L_{1,1}L_{2,2}}\sum_{i=1}^{k_1}\sum_{j=1}^{k_2}\omega_{1i}\omega_{2j}\exp\Bigl\{-\frac{z^{2}_{1i}+z^{2}_{2j}}{2}\Bigr\},\nonumber\\
=&\sum_{i=1}^{k_1}\sum_{j=1}^{k_2}\omega_{1i}\omega_{2j}g\bigl(z_{1i},z_{2i}\bigr),
\end{align}
where
\begin{align*}%\label{Gaussian-quadrature-multivariate-Cholesky-4}
z_{1i}=&\frac{\bigl(x_{1i}+1\bigr)\bigl(b_{1}-a_{1}\bigr)}{2L_{1,1}}+\frac{a_{1}-\mu_{1}}{L_{1,1}},\\
z_{2j}=&\frac{\bigl(x_{2j}+1\bigr)\bigl(b_2-a_2\bigr)}{2L_{2,2}}+\frac{a_{2}-\mu_{2}-L_{2,1}z_{1i}}{L_{2,2}}.
\end{align*}
The \verb+R+ code below approximates ${\cal{P}}$ given in (\ref{Gaussian-quadrature-multivariate-Cholesky-1}) using a $k$-point Gauss-Legendre rule using Cholesky decomposition of covariance matrix.
\vspace{5mm}
\begin{lstlisting}[style=deltaj]
R> Chol_quad <- function(a, b, Mu, Sigma, k)
+{
+ x <- matrix(0, nrow = k^2, ncol = 2)
+ ch <- t( chol(Sigma) )
+ threshold <- 5*sqrt( diag(Sigma) )
+ a0 <- a - Mu
+ b0 <- b - Mu
+ a0 <- ifelse( a0 < -threshold, -threshold, a0 )
+ b0 <- ifelse( b0 >  threshold,  threshold, b0 )
+ f1 <- function(x) ( x + 1 )*(b0[1] - a0[1])/(2*ch[1,1]) + a0[1]/ch[1,1]
+ f2 <- function(x, y) 1/(2*pi)*exp( -x^2/2 - y^2/2 )
+ out <- quad_rule(k, type = "LE", alpha = 0, beta = 0)
+ weight <- apply( expand.grid(out$weight, out$weight), 1, prod )
+ x1 <- f1( out$node )
+ x2 <- out$node 
+ x[, 1]<- rep(x1, each = k)
+ x[, 2] <- as.vector( t( sapply(1:k, function(j) (x2[j] + 1)*(b0[2] - a0[2])/
+                       (2*ch[2,2]) + (a0[2] - ch[2,1]*x1)/ch[2,2] ) ) )
+ I1 <- (b0[1] - a0[1])*(b0[2] - a0[2])/( 4*ch[1,1]*ch[2,2] )*
+        sum( weight*f2(x[, 1], x[, 2]) )
+ min( abs(I1), 1)
+}
\end{lstlisting}
\end{example}
\section{Computing the CDF of multivariate Gaussian distribution using Gaussian Quadrature}\label{Computing-CDF-multivariate-Gaussian-Gaussian-Quadrature} 
Recall form Subsection \ref{Computing-CDF-multivariate-Gaussian} that computes the CDF of Gaussian distribution using SOV technique proposed by \cite{genz1992numerical}. Herein, we are willing to compute
${\cal{P}}$ through the Gaussian quadrature based on spectral decomposition (\ref{spectral-decomposition}) of covariance matrix. \citep{forbes2014new} show that the PDF of each $p$-dimensional Gaussian distribution can be represented as the product of $p$ univariate Gaussian PDFs as
\begin{align}\label{Gaussian-quadrature-multivariate-spectral-1}
\boldsymbol{\phi}_{p}\bigl(\boldsymbol{x}\big \vert \boldsymbol{\mu}, \Sigma\bigr)=
\boldsymbol{\phi}_{p}\bigl(\boldsymbol{x}\big \vert \boldsymbol{\mu}, VDV^{\top}\bigr)=
\prod_{i=1}^{p}{\phi}\Bigl(\boldsymbol{x}\Big \vert \bigl[V^{\top}\boldsymbol{z}\big]_{i}, d_{i}\Bigr)
=
\prod_{i=1}^{p}{\phi}\Bigl(\bigl[V^{\top}\boldsymbol{x}\bigr]_{i}\Big \vert \bigl[V^{\top}\boldsymbol{\mu}\bigr]_{i}, d_{i}\Bigr),
\end{align}
where $\boldsymbol{z}=\boldsymbol{x}-\boldsymbol{\mu}$, $\bigl[V^{\top}\boldsymbol{z}\bigr]_{i}$ denotes the $i$th element of vector $V^{\top}\boldsymbol{z}$, and $d_{i}$ is the $i$th diagonal entry of the diagonal matrix $D$ whose main diagonal elements are eigen values of $\Sigma$. Let $p$-dimensional random vector $\boldsymbol{X}$ follows ${\cal{N}}(\boldsymbol{\mu}, \Sigma)$. For computing ${\cal{P}}=P(\boldsymbol{a}\leq \boldsymbol{X}\leq \boldsymbol{b})$, based on spectral decomposition, we have
\begin{align}\label{Gaussian-quadrature-multivariate-spectral-2}
{\cal{P}}=&\int_{a_{1}}^{b_{1}}\cdots \int_{a_{p}}^{b_{p}}
\prod_{i=1}^{p}{\phi}\Bigl(\bigl[V^{\top}\boldsymbol{x}\bigr]_{i}\Big \vert \bigl[V^{\top}\boldsymbol{\mu}\bigr]_{i}, d_{i}\Bigr) dx_{1}\cdots dx_{p}\nonumber\\
=&\int_{a_{1}-\mu_{1}}^{b_{1}-\mu_{1}}\cdots \int_{a_{p}-\mu_{p}}^{b_{p}-\mu_{p}}
\prod_{i=1}^{p}{\phi}\Bigl(\bigl[V^{\top}\boldsymbol{x}\bigr]_{i}\Big \vert 0, d_{i}\Bigr) dx_{1}\cdots dx_{p}\nonumber\\
=&\int_{-1}^{1}\cdots \int_{-1}^{1}
\prod_{i=1}^{p}\Bigl[\frac{b_{i}-a_{i}}{2}\Bigr]{\phi}\Bigl(\Bigl[V^{\top}\boldsymbol{y}\Bigr]_{i}\Big \vert 0, d_{i}\Bigr) dx_{1}\cdots dx_{p},
\end{align}
where $\boldsymbol{y}=(\boldsymbol{x}+1)(\boldsymbol{b}-\boldsymbol{a})/2+\boldsymbol{a}-\boldsymbol{\mu}$. Using the Gauss-Laguerre rule the quantity ${\cal{P}}$ can be approximated as
\begin{align}\label{Gaussian-quadrature-multivariate-spectral-3}
{\cal{P}}\approx&\prod_{i=1}^{p}\Bigl[\frac{b_{i}-a_{i}}{2\sqrt{2\pi d_{i}}}\Bigr]\times\sum_{i_{1}=1}^{k_{1}}\cdots\sum_{i_{p}=1}^{k_{p}}\omega_{1i_{1}}\cdots\omega_{pi_{p}}
\exp\Bigl\{-\frac{1}{2}\sum_{r=1}^{p}\frac{\bigl(\sum_{c=1}^{p}v_{r,c}\times z_{ci_{c}}\bigr)^{2}}{d_{r}} \Bigr\},
\end{align}
where $z_{ci_{c}}=\bigl(x_{ci_{c}}+1\bigr)\bigl(b_{c}-a_{c}\bigr)/2+a_{c}-\mu_{c}$ in which $x_{c i_{c}}$ is the $i_{c}$th node computed based on a $k_{c}$-point Gauss-Laguerre rule for the $c$th coordinate and furthermore $v_{r,c}$ denotes the  $(r,c)$th entry of matrix $V^{T}$. 
\begin{example}\label{Gaussian-quadrature-multivariate-spectral-exam-1}%\lipsum*[]
Herein, we want to compute ${\cal{P}}$ given in Example \ref{Gaussian-quadrature-multivariate-Cholesky-exam-1} based on spectral decomposition discussed earlier. For $p=2$ and $k_1=k_2=k$, it follows from (\ref{Gaussian-quadrature-multivariate-spectral-3}) that
 \begin{align}\label{Gaussian-quadrature-multivariate-spectral-4}
{\cal{P}}\approx&\prod_{i=1}^{2}\Bigl[\frac{b_{i}-a_{i}}{2\sqrt{2\pi d_{i}}}\Bigr]\times\sum_{i=1}^{k}\sum_{j=1}^{k}\omega_{1i}\omega_{2j}
\exp\Bigl\{-\frac{\bigl(v_{1,1} z_{1i}+v_{1,2} z_{2j}\bigr)^{2}}{2d_{1}}-\frac{\bigl(v_{2,1} z_{1i}+v_{2,2} z_{2j}\bigr)^{2}}{2d_{2}} \Bigr\},\nonumber\\
=&\sum_{i=1}^{k}\sum_{j=1}^{k}\omega_{1i}\omega_{2j}g\bigl(z_{1i},z_{2i}\bigr),
\end{align}
where $z_{ij}=\bigl(x_{ij}+1\bigr)\bigl(b_{i}-a_{i}\bigr)/2+a_{i}-\mu_{i}$ in which $x_{ij}$ (or $\omega_{ij}$) for $i=1,2$ and $j=1,\cdots,k$ is the $j$th node (or weight) computed based on a $k$-point Gauss-Laguerre rule for the $i$th coordinate. Table \ref{table-Gaussian-quadrature-multivariate-spectral-1} shows the performance of the $k$-point Gauss-Legendre rule for approximating ${\cal{P}}=P(\boldsymbol{a}\leq \boldsymbol{X}\leq \boldsymbol{b})$ when $\boldsymbol{\mu}=(0,0)^{\top}$ and covariance matrix $\Sigma=\bigl[(1,\rho)^{\top},(1,\rho)^{\top}\bigr]$. It is worth to note that both of the Cholesky and spectral decompositions show the same results in cases discussed in Table \ref{table-Gaussian-quadrature-multivariate-spectral-1}.
\vspace{5mm}
\par
\begin{table}[h]
\begin{threeparttable}
\caption{ARE for approximating ${\cal{P}}=P(\boldsymbol{a}\leq \boldsymbol{X}\leq \boldsymbol{b})$ using $k$-point Gauss-Legendre rule based on Cholesky and spectral decompositions.}
\label{table-Gaussian-quadrature-multivariate-spectral-1}
\begin{tabular}{lllccc} 
\cline{1-6} 
$\rho=0.99$                                              &              &\multicolumn{4}{c}{ARE}                   \\ \cline{1-6}
                            &                            &Exact value\tnote{1} & $k=10$     &$k=20$       & $k=40$  \\ \cline{1-6}
$a=(-\infty,-\infty)^{\top}$& $b=(2,2)^{\top}$           &0.97421138              &1.245136942  &0.20878929   &5.673800e-04\\
$a=(2,2)^{\top}$            & $b=(\infty, \infty)^{\top}$&0.01971164              &0.003664231  &0.00198487   &1.984895e-03 \\
$a=(-2,-2)^{\top}$          & $b=(2,2)^{\top}$           &0.94842275              &0.520591497  &0.01685183   &7.154119e-08 \\
\cline{1-6}
$\rho=0.50$                                              &              &\multicolumn{4}{c}{}                   \\ \cline{1-6}
                            &                            &       & $k=10$     &$k=20$       & $k=40$  \\ \cline{1-6}
$a=(-\infty,-\infty)^{\top}$& $b=(2,2)^{\top}$           &0.95855268              &6.519226e-05 &6.557075e-05 &6.557075e-05\\
$a=(2,2)^{\top}$            & $b=(\infty, \infty)^{\top}$&0.00405294              &8.493075e-03 &8.493075e-03 &8.493075e-03 \\
$a=(-2,-2)^{\top}$          & $b=(2,2)^{\top}$           &0.91711185              &2.453014e-09 &2.856401e-09 &2.856405e-09 \\
\cline{1-6}
$\rho=0.00$                                              &              &\multicolumn{4}{c}{}                   \\ \cline{1-6}
                            &                            &             & $k=10$     &$k=20$       & $k=40$  \\ \cline{1-6}
$a=(-\infty,-\infty)^{\top}$& $b=(2,2)^{\top}$           &0.95501730           &6.438875e-05 &6.481121e-05 &6.481121e-05\\
$a=(2,2)^{\top}$            & $b=(\infty, \infty)^{\top}$&0.00051756           &2.765946e-03 &2.765946e-03 &2.765946e-03 \\
$a=(-2,-2)^{\top}$          & $b=(2,2)^{\top}$           &0.91106974           &6.285136e-09 &6.829245e-09 &6.829248e-09 \\
\cline{1-6}                
\end{tabular}
\begin{tablenotes}
\item[1] The exact value has been computed using \verb+Maple+ software up to eight decimal places.
\end{tablenotes}
\end{threeparttable}
\end{table}
\vspace{8mm}
Figure \ref{fig-Gaussian-quadrature-multivariate-3} displays the address of nodes and product of weights using the Gauss-Legendre rule with $k_1=k_2=5$ points for approximating ${\cal{P}}$ when $\boldsymbol{\mu}=(0,0)^{\top}$ and $\Sigma=\bigl[(1,\rho)^{\top},(1,\rho)^{\top}\bigr]$. More investigations reveal that a $k$-point Gauss-Legendre rule based on spectral decomposition shows superior performance than the Cholesky decomposition since in the latter case the integration region is rectangular. In general, for approximating a $p$-dimensional integral one needs to compute $k^p$ nodes and weights that computationally is expensive when $p$ becomes large. In such case, the Gaussian quadrature rules are not applicable. Alternatively, we suggest the use of \verb+R+ package \verb+TruncatedNormal+ that has been developed by \citep{botev2017normal} and is available at CRAN. Our proposed function \verb+spec_quad+ works well for small $p$ ($p\leq 5$, say).
\begin{figure}[!h]
\center
\includegraphics[width=55mm,height=55mm]{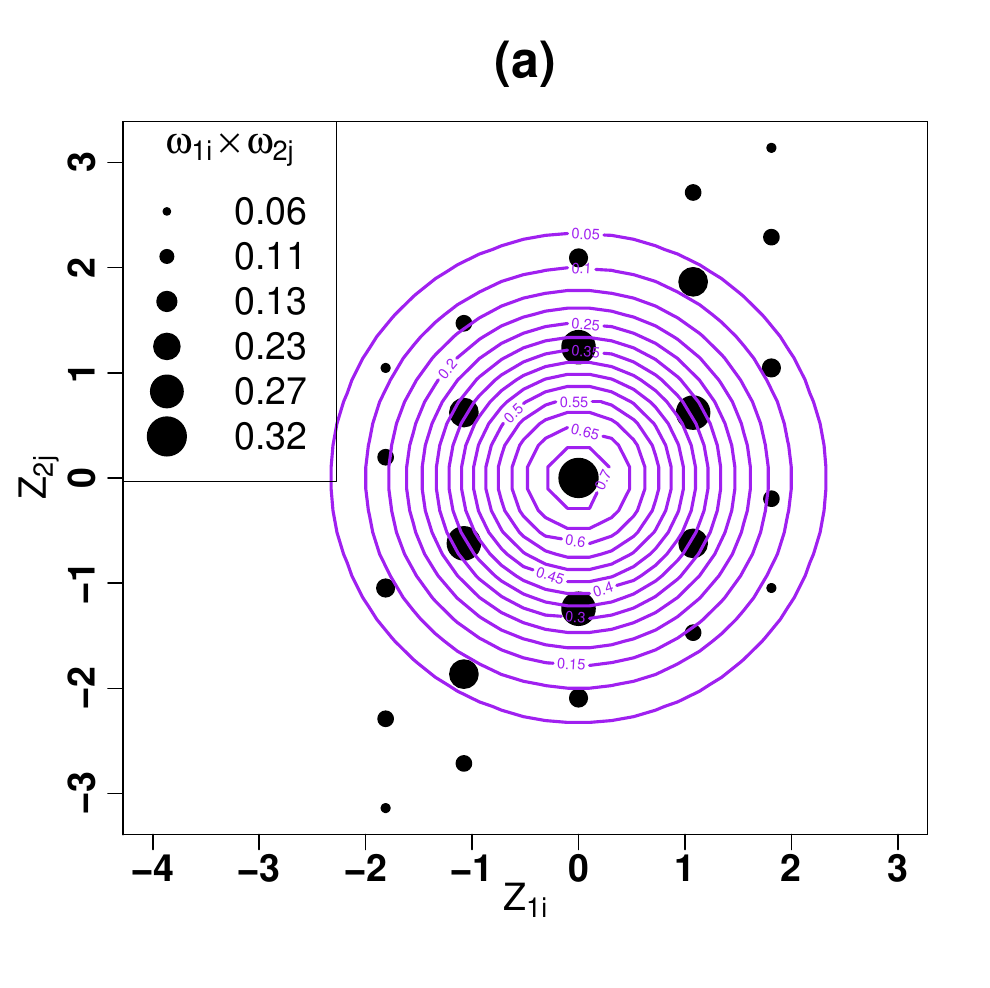}
\includegraphics[width=55mm,height=55mm]{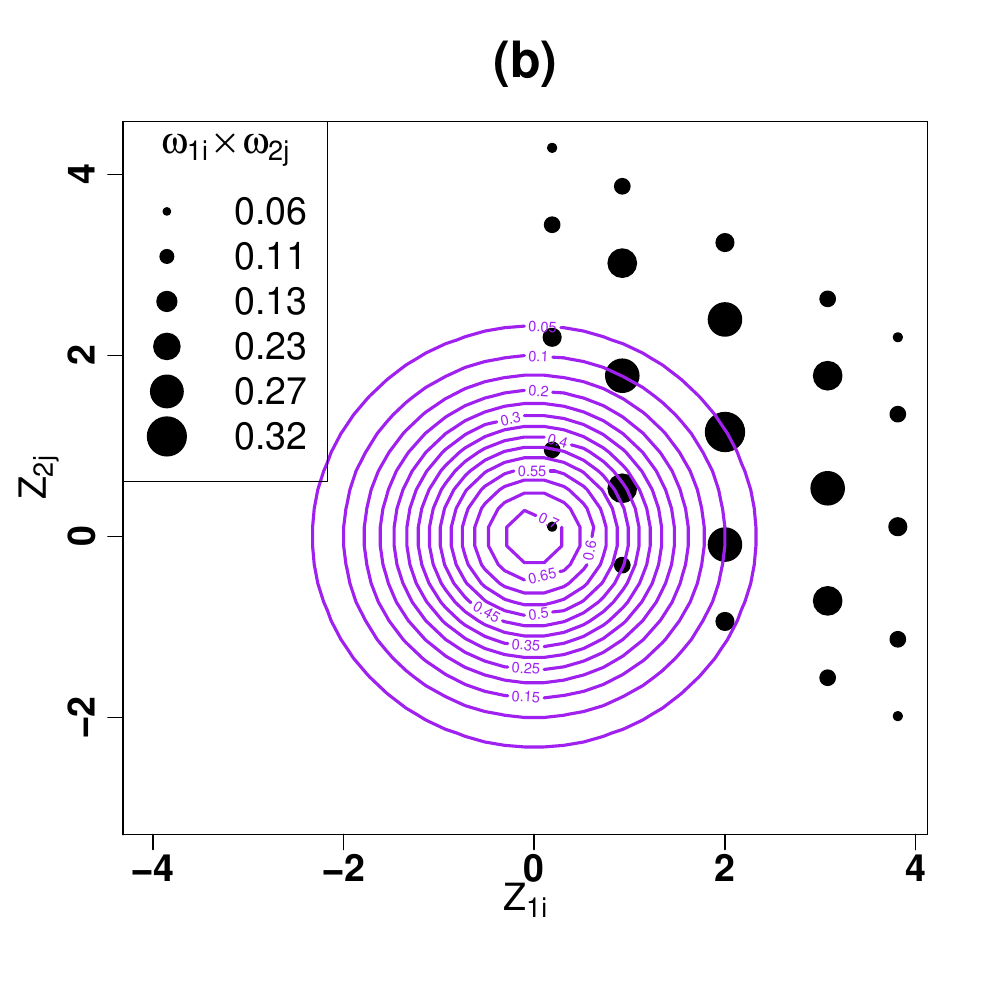}
\includegraphics[width=55mm,height=55mm]{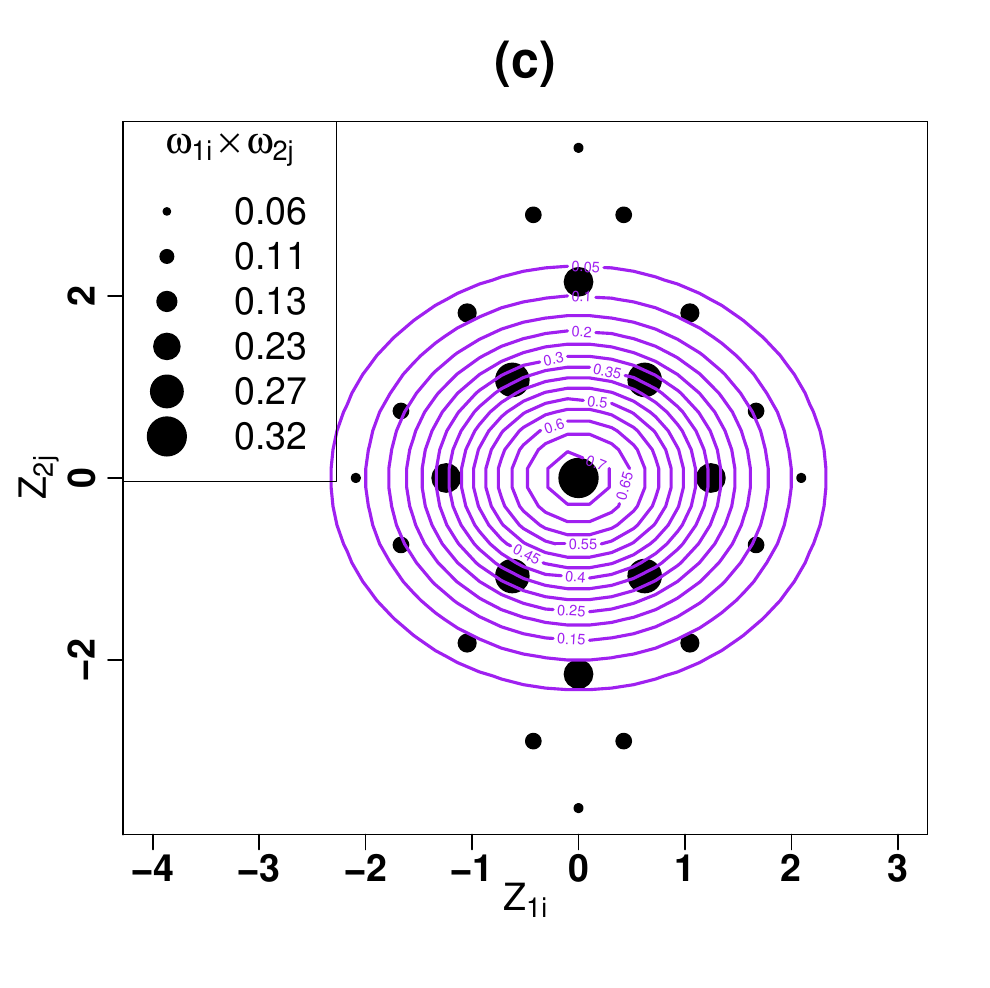}
\includegraphics[width=55mm,height=55mm]{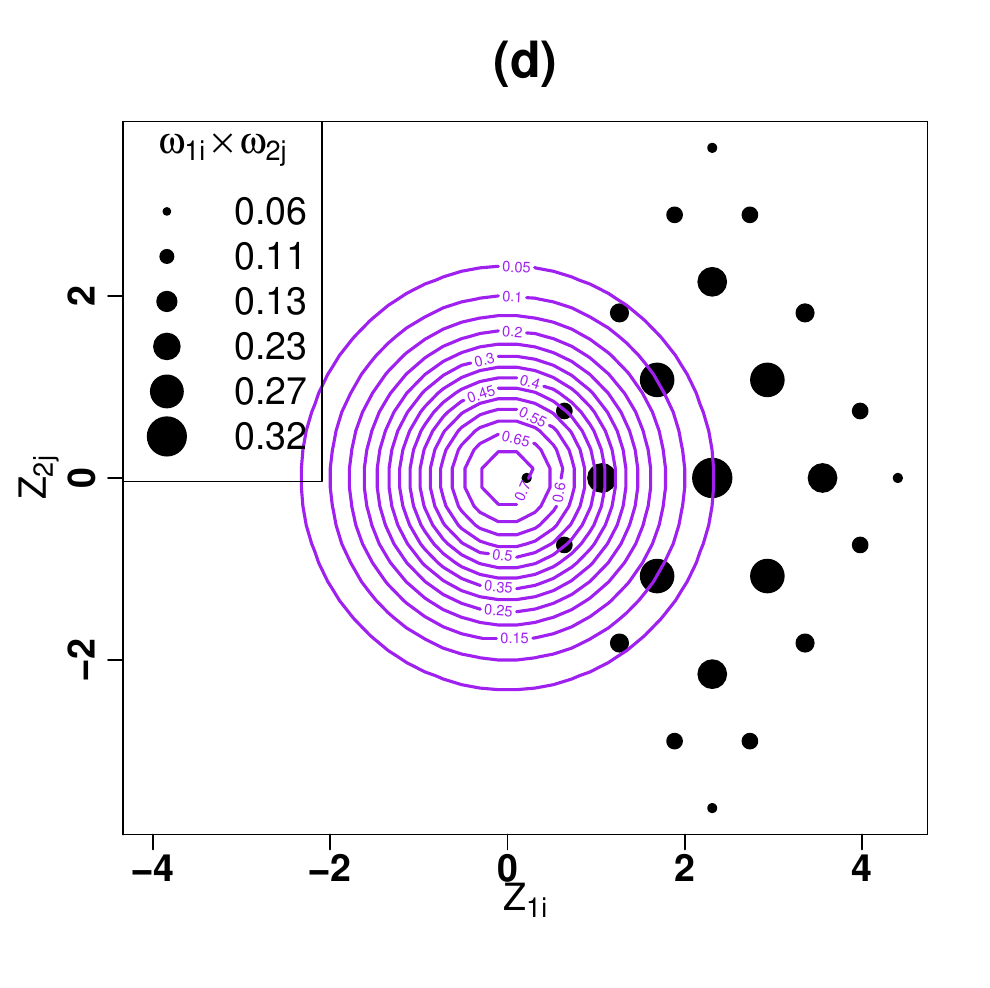}
\caption{First row: contour plot of $g\bigl(z_{1i}, z_{2j}\bigr)$ given in the RHS of (\ref{Gaussian-quadrature-multivariate-Cholesky-3}). The filled black points show the address of nodes based on the Cholesky decomposition for (a): $\boldsymbol{a}=(-2,-2)^{\top}$, $\boldsymbol{b}=(2,2)^{\top}$, and $\rho=-0.50$ and (b): $\boldsymbol{a}=(0,0)^{\top}$, $\boldsymbol{b}=(\infty,\infty)^{\top}$, and $\rho=0.50$. Second row: contour plot of $g\bigl(z_{1i}, z_{2j}\bigr)$ given in the RHS of (\ref{Gaussian-quadrature-multivariate-spectral-4}). The filled black points show the address of nodes based on the spectral decomposition for (c): $\boldsymbol{a}=(-2,-2)^{\top}$, $\boldsymbol{b}=(2,2)^{\top}$, and $\rho=-0.50$ and (d): $\boldsymbol{a}=(0,0)^{\top}$, $\boldsymbol{b}=(\infty,\infty)^{\top}$, and $\rho=0.50$.}
\label{fig-Gaussian-quadrature-multivariate-3}
\end{figure}
\vspace{5mm}
\begin{lstlisting}[style=deltaj]
R> spec_quad <- function(a, b, Mu, Sigma, k)
+ {
+  p <- length(Mu)
+  n <- k^p
+  threshold <- 5*sqrt( diag(Sigma) )
+  a0 <- a - Mu
+  b0 <- b - Mu
+  a0 <- ifelse( a0 < -threshold, -threshold, a0 )
+  b0 <- ifelse( b0 >    threshold,  threshold, b0 )
+  e <- eigen(Sigma)
+  ev <- e$vectors
+  out <- quad_rule(k, type = "LE", alpha = 0, beta = 0)
+  weight <- out$weight
+  node <- out$node
+  weight_matrix <- as.data.frame( matrix(rep(weight, p), nrow = k, ncol = p) )
+  node_matrix   <- as.data.frame( matrix(rep(node  , p), nrow = k, ncol = p) )
+  weight_grid <- as.matrix( as.vector( do.call( expand.grid, weight_matrix ) ),
+                         														nrow = n, ncol = p)
+  omega <- sapply( 1:n, function(i) prod( weight_grid[i, ] ) )
+  node_grid <- as.matrix( as.vector( do.call( expand.grid, node_matrix ) ),
+                         														nrow = n, ncol = p)
+  zij <- node_grid %*% diag((b0 - a0)/2) + matrix( (b0 + a0)/2 , n, p, byrow = T )
+  p1 <- apply( ( t( zij%*%sweep(ev, 2, sqrt(e$values), "/") ) )^2, 2, sum) 
+  I1 <- 1/sqrt( (2*pi)^p*prod(e$values) )*prod( (b0 - a0)/2 )*sum( omega*exp( -p1/2 ) )
+  min( abs(I1), 1)
+ }
\end{lstlisting}
\end{example}
%%%%%%%%%%%%%%%%%%%%%%%%
\section{Moments of truncated distributions}\label{trunc-mom}
Truncated distribution is widely used in different fields of statistics such as, survival analysis, regression analysis, and model-based clustering. We write ${\cal{X}}$ to denote the  truncated version of random variable $X$ on $(a,b)$. The random variable may take any value on real line, but ${\cal{X}}$ just takes values on $(a,b)$. The latter is known as the doubly truncated of $X$. There are two other types of truncated random variable including the left truncated random variables. If ${\cal{X}}$ is defined on $(b, \infty)$, then we may say that ${\cal{X}}$ is the left truncated version of $X$ at $b$. Likewise, if ${\cal{X}}$ is defined on $(-\infty, a)$, then we may say that ${\cal{X}}$ is the right truncated version of $X$ at $a$. The PDF of random variable ${\cal{X}}$ truncated on $(a,b)$ is given by
\begin{align}\label{truncated-moment-part1}
%f_{{\cal{X}}_{a,b}}(x)=
\frac{f(x)}{\int_{a}^{b}f(u)du},
\end{align} 
where $-\infty<a<x<b<\infty$. Each doubly random variable ${\cal{X}}$ is in fact a conditional version of random variable $X$. In other words
\begin{align}
{{\cal{X}}}_{a,b}=X \big \vert X \in (a,b).
\end{align} 
Let $F(\cdot)$ denotes the CDF of random variable $X$. Elementary calculations show that the CDF of ${{\cal{X}}}$ can be represented in terms of $F(\cdot)$ as follows.
\begin{align}
%F_{{\cal{X}}_{a,b}}(x)=
\frac{F(x)-F(a)}{F(b)-F(a)},~~-\infty<a<x<b<\infty.
\end{align} 
Hence, in univariate case, the $m$th moment of truncated random variable $X$ on $(a,b)$, for $m=1,2,3,\cdots$ is
\begin{align}\label{truncated-moment-part2}
\Omega^m= E\bigl(X^m\big \vert a<X<b\bigr)=\frac{\int_{a}^{b} u^{m} f(u)du}{F(b)-F(a)}, ~-\infty<a<x<b<\infty.
\end{align}
In multivariate case, the PDF of truncated random vector $\boldsymbol{X}$ on hyper-cube $(\boldsymbol{a}, \boldsymbol{b})\subset \mathbb{R}^{p}$ with lower bound $\boldsymbol{a}=(a_1,\cdots,a_p)^{\top}$ and upper bound $\boldsymbol{b}=(b_1,\cdots,b_p)^{\top}$ is given by 
\begin{align}\label{truncated-moment-part3}
\frac{f(\boldsymbol{x})}{\int_{\boldsymbol{a}}^{\boldsymbol{b}}f(\boldsymbol{u})d\boldsymbol{u}}=\frac{f(\boldsymbol{x})}{F(\boldsymbol{b})-F(\boldsymbol{a})},
\end{align} 
where $\boldsymbol{a}<\boldsymbol{x}<\boldsymbol{b}$ and $F(\cdot)$ is the CDF of $\boldsymbol{X}$. So, the first and second moments of truncated random vector $\boldsymbol{X}$, respectively, are
\begin{align*}
\boldsymbol{\Omega}=&E\bigl(\boldsymbol{X}\big \vert \boldsymbol{a}<\boldsymbol{X}<\boldsymbol{b}\bigr)=E\bigl(\boldsymbol{X}\big \vert a_1<X_1<b_1,\cdots,a_p<X_p<b_p\bigr),\nonumber\\
\boldsymbol{\Omega}^2=&E\bigl(\boldsymbol{X}\boldsymbol{X}^{\top}\big \vert \boldsymbol{a}<\boldsymbol{X}<\boldsymbol{b}\bigr)=E\bigl(\boldsymbol{X}\boldsymbol{X}^{\top}\big \vert a_1<X_1<b_1,\cdots,a_p<X_p<b_p\bigr).
\end{align*}  
%We have
%\begin{align}
%{\cal{\boldsymbol{X}}}_{\boldsymbol{a},\boldsymbol{b}}=\boldsymbol{X} \Big \vert a_1<X_1<b_1,a_2<X_2<b_2,\cdots,a_p<X_p<b_p.
%\end{align} 
Herein, the concern of study mainly is devoted to computing the first and second moments given by since these quantities have received more attention than other moments in practice.
% $\boldsymbol{\Omega}=E\bigl({\cal{X}}_{a,b}\bigr)$.
%and $E\bigl({\cal{X}}^{\top}_{a,b}{\cal{X}}_{a,b}\bigr)$. 
If $\boldsymbol{X}\sim f(\cdot)$, by definition, the first moment of $\boldsymbol{X}$ is computed as
\begin{align}\label{truncated-moment-part4}
\boldsymbol{\Omega}=&\frac{1}{F(\boldsymbol{b})-F(\boldsymbol{a})} \int_{\boldsymbol{a}}^{\boldsymbol{b}}\boldsymbol{u}f(\boldsymbol{u})d\boldsymbol{u}.
%E\bigl({\cal{X}}_{a,b}\bigr)=&\frac{1}{P(\boldsymbol{a}<\boldsymbol{X}<\boldsymbol{b})} \int_{\boldsymbol{a}}^{\boldsymbol{b}}  {\boldsymbol{u}}^{\top}{\boldsymbol{u}}   f(\boldsymbol{u})d\boldsymbol{u}.\nonumber
\end{align}
The Monte Carlo approximation of $\boldsymbol{\Omega}$ can be obtained by considering the instrumental distribution with PDFs $f(\cdot)/\bigl[F(a)-F(b)\bigr]$. Here, we can use the rejection approach for simulating from truncated random variable as follows. Suppose, $x_1,\cdots,x_N$, for sufficiently large $N$, are realizations from instrumental with PDF $f_1$, the Monte Carlo approximation of $\boldsymbol{\Omega}$ is obtained as
\begin{align}
\hat{\boldsymbol{\Omega}}=\frac{1}{N}\sum_{i=1}^{N}f_1\bigl(x_i\bigr).
\end{align}
The pseudo code for sampling from $f(\cdot)/\bigl[F(a)-F(b)\bigr]$ is given as follows.
\begin{enumerate}
\item Read ${a}, {b}$, and $N$;
\item Set $j=0$ and ${y}=0$;
\item Simulate ${x}\sim f(\cdot)$
\item If $a<x<b$, then accept ${x}$ as a generation from $f(\cdot)/\bigl[F(a)-F(b)\bigr]$, set ${y}={y}+x$ and $j = j+1$;
\item If $j=N$, then go to the next step, otherwise return to step 2;
\item Approximate $\Omega$ as $\hat{\Omega}=y/N$;
\item End.
\end{enumerate}
The focus of this chapter is placed on compute the the first and second moments of truncated Gaussian distribution in univariate and multivariate cases.
Herein, we proceed to compute the exact values of first and second moments of the truncated univariate and multivariate Gaussian distributions. 
\section{Moments of truncated Gaussian distribution in univariate case}\label{first-moment-truncated-univariate-Gaussian-distribution}
For simplicity, we use generic symbols $\phi\bigl({x}\big \vert {\mu}, \sigma^2\bigr)$ and ${\Phi}\bigl({x}\big \vert {\mu}, \sigma^2\bigr)$ to show, accordingly, the PDF and CDF of a Gaussian distribution at point ${x}$ for which ${\mu}$ and $\sigma$ are the location and scale parameters, respectively. We further write ${X} \sim {\cal{N}}({\mu}, \sigma^2)$ to show that random variable ${X}$ follows a Gaussian distribution with location parameter ${\mu}$ and variance  $\sigma^2$. The corresponding truncated family on $(a,b)$ is represented by $t{\cal{N}}\bigl({\mu}, \sigma^2, a, b\bigr)$. 
\subsection{First two moments of truncated Gaussian distribution}
Let $X \sim {\cal{N}}\bigl(\mu, \sigma^2\bigr)$. By definition, from (\ref{truncated-moment-part2}) for $m=1$, it follows that
\begin{align}\label{first-moment-truncated-uni variate-Gaussian-distribution-1}
\Omega=E\bigl(X \big \vert a<X<b\bigr)=\frac{1}{\Phi(b\vert \mu, \sigma^2)-\Phi(a\vert \mu, \sigma^2)}\int_{a}^{b}x\times \phi(x\vert \mu, \sigma^2)dx.
\end{align} 
It is easy to check that
\begin{align}\label{first-moment-truncated-univariate-Gaussian-distribution-2}
\frac{\partial }{\partial \mu}\phi(x\vert \mu, \sigma^2)=&\Bigl(\frac{x-\mu}{\sigma^2}\Bigr)\phi(x\vert \mu, \sigma^2).
\end{align} 
Integrating both sides of (\ref{first-moment-truncated-univariate-Gaussian-distribution-2}) and dividing by $P_{G}=\Phi(b\vert \mu, \sigma^2)-\Phi(a\vert \mu, \sigma^2)$ yields
\begin{align*}
\frac{1}{P_{G}}\int_{a}^{b} \frac{\partial }{\partial \mu} \phi(x\vert \mu, \sigma^2)dx=\frac{\Omega}{\sigma^2}-\frac{\mu}{\sigma^2}.
\end{align*} 
It follows that
\begin{align*}
\Omega&=\mu +\frac{\sigma^2}{P_{G}} \frac{\partial }{\partial \mu}\int_{a-\mu}^{b-\mu}\phi(u\vert 0, \sigma^2)du\nonumber\\
  &=\mu +\frac{\sigma^2}{P_{G}} \Bigl[ \phi(a\vert \mu, \sigma^2)-\phi(b\vert \mu, \sigma^2)\Bigr].
\end{align*} 
Hence,
\begin{align}\label{first-moment-truncated-univariate-Gaussian-distribution-3}
\Omega=E\bigl(X \big \vert a<X<b\bigr)=\mu +\sigma^2\frac{\phi(a\vert \mu, \sigma^2)-\phi(b\vert \mu, \sigma^2)}{\Phi(b\vert \mu, \sigma^2)-\Phi(a\vert \mu, \sigma^2)}.
\end{align} 
By definition, from (\ref{truncated-moment-part2}) for $m=2$, it follows that
\begin{align*}%\label{first-moment-truncated-univariate-Gaussian-distribution-1}
\Omega^2=E\bigl(X^2 \big \vert a<X<b\bigr)=\frac{1}{P_{G}}\int_{a}^{b}x^2\phi(x\vert \mu, \sigma^2)dx,
\end{align*} 
Taking the second order derivative from (\ref{first-moment-truncated-univariate-Gaussian-distribution-2}) with respect to $\mu$, we have 
\begin{align}\label{first-moment-truncated-univariate-Gaussian-distribution-4}
\frac{\partial^2 }{\partial \mu^2}\phi(x\vert \mu, \sigma^2)=&\Bigl(\frac{x-\mu}{\sigma^2}\Bigr)^2\phi(x\vert \mu, \sigma^2)-\frac{\phi(x\vert \mu, \sigma^2)}{\sigma^2}.
\end{align} 
Integrating both sides of (\ref{first-moment-truncated-univariate-Gaussian-distribution-4}), dividing by $P_{G}$, and then rearranging yields
\begin{align}\label{first-moment-truncated-univariate-Gaussian-distribution-5}
\frac{1}{P_{G}}\frac{\partial^2 }{\partial \mu^2} \int_{a-\mu}^{b-\mu} \phi(x\vert 0, \sigma^2)dx=
\frac{1}{P_{G}}\int_{a}^{b}\Bigl(\frac{x-\mu}{\sigma^2}\Bigr)^2\phi(x\vert \mu, \sigma^2)dx-\frac{1}{\sigma^2}.
\end{align} 
It is easy to check that the LHS of (\ref{first-moment-truncated-univariate-Gaussian-distribution-5}) becomes 
\begin{align}\label{first-moment-truncated-univariate-Gaussian-distribution-6}
\frac{1}{P_{G}}\frac{\partial^2 }{\partial \mu^2} \int_{a-\mu}^{b-\mu} \phi(x\vert 0, \sigma^2)dx=\frac{1}{P_{G}}\Bigl[\Bigl(\frac{a-\mu}{\sigma^2}\Bigr)\phi(a\vert \mu, \sigma^2)-\Bigl(\frac{b-\mu}{\sigma^2}\Bigr)\phi(b\vert \mu, \sigma^2)\Bigr],
\end{align} 
and further
\begin{align}\label{first-moment-truncated-univariate-Gaussian-distribution-7}
\frac{1}{P_{G}}\int_{a}^{b}\Bigl(\frac{x-\mu}{\sigma^2}\Bigr)^2\phi(x\vert \mu, \sigma^2)dx=&\frac{1}{\sigma^2}E\Bigl[\Bigl(\frac{X-\mu}{\sigma}\Bigr)^2\Big \vert a<X<b\Bigr]\nonumber\\
=&\frac{1}{\sigma^4}E\bigl(X^2\big \vert a<X<b\bigr)-2\frac{\mu}{\sigma^4} E\bigl(X\big \vert a<X<b\bigr)\Bigr]+\frac{\mu^2}{\sigma^4}.
\end{align} 
Hence, substituting the RHS of (\ref{first-moment-truncated-univariate-Gaussian-distribution-6}) and (\ref{first-moment-truncated-univariate-Gaussian-distribution-7}) into (\ref{first-moment-truncated-univariate-Gaussian-distribution-5}) and rearranging, we obtain
\begin{align*}
\Omega^2=\sigma^2\frac{(a-\mu)\phi(a\vert \mu, \sigma^2)-(b-\mu)\phi(b\vert \mu, \sigma^2)}{\Phi(b\vert \mu, \sigma^2)-\Phi(a\vert \mu, \sigma^2)}+\sigma^2-\mu^2+2\mu \Omega.
\end{align*} 
In what follows \verb+R+ function \verb+Ex+ is given for computing the first and second moments of truncated univariate Gaussian distribution.
\begin{lstlisting}[style=deltaj]
R> Ex <- function(mu, sigma, a, b)
+ {
+ 	pb <- pnorm(b, mu, sigma) 
+ 	pa <- pnorm(a, mu, sigma)
+ 	da <- dnorm(a, mu, sigma)
+ 	db <- dnorm(b, mu, sigma)
+ 	Omega <- mu + sigma^2*( da - db )/( pb - pa )
+ 	P1 <- pnorm(b, mu, sigma) - pnorm(a, mu, sigma) 
+ 	ex <- mu + sigma^2*( da - db )/( pb - pa )
+ 	Omega2 <- sigma^2*( (a-mu)*da - (b-mu)*db )/( pb - pa ) + sigma^2 -mu^2 + 2*mu*ex
+ 	out2 <- list("Ex" = Omega, "Exx" = Omega2)
+ 	return( out2 )
+ }
\end{lstlisting}
\section{First moment of truncated multivariate Gaussian distribution}\label{first-moment-truncated-multivariate-Gaussian-distribution}
For simplicity, we use generic symbols $\boldsymbol{\phi}_{p}\bigl(\boldsymbol{x}\big \vert \boldsymbol{\mu}, \Sigma\bigr)$ and $\boldsymbol{\Phi}_{p}\bigl(\boldsymbol{x}\big \vert \boldsymbol{\mu}, \Sigma\bigr)$ to denote, accordingly, the PDF and CDF of a $p$-dimensional Gaussian distribution at point $\boldsymbol{x}$ for which $\boldsymbol{\mu}$ and $\Sigma$ are the location and scale parameters, respectively. Also, we write $\boldsymbol{X} \sim {\cal{N}}_{p}\bigl(\boldsymbol{\mu}, \Sigma\bigr)$ to indicate that random vector $\boldsymbol{X}$ follows a Gaussian distribution with location vector $\boldsymbol{\mu}$ and scale matrix $\Sigma$. The corresponding family of Gaussian distributions truncated on on hyper-cube $(\boldsymbol{a}, \boldsymbol{b})\subset \mathbb{R}^{p}$ is shown by ${\cal{TN}}_{p}\bigl(\boldsymbol{\mu}, \Sigma, \boldsymbol{a}, \boldsymbol{b}\bigr)$. Furthermore, $\boldsymbol{\Omega}$ denotes the first moment of family ${\cal{TN}}_{p}\bigl(\boldsymbol{\mu}, \Sigma, \boldsymbol{a}, \boldsymbol{b}\bigr)$. Suppose $p$-dimensional random vector ${\boldsymbol{X}}=(X_1,\cdots,X_p)^{\top}$ follows ${\cal{N}}_{p}(\boldsymbol{\mu},\Sigma)$, we define the Mahalanobis distance as 
% One of the most important truncated random variable is $\vert X \vert$ that is a left truncated version of $X$ at point zero based on definition of truncation. The random variable $\vert X \vert$ received much attention in different fields of statistics both univariate and multivariate cases. 
\begin{align}\label{delta}
\delta(\boldsymbol{x},\Sigma)=\boldsymbol{{{x}}}^{\top}{\Sigma}^{-1}\boldsymbol{x},
\end{align} 
and
\begin{align}\label{vector-representation}
&\boldsymbol{v}_{i}[x]=\bigl({v}_{1},\cdots,{v}_{i-1},x,{v}_{i+1},\cdots,{v}_{p}\bigr),\\ 
&\boldsymbol{v}[-i]=\bigl({v}_{1},\cdots,{v}_{i-1},{v}_{i+1},\cdots,{v}_{p} \bigr),
\end{align}
in which $\boldsymbol{v}=\bigl(v_1,v_2,\cdots,v_p\bigr)^{\top}$ is a vector of real values. By definition, we can write
\begin{align}\label{first-truncated-moment-multivariate-Gaussian-part1}
\boldsymbol{\Omega}=&\frac{1}{{\cal{P}}_{G}} \int_{\boldsymbol{a}}^{\boldsymbol{b}}\boldsymbol{x}\times \boldsymbol{\phi}_{p}\bigl(\boldsymbol{x}\big \vert \boldsymbol{\mu}, \Sigma\bigr)d\boldsymbol{x}\nonumber\\
&\frac{1}{{\bf{C}}_{G}{\cal{P}}_{G}} \int_{\boldsymbol{a}}^{\boldsymbol{b}}\boldsymbol{x}\exp\Bigl\{-\frac{\delta(\boldsymbol{x}-\boldsymbol{\mu},\Sigma)}{2}\Bigr\}d\boldsymbol{x}\nonumber\\=&\frac{1}{{\bf{C}}_{G}{\cal{P}}_{G}}
                       \int_{\boldsymbol{a}}^{\boldsymbol{b}}(\boldsymbol{x}-\boldsymbol{\mu})\exp\Bigl\{-\frac{\delta(\boldsymbol{x}-\boldsymbol{\mu},\Sigma)}{2}\Bigr\}d\boldsymbol{x}\nonumber\\
                       &+\frac{\boldsymbol{\mu}}{{\bf{C}}_{G}{\cal{P}}_{G}}
\int_{\boldsymbol{a}}^{\boldsymbol{b}}\exp\Bigl\{-\frac{\delta(\boldsymbol{x}-\boldsymbol{\mu},\Sigma)}{2}\Bigr\}d\boldsymbol{x}\nonumber\\
=&  \Sigma \boldsymbol{I}+\boldsymbol{\mu},
%\Phi_{q}\Bigl(\sqrt{r}\boldsymbol{m}\Big|\boldsymbol{0},\boldsymbol{\Delta}\Bigr)
\end{align}
where ${\bf{C}}_{G}=(2\pi)^{p/2} \vert \Sigma \vert ^{1/2}$, ${\cal{P}}_{G}=\boldsymbol{\Phi}_{p}\bigl(\boldsymbol{b}\big\vert\boldsymbol{\mu}, {\Sigma}\bigr)-\boldsymbol{\Phi}_{p}\bigl(\boldsymbol{a}\big\vert\boldsymbol{\mu}, {\Sigma}\bigr)$.
Since
\begin{align*}
                    \frac{\partial}{\partial \boldsymbol{\mu}}\frac{\delta(\boldsymbol{x}-\boldsymbol{\mu},\Sigma)}{2}= - {\Sigma}^{-1}(\boldsymbol{x}-\boldsymbol{\mu}),
%\Phi_{q}\Bigl(\sqrt{r}\boldsymbol{m}\Big|\boldsymbol{0},\boldsymbol{\Delta}\Bigr)
\end{align*}
then
\begin{align*}%\label{first-truncated-moment-multivariate-Gaussian-part2}
\boldsymbol{I}=&\frac{1}{{\bf{C}}_{G}{\cal{P}}_{G}}\frac{\partial}{\partial \boldsymbol{\mu}} \int_{\boldsymbol{a}}^{\boldsymbol{b}}\exp\Bigl\{-\frac{\delta(\boldsymbol{x}-\boldsymbol{\mu},\Sigma)}{2}\Bigr\}d\boldsymbol{x}\nonumber\\
=&\frac{1}{{\bf{C}}_{G}{\cal{P}}_{G}}\frac{\partial}{\partial \boldsymbol{\mu}} \int_{\boldsymbol{a}-\boldsymbol{\mu}}^{\boldsymbol{b}-\boldsymbol{\mu}}\exp\Bigl\{-\frac{\delta(\boldsymbol{u},\Sigma)}{2}\Bigr\}d\boldsymbol{u}\nonumber\\
%=&\frac{\Sigma}{\alpha{\cal{P}}} \int_{\boldsymbol{a}_{-i}-\boldsymbol{\mu}_{-i}}^{\boldsymbol{b}_{-i}-\boldsymbol{\mu}_{-i}}\biggl[\exp\Bigl\{-\frac{d\bigl(\boldsymbol{u}_{-i(a_{i}-\mu_{i})}\bigr)}{2}\Bigr\}-\exp\Bigl\{-\frac{d\bigl(\boldsymbol{u}_{-i(b_{i}-\mu_{i})}\bigr)}{2}\Bigr\}\biggr]d\boldsymbol{u}_{-i}\nonumber\\
=&\bigl({{I}}_{1},{{I}}_{2},\cdots,{{I}}_{p}\bigr)^{\top},
\end{align*}
where 
\begin{align}\label{first-truncated-moment-multivariate-Gaussian-part3}
{{I}}_{i}=\frac{1}{{\bf{C}}_{G}{\cal{P}}_{G}}\int_{\boldsymbol{a}[-i]-\boldsymbol{\mu}[-i]}^{\boldsymbol{b}[-i]-\boldsymbol{\mu}[-i]}\biggl[\exp\Bigl\{-\frac{\delta\bigl(\boldsymbol{u}_{i}[a_{i}-\mu_{i}],\Sigma\bigr)}{2}\Bigr\}-\exp\Bigl\{-\frac{\delta\bigl(\boldsymbol{u}_{i}[b_{i}-\mu_{i}],\Sigma\bigr)}{2}\Bigr\}\biggr]d\boldsymbol{u}[-i],
\end{align}
%where, for $i=1,\cdots,p$, we have
for $i=1,\cdots,p$. We note that the quadratic form $\delta\bigl(\boldsymbol{u}_{i}[x],\Sigma\bigr)$ can be decomposed as
\begin{align}\label{quadratic-decomposition}
\delta\bigl(\boldsymbol{u}_{i}[x],\Sigma\bigr)=x^2\times\bigl(\Sigma_{i,i}\bigr)^{-1}+
\delta\Bigl(\boldsymbol{u}[-i]-\boldsymbol{\xi}(x),\Delta_{i}\Bigr),
\end{align}
where 
\begin{align}
\boldsymbol{\xi}(x)=&\Sigma_{i,-i}\times\bigl(\Sigma_{i,i}\bigr)^{-1}\times{x},\label{xi-uni-decomposition-1}\\
\Delta_{i}=&\Sigma_{-i, -i}-\bigl(\Sigma_{i,i}\bigr)^{-1} \times \bigl(\Sigma_{i,-i}\bigr)^{\top} \Sigma_{i, -i}\label{Delta-uni-decomposition-1},
\end{align}
in which  $\Sigma_{i,j}$ denotes the $(i,j)$th element of $\Sigma$, $\Sigma_{-i, -j}$ denotes the matrix $\Sigma$ when its $i$th row and $j$th column is eliminated, and $\Sigma_{i, -j}$ is the $i$th row of matrix $\Sigma$ when its $j$the column is eliminated, for $i,j=1,\cdots,p$.
Hence, using expression (\ref{quadratic-decomposition}), the integrand in the RHS of (\ref{first-truncated-moment-multivariate-Gaussian-part3}) can be rewritten as 
\begin{align*}\label{first-truncated-moment-multivariate-Gaussian-part4}
I_{i}=&\frac{1}{{\bf{C}}_{G}{\cal{P}}_{G}} \exp\Bigl\{-\frac{(a_{i}-\mu_{i})^2}{2\Sigma_{i,i}}\Bigr\}\int_{\boldsymbol{a}[-i]-\boldsymbol{\mu}[-i]}^{\boldsymbol{b}[-i]-\boldsymbol{\mu}[-i]}
\exp\Bigl\{-\frac{1}{2}\delta\Bigl(\boldsymbol{u}[-i]-\boldsymbol{\xi}(a_{i}-\mu_{i}),\Delta_{i}\Bigr) \Bigr\} d\boldsymbol{u}[-i]\nonumber\\
&-\frac{1}{{\bf{C}}_{G}{\cal{P}}_{G}} \exp\Bigl\{-\frac{(b_{i}-\mu_{i})^2}{2\Sigma_{i,i}}\Bigr\} \int_{\boldsymbol{a}[-i]-\boldsymbol{\mu}[-i]}^{\boldsymbol{b}[-i]-\boldsymbol{\mu}[-i]}
\exp\Bigl\{-\frac{1}{2}\delta\Bigl(\boldsymbol{u}[-i]-\boldsymbol{\xi}(b_{i}-\mu_{i}),\Delta_{i}\Bigr)\Bigr\} d\boldsymbol{u}[-i].
%=&\frac{\Sigma}{{\cal{P}}}\times\bigl({\cal{I}}_{1},{\cal{I}}_{2},\cdots,{\cal{I}}_{p}\bigr)^{\top},
\end{align*}
The quantity ${\bf{C}}_{G}$ can be represented as 
\begin{align}
{\bf{C}}_{G}=\sqrt{2\pi \Sigma_{i,i}} (2\pi)^{\frac{p-1}{2}}\big \vert \Delta_{i}\big  \vert^{\frac{1}{2}}.
\end{align}
It follows that
\begin{align}\label{first-truncated-moment-multivariate-Gaussian-part5}
I_{i}=&\frac{1}{{\cal{P}}_{G}} \phi\Bigl(a_{i}\Big \vert \mu_{i}, \Sigma_{i,i}\Bigr)
\biggl[\boldsymbol{\Phi}_{p-1}\Bigl(\boldsymbol{b}[-i]-\boldsymbol{\mu}[-i]\Big\vert\boldsymbol{\xi}(a_{i}-\mu_{i}),\Delta_{i}\Bigr)\nonumber\\
&-\boldsymbol{\Phi}_{p-1}\Bigl(\boldsymbol{a}[-i]-\boldsymbol{\mu}[-i]\Big\vert\boldsymbol{\xi}(a_{i}-\mu_{i}), \Delta_{i}\Bigr)\biggr]
\nonumber\\
&-\frac{1}{{\cal{P}}_{G}} \phi\Bigl(b_{i}\Big \vert \mu_{i}, \Sigma_{i,i}\Bigr)
\biggl[\boldsymbol{\Phi}_{p-1}\Bigl(\boldsymbol{b}[-i]-\boldsymbol{\mu}[-i]\Big\vert\boldsymbol{\xi}(b_{i}-\mu_{i}), \Delta_{i}\Bigr)\nonumber\\
&-\boldsymbol{\Phi}_{p-1}\Bigl(\boldsymbol{a}[-i]-\boldsymbol{\mu}[-i]\Big\vert\boldsymbol{\xi}(b_{i}-\mu_{i}), \Delta_{i}\Bigr)\biggr].
%=&\frac{\Sigma}{{\cal{P}}}\times\bigl({\cal{I}}_{1},{\cal{I}}_{2},\cdots,{\cal{I}}_{p}\bigr)^{\top},
\end{align}
Computing $I_{i}$ in (\ref{first-truncated-moment-multivariate-Gaussian-part5}), for $i=1,\cdots,p$, the first moment of truncated Gaussian distribution is then obtained by substituting the constructed vector $\boldsymbol{I}$ into the RHS of (\ref{first-truncated-moment-multivariate-Gaussian-part1}). 
\par
\begin{figure}[!h]
\center
\includegraphics[width=60mm,height=60mm]{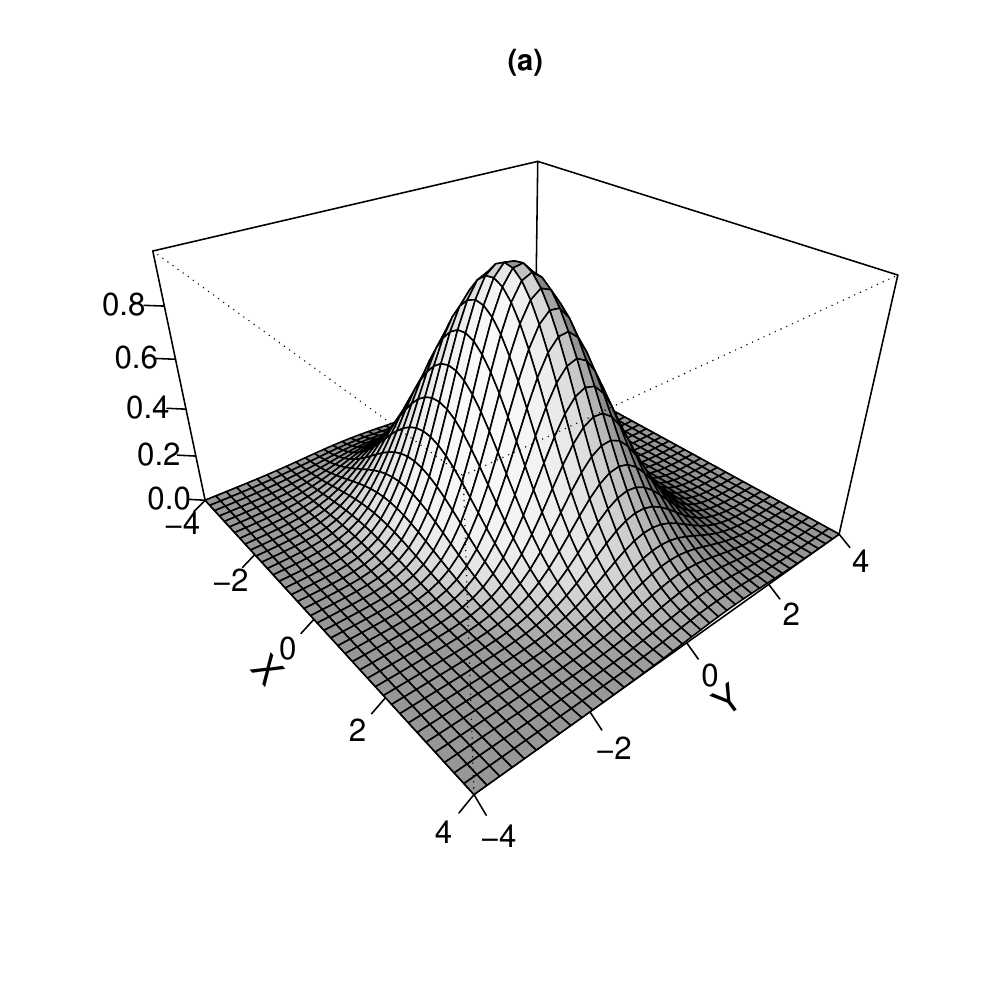}
\includegraphics[width=60mm,height=60mm]{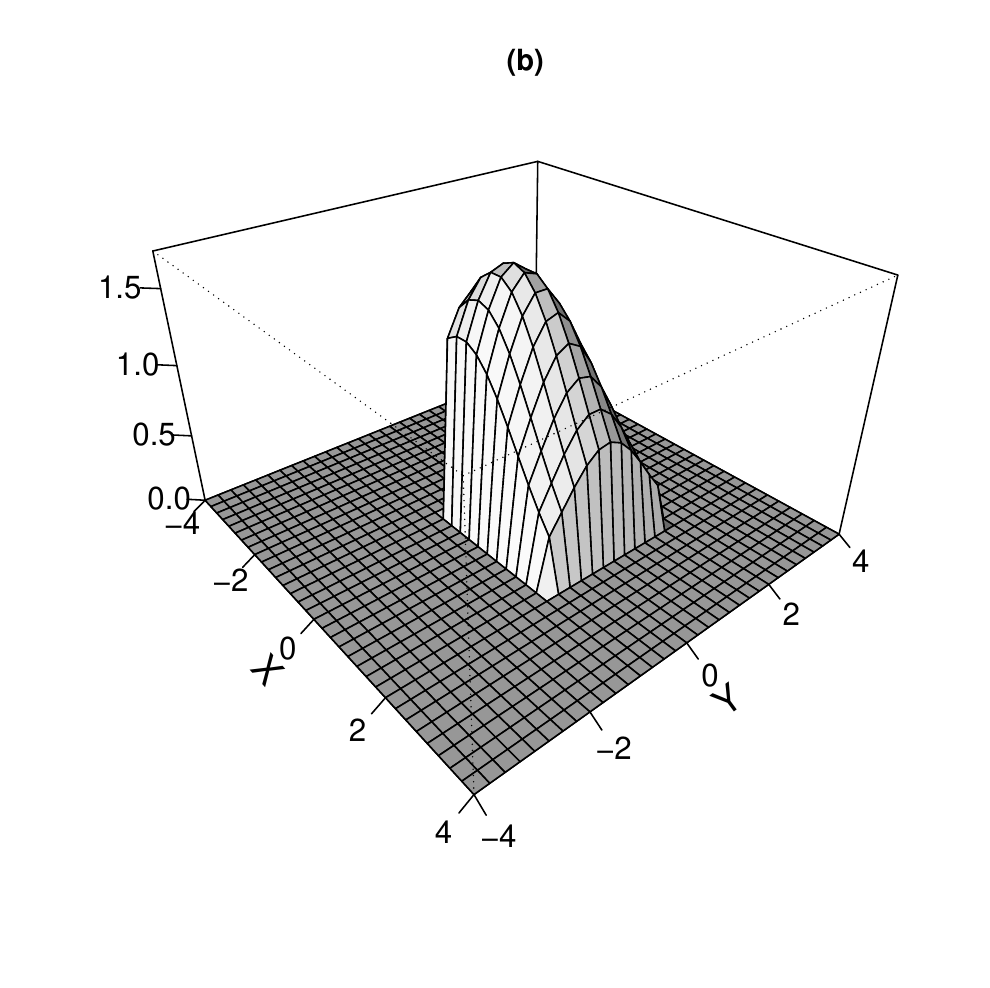}\\
\includegraphics[width=60mm,height=60mm]{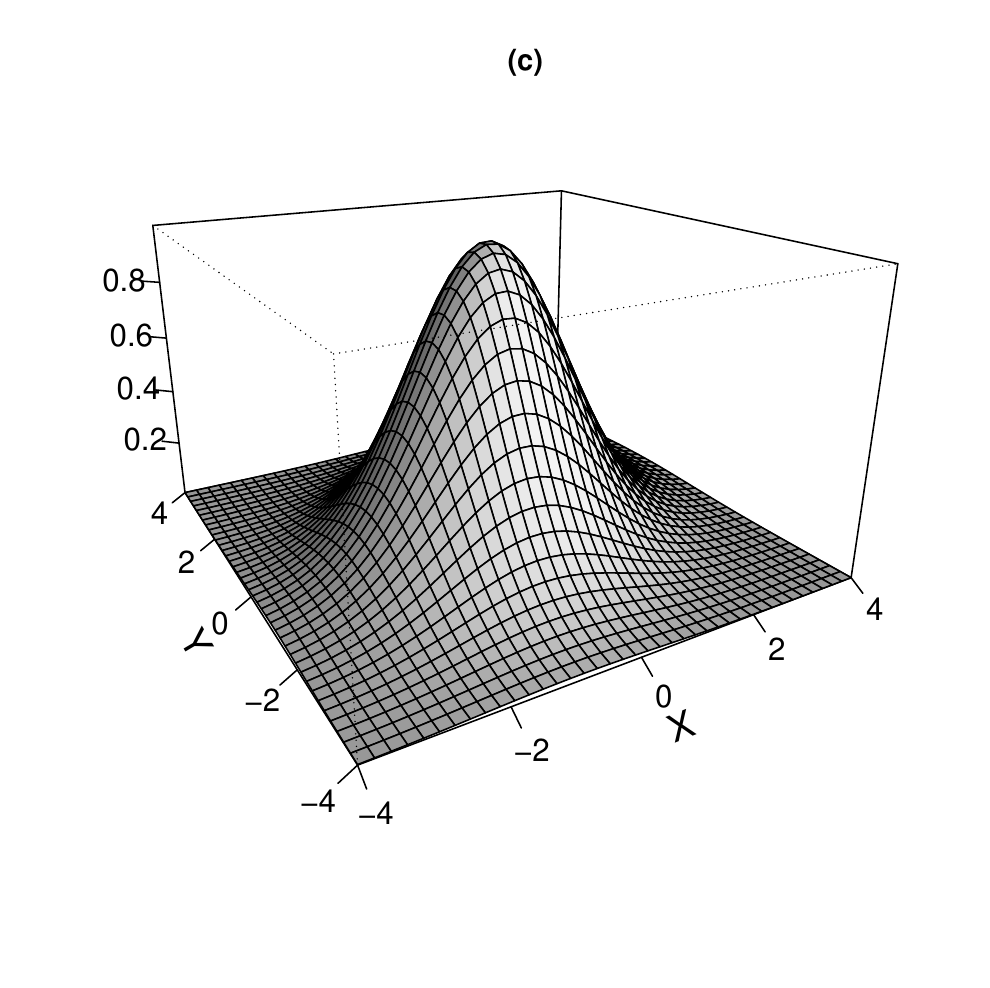}
\includegraphics[width=60mm,height=60mm]{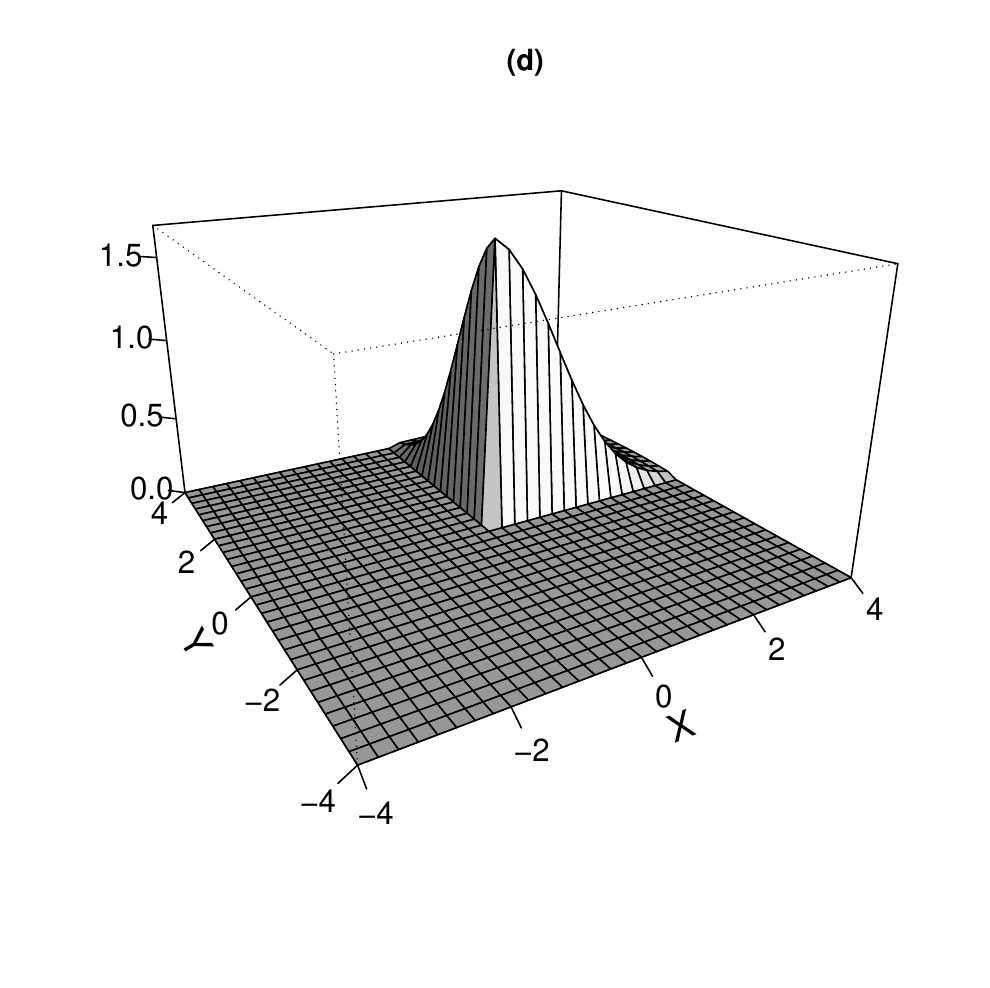}
\caption{(a): The PDF of bivariate Gaussian distribution with parameters $\mu_{1}=\mu_{2}=0$, 
$\sigma_{1}^{2}=1$, $\sigma_{2}^{2}=2$, and $\rho=0.3535$. (b): The PDF of the corresponding truncated bivariate Gaussian distribution on area with lower limits $\boldsymbol{a}=(-1,-1)^{\top}$ and upper limits $\boldsymbol{b}=(2,2)^{\top}$. (c) The PDF of bivariate Gaussian distribution with parameters $\mu_{1}=\mu_{2}=0$, 
$\sigma_{1}^{2}=2$, $\sigma_{2}^{2}=2$, and $\rho=0.25$. (b): The PDF of the corresponding truncated bivariate Gaussian distribution on area with lower limits $\boldsymbol{a}=(0,0)^{\top}$ and upper limits $\boldsymbol{b}=(+\infty,+\infty)^{\top}$.}
\label{plot-bivariate-truncated-Gaussian-PDF}
\end{figure}

 Here, we carry out a small simulation study for computing the first moment of vector $\boldsymbol{X} \sim{{\cal{TN}}}_{2}(\boldsymbol{0},\Sigma,\boldsymbol{a},\boldsymbol{b})$ under two scenarios. Under the first scenario, we assume that $\sigma_{1}^{2}=2, \sigma_{2}^{2}=1, \rho=0.3535$, $\boldsymbol{a}=(-1,-1)^{\top}$, and $\boldsymbol{b}=(2,2)^{\top}$. Under the second scenario it is assumed that $\sigma_{1}^{2}=2, \sigma_{2}^{2}=2, \rho=0.25$, $\boldsymbol{a}=(0,0)^{\top}$, and $\boldsymbol{b}=(+\infty,+\infty)^{\top}$. For both scenarios, we compute $\boldsymbol{\Omega}$ by generating samples of sizes $N=2000,5000$, and 20000 based on 5000 runs. The instrumental distribution is set to be ${\cal{N}}_{2}(\boldsymbol{\mu},\Sigma)$. For computing $\boldsymbol{\Omega}$ through the ${\cal{N}}_{2}(\boldsymbol{\mu},\Sigma)$, one can follow two ways. In the first way the quantity $\boldsymbol{\Omega}$ is estimated as
\begin{align}
\hat{\boldsymbol{\Omega}}=\frac{1}{N}\sum_{i=1}^{N} \vert \boldsymbol{x}_{i}\big \vert
\end{align}
where $\big \vert \boldsymbol{x}_{i}\big \vert=\bigl(\big \vert x_{1i}\big \vert, \big \vert x_{2i}\big \vert \bigr)^{\top}$ where $\boldsymbol{x}_{i}$s come independently from ${\cal{N}}_{2}(\boldsymbol{\mu},\Sigma)$, for sufficiently large $N$. The second method computes $\hat{\boldsymbol{\Omega}}$ based on sample of size $N$ when the lower and upper truncation bounds are $\boldsymbol{a}=\boldsymbol{0}$ and $\boldsymbol{b}=+\boldsymbol{\infty}$, respectively. Details for implementing the second method are given as follows.
\begin{enumerate}
\item Read $\boldsymbol{a}=\boldsymbol{0}, \boldsymbol{b}=+\boldsymbol{\infty}, \boldsymbol{\mu}$, $\Sigma$, and $N$;
\item Set $j=0$ and $\boldsymbol{y}=(0,0)^{\top}$;
\item Simulate $\boldsymbol{x}=(x_1,x_2)^{\top}\sim{\cal{N}}_{2}(\boldsymbol{0},\Sigma)$
\item If $a_1<x_1<b_1$ and $a_2<x_2<b_2$, then accept $\boldsymbol{x}$ as a generation from ${{\cal{TN}}}_{2}(\boldsymbol{0},\Sigma,\boldsymbol{a},\boldsymbol{b})$, set $\boldsymbol{y}=\boldsymbol{y}+\bigl(x_1,x_2\bigr)^{\top}$ and $j = j+1$;
\item If $j=N$, then go to the next step, otherwise return to step 2;
\item Compute an approximation of $\boldsymbol{\Omega}=\bigl({{\Omega}}_{1},{{\Omega}}_{2}\bigr)^{\top}$ as $\hat{\boldsymbol{\Omega}}=\bigl({y}_{1}/N,{y}_{2}/N\bigr)^{\top}$;
\item End.
\end{enumerate}
While the exact value of $\boldsymbol{\Omega}$ under both scenarios is computed using (\ref{first-truncated-moment-multivariate-Gaussian-part1}), Table \ref{table-importance-sampling-bivariate-truncated-Gaussian} shows the results of simulation study for approximating $\boldsymbol{\Omega}$ under both scenarios through the instrumental ${\cal{N}}_{2}(\boldsymbol{\mu},\Sigma)$.  
\begin{table}[h!]
\center
\caption{Summary statistics for computing $\boldsymbol{\Omega}$ through the exact and importance sampling with ${\cal{N}}_{2}(\boldsymbol{0},\Sigma)$ as the instrumental.} 
\begin{tabular}{lllcclcccc} 
\cline{1-10} 
&&&&&&\multicolumn{4}{c}{summary}\\ \cline{7-10} 
\rotatebox{90}{\parbox{.75cm}{\centering {\tiny{Scenario}}}}&$\boldsymbol{a}^{\top}$&$\boldsymbol{b}^{\top}$&$N$&${\boldsymbol{\Omega}}$ & $\hat{\boldsymbol{\Omega}}$&bias&SE&min.&max.\\ \cline{1-10}
1&(-1,-1)&(2,2)					    &2000 &0.3691&  $\hat{{\Omega}_{1}}$& -1.7e-04& 0.0174& 0.3021 &  0.4440\\
     &    &        					&     &0.2603&  $\hat{{\Omega}_{2}}$& 5.9e-05& 0.0158& 0.2072 &  0.3158\\
    &     &        					&5000 &0.3691&  $\hat{{\Omega}_{1}}$& 7.5e-06& 0.0113& 0.3268 &  0.4071\\
    &     &       					&     &0.2603&  $\hat{{\Omega}_{2}}$&-1.8e-04& 0.0101& 0.2255 &  0.2971\\
\cline{1-10}   
2&(0,0)&$(\infty,\infty)$	            &2000 &1.2150&$\hat{{\Omega}_{1}}$&-7.8e-05&0.0197&1.1287 &1.2802\\
	 &  &							&     &1.2150&$\hat{{\Omega}_{2}}$&-8.3e-04&0.0196&1.1462 &1.2827\\
	 &  &							&5000 &1.2150&$\hat{{\Omega}_{1}}$&-7.4e-04 &0.0124&1.1690 &1.2603\\
	 &  &							&     &1.2150&$\hat{{\Omega}_{2}}$&-4.7e-04 &0.0125&1.1701 &1.2774\\	
\cline{1-10}                                                             			                 
\end{tabular} 
%\tabnote{Hint: CI is short form for confidence interval.}                                                                                                           
\label{table-importance-sampling-bivariate-truncated-Gaussian}
\end{table}
The PDF of bivariate Gaussian distribution and the corresponding truncated version under two scenarios mentioned above is depicted in Figure \ref{plot-bivariate-truncated-Gaussian-PDF}.
The corresponding \verb+R+ function for computing $\boldsymbol{\Omega}$ is given by the following. 
\begin{lstlisting}[style=deltaj]
R> library(mvtnorm) # for computing multivariate Gaussian CDF 
R> a <- c(-1, -1); b <- c(2, 2); Mu <- c(0, 0); 
R> Sigma <- matrix( c(2, 0.5, 0.5, 2), nrow = length(Mu), ncol = length(Mu), byrow = TRUE) 
R> EX <- function(Mu, Sigma, a, b)
+{
+ if( any(a > b) == TRUE ) stop(message("lower bound must be smaller than the upper bound."))
+ if( length(a) != length(Mu) || length(b) != length(Mu) ) stop(message("truncation bounds and location parameter must be of the same size."))
+	p  <- length(Mu)
+	Omega <- numeric(p)
+	V <- numeric(p)
+	if(p == 2)
+	{
+	for(i in 1:p)
+	{
+	Mu_ia <- Sigma[i, -i]*(a[i] - Mu[i])/Sigma[i, i] 
+	Mu_ib <- Sigma[i, -i]*(b[i] - Mu[i])/Sigma[i, i]
+	Delta_i <- sqrt( Sigma[-i, -i] - Sigma[i, -i]^2/Sigma[i, i] )
+	V[i] <-  dnorm(a[i] - Mu[i], 0, sd = sqrt(Sigma[i, i]) )*(
+			pnorm(b[-i] - Mu[-i], mean = Mu_ia, sd = Delta_i)-
+			pnorm(a[-i] - Mu[-i], mean = Mu_ia, sd = Delta_i) )-
+			dnorm(b[i] - Mu[i], 0, sd = sqrt(Sigma[i, i]) )*(
+			pnorm(b[-i] - Mu[-i], mean = Mu_ib, sd = Delta_i)-
+			pnorm(a[-i] - Mu[-i], mean = Mu_ib, sd = Delta_i) )
+	}
+	}else{
+    for(i in 1:p)
+	{
+		Mu_ia <- Sigma[i, -i]*(a[i] - Mu[i])/Sigma[i, i] 
+		Mu_ib <- Sigma[i, -i]*(b[i] - Mu[i])/Sigma[i, i]
+		Delta_i <- Sigma[-i, -i] - c(Sigma[i, -i]%*%t(Sigma[i, -i])/(Sigma[i, i]))
+		V[i] <- dnorm(a[i] - Mu[i], 0, sqrt(Sigma[i, i]) )*
+          		pmvnorm(a[-i] - Mu[-i], b[-i] - Mu[-i], mean = Mu_ia, sigma = Delta_i)[1] -
+			dnorm(b[i] - Mu[i], 0, sqrt(Sigma[i, i]) )*
+          		pmvnorm(a[-i] - Mu[-i], b[-i] - Mu[-i], mean = Mu_ib, sigma = Delta_i)[1]
+    }
+	}
+  Omega <- Sigma%*%V/pmvnorm(a, b, mean = Mu, sigma = Sigma)[1] + Mu
+ return(Omega)
+ }
R>  EX(a, b, Mu, Sigma)
\end{lstlisting}
\section{Second moment of truncated Gaussian distribution}\label{second-moment-truncated-multivariate-Gaussian-distribution}
Let $\boldsymbol{v}=\bigl(v_1\cdots,v_p\bigr)^{\top}$ represent a vector of $p$ real-valued constants. Based on vector $\boldsymbol{v}$, for $i<j$, we define the following notations. 
\begin{align}
&\boldsymbol{v}_{i,j}[x,y]=\bigl({v}_{1},\cdots,{v}_{i-1},x,{v}_{i+1},\cdots,{v}_{j-1},y,{v}_{j+1},\cdots,{v}_{p}\bigr)^{},\label{vector-representation1}\\ 
&\boldsymbol{v}[-i,-j]=\bigl({v}_{1},\cdots,{v}_{i-1},{v}_{i+1},\cdots,{v}_{j-1},{v}_{j+1},\cdots,{v}_{p}\bigr)^{},\label{vector-representation2}\\
&\boldsymbol{v}[i,j]=\bigl({v}_{i},{v}_{j}\bigr)^{}\label{vector-representation3}.
\end{align}
Let $\boldsymbol{X}\sim{\cal{TN}}_{p}(\boldsymbol{\mu},\Sigma,\boldsymbol{a},\boldsymbol{b})$. By definition, the second (raw) moment of truncated Gaussian distribution is given by
\ref{second-moment-truncated-multivariate-Gaussian-distribution}
\begin{align}%\label{second-truncated-moment-multivariate-Gaussian-part1}
\boldsymbol{\Omega}^{2}=E\bigl(\boldsymbol{X}\boldsymbol{X}^{\top}\bigr)=&\frac{1}{{\bf{C}}_{G}{\cal{P}}_{G}} \int_{\boldsymbol{a}}^{\boldsymbol{b}}\boldsymbol{x}\boldsymbol{x}^{\top}\exp\Bigl\{-\frac{\delta(\boldsymbol{x}-\boldsymbol{\mu},\Sigma)}{2}\Bigr\}d\boldsymbol{x}.\nonumber
\end{align}
%=&\frac{1}{C_{G}{\cal{P}}_{G}}
%                       \int_{\boldsymbol{a}}^{\boldsymbol{b}}(\boldsymbol{x}-\boldsymbol{\mu})(\boldsymbol{x}-\boldsymbol{\mu})^{\top}\exp\Bigl\{-\frac{\delta(\boldsymbol{x}-\boldsymbol{\mu},\Sigma)}{2}\Bigr\}d\boldsymbol{x}\nonumber\\
%                       &+\frac{\boldsymbol{\mu}}{C_{G}{\cal{P}}_{G}}
%\int_{\boldsymbol{a}}^{\boldsymbol{b}}\exp\Bigl\{-\frac{\delta(\boldsymbol{x}-\boldsymbol{\mu},\Sigma)}{2}\Bigr\}d\boldsymbol{x}\nonumber\\
%=&  \boldsymbol{I}+\boldsymbol{\mu}.
%\Phi_{q}\Bigl(\sqrt{r}\boldsymbol{m}\Big|\boldsymbol{0},\boldsymbol{\Delta}\Bigr)
We recall that ${\bf{C}}_{G}=(2\pi)^{p/2} \vert \Sigma \vert ^{1/2}$, ${\cal{P}}_{G}=\boldsymbol{\Phi}_{p}\bigl(\boldsymbol{b}\big\vert\boldsymbol{\mu}, {\Sigma}\bigr)-\boldsymbol{\Phi}_{p}\bigl(\boldsymbol{a}\big\vert\boldsymbol{\mu}, {\Sigma}\bigr)$. Using matrix algebra, one can see that
\begin{align}\label{second-partial-mu}
 \frac{\partial^{2}}{\partial \boldsymbol{\mu}\partial \boldsymbol{\mu}^{\top}}\exp\Bigl\{-\frac{\delta(\boldsymbol{x}-\boldsymbol{\mu},\Sigma)}{2}\Bigr\}= 
 &\Bigl[ {\Sigma}^{-1}(\boldsymbol{x}-\boldsymbol{\mu})(\boldsymbol{x}-\boldsymbol{\mu})^{\top}\Sigma^{-1}-\Sigma^{-1}\Bigr]\nonumber\\
 &\times\exp\Bigl\{-\frac{\delta(\boldsymbol{x}-\boldsymbol{\mu},\Sigma)}{2}\Bigr\}.
\end{align}
%\Phi_{q}\Bigl(\sqrt{r}\boldsymbol{m}\Big|\boldsymbol{0},\boldsymbol{\Delta}\Bigr)
Integrating both sides of (\ref{second-partial-mu}), then multiplying from left and right by $\Sigma$, and finally rearranging, we have
\begin{align}\label{second-truncated-moment-multivariate-Gaussian-part2}
\boldsymbol{\Omega}^{2}=&\boldsymbol{\Omega}\boldsymbol{\mu}^{\top}+\boldsymbol{\mu}\boldsymbol{\Omega}^{\top}-\boldsymbol{\mu}\boldsymbol{\mu}^{\top}+\Sigma+
\frac{\Sigma}{{\bf{C}}_{G}{\cal{P}}_{G}} \int_{\boldsymbol{a}}^{\boldsymbol{b}}  \frac{\partial^{2}}{\partial \boldsymbol{\mu}\partial \boldsymbol{\mu}^{\top}}\exp\Bigl\{-\frac{\delta(\boldsymbol{x}-\boldsymbol{\mu},\Sigma)}{2}\Bigr\}d\boldsymbol{x} \Sigma \nonumber\\
=&\boldsymbol{\Omega}\boldsymbol{\mu}^{\top}+\boldsymbol{\mu}\boldsymbol{\Omega}^{\top}-\boldsymbol{\mu}\boldsymbol{\mu}^{\top}+\Sigma+\frac{\Sigma}{{\bf{C}}_{G}{\cal{P}}_{G}}   \frac{\partial^{2}}{\partial \boldsymbol{\mu}\partial \boldsymbol{\mu}^{\top}}\int_{\boldsymbol{a}-\boldsymbol{\mu}}^{\boldsymbol{b}-\boldsymbol{\mu}}\exp\Bigl\{-\frac{\delta(\boldsymbol{u},\Sigma)}{2}\Bigr\}d\boldsymbol{u} \Sigma \nonumber\\
%=&\frac{\Sigma}{\alpha{\cal{P}}} \int_{\boldsymbol{a}_{-i}-\boldsymbol{\mu}_{-i}}^{\boldsymbol{b}_{-i}-\boldsymbol{\mu}_{-i}}\biggl[\exp\Bigl\{-\frac{d\bigl(\boldsymbol{u}_{-i(a_{i}-\mu_{i})}\bigr)}{2}\Bigr\}-\exp\Bigl\{-\frac{d\bigl(\boldsymbol{u}_{-i(b_{i}-\mu_{i})}\bigr)}{2}\Bigr\}\biggr]d\boldsymbol{u}_{-i}\nonumber\\
=&\boldsymbol{\Omega}\boldsymbol{\mu}^{\top}+\boldsymbol{\mu}\boldsymbol{\Omega}^{\top}-\boldsymbol{\mu}\boldsymbol{\mu}^{\top}+\Sigma+\Sigma \boldsymbol{I}\Sigma,
\end{align}
where
\begin{align}\label{second-truncated-moment-multivariate-Gaussian-part3}
\boldsymbol{I}=\frac{1}{{\bf{C}}_{G}{\cal{P}}_{G}}\frac{\partial^{2}}{\partial \boldsymbol{\mu}\partial \boldsymbol{\mu}^{\top}}\int_{\boldsymbol{a}-\boldsymbol{\mu}}^{\boldsymbol{b}-\boldsymbol{\mu}}\exp\Bigl\{-\frac{\delta(\boldsymbol{u},\Sigma)}{2}\Bigr\}d\boldsymbol{u}.
\end{align}
As it may seen, $\boldsymbol{I}$ is a $ p \times p$ square matrix whose $(i,j)$th element is shown by $I_{i,j}$. In order to compute $I_{i,j}$, we consider three scenarios given by the following. 
\begin{itemize}
\item {\bf{Scenario 1}}: $i\neq j =1,\cdots,p$ and $p=2$
\item {\bf{Scenario 2}}: $i\neq j =1,\cdots,p$ and $p>2$
\item {\bf{Scenario 3}}: $i=j=1,\cdots,p$
\end{itemize}
In what follows, we proceed to compute $I_{i,j}$ under three scenarios mentioned above. 
\begin{itemize}
\item {\bf{Scenario 1}} ($i\neq j =1,\cdots,p$, and $p=2$): It follows from (\ref{second-truncated-moment-multivariate-Gaussian-part3}) that 
\begin{align}\label{second-truncated-moment-multivariate-Gaussian-case1-part1}
I_{1,2}=\frac{1}{{\bf{C}}_{G}{\cal{P}}_{G}}\frac{\partial^2}{\partial \mu_1\partial \mu_2}\int_{a_1-\mu_1}^{b_1-\mu_1}\int_{a_2-\mu_2}^{b_2-\mu_2}
\exp\Bigl\{-\frac{\delta(\boldsymbol{u},\Sigma)}{2}\Bigr\}du_{1}du_{2}.
\end{align}
Applying the Leibniz integral rule for the RHS of (\ref{second-truncated-moment-multivariate-Gaussian-case1-part1}), we have
\begin{align*}%\label{second-truncated-moment-multivariate-Gaussian-case1-part2}
I_{1,2}=\frac{1}{{\cal{P}}_{G}}\Bigl[\phi\bigl(\boldsymbol{b}\big \vert \boldsymbol{\mu},\Sigma\bigr) -\phi\bigl((b_1,a_2)\big \vert \boldsymbol{\mu},\Sigma\bigr)
+\phi\bigl(\boldsymbol{a}\big \vert \boldsymbol{\mu},\Sigma\bigr)-\phi\bigl((a_1,b_2)\big \vert \boldsymbol{\mu},\Sigma\bigr)\Bigr].
\end{align*}
Obviously $I_{2,1}=I_{1,2}$.
\item {\bf{Scenario 2}} ($i\neq j =1,\cdots,p$, and $p>2$):  
%It follows from (\ref{second-truncated-moment-Gaussian-part3}) that 
%Evidently $I_{1,2}=I_{2,1}$. Likewise, for computing $I_{2,2}$, we can write
%\begin{align}\label{second-truncated-moment-Gaussian-case1-part3}
%I_{2,2}&=\frac{1}{C_{G}{\cal{P}}_{G}}\frac{\partial^2}{\partial \mu_2\partial \mu_2}\int_{a_1-\mu_1}^{b_1-\mu_1}\int_{a_2-\mu_2}^{b_2-\mu_2}
%\exp\Bigl\{-\frac{\delta(\boldsymbol{u},\Sigma)}{2}\Bigr\}du_{1}du_{2}\nonumber\\
%&=\frac{1}{C_{G}{\cal{P}}_{G}}\int_{a_1-\mu_1}^{b_1-\mu_1} \Bigl[\exp\Bigl\{-\frac{\delta\bigl((u,a_2-\mu_2),\Sigma\bigr)}{2}\Bigr\}-\exp\Bigl\{-\frac{\delta\bigl((u,b_2-\mu_2),\Sigma\bigr)}{2}\Bigr\}\Bigr]du.
%\end{align}
%Since
%\begin{align}
%C_{G}=\frac{1}{\sqrt{2\pi \Sigma_{2,2}}}\frac{1}{\sqrt{2\pi \bigl[\Sigma_{1,1}-\Sigma^{-1}_{2,2}\Sigma^{2}_{1,2}\bigr]}},
%\end{align}
%we have
%\begin{align}\label{second-truncated-moment-Gaussian-case1-part4}
%I_{2,2}&=\frac{1}{{\cal{P}}_{G}}\phi\bigl({b}_2\big \vert {\mu}_2,\sqrt{\Sigma_{2,2}}\bigr)\Phi
%\end{align}
For a given vector $(x,y)^{\top}$, it can be seen that the quadratic form $\delta\bigl(\boldsymbol{u}_{i,j}[x,y],\Sigma\bigr)$, for which $\boldsymbol{u}_{i,j}[x,y]$ is defined in (\ref{vector-representation1}), can be decomposed as
\begin{align}\label{quadratic-ij-decomposition}
\delta\bigl(\boldsymbol{u}_{i,j}[x,y],\Sigma\bigr)= \delta_{ij}(x,y) + \delta\Bigl(\boldsymbol{u}[-i,-j]-\boldsymbol{\xi}(x,y),\Delta_{ij}\Bigr),
\end{align}
where $\delta_{ij}(x,y)=(x,y)^{\top}\Sigma^{-1}_{{\cal{I}},{\cal{I}}}(x,y)$ for ${\cal{I}}=\{i,j\}$ and
\begin{align}
\boldsymbol{\xi}(x,y)=&\Sigma_{-{\cal{I}},{\cal{I}}}\bigl(\Sigma_{{\cal{I}},{\cal{I}}}\bigr)^{-1}(x,y),\label{xi-multi-decomposition}\\
\Delta_{ij}=&\Sigma_{-{\cal{I}},-{\cal{I}}}-\Sigma_{-{\cal{I}},{\cal{I}}}^{}\bigl(\Sigma_{{\cal{I}},{\cal{I}}}\bigr)^{-1}\Sigma_{{\cal{I}}, -{\cal{I}}}.\label{Delta-multi-decomposition}
\end{align}
Herein, $\boldsymbol{\xi}(x,y)$ is a vector of length $p-2$, $\Delta_{ij}$ is a $(p-2)\times (p-2)$ positive definite scale matrix, and $2\times 2$ positive definite scale matrix matrix $\Sigma_{{\cal{I}},{\cal{I}}}$ is given by
\begin{align*}
\Sigma_{{\cal{I}},{\cal{I}}}= \biggl[\begin{matrix}\Sigma_{i,i} & \Sigma_{i,j}\\  \Sigma_{j,i} & \Sigma_{j,j}\end{matrix}\biggr].
\end{align*}
For constructing $(p-2)\times 2$ matrix $\Sigma_{-{\cal{I}},{\cal{I}}}$, first construct a  two-column matrix based on the $i$th and $j$th columns of matrix $\Sigma$ and then remove $i$th and $j$th rows of the constructed two-column matrix. We apply the Leibniz integral rule twice to the RHS of (\ref{second-truncated-moment-multivariate-Gaussian-part3}), and simultaneously utilize the property (\ref{quadratic-ij-decomposition}), to see that
\begin{align}\label{second-truncated-moment-multivariate-Gaussian-case2-part1}
I_{i,j}=&\frac{1}{{\bf{C}}_{G}{\cal{P}}_{G}} \exp\Bigl\{-\frac{\delta_{ij}(\boldsymbol{b}[i,j]-\boldsymbol{\mu}[i,j])}{2}\Bigr\} 
\times I_{1}\bigl(\boldsymbol{b}[i,j]\bigr)\nonumber\\
&-\frac{1}{{\bf{C}}_{G}{\cal{P}}_{G}} \exp\Bigl\{-\frac{\delta_{ij}\bigl((\boldsymbol{a}[i],\boldsymbol{b}[j])-\boldsymbol{\mu}[i,j]\bigr)}{2}\Bigr\}
\times I_{1}\bigl(\boldsymbol{a}[i],\boldsymbol{b}[j]\bigr)\nonumber\\
&-\frac{1}{{\bf{C}}_{G}{\cal{P}}_{G}} \exp\Bigl\{-\frac{\delta_{ij}\bigl((\boldsymbol{b}[i],\boldsymbol{a}[j])-\boldsymbol{\mu}[i,j]\bigr)}{2}\Bigr\}
 \times I_{1}\bigl(\boldsymbol{b}[i],\boldsymbol{a}[j]\bigr)\nonumber\\
&+\frac{1}{{\bf{C}}_{G}{\cal{P}}_{G}} \exp\Bigl\{-\frac{\delta_{ij}(\boldsymbol{a}[i,j]-\boldsymbol{\mu}[i,j])}{2}\Bigr\}
\times I_{1}\bigl(\boldsymbol{a}[i,j]\bigr),
\end{align}
where 
 \begin{align}\label{second-truncated-moment-multivariate-Gaussian-case2-part2}
 I_{1}(\boldsymbol{z})=&\int_{\boldsymbol{a}[-i,-j]-\boldsymbol{\mu}[-i,-j]}^{\boldsymbol{b}[-i,-j]-\boldsymbol{\mu}[-i,-j]}
\exp\Bigl\{-\frac{1}{2}\delta\Bigl(\boldsymbol{u}[-i,-j]-\boldsymbol{\xi}(\boldsymbol{z}),\Delta_{ij}\Bigr) \Bigr\} d\boldsymbol{u}[-i,-j].
 \end{align}
We further notice that the quantity ${\bf{C}}_{G}$ can be represented as 
\begin{align}\label{second-truncated-moment-multivariate-Gaussian-case2-part3}
{\bf{C}}_{G}=(2\pi)^{\frac{p}{2}} \vert \Sigma \vert ^{\frac{1}{2}}={2\pi \big \vert \Sigma_{{\cal{I}},{\cal{I}}}\big \vert^{\frac{1}{2}} } (2\pi)^{\frac{p-2}{2}}\big \vert \Delta_{ij}\big  \vert^{\frac{1}{2}}.
\end{align}
Substituting (\ref{second-truncated-moment-multivariate-Gaussian-case2-part3}) into the RHS of (\ref{second-truncated-moment-multivariate-Gaussian-case2-part1}), it follows that
\begin{align}\label{second-truncated-moment-multivariate-Gaussian-case2-part4}
I_{i,j}=&\frac{1}{{\cal{P}}_{G}} \boldsymbol{\phi}_{2}\Bigl(\boldsymbol{b}[i,j]\Big \vert \boldsymbol{\mu}[i,j], \Sigma_{{\cal{I}},{\cal{I}}}\Bigr)
\times I_{2}(\boldsymbol{b}[i,j])\nonumber\\
&-\frac{1}{{\cal{P}}_{G}} \boldsymbol{\phi}_{2}\Bigl((\boldsymbol{a}[i],\boldsymbol{b}[j])\Big \vert \boldsymbol{\mu}[i,j], \Sigma_{{\cal{I}},{\cal{I}}}\Bigr)
\times I_{2}(\boldsymbol{a}[i],\boldsymbol{b}[j])\nonumber\\
&-\frac{1}{{\cal{P}}_{G}} \boldsymbol{\phi}_{2}\Bigl((\boldsymbol{b}[i],\boldsymbol{a}[j])\Big \vert \boldsymbol{\mu}[i,j], \Sigma_{{\cal{I}},{\cal{I}}}\Bigr)
\times I_{2}(\boldsymbol{b}[i],\boldsymbol{a}[j])\nonumber\\
&+\frac{1}{{\cal{P}}_{G}}\boldsymbol{\phi}_{2}\Bigl(\boldsymbol{b}[i,j]\Big \vert \boldsymbol{\mu}[i,j], \Sigma_{{\cal{I}},{\cal{I}}}\Bigr)\times I_{2}(\boldsymbol{a}[i,j]),
\end{align}
where
\begin{align*}%\label{second-truncated-moment-multivariate-Gaussian-case2-part5}
 I_{2}(\boldsymbol{z})=&\boldsymbol{\Phi}_{p-2}\Bigl(\boldsymbol{b}[-i,-j]-\boldsymbol{\mu}[-i,-j]\Big\vert\boldsymbol{\xi}(\boldsymbol{z}),\Delta_{ij}\Bigr)\nonumber\\
&-\boldsymbol{\Phi}_{p-2}\Bigl(\boldsymbol{b}[-i,-j]-\boldsymbol{\mu}[-i,-j]\Big\vert\boldsymbol{\xi}(\boldsymbol{z}),\Delta_{ij}\Bigr).
 \end{align*}
The matrix $\boldsymbol{I}$ is constructed by computing $I_{i,j}$ in (\ref{second-truncated-moment-multivariate-Gaussian-case2-part4}) for all $i\neq j =1,\cdots,p$. The second moment of truncated Gaussian distribution is then obtained by substituting $\boldsymbol{I}$ into the RHS of (\ref{second-truncated-moment-multivariate-Gaussian-part2}).
%%%%%%%%%%%%%%%%%%%%%%%%%%%%%%%%%%
%%%%%%%%%%%%%%%%%%%%%%%%%%%%%%%%%%
\item {\bf{Scenario 3}} ($i= j =1,\cdots,p$): Using the Leibniz integral rule, the first order derivative of the RHS of (\ref{second-truncated-moment-multivariate-Gaussian-part3}) is

\begin{align}\label{second-truncated-moment-multivariate-Gaussian-case3-part1}
\frac{\partial}{\partial {\mu}_{i}}\boldsymbol{I}=&\frac{1}{{\bf{C}}_{G}{\cal{P}}_{G}} \frac{\partial}{\partial {\mu}_{i}} \int_{\boldsymbol{a}-\boldsymbol{\mu}}^{\boldsymbol{b}-\boldsymbol{\mu}}\exp\Bigl\{-\frac{\delta(\boldsymbol{u},\Sigma)}{2}\Bigr\}d\boldsymbol{u}\nonumber\\
=&\frac{1}{{\bf{C}}_{G}{\cal{P}}_{G}}\int_{\boldsymbol{a}[-i]-\boldsymbol{\mu}[-i]}^{\boldsymbol{b}[-i]-\boldsymbol{\mu}[-i]}\biggl[\exp\Bigl\{-\frac{\delta\bigl(\boldsymbol{u}_{i}[a_{i}-\mu_{i}],\Sigma\bigr)}{2}\Bigr\}\nonumber\\
&-\exp\Bigl\{-\frac{\delta\bigl(\boldsymbol{u}_{i}[b_{i}-\mu_{i}],\Sigma\bigr)}{2}\Bigr\}\biggr]d\boldsymbol{u}[-i],
\end{align}
%	Mu_ia <- Sigma[i, -i]*(a[i] - Mu[i])/Sigma[i, i] 
%	Mu_ib <- Sigma[i, -i]*(b[i] - Mu[i])/Sigma[i, i]
%	Delta_i <- Sigma[-i, -i] - Sigma[i, -i]%*%t(Sigma[i, -i])/Sigma[i, i]
for $i=1,\cdots,p$. We note that the quadratic form $\delta\bigl(\boldsymbol{u}_{i}[x],\Sigma\bigr)$, for which $\boldsymbol{u}_{i}[x]$ is defined in (\ref{vector-representation1}), can be decomposed as
\begin{align}\label{quadratic-ij-decomposition-case3}
\delta\bigl(\boldsymbol{u}_{i}[x],\Sigma\bigr)= \delta_{ii}(x,x) + \delta\Bigl(\boldsymbol{u}[-i]-\boldsymbol{\xi}(x),\Delta_{ii}\Bigr),
\end{align}
where $\delta_{ii}(x,x)=\bigl(x_i-\mu_i\bigr)^{2}\times\bigl(\Sigma_{i,i}\bigr)^{-1}$ is a scalar and
\begin{align}
\boldsymbol{\xi}(x)=&\Sigma_{i,-i}\times \bigl(\Sigma_{i,i}\bigr)^{-1}\times x,\label{xi-uni-decomposition}\\
\Delta_{ii}=&\Sigma_{-i,-i}-\Sigma_{i,-i}\bigl(\Sigma_{i,-i}\bigr)^{\top}\times\bigl(\Sigma_{i,i}\bigr)^{-1}\label{Delta-uni-decomposition}.
\end{align}
We note that $\boldsymbol{\xi}(x)$ is a vector of length $p-1$ and $\Delta_{ii}$ is a $(p-1)\times (p-1)$ positive definite scale matrix. We further notice that the quantity ${\bf{C}}_{G}$ can be represented as 
\begin{align}\label{CG-decomposition-case3}
{\bf{C}}_{G}={(2\pi)^{\frac{1}{2}}\big \vert \Sigma_{i,i}\big \vert^{\frac{1}{2}}}(2\pi)^{\frac{p-1}{2}}\big \vert \Delta_{i}\big  \vert^{\frac{1}{2}}.
\end{align}
Using information given in (\ref{quadratic-ij-decomposition-case3}), (\ref{xi-uni-decomposition}), (\ref{Delta-uni-decomposition}), and (\ref{CG-decomposition-case3}), the RHS of (\ref{second-truncated-moment-multivariate-Gaussian-case3-part1}) can be represented as
\begin{align}\label{second-truncated-moment-multivariate-Gaussian-case3-part2}
\frac{\partial}{\partial {\mu}_{i}}\boldsymbol{I}=&\frac{1}{{\cal{P}}_{G}} {\phi}\Bigl(a_{i}\Big \vert {\mu}_{i}, \Sigma_{i,i}\Bigr)\times \biggl[\boldsymbol{\Phi}_{p-1}\Bigl(\boldsymbol{b}[-i]-\boldsymbol{\mu}[-i]\Big\vert\boldsymbol{\xi}({a}_{i}-{\mu}_{i}),\Delta_{ii}\Bigr)\nonumber\\
&-\boldsymbol{\Phi}_{p-1}\Bigl(\boldsymbol{a}[-i]-\boldsymbol{\mu}[-i]\Big\vert\boldsymbol{\xi}({a}_{i}-{\mu}_{i}), \Delta_{ii}\Bigr)\biggr]-
\nonumber\\
&\frac{1}{{\cal{P}}_{G}} {\phi}\Bigl({b}_{i}\Big \vert {\mu}_{i}, \Sigma_{i,i}\Bigr)\times \biggl[\boldsymbol{\Phi}_{p-1}\Bigl(\boldsymbol{b}[-i]-\boldsymbol{\mu}_{i}\Big\vert\boldsymbol{\xi}({b}_{i}-{\mu}_{i}),\Delta_{ii}\Bigr)\nonumber\\
&-\boldsymbol{\Phi}_{p-1}\Bigl(\boldsymbol{a}[-i]-\boldsymbol{\mu}[-i]\Big\vert\boldsymbol{\xi}({b}_{i}-{\mu}_{i}), \Delta_{ii}\Bigr)\biggr]\nonumber\\
=&{I}_{ia}-{I}_{ib}.
\end{align}
Now, taking the second partial derivative with respect to $\mu_i$ from RHS of (\ref{second-truncated-moment-multivariate-Gaussian-case3-part2}), we obtain
\begin{align}\label{second-truncated-moment-multivariate-Gaussian-case3-part3}
\frac{\partial {I}_{ia}}{\partial {\mu}_{i}}=&\frac{\bigl(a_{i}- {\mu}_{i}\bigr)}{\Sigma_{i,i}\times{\cal{P}}_{G}} {\phi}\Bigl(a_{i}\Big \vert {\mu}_{i}, \Sigma_{i,i}\Bigr) \biggl[\boldsymbol{\Phi}_{p-1}\Bigl(\boldsymbol{b}[-i]-\boldsymbol{\mu}[-i]\Big\vert\boldsymbol{\xi}({a}_{i}-{\mu}_{i}),\Delta_{ii}\Bigr)-\nonumber\\
&\boldsymbol{\Phi}_{p-1}\Bigl(\boldsymbol{a}[-i]-\boldsymbol{\mu}[-i]\Big\vert\boldsymbol{\xi}({a}_{i}-{\mu}_{i}), \Delta_{ii}\Bigr)\biggr]-\frac{{\phi}\bigl(a_{i}\big \vert {\mu}_{i}, \Sigma_{i,i}\bigr)}{\Sigma_{i,i}\times{\cal{P}}_{G}} \times\Sigma_{-i,i}\bigl(\Delta_{ii}\bigr)^{-1}\times \nonumber\\ 
& \int_{\boldsymbol{a}[-i]-\boldsymbol{\mu}[-i]}^{\boldsymbol{b}[-i]-\boldsymbol{\mu}[-i]} \Bigl(\boldsymbol{u}[-i]-\boldsymbol{\xi}({a}_{i}-{\mu}_{i})\Bigr)\exp\Bigl\{-\frac{\delta\bigl(\boldsymbol{u}[-i]-\boldsymbol{\xi}({a}_{i}-{\mu}_{i}), \Delta_{ii}\bigr)}{2}\Bigr\}d\boldsymbol{u}[-i].
\end{align}
Clearly, the RHS of (\ref{second-truncated-moment-multivariate-Gaussian-case3-part3}) can be represented in terms of the first moment of a truncated Gaussian distribution. That means
\begin{align}\label{second-truncated-moment-multivariate-Gaussian-case3-part5}
\frac{\partial {I}_{ia}}{\partial {\mu}_{i}}=&\frac{1}{\Sigma_{i,i}\times{\cal{P}}_{G}} {\phi}\Bigl(a_{i}\Big \vert {\mu}_{i}, \Sigma_{i,i}\Bigr) \biggl[\boldsymbol{\Phi}_{p-1}\Bigl(\boldsymbol{b}[-i]-\boldsymbol{\mu}[-i]\Big\vert\boldsymbol{\xi}({a}_{i}-{\mu}_{i}),\Delta_{ii}\Bigr)-\nonumber\\
&\boldsymbol{\Phi}_{p-1}\Bigl(\boldsymbol{a}[-i]-\boldsymbol{\mu}[-i]\Big\vert\boldsymbol{\xi}({a}_{i}-{\mu}_{i}), \Delta_{ii}\Bigr)\biggr]\times \Bigl[a_i-\mu_i-  \Sigma_{-i,i}\bigl(\Delta_{ii}\bigr)^{-1}\boldsymbol{\Omega}_{a}\Bigr],
\end{align}
where 
\begin{align*}%\label{second-truncated-moment-multivariate-Gaussian-case3-part6}
\boldsymbol{\Omega}_{a}=E\Bigl(\boldsymbol{Y}_{a}\Big \vert \boldsymbol{a}[-i]-\boldsymbol{\mu}[-i]-\boldsymbol{\xi}({a}_{i}-{\mu}_{i})<\boldsymbol{Y}_{a}<\boldsymbol{b}[-i]-\boldsymbol{\mu}[-i]-\boldsymbol{\xi}({a}_{i}-{\mu}_{i})\Bigr),
\end{align*}
in which $\boldsymbol{Y}_{a}\sim {\cal{N}}_{p-1}\bigl(\boldsymbol{0}, \Delta_{ii}\bigr)$. For computing $\partial {I}_{ib}/\partial {\mu}_{i}$, we likewise have
%\begin{align}\label{second-truncated-moment-multivariate-Gaussian-case3-part6}
%\frac{\partial {I}_{ib}}{\partial {\mu}_{i}}=&\frac{\bigl(b_{i}- {\mu}_{i}\bigr)}{\Sigma_{i,i}\times{\cal{P}}_{G}} {\phi}\Bigl(b_{i}\Big \vert {\mu}_{i}, \Sigma_{i,i}\Bigr) \biggl[\boldsymbol{\Phi}_{p-1}\Bigl(\boldsymbol{b}[-i]-\boldsymbol{\mu}[-i]\Big\vert\boldsymbol{\xi}({b}_{i}-{\mu}_{i}),\Delta_{ii}\Bigr)-\nonumber\\
%&\boldsymbol{\Phi}_{p-1}\Bigl(\boldsymbol{a}[-i]-\boldsymbol{\mu}[-i]\Big\vert\boldsymbol{\xi}({b}_{i}-{\mu}_{i}), \Delta_{ii}\Bigr)\biggr]\nonumber\\
%&-\frac{P(\boldsymbol{a}_{iU}<\boldsymbol{Y}<\boldsymbol{b}_{iU})}{\Sigma_{i,i}\times{\cal{P}}_{G}} {\phi}\Bigl(b_{i}\Big \vert {\mu}_{i}, \Sigma_{i,i}\Bigr)\times  \Sigma_{-i,i}\bigl(\Delta_{ii}\bigr)^{-1}\boldsymbol{\Omega}_{U},
%\end{align}
\begin{align}\label{second-truncated-moment-multivariate-Gaussian-case3-part7}
\frac{\partial {I}_{ib}}{\partial {\mu}_{i}}=&\frac{1}{\Sigma_{i,i}\times{\cal{P}}_{G}} {\phi}\Bigl(b_{i}\Big \vert {\mu}_{i}, \Sigma_{i,i}\Bigr) \biggl[\boldsymbol{\Phi}_{p-1}\Bigl(\boldsymbol{b}[-i]-\boldsymbol{\mu}[-i]\Big\vert\boldsymbol{\xi}({b}_{i}-{\mu}_{i}),\Delta_{ii}\Bigr)-\nonumber\\
&\boldsymbol{\Phi}_{p-1}\Bigl(\boldsymbol{a}[-i]-\boldsymbol{\mu}[-i]\Big\vert\boldsymbol{\xi}({b}_{i}-{\mu}_{i}), \Delta_{ii}\Bigr)\biggr]\times \Bigl[b_i-\mu_i-  \Sigma_{-i,i}\bigl(\Delta_{ii}\bigr)^{-1}\boldsymbol{\Omega}_{b}\Bigr],
\end{align}
where
\begin{align*}%\label{second-truncated-moment-multivariate-Gaussian-case3-part8}
\boldsymbol{\Omega}_{b}=E\Bigl(\boldsymbol{Y}_{b}\Big \vert \boldsymbol{a}[-i]-\boldsymbol{\mu}[-i]-\boldsymbol{\xi}({b}_{i}-{\mu}_{i})<\boldsymbol{Y}_{b}<\boldsymbol{b}[-i]-\boldsymbol{\mu}[-i]-\boldsymbol{\xi}({b}_{i}-{\mu}_{i})\Bigr),
\end{align*}
in which $\boldsymbol{Y}_{b}\sim {\cal{N}}_{p-1}\bigl(\boldsymbol{0}, \Delta_{ii}\bigr)$. Once we have computed $\partial {I}_{ia}/\partial {\mu}_{i}$ in (\ref{second-truncated-moment-multivariate-Gaussian-case3-part5}) and $\partial {I}_{ib}/\partial {\mu}_{i}$ in (\ref{second-truncated-moment-multivariate-Gaussian-case3-part7}), for $i=1,\cdots,p$, then the diagonal elements of $\boldsymbol{I}$ given by (\ref{second-truncated-moment-multivariate-Gaussian-part2}) are available.
\end{itemize}
Once we have construct matrix $\boldsymbol{I}$, we can compute $\boldsymbol{\Omega}^2$ by substituting computed $\boldsymbol{I}$ at the RHS of (\ref{second-truncated-moment-multivariate-Gaussian-part2}). 
\par Herein, we carry out a small simulation study for computing the first moment of vector $(X,Y)^{\top}$ under two scenarios. Within the first scenario, we assume that $(X,Y)^{\top}$ follows truncated bivariate Gaussian distribution with parameters $\boldsymbol{\mu}=\boldsymbol{0}$, $\Sigma~\bigl(\sigma_{1}^{2}=2, \sigma_{2}^{2}=1, \rho=0.3535\bigr)$ truncated on area with lower limits $\boldsymbol{a}=(-1,-1)^{\top}$ and upper limits $\boldsymbol{b}=(2,2)^{\top}$. Under the second scenario it is assumed that $(X,Y)^{\top}$ follows truncated bivariate Gaussian distribution with parameters $\boldsymbol{\mu}=\boldsymbol{0}$, $\Sigma$ $\bigl(\sigma_{1}^{2}=2, \sigma_{2}^{2}=2, \rho=0.25\bigr)$ truncated on area with lower limits $\boldsymbol{a}=(0,0)^{\top}$ and upper limits $\boldsymbol{b}=(+\infty,+\infty)^{\top}$. We note that the concern of the second scenario is placed on computing the absolute moment, that is $\boldsymbol{\Omega}=E(\vert\boldsymbol{X}\vert)$. For both scenario, we compute $\boldsymbol{\Omega}$ by generating samples of sizes $N=2000, 5000$, and 20000 based on 5000 runs. The instrumental distribution is set to be ${\cal{N}}_{2}(\boldsymbol{\mu},\Sigma)$. For computing $\boldsymbol{\Omega}=E\bigl(\boldsymbol{{\cal{X}}}_{a,b}\bigr)$ through the ${\cal{N}}_{2}(\boldsymbol{\mu},\Sigma)$, one can follow two ways. In the first way the quantity $\boldsymbol{\Omega}$ is estimated as
\begin{align}
\hat{\boldsymbol{\Omega}}=\frac{1}{N}\sum_{i=1}^{N} \vert \boldsymbol{x}_{i}\big \vert
\end{align}
where $\big \vert \boldsymbol{x}_{i}\big \vert=\bigl(\big \vert x_{1i}\big \vert, \big \vert x_{2i}\big \vert \bigr)^{\top}$ where $\boldsymbol{x}_{i}$s come independently from ${\cal{N}}_{2}(\boldsymbol{\mu},\Sigma)$, for sufficiently large $N$. The second method proceeds to compute $\hat{\boldsymbol{\Omega}}$ based on sample of size $N$ when the lower and upper truncation bounds are $\boldsymbol{a}=\boldsymbol{0}$ and $\boldsymbol{b}=+\boldsymbol{\infty}$, respectively. Details for implementing the second method are given as follows. The corresponding \verb+R+ function \verb+EXX+ for computing $\boldsymbol{\Omega}$ is given by the following. 
\begin{lstlisting}[style=deltaj]
R> library(mvtnorm)  # for computing multivariate Gaussian CDF 
R> EXX <- function(Mu, Sigma, a, b)
+{
+ if(any(a>b) == TRUE) stop(message("vector a must be smaller than vector b."))
+ p <- length(Mu)
+ PG <- pmvnorm(lower = a, upper = b, mean = Mu, sigma = Sigma)[1]
+ II <- matrix(0, nrow = p, ncol = p)
+ if( p == 2 )
+ {
+ for(i in 1:p){
+ 	Mu_ia <- Sigma[i, -i]*(a[i] - Mu[i])/Sigma[i, i] 
+ 	Mu_ib <- Sigma[i, -i]*(b[i] - Mu[i])/Sigma[i, i]
+ 	Delta_i <- Sigma[-i, -i] - Sigma[i, -i]*Sigma[i, -i]/Sigma[i, i]
+ 	La_i <- a[-i] - Mu[-i] - Mu_ia	
+ 	Ua_i <- b[-i] - Mu[-i] - Mu_ia	
+ 	Lb_i <- a[-i] - Mu[-i] - Mu_ib	
+ 	Ub_i <- b[-i] - Mu[-i] - Mu_ib	
+  PG_ia <- pnorm( Ua_i, mean = rep(0, p-1), sd = sqrt( Delta_i) )-
+           pnorm( La_i, mean = rep(0, p-1), sd = sqrt( Delta_i) )
+  PG_ib <- pnorm( Ub_i, mean = rep(0, p-1), sd = sqrt( Delta_i) )-
+           pnorm( Lb_i, mean = rep(0, p-1), sd = sqrt( Delta_i) )
+	Ex_ia <- Delta_i*( dnorm(La_i, 0, sqrt(Delta_i)) - 
+  										dnorm(Ua_i, 0, sqrt(Delta_i)))/PG_ia
+	Ex_ib <- Delta_i*( dnorm(Lb_i, 0, sqrt(Delta_i)) - 
+ 										dnorm(Ub_i, 0, sqrt(Delta_i)))/PG_ib
+  p1 <-(a[i]-Mu[i])/(PG*Sigma[i, i])*dnorm(a[i]-Mu[i],0,sqrt(Sigma[i, i]) )*(
+        pnorm(b[-i] - Mu[-i], mean = Mu_ia, sd = sqrt( Delta_i) )-
+        pnorm(a[-i] - Mu[-i], mean = Mu_ia, sd = sqrt( Delta_i) ) )-
+  PG_ia/PG*dnorm(a[i] - Mu[i], 0, sqrt(Sigma[i, i]) )*
+  Sigma[i,-i]*Ex_ia/(Delta_i*Sigma[i, i])
+  p2 <- -(b[i]-Mu[i])/(PG*Sigma[i, i])*dnorm(b[i]-Mu[i],0,sqrt(Sigma[i, i]))*(
+        pnorm(b[-i] - Mu[-i], mean = Mu_ib, sd = sqrt( Delta_i) )-
+        pnorm(a[-i] - Mu[-i], mean = Mu_ib, sd = sqrt( Delta_i) ) ) +
+ 	PG_ib/PG*dnorm(b[i] - Mu[i], 0, sqrt(Sigma[i, i]) )*
+ 	Sigma[i,-i]*Ex_ib/(Delta_i*Sigma[i, i])
+  II[i, i] <- p1 + p2
+ }
+ x_aa <- a - Mu
+ x_ab <- c(a[1] - Mu[1], b[2] - Mu[2])
+ x_ba <- c(b[1] - Mu[1], a[2] - Mu[2])
+ x_bb <- b - Mu
+ II[1, 2] <-( dmvnorm(x_bb, mean = rep(0, p), sigma =  Sigma )-
+       	dmvnorm(x_ab, mean = rep(0, p), sigma =  Sigma )-
+ 	dmvnorm(x_ba, mean = rep(0, p), sigma =  Sigma )+
+       	dmvnorm(x_aa, mean = rep(0, p), sigma =  Sigma ))/PG
+ II[2, 1] <- II[1, 2]
+ }
+ if( p == 3 )
+ {
+ for(i in 1:(p - 1)){
+ for(j in (i + 1):p){
+ I_ij <- c(i, j)
+ Sigma_ij <- Sigma[I_ij, I_ij]
+ I0 <- Sigma[I_ij, -I_ij]%*%solve(Sigma_ij)
+ x_aa <- c(a[i] - Mu[i], a[j] - Mu[j])
+ x_ab <- c(a[i] - Mu[i], b[j] - Mu[j])
+ x_ba <- c(b[i] - Mu[i], a[j] - Mu[j])
+ x_bb <- c(b[i] - Mu[i], b[j] - Mu[j])
+ Mu_aa <- (I0%*%x_aa)[1]
+ Mu_ab <- (I0%*%x_ab)[1]
+ Mu_ba <- (I0%*%x_ba)[1]
+ Mu_bb <- (I0%*%x_bb)[1]
+ Delta_ij <- Sigma[-I_ij, -I_ij] - ( I0%*%Sigma[-I_ij, I_ij] )[1]
+ p1 <- dmvnorm(x_bb, mean = rep(0, 2), sigma = Sigma_ij)*(
+ pnorm(b[-I_ij]  - Mu[-I_ij], mean = Mu_bb, sd = sqrt( Delta_ij ) )-
+ pnorm(a[-I_ij]  - Mu[-I_ij], mean = Mu_bb, sd = sqrt( Delta_ij ) ) )
+ p2 <- dmvnorm(x_ab, mean = rep(0, 2), sigma = Sigma_ij)*(
+ pnorm(b[-I_ij]  - Mu[-I_ij], mean = Mu_ab, sd = sqrt( Delta_ij ) )-
+ pnorm(a[-I_ij]  - Mu[-I_ij], mean = Mu_ab, sd = sqrt( Delta_ij ) ) )
+ p3 <- dmvnorm(x_ba, mean = rep(0, 2), sigma = Sigma_ij)*(
+ pnorm(b[-I_ij]  - Mu[-I_ij], mean = Mu_ba, sd = sqrt( Delta_ij ) )-
+ pnorm(a[-I_ij]  - Mu[-I_ij], mean = Mu_ba, sd = sqrt( Delta_ij ) ) )
+ p4 <- dmvnorm(x_aa, mean = rep(0, 2), sigma = Sigma_ij)*(
+ pnorm(b[-I_ij]  - Mu[-I_ij], mean = Mu_aa, sd = sqrt( Delta_ij ) )-
+ pnorm(a[-I_ij]  - Mu[-I_ij], mean = Mu_aa, sd = sqrt( Delta_ij ) ) )
+ II[i, j] <- (p1 - p2 - p3 + p4)/PG
+ II[j, i] <- II[i, j]
+ }
+ }
+ for(i in 1:p){
+ 	Mu_ia <- Sigma[i, -i]*(a[i] - Mu[i])/Sigma[i, i] 
+ 	Mu_ib <- Sigma[i, -i]*(b[i] - Mu[i])/Sigma[i, i]
+ 	Delta_i <- Sigma[-i, -i] - Sigma[i, -i]%*%t(Sigma[i, -i])/Sigma[i, i]
+ 	La_i <- a[-i] - Mu[-i] - Mu_ia
+ 	Ua_i <- b[-i] - Mu[-i] - Mu_ia	
+ 	Lb_i <- a[-i] - Mu[-i] - Mu_ib	
+ 	Ub_i <- b[-i] - Mu[-i] - Mu_ib	
+ PG_ia <- pmvnorm(lower=La_i, upper=Ua_i, mean=rep(0, p-1), sigma=Delta_i)[1]
+ PG_ib <- pmvnorm(lower=Lb_i, upper=Ub_i, mean=rep(0, p-1), sigma=Delta_i)[1]
+ p5 <- (a[i]-Mu[i])/(PG*Sigma[i, i])*dnorm(a[i]-Mu[i],0,sqrt(Sigma[i, i]) )*
+       pmvnorm(a[-i] - Mu[-i], b[-i] - Mu[-i], mean = Mu_ia, sigma = Delta_i)[1] -
+ PG_ia/PG*dnorm(a[i]-Mu[i],0,sqrt(Sigma[i, i]) )*(t(Sigma[i, -i])%*%
+ solve(Delta_i)%*%EX( rep(0,p-1), Delta_i, La_i, Ua_i)/Sigma[i, i])[1]
+ p6 <- -(b[i]-Mu[i])/(PG*Sigma[i, i])*dnorm(b[i]-Mu[i],0,sqrt(Sigma[i, i]) )*
+        pmvnorm(a[-i] - Mu[-i], b[-i] - Mu[-i], mean = Mu_ib, sigma = Delta_i)[1] +
+ PG_ib/PG*dnorm(b[i] - Mu[i],0, sqrt(Sigma[i, i]) )*(t(Sigma[i, -i])%*%
+ solve(Delta_i)%*%EX(rep(0,p-1), Delta_i, Lb_i, Ub_i)/Sigma[i, i])[1]
+ II[i, i] <- p5 + p6
+ }
+ }
+ if ( p > 3 )
+ {
+ for(i in 1:(p - 1)){
+ for(j in (i + 1):p){
+ I_ij <- c(i, j)
+ Sigma_ij <- Sigma[I_ij, I_ij]
+ I0 <- Sigma[-I_ij, I_ij]%*%solve(Sigma_ij)
+ x_aa <- c( a[i] - Mu[i], a[j] - Mu[j] )
+ x_ab <- c( a[i] - Mu[i], b[j] - Mu[j] )
+ x_ba <- c( b[i] - Mu[i], a[j] - Mu[j] )
+ x_bb <- c( b[i] - Mu[i], b[j] - Mu[j] )
+ Mu_aa <- c( I0%*%x_aa )
+ Mu_ab <- c( I0%*%x_ab )
+ Mu_ba <- c( I0%*%x_ba )
+ Mu_bb <- c( I0%*%x_bb )
+ Delta_ij <- Sigma[-I_ij, -I_ij] - ( I0%*%Sigma[-I_ij, I_ij] )[1]
+ p1 <- dmvnorm(x_bb, mean = rep(0, 2), sigma = Sigma_ij)*
+ pmvnorm(a[-I_ij]-Mu[-I_ij], b[-I_ij]-Mu[-I_ij], mean = Mu_bb, sigma=Delta_ij )
+ p2 <- dmvnorm(x_ab, mean = rep(0, 2), sigma = Sigma_ij)*
+ pmvnorm(a[-I_ij]-Mu[-I_ij], b[-I_ij]-Mu[-I_ij], mean = Mu_ab, sigma=Delta_ij )
+ p3 <- dmvnorm(x_ba, mean = rep(0, 2), sigma = Sigma_ij)*
+ pmvnorm(a[-I_ij]-Mu[-I_ij], b[-I_ij]-Mu[-I_ij], mean = Mu_ba, sigma=Delta_ij )
+ p4 <- dmvnorm(x_aa, mean = rep(0, 2), sigma = Sigma_ij)*
+ pmvnorm(a[-I_ij]-Mu[-I_ij], b[-I_ij]-Mu[-I_ij], mean = Mu_aa, sigma=Delta_ij )
+ II[i, j] <- (p1 - p2 - p3 + p4)/PG
+ II[j, i] <- II[i, j]
+ }
+ }
+ for(i in 1:p){
+ 	Mu_ia <- Sigma[i, -i]*(a[i] - Mu[i])/Sigma[i, i] 
+ 	Mu_ib <- Sigma[i, -i]*(b[i] - Mu[i])/Sigma[i, i]
+ 	Delta_i <- Sigma[-i, -i] - Sigma[i, -i]%*%t(Sigma[i, -i])/Sigma[i, i]
+ 	La_i <- a[-i] - Mu[-i] - Mu_ia
+ 	Ua_i <- b[-i] - Mu[-i] - Mu_ia	
+ 	Lb_i <- a[-i] - Mu[-i] - Mu_ib	
+ 	Ub_i <- b[-i] - Mu[-i] - Mu_ib	
+ PG_ia <- pmvnorm(lower=La_i, upper=Ua_i, mean=rep(0, p-1), sigma=Delta_i)[1]
+ PG_ib <- pmvnorm(lower=Lb_i, upper=Ub_i, mean=rep(0, p-1), sigma=Delta_i)[1]
+ p5 <- (a[i]-Mu[i])/(PG*Sigma[i, i])*dnorm(a[i] - Mu[i], 0, sqrt(Sigma[i, i]))*
+       pmvnorm(a[-i] - Mu[-i], b[-i] - Mu[-i], mean=Mu_ia, sigma=Delta_i)[1] -
+ PG_ia/PG*dnorm(a[i] - Mu[i], 0, sqrt(Sigma[i, i]) )*(t(Sigma[i, -i])%*%
+ solve(Delta_i)%*%EX( rep(0, p - 1), Delta_i, La_i, Ua_i)/Sigma[i, i])[1]
+ p6 <- -(b[i]-Mu[i])/(PG*Sigma[i, i])*dnorm(b[i]-Mu[i],0,sqrt(Sigma[i, i]) )*
+        pmvnorm(a[-i] - Mu[-i], b[-i] - Mu[-i], mean = Mu_ib, sigma = Delta_i)[1] +
+ PG_ib/PG*dnorm(b[i] - Mu[i], 0, sqrt(Sigma[i, i]))*(t(Sigma[i, -i])%*%
+ solve(Delta_i)%*%EX(rep(0, p - 1), Delta_i, Lb_i, Ub_i)/Sigma[i, i])[1]
+ II[i, i] <- p5 + p6
+ }
+ }
+ Omega <- EX(Mu, Sigma, a, b)
+ EXX <- Omega%*%Mu + t(Omega%*%Mu) - Mu%*%t(Mu) + Sigma + Sigma%*%II%*%Sigma
+ return(EXX)
+ }
\end{lstlisting}
%%%%%%%%%%%%%%%%%%%%%
\section{Truncated skew Gaussian distribution}

Here, 
%we consider to draw Bayesian inference on parameters of a model which contains two latent variable. To this end, first, 
we introduce the class of skew Gaussian distributions that is known in the literature as the canonical fundamental unrestricted skew Gaussian distribution, see \cite{arellano2006unification}.
We write $\boldsymbol{X}\sim \text{SG}_{p,q}(\boldsymbol{\mu},\Sigma,{\Lambda})$ to denote that $p$-dimensional random vector $\boldsymbol{X}$ follows a canonical fundamental unrestricted skew Gaussian distribution with PDF given by  
\begin{align}\label{ursn}
f_{}(\boldsymbol{x}\big \vert \boldsymbol{\mu},{\Sigma},{{\Lambda}}) =2^q \boldsymbol{\phi}_{p}\bigl(\boldsymbol{x}\vert\boldsymbol{\mu},{\Omega}\bigr)
\boldsymbol{\Phi}_{q}\bigl(\boldsymbol{m}\vert\boldsymbol{0}_{q},{\Delta}\bigr),
\end{align}
where ${\Omega}={\Sigma}+{{\Lambda}}{{\Lambda}}^{\top}$, ${\Delta}=\boldsymbol{I}_q-{{\Lambda}}^{\top}{\Omega}^{-1}{{\Lambda}}$, $\boldsymbol{m}={\Lambda}^{\top}{{\Omega}}^{-1}(\boldsymbol{y}-\boldsymbol{\mu})$, and $\boldsymbol{I}_{q}$ denotes the
$q \times q$ identity matrix. Further, $\boldsymbol{\phi}_{p}\bigl(.\big \vert \boldsymbol{\mu},{\Omega}\bigr)$ denotes the PDF of a $p$-dimensional Gaussian distribution with location vector $\boldsymbol{\mu}$ and dispersion matrix $\Sigma$, and $\boldsymbol{\Phi}_{q}\bigl(.\big\vert\boldsymbol{0}_{q}, {\Delta}\bigr)$ is the CDF of a $q$-dimensional Gaussian distribution with location vector $\boldsymbol{0}_{q}$ (a vector of zeros of length $q$) and dispersion matrix ${\Delta}$. The random vector $\boldsymbol{X}$ admits stochastic representation as follows, see \citep{arellano2005fundamental,arellano2006unification,arellano2007bayesian}.
\begin{align}\label{ursnrep0}
\boldsymbol{X} \mathop=\limits^d\boldsymbol{\mu}+{\Lambda}\big \vert \boldsymbol{Z}_0 \big \vert+ {\Sigma}^{\frac{1}{2}}\boldsymbol{Z}_1,
\end{align}
where $\mathop=\limits^d$ means ``distributed as'' and random vectors $\boldsymbol{Z}_0\sim {\cal{N}}_{q}\bigl(\boldsymbol{0},\boldsymbol{I}_{q}\bigr)$ and $\boldsymbol{Z}_1\sim {\cal{N}}_{q}\bigl(\boldsymbol{0},\boldsymbol{I}_{p}\bigr)$ are independent.
%\begin{align}\label{ursnrep}
%\biggl[\begin{matrix}
%\boldsymbol{Z}_0\\ \boldsymbol{Z}_1
%\end{matrix}\biggr]
%\sim {\cal{N}}_{q+p}\biggl(\biggl[\begin{matrix}
%\boldsymbol{0}_{q}\\ \boldsymbol{0}_{p}
%\end{matrix}\biggr],
%\biggl[\begin{matrix}
%\boldsymbol{I}_{q}&\boldsymbol{0}_{q\times p}\\
%\boldsymbol{0}_{p\times q}& {\Sigma}\\
%\end{matrix}\biggr]
%\biggr).
%\end{align}
Furthermore, it can be seen that \citep{lee2016afinite,maleki2019robust,morales2022moments}  
\begin{align}
\boldsymbol{X}\mathop=\limits^d \boldsymbol{\mu} + \bigl(\boldsymbol{W}\big \vert \boldsymbol{W}_{0}>\boldsymbol{0}\bigr),
\end{align}
in which
\begin{align}\label{ursnrep}
\boldsymbol{Y}=\biggl[\begin{matrix}
\boldsymbol{W}_0\\ \boldsymbol{W}_{1}
\end{matrix}\biggr]
\sim {\cal{N}}_{q+p}\biggl(\biggl[\begin{matrix}
\boldsymbol{0}_{q}\\ \boldsymbol{0}_{p}
\end{matrix}\biggr],
\biggl[\begin{matrix}
\boldsymbol{I}_{q}&\boldsymbol{\Lambda}^{\top}\\
\boldsymbol{\Lambda}& {\Sigma}+\boldsymbol{\Lambda}\boldsymbol{\Lambda}^{\top}\\
\end{matrix}\biggr]
\biggr).
\end{align}
By the property (\ref{ursnrep}), we can construct a method for simulating from $\boldsymbol{X}\big \vert \boldsymbol{a}<\boldsymbol{X}<\boldsymbol{b}$ when $\boldsymbol{X}$ admits relation (\ref{ursnrep0}). To this end, we need to simulate from $(p+q)$-dimensional Gaussian random vector $\boldsymbol{Y}$ given by (\ref{ursnrep}). Some algebra show
\begin{align*}
\boldsymbol{X}\big \vert \bigl(\boldsymbol{a}<\boldsymbol{X}<\boldsymbol{b}\bigr) \mathop=\limits^d
\boldsymbol{\mu}^{*}+ \boldsymbol{Y}\big \vert \bigl(\boldsymbol{a}^{*}-\boldsymbol{\mu}^{*}<\boldsymbol{Y}<\boldsymbol{b}^{*}-\boldsymbol{\mu}^{*}\bigr),
\end{align*}
where $a^{*}=\bigl(\boldsymbol{0}^{\top}_{q}, \boldsymbol{a}^{\top}\bigr)^{\top}$, $b^{*}=\bigl( \boldsymbol{\infty}^{\top}_{q},\boldsymbol{b}^{\top}\bigr)^{\top}$, and $\boldsymbol{\mu}^{*}=\bigl( \boldsymbol{0}^{\top}_{q},\boldsymbol{\mu}^{\top}\bigr)^{\top}$ in which $\boldsymbol{\infty}_{q}$ is a vector of length $q$ of infinities.
% =(\infty,\cdots,\infty_\text{q})$. 
Hence, using the methodology given in Sections \ref{first-moment-truncated-multivariate-Gaussian-distribution} and \ref{second-moment-truncated-multivariate-Gaussian-distribution} for computing the first and second moments of multivariate Gaussian distribution, we can compute the first and second moments of truncated skew Gaussian distribution by computing $E\bigl( \boldsymbol{Y}\big \vert \boldsymbol{a}^{*}<\boldsymbol{Y}<\boldsymbol{b}^{*}\bigr)$ and $E\bigl(\boldsymbol{Y}\boldsymbol{Y}^{\top}\big \vert \boldsymbol{a}^{*}<\boldsymbol{Y}<\boldsymbol{b}^{*}\bigr)$, respectively. We note that an \verb+R+ package called \verb+MomTrunc+ uploaded at \verb+https://cran.r-project.org/web/packages/MomTrunc/inde+\\ \verb+x.html+ developed for this aim \citep{morales2022moments}. The following \verb+R+ code can be used for computing first two moments of the truncated random vector $\boldsymbol{X}$ where $\boldsymbol{X}\sim \text{SG}_{2,2}(\boldsymbol{\mu},\Sigma, {\Lambda})$ truncated on $(\boldsymbol{a},\boldsymbol{b})$ in which $\boldsymbol{\mu}=(0,0)^{\top}$, $\boldsymbol{a}=(-3, -3)^{\top}$, $\boldsymbol{b}=(2, 2)^{\top}$,
\begin{align*}
\Sigma=\left[\begin{matrix} 
2&0.5\\
0.5&2\\
\end{matrix}\right],
\Lambda=\left[\begin{matrix} 
4&2\\
2&-5\\
\end{matrix}\right].
\end{align*}
\begin{lstlisting}[style=deltaj]
library("MomTrunc")
a <- c(-3, -3); b <- c(2, 2); Mu <- c(0, 0); p <- 2; q<- 2
a_star <- c(0, 0, a); b_star  <- c(Inf, Inf, b); Mu_star  <- c(0, 0, Mu)
Sigma <- matrix( c(2, 0.5, 0.5, 2), nrow = p , ncol = p)
ch <- t(chol(Sigma));  Lambda <- matrix( c(4, 2 ,2, -5 ), nrow = p , ncol = q) 
p1 <- diag(q); p2 <- t(Lambda);  p3 <- t(p2); p4 <- Sigma + p3%*%t(p3)
Sigma_star <- matrix( 
      c(p1[1,1],p1[1,2], p2[1,1], p2[1,2],
        p1[2,1],p1[2,2], p2[2,1], p2[2,2],
        p3[1,1],p3[1,2], p4[1,1], p4[1,2],
	p3[2,1],p3[2,2], p4[2,1], p4[2,2]), nrow = p + q, ncol = p + q)
out <- MCmeanvarTMD(a_star, b_star, rep(0, p + q), Sigma_star, dist = "normal")
out$mean[(p+1):(p+q)]
out$EYY[(p+1):(p+q), (p+1):(p+q)]
\end{lstlisting}

%%%%%%%%%%%%%%%%%%%%%
\section{Gaussian scale mixture model}\label{GSM}
Here, we introduce the useful property known in the literature as {\it{Gaussian scale mixture model}}. Based on this property, one can represent the PDF and other characteristics of a wide range of distributions in terms of the Gaussian model.
\begin{dfn}\label{dfn-Gaussian-scale-mixture-model-univariate}
Let random variable $Z$  follows a standard Gaussian distribution. For given positive valued function $h(\cdot)$ and positive random variable $G$, the random variable $X$ is said to follow a Gaussian scale mixture model if we can write
\begin{align}
X=\mu + \sigma \frac{ Z}{h(G)},
\end{align}
where $\mu \in \mathbb{R}$ and $\sigma\in \mathbb{R}^{+}$ are the location and scale parameters, respectively. Herein, and random variable $G$ is the mixing variable.
\end{dfn}
Likewise, the multivariate version of the Gaussian scale mixture model (representation) is given as follows.
\begin{dfn}\label{dfn-Gaussian-scale-mixture-model-multivariate} Let random variable $\boldsymbol{Z}$ follows a $p$-dimensional standard Gaussian distribution. For given positive valued function $h(\cdot)$ and positive random variable $G$. The $p$-dimensional random vector $\boldsymbol{X}$ is said to follow a Gaussian scale mixture model if we have
\begin{align}
\boldsymbol{X}=\boldsymbol{\mu} + A \frac{\boldsymbol{Z}}{h(G)}
\end{align}
where $\boldsymbol{\mu} \in \mathbb{R}^{p}$ and $p \times p$ lower triangular matrix $A$ is called the Cholesky decomposition of positive definite scale matrix $\Sigma$ such that $\Sigma= AA^{\top}$. Herein, the univariate random variable $G$ plays the role of the mixing variable. 
\end{dfn}
The well-known example of Gaussian scale mixture model is the Student's $t$ distribution, with PDF given by (\ref{PDF-t}), for which $h(G)=\sqrt{G}$ where $G\sim {\cal{G}}(\nu/2,\nu/2)$.

\section{Moment of truncated Student's $t$ distribution}
Let random vector $\boldsymbol{X}=(X_1,\cdots,X_p)^{\top}$ follows a $p$-dimensional Student's $t$ distribution with $\nu>0$ degrees of freedom with PDF given by
\begin{align}\label{PDF-t}
f(\boldsymbol{x} \big \vert \boldsymbol{\Psi}\bigr)=&\frac{\Gamma \bigl(\frac{\nu+p}{2}\bigr)}{(\nu \pi)^{\frac{p}{2}}\Gamma \bigl(\frac{\nu}{2}\bigr)\vert \Sigma \vert ^{\frac{1}{2}}} \biggl[ 1 +\frac{\delta(\boldsymbol{x}-\boldsymbol{\mu},\Sigma)}{\nu}\biggr]^{-\frac{\nu+p}{2}},
\end{align}
where $\boldsymbol{\Psi}=(\boldsymbol{\mu},\Sigma, \nu)$ is the parameter vector and Mahalanobis distance $\delta(\cdot)$ is defined in (\ref{delta}). By definition
\begin{align}\label{moment-t-distribution1}
E(\boldsymbol{X} )=&\int_{-\boldsymbol{\infty}}^{\boldsymbol{\infty}}\frac{\boldsymbol{x} \Gamma \bigl(\frac{\nu+p}{2}\bigr)}{(\nu \pi)^{\frac{p}{2}}\Gamma \bigl(\frac{\nu}{2}\bigr)\vert \Sigma \vert ^{\frac{1}{2}}} \biggl[ 1 +\frac{\delta(\boldsymbol{x}-\boldsymbol{\mu},\Sigma)}{\nu}\biggr]^{-\frac{\nu+p}{2}} d \boldsymbol{x},
\end{align}
for $\nu >1$. The first moment of the truncated Student's $t$ distribution (shown by ${\cal{TT}}_{p}(\boldsymbol{\mu},\Sigma,\nu,\boldsymbol{a},\boldsymbol{b})$) can be evaluated through relation (\ref{truncated-moment-part4}) in which $f(\boldsymbol{u})$ is replaced with $f(\boldsymbol{x} \big \vert \boldsymbol{\Psi}\bigr)$ given in (\ref{PDF-t}) and $F(\boldsymbol{b})-F(\boldsymbol{a})=P(\boldsymbol{a}<\boldsymbol{X}<\boldsymbol{b})=\int_{\boldsymbol{a}}^{\boldsymbol{b}}f(\boldsymbol{x} \big \vert \boldsymbol{\Psi}\bigr)d\boldsymbol{x}$. However, herein, we proceed to compute $\boldsymbol{\Omega}=E\bigl(\boldsymbol{{{X}}}\big \vert \boldsymbol{a}<\boldsymbol{X}<\boldsymbol{b}\bigr)$ based on the property of {\it{Gaussian scale mixture model}} introduced in Section \ref{GSM}.
\subsection{First two moments of truncated Student's $t$ distribution distribution in univariate case}\label{first-moment-truncated-univariate-Gaussian-distribution}
Herein, we proceed to compute the first and second moments of an univariate Student's $t$ distribution. For simplicity, we use generic symbols ${t}\bigl({x}\big \vert {\mu}, \sigma^2, \nu\bigr)$ and ${T}\bigl({x}\big \vert {\mu}, \sigma^2, \nu\bigr)$ to denote, accordingly, the PDF and CDF of a Student's $t$ distribution with $\nu$ degrees of freedom at point ${x}$ for which ${\mu}$, $\sigma$, and $\nu$ play the role of location, scale, and degrees of freedom parameters, respectively. Furthermore, we write ${X} \sim {t}({\mu}, \sigma^2, \nu)$ to show that random variable ${X}$ follows a Student's $t$ distribution with location parameter ${\mu}$, scale parameter $\sigma$, and $\nu$ degrees of freedom. The corresponding truncated family on $(a,b)$ is represented by ${\cal{TT}}\bigl({\mu}, \sigma^2, \nu, a, b\bigr)$. 
\par If $X \sim t\bigl(\mu, \sigma^2, \nu\bigr)$, it can be seen from (\ref{truncated-moment-part2}) for $m=1$ that
\begin{align}\label{first-truncated-moment-univariate-t-part1}
\Omega=&\frac{1}{{\cal{P}}_{t}}\int_{a}^{b}x \times {t}\bigl({x}\big \vert {\mu}, \sigma^2, \nu\bigr)dx,\nonumber\\
=&\frac{1}{{\cal{P}}_{t}}\int_{a}^{b}\frac{x \Gamma \bigl(\frac{\nu+1}{2}\bigr)}{\sqrt{\nu \pi}\Gamma \bigl(\frac{\nu}{2}\bigr) \sigma } \biggl[ 1 +\frac{({x}-{\mu})^2}{\nu\sigma^2}\biggr]^{-\frac{\nu+1}{2}} d{x},
\end{align}
where ${\cal{P}}_{t}=T\bigl({b}\big \vert {\mu}, \sigma^2, \nu\bigr)-{T}\bigl({a}\big \vert {\mu}, \sigma^2, \nu\bigr)$. Rather computing $\Omega$ through (\ref{first-truncated-moment-univariate-t-part1}), we prefer to use Definition \ref{dfn-Gaussian-scale-mixture-model-univariate} that states each Student's $t$ distribution is a Gaussian scale mixture model. In fact ${X}$ admits the hierarchy given by
\begin{align*}%\label{hierarchy-t-distribution}
{X}\vert G &\sim {\cal{N}}\Bigl({\mu}, \frac{\sigma^2}{{G}}\Bigr),\nonumber\\
G&\sim{\cal{G}}\Bigl(\frac{\nu}{2},\frac{\nu}{2}\Bigr).
\end{align*}
Using above hierarchy, the quantity $\Omega$ can be computed as follows.
\begin{align*}%\label{moment-t-distribution2}
\Omega&=\frac{1}{{\cal{P}}_{t}}\int_{a}^{b}{x}
\int_{0}^{\infty}{\phi}\Bigl({x}\Big \vert \boldsymbol{\mu},\frac{\sigma^2}{g}\Bigr)
{\cal{G}}\Bigl(g\Big \vert \frac{\nu}{2},\frac{\nu}{2}\Bigr)dgd {x},\nonumber\\
&=\frac{1}{C_{G}{\cal{P}}_{t}}\int_{0}^{\infty}g^{\frac{1}{2}}\int_{a}^{b}{x} \exp\Bigl\{-\frac{g(x-\mu)^2}{2\sigma^2}\Bigr\}dx{\cal{G}}\Bigl(g\Big \vert \frac{\nu}{2},\frac{\nu}{2}\Bigr)dg,\nonumber\\
&=\frac{\sigma^2}{C_{G}{\cal{P}}_{t}}\int_{0}^{\infty}g^{-\frac{1}{2}}\int_{a}^{b}\frac{g(x-\mu)}{\sigma^2}\exp\Bigl\{-\frac{g(x-\mu)^2}{2\sigma^2}\Bigr\}dx{\cal{G}}\Bigl(g\Big \vert \frac{\nu}{2},\frac{\nu}{2}\Bigr)dg+\mu,\nonumber
%&=\frac{1}{C_{G}{\cal{P}}_{t}\Gamma\bigl(\frac{\nu}{2}\bigr)}\Bigl(\frac{\nu}{2}\Bigr)^{\frac{\nu}{2}}\int_{0}^{\infty}g^{\frac{\nu-1}{2}}\int_{a}^{b}{x} \exp\Bigl\{-g\Bigl[\frac{(x-\mu)^2}{2\sigma^2}+2\beta\Bigr]\Bigr\}dx dg,
\end{align*}
where $C_{G}=\sqrt{2\pi} \sigma$. On the other hand, since
\begin{align}\label{first-truncated-moment-univariate-t-part2}
\frac{\partial }{\partial \mu}  \exp\Bigl\{-\frac{g(x-\mu)^2}{2\sigma^2}\Bigr\}=&g\times \Bigl(\frac{x-\mu}{\sigma^2}\Bigr) \exp\Bigl\{-\frac{g(x-\mu)^2}{2\sigma^2}\Bigr\},
\end{align} 
it turns out that
\begin{align}\label{first-truncated-moment-univariate-t-part3}
\Omega=&\frac{\sigma^2}{C_{G}{\cal{P}}_{t}}\int_{0}^{\infty}g^{-\frac{1}{2}}\frac{\partial }{\partial \mu} \int_{a}^{b} \exp\Bigl\{-\frac{g(x-\mu)^2}{2\sigma^2}\Bigr\}dx{\cal{G}}\Bigl(g\Big \vert \frac{\nu}{2},\frac{\nu}{2}\Bigr)dg+\mu,\nonumber\\
=&\frac{\sigma^2}{C_{G}{\cal{P}}_{t}}\int_{0}^{\infty}g^{-\frac{1}{2}}\frac{\partial }{\partial \mu} \int_{a-\mu}^{b-\mu} \exp\Bigl\{-\frac{g u^2}{2\sigma^2}\Bigr\}du{\cal{G}}\Bigl(g\Big \vert \frac{\nu}{2},\frac{\nu}{2}\Bigr)dg+\mu,\nonumber\\
=&\frac{\sigma^2}{C_{G}{\cal{P}}_{t}}\int_{0}^{\infty}g^{-\frac{1}{2}} \Bigl[\exp\Bigl\{-\frac{g (a-\mu)^2}{2\sigma^2}\Bigr\}-\exp\Bigl\{-\frac{g (b-\mu)^2}{2\sigma^2}\Bigr\}\Bigr]{\cal{G}}\Bigl(g\Big \vert \frac{\nu}{2},\frac{\nu}{2}\Bigr)dg+\mu,\nonumber\\
=&\frac{\sigma^2}{C_{G}{\cal{P}}_{t}\Gamma\bigl(\frac{\nu}{2}\bigr)}\Bigl(\frac{\nu}{2}\Bigr)^{\frac{\nu}{2}}\int_{0}^{\infty}g^{\frac{\nu-1}{2}-1} \exp\Bigl\{-g\Bigl[\frac{(a-\mu)^2}{2\sigma^2}+\frac{\nu}{2}\Bigr]\Bigr\} dg\nonumber\\
&-\frac{\sigma^2}{C_{G}{\cal{P}}_{t}\Gamma\bigl(\frac{\nu}{2}\bigr)}\Bigl(\frac{\nu}{2}\Bigr)^{\frac{\nu}{2}}\int_{0}^{\infty}g^{\frac{\nu-1}{2}-1} \exp\Bigl\{-g\Bigl[\frac{(b-\mu)^2}{2\sigma^2}+\frac{\nu}{2}\Bigr]\Bigr\} dg+\mu\nonumber\\
=&\frac{\sigma^2\Gamma\bigl(\frac{\nu-1}{2}\bigr)}{C_{G}{\cal{P}}_{t}\Gamma\bigl(\frac{\nu}{2}\bigr)}\Bigl(\frac{\nu}{2}\Bigr)^{\frac{\nu}{2}}\biggl[\Bigl[\frac{(a-\mu)^2}{2\sigma^2}+\frac{\nu}{2}\Bigr]^{-\frac{\nu-1}{2}}+
     \Bigl[\frac{(b-\mu)^2}{2\sigma^2}+\frac{\nu}{2}\Bigr]^{-\frac{\nu-1}{2}}\biggr]+\mu.
\end{align}
Rearranging the RHS of (\ref{first-truncated-moment-univariate-t-part3}) in terms of Student's $t$ PDF, we obtain
\begin{align}\label{first-truncated-moment-univariate-t-part4}
\Omega=\frac{\sigma_{\nu}}{{\cal{P}}_{t}}\biggl[t\Bigl(\frac{a-\mu}{\sigma_{\nu}}\Big \vert 0, 1, \nu-2 \Bigr)
-t\Bigl(\frac{b-\mu}{\sigma_{\nu}}\Big \vert 0, 1, \nu-2 \Bigr)\biggr]+\mu,
\end{align}
where $\sigma_{\nu}=\sigma\sqrt{\nu}/\sqrt{\nu-2}$ for $\nu>2$. Applying the above argument to computing $\Omega$, we have
\begin{align}%\label{first-truncated-moment-univariate-t-part6}
\Omega^2=&\frac{1}{{\cal{P}}_{t}}\int_{a}^{b}{x^2}\int_{0}^{\infty}{\phi}\Bigl({x}\Big \vert \boldsymbol{\mu},\frac{\sigma^2}{g}\Bigr)
{\cal{G}}\Bigl(g\Big \vert \frac{\nu}{2},\frac{\nu}{2}\Bigr)dgd {x}.\nonumber
\end{align} 
For computing $\Omega^2$, we take into account the fact that 
\begin{align}\label{first-truncated-moment-univariate-t-part7}
\frac{\partial^2}{\partial \mu^2} \exp\Bigl\{-g\frac{(x-\mu)^2}{2\sigma^2}\Bigr\}=&\Bigl[-\frac{g}{\sigma^2}+ g^2\times \Bigl(\frac{x-\mu}{\sigma^2}\Bigr)^2\Bigr] 
\exp\Bigl\{-g\frac{(x-\mu)^2}{2\sigma^2}\Bigr\}.
\end{align} 
Rearranging equation (\ref{first-truncated-moment-univariate-t-part7}) yields
\begin{align}\label{first-truncated-moment-univariate-t-part8}
 \frac{g^2x^2}{\sigma^4} \exp\Bigl\{-g\frac{(x-\mu)^2}{2\sigma^2}\Bigr\}=&\frac{\partial^2 }{\partial \mu^2}  \exp\Bigl\{-g\frac{(x-\mu)^2}{2\sigma^2}\Bigr\}\nonumber\\
 &+\biggl[\Bigl(\frac{g}{\sigma^2}-\frac{\mu^2}{\sigma^4}g^2+2\frac{\mu}{\sigma^4}g^2x\Bigr)\exp\Bigl\{-g\frac{(x-\mu)^2}{2\sigma^2}\Bigr\}\biggr].
\end{align} 
Multiplying both sides of (\ref{first-truncated-moment-univariate-t-part8}) by $\sigma^4 g^{-3/2}{\cal{G}}\Bigl(g\Big \vert \frac{\nu}{2},\frac{\nu}{2}\Bigr)$, then integrating over $x$ and $g$, and finally dividing by $C_{G}{\cal{P}}_{t}$, we have 
\begin{align}\label{first-truncated-moment-univariate-t-part9}
\Omega^2=&\frac{\sigma^4}{C_{G}{\cal{P}}_{t}}\int_{0}^{\infty} g^{-\frac{3}{2}}\frac{\partial^2}{\partial \mu^2} \int_{a}^{b} \exp\Bigl\{-g\frac{(x-\mu)^2}{2\sigma^2}\Bigr\} {\cal{G}}\Bigl(g\Big \vert \frac{\nu}{2},\frac{\nu}{2}\Bigr)dgd {x}\nonumber\\
&+ \biggl[\frac{\sigma^2}{C_{G}{\cal{P}}_{t}}{\cal{R}}_{-\frac{1}{2},0} -\frac{\mu^2}{C_{G}{\cal{P}}_{t}}{\cal{R}}_{\frac{1}{2},0} +2\frac{\mu}{C_{G}{\cal{P}}_{t}}{\cal{R}}_{\frac{1}{2},1}\biggr],
\end{align}
where
\begin{align}\label{first-truncated-moment-univariate-t-part10}
{\cal{R}}_{i,j}=& \int_{a}^{b}\int_{0}^{\infty}g^{i}x^{j} \exp\Bigl\{-g\frac{(x-\mu)^2}{2\sigma^2}\Bigr\} {\cal{G}}\Bigl(g\Big \vert \frac{\nu}{2},\frac{\nu}{2}\Bigr)dgdx.
\end{align}
For integral in the RHS of (\ref{first-truncated-moment-univariate-t-part9}), it is not hard to check that
\begin{align}\label{first-truncated-moment-univariate-t-part11}
&\frac{\sigma^4}{C_{G}{\cal{P}}_{t}}\int_{0}^{\infty} g^{-\frac{3}{2}}\frac{\partial^2}{\partial \mu^2} \int_{a}^{b} \exp\Bigl\{-g\frac{(x-\mu)^2}{2\sigma^2}\Bigr\} {\cal{G}}\Bigl(g\Big \vert \frac{\nu}{2},\frac{\nu}{2}\Bigr)dgdx\nonumber\\
&=\frac{\sigma_{\nu}}{{\cal{P}}_{t}}(a-\mu)t\Bigl(\frac{a-\mu}{\sigma_{\nu}}\Big \vert 0, 1, \nu-2 \Bigr)-\frac{\sigma_{\nu}}{{\cal{P}}_{t}}(b-\mu)t\Bigl(\frac{b-\mu}{\sigma_{\nu}}\Big \vert 0, 1, \nu-2 \Bigr),
\end{align}
where $\sigma_{\nu}=\sigma\sqrt{\nu}/\sqrt{\nu-2}$ for $\nu>2$. Hence, replacing the RHS of (\ref{first-truncated-moment-univariate-t-part11}) with integral in RHS of (\ref{first-truncated-moment-univariate-t-part9}), we have 
\begin{align}\label{first-truncated-moment-univariate-t-part12}
\Omega^2=&\frac{\sigma_{\nu}}{{\cal{P}}_{t}}(a-\mu)t\Bigl(\frac{a-\mu}{\sigma_{\nu}}\Big \vert 0, 1, \nu-2 \Bigr)-\frac{\sigma_{\nu}}{{\cal{P}}_{t}}(b-\mu)t\Bigl(\frac{b-\mu}{\sigma_{\nu}}\Big \vert 0, 1, \nu-2 \Bigr)\nonumber\\
&+ \frac{\sigma^2}{C_{G}{\cal{P}}_{t}}{\cal{R}}_{-\frac{1}{2},0} -\frac{\mu^2}{C_{G}{\cal{P}}_{t}}{\cal{R}}_{\frac{1}{2},0} +2\frac{\mu}{C_{G}{\cal{P}}_{t}}{\cal{R}}_{\frac{1}{2},1}.
\end{align}
Moreover, it may seen, for $i>(1-\nu)/2$ and $j=\{0,1\}$, that
\begin{eqnarray}\label{first-truncated-moment-univariate-t-part13}
\displaystyle
{\cal{R}}_{i,j}=\left\{\begin{array}{c}
\displaystyle
\frac{2^{i}\Gamma\bigl(\frac{\nu(i)}{2}\bigr)\sigma_{i}\nu^{-i} }{\bigl[\pi\nu(i)\bigr]^{-\frac{1}{2}}\Gamma\bigl(\frac{\nu}{2}\bigr)}
\biggl[T\Bigl( \frac{b-\mu}{\sigma_{i}}\Big \vert 0, 1, \nu(i)\Bigr)-T\Bigl( \frac{a-\mu}{\sigma_{i}}\Big \vert 0, 1, \nu(i)\Bigr)\biggr],
~~~\mathrm{{if}}
\
j = 0,
\\
\displaystyle
\frac{2^{i}\Gamma\bigl(\frac{\nu(i)}{2}\bigr)\sigma_{i}\nu^{-i}}{ \bigl[\pi\nu(i)\bigr]^{-\frac{1}{2}} \Gamma\bigl(\frac{\nu}{2}\bigr)}
\biggl[T\Bigl( \frac{b-\mu}{\sigma_{i}}\Big \vert 0, 1, \nu(i)\Bigr)-T\Bigl( \frac{a-\mu}{\sigma_{i}}\Big \vert 0, 1, \nu(i)\Bigr)\biggr]\Omega_{*}^j,
\mathrm{if}
\
j>0,
\end{array} \right.
\end{eqnarray}
where $\sigma_{i}=\sigma\sqrt{\nu}/\sqrt{\nu+2i-1}$ (for $i>(1-\nu)/2$), $\nu(i)=\nu+2i-1$, and $\Omega_{*}^j=E\bigl(Y^j \big \vert a<Y<b\bigr)$ is the $j$th moment of $Y\sim {\cal{TT}}\bigl(\mu, \sigma_{i}^{2}, \nu(i), a,b\bigr)$. Computing the quantity ${\cal{R}}_{i,j}$ represented in (\ref{first-truncated-moment-univariate-t-part12}) using (\ref{first-truncated-moment-univariate-t-part13}), for $i=\{-1/2,1/2\}$ and $j=\{0,1\}$, and then replacing the results in the RHS of (\ref{first-truncated-moment-univariate-t-part9}), one can compute the quantity $\Omega^2$. The \verb+R+ function given by the following is developed for computing first two moments of the Student's $t$ distribution truncated on $(a,b)$.  
\begin{lstlisting}[style=deltaj]
R> Ex <- function(mu, sigma, nu, a, b)
+ {
+ 	CG <- sqrt(2*pi*sigma^2)
+ 	za <- (a - mu)/sigma; zb <- (b - mu)/sigma
+ 	Pt <- pt(zb, df = nu ) - pt(za, df = nu )
+ 			ex <- function(m, s, nu, a, b)
+      			{
+ 				za <- (a - m)/s; zb <- (b - m)/s
+ 				Pt <- pt(zb, df = nu ) - pt(za, df = nu )
+ 				s_nu <- s*sqrt( nu/(nu - 2) )
+ 				za_nu <- (a - m)/s_nu
+ 				zb_nu <- (b - m)/s_nu
+ 				p1 <- s_nu/Pt*dt(za_nu, df = nu - 2)
+ 				p2 <- s_nu/Pt*dt(zb_nu, df = nu - 2)
+ 				return( list("p1" = p1, "p2" = p2, "Omega" = p1 - p2 + m) ) 
+ 				}
+ 		out1 <- ex(mu, sigma, nu, a, b)
+ 		p3 <- (a - mu)*out1$p1 - (b - mu)*out1$p2
+ 			Rij <- function(i, j)
+ 			{
+ 				nu_i <- nu + 2*i -1 
+ 				sigma_i <- sigma*sqrt( nu/nu_i )
+ 				zb_i <- (b - mu)/sigma_i
+ 				za_i <- (a - mu)/sigma_i
+ 				C_ij <- (2/nu)^(i)*sqrt( pi*nu_i )*gamma(nu_i/2)/gamma(nu/2)*sigma_i
+ 				Pt_i <- pt(zb_i, df = nu_i ) - pt(za_i, df = nu_i )
+ 					if(j == 0)
+ 					{
+ 					R_ij <- C_ij*Pt_i
+ 					}else{
+ 					R_ij <- C_ij*Pt_i*ex(mu, sigma_i, nu_i, a, b)$Omega
+ 					}
+	 				R_ij
+ 				}
+ Omega2 <- p3 + sigma^2/(CG*Pt)*Rij(-1/2, 0) - mu^2/(CG*Pt)*Rij(1/2, 0) +
+  2*mu/(CG*Pt)*Rij(1/2, 1) 
+ out2 <- list("Ex" = out1$Omega, "Exx" = Omega2)
+ return( out2 )
+ }
\end{lstlisting}

\section{First moment of truncated Student's $t$ distribution in multivariate case}\label{first-moment-truncated-multivariate-t}
For simplicity, we use generic symbols $\boldsymbol{t}_{p}\bigl(\boldsymbol{x}\big \vert \boldsymbol{\mu}, \Sigma, \nu\bigr)$ and $\boldsymbol{T}_{p}\bigl(\boldsymbol{x}\big \vert \boldsymbol{\mu}, \Sigma, \nu\bigr)$ to denote, accordingly, the PDF and CDF of a $p$-dimensional Student's $t$ distribution with $\nu$ degrees of freedom at point $\boldsymbol{x}$ for which $\boldsymbol{\mu}$, $\Sigma$, and $\nu$ play the role of location, scale, and degrees of freedom parameters, respectively. Furthermore, we write $\boldsymbol{X} \sim \boldsymbol{t}_{p}\bigl(\boldsymbol{\mu}, \Sigma, \nu\bigr)$ to indicate that random vector $\boldsymbol{X}$ follows a Student's $t$ distribution with location vector $\boldsymbol{\mu}$, scale matrix $\Sigma$, and $\nu$ degrees of freedom. The corresponding family of Student's $t$ distributions truncated on on hyper-cube $(\boldsymbol{a}, \boldsymbol{b})\subset \mathbb{R}^{p}$ is shown by ${\cal{TT}}_{p}\bigl(\boldsymbol{\mu}, \Sigma, \nu, \boldsymbol{a}, \boldsymbol{b}\bigr)$.
\par Recalling from Definition \ref{dfn-Gaussian-scale-mixture-model-multivariate} that each Student's $t$ distribution is a Gaussian scale mixture model. In fact if $\boldsymbol{X}\sim \boldsymbol{t}_{p}\bigl(\boldsymbol{\mu}, \Sigma, \nu\bigr)$, then it admits the hierarchy given by
\begin{align}\label{hierarchy-t-distribution}
\boldsymbol{X}\vert G &\sim {\cal{N}}_{p}\Bigl(\boldsymbol{\mu}, \frac{\Sigma}{{G}}\Bigr),\nonumber\\
G&\sim{\cal{G}}\Bigl(\frac{\nu}{2},\frac{\nu}{2}\Bigr).
\end{align}
Using above hierarchy, the quantity $E(\boldsymbol{X})$ in (\ref{moment-t-distribution1}) can be represented as
\begin{align*}%\label{moment-t-distribution2}
E(\boldsymbol{X} )&=\int_{-\boldsymbol{\infty}}^{\boldsymbol{\infty}}\boldsymbol{x}
\int_{0}^{\infty}\boldsymbol{\phi}_{p}\Bigl(\boldsymbol{x}\Big \vert \boldsymbol{\mu},\frac{\Sigma}{g}\Bigr)
{\cal{G}}\Bigl(g\Big \vert \frac{\nu}{2},\frac{\nu}{2}\Bigr)dgd \boldsymbol{x}.
\nonumber\\
&=\frac{1}{{\bf{C}}_{G}}\int_{0}^{\infty}g^{\frac{p}{2}}\int_{-\boldsymbol{\infty}}^{\boldsymbol{\infty}}\boldsymbol{x} \exp\Bigl\{-\frac{g}{2}\delta(\boldsymbol{x}-\boldsymbol{\mu},\Sigma)\Bigr\}d \boldsymbol{x}{\cal{G}}\Bigl(g\Big \vert \frac{\nu}{2},\frac{\nu}{2}\Bigr)dg,
\end{align*}
where ${\bf{C}}_{G}=(2\pi)^{p/2}\big \vert \Sigma \big  \vert^{1/2}$. The first moment of truncated Student's $t$ distribution is
\begin{align*}%\label{truncated-moment-t-part1}
\boldsymbol{\Omega}=E(\boldsymbol{X}\vert \boldsymbol{a}<\boldsymbol{X}<\boldsymbol{b})=\frac{1}{{\bf{C}}_{G}{\cal{P}}_{t}}\int_{0}^{\infty}\int_{\boldsymbol{a}}^{\boldsymbol{b}}g^{\frac{p}{2}}\boldsymbol{x} \exp\Bigl\{-\frac{g}{2}\delta(\boldsymbol{x}-\boldsymbol{\mu},\Sigma)\Bigr\}d \boldsymbol{x}{\cal{G}}\Bigl(g\Big \vert \frac{\nu}{2},\frac{\nu}{2}\Bigr)dg,
\end{align*}
where ${\cal{P}}_{t}=P(\boldsymbol{a}<\boldsymbol{X}<\boldsymbol{b})$.
We can write
\begin{align}\label{truncated-moment-t-part2}
\boldsymbol{\Omega}=&\frac{1}{{\bf{C}}_{G}{\cal{P}}_{t}}\int_{0}^{\infty}\int_{\boldsymbol{a}}^{\boldsymbol{b}}g^{\frac{p}{2}} (\boldsymbol{x}-\boldsymbol{\mu})\exp\Bigl\{-\frac{g}{2}\delta(\boldsymbol{x}-\boldsymbol{\mu},\Sigma)\Bigr\}d\boldsymbol{x}{\cal{G}}\Bigl(g\Big \vert \frac{\nu}{2},\frac{\nu}{2}\Bigr)dg\nonumber\\
 &+\frac{\boldsymbol{\mu}}{{\bf{C}}_{G}{\cal{P}}_{t}}\int_{0}^{\infty} \int_{\boldsymbol{a}}^{\boldsymbol{b}}g^{\frac{p}{2}}\exp\Bigl\{-\frac{g}{2}\delta(\boldsymbol{x}-\boldsymbol{\mu},\Sigma)\Bigr\} d\boldsymbol{x}{\cal{G}}\Bigl(g\Big \vert \frac{\nu}{2},\frac{\nu}{2}\Bigr)dg\nonumber\\
=&\frac{1}{{\bf{C}}_{G}{\cal{P}}_{t}}\int_{0}^{\infty} \int_{\boldsymbol{a}}^{\boldsymbol{b}}g^{\frac{p}{2}} (\boldsymbol{x}-\boldsymbol{\mu})\exp\Bigl\{-\frac{g}{2}\delta(\boldsymbol{x}-\boldsymbol{\mu},\Sigma)\Bigr\} d\boldsymbol{x}{\cal{G}}\Bigl(g\Big \vert \frac{\nu}{2},\frac{\nu}{2}\Bigr)dg+\boldsymbol{\mu}\nonumber\\
=&\boldsymbol{I}+\boldsymbol{\mu}.
%\Phi_{q}\Bigl(\sqrt{r}\boldsymbol{m}\Big|\boldsymbol{0},\boldsymbol{\Delta}\Bigr)
\end{align}
Since
\begin{align*}
\frac{\partial}{\partial \boldsymbol{\mu}}\frac{g}{2}\delta(\boldsymbol{x}-\boldsymbol{\mu},\Sigma)=-g {\Sigma}^{-1}(\boldsymbol{x}-\boldsymbol{\mu}),
%\Phi_{q}\Bigl(\sqrt{r}\boldsymbol{m}\Big|\boldsymbol{0},\boldsymbol{\Delta}\Bigr)
\end{align*}
we have
\begin{align*}
\boldsymbol{I}=&\frac{\Sigma}{{\cal{P}}_{t}}\int_{0}^{\infty} \frac{g^{\frac{p}{2}-1}}{{\bf{C}}_{G}}\frac{\partial}{\partial \boldsymbol{\mu}} \int_{\boldsymbol{a}}^{\boldsymbol{b}}\exp\Bigl\{-\frac{g}{2}\delta(\boldsymbol{x}-\boldsymbol{\mu},\Sigma)\Bigr\}d\boldsymbol{x}{\cal{G}}\Bigl(g\Big \vert \frac{\nu}{2},\frac{\nu}{2}\Bigr) dg.
\end{align*}
Applying the change of variable $\boldsymbol{u}=\boldsymbol{x}-\boldsymbol{\mu}$, it follows that
\begin{align*}
\boldsymbol{I}=&\frac{\Sigma}{{\cal{P}}_{t}}\int_{0}^{\infty} \frac{g^{\frac{p}{2}-1}}{{\bf{C}}_{G}}\frac{\partial}{\partial \boldsymbol{\mu}} \int_{\boldsymbol{a}-\boldsymbol{\mu}}^{\boldsymbol{b}-\boldsymbol{\mu}}\exp\Bigl\{-\frac{g}{2}\delta(\boldsymbol{u},\Sigma)\Bigr\} d\boldsymbol{u}{\cal{G}}\Bigl(g\Big \vert \frac{\nu}{2},\frac{\nu}{2}\Bigr)dg\nonumber\\
=&\bigl({{I}}_{1},{{I}}_{2},\cdots,{{I}}_{p}\bigr)^{\top},
\end{align*}
where 
\begin{align}\label{truncated-moment-t-part4}
{{I}}_{i}=&\frac{\Sigma}{{\cal{P}}_{t}}\int_{0}^{\infty} \frac{g^{\frac{p}{2}-1}}{{\bf{C}}_{G}}\int_{\boldsymbol{a}[-i]-\boldsymbol{\mu}[-i]}^{\boldsymbol{b}[-i]-\boldsymbol{\mu}[-i]}\Biggl[\exp\Bigl\{-\frac{g\delta\bigl(\boldsymbol{u}_{i}[a_{i}-\mu_{i}],\Sigma\bigr)}{2}\Bigr\}\nonumber\\
&-\exp\Bigl\{-\frac{g\delta\bigl(\boldsymbol{u}_{i}[b_{i}-\mu_{i}],\Sigma\bigr)}{2}\Bigr\}\Biggr] d\boldsymbol{u}[-i]{\cal{G}}\Bigl(g\Big \vert \frac{\nu}{2},\frac{\nu}{2}\Bigr)dg,
\end{align}
%where, for $i=1,\cdots,p$, we have
for $i=1,\cdots,p$. Using identity (\ref{quadratic-decomposition}) and the fact that
\begin{align*}
\frac{ g^{\frac{p}{2}-1}}{{\bf{C}}_{G}}=\Bigl[\frac{1}{\sqrt{2\pi \Sigma_{i,i}}}\frac{1}{(2\pi)^{\frac{p-1}{2}}\big \vert \Delta_{ii}\big  \vert^{\frac{1}{2}}}\Bigr]\times g^{\frac{p}{2}-1}=\frac{g^{-1}}{\sqrt{2\pi }}\Bigl[\frac{\Sigma_{i,i}}{g}\Bigr]^{-\frac{1}{2}}\frac{1}{(2\pi)^{\frac{p-1}{2}}}\Big \vert\frac{ \Delta_{ii}}{g}\Big  \vert^{-\frac{1}{2}},
\end{align*}
the integrand in the RHS of (\ref{truncated-moment-t-part4}) can be rewritten as 
\begin{align}\label{truncated-moment-t-part5}
I_{i}=&\frac{\Sigma}{{\cal{P}}_{t}}\int_{0}^{\infty} \frac{g^{-1}}{\sqrt{2\pi }}\Bigl[\frac{\Sigma_{i,i}}{g}\Bigr]^{-\frac{1}{2}} \exp\Bigl\{-\frac{g(a_{i}-\mu_{i})^2}{2\Sigma_{i,i}}\Bigr\} {\cal{G}}\Bigl(g\Big \vert \frac{\nu}{2},\frac{\nu}{2}\Bigr)\nonumber\\
&\times\int_{\boldsymbol{a}[-i]-\boldsymbol{\mu}[-i]}^{\boldsymbol{b}[-i]-\boldsymbol{\mu}[-i]}
\frac{1}{(2\pi)^{\frac{p-1}{2}}}\Big \vert\frac{ \Delta_{ii}}{g}\Big  \vert^{-\frac{1}{2}}\exp\Bigl\{-\frac{g}{2}\delta\Bigl(\boldsymbol{u}[-i]-\boldsymbol{\xi}(a_{i}-\mu_{i}),\Delta_{ii}\Bigr)\Bigr\} d\boldsymbol{u}[-i]dg\nonumber\\
&-\frac{\Sigma}{{\cal{P}}_{t}} \int_{0}^{\infty}\frac{g^{-1}}{\sqrt{2\pi }}\Bigl[\frac{\Sigma_{i,i}}{g}\Bigr]^{-\frac{1}{2}}\exp\Bigl\{-\frac{g(b_{i}-\mu_{i})^2}{2\Sigma_{i,i}}\Bigr\}{\cal{G}}\Bigl(g\Big \vert \frac{\nu}{2},\frac{\nu}{2}\Bigr)\nonumber\\
&\times \int_{\boldsymbol{a}[-i]-\boldsymbol{\mu}[-i]}^{\boldsymbol{b}[-i]-\boldsymbol{\mu}[-i]}
\frac{1}{(2\pi)^{\frac{p-1}{2}}}\Big \vert\frac{ \Delta_{ii}}{g}\Big  \vert^{-\frac{1}{2}}\exp\Bigl\{-\frac{g}{2}\delta\Bigl(\boldsymbol{u}[-i]-\boldsymbol{\xi}(b_{i}-\mu_{i}),\Delta_{ii}\Bigr)\Bigr\} d\boldsymbol{u}[-i]dg,
\end{align}
%It follows that
%\begin{align}\label{I14}
%I_{i}=&\frac{\Sigma}{\alpha{\cal{P}}_{t}}\int_{0}^{\infty} g^{\frac{p}{2}-1} \phi\Bigl(a_{i}\Big \vert \mu_{i}, \frac{\Sigma_{i,i}}{g}\Bigr)
%\biggl[\boldsymbol{\Phi}_{p-1}\Bigl(\boldsymbol{b}[-i]-\boldsymbol{\mu}[-i]\Big\vert\boldsymbol{\xi}(a_{i}-\mu_{i}),\frac{\Delta_{i}}{g}\Bigr)\nonumber\\
%&-\boldsymbol{\Phi}_{p-1}\Bigl(\boldsymbol{a}[-i]-\boldsymbol{\mu}[-i]\Big\vert\boldsymbol{\xi}(a_{i}-\mu_{i}),\frac{\Delta_{i}}{g}\Bigr)\biggr]{\cal{G}}\Bigl(g\Big \vert \frac{\nu}{2},\frac{\nu}{2}\Bigr)dg
%\nonumber\\
%&-\frac{\Sigma}{\alpha{\cal{P}}_{t}}\int_{0}^{\infty} g^{\frac{p}{2}-1} \phi\Bigl(b_{i}\Big \vert \mu_{i}, \frac{\Sigma_{i,i}}{g}\Bigr)
%\biggl[\boldsymbol{\Phi}_{p-1}\Bigl(\boldsymbol{b}[-i]-\boldsymbol{\mu}[-i]\Big\vert\boldsymbol{\xi}(b_{i}-\mu_{i}), \frac{\Delta_{i}}{g}\Bigr)\nonumber\\
%&-\boldsymbol{\Phi}_{p-1}\Bigl(\boldsymbol{a}[-i]-\boldsymbol{\mu}[-i]\Big\vert\boldsymbol{\xi}(b_{i}-\mu_{i}), \frac{\Delta_{i}}{g}\Bigr)\biggr]{\cal{G}}\Bigl(g\Big \vert \frac{\nu}{2},\frac{\nu}{2}\Bigr)dg.
%%=&\frac{\Sigma}{{\cal{P}}}\times\bigl({\cal{I}}_{1},{\cal{I}}_{2},\cdots,{\cal{I}}_{p}\bigr)^{\top},
%\end{align}
where $\boldsymbol{\xi}(x)$ and $\Delta_{ii}$ are defined in (\ref{xi-uni-decomposition}) and (\ref{Delta-uni-decomposition}), respectively. Moreover, it is not hard to check that
\begin{align}\label{truncated-moment-t-part6}
&\frac{g^{-1}}{\sqrt{2\pi }}\Bigl[\frac{\Sigma_{i,i}}{g}\Bigr]^{-\frac{1}{2}} \exp\Bigl\{-\frac{g(a_{i}-\mu_{i})^2}{2\Sigma_{i,i}}\Bigr\}{\cal{G}}\Bigl(g\Big \vert \frac{\nu}{2},\frac{\nu}{2}\Bigr)\nonumber\\
=&\sqrt{\frac{\nu_2}{\Sigma_{i,i}}}
 t\Bigl(a_{i}\Big \vert \mu_{i},\nu_2\Sigma_{i,i},\nu-2\Bigr){\cal{G}}\Bigl(g\Big \vert\frac{\nu-1}{2},\lambda_{a} \Bigr),
\end{align}
where $\nu_2=\nu/(\nu-2)$, $\lambda_{a}=(a_{i}-\mu_{i})^{2}/(2\Sigma_{i,i})+\nu/2$. Further, for given vectors $\boldsymbol{c}_{1}$ and $\boldsymbol{c}_{2}$ of length $p-1$ such that $\boldsymbol{c}_{1}<\boldsymbol{c}_{2}$, we have
\begin{align}\label{truncated-moment-t-part7}
&\int_{0}^{\infty} \int_{\boldsymbol{c}_{1}}^{\boldsymbol{c}_{2}}\frac{1}{(2\pi)^{\frac{p-1}{2}}}\Big \vert\frac{ \Delta_{ii}}{g}\Big  \vert^{-\frac{1}{2}}
\exp\Bigl\{-\frac{g}{2}\delta\bigl(\boldsymbol{y}-\boldsymbol{m},\Delta_{ii}\bigr)\Bigr\}d\boldsymbol{y}{\cal{G}}\Bigl(g\Big \vert\frac{\nu-1}{2}, \lambda_{a} \Bigr)dg\nonumber\\
&= \int_{\boldsymbol{c}_{1}}^{\boldsymbol{c}_{2}}\frac{\lambda^{\frac{\nu-1}{2}}_{a}}{(2\pi)^{\frac{p-1}{2}}\big \vert \Delta_{ii}\big  \vert^{\frac{1}{2}} }
\frac{\Gamma\bigl(\frac{\nu+p-2}{2}\bigr)}{\Gamma\bigl(\frac{\nu-1}{2}\bigr)}
\Bigl[\frac{\delta\bigl(\boldsymbol{y}-\boldsymbol{m},\Delta_{ii}\bigr)}{2}+\lambda_{a}\Bigr]^{-\frac{\nu+p-2}{2}}d\boldsymbol{y}\nonumber\\
&=\boldsymbol{T}_{p-1}\Bigl(\boldsymbol{c}_{2}\Big \vert \boldsymbol{m}, \frac{2\lambda_{a}}{\nu-1}\Delta_{ii},\nu-1\Bigr)-\boldsymbol{T}_{p-1}\Bigl(\boldsymbol{c}_{1}\Big \vert \boldsymbol{m}, \frac{2\lambda_{a}}{\nu-1}\Delta_{ii},\nu-1\Bigr),
\end{align}
where $\nu >1$. 
%&=\int_{0}^{\infty}P\Bigl(\boldsymbol{c}_{1}-\boldsymbol{m}<\frac{\boldsymbol{Y}}{\sqrt{g}}<\boldsymbol{c}_{2}-\boldsymbol{m}\Bigr){\cal{G}}\Bigl(g\Big \vert\frac{\nu-1}{2}, \lambda_{a} \Bigr)dg,\nonumber\\
%&=P\Bigl(\boldsymbol{c}_{1}-\boldsymbol{m}<\frac{\boldsymbol{Y}}{\sqrt{G}}<\boldsymbol{c}_{2}-\boldsymbol{m}\Bigr),
%\end{align}
%%$\boldsymbol{Y}\sim {\cal{N}}_{p-1}(\boldsymbol{0},\Delta_{i})$ and $G\sim {\cal{G}}\bigl((\nu-1)/2, \lambda_{a} \bigr)$ in which
% $\lambda_{a}=(a_{i}-\mu_{i})^{2}/(2\Sigma_{i,i})+\nu/2$. Evidently the RHS of (\ref{truncated-moment-t-part3}) is 
%\begin{align}
%P_{2}=\boldsymbol{T}_{\frac{\nu-1}{2}}\Bigl(\boldsymbol{c}_{2}-\boldsymbol{m}\Big \vert \boldsymbol{0}, \Delta_{i}\Bigr)-\boldsymbol{T}_{\frac{\nu-1}{2}}\Bigl(\boldsymbol{c}_{1}-\boldsymbol{m}\Big \vert \boldsymbol{0}, \Delta_{i}\Bigr),
%\end{align}
Using the facts given in (\ref{truncated-moment-t-part6}) and (\ref{truncated-moment-t-part7}), the RHS of (\ref{truncated-moment-t-part5}) becomes
\begin{align}\label{truncated-moment-t-part8}
I_{i}=\sqrt{\frac{\nu_{2}}{\Sigma_{i,i}}}\frac{\Sigma}{{\cal{P}}_{t}}&t\Bigl(a_{i}\Big \vert \mu_{i},\nu_{2}\Sigma_{i,i},\nu-2\Bigr)\biggl[\boldsymbol{T}_{p-1}\Bigl(\boldsymbol{b}[-i]-\boldsymbol{\mu}[-i]\Big \vert \boldsymbol{\xi}(a_{i}-\mu_{i}), \frac{2\lambda_{a}}{\nu-1}\Delta_{ii},\nu-1\Bigr)\nonumber\\
&-\boldsymbol{T}_{p-1}\Bigl(\boldsymbol{a}[-i]-\boldsymbol{\mu}[-i]\Big \vert \boldsymbol{\xi}(a_{i}-\mu_{i}), \frac{2\lambda_{a}}{\nu-1}\Delta_{ii},\nu-1\Bigr)\biggr]-\nonumber\\
\sqrt{\frac{\nu_{2}}{\Sigma_{i,i}}}\frac{\Sigma}{{\cal{P}}_{t}}&t\Bigl(b_{i}\Big \vert \mu_{i},\nu_{2}\Sigma_{i,i},\nu-2\Bigr)\biggl[\boldsymbol{T}_{p-1}\Bigl(\boldsymbol{b}[-i]-\boldsymbol{\mu}[-i]\Big \vert \boldsymbol{\xi}(b_{i}-\mu_{i}), \frac{2\lambda_{b}}{\nu-1}\Delta_{ii},\nu-1\Bigr)\nonumber\\
&-\boldsymbol{T}_{p-1}\Bigl(\boldsymbol{a}[-i]-\boldsymbol{\mu}[-i]\Big \vert \boldsymbol{\xi}(b_{i}-\mu_{i}), \frac{2\lambda_{b}}{\nu-1}\Delta_{ii},\nu-1\Bigr)\biggr],
\end{align}
where $\nu>2$ and $\lambda_{b}=(b_{i}-\mu_{i})^{2}/(2\Sigma_{i,i})+\nu/2$. Computing $I_{i}$ in (\ref{truncated-moment-t-part8}), for $i=1,\cdots,p$, the first moment of truncated Student's $t$ distribution is then obtained by substituting the constructed vector $\boldsymbol{I}$ into the RHS of (\ref{truncated-moment-t-part2}). In what follows, we perform a small simulation study for computing $\boldsymbol{\tau}$ when $\boldsymbol{X}=(X_1,X_2,X_3)^{\top}$ follows three-dimensional Student's $t$ distribution with parameters  $\nu=4$, $\boldsymbol{\mu}=(1,1,1)^{\top}$ and
\begin{align}
\Sigma=\left[\begin{matrix} 
2&0.5&0.4\\
0.5&1.5&0.3\\
0.4&0.3&1
\end{matrix}\right].
\end{align}
The truncation bounds are $\boldsymbol{a}=(0,0,0)^{\top}$ and $\boldsymbol{b}=(10,10,10)^{\top}$. To this end, we compute $\boldsymbol{\Omega}$ using the exact method described above and the importance sampling in which the instrumental distribution is $\boldsymbol{t}_{3}(\boldsymbol{\mu},\Sigma,\nu)$. For computing $\boldsymbol{\Omega}$ through the importance sampling, we generate a sample of size $N$ through the rejection method. Table \ref{table-importance-sampling-bivariate-truncated-t} shows the results of simulation study for approximating $\boldsymbol{\Omega}$ under both scenarios through the instrumental $t_{2}(\boldsymbol{\mu},\Sigma,\nu)$.  
\begin{table}[h!]
\center
\caption{Summary statistics for computing $\boldsymbol{\Omega}=E(  \boldsymbol{X})$ where $\boldsymbol{X} \sim {\cal{TT}}_{2}(\boldsymbol{0},\Sigma,\nu,\boldsymbol{a},\boldsymbol{b})$ through the exact and importance sampling with $t_{2}(\boldsymbol{0},\Sigma,\nu)$ as the only instrumental.} 
\begin{tabular}{lllcclcccc} 
\cline{1-10} 
&&&&&&\multicolumn{4}{c}{summary}\\ \cline{7-10} 
&$\boldsymbol{a}^{\top}$&$\boldsymbol{b}^{\top}$&$N$&${\boldsymbol{\Omega}}$ & $\hat{\boldsymbol{\Omega}}$&bias&SE&min.&max.\\ \cline{1-10}
1&(-1,-1)&(2,2)					    &2000 &0.3691&  $\widehat{{\Omega}_{1}}$& -1.7e-04& 0.0174& 0.3021 &  0.4440\\
     &    &        					&     &0.2603&  $\widehat{{\Omega}_{2}}$& 5.9e-05& 0.0158& 0.2072 &  0.3158\\
    &     &        					&5000 &0.3691&  $\widehat{{\Omega}_{1}}$& 7.5e-06& 0.0113& 0.3268 &  0.4071\\
    &     &       					&     &0.2603&  $\widehat{{\Omega}_{2}}$&-1.8e-04& 0.0101& 0.2255 &  0.2971\\
\cline{1-10}   
2&(0,0)&$(\infty,\infty)$	            &2000 &1.2150&$\widehat{{\Omega}_{1}}$&-7.8e-05&0.0197&1.1287 &1.2802\\
	 &  &							&     &1.2150&$\widehat{{\Omega}_{2}}$&-8.3e-04&0.0196&1.1462 &1.2827\\
	 &  &							&5000 &1.2150&$\widehat{{\Omega}_{1}}$&-7.4e-04 &0.0124&1.1690 &1.2603\\
	 &  &							&     &1.2150&$\widehat{{\Omega}_{2}}$&-4.7e-04 &0.0125&1.1701 &1.2774\\	
\cline{1-10}                                                             			                 
\end{tabular} 
%\tabnote{Hint: CI is short form for confidence interval.}                                                                                                           
\label{table-importance-sampling-bivariate-truncated-t}
\end{table}
%\multicolumn{1}{l}{\multirow{3}{*}{}}
%$\left[\begin{matrix} 2&0.5\\ 0.5&1\end{matrix}\right]$(-1,-1)&(2,2)&2000 &0.3691& $\hat{{\tau}}_{1}}$         &-1.7e-04& 0.0174& 0.3021 & 0.3689 & 0.4440\\
\begin{lstlisting}[style=deltaj]
R> library(MomTrunc) #for computing the CDF of multivariate Student's t 
R> a  <- rep(0, p); Mu <- rep(1, p); b <- rep(100, p); nu <- 3;
R> Sigma <- matrix( c(2,0.5,0.4,0.5,1.5,0.3,0.4,0.3,1), nrow = 3, ncol = 3)
R> EX <- function(Mu, Sigma, nu, a, b)
+ {
+ p  <- length(Mu)
+ V  <- numeric(p)
+ for(i in 1:p){
+ p1 <- (a[i] - Mu[i])/sqrt(nu/(nu-2)*Sigma[i, i])
+ p2 <- (b[i] - Mu[i])/sqrt(nu/(nu-2)*Sigma[i, i])
+ lambda_a <- (a[i] - Mu[i])^2/(2*Sigma[i, i])+ nu/2
+ lambda_b <- (b[i] - Mu[i])^2/(2*Sigma[i, i])+ nu/2
+ da <- sqrt(nu)/sqrt( (nu-2)*Sigma[i, i] )*dt(p1, df = nu - 2)
+ db <- sqrt(nu)/sqrt( (nu-2)*Sigma[i, i] )*dt(p2, df = nu - 2)
+ xi <- function(x) Sigma[i, -i]*x/Sigma[i, i]
+ 	if(p == 2)
+ 	{
+ 	Delta_i  <- Sigma[-i, -i] - Sigma[-i, i]^2/Sigma[i, i]
+ 	}else{
+ 	Delta_i  <- Sigma[-i, -i] - Sigma[-i, i]%*%t(Sigma[-i, i])/Sigma[i, i]
+ 	}
+ Delta_ia <- 2*lambda_a/(nu - 1)*Delta_i
+ Delta_ib <- 2*lambda_b/(nu - 1)*Delta_i
+ p3 <- da*pmvEST( a[-i] - Mu[-i], b[-i] - Mu[-i], mu = xi(a[i] - Mu[i]), 
+ 						Sigma = Delta_ia, lambda = rep(0, p-1), tau = 0, nu = nu - 1 )
+ p4 <- db*pmvEST( a[-i] - Mu[-i], b[-i] - Mu[-i], mu = xi(b[i] - Mu[i]), 
+							Sigma = Delta_ib, lambda = rep(0, p-1), tau = 0, nu = nu - 1 )
+ V[i] <- p3 - p4
+ }
+ EX <- Sigma%*%V/pmvEST(a, b, Mu, Sigma, rep(0, p), tau = 0, nu = nu) + Mu
+ return(EX)
+ }
\end{lstlisting}

\chapter{Bayesian inference}\label{sec33}
\begin{chapquote}{Bradley Efron, Stanford University}
% \textit{\cite{fernholz2000conversation}}}
A Bayesian prior is an
assumption of an infinite
amount of past relevant
experience. But you cannot
forget that you have just made
up a whole bunch of data \footnote{\url{https://rss.onlinelibrary.wiley.com/doi/pdf/10.1111/j.1740-9713.2010.00460.x}}
\end{chapquote}

In the framework of classical (or frequentist) statistics, the unknown parameter is assumed to be fixed and there is not any guess or belief about its true value. The only source to estimate the unknown parameter is the sample evidence (information). Contrary to this school of thinking, the Bayesian statistics permits to involve our {\it{prior}} knowledge or belief in process of estimating of unknown parameter. For this reason, the classical and Bayesian statistics are sometimes called {\it{objective}} and {\it{subjective}}\footnote{The belief about unknown parameter differs from individual to another one, so it is {\it{subjective}}.} statistics, respectively. In other words, the subjectivity of belief about unknown parameter leads us to assume that unknown parameter is a random variable with a known distribution, called in the literature {\it{prior}} distribution. The Bayesian statistics would make a revision on the experimenter belief about the unknown parameter using the sample evidence. Let the PDF corresponds to the prior distribution of unknown parameter $\theta$ is shown by $\pi(\theta)$. Furthermore suppose the revised version of $\pi(\theta)$ given sample evidence $\boldsymbol{x}=\{x_1,x_2,\cdots,x_n\}$, is represented as $\pi(\theta \vert \boldsymbol{x})$ that is known in the literature as the {\it{posterior}} PDF. The elementary Bayesian formula links the {\it{prior}} with {\it{posterior}} as 
\begin{align}\label{bayes}
\pi(\theta \vert\boldsymbol{x})= 
\frac{f(\boldsymbol{x}\vert\theta)}{\int_{\theta}f(\boldsymbol{x}\vert\theta)\pi(\theta)d\theta}\pi(\theta)=
\frac{f(\boldsymbol{x}\vert\theta)}{m(\boldsymbol{x})}\pi(\theta)
\propto f(\boldsymbol{x}\vert\theta)\pi(\theta)
= L(\theta\vert \boldsymbol{x}) \pi(\theta),
\end{align}
where $
L(\theta\vert\boldsymbol{x})=f(\boldsymbol{x}\vert \theta)=f(x_1,x_2,\cdots,x_n\vert \theta)=\Pi_{i=1}^{n}f(x_{i}\vert\theta)$ is the likelihood function. Hence,
\begin{align}\label{bayes2}
\pi(\theta \vert\boldsymbol{x}) \propto \Pi_{i=1}^{n}f(x_{i}\vert\theta) \pi(\theta),
\end{align}
Notice that joint PDF of sample evidence in (\ref{bayes2}), is represented as the product of marginal PDFs since the sample evidence $x_1,x_2,\cdots,x_n$ are assumed to be independent. Notice that in (\ref{bayes}) the {\it{prior}} may be either {\it{proper}} or {\it{improper}}\footnote{The {\it{prior}} $\pi(\theta)$ is called improper if $\int_{\theta} \pi(\theta)d\theta$ diverges; otherwise $\pi(\theta)$ is called {\it{proper}}.}. If the choosen {\it{prior}} $\pi(\theta)$ is {\it{improper}}, then the formal Bayes formula (\ref{bayes}) is valid if we have $\pi(\theta \vert\boldsymbol{x})$ is {\it{proper}} or equivalently $m(\boldsymbol{x})=\int_{\theta}f(\boldsymbol{x}\vert\theta)\pi(\theta)d\theta<\infty$. Furthermore, we note that the parameter space may be $d$-dimensional (multi-parameter case), represented as $\boldsymbol{\theta}=(\theta_1,\cdots,\theta_{d})^{\top}$.

%\begin{align}\label{mx1}
%{\color{black}{
%\pi(\Psi \vert\boldsymbol{x},\boldsymbol{u})= 
%\frac{f(\boldsymbol{x}, \boldsymbol{u}\vert\Psi)\pi(\Psi)}{\int_{\Psi}f(\boldsymbol{x}, \boldsymbol{u}\vert\Psi)\pi(\Psi)d\Psi}=
%\frac{f(\boldsymbol{x}\vert \boldsymbol{u}, \Psi)f(\boldsymbol{u}\vert \Psi)\pi(\Psi)}{m(\boldsymbol{x},\boldsymbol{u})}
%\propto L_{c}(\Psi\vert \boldsymbol{x},\boldsymbol{u}) \pi(\Psi),
%}}
%\end{align} 
%where the {\it{complete}} data likelihood function 
%\begin{align*}
%L_{c}(\Psi\vert\boldsymbol{x})=f(\boldsymbol{x}\vert \boldsymbol{u}, \Psi)f( \boldsymbol{u}\vert \Psi)=\Pi_{i=1}^{n}f(x_{i}\vert u_{i},\Psi)f(u_{i}\vert\Psi)
%\end{align*}
\section{Prior selection}\label{sec4}
Although the prior reflects experimenter's belief or {\it{prior}} knowledge on $\theta$, but the term {\it{subjectivity}} does not mean that experimenter is free to choose any candidate for distribution of {\it{prior}}. In practice, there are a veriety of rules for this mean that makes the process of {\it{prior}} selection a quite {\it{objective}}\footnote{The term {\it{objective}} means that experimenter has some plausible guideline for selecting the distribution of {\it{prior}}.} process that may need comprehensive study or knowledge about the unknown parameter. The structure of {\it{prior}} may also depend on some extra parameter(s) that is known in the literature as {\it{hyperparameter}}. In what follows, we cite some known classes of {\it{objective prior}}. 
\subsection{Uniform prior}
The first class is the uniform prior $\pi_{U}(\theta)$, that is useful when the parameter space is compact interval and experimenter believes that all points in parameter space are equally likely outcomes. The famous example of this type may happen when one is interested in Bayesian inference for the success probability in a Bernoulli process for which we may have $\pi_{U}(\theta)\propto 1$. Considering a beta distribution for the law of {\it{prior}}, 
%that is $\pi_{U}(\theta)\sim \text{Beta}(a,b)$, 
then $\pi_{U}(\theta\vert a,b)$, for which $a$ and $b$ play the role of {\it{hyperparameters}}, is no longer uniform. 
\cite{doob1949application}
%For example, a uniform distribution on (0, 2) is a suitable prior for the tail thickness in this work. 
 %\footnote{In the case of single parameter, we have ${I}({\theta})=-E\bigl[\partial^2/\partial \theta^2 \log L({\theta}\vert \boldsymbol{x}) \bigr]$.}. %The third class of priors is the maximum entropy prior $\pi_{M}(\theta)$, introduced by \cite{jaynes1968prior}. This type of prior is obtained by maximizing the entropy of the distribution subject to some constraints, representing available information on $\theta$. 
\subsection{Reference prior}
The first class of {\it{priors}} is known as the reference {\it{prior}} $\pi_{R}(\theta)$, 
proposed by \cite{bernardo1979reference}. For convenience let us to give some useful definitions as follows.
\begin{dfn} Let $F(x\vert \boldsymbol{\theta})$ and $f(x\vert\boldsymbol{\theta})$ denote, respectively, the CDF and PDF of the random variable $X$ with unknown parameter $\theta \in \boldsymbol{\theta}$. 
\hfill
\begin{itemize}
\item[(i)] {\bf{location family}}: The family of random variable $X$ is a location family if ${{F}}(x-\theta\vert\boldsymbol{\theta})$ or $f(x-\theta \vert \boldsymbol{\theta})$ for $\theta \in {\mathbb{R}}$ no longer depends on $\theta$. Statistically speaking, distribution of transformation $X-\theta$ does not depend on $\theta$. Herein, $\theta$ is known in the literature as the location parameter.
\item[(ii)] {\bf{scale family}}: The family of random variable $X$ is a scale family if ${{F}}(x/\theta\vert\boldsymbol{\theta})$ or $1/\theta \times f(x/\theta \vert \boldsymbol{\theta})$ for $\theta \in {\mathbb{R}^{+}}$ no longer depends on $\theta$. Statistically speaking, distribution of transformation $X/\theta$  does not depend on $\theta$. Herein, $\theta$ is known in the literature as the scale parameter.
\item[(iii)] {\bf{sufficient statistic}} Let $\boldsymbol{x}=\{x_1,x_2,\cdots,x_n\}$ denote a sequence of observations of random variables $X_1,X_2,\cdots,X_n$ that follow independently a family of distributions with PDF $f(x \vert \boldsymbol{\theta})$ . A function of random variables $\boldsymbol{X}=\{X_1,X_2,\cdots,X_n\}$ such as $S(\boldsymbol{X})$ is called {\it{sufficient statistic}} if $S(\boldsymbol{X})$ provides information enough about unknown parameter $\theta \in \boldsymbol{\theta}$. Theoretically speaking, the conditional distribution of $\boldsymbol{X}$ given $S(\boldsymbol{X})$ does not depend on $\theta$.
\end{itemize}
\end{dfn}
The reference {\it{prior}} is obtained by maximizing the expected value of the Kullback-Leibler (or logarithmic) divergence between the {\it{prior}} and the posterior. The logarithmic divergence of PDF $\pi(\theta)$ from PDF $\tilde{\pi}(\theta)$ is represented as
\begin{align}\label{kl}
{\cal{K}}(\pi \vert\vert \tilde{\pi})= \int_{\theta}{\pi}(\theta)\log \frac{{\pi}(\theta)}{\tilde{\pi}(\theta)} d\theta,
\end{align}
provided that integral (\ref{kl}) is finite. The reference {\it{prior}} must justify two convincing rationales including {\it{permissibility}} and {\it{expected logarithmically convergence}} \cite[Definition 5]{berger2009formal}. There follows a listing of nice properties of the reference prior including: 
\begin{itemize}
\item [(i)] Consistency under reparameterization of the unknown parameter $\theta$
\item[(ii)] Independence of sample size when observations are independent
\item[(iii)] Compatibility with sufficient statistic
\item[(iiii)] For a distribution for which nothing is known about unknown parameter $\theta$, the reference prior follows the uniform distribution 
\end{itemize}
For a proof of properties noted above, we refer the reader to \cite{berger2009formal}. The property (i) states, for a location family, that the reference {\it{prior}} for location parameter is flat (equally likely over $\mathbb{R}$) and within a scale family, the reference {\it{prior}} of the scale parameter $\theta$ is proportional to $1/\theta$ that is the same of Jeffreys {\it{prior}}, that is $\pi_{R}({\theta})=\pi_{J}({\theta})\propto 1/\theta$ \citep{bernardo2005reference}. Although the reference prior is an objective prior and inherits nice properties among them some noted above, but it suffers from some drawbacks as follows. 
\begin{itemize}
\item [(i)] The reference prior must be computed point-by-point and then approximated through interpolation techniques.
\item[(ii)] The burden of computations grows when sample size increases and it may be critical if $f(x\vert \theta)$ lacks closed form such as $\alpha$-stable distribution. 
\item[(iii)] Although the conditions for existence of reference prior for multi-parameter case is guaranteed, but obviously demands for more computational complexity. 
\end{itemize}
In general, a reference prior is obtained by following the steps of the algorithm given below  \citep{berger2009formal}.
\begin{algorithm}
\caption{Computing the reference prior}
\label{Computing the reference prior}
\begin{algorithmic}[1]
\State Choose a moderate integer value for $k$;
\State Choose an arbitrary positive function $\pi^{*}(\theta)$ such as $\pi^{*}(\theta)=1$;
\State Consider the sequence $\theta_1,\cdots, \theta_m$ to be $m$ arbitrary distinct points that $\theta$ can take on;
\State Set $j=1$;
\While{$j \leq m$}  %\Comment{put some comments here}
\State Generate realizations $x_{1},x_{2},\cdots,x_{k}$ from PDF $f\bigl(\cdot \big \vert \theta_{j}\bigr)$;
\State Compute the integral $c_{j}=\int_{\mathbb{R}}\prod_{i=1}^{k}f\bigl(x_{i} \big \vert u\bigr)\pi^{*}\bigl(u\bigr) du$;
\State Compute $r_{j}=\log \Bigl(\prod_{i=1}^{k}f\bigl(x_{j} \big \vert \theta_{j}\bigr)\pi^{*}\bigl(\theta_{j}\bigr) / c_{j}\Bigr)$;
\State  Compute $\pi\bigl(\theta_{j}\bigr)=\exp\bigl\{m^{-1}\sum_{j=1}^{m}r_{j}\bigr\}$ and store the pair $\bigl(\theta_{j},\pi(\theta_{j})\bigr)$
        \State Set $j \leftarrow j+1$;
        %\Comment{another comment}
        %\State $var3 \leftarrow var4$
     \EndWhile  %\label{roy's loop}
     \State {\bf{end}}
     \State The set $\bigl\{\bigl(\theta_{j},\pi(\theta_{j})\bigr)\bigr\}_{j=1}^{m}$ is the sequence of $m$ paired from which $\pi(\theta)$ can be obtained through interpolation. 
%\EndProcedure
\end{algorithmic}
\end{algorithm}
\subsection{Jeffreys prior}
The second class of objective {\it{prior}} is known as the Jeffreys {\it{prior}} $\pi_{J}(\boldsymbol{\theta})$ that was proposed by \cite{jeffreys1946invariant}. We have
%The pdf of this prior is proportional to the square root of the determinant of the Fisher information matrix represented as
\begin{align}\label{Jeffreys}
\pi_{J}(\boldsymbol{\theta}) \propto \sqrt{ {\text{det}} I(\boldsymbol{\theta})},
\end{align}
where ${I}(\boldsymbol{\theta})=-E\bigl[\partial^2/\partial \theta_{i} \partial \theta_{j} \log L(\boldsymbol{\theta}\vert \boldsymbol{x}) \bigr]$, for $i,j=1,\cdots,d$ is the Fisher information matrix.
\subsection{Conjugate prior}
The last class of priors is the conjugate prior $\pi_{C}(\theta)$. If prior and posterior follow the same family of distributions, then the corresponding {\it{prior}} is called a conjugate {\it{prior}}. For example, let $x_1,x_2,\cdots,x_n$ independently follow ${\cal{N}}(\mu,\sigma^2)$. Considering a {\it{prior}} with PDF ${\cal{N}}(\cdot \vert \mu_{0},\sigma_{0}^2)$, it is easy to see that {\it{posterior}} of $\mu$ is
\begin{align}\label{gausssianconjugacy}
\pi( \mu \vert \boldsymbol{x}) \sim {\cal{N}}\biggl(
\frac{\frac{\sigma^{2}}{n}\mu_{0}+\sigma_{0}^{2}\bar{x}}{
          \frac{\sigma^{2}}{n}+\sigma_{0}^{2}},
      \Bigl(\frac{n}{\sigma^{2}}+\frac{1}{\sigma_{0}^{2}}\Bigr)^{-1}
\biggr),
\end{align}
where $\bar{x}$ is the sample mean and $\boldsymbol{\theta}_{0}=(\mu_{0},\sigma_{0})^{\top}$ is vector of {\it{hyperparameters}}. As it is seen from (\ref{gausssianconjugacy}), the {\it{posterior}} follows a Gaussian distribution too. 
%follows an inverse-gamma (IG) is a conjugate prior for $\delta=\sigma^2$ since the pertaining posterior follows an IG distribution. 
We note that the conjugate prior is an {\it{subjective}} prior since it is chosen by experimenter without rationale and just for the sake of simplicity in form and specially sampling from. The main advantage of the conjugate {\it{prior}} is that sampling from the corresponding posterior is an easy task.
\par
If the PDF of family under study has not closed form, finding $\pi_{J}(\theta)$, $\pi_{R}(\theta)$, or even $\pi_{C}(\theta)$ may be difficult. As we will see in the next section, uniform and conjugate priors will be proposed for $\alpha$ and $\delta=\sigma^2$, respectively. These types were chosen by \citep{buckle1995bayesian,lombardi2007bayesian,godsill1999bayesian} for Bayesian estimation of tail thickness and scale parameters of $\alpha$-{\it{stable}} distribution. 
%If the Bayesian inference about scale parameter is of interest, then a common prior is the IG conjugate prior \cite{gelman2006prior} or Jeffreys (improper) prior $\pi_{J}(\sigma) \propto 1/\sigma$ \cite{karakucs2020modelling}. 
However, some drawbacks of IG conjugate {\it{prior}} were addressed by \citep{gelman2006prior,teimouri2024bayesian}. 
%In this work, as we will see, the IG conjugate prior does not work well when the sample size is small. 
Indeed, investigating advantages and disadvantages of above-noted {\it{priors}} in more details needs more space that is out of scope of this work. We refer the reader to \citep{jeffreys1946invariant,jaynes1968prior,gelman1995bayesian,berger2009formal,box2011Bayesian} for further specific information. 
%%%%%%%%%%%%%%%%%%%%%%%
%%%%%%%%%%%%%%%%%%%%%%%
\section{Empirical Bayes}\label{empiricalbayes}
As noted earlier, the prior itself may depend on hyperparameter(s). Hence, sometimes we use the notation $\pi(\theta\vert \boldsymbol{\theta}_{0})$ rather than $\pi(\theta)$ to emphasize the fact that prior depends on hyperparameter $\boldsymbol{\theta}_{0}$. Evidently, for characterizing distribution of the unknown parameter $\theta$ through the posterior in (\ref{bayes}), it may be seen
\begin{align}\label{bayes3}
\pi\bigl(\theta \big \vert\boldsymbol{x},\boldsymbol{\theta}_{0}\bigr) =\frac{f(\boldsymbol{x}\vert\theta)}{m\bigl(\boldsymbol{x}\vert \boldsymbol{\theta}_{0}\bigr)}\pi\bigl(\theta\vert \boldsymbol{\theta}_{0}\bigr),
\end{align}
where the unknown {\it{hyperparameter}} $\boldsymbol{\theta}_{0}$ plays the role of a {\it{nuisance}} parameter. The empirical Bayes is essentially an inferential approach, for estimating $\boldsymbol{\theta}_{0}$, that was introduced by \cite{herbert1956empirical}. It takes two {\it{non-parametric}} and {\it{parametric}} types \citep{casella1985introduction}. The empirical Bayes proposed by \cite{herbert1956empirical} is a {\it{non-parametric}} approach under which experimenter has no any assumption about specification(s) of the prior's distribution while within a {\it{parametric}} framework the family of prior is fully specified by experimenter \citep{morris1983parametric,casella1985introduction}. More precisely, the {\it{parametric}} empirical Bayes takes into account information provided by sample evidence $\boldsymbol{x}$ for estimating {\it{hyperparameter}} $\boldsymbol{\theta}_{0}$. Herein, we assume that $\boldsymbol{\theta}_{0}$ is fixed, but unknown. If $\boldsymbol{\theta}_{0}$ itself follows some distribution, then the procedure of estimating $\theta$ within Bayesian framework is called {\it{hierarchical Bayesian inference}}. For a single observation $x_i$, we recall from (\ref{bayes3}) that 
\begin{align}\label{empirical-marginal}
\pi\bigl(\theta \big \vert{x}_{i},\boldsymbol{\theta}_{0}\bigr) =\frac{f({x}_{i}\vert\theta)}{m\bigl({x}_{i}\vert \boldsymbol{\theta}_{0}\bigr)}\pi\bigl(\theta\vert \boldsymbol{\theta}_{0}\bigr).
\end{align}
If $\pi\bigl(\theta \big \vert{x}_{i},\boldsymbol{\theta}_{0}\bigr)$ is proper, then $m\bigl({x}_{i}\vert \boldsymbol{\theta}_{0}\bigr)$ in denominator of (\ref{empirical-marginal}) itself is a PDF since $0<m\bigl({x}_{i}\vert \boldsymbol{\theta}_{0}\bigr)<\infty$ and $\int_{\mathbb{R}}m\bigl({x}_{i}\vert \boldsymbol{\theta}_{0}\bigr)d\boldsymbol{x}=1$. In fact, $m\bigl({x}_{i}\vert \boldsymbol{\theta}_{0}\bigr)$ plays the role of {\it{marginal}} PDF (likelihood function) when $\boldsymbol{\theta}$ is integrated out. The unknown hyperparameter $\boldsymbol{\theta}_{0}$ is then estimated through the maximum likelihood (ML) by maximizing $\prod_{i=1}^{n}m\bigl({x}_{i}\vert \boldsymbol{\theta}_{0}\bigr)$ with respect to $\boldsymbol{\theta}_{0}$ or using the method of moments \citep{teimouri2024bayesian}. In what follows, we proceed to compute the {\it{parametric}} empirical Bayes through an example.
%To do this, notice that the function $m\bigl(\boldsymbol{x}\vert \boldsymbol{\theta}_{0}\bigr)$ in denominator of (\ref{bayes3}) itself if the pdf of joint distribution. This is because $0<m\bigl(\boldsymbol{x}\vert \boldsymbol{\theta}_{0}\bigr)<\infty$ and $\int_{\mathbb{R}^{n}}m\bigl(\boldsymbol{x}\vert \boldsymbol{\theta}_{0}\bigr)d\boldsymbol{x}=1$. Herein, $\boldsymbol{\theta}_{0}$ can be seen as the parameters of the distribution and $m\bigl(\boldsymbol{x}\vert \boldsymbol{\theta}_{0}\bigr)$ plays the role of {\it{marginal}} pdf (likelihood function) since $\boldsymbol{\theta}$ is integrating out. To this end, one may use the method of maximum likelihood (ML) or method of moments \citep{teimouri2024bayesian} for estimating (recovering) $\boldsymbol{\theta}_{0}$ through the {\it{marginal}} likelihood. In what follows, we proceed to compute the {\it{parametric}} empirical Bayes through an example.
\begin{example}\label{exam-empiricalgaussian}%\lipsum*[]
Let $x_1,x_2,\cdots,x_n$ independently follow ${\cal{N}}(\mu,\sigma^2)$ in which scale parameter $\sigma>0$ is assumed to be known. Furthermore, a Gaussian prior  $\pi(\mu)={\cal{N}}\bigl(\mu \big \vert \mu_{0},\sigma_{0}^2\bigr)$, is assumed for $\mu$. For estimating $\boldsymbol{\theta}_{0}=(\mu_{0},\sigma_{0})^{\top}$, we can write
\begin{align*}%\label{bayes3}
m\bigl({x}_{i}\vert \boldsymbol{\theta}_{0}\bigr)=\int_{-\infty}^{\infty}f\bigl({x}_{i} \big \vert \mu\bigr)\pi\bigl(\mu \big \vert \boldsymbol{\theta}_{0}\bigr)d\mu=\int_{-\infty}^{\infty}f\bigl(x_i \big \vert \mu\bigr)\pi\bigl(\mu \big \vert \boldsymbol{\theta}_{0}\bigr)d\mu,
\end{align*} 
%\Pi_{i=1}^{n}
where $f\bigl(x_i \big \vert \mu\bigr)={\cal{N}}\big(x_{i}\big \vert\mu,\sigma^2\big)$. We have
\begin{align*}%\label{bayes3}
m\bigl({x}_{i}\vert \boldsymbol{\theta}_{0}\bigr)=\int_{-\infty}^{\infty}\frac{1}{\sqrt{2\pi} \sigma}\exp\Bigl\{-\frac{\bigl(x_{i}-\mu\bigr)^{2}}{2\sigma^{2}}\Bigr\}\frac{1}{\sqrt{2\pi} \sigma_{0}}\exp\Bigl\{-\frac{\bigl(\mu-\mu_{0}\bigr)^{2}}{2\sigma_{0}^{2}}\Bigr\}d\mu.
\end{align*} 
Simplifying and some algebra leads to
\begin{align*}%\label{bayes3}
m\bigl({x}_{i}\vert \boldsymbol{\theta}_{0}\bigr)={\cal{N}}\big(x_{i}\big \vert\mu_{0},\sigma^2+\sigma_{0}^{2}\big).
\end{align*} 
It is well known that the ML estimator of mean and variance of each Gaussian distribution are the sample counterparts. So, the empirical Bayesian estimator of $\mu_{0}$ and $\sigma_{0}$ are $\widehat{\mu_{0}}_{EB}=\bar{x}$ and $\widehat{\sigma_{0}}_{EB}=\sqrt{\vert S^2-\sigma^2\vert}$ where $\bar{x}$ and $S^2$ are the sample mean and variance, respectively. It is noteworthy that the quantity $S^2-\sigma^2$ may be negative especially when sample size is small. Hence, the absolute sign is used to avoid negative value for $\widehat{\sigma_{0}}_{EB}$. 
%
%\begin{align}\label{bayes5}
%l\bigl(\boldsymbol{\theta}_{0}\big \vert \boldsymbol{x}\bigr) =\text{C} -\frac{n}{2} \log \bigl(\sigma_{0}^{2}+\sigma^{2}\bigr)-\frac{\sum_{i=1}^{n}x_{i}^{2}}{2\sigma^{2}}-\frac{n\mu^{2}_{0}}{2\sigma_{0}^{2}} +
%\frac{1}{2}\Bigl(\frac{\sigma_{0}^{2}\sigma^{2}}{\sigma_{0}^{2}+\sigma^{2}}\Bigr)\sum_{i=1}^{n}\Bigl(
%\frac{ x_{i}}{\sigma^{2}}+\frac{\mu_{0}}{\sigma_{0}^{2}}\Bigr)^{2},
%\end{align}
%where constant $\text{C}$ is independent of $\boldsymbol{\theta}_{0}$. Differentiating the RHS of (\ref{bayes5}) with respect to $\boldsymbol{\theta}_{0}$ and then solving resultant in terms of $\boldsymbol{\theta}_{0}$ yields {parametric} empirical Bayes estimators for $\mu_{0}$ and $\sigma^{2}_{0}$, that is, $\widehat{\mu_{0}}$ and $\widehat{\sigma^{2}_{0}}$, as $\bar{x}$ and $1/n\sum_{i=1}^{n}(x_{i}-\bar{x})^2-\sigma^2$, respectively. 
\end{example}
It is immediate from Example \ref{exam-empiricalgaussian} above that the Bayesian estimator of $\mu$, that is $\hat{\mu}_{B}$, becomes the average of the posterior (\ref{gausssianconjugacy}). Hence, substituting $\widehat{\mu_{0}}_{EB}$ and $\widehat{\sigma_{0}}_{EB}$ obtained above in RHS of (\ref{gausssianconjugacy}) yields $\hat{\mu}_{B}=\bar{x}$ with standard error
\begin{align*}%\label{muB}
SE\bigl(\hat{\mu}_{B}\bigr)= \Bigl(\frac{n}{\sigma^{2}}+\frac{1}{\vert S^2-\sigma^2\vert}\Bigr)^{-\frac{1}{2}}.
\end{align*}
Evidently, $\hat{\mu}_{B}$ is more efficient than the ML counterpart $\hat{\mu}_{ML}=\bar{x}$ since
\begin{align*}%\label{SE} 
SE\bigl(\hat{\mu}_{B}\bigr)<SE\bigl(\hat{\mu}_{ML}\bigr)=\frac{\sigma}{\sqrt{n}}.
\end{align*}
A different version of Example \ref{exam-empiricalgaussian} in which $x_i$s independently, for $i=1,\cdots,n$, follow ${\cal{N}}(\mu_{i},\sigma^2)$ is discussed by \cite{casella1985introduction}.
\begin{example}\label{exam-empiricalbernoulli}%\lipsum*[]
Suppose $x_1,x_2,\cdots,x_n$ follow independently a Bernoulli distribution with success probability $y$, that is $X_i \sim {\cal{BER}}(y)$ for $i=1,\cdots,n$ and $0<y<1$. For computing the Bayes estimation of $y$, that is $\hat{y}_{B}$, we may consider a Beta distribution with PDF $\pi\bigl(y \big \vert \boldsymbol{\theta}_{0}\bigr)=1/{\cal{B}}(a,b)y^{a-1}(1-y)^{b-1}$ for the prior of $y$ where $\boldsymbol{\theta}_{0}=(a,b)^{\top}$ is the vector of hyperparameters. We can write
\begin{align}\label{posteriorbernoulli}
\pi\bigl(y \big \vert \boldsymbol{x},\boldsymbol{\theta}_{0}\bigr )&= y^{\sum_{i=1}^{n}x_{i}}(1-y)^{n-\sum_{i=1}^{n}x_{i}}\pi(y\vert a,b)\nonumber\\
&\propto y^{\sum_{i=1}^{n}x_{i}+a-1}(1-y)^{n-\sum_{i=1}x_{i}+b-1},
\end{align}
where $\boldsymbol{x}=\{x_1,x_2,\cdots,x_n\}$. Obviously, $y \big \vert \boldsymbol{x}$ follows a beta distribution. So, we can consider the mean of posterior, that is $(\sum_{i=1}^{n}x_{i}+a-1)/(n+a+b-2)$, as the Bayes estimator for the success probability $y$. Recalling from Subsection \ref{empiricalbayes}, for computing the vector of hyperparameters $\boldsymbol{\theta}_{0}=(a,b)^{\top}$, we can write
\begin{align*}%\label{empiricalbernoulli}
m\bigl({x}_{i}\vert \boldsymbol{\theta}_{0}\bigr)=\int_{0}^{1}f\bigl({x}_{i} \big \vert y\bigr)\pi\bigl(y \big \vert \boldsymbol{\theta}_{0}\bigr)dy
&=\int_{0}^{1} \frac{y^{x_{i}+a-1}(1-y)^{b-x_{i}}}{{\cal{B}}(a,b)}dy\nonumber\\
&=\frac{{\cal{B}}(x_{i}+a,b-x_{i}+1)}{{\cal{B}}(a,b)}\nonumber\\
&=\frac{\Gamma(x_i+a)\Gamma(b-x_i+1)}{(a+b)\Gamma(a)\Gamma(b)}.
\end{align*} 
Hence the likelihood function becomes
\begin{align}\label{empirical-bernoulli}
m\bigl(\boldsymbol{x}\vert \boldsymbol{\theta}_{0}\bigr)=\frac{\Pi_{i=1}^{n}\Gamma(x_i+a)\Gamma(b-x_i+1)}{(a+b)^n[\Gamma(a)\Gamma(b)]^n}.
\end{align} 
Differentiating the RHS of (\ref{empirical-bernoulli}) with respect to $\boldsymbol{\theta}_{0}$ and some algebra leads to the empirical Bayes estimator of $\boldsymbol{\theta}_{0}$ as $\widehat{\boldsymbol{\theta}_{0}}_{EB}=\bigl(\sum_{i=1}^{n}x_{i}/n, (n-\sum_{i=1}^{n}x_{i})/n\bigr)^{\top}$. 
\end{example}
If in Example \ref{exam-empiricalbernoulli}, the Bayesian estimator $\hat{y}_{B}$, of success probability $y$ is desired, one could assume $\hat{y}_{B}$ is equal to the expected value of posterior as usual. To obtain $\hat{y}_{B}$, we recall from (\ref{posteriorbernoulli}) that 
\begin{align} 
\pi(y \big \vert \boldsymbol{x} )= \frac{y^{\sum_{i=1}^{n}x_{i}+a-1}(1-y)^{n-\sum_{i=1}x_{i}+b-1}}{{\cal{B}}\bigl(\sum_{i=1}^{n}x_{i}+a-1, n-\sum_{i=1}x_{i}+b-1\bigr)}.
\end{align} 
It turns out that $y \big \vert \boldsymbol{x}$ follows a Beta distribution with parameters $\sum_{i=1}^{n}x_{i}+a$ and $n-\sum_{i=1}x_{i}+b$. Hence,
\begin{align}\label{bayesbernoulli1}
\hat{y}_{B}=E\bigl(y\big \vert \boldsymbol{x}\bigr)&=\int_{0}^{1} y \pi(y \big \vert \boldsymbol{x}) dy=\frac{\sum_{i=1}^{n}x_{i}+a}{n+a+b}.
\end{align} 
Substituting $\boldsymbol{\theta}_{0}$ computed above into the RHS of (\ref{bayesbernoulli1}) yields 
\begin{align*}%\label{bayesbernoulli2}
\hat{y}_{B}=\frac{\sum_{i=1}^{n}x_{i}+\frac{\sum_{i=1}^{n}x_{i}}{n}}{n+1}.
\end{align*}
As it is see, the Bayesian estimator $\hat{y}_{B}$ is asymptotically (that is, $n \rightarrow \infty$) equal to the maximum likelihood (ML) estimator $\hat{y}_{ML}=\sum_{i=1}^{n}x_{i}/n$ of success probability $y$. 
%\frac{\sigma_{0}^{2}\sigma^{2}}{\sigma_{0}^{2}+\sigma^{2}}
\begin{example}\label{exam-empirical-gamma}%\lipsum*[]
Let $x_1,x_2,\cdots,x_n$ come independently from ${\cal{G}}(\alpha,\beta)$ with unknown $\alpha$ and known $\beta$. For computing the Bayesian estimator of $\alpha$, we suppose that $\pi\bigl(\alpha\big \vert \boldsymbol{\theta}_{0}\bigr)={\cal{G}}\bigl(\alpha \big \vert \alpha_{0},\beta_{0}\bigr)$.
%For constructing an empirical confidence interval for $\alpha$, 
But, first we need to determine hyperparameter $\boldsymbol{\theta}_{0}=(\alpha_{0},\beta_{0})^{\top}$. To this end, we proceed through the empirical Bayes as follows.
\begin{align*}
\pi\bigl(\alpha \big \vert \boldsymbol{x},\boldsymbol{\theta}_{0}\bigr)= \frac{f(\boldsymbol{x} \vert \alpha)\pi\bigl(\alpha\big \vert \boldsymbol{\theta}_{0}\bigr)}{m\bigl(\boldsymbol{x}\vert \boldsymbol{\theta}_{0}\bigr)},
\end{align*} 
where
\begin{align*}
f(\boldsymbol{x} \vert \alpha)\pi\bigl(\alpha\big \vert \boldsymbol{\theta}_{0}\bigr)=&\Bigl[\prod_{i=1}^{n}\frac{\beta^{\alpha}x_{i}^{\alpha-1}}{\Gamma(\alpha)}\exp
\bigl\{-\beta x_{i}\bigr\}\Bigr]\frac{\beta_{0}^{\alpha_{0}}\alpha^{\alpha_{0}-1}}{\Gamma\bigl(\alpha_{0}\bigr)}\exp\bigl\{-\beta_{0} \alpha\bigr\},\nonumber\\
=&\frac{\beta_{0}^{\alpha_{0}}\beta^{n\alpha}\alpha^{\alpha_{0}-1}}{\Gamma\bigl(\alpha_{0}\bigr)\bigl[\Gamma(\alpha)\bigr]^{n}}\Bigl[\prod_{i=1}^{n}x_{i}\Bigr]^{\alpha-1}\exp\bigl\{-\beta \sum_{i=1}^{n}x_{i}-\beta_{0}\alpha\bigr\},\nonumber
%\propto&\frac{\beta^{n\alpha}\alpha^{\alpha_{0}-1}}{\bigl[\Gamma(\alpha)\bigr]^{n}}\Bigl[\prod_{i=1}^{n}x_{i}\Bigr]^{\alpha-1}\exp\bigl\{-\beta \sum_{i=1}^{n}x_{i}-\beta_{0}\alpha\bigr\}.
\end{align*}
and
\begin{align}\label{empiricalgamma}
m\bigl({x}_{i}\vert \boldsymbol{\theta}_{0}\bigr)&=\int_{0}^{\infty}f\bigl({x}_{i} \big \vert \alpha\bigr)\pi\bigl(\alpha \big \vert \boldsymbol{\theta}_{0}\bigr)d\alpha\nonumber\\
&=\int_{0}^{\infty} \beta^{\alpha}\frac{x_{i}^{\alpha-1}}{\Gamma(\alpha)}\exp
\bigl\{-\beta x_{i}\bigr\}\beta_{0}^{\alpha_{0}}\frac{\alpha^{\alpha_{0}-1}}{\Gamma\bigl(\alpha_{0}\bigr)}\exp\bigl\{-\beta_{0} \alpha\bigr\}d\alpha.
\end{align} 
Obviously, the integral in the RHS of (\ref{empiricalgamma}) has no closed form and so computing $\widehat{\boldsymbol{\theta}_{0}}_{EB}$ using the ML approach is computationally cumbersome. Herein, we proceed to find the moment-based estimator of ${\boldsymbol{\theta}_{0}}$. First note that  
\begin{align}\label{empiricalgamma2}
E\bigl(X^{s} \big \vert \boldsymbol{\theta}_{0}\bigr)&=\int_{0}^{\infty} x^{s}\int_{0}^{\infty}f\bigl(x \big \vert \alpha\bigr)\pi\bigl(\alpha \big \vert \boldsymbol{\theta}_{0}\bigr)d\alpha dx\nonumber\\
&=\int_{0}^{\infty} x^{s}\int_{0}^{\infty} \beta^{\alpha}\frac{x^{\alpha-1}}{\Gamma(\alpha)}\exp
\bigl\{-\beta x\bigr\}\beta_{0}^{\alpha_{0}}\frac{\alpha^{\alpha_{0}-1}}{\Gamma\bigl(\alpha_{0}\bigr)}\exp\bigl\{-\beta_{0} \alpha\bigr\}d\alpha dx\nonumber\\
&=\int_{0}^{\infty}\frac{ \beta^{\alpha}\beta_{0}^{\alpha_{0}}\alpha^{\alpha_{0}-1}\exp\bigl\{-\beta_{0} \alpha\bigr\}}{\Gamma(\alpha)\Gamma\bigl(\alpha_{0}\bigr)}\int_{0}^{\infty} x^{s+\alpha-1}\exp
\bigl\{-\beta x\bigr\} dxd\alpha\nonumber\\
&=\int_{0}^{\infty}\frac{ \beta^{\alpha}\beta_{0}^{\alpha_{0}}\alpha^{\alpha_{0}-1}\Gamma(s+\alpha)\exp\bigl\{-\beta_{0} \alpha\bigr\}}{\Gamma(\alpha)\Gamma\bigl(\alpha_{0}\bigr)\beta^{s+\alpha}}d\alpha,
\end{align} 
where $s \in \mathbb{N}$. It is easy to check that 
\begin{align}\label{empirical-gamma-moment1}
E\bigl(X \big \vert \boldsymbol{\theta}_{0}\bigr)&=\frac{\alpha_{0}}{\beta_{0} \beta},
\end{align} 
and
\begin{align}\label{empirical-gamma-moment2}
E\bigl(X^2 \big \vert \boldsymbol{\theta}_{0}\bigr)&=\frac{\alpha_{0}}{\beta_{0} \beta^2} + \frac{\alpha_{0}\bigl(\alpha_{0}+1\bigr)}{\beta^{2}_{0} \beta^2}. 
\end{align} 
Using (\ref{empirical-gamma-moment1}) and (\ref{empirical-gamma-moment2}), and some simplifications yields 
\begin{align}\label{empirical-gamma-alpha0}
\alpha_{0}=\frac{\beta\bigl[E\bigl(X \big \vert \boldsymbol{\theta}_{0}\bigr)\bigr]^2}{\beta \text{var}\bigl(X \big \vert \boldsymbol{\theta}_{0}\bigr)-E\bigl(X \big \vert \boldsymbol{\theta}_{0}\bigr)}.
\end{align} 
So, based on sample evidence $\{x_{1},x_{2},\cdots,x_{n}\}$, the moment-based estimator $\widehat{\alpha_{0}}_{EB}$ is given by
\begin{align}\label{empirical-gamma-alpha0hat}
\widehat{\alpha_{0}}_{EB}=\frac{\beta \bar{x}^{2}}{\big \vert \beta S^2-\bar{x}\big \vert}.
\end{align} 
Consequently, the moment-based estimator $\widehat{\beta_{0}}_{EB}$ is then given by
\begin{align}\label{empirical-gamma-beta0hat}
\widehat{\beta_{0}}_{EB}=\frac{\widehat{\alpha_{0}}_{EB}}{\beta \bar{x}}.
\end{align} 
As it is seen, theoretically, denominator in the RHS of (\ref{empirical-gamma-alpha0}) is zero. This means that $\widehat{\alpha_{0}}_{EB}$ must be large. Moreover, the absolute sign is used in denominator of (\ref{empirical-gamma-alpha0hat}) to avoid negative value for $\widehat{\alpha_{0}}_{EB}$. 
\begin{figure}[h]
\center
\includegraphics[width=55mm,height=55mm]{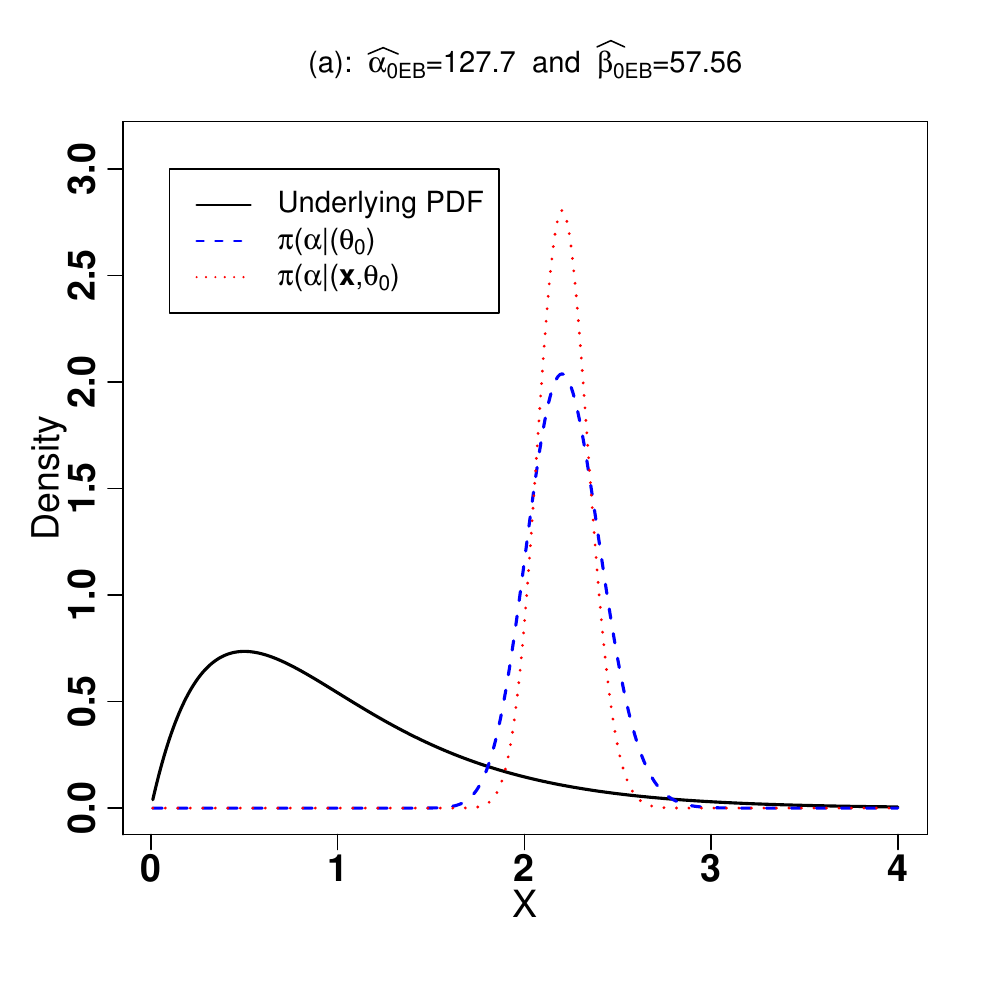}
\includegraphics[width=55mm,height=55mm]{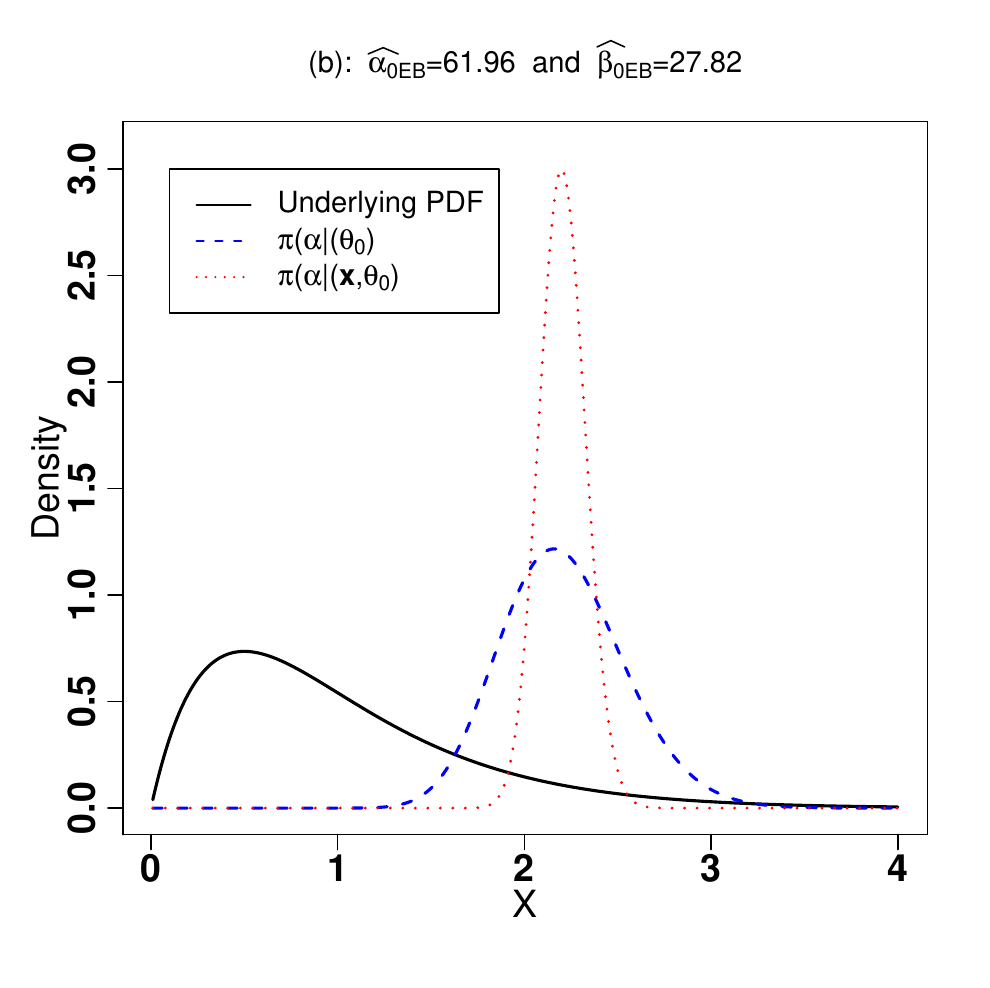}
\caption{Schematic diagram for PDF of underlying, prior, and posterior when (a): $n=50$ and (b): $n=100$.}
\label{empiricalgamma}
\end{figure}
Figure \ref{empiricalgamma} shows the visual comparison between prior and posterior distributions when sample evidence are generated from ${\cal{G}}(\alpha=2,\beta=2)$ as underlying distribution for two data sets of sizes $n=50$ and $n=100$.
\end{example}
%%%%%%%%%%%%%%%%%%%%%%%%%%%%%%%%%%%%%%%%%%%
\section{Credible interval}
One common type of inference in frequentist statistics is {\it{confidence interval}}. The Bayesian analogue of confidence interval is known as {\it{credible interval}} and defined as follows.
\begin{dfn}\label{credibleinterval}
Given the observed sample evidence $\boldsymbol{x}=\{x_1,\cdots,x_n\}$, the interval $(L,U)$ is a $100(1-\gamma)\%$ credible interval for unknown parameter $\theta$ if
\begin{align}
P(L<\theta <U\vert \boldsymbol{x}) =\int_{L}^{U} \pi(\theta \vert \boldsymbol{x}) d\theta \geq 1-\gamma,
\end{align}
where $0<\gamma<1$. 
\end{dfn}
Indeed, Definition \ref{credibleinterval} states that there is at least $100(1-\gamma)\%$ confidence that interval $(L,U)$ includes unknown parameter $\theta$. Figure \ref{credibleinterval} provides a schematic diagram for credible interval.
\begin{figure}[h]
\center
\includegraphics[width=55mm,height=55mm]{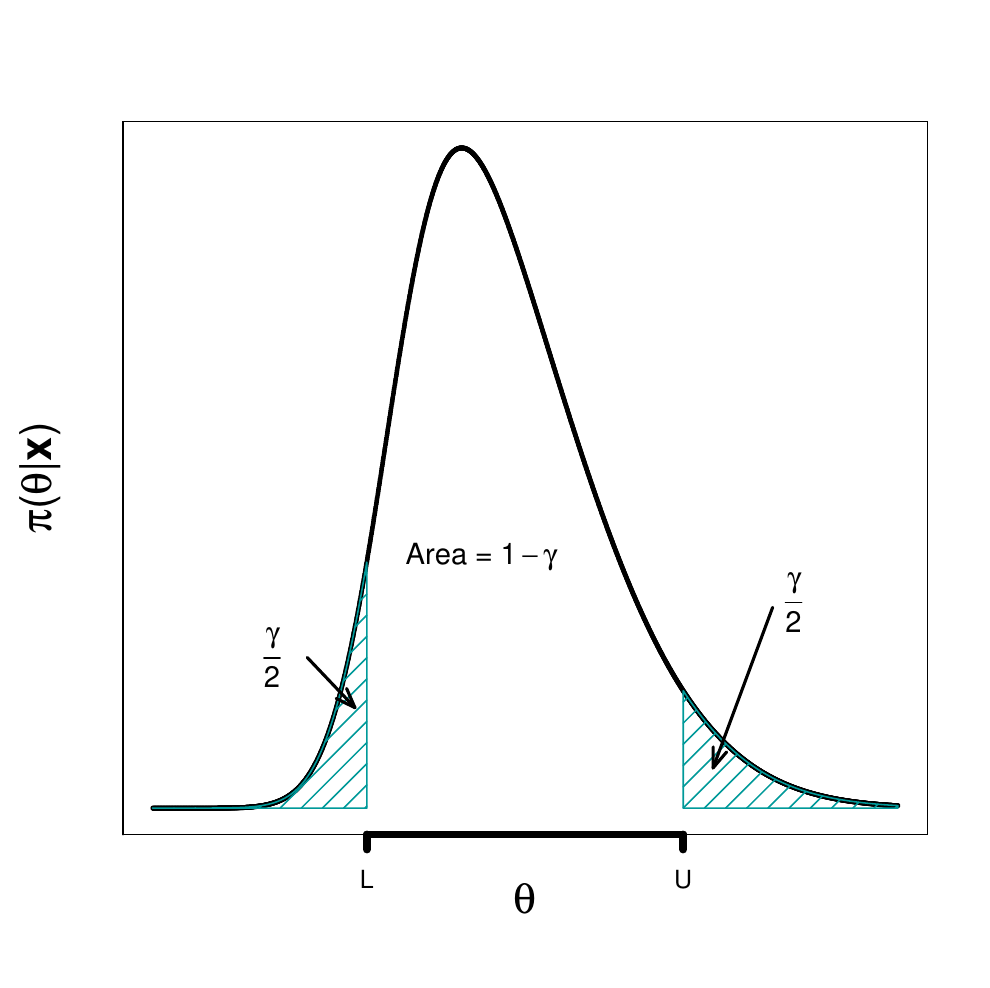}
\caption{Schematic diagram for credible interval.}
\label{credibleinterval}
\end{figure}
\begin{example}\label{exam-credibleinterval}%\lipsum*[]
Let $x_1,x_2,\cdots,x_n$ come independently from ${\cal{N}}(\mu,\sigma^2)$ with known variance. By considering a conjugate prior ${\cal{N}}(\cdot \vert \mu_{0},\sigma_{0}^2)$ for $\mu$, we recall from (\ref{gausssianconjugacy}) that $\mu \vert \boldsymbol{x}$ follows a Gaussian distribution. Hence
\begin{align}\label{conjugacy1}
\sqrt{\frac{n}{\sigma^{2}}+\frac{1}{\sigma_{0}^{2}}}
\biggl(\mu -\frac{n^{-1}\sigma^{2}\mu_{0}+\sigma_{0}^{2}\bar{x}}{n^{-1}\sigma^{2}+\sigma_{0}^{2} }\biggr)
          \sim{\cal{N}}(0,1).
\end{align}
Therefore, $100(1-\gamma)\%$ credible interval for $\mu$ is given by
\begin{align}\label{conjugacy1}
\frac{n^{-1}\sigma^{2}\mu_{0}+\sigma_{0}^{2}\bar{x}}{n^{-1}\sigma^{2}+\sigma_{0}^{2} } \pm z_{\gamma/2}\sqrt{\frac{n}{\sigma^{2}}+\frac{1}{\sigma_{0}^{2}}},
\end{align}
where $z_{\gamma/2}$ is the upper $\gamma/2$-quantile of the standard Gaussian distribution, that is, $P\bigl(Z>z_{\gamma/2}\bigr)=\gamma/2$ in which $Z\sim{\cal{N}}(0,1)$.
% represents the standard Gaussian $\alpha/2$th upper tail quantile, that is $P(.
\end{example}
As it is seen from Example \ref{exam-credibleinterval}, for constructing credible interval, the upper and lower $\gamma/2$ of the posterior must be known. If the posterior gets complicated from the latter quantities cannot be found easily. In such cases, credible interval is constructed based on upper $\gamma/2$-quantile of the samples generated from the posterior. Let $\pi_{\gamma/2}$ denote the posterior's upper $\gamma/2$-quantile, that is 
\begin{align}
\int_{\pi_{\gamma/2}}^{\infty}\pi(\theta \vert \boldsymbol{x})d\theta=\frac{\gamma}{2}.
\end{align}
Furthermore, suppose $\lceil{u}\rceil$ denotes the smallest integer value equal or greater than real value $u$ and $\theta_{(r)}$ is $r$th order statistic of sample $\{\theta_{1},\theta_{2},\cdots,\theta_{N}\}$ generated from posterior with PDF $\pi(\theta \vert \boldsymbol{x})$. When $N \rightarrow \infty$, then strong law of large numbers states that $\pi_{\gamma/2} \approx \theta_{(v)}$ where $v=\lceil{(1-\gamma/2)N}\rceil$. The definition below gives the empirical credible interval. 
\begin{dfn}\label{empiricalcredibleinterval}
Let $\{\theta_{1},\theta_{2},\cdots,\theta_{N}\}$, for large $N$, denote generated samples form posterior with PDF $\pi(\theta \vert \boldsymbol{x})$. A $100(1-\gamma)\%$ empirical credible interval for unknown parameter $\theta$ is then give by
\begin{align}
\bigl(L_{e},U_{e}\bigr)=\bigl(\theta_{(v)}, \theta_{(N-v)}  \bigr),
\end{align}
where $v=\lceil{(1-\gamma/2)N}\rceil$. 
\end{dfn}
\begin{example}[continued]\label{exam-credibleinterval-gamma}%\lipsum*[]
Recalling from Example \ref{exam-empirical-gamma}, suppose we observed $n=200$ observations, represented as $\boldsymbol{x}=\bigl\{x_1,x_2,\cdots,x_{200}\bigr\}$ from ${\cal{G}}(\alpha=2,\beta=2)$. For known $\beta$, using the moment-based method described earlier in Example \ref{exam-empirical-gamma}, we obtain $\boldsymbol{\theta}_{0}=\bigl(\widehat{\alpha_{0}}_{EB}, \widehat{\beta_{0}}_{EB}\bigr)^{\top}=(62.179,30.083)^{\top}$. The corresponding posterior is given by
%t $\pi\bigl(\alpha\big \vert \boldsymbol{\theta}_{0}\bigr)={\cal{G}}\bigl(\alpha \big \vert \alpha_{0},\beta_{0}\bigr)$.
%For constructing an empirical confidence interval for $\alpha$, 
%But, first we need to determine hyperparameter $\boldsymbol{\theta}_{0}=(\alpha_{0},\beta_{0})^{\top}$. To this end, we proceed through the empirical Bayes as follows.
\begin{align}\label{credibleinterval-gamma-postalpha}
\pi\bigl(\alpha \big \vert \boldsymbol{x},\boldsymbol{\theta}_{0}\bigr)\propto&\Bigl[\prod_{i=1}^{n}\frac{\beta^{\alpha}x_{i}^{\alpha-1}}{\Gamma(\alpha)}\exp
\bigl\{-\beta x_{i}\bigr\}\Bigr]\frac{\beta_{0}^{\alpha_{0}}\alpha^{\alpha_{0}-1}}{\Gamma\bigl(\alpha_{0}\bigr)}\exp\bigl\{-\beta_{0} \alpha\bigr\},\nonumber\\
\propto&\frac{\beta^{n\alpha}\alpha^{\alpha_{0}-1}}{\bigl[\Gamma(\alpha)\bigr]^{n}}\Bigl[\prod_{i=1}^{n}x_{i}\Bigr]^{\alpha-1}\exp\bigl\{-\beta_{0}\alpha\bigr\}.
\end{align}
In order to construct a 95\% empirical credible interval for $\alpha$, we need to sample from posterior (\ref{credibleinterval-gamma-postalpha}) using methods such as MH or ARS. Taking into account into the fact that $\widehat{\alpha_{0}}_{EB}=62.179$ and it is simple to see that the posterior (\ref{credibleinterval-gamma-postalpha}) is log-concave for $\alpha_{0}\geq 1$. So, we proceed for simulating from this posterior using the ARS approach. Based on a sample of size 5000 generated from posterior (\ref{credibleinterval-gamma-postalpha}), we obtain quantities $L_e=1.778$, $\hat{\alpha}_{B}=1.945$, and $U_e=2.121$. The scatterplot of generated samples as well as the latter quantities are shown in Figure \ref{empirical-credibleinterval-gamma}. 
\begin{figure}[h]
\center
\includegraphics[width=55mm,height=55mm]{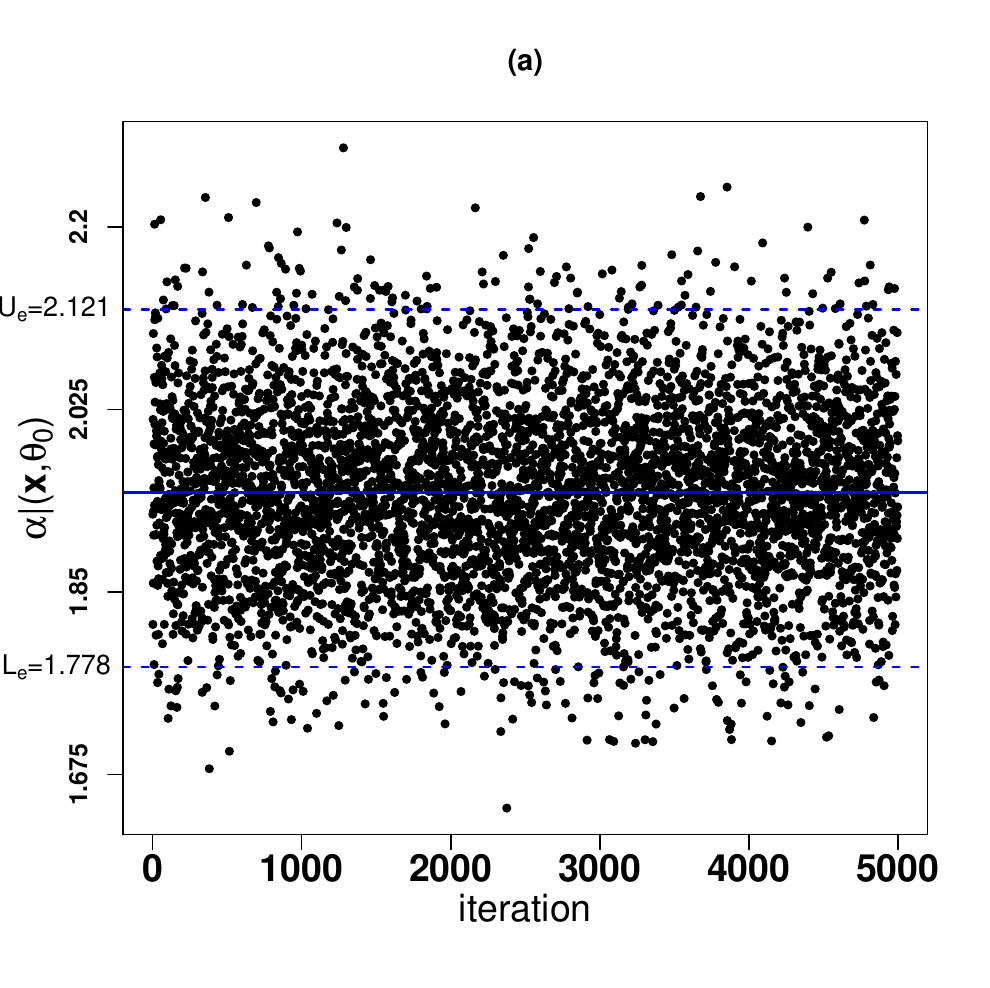}
\includegraphics[width=55mm,height=55mm]{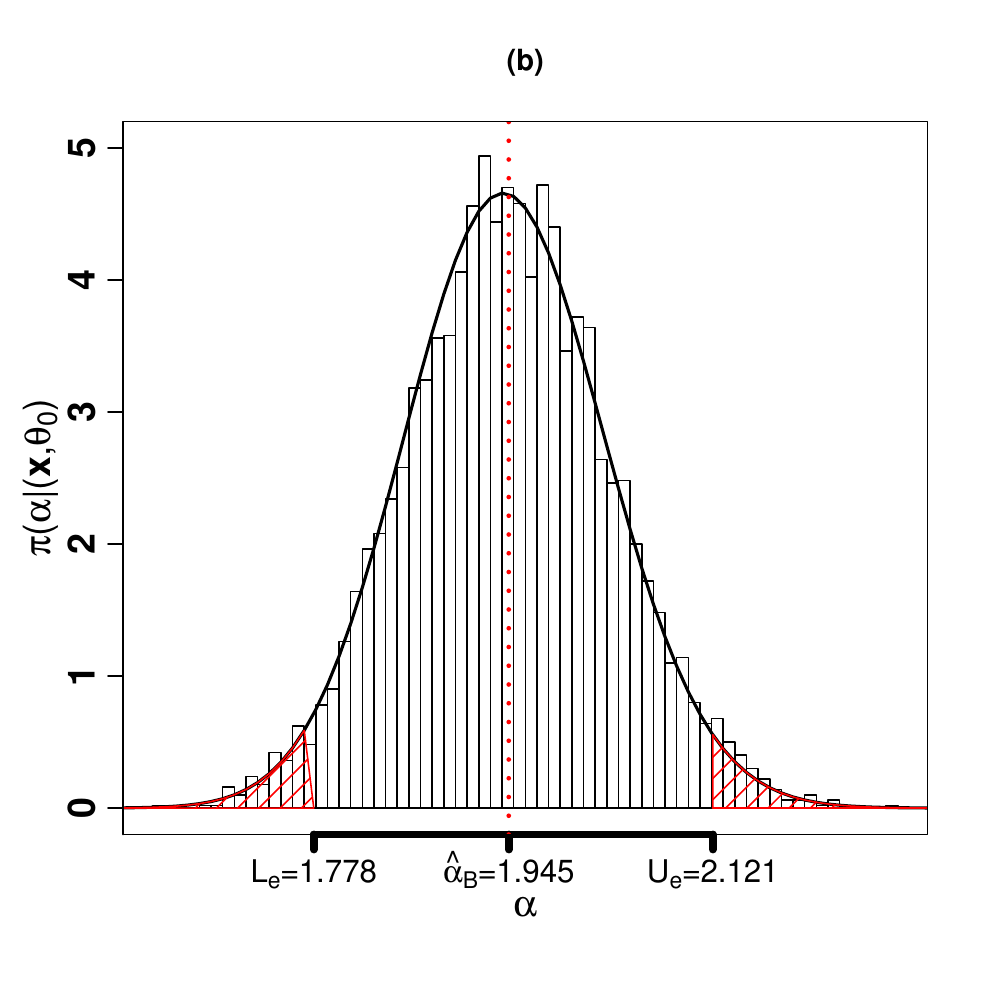}
\caption{(a): the 95\% empirical credible interval for parameter $\alpha$ based on a sample of $n=200$ realizations generated from ${\cal{G}}(\alpha=2,\beta=2)$ distribution and (b): histogram of the generated samples with fitted posterior $\pi\bigl(\alpha \big \vert \boldsymbol{x},\boldsymbol{\theta}_{0}\bigr)$.}
\label{empirical-credibleinterval-gamma}
\end{figure}
The \verb+R+ code for constructing the 95\% empirical credible interval for $\alpha$ is given below.
\begin{lstlisting}[style=deltaj]
R> f.a <- function(x, beta, alpha0, beta0, y)
+ 	{
+		 n=length(y); n*x*log(beta)-n*lgamma(x)+(alpha0-1)*log(x)+
+			(x-1)*sum(y)-beta0*x
+ 	}
R> fprim.a <- function(x, beta, alpha0, beta0, y)
+ 	{
+ 		 	 n=length(y); n*log(beta)-n*digamma(x)+(alpha0-1)/x+sum(y)-beta0
+ 	}
R> library(ars)
R> set.seed(20240608) 
R> n <- 200; alpha <-2; beta <- 2
R> y <- rgamma(n, shape = alpha, rate = beta)
R> alpha0 <- beta*mean(y)^2/abs( beta*var(y) - mean(y) )
R> beta0 <- alpha0/(beta*mean(y))
R> N <- 5000  # number of generations
R> M <- 0  # size of burn-in period
R> theta <- rep(0, N)
R> for(j in 1:N)
+	{
+			 theta[j] <- ars(1, f.a, fprim.a, x = c(0.1, 4, 10, 30), m = 4,
+				 lb = TRUE, xlb = 0,  ub = FALSE, beta = beta, alpha0 = alpha0, 
+ 			beta0 = beta0, y = log(y) )
+ 	}
R> Le <- quantile(theta[(M+1):N], 0.025)[[1]]
R> Ue <- quantile(theta[(M+1):N], 0.975)[[1]]
R> alphahat_B <- mean(theta)
R> list("Lower_bound"=Le, "Bayes estimator"=alphahat_B, "Upper_bound"=Ue)
\end{lstlisting}
\end{example}
\chapter{{\color{black}{Gibbs sampling}}}\label{sec3}
\begin{chapquote}{Bradley Efron, Stanford University}
% \textit{\cite{fernholz2000conversation}}}
``Statistics is the science of information gathering, especially when the information arrives in little pieces instead of big ones.''
\end{chapquote}
This section has three parts. First, we describe the concept of Gibbs sampling as a general framework for generating independent samples from the given multivariate target distribution. Second, we discuss the situation that the Gibbs sampling is utilized to estimate the model's unknown parameter(s) through the Bayesian paradigm.   
\section{Gibbs sampling in broad sense}
The Gibbs sampling (or sampler) is an iterative Markov chain Monte Carlo (MCMC) sampling approach
was developed by \cite{geman1984stochastic} for image reconstruction, but its application in statistical analyses were shown by \cite{gelfand1990sampling}. Assuming that we are at $t$th iteration of the Gibbs sampling and $\pi(\boldsymbol{x})$ denote the PDF of target distribution from which generating sample $\boldsymbol{x}=(x_1,\cdots,x_p)^{p}$ is of interest. For convenience, there follows a set of generic symbols as
%\begin{dfn} \\
\begin{align}
%\item 
&\boldsymbol{x}^{(t)}=\bigl({x}_{1}^{(t)},{x}_{2}^{(t)},\cdots,{x}_{p}^{(t)} \bigr),\nonumber\\
%\item 
&\boldsymbol{x}_{i}^{(t)}[y]=\bigl({x}_{1}^{(t)},\cdots,{x}_{i-1}^{(t)},y,{x}_{i+1}^{(t)},\cdots,{x}_{p}^{(t)} \bigr),\nonumber\\ 
&\boldsymbol{x}^{(t)}[-i]=\bigl({x}_{1}^{(t)},\cdots,{x}_{i-1}^{(t)},{x}_{i+1}^{(t)},\cdots,{x}_{p}^{(t)} \bigr).\label{minus}
%\item full conditional distribution: the conditional distribution of each ${\theta}_{i}$(for $i=1,\cdots,p$) given others, that is
%${\theta}_{i}\big \vert {\theta}_{1},{\theta}_{2},\cdots,{\theta}_{i-1},{\theta}_{i+1},\cdots,{\theta}_{p}$ is known in the literature as the {\it{full conditional}} of ${\theta}_{i}$
\end{align}
It is noteworthy that sampling from a $p$-dimensional distribution using the methods such as MH or AMH discussed earlier may involve undesired computational burden. In such cases, the Gibbs sampling technique that works efficiently is becoming increasingly popular in a wide range of applications. Indeed, each Gibbs sampling scheme works by generating sample from conditional distribution of vector $\boldsymbol{x}=(x_1,\cdots,x_p)^{p}$ that is known in the literature as the {\it{full conditional}}. For example, for generating $\boldsymbol{x}=(x_1,\cdots,x_p)^{p}$ form target PDF $\pi(\boldsymbol{x})$, the conditional distribution (or PDF) $\pi\bigl(x_{i}\big \vert \boldsymbol{x}[-i]\bigr)$ is called {\it{full conditional}} of $x_i$ (for $i=1,\cdots,p$). It is proved \citep{geman1984stochastic} that, after {\it{burn-in}} period \citep{roberts1994simple}, the generated sample comes from the target distribution whose PDF is $\pi(\boldsymbol{x})$. Each Gibbs sampling technique is summarized by the following algorithm.
\begin{algorithm}
\caption{Gibbs sampling Technique}
\begin{algorithmic}[1]\label{Gibbs}
%\Procedure{Roy}{$a,b$}     %  \Comment{This is a test}
 %   \State System Initialization
 %   \State Read the value 
%    \If{$condition = True$}
%        \State Do this
%        \If{$Condition \geq 1$}
%        \State Do that
%        \ElsIf{$Condition \neq 5$}
%        \State Do another
%        \State Do that as well
%        \Else
%        \State Do otherwise
%        \EndIf
%    \EndIf
 \State Set $t=1$, read $M$ ({\it{burn-in}} period), $N$ (total number of iterations), and 
 \State initiate algorithm with randomly selected sample $\boldsymbol{x}^{(0)}$;
    \While{$t \leq N$}  %\Comment{put some comments here}
    \While{$i \leq p$}  %\Comment{put some comments here}
        \State  Generate sample $y$ from full conditional with PDF $\pi\bigl(x^{(t)}_{i}\big \vert \boldsymbol{x}^{(t-1)}[-i]\bigr)$;
        \State Set $\boldsymbol{x}^{(t+1)}\leftarrow \boldsymbol{x}_{i}^{(t)}[y]$ (that is 
        \State ${x}_{i}^{(t+1)} \leftarrow y$ and $\boldsymbol{x}^{(t+1)}[-i] \leftarrow\boldsymbol{x}^{(t)}[-i]$);
        \State Set $i \leftarrow i+1$;
        %\Comment{another comment}
        %\State $var3 \leftarrow var4$
     \EndWhile  %\label{roy's loop}
     \State {\bf{end}}
         \State Set $t \leftarrow t+1$;
    \EndWhile  %\label{roy's loop}
                  \State {\bf{end}}
   \State  Sequence $\bigl\{\boldsymbol{x}_{M},\boldsymbol{x}_{M+1},\cdots,\boldsymbol{x}_{N}\bigr\}$ is a sample of size $N-M+1$ from 
    \State   target  distribution with PDF $\pi\bigl(\boldsymbol{x})$.          
%\EndProcedure
\end{algorithmic}
\end{algorithm}
\begin{example}\label{exam-bivariategaussian-simulation}%\lipsum*[]
Let $\boldsymbol{X}=\bigl(X_1,X_2\bigr)^{\top}$ follows a bivariate Gaussian distribution with PDF
\begin{align}\label{bivariatenormmal}
\pi\bigl(\boldsymbol{x}\big \vert \rho \bigr)=\frac{1}{2\pi \sqrt{1-\rho^2}}\exp\Bigl\{ -\frac{x_{1}^{2}+x_{2}^2-2\rho x_1x_2}{2(1-\rho^2)}\Bigr\}.
\end{align}
where $\boldsymbol{x}=\bigl(x_1, x_2 \bigr)^{\top}$ and $-1<\rho<1$. Since we have just two variables $X_1$ and $X_2$, so for simulating from $\boldsymbol{X}$ using the Gibbs sampling technique, we need to simulate from full conditionals $X_1 \big \vert \bigl(X_2, \rho \bigr)$ and $X_2 \big \vert \bigl(X_1, \rho \bigr)$. It can be easily seen that
\begin{align}\label{full1}
\pi\bigl(x_1 \big \vert x_2 ,\rho \bigr)\propto \exp\Bigl\{ -\frac{x_{1}^{2}-2\rho x_1x_2}{2(1-\rho^2)}\Bigr\}.
\end{align}
Since $x_{1}^{2}-2\rho x_1x_2=(x_1-\rho x_2)^2-(\rho x_2)^2$, then it follows that
\begin{align}\label{full2}
\pi\bigl(x_1 \big \vert x_2,\rho \bigr)\propto \exp\Bigl\{ -\frac{(x_1-\rho x_2)^2}{2(1-\rho^2)}\Bigr\}.
\end{align}
This means that
$X_1 \big \vert \bigl(X_2, \rho\bigr) \sim {\cal{N}}(\rho x_2, 1-\rho^2)$. Likewise, we can write $X_2 \big \vert \bigl(X_1, \rho\bigr) \sim {\cal{N}}(\rho x_1, 1-\rho^2)$. Hence, given $\boldsymbol{x}^{(0)}=\bigl(x^{(0)}_1,x^{(0)}_2\bigr)^{\top}$, for simulating a sample of $n=3000$ observations when $M=2000$, the steps of the Gibbs sampling given in Algorithm (\ref{Gibbs}) are as follows.
\begin{enumerate}
\item Set $t=1$, and generate $\boldsymbol{x}^{(0)}=\bigl(x^{(0)}_{1},x^{(0)}_{2}\bigr)^{\top}$ in which $x^{(0)}_{1} \sim {\cal{N}}\bigl(0, 1-\rho^2\bigr)$ and $x^{(0)}_{2} \sim {\cal{N}}\bigl(0, 1-\rho^2\bigr)$;
\item Set $t=0$;
\item Simulate $x^{(t+1)}_{1}$ from the full conditional $x^{(t+1)}_{1} \big \vert \bigl(x^{(t)}_{2},\rho\bigr) \sim {\cal{N}}\bigl(\rho x^{(t)}_{2}, 1-\rho^2\bigr)$;
\item Having $x^{(t+1)}_{1}$, simulate $x^{(t+1)}_{2}$ from the full conditional $x^{(t+1)}_{2} \big \vert \bigl(x^{(t)}_{1},\rho\bigr)  \sim {\cal{N}}\bigl(\rho x^{(t+1)}_{1}, 1-\rho^2\bigr)$;
\item If $t=5000$, then go to the next step, otherwise $t \leftarrow t+1$ and return step 3;
\item Sequence $\bigl\{\boldsymbol{x}_{2001},\boldsymbol{x}_{2002},\cdots,\boldsymbol{x}_{5000}\bigr\}$ constitutes a sample of size 3000 from target distribution with PDF $\pi\bigl(\boldsymbol{x} \big \vert \rho \bigr)$.    
 \end{enumerate}
The associated \verb+R+ code for generating 3000 realizations from ${\cal{N}}_{2}(\mu_{1}=0, \mu_{2}=0, \sigma_{1}^{2}=1, \sigma_{2}^{2}=1, \rho=0.50)$ is given as follows.
\begin{lstlisting}[style=deltaj]
R > set.seed(20240520)
R > N <- 5000  # number of generations
R > M <- 2000  # size of burn-in period
R > rho <- 0.5
R > X <- matrix(0, nrow = N, ncol = 2)
R > sigma1 <- sigma2 <- 1
R > Sigma <- matrix( c(1, rho*sigma1*sigma2, rho*sigma1*sigma2, 1), 2, 2)
R > X[1, ] <- rnorm (2, mean = 0, sd = 1 - rho^2) # intitial generation
R > j <- 1
R > while(j < N)
+	{
+  			                  X[j + 1, 1] <- rnorm (1, mean = rho*sqrt(Sigma[1,1]/Sigma[2,2])*X[j    , 2], 
+	            			sd = 1 - rho^2)
+  			X[j + 1, 2] <- rnorm (1, mean = rho*sqrt(Sigma[2,2]/Sigma[1,1])*X[j + 1, 1],
+			   	sd= 1 - rho^2)
+		       		j <- j + 1
+	             		}
R > plot(X[(N-M+1):N,])
\end{lstlisting}
%^{\top}\sim {\cal{N}}_{2}(\boldsymbol{0}_2,\Sigma)$ where
%  In Example \ref{eam2} it is easy to check, for $t \geq 1$ that
%\begin{equation}
%\biggl(\begin{array}{c}
%x^{(t)}_{1}\\
%x^{(t)}_{2} \end{array}\biggr) \sim {\cal{N}}_{2}\Bigl(\biggl(\begin{array}{c}
%\rho^{(2t-1)}x^{(0)}_{2}\\
%\rho^{(2t)}x^{(0)}_{2} \end{array}\biggr) \Bigr)
%\end{equation}
 \end{example}
 \begin{figure}[h]
\center
\includegraphics[width=55mm,height=55mm]{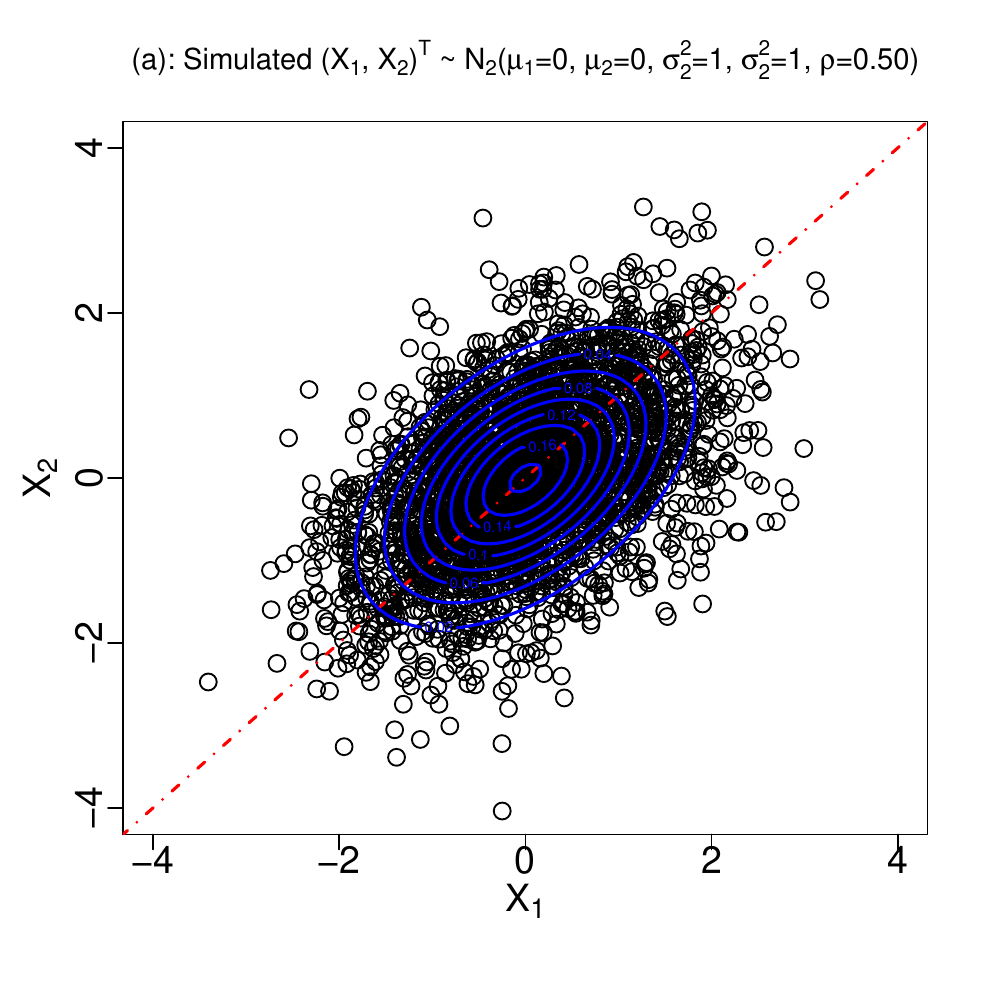}
\includegraphics[width=55mm,height=55mm]{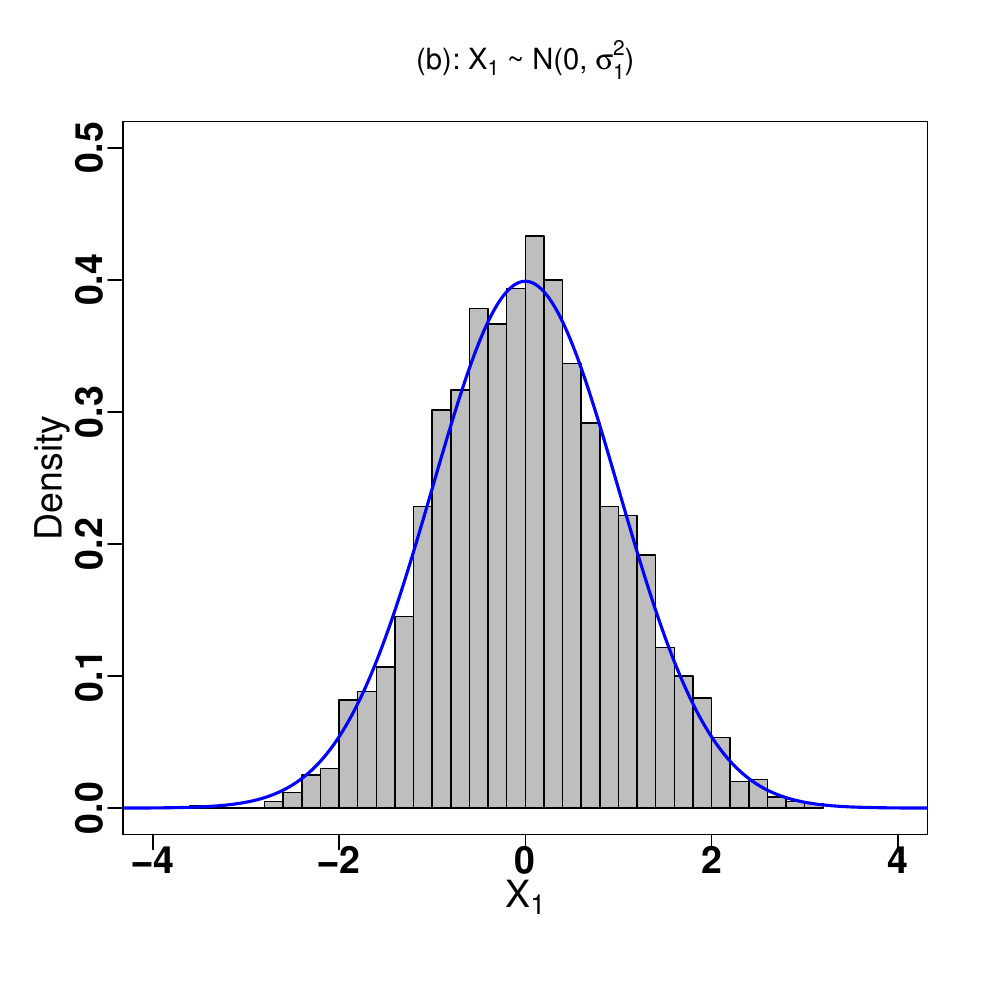}
\includegraphics[width=55mm,height=55mm]{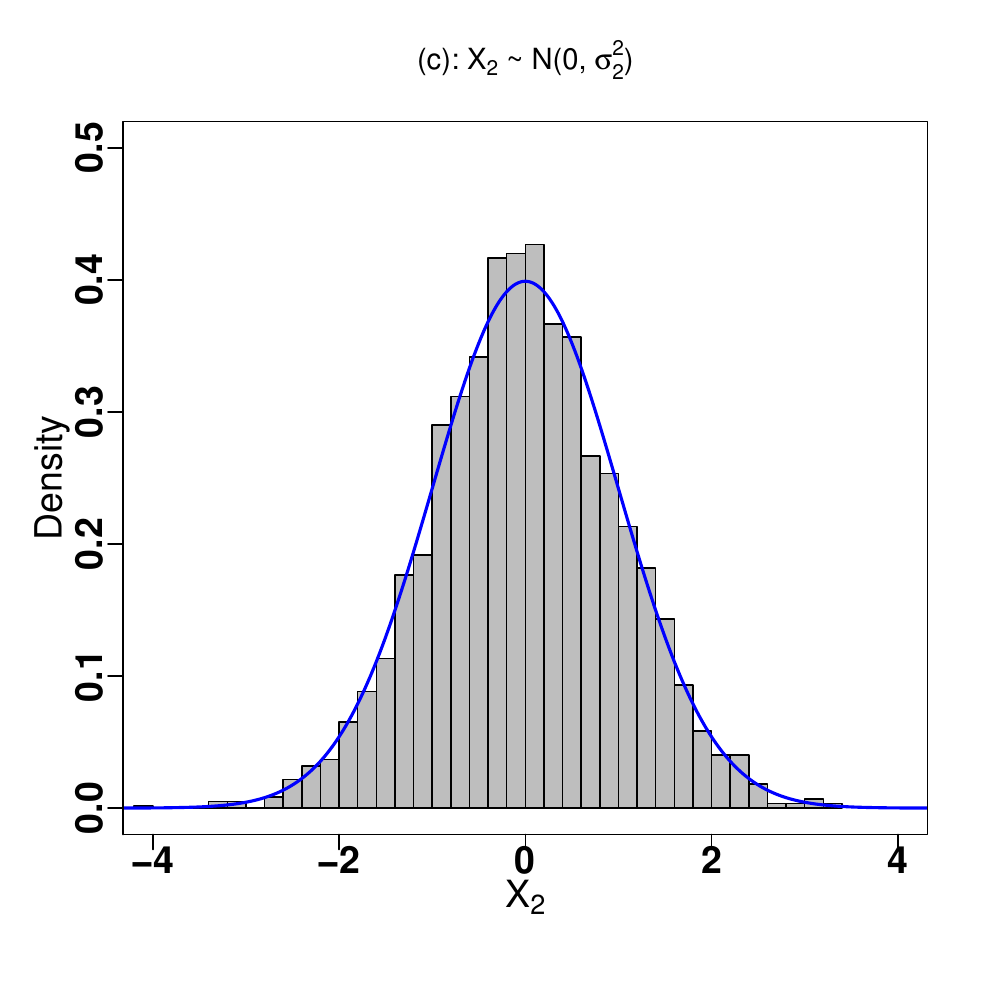}
\caption{(a): Scatterplot with fitted contours based on 3000 samples generated from ${\cal{N}}_{2}(\mu_{1}=0, \mu_{2}=0, \sigma_{1}^{2}=1, \sigma_{2}^{2}=1, \rho=0.50)$, (b): histogram of generated samples from marginal $X_{1}$, and (c): histogram of generated samples from marginal $X_{2}$.}
\label{fig3}
\end{figure}
%%%%%%%%%%%%%%%%%%%%%%%%%%%%%%%%%%%%%%%%%%%%%%%%%
\section{Gibbs sampling in Bayesian paradigm}
Herein, we show how a Gibbs sampling technique described in the previous section can be used for computing the Bayesian estimator of the unknown parameter $\boldsymbol{\theta}=(\theta_1,\cdots,\theta_p)^{\top}$. Notice that, based on random sample $\{x_1,\cdots,x_n\}$ following a distribution with PDF $f(\cdot\vert \boldsymbol{\theta})$, the Bayes estimator of $\boldsymbol{\theta}$ usually becomes the mean\footnote{Statistically speaking, the form of Bayesian estimator depends on the type of the loss function. For example, if the loss function takes either {\it{square}} or {\it{absolute}} form, then the Bayesian estimator is {\it{mean}} or {\it{median}} of posterior, respectively. But, the most commonly used summary representing the central tendency such as mean, median, or mode of the posterior would be considered as the Bayesian estimator.} of the posterior defined as
\begin{align}\label{bayesestim}
E\bigl[\boldsymbol{\theta}\big \vert \boldsymbol{x}\bigr]=\int_{\mathbb{R}^{p}}\boldsymbol{\theta}\pi(\boldsymbol{\theta}\vert \boldsymbol{x})d \boldsymbol{\theta}.
\end{align}
Computing integration (\ref{bayesestim}) may involve huge computational burden especially when number of parameters $p$, is large. The law of large numbers guarantees that the sample mean converges with probability one to the population mean when sample size is sufficiently large, that is
\begin{align*}%\label{bayesestim1}
\hat{\boldsymbol{\theta}}=\underset{N \rightarrow \infty}{\operatorname{limit}}~\frac{\boldsymbol{\theta}^{(1)}+\cdots+\boldsymbol{\theta}^{(N)}}{N}=E\bigl[\boldsymbol{\theta}\big \vert \boldsymbol{x}\bigr].
\end{align*}
Hence, if we could generate sample enough from the posterior $\pi(\boldsymbol{\theta}\vert \boldsymbol{x})$, then $\boldsymbol{\theta}$ can be estimated via $\hat{\boldsymbol{\theta}}$ in which $\{\boldsymbol{\theta}^{(1)},\cdots,\boldsymbol{\theta}^{(N)}\}$ is generated from posterior  $\pi(\boldsymbol{\theta}\vert \boldsymbol{x})$ when $N$ is sufficiently large (here $\boldsymbol{\theta}^{(i)}$ denotes the $i$th sample generated from posterior $\pi(\boldsymbol{\theta}\vert \boldsymbol{x})$). Generating sample from the target distribution (that is here $\pi(\boldsymbol{\theta}\vert \boldsymbol{x})$) using the MCMC techniques described in the previous section appears to be challenging when $p$ is large. In such cases, the Gibbs sampling is guaranteed to work well. 

\begin{figure}[h]
\center
\includegraphics[width=55mm,height=55mm]{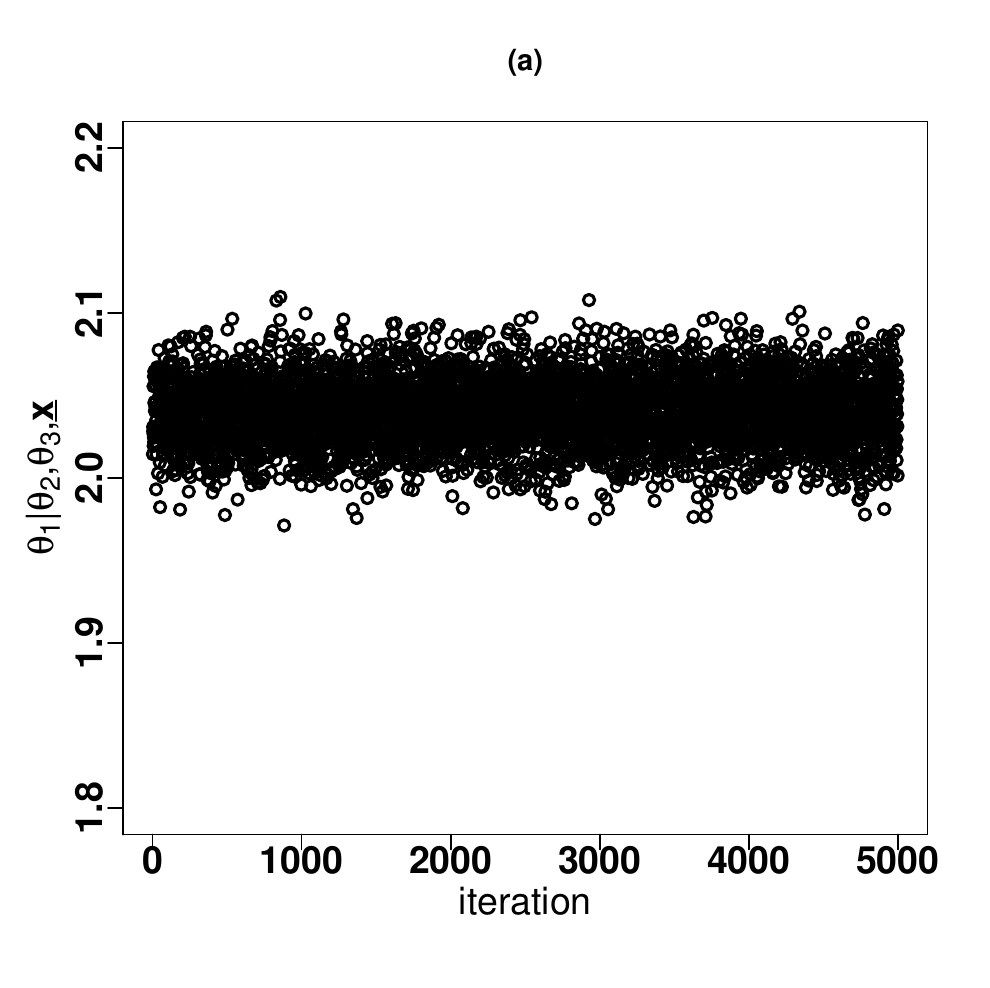}
\includegraphics[width=55mm,height=55mm]{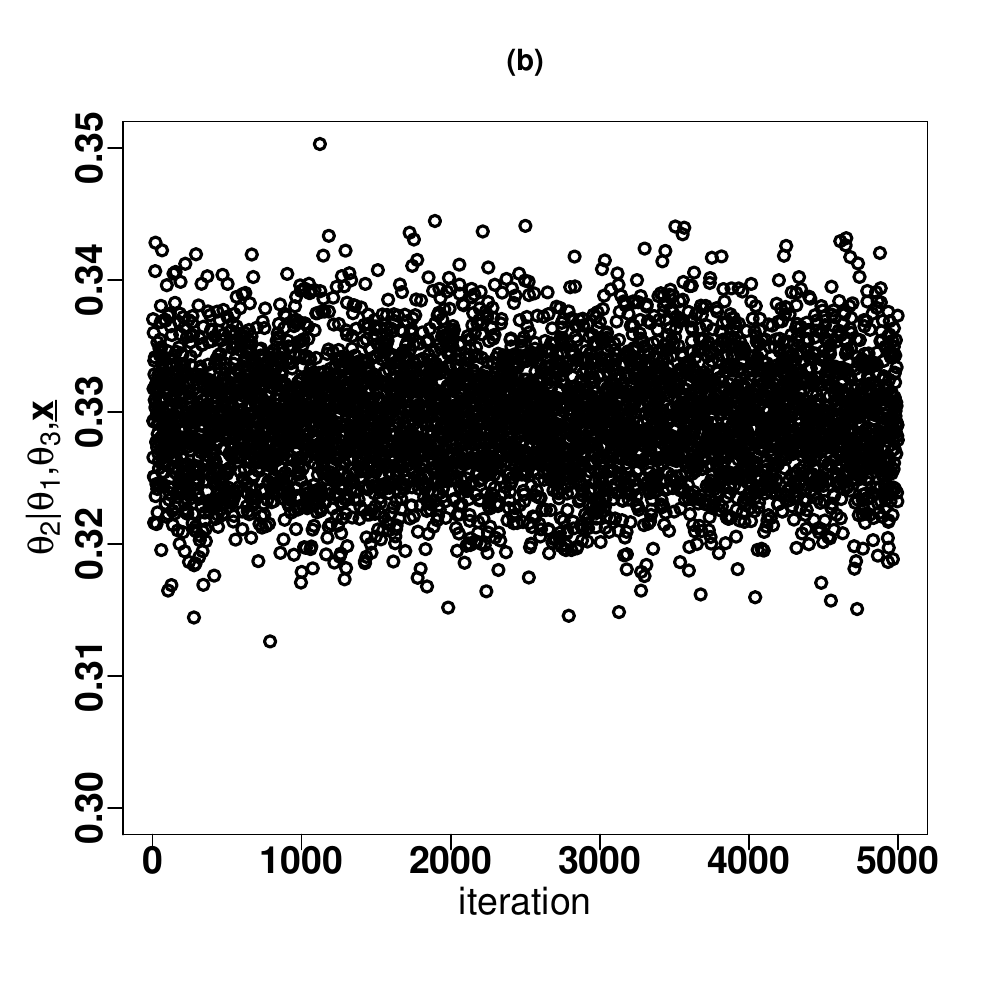}
\includegraphics[width=55mm,height=55mm]{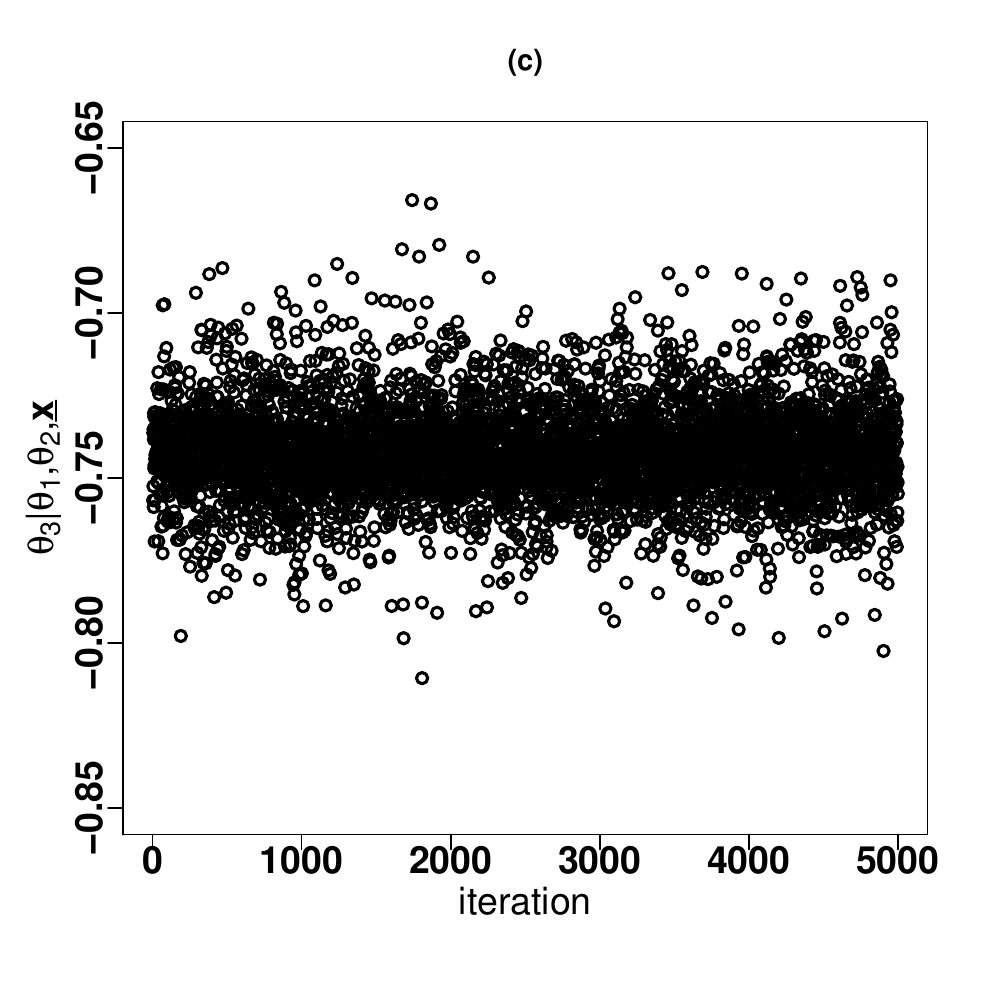}
\caption{Scatterplot of 5000 samples generated from full conditionals: (a) 
${\theta}_{1} \big \vert (\theta_{2},\theta_{3},{{\boldsymbol{x}}})$, (b): 
${\theta}_{2} \big \vert (\theta_{1},\theta_{3},{{\boldsymbol{x}}})$, and (c): 
${\theta}_{3} \big \vert (\theta_{1},\theta_{2},{{\boldsymbol{x}}})$.
}
\label{fig5}
\end{figure}

\begin{figure}[h]
\center
\includegraphics[width=55mm,height=55mm]{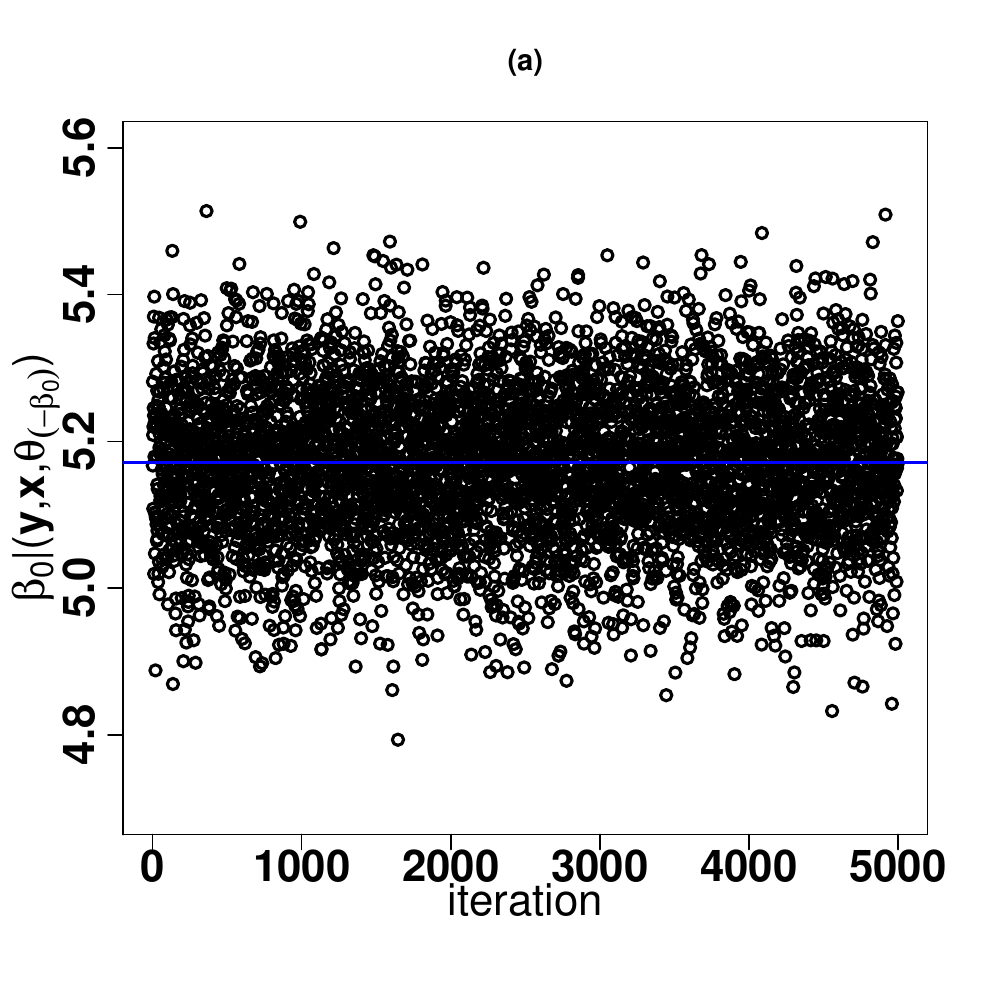}
\includegraphics[width=55mm,height=55mm]{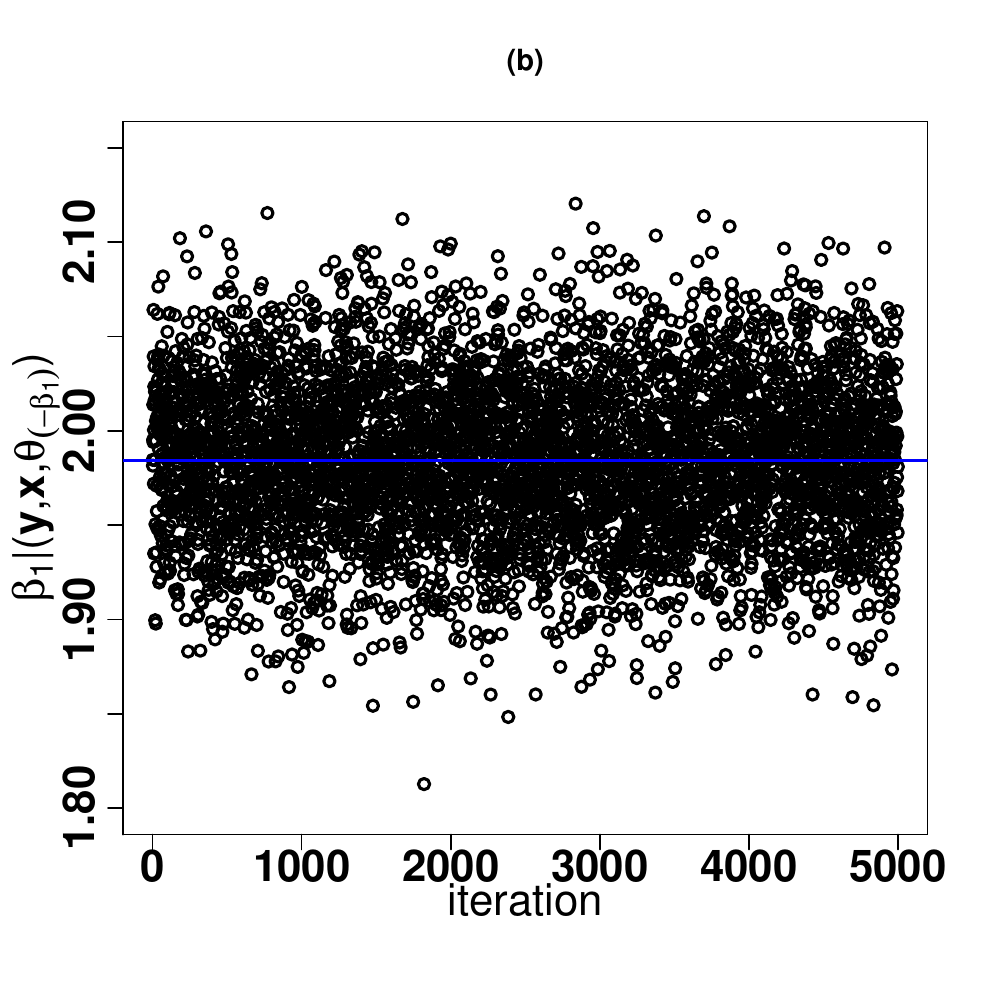}
\includegraphics[width=55mm,height=55mm]{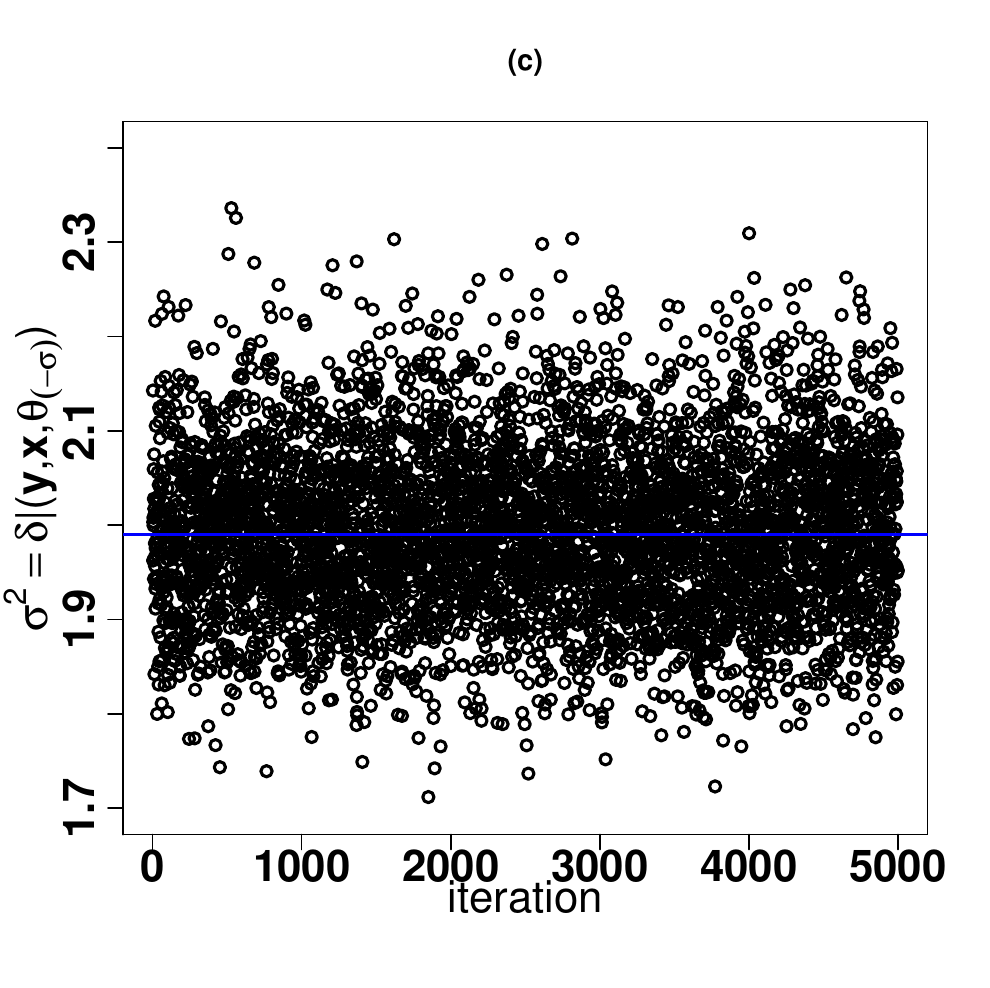}
\caption{
Scatterplot of 5000 samples generated from full conditionals: (a) $\beta_{0} \big \vert \bigl(\boldsymbol{y},\boldsymbol{x},\boldsymbol{\theta}_{(-\beta_{0})} \bigr)$
, (b): $\beta_{1} \big \vert \bigl(\boldsymbol{y},\boldsymbol{x},\boldsymbol{\theta}_{(-\beta_{1})} \bigr)$, and (c): $\sigma \big \vert \bigl(\boldsymbol{y},\boldsymbol{x},\boldsymbol{\theta}_{(-\sigma)} \bigr)$. The blue line, in each subfigure, shows the Bayesian estimator.}
\label{simplereg}
\end{figure}

\begin{figure}[h]
\center
\includegraphics[width=40mm,height=50mm]{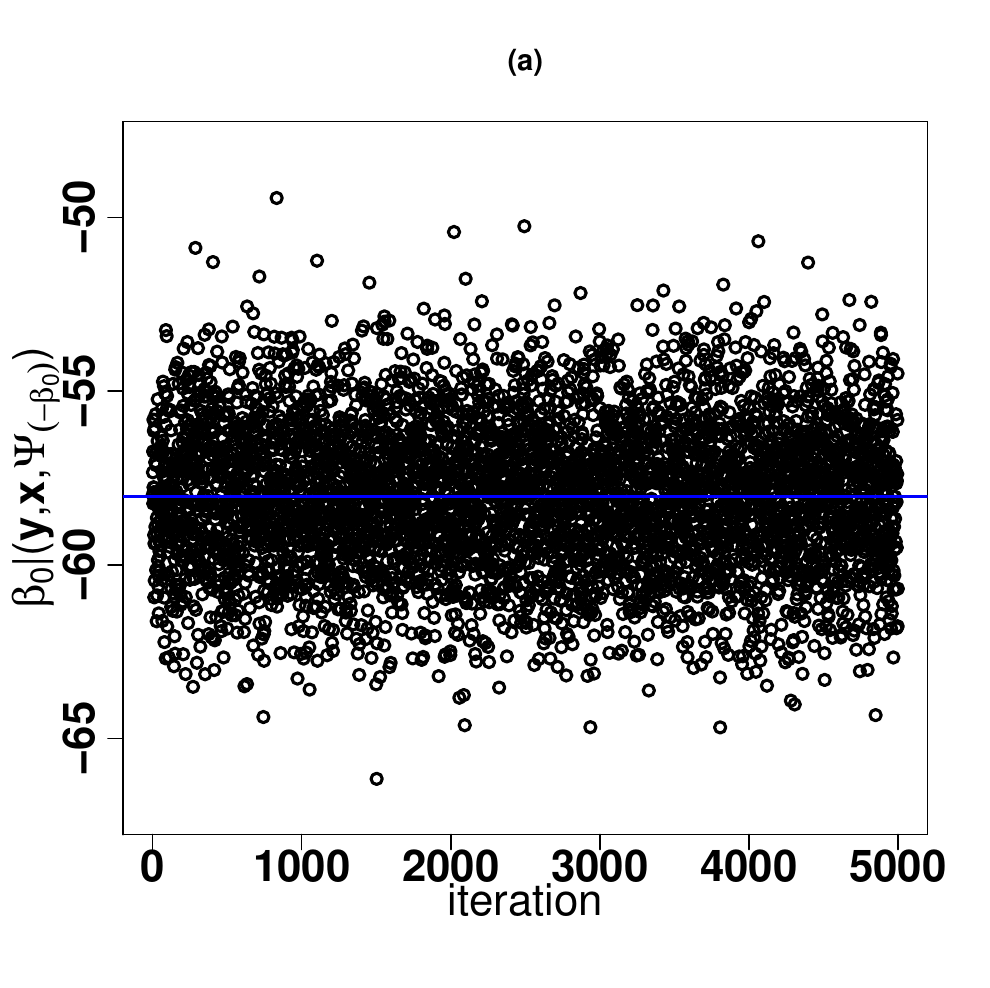}
\includegraphics[width=40mm,height=50mm]{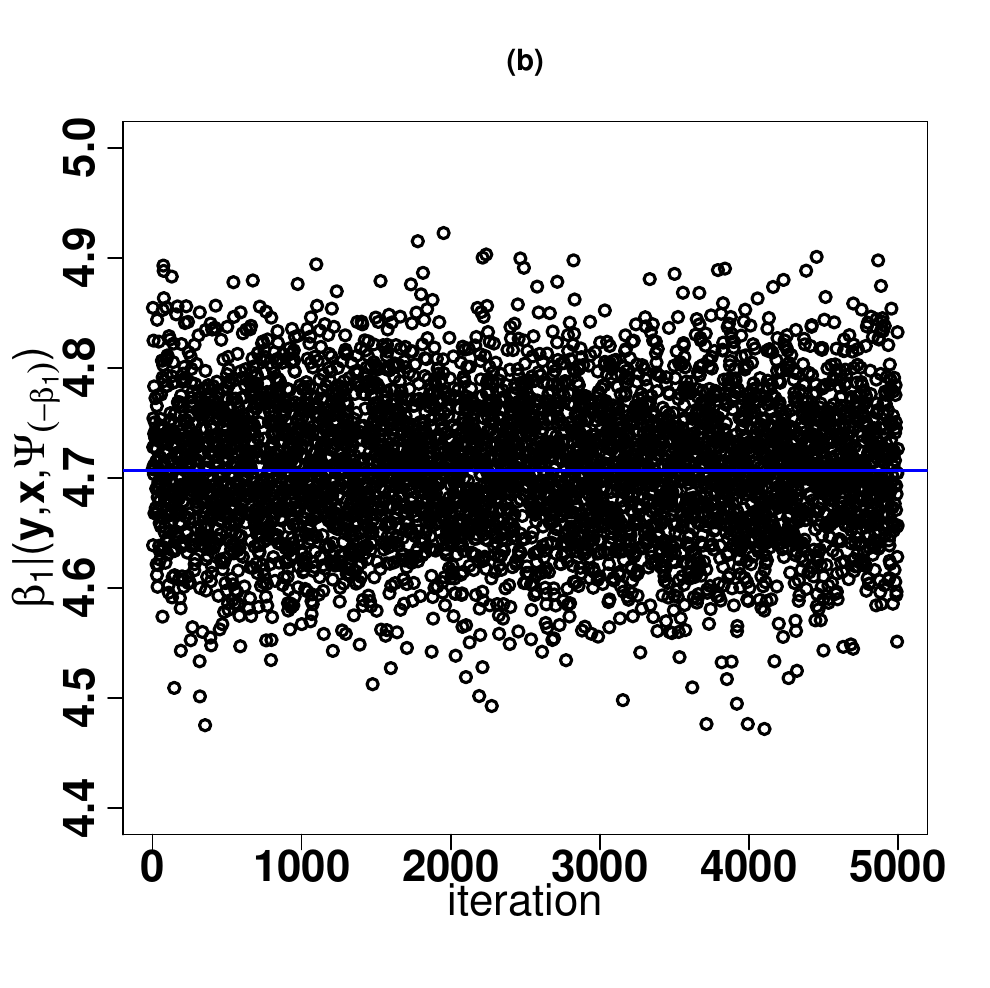}\\
\includegraphics[width=40mm,height=50mm]{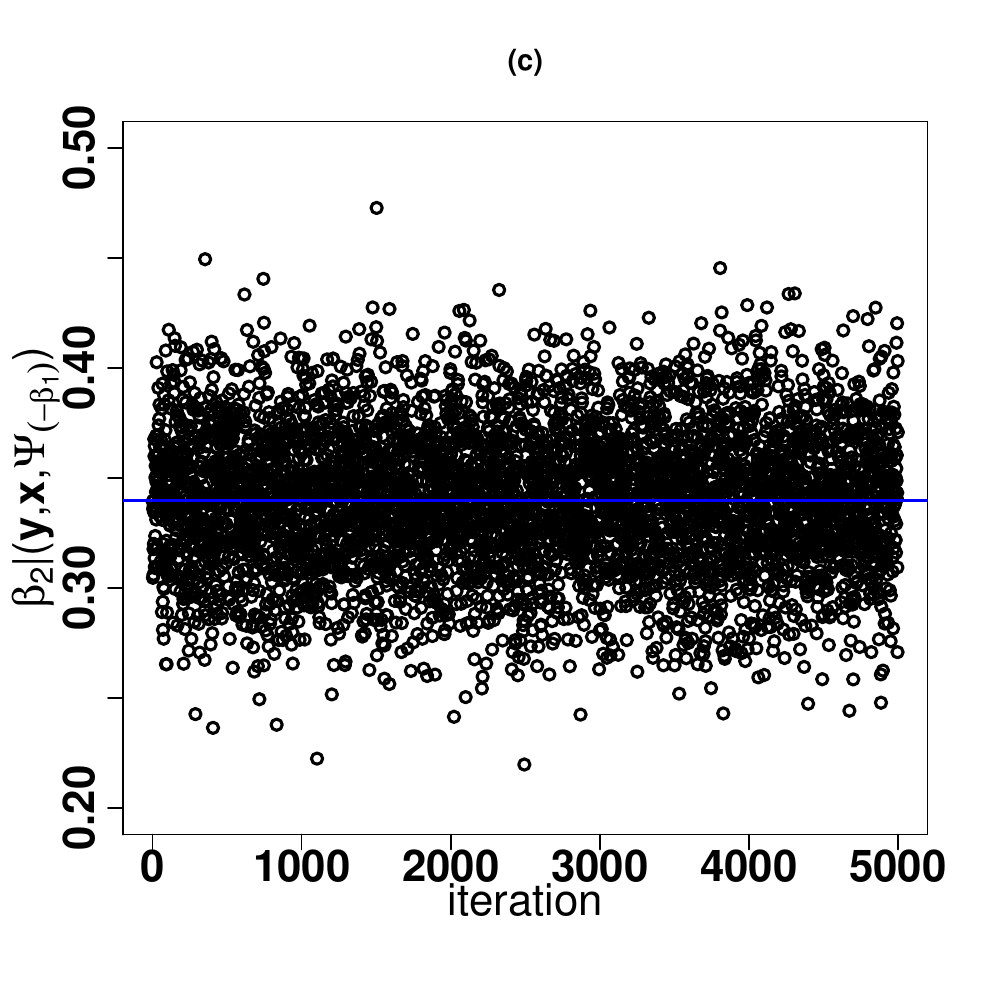}
\includegraphics[width=40mm,height=50mm]{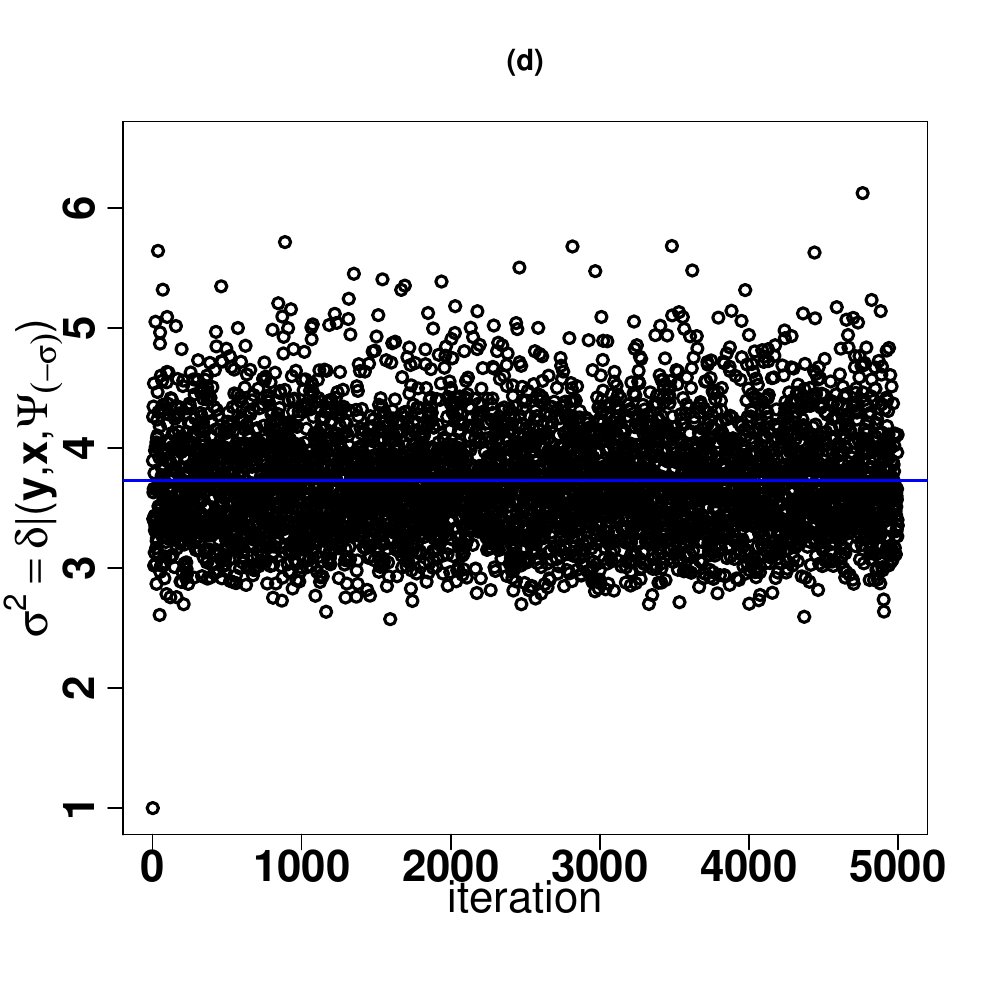}
\caption{
Scatterplot of 5000 samples generated from full conditionals: (a) $\beta_{0} \big \vert \bigl(\boldsymbol{y},\boldsymbol{x},\boldsymbol{\Psi}_{(-\beta_{0})} \bigr)$
, (b): $\beta_{1} \big \vert \bigl(\boldsymbol{y},\boldsymbol{x},\boldsymbol{\Psi}_{(-\beta_{1})} \bigr)$, (c): $\beta_{2} \big \vert \bigl(\boldsymbol{y},\boldsymbol{x},\boldsymbol{\Psi}_{(-\beta_{2})} \bigr)$, and (d): $\sigma^2 =\delta\big \vert \bigl(\boldsymbol{y},\boldsymbol{x},\boldsymbol{\Psi}_{(-\delta)} \bigr)$. The blue line, in each subfigure, shows the Bayesian estimator.}
\label{multiplereg}
\end{figure}

It should be noted that since in each Bayesian framework the unknown parameter $\boldsymbol{\theta}=(\theta_1,\cdots,\theta_p)^{\top}$ itself is a random variable, in addition of the model's variables, herein we must consider an extra number of $p$ additional full conditionals within a Gibbs sampler schemes. For instance, recalling from the Example (3), we must consider three full conditionals given by variables $X_1\big \vert X_2,\rho$, $X_2\big \vert X_1,\rho$, and $\rho \big \vert X_1, X_2$. In what follows, we consider the foregoing example when Bayesian inference on $\rho$ is desired.  
\begin{example}\label{exam-bivariategaussian-Bayesian}%\lipsum*[]
Let ${\underline{\boldsymbol{x}}}=\{\boldsymbol{x}_1,\boldsymbol{x}_2,\cdots, \boldsymbol{x}_n$\} account for a sequence of $n$ independent observations in which $\boldsymbol{x}_i=\bigl(x_{i1},x_{i2}\bigr)^{\top}$. Furthermore, assume $\boldsymbol{x}_i$ (for $i=1,2,\cdots,n$) follows a bivariate Gaussian distribution with PDF given by \begin{align}\label{bivariatenormmal1}
\pi\bigl(\boldsymbol{x}_{i}\big \vert \boldsymbol{\theta}\bigr)=\frac{1}{2\pi \sigma_1 \sigma_2 \sqrt{1-\rho^2}}\exp\Bigl\{-\frac{1}{2(1-\rho^2)}\Bigl[\Bigl(\frac{x_{i1}}{\sigma_1}\Bigr)^{2}+\Bigl(\frac{x_{i2}}{\sigma_2}\Bigr)^{2}-2\rho \frac{x_{i1}x_{i2}}{\sigma_1\sigma_2}\Bigr]\Bigr\},
\end{align}
where $\boldsymbol{\theta}=(\sigma_1,\sigma_2,\rho)^{\top}$ is the unknown parameter vector. Using a reparameterization $\boldsymbol{\theta}=(\theta_1,\theta_2,\theta_3)^{\top}$ in which $\theta_1=\sigma_1/\sigma_2$, $\theta_2=\sigma_1 \sigma_2 \sqrt{1-\rho^2}$, and $\theta_3=\rho$, the bivariate Gaussian PDF is rewritten as
 by \begin{align}\label{bivariatenormmal2}
\pi\bigl(\boldsymbol{x}_{i}\big \vert \boldsymbol{\theta}\bigr)=\frac{1}{2\pi \theta_2}\exp\Bigl\{-\frac{1}{2(1-\theta_{3}^2)^{1/2}\theta_2}\Bigl[\frac{x^{2}_{i1}}{\theta_1}+\theta_1 x^{2}_{i2}-2\theta_3 x_{i1}x_{i2}\Bigr]\Bigr\}.
\end{align}
For computing the Bayesian estimator of $\boldsymbol{\theta}$, we first note that
\begin{align}\label{bayesestim2}
\pi\bigl(\rho \big \vert {\underline{\boldsymbol{x}}} \bigr) &\propto \prod_{i=1}^{n}\pi\bigl(\boldsymbol{x}_{i}\big \vert \boldsymbol{\theta})\pi(\boldsymbol{\theta}\bigr).
%&\propto\frac{1}{\bigl(1-\rho^2\bigr)^{\frac{n}{2}}}\exp\Bigl\{ -\frac{\sum_{i=1}^{n}x_{i1}^{2}+\sum_{i=1}^{n}x_{i2}^2-2\rho \sum_{i=1}^{n}x_{i1}x_{i2}}{2(1-\rho^2)}\Bigr\}.
%%=\frac{1}{2\pi \sqrt{1-\rho^2}}\exp\Bigl\{ -\frac{x_{1}^{2}+x_{2}^2-2\rho x_1x_2}{2(1-\rho^2)}\Bigr\}.
\end{align}
The marginals of joint prior $\pi(\boldsymbol{\theta})$ are assumed to be independent, that is, $\pi(\boldsymbol{\theta})=\pi(\theta_1)\pi(\theta_2)\pi(\theta_3)$. Also, we consider gamma prior for both of $\theta_1$, inverse gamma prior for $\theta_2$, and an uniform on (-1,1) prior for $\theta$ represented, respectively, as follows.
\begin{align}\label{bayesestim3}
\pi(\theta_1\big \vert a_1, b_1)&=\frac{b_{1}^{a_{1}}\theta_{1}^{a_{1}-1}}{\Gamma(a_{1})}\exp\bigl\{-b_{1}\theta_1\bigr\},~~\theta_1>0,\nonumber\\
\pi(\theta_2\big \vert a_2, b_2)&=\frac{b_{2}^{a_{2}}\theta_{2}^{-a_{2}-1}}{\Gamma(a_{2})}\exp\bigl\{-\frac{b_{2}}{\theta_2}\bigr\},~~\theta_2>0,\nonumber\\
\pi(\theta_3)&=\frac{1}{2},~~-1<\theta_3<1.
\end{align}
We note that, here we consider above priors just for the simplicity of computations. It follows that
\begin{align}\label{bayesestim2}
\pi\bigl(\boldsymbol{\theta} \big \vert {\underline{\boldsymbol{x}}}\bigr) \propto &\frac{1}{\theta_{2}^{n}}
\exp\Bigl\{ -\frac{1}{2\theta_2 \sqrt{1-\theta_{3}^{2}}}\Bigl[\frac{\sum_{i=1}^{n}x_{i1}^{2}}{\theta_1}+\theta_1\sum_{i=1}^{n}x_{i2}^{2}-2\theta_{3} \sum_{i=1}^{n}x_{i1}x_{i2}{
}\Bigr]\Bigr\}\nonumber\\
&\times \frac{b_{1}^{a_{1}}\theta_{1}^{a_{1}-1}}{\Gamma(a_{1})}\exp\bigl\{-b_{1}\theta_1\bigr\}\times\frac{b_{2}^{a_{2}}\theta_{2}^{-a_{2}-1}}{\Gamma(a_{2})}\exp\bigl\{-\frac{b_{2}}{\theta_2}\bigr\}\times \frac{1}{2}
%%=\frac{1}{2\pi \sqrt{1-\rho^2}}\exp\Bigl\{ -\frac{x_{1}^{2}+x_{2}^2-2\rho x_1x_2}{2(1-\rho^2)}\Bigr\}.
\end{align}
We note that, herein, the model's variables are $\theta_{1}$, $\theta_{2}$, and 
$\theta_{3}$. So, by dropping out all terms that do not depend on $\theta_1$ from the posterior (\ref{bayesestim2}), then the full conditional ${\theta}_{1} \big \vert (\theta_2,\theta_3,{\underline{\boldsymbol{x}}})$, after some algebra, is shown to be\footnote{It is noteworthy that ${\underline{\boldsymbol{x}}}$ has been observed and we no longer look at it as variable. So, the model's variables are $\theta_{1}$, $\theta_{2}$, and 
$\theta_{3}$.}
\begin{align}\label{bayesestim4}
\pi\bigl({\theta}_{1} \big \vert \theta_2,\theta_3,{\underline{\boldsymbol{x}}}\bigr) \propto &\theta_{1}^{-a_{1}-1}
\exp\Bigl\{ -\frac{\theta_1}{2}\Bigl[2b_{1}+\frac{\sum_{i=1}^{n}x_{i2}^{2}}{ \theta_{2}(1-\theta_{3})^{1/2}}\Bigr]-\frac{1}{2\theta_1}\Bigl[ \frac{\sum_{i=1}^{n}x_{i1}^{2}}{\theta_{2}(1-\theta_{3})^{1/2}}\Bigr]\Bigr\}.
%=\frac{1}{2\pi \sqrt{1-\rho^2}}\exp\Bigl\{ -\frac{x_{1}^{2}+x_{2}^2-2\rho x_1x_2}{2(1-\rho^2)}\Bigr\}.
\end{align}
Comparing the RHSs of (\ref{bayesestim4}) and (\ref{pdf-gig}), it turns out that (\ref{bayesestim4}) is the pseudo PDF corresponds to a ${\cal{GIG}}(a,b,c)$ distribution where $a=a_1$, $b=2b_{1}+\sum_{i=1}^{n}x_{i2}^{2}/\bigl[\theta_{2}(1-\theta_{3})^{1/2}\bigr]$, $c=\sum_{i=1}^{n}x_{i1}^{2}/\bigl[\theta_{2}(1-\theta_{3})^{1/2}\bigr]$. So,
\begin{align}
{\theta}_{1} \big \vert (\theta_2,\theta_3,{\underline{\boldsymbol{x}}}) \sim {\cal{GIG}}\Bigl(a_1, 2b_{1}+\sum_{i=1}^{n}x_{i2}^{2}/\bigl[\theta_{2}(1-\theta_{3})^{1/2}\bigr],  \sum_{i=1}^{n}x_{i1}^{2}/\bigl[\theta_{2}(1-\theta_{3})^{1/2}\bigr]\Bigr).
\end{align}  
Likewise, we can see that the full conditional ${\theta}_{2} \big \vert (\theta_1,\theta_3,{\underline{\boldsymbol{x}}})$ follows an inverse gamma distribution. We have
\begin{align}
{\theta}_{2} \big \vert (\theta_1,\theta_3,{\underline{\boldsymbol{x}}}) \sim &{\cal{IG}}\biggl(n+a_2, b_{2}+ \frac{1}{2\theta_{2}(1-\theta_{3})^{1/2}}\nonumber\\
&\times\Bigl[\theta^{-1}_{1}\sum_{i=1}^{n}x_{i1}^{2}+\theta_{1} \sum_{i=1}^{n}x_{i2}^{2}-2\theta_3\sum_{i=1}^{n}x_{i1}x_{i2}\Bigr] \biggr),
\end{align}
and finally 
\begin{align}
{\theta}_{3} \big \vert (\theta_1,\theta_2,{\underline{\boldsymbol{x}}}) \propto
\exp\biggl\{ -\frac{1}{2\theta_2 \sqrt{1-\theta_{3}^{2}}}\Bigl[\theta_{1}^{-1}\sum_{i=1}^{n}x_{i1}^{2}+\theta_1\sum_{i=1}^{n}x_{i2}^{2}-2\theta_{3} \sum_{i=1}^{n}x_{i1}x_{i2}\Bigr]\biggr\}.
\end{align}
As it is seen, the full conditional ${\theta}_{3} \big \vert (\theta_1,\theta_2,{\underline{\boldsymbol{x}}})$ has no closed form. Hence, we can use the {MH within Gibbs sampling} technique. Recall from Subsection \ref{MHsection}, considering a symmetric uniform proposal on (-1,1), the transition probability $p\bigl(\theta_{3(n)} \rightarrow \theta_{3(n+1)}\bigr)=\min\bigl\{1, \exp\{R_1-R_2\}\bigr\}$, is
\begin{align*}%\label{ratiotheta3}
R_1= -\frac{1}{2\theta_{2} \sqrt{1-\theta_{3(n+1)}^{2}}}\Bigl[\theta_{1}^{-1}\sum_{i=1}^{n}x_{i1}^{2}+
\theta_{1}
\sum_{i=1}^{n}x_{i2}^{2}-2\theta_{3(n+1)} \sum_{i=1}^{n}x_{i1}x_{i2}{
}\Bigr]
\end{align*}
\begin{align}
R_2=-\frac{1}{2\theta_2 \sqrt{1-\theta_{3(n)}^{2}}}\Bigl[\theta_{1}^{-1}\sum_{i=1}^{n}x_{i1}^{2}+\theta_1\sum_{i=1}^{n}x_{i2}^{2}-2\theta_{3(n)} \sum_{i=1}^{n}x_{i1}x_{i2}{
}\Bigr]
\end{align}
In what follows, we give the steps for computing Bayesian estimation of ${\boldsymbol{\theta}}$. 
\begin{enumerate}
\item Read $N, M, a_1, b_1, a_2, b_2$, and suggest $\boldsymbol{\theta}^{(0)}=\bigl(\theta^{(0)}_{1},\theta^{(0)}_{2},\theta^{(0)}_{3}\bigr)^{\top}$ arbitrarily;
\item Set $t=0$;
\item Simulate $\theta^{(t+1)}_{1}$ from the full conditional $\theta^{(t+1)}_{1} \big \vert \bigl(\theta^{(t)}_{2},\theta^{(t)}_{3}, {\underline{\boldsymbol{x}}}\bigr)$ that follows 
\begin{align*}
{\cal{GIG}}\Bigl(a_1, 2b_{1}+\sum_{i=1}^{n}x_{2i}^{2}/\bigl[\theta^{(t)}_{2}(1-\theta^{(t)}_{3})^{1/2}\bigr],  \sum_{i=1}^{n}x_{1i}^{2}/\bigl[\theta^{(t)}_{2}(1-\theta^{(t)}_{3})^{1/2}\bigr]\Bigr);
\end{align*}  
\item Having $\theta^{(t+1)}_{1}$, simulate $\theta^{(t+1)}_{2}$ from the full conditional $\theta^{(t)}_{2} \big \vert \bigl(\theta^{(t+1)}_{1},\theta^{(t)}_{3},{\underline{\boldsymbol{x}}}\bigr)$ that follows 
\begin{align*}
&{\cal{GIG}}\biggl(n+a_2, 2b_{2}, \Bigl[\frac{1}{\theta^{(t+1)}_{1}}\sum_{i=1}^{n}x_{i1}^{2}+\theta^{(t+1)}_{1} \sum_{i=1}^{n}x_{i2}^{2}-2\theta^{(t)}_3\sum_{i=1}^{n}x_{i1}x_{i2}\Bigr] \nonumber\\
&\Bigl[\theta^{(t)}_{2}(1-\theta^{(t)}_{3})^{1/2}\Bigr]^{-1}\biggr);
\end{align*}
\item Having $\theta^{(t+1)}_{1}$ and $\theta^{(t+1)}_{2}$, simulate $\theta^{(t+1)}_{3}$ from the full conditional $\theta^{(t)}_{3} \big \vert \bigl(\theta^{(t+1)}_{2},\theta^{(t)}_{3},{\underline{\boldsymbol{x}}}\bigr)$ using MH-within-Gibbs sampling technique as follows.  
\begin{enumerate}
\item Set $k=0$, $K=100$, and choose the initial state as
$y_{0}\sim {\cal{U}}(-1,1)$;
\item Generate $y^{(k+1)}\sim {\cal{U}}(-1,1)$
\item Compute $p\bigl(y^{(k)} \rightarrow y^{(k+1)}\bigr)=\min\bigl\{1,\exp\{ R_{1}- R_{2}\}\bigr\}$
where
\begin{equation*}
R_1=  -\frac{1}{2\theta^{(t+1)}_{2} \sqrt{1-\bigl(y^{(k+1)}\bigr)^{2}}}
\Bigl[\frac{1}{\theta^{(t+1)}_{1}}\sum_{i=1}^{n}x_{i1}^{2}+
\theta_{1}\sum_{i=1}^{n}x_{i1}^{2}-2y^{(k+1)} \sum_{i=1}^{n}x_{i1}x_{i2}\Bigr]
\end{equation*}
and
\begin{equation*}
R_2=-\frac{1}{2\theta^{(t+1)}_2 \sqrt{1-\bigl(y^{(k)}\bigr)^{2}}}
\Bigl[\frac{1}{\theta^{(t+1)}_{1}}\sum_{i=1}^{n}x_{i1}^{2}+\theta^{(t+1)}_1 \sum_{i=1}^{n}x_{i2}^{2}-2y^{(k)} \sum_{i=1}^{n}x_{i1}x_{i2}\Bigr]
\end{equation*}
\item Generate $u\sim {\text{U}}(0, 1)$;
\item If $u<p\bigl(y^{(k)}, y^{(k+1)}\bigr)$ then $y^{(k)} \rightarrow y^{(k+1)}$ and $k \rightarrow k+1$; 
\item Repeat algorithm from step (b);
\item If $k=K$ then $ y^{(k+1)} \rightarrow\theta^{(t+1)}_{3}$ and stop the algorithm. 
\end{enumerate}
\item If $t=N$, then go to the next step, otherwise $t+1 \rightarrow t$ and return step 3;
\item The Bayesian estimator $\hat{\boldsymbol{\theta}}_{B}$, is the average of $N-M+1$ samples, that is, $\boldsymbol{\theta}^{(M+1)},\boldsymbol{\theta}^{(M+2)},\cdots,\boldsymbol{\theta}^{(M+N)}$ generated from target PDF $\pi\bigl(\boldsymbol{\theta}\vert {\underline{\boldsymbol{x}}}\bigr)$.
 \end{enumerate}
We note that, in practice, hyperparameters $a_1,b_1,a_2$ and $b_2$ are determined based on experimenter's belief or the empirical Bayes. If sample size is small, then the improper Jeffreys {\it{prior}} is suggested. It is noteworthy that for simulating from GIG distribution, one can use the method proposed by \cite{hormann2014generating}. 
\end{example}
The pertaining \verb+R+ code for implementing example above is given as follows. To this end, a sample of $N=5000$ realizations have been generated from a zero-mean bivariate Gaussian when $\sigma_1=0.5$, $\sigma_2=1$, and $\rho=-0.75$. The output of the Gibbs sampler is displayed in Figure \ref{fig5}.
\begin{lstlisting}[style=deltaj]
R > set.seed(20240522)
R > library(MASS)
R > library(GIGrvg)
R > n <- 5000  # sample size
R > Mu <- c(0, 0)
R > X <- matrix(0, nrow = n, ncol = 2)
R > sigma1 <- 1; sigma2 <- .5;  rho <- -0.75
R > Sigma <- matrix( c(sigma1^2,rho*sigma1*sigma2,rho*sigma1*sigma2,sigma2^2), 2, 2)
R > X <- mvrnorm(n, Mu, Sigma)
R > n.sim <- 5000		    # number of Gibbs sampling iterations
R > n.burn <- 2000  # length of burn-in period
R > K <- 100							# number of MH algorithm iterations
R > theta3i <- rep(0, K)
R > a1 <- 0.5; b1<- 0.5; a2 <- 0.5; b2 <- 0.5;
R > theta <- matrix(0, nrow = n.sim, ncol = 3)
R > theta[1, ] <- c(1, 1, 0.75)
R >  sx1 <- sum(X[, 1]^2);  sx2 <- sum(X[, 2]^2); sx12 <- sum(X[, 1]*X[, 2])
R > j <- 1
R > 	while(j < n.sim)
+			{
+			c0   <- ( theta[j, 2]*sqrt( 1 - theta[j, 3]^2 ) )
+			rate1 <- function(x) (sx1/theta[j, 1] + theta[j, 1]*sx2 - 2*x*sx12)/sqrt(1-x^2)
+			theta[j + 1, 1] <- rgig(n = 1, lambda = a1, chi = sx1/c0, psi = 2*b1 + sx2/c0  )
+			theta[j + 1, 2] <- 1/rgamma(n=1, shape=n + a2, rate=rate1(theta[j, 3])/2 + b2 )
#		 	start of MH-within-Gibbs sampling for sampling
#		 			           from full conditional of theta3
+			theta3i[1] <- runif(1, -1, 1)
+				for(k in 2:K)
+				{
+					theta3.n <- runif(1, -1, 1)
+					dif <- -( rate1(theta3.n)  -  rate1(theta3i[k - 1]) )/(2*theta[j + 1, 2])
+						if( runif(1) < exp( dif ) )
+						{
+							theta3i[k] <- theta3.n           # theta3.n is accepted
+						}else{
+							theta3i[k] <- theta3i[ k - 1] # theta3.n is rejected
+						}
+				}
#		 	start of MH-within-Gibbs sampling for sampling
#		 			           from full conditional of theta3
+ 		theta[j + 1, 3] <- theta3i[k]
+ 		j <- j + 1
+			}
R > theta.hat <- apply(theta[(n.sim - n.burn + 1):n.sim, ], 2, mean)
R > rho <- theta.hat[3]  #estimator of rho
R > sigma1 <- sqrt(theta.hat[1]*theta.hat[2]/sqrt(1-rho.hat^2)) #estimator of sigma1
R > sigma2 <- sigma1.hat/theta.hat[1] #estimator of sigma2
\end{lstlisting}
\begin{example}\label{exam-simplereg}%\lipsum*[]
Suppose response variable $Y$ depends linearly on independent variable (covariate) $X$ through a simple linear model as 
\begin{align}\label{regressionmodel2}
y_{i}=&\beta_0+\beta_{1}x_{i}+\epsilon_{i};~~ i=1,2,3,\cdots,n.
\end{align}
Recording $n$ measurements on $Y$ represented as $y_1,y_2,y_3,\cdots,y_n$ at the corresponding levels of covariate $X$ shown by $x_{1},x_{2},\cdots,x_{n}$, the focus is placed on estimating the regression coefficients $\beta_0$ and $\beta_{1}$ through the Bayesian approach. Herein, $\epsilon_{i}$s are the simple linear model's theoretical errors, independently (that is, ${\text{cov}}\bigl(\epsilon_{i},\epsilon_{j}\bigr)=0$ for $i\neq j =1,2,3,\cdots,n$) assumed to follow ${\cal{N}}\bigl(0,\sigma^{2}\bigr)$. Of course the skewed Student's $t$ \citep{teimouri2020forestfit} and $\alpha$-stable \citep{nolan2013linear} may be reasonable candidates for distribution of theoretical error in presence of outliers that yield more robust estimation of ${\boldsymbol{\beta}}=(\beta_{0},\beta_{1})^{\top}$ than the Gaussian one. It is known that the least-square (LS) estimator $\hat{\boldsymbol{\beta}}_{ls}$ and ML estimator $\hat{\boldsymbol{\beta}}_{ml}$, of the regression coefficients $\boldsymbol{\beta}$ are the same and given by 
\begin{align}\label{betaml}
\hat{{\beta}_{1}}_{ML}=\hat{{\beta}_{1}}_{LS}&={\frac {\sum _{i=1}^{n}(x_{i}-{\bar {x}})(y_{i}-{\bar {y}})}{\sum _{i=1}^{n}(x_{i}-{\bar {x}})^{2}}},\nonumber\\
\hat{{\beta}_{0}}_{ML}=\hat{{\beta}_{0}}_{LS}&={\bar {y}}-\hat{{\beta}_{0}}_{ML} \times\bar {x},
\end{align}
 where $\bar{x}=\bigl(x_{1}+x_{2}+\cdots+x_{n}\bigr)/n$. Assuming independent Gaussian priors for elements of $\boldsymbol{\beta}$ and a mutually independent conjugate inverse gamma prior for $\delta=\sigma^2$ as
\begin{align}
\pi(\beta_{0})=&{\cal{N}}\bigl(\beta_{0}\big \vert \mu_{0},\sigma^{2}_{0}\bigr),\nonumber\\
\pi(\beta_{1})=&{\cal{N}}\bigl(\beta_{1}\big \vert \mu_{1},\sigma^{2}_{1}\bigr),\nonumber\\
\pi(\delta)=&{\cal{IG}}\Bigl(\delta\Big \vert \frac{a_{0}}{2},\frac{b_{0}}{2}\Bigr).\nonumber
\end{align}
It follows that
\begin{align}\label{bayes}
\pi(\boldsymbol{\theta} \vert\boldsymbol{y},\boldsymbol{x})=& L\bigl(\boldsymbol{\theta} \big \vert \boldsymbol{y},\boldsymbol{x}\bigr) \pi(\beta_{0}) \pi(\beta_{1})\pi(\delta)\nonumber\\
\propto& \frac{1}{\delta^{\frac{n}{2}}}\exp\Bigl\{-\frac{1}{2\delta}\sum_{i=1}^{n}\bigl(y_{i}-\beta_0-\beta_{1}x_{i}\bigr)^2\Bigr\}\frac{1}{\sqrt{2\pi \sigma^{2}_{0}}}\exp\Bigl\{-\frac{\bigl(\beta_{0}-\mu_{0}\bigr)^2}{2\sigma^{2}_{0}}\Bigr\}\nonumber\\
&\times\frac{1}{\sqrt{2\pi \sigma^{2}_{1}}}\exp\Bigl\{-\frac{\bigl(\beta_{1}-\mu_{1}\bigr)^2}{2\sigma^{2}_{1}}\Bigr\}\Bigl(\frac{b_{0}}{2}\Bigr)^{\frac{a_{0}}{2}}
\frac{\delta^{-\frac{a_{0}}{2}-1}}{\Gamma\bigl(\frac{a_{0}}{2}\bigr)}\exp\Bigl\{-\frac{b_{0}}{2\delta}\Bigr\},
\end{align}
where $\delta=\sigma^2$ and $\boldsymbol{\theta}=({\beta}_{0},{\beta}_{1},\delta)^{\top}$. Then more algebra shows
\begin{align}\label{coefbeta0}
\pi\bigl({\beta}_{0} \big \vert \boldsymbol{y},\boldsymbol{x},\boldsymbol{\theta}_{(-\beta_{0})}\bigr) = {\cal{N}}\biggl( {\beta}_{0}\bigg \vert D_{0} \biggl[\frac{\mu_{0}}{\sigma^{2}_{0}}+ \frac{\sum_{i=1}^{n} \bigl(y_{i} -\beta_{1}x_{i}\bigr)}{\sigma^{2}}\biggr], D_{0}\biggr),
\end{align}
and
\begin{align}\label{coefbeta1}
\pi\bigl({\beta}_{1} \big \vert \boldsymbol{y},\boldsymbol{x},\boldsymbol{\theta}_{(-\beta_{1})}\bigr) = {\cal{N}}\biggl( {\beta}_{1}\bigg \vert D_{1} \biggl[\frac{\mu_{1}}{\sigma^{2}_{1}}+ \frac{\sum_{i=1}^{n} x_{i}\bigl(y_{i}-\beta_{0}\bigr)}{\sigma^{2}}\biggr], D_{1}\biggr),
\end{align}
where $D_{0}=\sigma_{0}^{2} \delta/\bigl(n\sigma_{0}^{2}+\delta\bigr)$ and $D_{1}=\sigma_{1}^{2} \delta/\bigl(\sigma_{1}^{2}\sum_{i=1}^{n}x^{2}_{i}+\delta\bigr)$. Furthermore, by considering transformation $\delta=\sigma^2$, it can be seen easily that
  \begin{align}\label{coefsigma}
\pi\bigl({\delta} \big \vert \boldsymbol{y},\boldsymbol{x}, \boldsymbol{\theta}_{(-\boldsymbol{\beta})}\bigr) = {\cal{IG}}\biggl( \delta \bigg \vert \frac{a_{0}+n}{2}, \frac{a_{0}+\sum_{i=1}^{n}\bigl(y_{i}-{x}_{i}{\beta}\bigr)^{2}}{2}\biggr]\biggr).
\end{align}
Applying a monotone transformation such used as above is a common trick in each Bayesian paradigm to yield a conjugate prior. Finally, the simple reverse transformation gives back the Bayesian estimator of $\sigma$.  
There follows the steps for computing Bayesian estimation of ${\boldsymbol{\theta}}$. 
\begin{enumerate}
\item Read $N, M$, determine hyperparameters $\mu_{0},\mu_{1},\sigma_{0},\sigma_{1}, a_0, b_{0}$, and propose $\boldsymbol{\theta}^{(0)}=\bigl(\beta^{(0)}_{0},\beta^{(0)}_{1},\delta^{(0)}\bigr)^{\top}$;
\item Set $t=0$;
\item Simulate $\beta^{(t+1)}_{0}$ from the full conditional $\beta_{0} \big \vert \bigl({\boldsymbol{y}},{\boldsymbol{x}}, \boldsymbol{\theta}_{(-{\beta}_{0})}\bigr)$ that follows 
\begin{align*}
{\cal{N}}\biggl( D^{(t)}_{0} \biggl[\frac{\mu_{0}}{\sigma^{2}_{0}}+ \frac{\sum_{i=1}^{n} \bigl(y_{i} -\beta^{(t)}_{1}x_{i}\bigr)}{\delta^{(t)}}\biggr], D^{(t)}_{0}\biggr),
\end{align*}  
where $D^{(t)}_{0}=\sigma_{0}^{2} \delta^{(t)}/\bigl(n\sigma_{0}^{2}+\delta^{(t)}\bigr)$.
\item Having ${\beta}^{(t+1)}_{0}$, simulate $\beta^{(t+1)}_{1}$ from the full conditional $\beta_{1} \big \vert \bigl({\boldsymbol{y}},{\boldsymbol{x}}, \boldsymbol{\theta}_{(-{\beta}_{1})}\bigr)$ that follows 
\begin{align*}
{\cal{N}}\biggl( D^{(t)}_{1} \biggl[\frac{\mu_{1}}{\sigma^{2}_{1}}+ \frac{\sum_{i=1}^{n} x_{i}\bigl(y_{i}-\beta^{(t+1)}_{0}\bigr)}{{\delta^{(t)}}}\biggr], D^{(t)}_{1}\biggr),
\end{align*}  
where $D^{(t)}_{1}=\sigma_{1}^{2} {\delta^{(t)}}/\bigl(\sigma_{1}^{2}\sum_{i=1}^{n}x^{2}_{i}+{\delta^{(t)}}\bigr)$.
\item Having $\boldsymbol{\beta}^{(t+1)}$, simulate ${\delta}^{(t+1)}$ from the full conditional $\delta \big \vert \bigl({\boldsymbol{y}},{\boldsymbol{x}}, \boldsymbol{\theta}_{(-\delta)}\bigr)$ that follows 
\begin{align*}
{\cal{IG}}\biggl( \delta \bigg \vert \frac{a_{0}+n}{2}, \frac{b_{0}+\sum_{i=1}^{n}\bigl(y_{i}-{\beta}^{(t+1)}_{0}-{x}_{i}{\beta}^{(t+1)}_{1}\bigr)^{2}}{2}\biggr]\biggr).
\end{align*} 
\item If $t=N$, then go to the next step. Otherwise set $t+1 \rightarrow t$ and return to step 3;
\item The average of sequence $\bigl\{\boldsymbol{\theta}^{(M+1)},\boldsymbol{\theta}^{(M+2)},\cdots,\boldsymbol{\theta}^{(M+N)}\bigr\}$ sampled from target distribution with PDF $\pi\bigl(\boldsymbol{\theta}\vert \boldsymbol{y},\boldsymbol{x}\bigr)$ is the Bayesian estimator $\hat{\boldsymbol{\theta}}_{B}=\bigl(\hat{\beta_{0}}_{B},\hat{\beta_{1}}_{B},\hat{\sigma}_{B}\bigr)^{\top}$.
 \end{enumerate}
Herein, we set the hyperparameters $\{\mu_{0},\mu_{1}\}$ are supposed to be $\widehat{\boldsymbol{\beta}_{ml}}$ and also, we set $\sigma_{0}=1$, $\sigma_{1}=1$, and $\sigma_{2}=1$. Furthermore, $a_0$ and $b_0$ are determined through the empirical Bayes as follows.
\begin{align}\label{empiricalregression}
m\bigl(\boldsymbol{y}\vert {a}_{0},{b}_{0}\bigr)=&\Pi_{i=1}^{n}m\bigl({y}_{i}\vert {a}_{0},{b}_{0}\bigr)\nonumber\\
=&\Pi_{i=1}^{n}\int_{0}^{\infty}f({y}_{i} \vert \delta )\pi\bigl(\delta \big \vert {a}_{0},{b}_{0}\bigr)d\delta\nonumber\\
=&\Pi_{i=1}^{n}\int_{0}^{\infty}
\frac{\delta^{-\frac{1}{2}}}{\sqrt{2 \pi}}\exp\Bigl\{-\frac{1}{2\delta}\bigl(y_{i}-\beta_{0}-\beta_{1}x_{i}\bigr)^2\Bigr\} \nonumber\\
&\times \bigl(\frac{b_{0}}{2}\bigr)^{\frac{a_{0}}{2}} \delta^{-\frac{a_{0}}{2}-1}{\Gamma^{-1}\bigl(\frac{a_{0}}{2}\bigr)}\exp\Bigl\{-\frac{b_{0}}{2\delta}\Bigr\} d\delta\nonumber\\
=&\Pi_{i=1}^{n}\bigl(\frac{b_{0}}{2}\bigr)^{\frac{a_{0}}{2}}\int_{0}^{\infty}
\frac{\delta^{-\frac{a_{0}+1}{2}-1}}{\sqrt{2 \pi}\Gamma\bigl(\frac{a_{0}}{2}\bigr)}\exp\Bigl\{-\frac{1}{2\delta}\bigl[\bigl(y_{i}-\beta_{0}-\beta_{1}x_{i}\bigr)^2+b_{0}\bigr]\Bigr\}d\delta\nonumber\\
=&\Pi_{i=1}^{n} \frac{\Gamma\bigl(\frac{a_{0}+1}{2}\bigr)}{\sqrt{b_{0} \pi}\Gamma\bigl(\frac{a_{0}}{2}\bigr) }\Bigl[1+\frac{\bigl(y_{i}-\beta_{0}-\beta_{1}x_{i}\bigr)^2}{b_{0}}\Bigr]^{-\frac{a_{0}+1}{2}},
%=&\Pi_{i=1}^{n}{\cal{T}}\bigl(y_{i}-\boldsymbol{x}_{i}^{\top}\boldsymbol{\beta}\big \vert a_{0}\bigr),
\end{align} 
where $f({y}_{i} \vert \delta )={\cal{N}}\bigl({y}_{i}\big \vert \beta_{0}+\beta_{1}x_{i}, \delta=\sigma^2\bigr)$. Further, taking into account the fact that if $a_0=b_0$, the RHS of (\ref{empiricalregression}) becomes $\Pi_{i=1}^{n}{{t}}_{}(y_{i} \vert \beta_{0}+\beta_{1}x_{i}, b_{0}, a_{0})$. Assuming $\boldsymbol{\beta}$ is known, the ML estimator of $a_0$ and $b_0$ is obtained by maximizing the RHS of (\ref{empiricalregression})
% is the likelihood function of Student's $t$ distribution with $a_{0}$ degrees of freedom 
based on the sequence of $n$ observed data $\bigl\{y_{i}-\beta_{0}-\beta_{1}x_{i}\bigr\}_{i=1}^{n}$. Moreover, the hyperparameters can be estimated through the the moment-based (MO) approach. While the likelihood function in the RHS of (\ref{empiricalregression}) is complicated in terms of $a_0$, a simple argument shows that this function gets its maximum when $b_0$ is large. We consider this fact when finding the ML estimators of $a_0$ and $b_0$. 
%Herein, the moment estimator of $a_{0}$ is obtained as $\widehat{a_{0}}_{MO}=2S^{2}_{y}/\bigl(S^{2}_{y}-1\bigr)$ where $S^{2}_{y}$ denotes the sample variance of vector $\boldsymbol{y}-\boldsymbol{x}^{\top}\boldsymbol{\beta}$.
To illustrate how one can estimate the simple regression coefficients within a Bayesian framework, we generate $n=200$ realizations $x_1,x_2,\cdots,x_n$ form uniform distribution on (-1,1) and then responses $y_1,y_2,\cdots,y_n$ are obtained using the linear relationship $y_i=\beta_0+\beta_1 x_i + \epsilon_{i}=5+2 x_i + \epsilon_{i}$ where $\epsilon_{i} \sim {\cal{N}}(0,1)$ for $i=1,2,\cdots,n$. 
\end{example}
The corresponding \verb+R+ code for implementing this example is given as follows. The output of the Gibbs sampler is displayed in Figure \ref{simplereg}.
\begin{lstlisting}[style=deltaj]
R> set.seed(20240530)
R> n <- 200
R> x <- runif( n, -4, 4)
R> y <- 5 + 2*x + rnorm(n, mean = 0, sd = 2)
R> out <- lm(y ~ x)
R> ML <- as.vector( coefficients(out) )
R> error <- as.vector( y - ML[1] - ML[2]*x )
R> param0 <- var(error)
R> obj <- function(par) -sum( -log(par[2])/2 + lgamma( (par[1]+1)/2 ) -
+									 lgamma( par[1]/2 ) - (par[1]+1)/2*log( 1 + error^2/par[2] )) 
R> hyper0 <- optim( rep(param0, 2), fn = obj, method = "L-BFGS-B", 
+												lower = rep(0.5, 2), upper = rep(n, 2) )$par
R> N <- 5000  # number of generations
R> M <- 2000  # size of burn-in period
R> theta <-matrix(0, nrow = N, ncol = 3 )
R> sigma0 <- 1; sigma1 <- 1;
R> Sigma0 <- c(sigma0, sigma1)
R> mu0 <- ML
R> delta <- 1
R> theta[1, ] <- c(ML, delta )
R> s.x2 <- c(n, sum(x^2) )
R> j <- 1
R> while(j < N)
+		{
+		 		Di <- Sigma0^2*theta[j, 3]/(Sigma0^2*s.x2 + theta[j, 3])
+				A0 <- sum( y - x*theta[j, 2] )
+				Mu0 <- Di[1]*(mu0[1]/Sigma0[1]^2 + sum(A0)/delta)
+				hat0 <- rnorm(1, Mu0, sqrt(Di[1]) )
+				theta[j + 1, 1] <- hat0
+				A1 <- sum( x*(y - theta[j + 1, 1]) )
+				Mu1 <- Di[2]*(mu0[2]/Sigma0[2]^2 + sum(A1)/delta)
+				hat1 <- rnorm(1, Mu1, sqrt(Di[2]) )
+				theta[j + 1, 2] <- hat1
+				SSE <- sum(error^2)
+				delta <- 1/sqrt( rgamma( 1, shape = (theta0[1] + n)/2, 
+													rate = (theta0[2] + SSE)/2 ) )
+				theta[j + 1, 3] <- delta 
+				j <- j + 1
+		}
R> Bayes <- apply(theta[(N - M + 1):N, ], 2, mean)
R> list(beta0 = Bayes[1], beta1 = Bayes[2], sigma = Bayes[3]) 
\end{lstlisting}
%%%%%%%%%%%%%%%%%%%%%%% muktiple regression 
\begin{example}\label{exam-multiplereg}%\lipsum*[]
Suppose response variable $Y$ depends linearly on independent variable (covariate) $X$ through a simple linear model as 
\begin{align*}%\label{regressionmodel2}
y_{i}=&\beta_0+\beta_{1}x_{i1}+\beta_{2}x_{i2}+\cdots+\beta_{K}x_{iK}+\epsilon_{i}\nonumber\\
=&\boldsymbol{x}_{i}^{\top}\boldsymbol{\beta}+\epsilon_{i};~~ i=1,2,3,\cdots,n,
\end{align*}
where $\boldsymbol{x}_{i}=\bigl(1,x_{i1},\cdots,x_{iK}\bigr)^{\top}$, $\boldsymbol{\beta}=\bigl(\beta_{0},\beta_{1},\cdots,\beta_{K}\bigr)^{\top}$ is the vector of unknown regression coefficients, and $\epsilon_{i}$s are the model's theoretical error assumed to follow ${\cal{N}}\bigl(0,\sigma^{2}\bigr)$ independently (that is, ${\text{cov}}\bigl(\epsilon_{i},\epsilon_{j}\bigr)=0$ for $i\neq j =1,2,3,\cdots,n$). Of course the skewed Student's $t$ \citep{teimouri2020forestfit} and $\alpha$-stable \citep{nolan2013linear} may be reasonable candidates for distribution of theoretical error in presence of outliers that yield more robust estimation of ${\boldsymbol{\beta}}$ than the Gaussian one. It is known that the least-square (LS) estimator $\hat{\boldsymbol{\beta}}_{ls}$ and ML estimator $\hat{\boldsymbol{\beta}}_{ml}$, of the regression coefficients $\boldsymbol{\beta}$ are the same and given by 
\begin{align}\label{ML-multiplereg}
\hat{\boldsymbol{\beta}}_{ML}=\hat{\boldsymbol{\beta}}_{LS}=\bigl(\boldsymbol{X}^{\top} \boldsymbol{X}\bigr)^{-1}\boldsymbol{X}^{\top}\boldsymbol{y},
\end{align}
 where $\boldsymbol{y}$ (vector of responses) and $\boldsymbol{X}$ is $n\times (K+1)$ design matrix are
\begin{align}\label{design-matrix} 
\boldsymbol{y}=
\left[\begin{matrix}
y_{1}\\
y_{2}\\
\vdots\\
y_{n}
\end{matrix}\right],~
\boldsymbol{X}=
\left[\begin{matrix}
\boldsymbol{x}_{1}\\
\boldsymbol{x}_{2}\\
\vdots\\
\boldsymbol{x}_{n}
\end{matrix}\right]=
\left[\begin{matrix}
1& x_{11}& x_{12}&\cdots&x_{1K}\\
1& x_{21}& x_{22}&\cdots&x_{2K}\\
\vdots&\vdots&\vdots&\vdots&\vdots\\
1& x_{n1}& x_{n2}&\cdots&x_{nK}\\
\end{matrix}\right].
\end{align}
Although it is possible to consider a Gaussian conjugate prior for each regression coefficient separately, but we prefer to consider a $(K+1)$-dimensional Gaussian prior for vector of regression coefficients $\boldsymbol{\beta}$ and a mutually independent inverse gamma conjugate prior for $\delta=\sigma^2$ as
\begin{align}\label{prior-multipleregression}
\pi(\boldsymbol{\beta})=&{\cal{N}}_{K+1}\bigl(\boldsymbol{\beta} \big \vert \boldsymbol{\mu}_{0}, \Sigma_{0}\bigr),\nonumber\\
\pi(\delta)=&{\cal{IG}}\Bigl(\delta\Big \vert \frac{a_{0}}{2},\frac{b_{0}}{2}\Bigr).
\end{align}
It follows that
\begin{align}\label{bayes}
\pi(\boldsymbol{\Psi} \vert\boldsymbol{y},\boldsymbol{x})=& L\bigl(\boldsymbol{\Psi} \big \vert \boldsymbol{y},\boldsymbol{x}\bigr) \pi(\boldsymbol{\beta})\nonumber\\
\propto& \delta^{-\frac{n}{2}}\exp\Bigl\{-\frac{1}{2\delta}\sum_{i=1}^{n}\bigl(y_{i}-\boldsymbol{x}_{i}^{\top}\boldsymbol{\beta}\bigr)^2\Bigr\}
 \exp\Bigl\{-\frac{1}{2}\bigl(\boldsymbol{\beta}-\boldsymbol{\mu}_{0}\bigr)^{\top}\Sigma_{0}^{-1}\bigl(\boldsymbol{\beta}-\boldsymbol{\mu}_{0}\bigr)\Bigr\}\nonumber\\
&\times \Bigl(\frac{a_{0}}{2}\Bigr)^{\frac{a_{0}}{2}}
\frac{\delta^{-a_{0}/2-1}}{\Gamma\bigl(\frac{a_{0}}{2}\bigr)}\exp\Bigl\{-\frac{a_{0}}{2\delta}\Bigr\},
\end{align}
where $\boldsymbol{\Psi}=(\boldsymbol{\beta}^{\top},\delta)^{\top}$. For computing the full conditional $\boldsymbol{\beta} \big \vert \bigl(\boldsymbol{y},\boldsymbol{x},\boldsymbol{\Psi}_{(-\boldsymbol{\beta})}\bigr)$, it is worth taking into account to the fact that
\begin{align*}
\sum_{i=1}^{n}\bigl(y_{i}-\boldsymbol{x}_{i}^{\top}\boldsymbol{\beta}\bigr)^2=
\sum_{i=1}^{n}y^{2}_{i} -2 \boldsymbol{\beta}^{\top}\boldsymbol{X}^{\top}\boldsymbol{y}+
\boldsymbol{\beta}^{\top}\boldsymbol{X}^{\top}\boldsymbol{X}\boldsymbol{\beta}.
\end{align*}
%where 
%\begin{align}\label{Xdot}
%X_{\bullet}=\sum_{i=1}^{n}\boldsymbol{x}_{i}^{\top}=\Bigl[n,\sum_{i=1}^{n}x_{i1},\cdots,\sum_{i=1}^{n}x_{iK}\Bigr].
%\end{align}
Then more algebra shows
\begin{align}\label{bayes-beta-multiplereg}
\pi\bigl(\boldsymbol{\beta} \big \vert \boldsymbol{y},\boldsymbol{x},\boldsymbol{\Psi}_{(-\boldsymbol{\beta})}\bigr)
\propto& \exp\Bigl\{-\frac{1}{2}\bigl(\boldsymbol{\beta}-\boldsymbol{\mu}_{\bullet}\bigr)^{\top}\Sigma^{-1}_{\bullet}\bigl(\boldsymbol{\beta}-\boldsymbol{\mu}_{\bullet}\bigr)\Bigr\}\nonumber,
\end{align}
where $\boldsymbol{\mu}_{\bullet}=\Sigma_{\bullet}\bigl[\Sigma_{0}^{-1}\boldsymbol{\mu}_{0}+\boldsymbol{X}^{\top}\boldsymbol{y}\bigr]$ and $\Sigma_{\bullet}=\bigl[\Sigma_{0}^{-1}+\boldsymbol{X}^{\top}\boldsymbol{X}\bigr]^{-1}$.
So,
\begin{align*}%\label{coef-beta-multiplereg}
\pi\bigl(\boldsymbol{\beta} \big \vert \boldsymbol{y},\boldsymbol{x},\boldsymbol{\Psi}_{(-\boldsymbol{\beta})}\bigr) = {\cal{N}}_{K+1}\biggl( \boldsymbol{\beta} \Big \vert \Sigma_{\bullet}\Bigl[\Sigma_{0}^{-1}\boldsymbol{\mu}_{0}+\boldsymbol{X}^{\top}\boldsymbol{y}\Bigr], \Sigma_{\bullet}\biggr).
\end{align*}
Furthermore, it can be seen easily that
  \begin{align*}%\label{coef-sigma-multiplereg}
\pi\bigl({\delta} \big \vert \boldsymbol{y},\boldsymbol{x}, \boldsymbol{\Psi}_{(-{\delta})}\bigr) = {\cal{IG}}\biggl( \delta \bigg \vert \frac{a_{0}+n}{2}, \frac{b_{0}+\sum_{i=1}^{n}\bigl(y_{i}-\boldsymbol{x}_{i}^{\top}\boldsymbol{\beta}\bigr)^{2}}{2}\biggr]\biggr).
\end{align*}
The hyperparameters $a_0$ and $b_0$ can be estimated in the same way as in Example \ref{exam-simplereg}, $\boldsymbol{\mu}_{0}$ is assumed to be (\ref{ML-multiplereg}), and $\Sigma_{0}=\text{cov}\bigl(\hat{\boldsymbol{\beta}}_{ML}\bigr)=\sigma^2\bigl(\boldsymbol{X}^{\top} \boldsymbol{X}\bigr)^{-1}$.
% where 
%\begin{align}\label{cov-ML-multiplereg}
%\boldsymbol{X}^{\top} \boldsymbol{X}=
%\left[\begin{matrix}
%n                                & \sum_{i=1}^{n}x_{i1}          &\sum_{i=1}^{n}x_{i2}           &\cdots&\sum_{i=1}^{n}x_{iK}\\
%\sum_{i=1}^{n}x_{i1}& \sum_{i=1}^{n}x^{2}_{i1}  &\sum_{i=1}^{n}x_{i1}x_{i2}  &\cdots&\sum_{i=1}^{n}x_{i1}x_{iK}\\
%\vdots&\vdots&\vdots&\vdots&\vdots\\
%\sum_{i=1}^{n}x_{iK}& \sum_{i=1}^{n}x_{iK}x_{i1}&  \sum_{i=1}^{n}x_{iK}x_{i2}&\cdots& \sum_{i=1}^{n}x^{2}_{iK}\\
%\end{matrix}\right].
%\end{align}
In what follows, we give the steps for computing Bayesian estimation of ${\boldsymbol{\Psi}}$ in Example \ref{exam-multiplereg}. 
\begin{enumerate}
\item Read $N, M$, and determine hyperparameters $a_0$, $b_{0}$, $\boldsymbol{\mu}_{0}$, and $\Sigma_{0}$;
\item Set $t=0$;
\item Simulate $\boldsymbol{\beta}^{(t+1)}$ from the full conditional $\boldsymbol{\beta} \big \vert \bigl({\boldsymbol{y}},{\boldsymbol{x}}, \boldsymbol{\Psi}_{(-\boldsymbol{\beta})}\bigr)$ with PDF 
\begin{align*}
\pi\bigl(\boldsymbol{\beta} \big \vert \boldsymbol{y},\boldsymbol{x},\boldsymbol{\Psi}_{(-\boldsymbol{\beta})}\bigr) = {\cal{N}}_{K+1}\Bigl( \boldsymbol{\beta} \Big \vert \boldsymbol{\mu}_{\bullet}, \Sigma_{\bullet}\Bigl),
\end{align*}
where $\boldsymbol{\mu}_{\bullet}=\Sigma_{\bullet}\bigl[\Sigma_{0}^{-1}\boldsymbol{\mu}_{0}+\boldsymbol{X}^{\top}\boldsymbol{y}\bigr]$ and $\Sigma_{\bullet}=\bigl[\Sigma_{0}^{-1}+\boldsymbol{X}^{\top}\boldsymbol{X}\bigr]^{-1}$.
So,
\item Having $\boldsymbol{\beta}^{(t+1)}$, set ${\delta}^{(t+1)}=\sqrt{z}$ where $z$ follows ${\cal{IG}}$ distribution with PDF
\begin{align*}
{\cal{IG}}\biggl( z \bigg \vert \frac{a_{0}+n}{2}, \frac{a_{0}+\sum_{i=1}^{n}\bigl(y_{i}-\boldsymbol{x}_{i}^{\top}\boldsymbol{\beta}^{(t+1)}\bigr)^{2}}{2}\biggr]\biggr).
\end{align*} 
\item If $t=N$, then go to the next step. Otherwise set $t+1 \rightarrow t$ and return to step 3;
\item The average of sequence $\bigl\{\boldsymbol{\Psi}^{(M+1)},\boldsymbol{\Psi}^{(M+2)},\cdots,\boldsymbol{\Psi}^{(M+N)}\bigr\}$ sampled from target distribution with PDF $\pi\bigl(\boldsymbol{\Psi}\vert \boldsymbol{y},\boldsymbol{x}\bigr)$ is the Bayesian estimator $\hat{\boldsymbol{\Psi}}_{B}=\bigl(\hat{\boldsymbol{\beta}}_{B},\hat{\sigma}_{B}\bigr)^{\top}$.      
 \end{enumerate}
\end{example}
Herein, Example \ref{exam-multiplereg} will be illustrated with computing the Bayesian estimator of regression plane coefficients fitted to \verb+trees+ data. This data set is available in \verb+R+ environment. The corresponding \verb+R+ code for implementing this example is given as follows. The output of the Gibbs sampler is displayed in Figure \ref{multiplereg}.
\begin{lstlisting}[style=deltaj]
R> set.seed(20240601)
R> data(trees)
R> y <- trees$Volume;
R> n <- length(y)
R> x <- cbind(trees$Girth, trees$Height)
R> K <- 3 # number of regression coefficients
R> X <- as.matrix( cbind(1, x), nrow = n, ncol = K)
R> out <- lm(y ~ X[, 2] + X[, 3])
R> ML <- as.vector( coefficients(out) )
R> error <- as.vector(y - X%*%ML)
R> param0 <- var(error)
R> obj <- function(par) -sum( -log(par[2])/2 + lgamma( (par[1]+1)/2 ) -
+												 lgamma( par[1]/2 ) - (par[1]+1)/2*log( 1 + error^2/par[2] )) 
R> theta0 <- optim( rep(param0, 2), fn = obj, method = "L-BFGS-B",
+												  lower = rep(0.5, 2),	 upper = rep(n, 2) )$par
R> N <- 5000  # number of generations
R> M <- 2000  # size of burn-in period
R> Psi <- matrix(0, nrow = N, ncol = K + 1)
R> delta <- 1
R> Psi[1, ] <- c(ML, delta )
R> hat <- Psi[1, ]
R> Mu0 <- ML
R> sigma2_hat <- theta0[2]/abs(theta0[1] - 1)
R> Sigma0_inv <- solve( sigma2_hat*solve(t(X)%*%X) )
R> j <- 1
R> while(j < N)
+		 {
+		 Sigma_bullet <- solve( Sigma0_inv + t(X)%*%X )
+		 Mu_bullet <- Sigma_bullet%*%( Sigma0_inv%*%Mu0 + t(X)%*%y )
+		 Psi[j+1, 1:K] <- mvrnorm(1, mu = Mu_bullet, Sigma = Sigma_bullet )
+		 error <- as.vector( y - X%*%Psi[j+1, 1:K] )
+		 SSE <- sum(error^2)
+		 delta <- 1/sqrt(rgamma( 1, shape = (theta0[1] + n)/2,
+												 rate = (theta0[2] + SSE)/2 ) )
+		 Psi[j + 1, 4] <- delta 
+		 j <- j + 1
+		 }
R> Bayes <- apply(Psi[(N - M + 1):N, ], 2, mean)
R> list(beta0=Bayes[1],beta1=Bayes[2],beta2=Bayes[3],sigma=sqrt(Bayes[4]))
\end{lstlisting}
\begin{figure}[!h]
\center
\includegraphics[width=90mm,height=60mm]{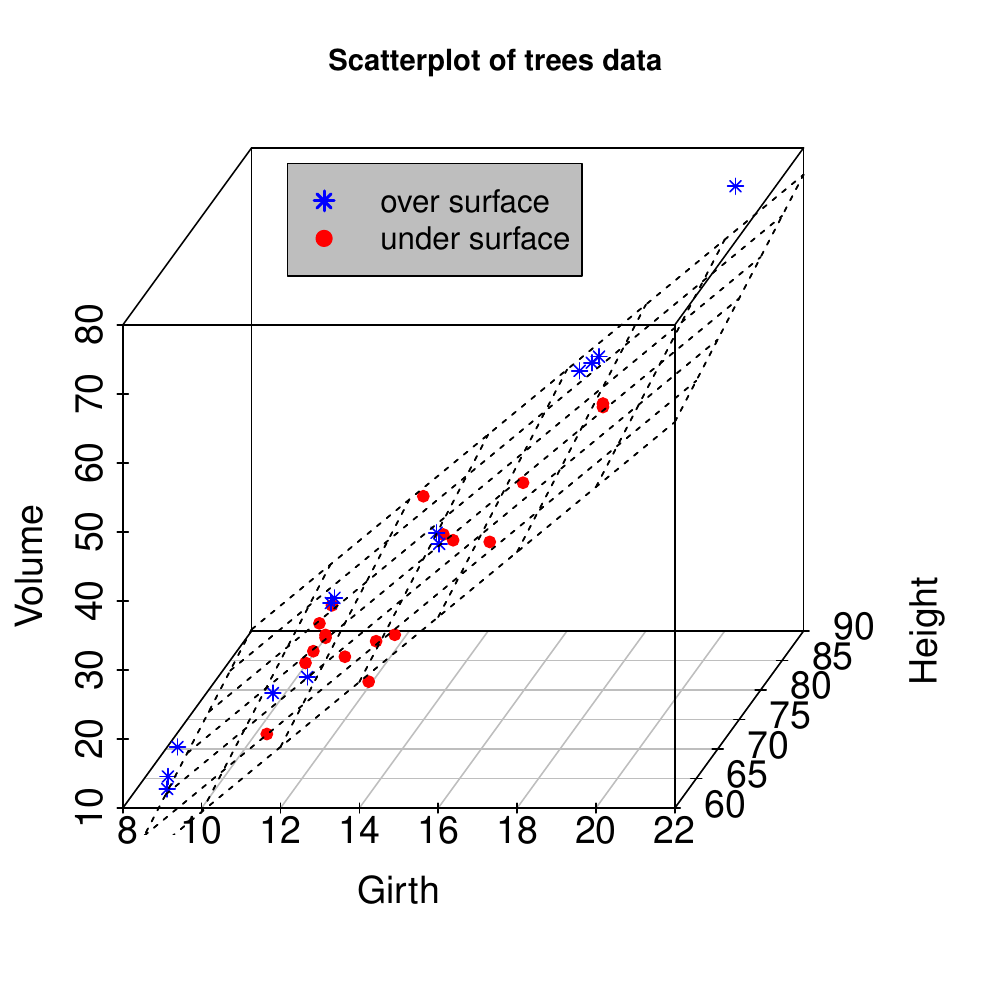}
\caption{
Scatterplot of trees data. Superimposed is the fitted regression plane whose coefficients are estimated through the Bayesian paradigm.}
\label{multiplereg-surface}
\end{figure}
%%%%%%%%%%%%%%%%%%%%%%%%%%%%%%%%%%%%%%%%%%%%%%%%
\section{Gibbs sampling in Bayesian paradigm with one latent variable}
\begin{figure}[h]
\center
\includegraphics[width=55mm,height=55mm]{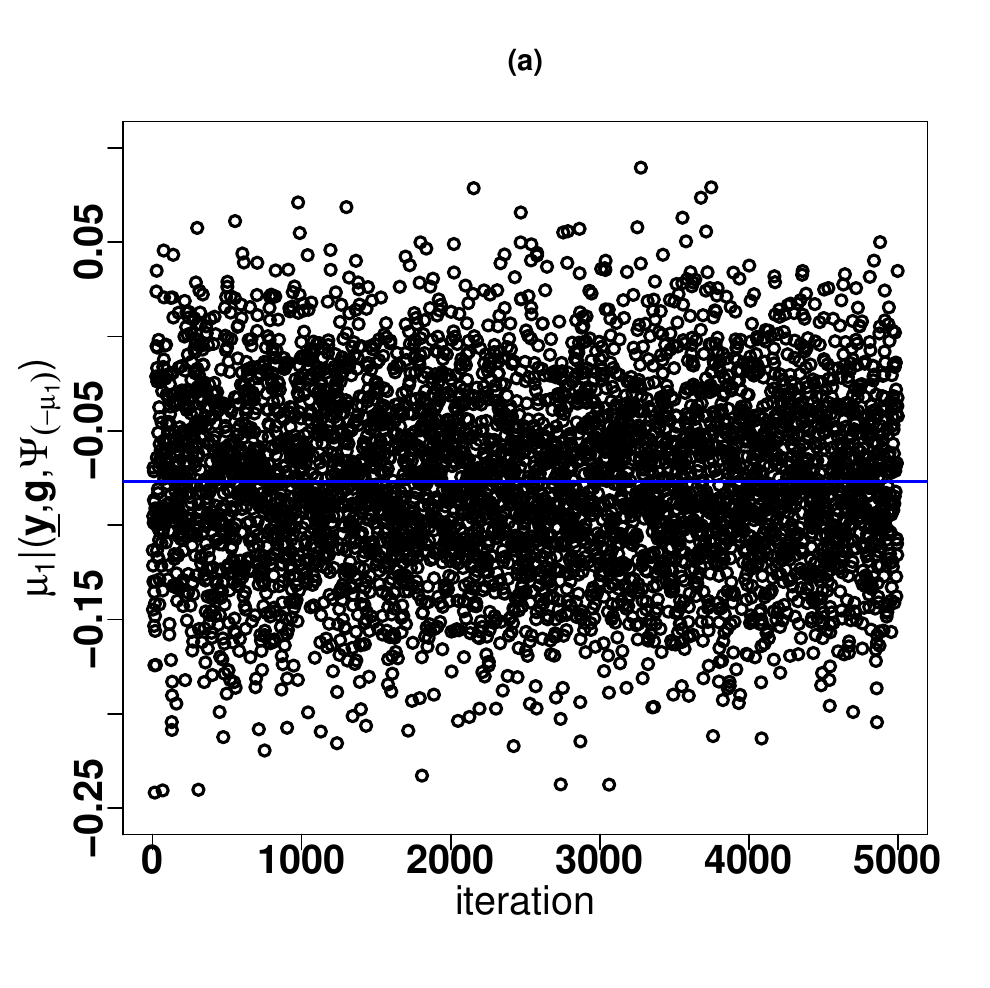}
\includegraphics[width=55mm,height=55mm]{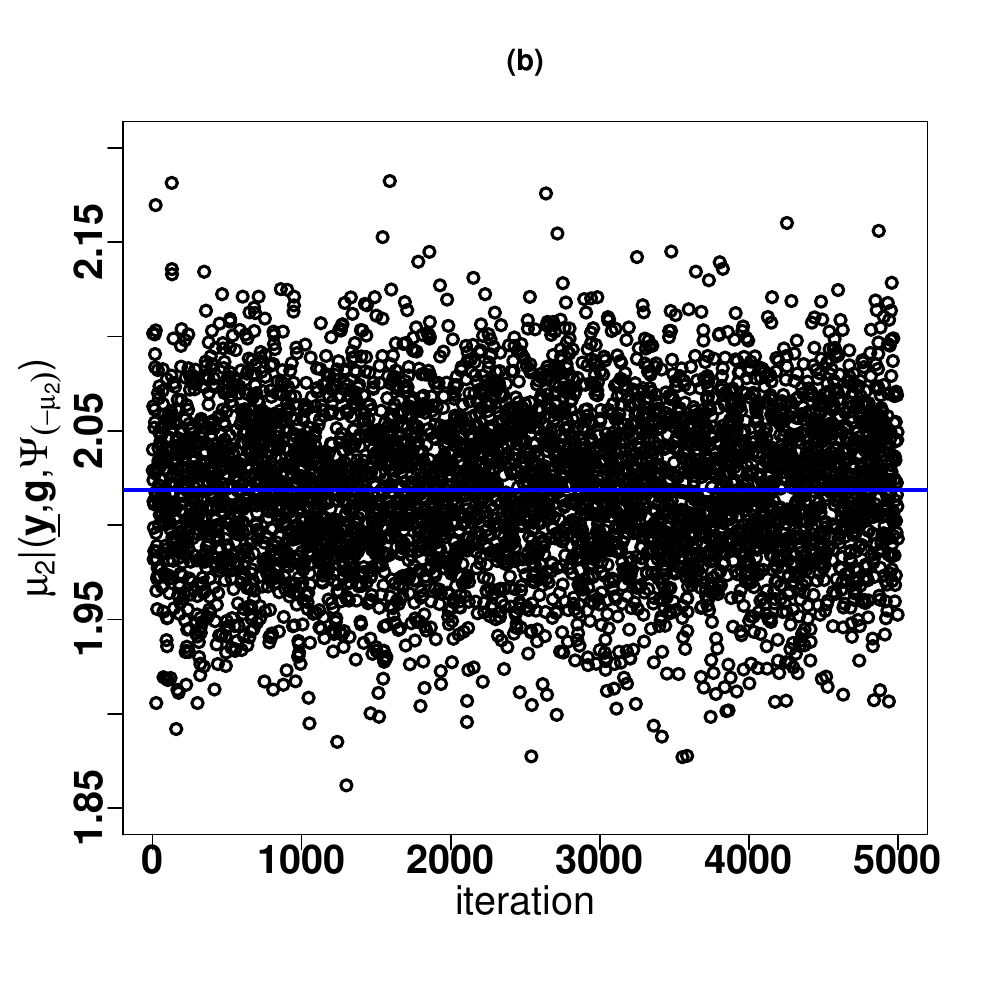}
\includegraphics[width=55mm,height=55mm]{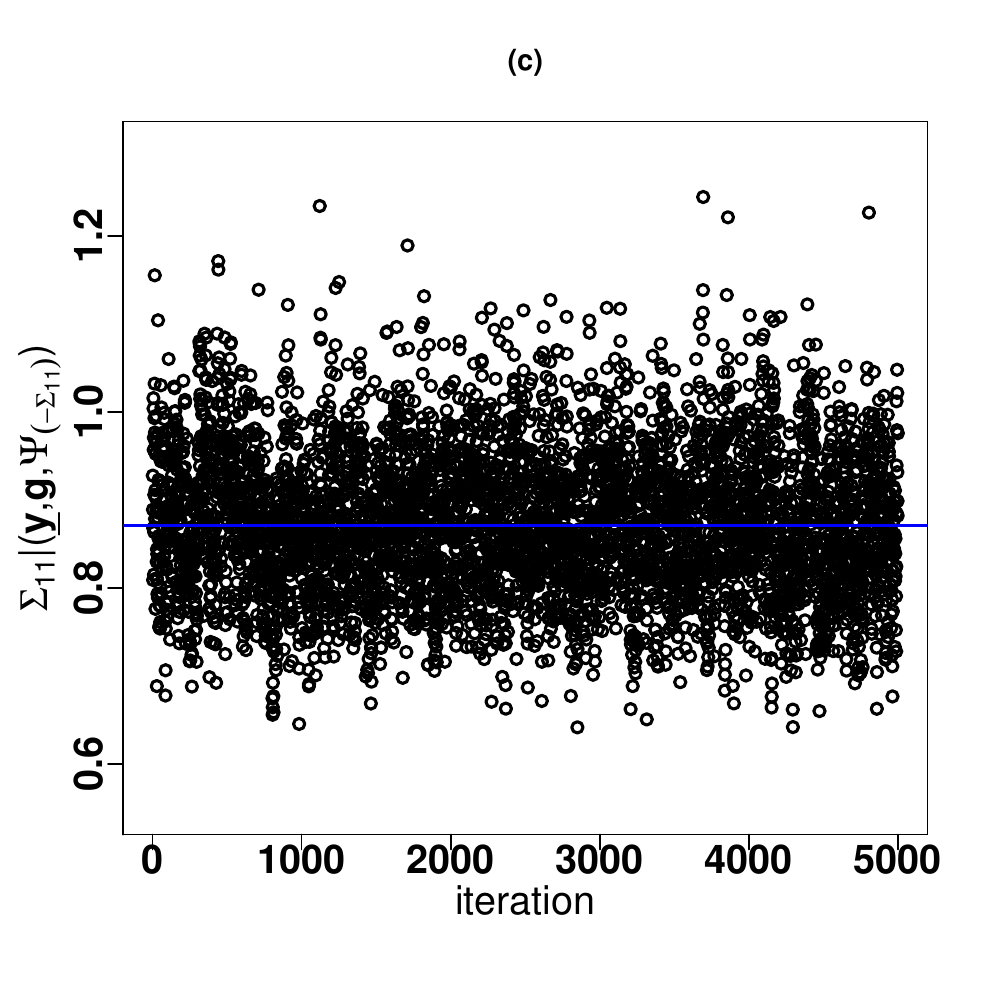}
\includegraphics[width=55mm,height=55mm]{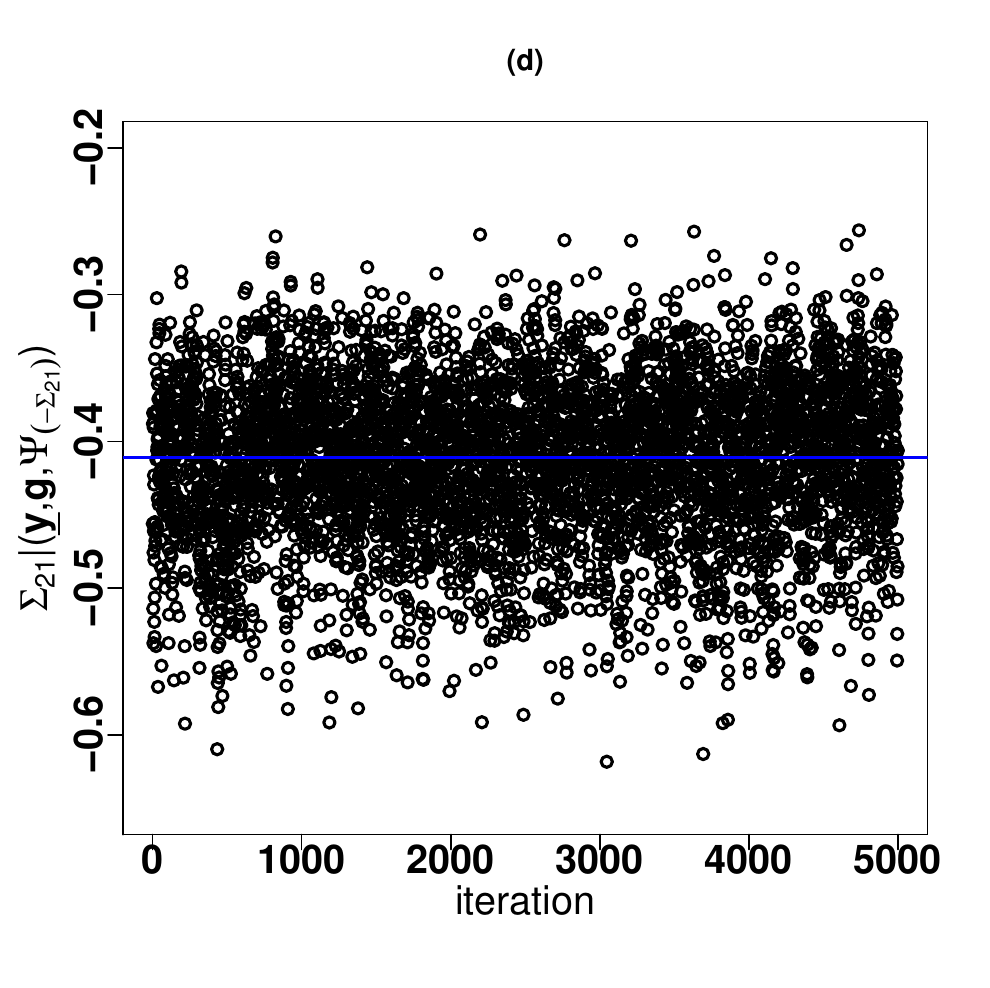}
\includegraphics[width=55mm,height=55mm]{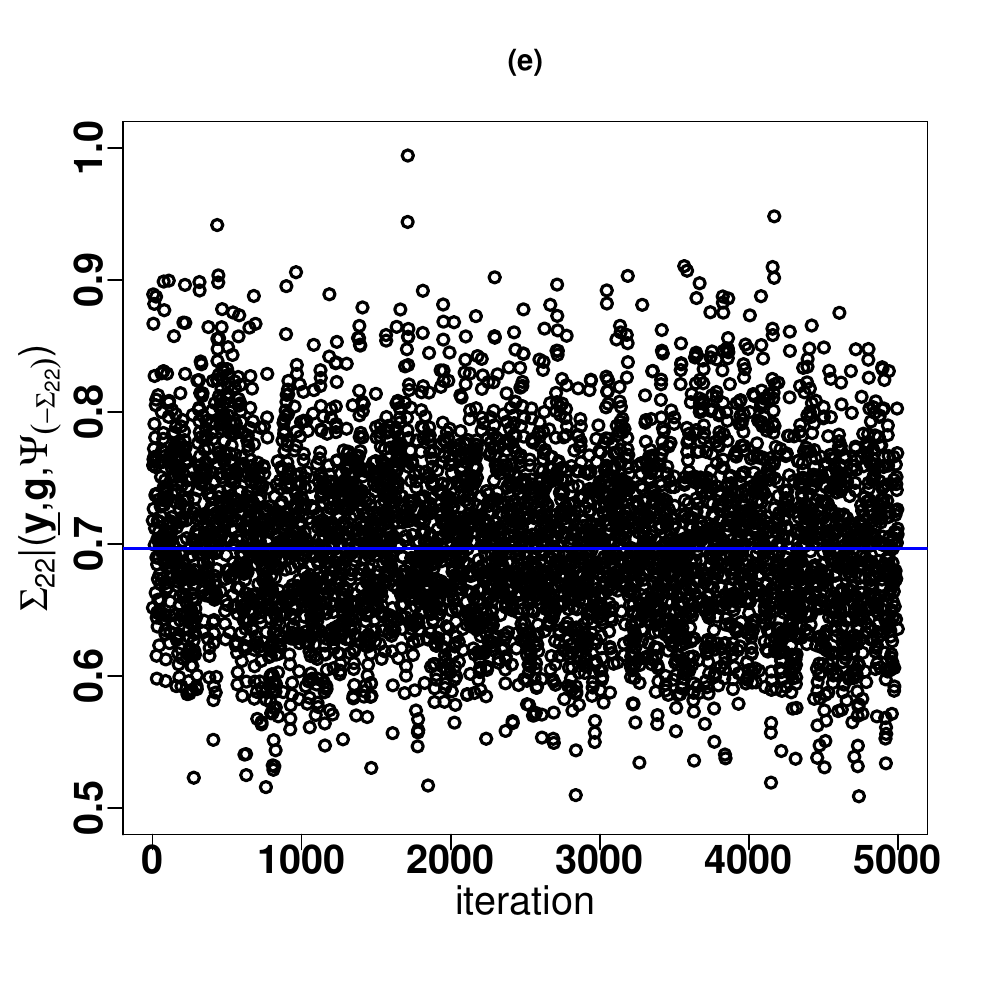}
\includegraphics[width=55mm,height=55mm]{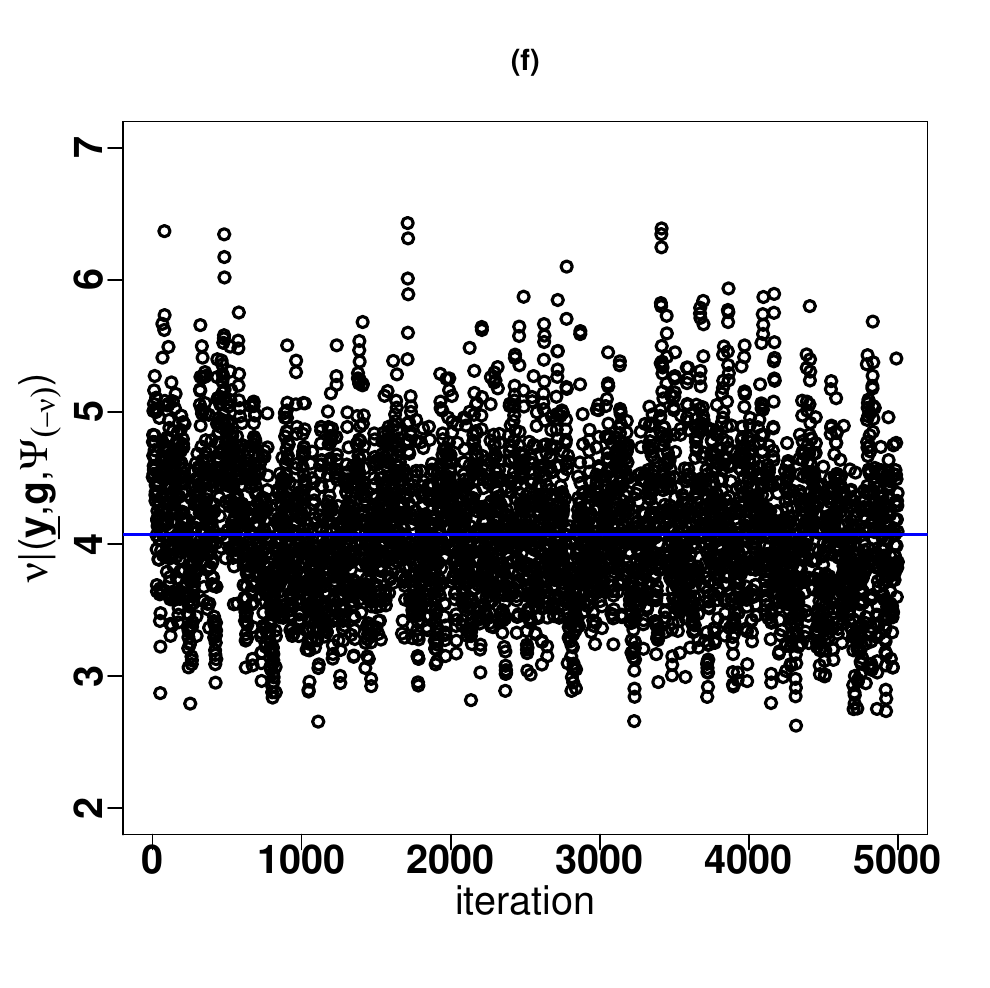}
\caption{Scatterplot of 5000 samples generated from full conditionals: (a) $\mu_{1} \big \vert \bigl({{\boldsymbol{y}}},{\boldsymbol{g}},\Psi_{(-{\mu}_{1})} \bigr)$, (b): ${\mu}_{2} \big \vert \bigl({{\boldsymbol{y}}},{\boldsymbol{g}},\Psi_{(-{\mu}_{2})} \bigr)$, (c): ${\Sigma}_{11} \big \vert \bigl({{\boldsymbol{y}}},{\boldsymbol{g}},\Psi_{(-\Sigma_{11})} \bigr)$, (d) ${\Sigma}_{12} \big \vert \bigl({{\boldsymbol{y}}},{\boldsymbol{g}},\Psi_{(-\Sigma_{12})} \bigr)$, (e) ${\Sigma}_{22} \big \vert \bigl({{\boldsymbol{y}}},{\boldsymbol{g}},\Psi_{(-\Sigma_{22})} \bigr)$, and (f) ${\nu} \big \vert \bigl({{\boldsymbol{y}}},{\boldsymbol{g}},\Psi_{(-\nu)} \bigr)$. The blue line, in each subfigure, shows the Bayesian estimator.}
\label{fig6}
\end{figure}

Sometimes implementing the MCMC techniques such as MH or Gibbs sampling for drawing the Bayesian inference is computationally inefficient due to the complex form of the posterior. In such cases, inference about the unknown parameter(s) may be drawn by involving some latent ({\it{unobservable}} or {\it{augmented}}) variable(s). The rationale behind using the term ``latent'' traced back to this fact that  the experimenter cannot observe some extra variable(s) existing in the model at quick glance, but there exists. For example, let random vector $\boldsymbol{X}=(X_1,\cdots,X_p)^{\top}$ follows a $p$-dimensional Student's $t$ distribution with $\nu>0$ degrees of freedom with PDF given by (\ref{PDF-t}). Evidently implementing the MH or Gibbs sampling in which the full conditionals are constructed based on PDF (\ref{PDF-t}) is quite difficult. On the other hand, it is easy to see that $\boldsymbol{X}$ admits the stochastic representation 
%$\boldsymbol{Y}\mathop=\limits^d \boldsymbol{\mu}+ \Sigma^{1/2} \boldsymbol{X}$ in which $\boldsymbol{X}=(X_1,\cdots,X_p)^{\top}$ follows a standard $p$-dimensional Student's $t$ distribution with location parameter $\boldsymbol{\mu}$, scale matrix $\Sigma$, $\nu$ degrees of freedom. To be more precise, the random vector $\boldsymbol{Y}$ can be represented as 
$\boldsymbol{X}\mathop=\limits^d \boldsymbol{\mu}+ \Sigma^{1/2} \boldsymbol{Z}/\sqrt{G}$ where $\boldsymbol{Z}\sim {\cal{N}}_{p}\bigl(\boldsymbol{0}_{p},\boldsymbol{I}_{p}\bigr)$ and $G\sim {\cal{G}}(\nu/2,\nu/2)$ (${\cal{G}}(a,b)$ accounts for a gamma distribution with shape and rate parameters, $a$ and $b$, respectively.). Herein, random variable $G$ plays the role of a latent variable. We also can consider more than one latent variable in each statistical model. Nevertheless, the Gibbs sampling is a very useful approach in presence of latent variable in the case of either sampling or Bayesian inference. In what follows, we give a definition that plays a main role in understanding Bayesian inference for the parameters of a statistical model that includes latent variable(s). 
\begin{dfn}\label{def4}
Let $x_1,\cdots,x_n$ are a sample of $n$ independent {observed} values from a model with PDF $f(x\vert \boldsymbol{\theta})$ where $\boldsymbol{\theta}$ is the unknown parameter vector. If model includes latent variable $g$ whose PDF is $f(g\vert \boldsymbol{\theta}_{g})$ in which $\boldsymbol{\theta}_{g}$ is the unknown parameter vector of distribution of $g$. Furthermore, let $f(x,g\vert \boldsymbol{\Psi})$ is the joint PDF of vector $(x,g)^{\top}$ where $\boldsymbol{\Psi}=\boldsymbol{\theta}\cup \boldsymbol{\theta}_{g}$. Then we have:
\begin{itemize}
\item The sequence $\bigl\{x_1,\cdots,x_n\bigr\}$ and $\bigl\{g_1,\cdots,g_n\bigr\}$, are called the {\it{observed}} and {\it{latent}} data, respectively.
\item The sequence of paired-data $\{\bigl(x_1,g_1\bigr),\cdots,\bigl(x_n,g_n\bigr)\}$ is called {\it{complete data}} that evidently is the union of {\it{observed}} and {\it{latent}} data. 
\item The {\it{complete data}} likelihood function becomes
\begin{equation}\label{conmar}
L_{c}(\boldsymbol{\Psi}\vert \boldsymbol{x},\boldsymbol{g})=f(\boldsymbol{x},\boldsymbol{g}\vert\boldsymbol{\Psi})=\prod_{i=1}^{n}f\bigl(x_{i}\big \vert g_{i}, \boldsymbol{\Psi}\bigr)=\prod_{i=1}^{n}f\bigl(x_{i}\big \vert g_{i}, \boldsymbol{\theta}\bigr)f\bigl(g_{i} \big \vert \boldsymbol{\theta}_{g}\bigr).
\end{equation} 
\end{itemize}
\end{dfn}
Definition \ref{def4} is based on just one latent variable, but can be easily extended and is valid for two or more latent variables. It is noteworthy that for a Bayesian paradigm in which we have latent variable(s), we rewrite the posterior (\ref{bayes}) in terms of the {\it{complete data}} likelihood function as follows. 
\begin{align}\label{bayesc}
\pi(\boldsymbol{\Psi} \vert\boldsymbol{x}, \boldsymbol{g}) 
\propto L_{c}(\boldsymbol{\Psi}\vert \boldsymbol{x},\boldsymbol{g}) \pi(\boldsymbol{\Psi}),
\end{align}
where $L_{c}(\boldsymbol{\Psi}\vert \boldsymbol{x},\boldsymbol{g})$ is given by Definition \ref{def4}. To implement the Bayesian paradigm, notice that the conditional PDF $f\bigl(x_{i}\big \vert g_{i}, \boldsymbol{\theta}\bigr)$ and marginal PDF $f\bigl(g_{i}\big \vert \boldsymbol{\theta}_{g}\bigr)$ given in RHS of (\ref{conmar}) must be known and possibly in closed form. In what follows, we give an example in which Bayesian inference on the parameters of a Student's $t$ distribution is desired.
\begin{example}\label{exam-t}%\lipsum*[]
\cite{he2021objective}
Suppose ${\underline{\boldsymbol{x}}}=\bigl(\boldsymbol{x}_{1},\cdots,\boldsymbol{x}_{n}\bigr)$ is vector of $n$ independent observations from a $p$-dimensional Student's $t$ distribution. Recall that for each Student's $t$ random vector $\boldsymbol{X}$ we can write $\boldsymbol{X}\mathop=\limits^d \boldsymbol{\mu}+ \Sigma^{1/2} \boldsymbol{Z}/\sqrt{G}$. Evidently $\boldsymbol{X}$ admits the hierarchy given by
\begin{align}\label{hierarchicalt}
\boldsymbol{X}\vert G &\sim f\bigl(\boldsymbol{x}_{i}\vert g_{i},\boldsymbol{\theta}=(\boldsymbol{\mu}, \Sigma)\bigr) = {\cal{N}}_{p}\Bigl(\boldsymbol{x}_{i}\Big \vert \boldsymbol{\mu}, \frac{\Sigma}{{G}}\Bigr),\nonumber\\
G & \sim f\bigl(g_{i}\big \vert \boldsymbol{\theta}_{g}=\nu \bigr) = {\cal{G}}\bigl(g_{i}\big \vert \nu/2, \nu/2\bigr),
\end{align}
where ${\cal{N}}_{p}\bigl(.\big \vert \boldsymbol{\mu},{\Sigma}\bigr)$ denotes the PDF of a $p$-dimensional Gaussian distribution with location vector $\boldsymbol{\mu}$ and scale matrix ${\Sigma}$ and ${\cal{G}}(\cdot \big \vert a, b)$ accounts for the PDF of a gamma distribution with shape parameter $a$ and rate parameter $b$. Hence, the complete data likelihood is
\begin{align}\label{cd12}
L_{c}(\boldsymbol{\Psi}\vert {\underline{\boldsymbol{x}}}, \boldsymbol{g})=& \prod_{i=1}^{n}\Bigl[{\cal{N}}_{p}\bigl(\boldsymbol{x}_i\big \vert\boldsymbol{\mu}, [g_{i}]^{-1} \Sigma \bigr) f\bigl(g_{i} \big \vert\boldsymbol{\theta}_{g}\bigr)\Bigr]\nonumber\\
=&\Bigl[\Gamma \Bigl(\frac{\nu}{2}\Bigr)\Bigr]^{-n}\Bigl(\frac{\nu}{2}\Bigr)^{\frac{n\nu}{2}}\nonumber\\
&\times \prod_{i=1}^{n}\biggl\{\bigl[g_{i}\bigr]^{\frac{\nu +p}{2}-1}\exp\Bigl\{- \frac{g_{i}}{2} \Bigl[ \nu +\bigl(\boldsymbol{x}_i-\boldsymbol{\mu}\bigr)^{\top}\Sigma^{-1}\bigl(\boldsymbol{x}_i-\boldsymbol{\mu}\bigr)\Bigr] \Bigr\}\biggr\},
\end{align}
where $\boldsymbol{\Psi}=\boldsymbol{\theta}\cup \boldsymbol{\theta}_{g}=(\boldsymbol{\mu},\Sigma,\nu)$ is the model's parameter vector. 
%More algebra shows
%\begin{align}\label{cdll1}
%\pi(\boldsymbol{\Psi}\vert \underline{\boldsymbol{x}}, \boldsymbol{g})
%\propto& \pi(\boldsymbol{\Psi}).
%\end{align}
For the prior, we proceed to assume independence, that is $\pi(\boldsymbol{\Psi})=\pi(\boldsymbol{\mu})\pi(\Sigma)\pi(\nu)$ in which $\pi(\boldsymbol{\mu})$ and $\pi(\Sigma)$ are conjugate Gaussian and inverse Wishart priors. Furthermore, we let   $\pi(\nu)$ to be the PDF of a gamma distribution. We have
\begin{align*}
\pi(\boldsymbol{\mu})&\sim {\cal{N}}_{p}({\cal{M}}_{0}, {\cal{S}}_{0}),\nonumber\\
\pi(\Sigma)&\sim {\cal{IW}}({\cal{D}}_{0}, \nu_0),\nonumber\\
\pi(\nu)&\sim {\cal{G}}\bigl(a_0,b_0\bigr).
\end{align*}
We note that in above an objective prior such as Jeffreys' prior can be chosen, however, the most concern is placed on $\pi(\nu)$ that was proposed by \cite{fernandez1999multivariate} to be $\pi(\nu)=10\exp\{-10\nu\}$.
% or $\pi(\nu)=10\exp\{-10\nu\}$ by \cite{wang2018bayesian}.   in type of prior   chhose
It can be seen that the jeffreys' prior  becomes $\pi(\boldsymbol{\mu},\Sigma,\nu)=\vert \Sigma \vert^{-(p+1)/2}$ \citep{fernandez1999multivariate}, and so considering a flat prior for both of $\boldsymbol{\mu}$ and $\nu$ is acceptable. We note that for drawing inference about $\boldsymbol{\Psi}$, one needs to generate from full conditional of the latent variable too. So, by considering $\boldsymbol{\mu}$, $\Sigma$, $\nu$, and $g$ (latent variable) as the model's variables, within a Gibbs sampling framework, we proceed to sample from full conditionals $\boldsymbol{\mu}\vert ({\Sigma}, \nu)$, $\Sigma \vert (\boldsymbol{\mu}, \nu)$, $\nu\vert (\boldsymbol{\mu}, {\Sigma})$. But, keep in mind that for each observed data $\boldsymbol{y}_{i}$, $g_i$ (for $i=1,\cdots,n$) is also latent 
unobservable). In order to $g_i$ being known, we proceed to simulate $g_i$ (for $i=1\cdots,n$) for each $\boldsymbol{y}_{i}$. To do this, in each iteration of the Gibbs sampling, we simulate (for each $\boldsymbol{y}_{i}$) from full conditional $g_i\big \vert \bigl(\boldsymbol{\mu}, {\Sigma}, \nu, {\boldsymbol{y}}_{i} \bigr)$ as follows. From RHS of (\ref{cd12}), for a single observed value $\boldsymbol{y}_{i}$, it follows that
\begin{align*}%\label{cdll2}
\pi\bigl(g_i\big \vert \boldsymbol{\Psi}, \boldsymbol{y}_{i}\bigr)
%\propto& 
%\bigl[g_{i}\bigr]^{\frac{p+1}{2}}\exp\Bigl\{- \frac{g_{i}}{2} \Bigl[ \bigl(\boldsymbol{y}_i-\boldsymbol{\mu}\bigr)^{\top}\Sigma^{-1}\bigl(\boldsymbol{y}_i-\boldsymbol{\mu}\bigr)\Bigr] \Bigr\}
%\frac{1}{\Gamma \Bigl(\frac{\nu}{2}\Bigr)}
%\Bigl(\frac{\nu}{2}\Bigr)^{\frac{\nu}{2}}
%g_{i}^{\frac{\nu}{2}-1}\exp\bigl\{-\frac{g_{i}\nu}{2}\bigr\}\nonumber\\
&\propto \bigl[g_{i}\bigr]^{\frac{\nu +p}{2}-1}\exp\Bigl\{- \frac{g_{i}}{2} 
\Bigl[ \nu+ \bigl(\boldsymbol{y}_i-\boldsymbol{\mu}\bigr)^{\top}\Sigma^{-1}\bigl(\boldsymbol{y}_i-\boldsymbol{\mu}\bigr)\Bigr] \Bigr\}.
\end{align*}
This means that 
\begin{align*}
\pi\bigl(g_i\big \vert \boldsymbol{\Psi}, \boldsymbol{y}_{i}\bigr)\sim {\cal{G}}\Bigl((\nu +p)/2,  \nu/2+ \bigl(\boldsymbol{y}_i-\boldsymbol{\mu}\bigr)^{\top}\Sigma^{-1}\bigl(\boldsymbol{y}_i-\boldsymbol{\mu}\bigr)/2\Bigr).
\end{align*}
Let $\boldsymbol{\Psi}_{(-\theta)}$ denote the whole parameter vector excluding its element $\theta$, once we have generated a whole vector of latent variable as $\boldsymbol{g}=(g_1,\cdots,g_n)^{\top}$, the complete data is known and we proceed to simulate from other full conditionals as follows.
\begin{itemize}
%\item  $\pi(g_{i}\big \vert \bigl(\boldsymbol{\mu}, \boldsymbol{\Sigma}, \nu, \boldsymbol{y}_{i}\bigr)\propto$
%\item  {\bf{full conditional of $\boldsymbol{\mu} \big \vert (\boldsymbol{Y}=\boldsymbol{y},\boldsymbol{U}=\boldsymbol{u}, \boldsymbol{G}=\boldsymbol{g}, \Psi_{(-\boldsymbol{\mu})})$:}}$ By considering a conjugate prior, that is $\pi(\boldsymbol{\mu}) \sim {\cal{N}}_{p}({\cal{M}}_{0}, {\cal{S}}_{0})$, more 
\item {\bf{full conditional of $\boldsymbol{\mu} \big \vert \bigl({{\underline{\boldsymbol{y}}}},\boldsymbol{g}, \Psi_{(-\boldsymbol{\mu})}\bigr)$:}} By considering a conjugate prior, shown earlier by $\pi(\boldsymbol{\mu}) \sim {\cal{N}}_{p}({\cal{M}}_{0}, {\cal{S}}_{0})$, we have
\begin{align*}
\pi\bigl(\boldsymbol{\mu} \big \vert {{\underline{\boldsymbol{y}}}},\boldsymbol{g}, \Psi_{(-\boldsymbol{\mu})}\bigr)  \propto L_{c}(\boldsymbol{\Psi}\vert {\underline{\boldsymbol{x}}}, \boldsymbol{g}) {\cal{N}}_{p}\bigl( \boldsymbol{\mu}\big \vert {\cal{M}}_{0}, {\cal{S}}_{0}\bigr).
\end{align*}
More algebra shows that
\begin{align*}
\boldsymbol{\mu} \big \vert \bigl({{\underline{\boldsymbol{y}}}},\boldsymbol{g}, \Psi_{(-\boldsymbol{\mu})}\bigr) \sim {\cal{N}}_{p}\Bigl({\cal{Q}}_{0}\Bigl[{\cal{S}}^{-1}_{0}{\cal{M}}_{0}+\Sigma^{-1}\sum_{i=1}^{n}g_{i}\boldsymbol{y}_{i}\Bigr],{\cal{Q}}_{0}\Bigr),
\end{align*}
where
\begin{align*}
{\cal{Q}}_{0}=\Bigl[{\cal{S}}^{-1}_{0}+\Sigma^{-1}\sum_{i=1}^{n}g_{i}\Bigr]^{-1}.
\end{align*}
\item {\bf{full conditional of 
${\Sigma} \big \vert \bigl({{\underline{\boldsymbol{y}}}},\boldsymbol{g}, \Psi_{(-{\Sigma})}\bigr)$:}} By considering a conjugate prior, that is ${\Sigma}\sim {\cal{IW}}\bigl({\cal{D}}_{0}, \nu_{0}\bigr)$, we can write
\begin{align*}
\pi\bigl(\Sigma \big \vert {{\underline{\boldsymbol{y}}}},\boldsymbol{g}, \Psi_{(-\Sigma)}\bigr)  \propto L_{c}(\boldsymbol{\Psi}\vert {\underline{\boldsymbol{x}}}, \boldsymbol{g}) {\cal{IW}}(\Sigma\big \vert {\cal{D}}_{0}, \nu_{0}\bigr),
\end{align*}
where ${\cal{IW}}(\Sigma\big \vert {\cal{D}}_{0}, \nu_{0}\bigr)$ denotes the PDF of an inverse Wishart distribution.
More algebra shows that
\begin{align*}
{\Sigma} \big \vert \bigl({{\underline{\boldsymbol{y}}}},\boldsymbol{g}, \Psi_{(-{\Sigma})}\bigr) \sim  {\cal{IW}}\bigl({\cal{R}}_{0}, \nu_{0}+n\bigr),
\end{align*}
where ${\cal{R}}_{0}={\cal{D}}_{0} +\sum_{i=1}^{n} g_{i} \bigl(\boldsymbol{y}_i-\boldsymbol{\mu}\bigr)\bigl(\boldsymbol{y}_i-\boldsymbol{\mu}\bigr)^{\top}$.
\item {\bf{full conditional of 
${\nu} \big \vert \bigl({\underline{\boldsymbol{y}}}
,\boldsymbol{g}, \Psi_{(-{\nu})}\bigr)$:
}} By considering a truncated exponential distribution with PDF 
$\pi(\nu)= {\cal{G}}\bigl(\nu-2 \big \vert 1, b_{0}\bigr)$ for $\nu \geq 2$, we can write
\begin{align}\label{fullnu}
\pi\bigl(\nu \big \vert {\underline{\boldsymbol{y}}},\boldsymbol{g}, \Psi_{(-{\nu})}\bigr)& \propto L_{c}(\boldsymbol{\Psi}\vert \underline{\boldsymbol{x}}, \boldsymbol{g}) \pi(\nu) \nonumber\\
&\propto \Bigl[\Gamma \Bigl(\frac{\nu}{2}\Bigr)\Bigr]^{-n}
\Bigl(\frac{\nu}{2}\Bigr)^{\frac{n\nu}{2}}b_{0}
\exp\Bigl\{-\frac{\nu}{2}\bigl[2b_{0}+\sum_{i=1}^{n}g_{i}\bigr]\Bigr\}\prod_{i=1}^{n}\bigl[g_{i}\bigr]^{\frac{\nu}{2}}.
\end{align}
As it is seen, the full conditional (\ref{fullnu}) has not closed form. The MH algorithm can be suggested to draw sample from this full conditional. It is easy to check that $\pi\bigl(\nu \big \vert {\underline{\boldsymbol{y}}},\boldsymbol{g}, \Psi_{(-{\nu})}\bigr)$ is log-concave and hence the ARS-within-Gibbs sampling algorithm can be suggested for simulating from this full conditional. For computing the vector of hyperparameter $\boldsymbol{\theta}_{0}=b_0$, it follows from hierarchy (\ref{hierarchicalt}) that
\begin{align}\label{bayes4}
%m\bigl(\boldsymbol{x}\vert \boldsymbol{\theta}_{0}\bigr)&=\int_{0}^{\infty} f\bigl(\boldsymbol{x}_{i}\big \vert g_{i},\boldsymbol{\theta}\bigr) f\bigl(g_{i}\big \vert \boldsymbol{\theta}_{g}=\nu \bigr)\pi\bigl(\nu \big \vert \boldsymbol{\theta}_{0}\bigr) d\nu \nonumber\\
%&=\frac{(2\pi)^{}}{\vert \Sigma \vert^{\frac{1}{2}}}\int_{0}^{\infty} \Bigl[\Gamma \Bigl(\frac{\nu}{2}\Bigr)\Bigr]^{-1}\Bigl(\frac{\nu}{2}\Bigr)^{\frac{\nu}{2}}b_{0}\exp\{-b_{0}\nu\}\nonumber\\
%&\times \bigl[g_{i}\bigr]^{\frac{\nu +p+1}{2}-1}\exp\Bigl\{- \frac{g_{i}}{2} \Bigl[ \nu +\bigl(\boldsymbol{x}_i-\boldsymbol{\mu}\bigr)^{\top}\Sigma^{-1}\bigl(\boldsymbol{x}_i-\boldsymbol{\mu}\bigr)\Bigr] \Bigr\},
m\bigl(\underline{\boldsymbol{x}}\vert {b}_{0}\bigr)&=\Pi_{i=1}^{n}\int_{2}^{\infty}
f\bigl(\boldsymbol{x}_{i}\big \vert\nu\bigr) \pi\bigl(\nu \big \vert {b}_{0}\bigr) d\nu \nonumber\\
&=\Pi_{i=1}^{n}\int_{2}^{\infty}\frac{\Gamma \bigl(\frac{\nu+p}{2}\bigr)\pi\bigl(\nu \big \vert {b}_{0}\bigr)d\nu}{(\nu \pi)^{\frac{p}{2}}\Gamma \bigl(\frac{\nu}{2}\bigr)\vert \Sigma \vert ^{\frac{1}{2}} \bigl[ 1 +\frac{(\boldsymbol{x}-\boldsymbol{\mu})^{\top}\Sigma^{-1}(\boldsymbol{x}-\boldsymbol{\mu})}{\nu}\bigr]^{\frac{\nu+p}{2}}},
\end{align}
where $\underline{\boldsymbol{x}}=(\boldsymbol{x}_{1},\cdots,\boldsymbol{x}_{n})$. Maximizing the RHS of (\ref{bayes4}) with respect to $b_0$ is not a simple task, an so we proceed to compute the moment-type estimator of $b_0$ as follows. First notice that
\begin{align*}
E\bigl[(\boldsymbol{X}-\boldsymbol{\mu}) (\boldsymbol{X}-\boldsymbol{\mu}) ^{\top}\big \vert b_{0}\bigr)=& \int_{2}^{\infty} \int_{\mathbb{R}^{p}} (\boldsymbol{x}-\boldsymbol{\mu}) (\boldsymbol{x}-\boldsymbol{\mu}) ^{\top} f\bigl(\boldsymbol{x}\big \vert\nu\bigr) \pi\bigl(\nu \big \vert {b}_{0}\bigr) d\boldsymbol{x} d\nu\nonumber\\
=& \int_{2}^{\infty} \int_{\mathbb{R}^{p}}\frac{(\boldsymbol{x}-\boldsymbol{\mu}) (\boldsymbol{x}-\boldsymbol{\mu}) ^{\top}\Gamma \bigl(\frac{\nu+p}{2}\bigr)}{(\nu \pi)^{\frac{p}{2}}\Gamma \bigl(\frac{\nu}{2}\bigr)\vert \Sigma \vert ^{\frac{1}{2}}\bigl[ 1 +\frac{(\boldsymbol{x}-\boldsymbol{\mu})^{\top}\Sigma^{-1}(\boldsymbol{x}-\boldsymbol{\mu})}{\nu}\bigr]^{\frac{\nu+p}{2}}}
\nonumber\\
&\times \frac{b^{b_0-1}_{0}}{\Gamma(b_0)}(\nu-2)^{b_0-1}\exp\{-b_0(\nu-2)\} d\boldsymbol{x}d\nu\nonumber\\
=&  \int_{2}^{\infty} \Bigl(\frac{\nu}{\nu-2}{\Sigma}\Bigr)\frac{b^{b_0-1}_{0}}{\Gamma(b_0)}(\nu-2)^{b_0-1}\exp\{-b_0(\nu-2)\}d\nu\nonumber\\
=&3-\frac{1}{b_0}.
\end{align*}
Let AD denote the average of diagonal elements of matrix $\big[\sum_{i=1}^{n}(\boldsymbol{x}^{2}_{i}-\boldsymbol{\mu})(\boldsymbol{x}^{2}_{i}-\boldsymbol{\mu})^{\top}/n\bigr]\Sigma^{-1}$, then $b_0$ can be exploited as $b_{0}=1/(3-AD)$.
%\footnote{The second moment of Student's $t$ distribution exists if $\nu>2$.} 
\end{itemize}
Furthermore, one can employ the MH-within-Gibbs sampling technique for drawing sample from full conditional (\ref{fullnu}). To this end, we assume gamma distribution for the prior with PDF $\pi(\nu)= {\cal{G}}\bigl(. \big \vert b_0, b_0\bigr)$. Herein there is just one hyperparameter that we prefer to compute it as $b_{0}=1/(3-AD)$. We further choose $gamma\bigl(b_0, b_0\bigr)$ as the candidate. For other hyperparameters including ${\cal{M}}_{0}, {\cal{S}}_{0}, {\cal{D}}_{0}$, and $\nu_0$, in example above, we set ${\cal{M}}_{0}$ to be the mean of ${\underline{\boldsymbol{y}}}$ and $\nu_0=1$. For ${\cal{S}}_{0}$ and ${\cal{D}}_{0}$ we suggest to consider $\boldsymbol{I}_{p}$. The steps for computing the Bayesian estimator of ${\boldsymbol{\Psi}}$ are given by Algorithm \ref{Bayesian inference for Student's $t$ distribution}. 
\end{example}
\begin{algorithm}
\caption{Bayesian inference for Student's $t$ distribution}
\label{Bayesian inference for Student's $t$ distribution}
\begin{algorithmic}[1]
%\Procedure{}{}     %  \Comment{This is a test}
 %   \State System Initialization
 %   \State Read the value 
%    \If{$condition = True$}
%        \State Do this
%        \If{$Condition \geq 1$}
%        \State Do that
%        \ElsIf{$Condition \neq 5$}
%        \State Do another
%        \State Do that as well
%        \Else
%        \State Do otherwise
%        \EndIf
%    \EndIf
%\State Set $t=1$, read $M$ as the {\it{burn-in}} period, and initiate algorithm with 
%\State randomly selected sample $\boldsymbol{x}^{(0)}$
\State Read $N$, $M$, ${{K}}$, ${\cal{M}}_{0}$, ${\cal{S}}_{0}$, ${\cal{D}}_{0}$, and $\nu_0$;
% , and suggest $\boldsymbol{\theta}^{(0)}=\bigl(\theta^{(0)}_{1},\theta^{(0)}_{2},\theta^{(0)}_{3}\bigr)^{\top}$ arbitrarily;
\State Set $t \leftarrow 0$;
\State Set $\boldsymbol{\mu}^{(0)} \leftarrow {\cal{M}}_{0}$, $\Sigma^{(0)}\leftarrow 1/n\sum_{i=1}^{n}\bigl(\boldsymbol{y}_i-\boldsymbol{\mu}^{(0)} \bigr)\bigl(\boldsymbol{y}_i-\boldsymbol{\mu}^{(0)} \bigr)^{\top}$, ${\nu}^{(0)}=2$,
\State and set $\boldsymbol{\Psi}^{(0)}=\bigl(\boldsymbol{\mu}^{(0)}, \Sigma^{(0)}, \nu^{(0)}\bigr)$;
    \While{$t \leq N$}  %\Comment{put some comments here}
    \State Set $i=1$;
    \While{$i \leq n$}  %\Comment{put some comments here}
    \State Simulate $g_i$ from ${\cal{G}}(a,b)$ where $a=(\nu^{(t)} +p+1)/2$
     and
      $b=$
      \State $\Bigl[\nu^{(t)} /2+ \bigl(\boldsymbol{y}_i-\boldsymbol{\mu}^{(t)} \bigr)^{\top}\bigl[\Sigma^{(t)} \bigr]^{-1}\bigl(\boldsymbol{y}_i-\boldsymbol{\mu}^{(t)} \bigr)\Bigr]/2$;
    \State Set $i \leftarrow i+1$;
        %\Comment{another comment}
        %\State $var3 \leftarrow var4$
             \EndWhile  %\label{roy's loop}
             \State {\bf{end}}
        \State  Generate $\boldsymbol{\mu}^{(t+1)}$ from 
${\cal{N}}_{p}\Bigl({\cal{Q}}_{0}\Bigl[{\cal{S}}^{-1}_{0}{\cal{M}}_{0}+\bigl[\Sigma^{(t)} \bigr]^{-1}\sum_{i=1}^{n}g_{i}\boldsymbol{y}_{i}\Bigr],{\cal{Q}}_{0}\Bigr)$
where
\State ${\cal{Q}}_{0}=\Bigl[{\cal{S}}^{-1}_{0}+\bigl[\Sigma^{(t)} \bigr]^{-1}\sum_{i=1}^{n}g_{i}\Bigr]^{-1}$;
\State  Generate $\Sigma^{(t+1)}$ from
${\cal{IW}}\bigl({\cal{R}}_{0}, \nu_{0}+n\bigr)$ where ${\cal{R}}_{0}={\cal{D}}_{0}$ +
\State $\sum_{i=1}^{n} g_{i} \bigl(\boldsymbol{y}_{i}-\boldsymbol{\mu}^{(t+1)}\bigr)\bigl(\boldsymbol{y}_{i}-\boldsymbol{\mu}^{(t+1)}\bigr)^{\top}$;
        %\Comment{another comment}
        %\State $var3 \leftarrow var4$
    %%%%%%%%%%%%%%%%%%%%%%%%%    
      \State Use the MH-within-Gibbs sampling technique for simulating $\nu^{(t+1)}$
        \State as follows;
     \State Set $k=0$ and $x_{0} \leftarrow \nu_0$
   \While{$k \leq {{K}}$}  %\Comment{put some comments here}
\State $x_{0} \sim {\cal{G}}\bigl(a_{0},b_{0}\bigr)$;
\State Generate $x^{(k+1)}\sim {\cal{G}}\bigl(a_{0},b_{0}\bigr)$;
\State Compute $p\bigl(x^{(k)} \rightarrow x^{(k+1)}\bigr)=\min\bigl\{1,\exp\{ R_{1}- R_{2}\}\bigr\}$
where 
\State $R_1= -n \log \Gamma \bigl(\frac{x^{(k+1)}}{2}\bigr)+\frac{n x^{(k+1)}}{2} \log\bigl(\frac{x^{(k+1)}}{2}\bigr)+ (b_0-1)\log x^{(k+1)}$
\State $-\frac{x^{(k+1)}}{2}\bigl[2b_{0}+\sum_{i=1}^{n}g_{i}\bigr] +\frac{x^{(k+1)}}{2}\sum_{i=1}^{n} \log g_{i}$
\State and
\State $R_2= -n \log \Gamma \bigl(\frac{x^{(k)}}{2}\bigr)+\frac{n x^{(k)}}{2}\log\bigl(\frac{x^{(k)}}{2}\bigr)+ (b_0-1)\log x^{(k)}$
\State $-\frac{x^{(k)}}{2}\bigl[2b_{0}+\sum_{i=1}^{n}g_{i}\bigr] +\frac{x^{(k)}}{2}\sum_{i=1}^{n} \log g_{i}$;
\State Generate $u\sim {\cal{U}}(0, 1)$;
\State If $u<p\bigl(x^{(k)}, x^{(k+1)}\bigr)$ then $x^{(k)} \rightarrow x^{(k+1)}$ and $k \rightarrow k+1$; 
\State Repeat algorithm from line (21);
\State If $k={{K}}$ then $\nu^{(t+1)} \leftarrow x^{(k+1)}$ and stop the algorithm; 
             \EndWhile  %\label{roy's loop}
                       \State {\bf{end}}
    \EndWhile  %\label{roy's loop}
            \State Set $\boldsymbol{\Psi}^{(t+1)} \leftarrow \bigl(\boldsymbol{\mu}^{(t+1)}, \Sigma^{(t+1)}, \nu^{(t+1)}\bigr)$ and $t \leftarrow t+1$;
                                   \State {\bf{end}}
   \State  Sequence $\bigl\{\boldsymbol{\Psi}^{(M)},\boldsymbol{\Psi}^{(M+1)},\cdots,\boldsymbol{\Psi}^{(M)}\bigr\}$ is a sample of size $N-M+1$ from 
   \State posterior with PDF $\pi\bigl(\boldsymbol{\Psi}\vert\underline{\boldsymbol{x}})$;
  \State The Bayesian estimator of $\hat{\boldsymbol{\Psi}}_{B}$ is given by $\frac{1}{N-M+1}\sum_{t=1}^{N-M+1}\boldsymbol{\Psi}^{(t)}$.
  %%\State {\bf{end procedure}}
%\EndProcedure
\end{algorithmic}
\end{algorithm}
The pertaining \verb+R+ code for implementing example above is given as follows. The \verb+R+ code is applied to a sample of $n=500$ realizations have been generated from a bivariate Student's $t$ distribution with parameters $\boldsymbol{\mu}=(0,2)^{\top}$, $\sigma_{11}=1$, $\sigma_{12}=-0.50$, $\sigma_{22}=0.75$, and $\nu=5$. The output of the Gibbs sampler is displayed in Figure \ref{fig6}.
\pagebreak{}
\begin{lstlisting}[style=deltaj]
R > rssg<-function(n, nu, Mu, Sigma)
+		{
+			Dim <- length(Mu)
+			Y <- matrix(NA, nrow = n, ncol = Dim)
+			for (i in 1:n)
+			{
+				Z <- rgamma(1, shape = nu/2, rate = nu/2) 
+				X <- mvrnorm(n = 1, mu = rep(0, Dim), Sigma = Sigma )
+				Y[i, ] <- Mu + X/sqrt(Z)
+			}
+ 			Y
+		}
R >  f.a <- function(x, b0, y)
+		{
+			n <- length(y)
+			-n*lgamma(x/2) + n*(x/2)*log(x/2) - x/2*( 2*b0 + sum(y) ) + x/2*sum( log(y) ) + (b0-1)*log(x)
+		}
R >   fprim.a <- function(x, b0, y)
+		{
+			n <- length(y)
+			-n*digamma(x/2)/2 + n/2 + n/2*log(x/2) - ( 2*b0 + sum(y) )/2 + sum( log(y) )/2 + (b0-1)/x
+		}
R > set.seed(20240529)
R > library(MASS); library(ars)
R > n <- 500; nu <- 5; Mu <- c(0, 2) 
R > Sigma <- matrix( c(1,-0.5,-0.5,0.75), 2, 2)
R > Y <- rssg(n, nu, Mu, Sigma)
R > N <- 5000; M <- 2000; p <- length(Mu)
R > Psi <- matrix(0, nrow = N, ncol = p + 1 + p^2)
R > g <- rate <- rep(0, n)
R > S0Mu <- diag(2); M0Mu <- rep(0, 2); nu0 <- 2 
R > D0 <- matrix(diag(2), 2, 2)
R > for(j in 1:N)
+			{
+				rate <- mahalanobis(x = Y, center = Mu, cov = Sigma )
+				for(i in 1:n) g[i] <- rgamma( 1, shape = (nu + p)/2, rate = rate[i]/2 + nu/2 ) 
+				Sigmainv <- solve( Sigma )
+				dy <- mahalanobis( Y, center = Mu, cov = Sigma )
+				Sum.g <- sum(g)
+				 M_sum <- rowSums( sapply(1:n, function(i) g[i]*c( Y[i, ] ) ) )
+				Q <- solve( solve( S0Mu ) + Sigmainv*Sum.g )
+				Mu <- c( mvrnorm(1, mu = Q%*%( solve( S0Mu )%*%M0Mu + 
+											 Sigmainv%*%M_sum ) , Sigma = Q) )
+				Psi[j, 1:p] <- Mu
+				R0 <- matrix(0, nrow = p, ncol = p) 
+				for(i in 1:n){R0 <- R0 + g[i]*c( Y[i, ] - Mu )%*%t(c( Y[i, ] - Mu ))} 
+				Sigma <- solve(rWishart(1, df = nu0 + n, Sigma = solve(D0 + R0) )[,,1])
+				Psi[j, (p+1):(p + p^2)] <- as.numeric(Sigma)
+				b0 <- 1/(3-(cov(Y)%*%solve(Sigma))[1,1])
+				nu <- ars(1, f.a, fprim.a, x = c(2, 5, 10, 50), m = 4, 
+										 lb = TRUE, xlb = 2, b0 = b0, y = g)
+				Psi[j, p + 1 + p^2] <- nu
+			 }
R > Psi.hat <- apply(Psi[(N - M + 1):N, ], 2, mean)
R > Mu <- Psi.hat[1:p]
R > Sigma <- Psi.hat[(p+1):(p + p^2)]
R > nu <- Psi.hat[p + 1 + p^2]
\end{lstlisting}
%%%%%%%%%%%%%%%%%%%%%%%%%%%%%%%%%%
%%%%%%%%%%%%%%%%%%%%%%%%%%%%%%%%%%
\section{Gibbs sampling in Bayesian paradigm with two latent variables}
\subsection{Heavy-tailed skew models}
The Bayesian paradigm for $\text{SG}_{p,q}(\boldsymbol{\mu},\Sigma,{\Lambda})$ distribution with representation (\ref{ursnrep0}) has been developed by \cite{liseo2013bayesian}. The skew Gaussian model introduced in Section \ref{skew-Gaussian} can be developed for modelling the heavy tailed skewed data. This can be accomplished by involving an extra latent variable $G$. Let 
%$\boldsymbol{X} \sim \text{SG}_{p,q}(\boldsymbol{0}_{p},\Sigma,\boldsymbol{\Lambda})$, then a heavy tailed distribution then random vector
% is said to have skew normal G (SNG) distribution, denoted by $\boldsymbol{Y} \sim \text{SNG}_{p,q}(\boldsymbol{\theta},\boldsymbol{\mu},\Sigma,\Lambda)$, if we have
\begin{align}\label{sngl1}
\boldsymbol{Y} \mathop=\limits^d\boldsymbol{\mu}+\frac{\boldsymbol{X}}{\sqrt{h(G)}},
\end{align}
in which $\boldsymbol{X}\sim \text{SG}_{p,q}(\boldsymbol{0}_{p},\Sigma,{\Lambda})$ and $h(\cdot)$ is a continuous monotone real function. Equivalently,
%%%%This model that is a location-scale $\boldsymbol{X}$. 
\begin{align}\label{sngl2}
\boldsymbol{Y} \mathop=\limits^d \boldsymbol{\mu}+\frac{{\Lambda}\big \vert \boldsymbol{Z}_{0}\big \vert}{\sqrt{h(G)}}+{\Sigma}^{\frac{1}{2}} \frac{\boldsymbol{Z}_1}{\sqrt{h(G)}}.
\end{align}
It is easy to see that random vector $\boldsymbol{Y}$ in (\ref{sngl2}) admits the hierarchy given by   
 \begin{align}\label{hierarchical1}
\boldsymbol{Y}\vert {\boldsymbol{U}}, G &\sim {\cal{N}}_{p}\Bigl(\boldsymbol{\mu}+{\Lambda}{\boldsymbol{U}}, \frac{\Sigma}{h(G)}\Bigr),\nonumber\\
{\boldsymbol{U}}\vert G &\sim {\cal{HN}}_{q}\Bigl(\boldsymbol{0}_{q}, \frac{\boldsymbol{I}_{q}}{h(G)}\Bigr),\nonumber\\
G & \sim f(g \vert\boldsymbol{\theta}),
\end{align}
where the short form ${\cal{HN}}_{q}(\boldsymbol{0}_{q}, \Sigma)$ accounts for the half-normal distribution truncated on ${{\mathbb{R}}}^{q+}$ with a zero-location vector of length $q$ and scale matrix $\Sigma$. 
Regarding $\bigl\{\boldsymbol{y}_1,\cdots,\boldsymbol{y}_n\bigr\}$ as the sequence of observed data and two sequences of latent variables are represented as $\bigl\{\boldsymbol{u}_{1},\cdots,\boldsymbol{u}_{n}\bigr\}$ and $\bigl\{g_{1},\cdots,g_{n}\bigr\}$. Based on hierarchy (\ref{hierarchical1})
% l representation for the complete data under mechanism one as follows.
%\begin{align*}
%\boldsymbol{Y} \big \vert \boldsymbol{U}=\boldsymbol{u}_i, {G}=g_i,&\sim {\cal{N}}_{p}\bigl(\boldsymbol{\mu}+\boldsymbol{\Lambda}\boldsymbol{u}_{i}, h(g_{i})\Sigma  \bigr),\\
%\boldsymbol{U} \big \vert G=g_{i} &\sim {\cal{HN}}_{q}\bigl(\boldsymbol{0}_{q}, h(g_{i})\boldsymbol{I}_{q}\bigr),\\
%G &\sim f_{G}\bigl(g_i \big \vert \boldsymbol{\theta}),\\
%%\boldsymbol{H}_i&\sim\text{Mult}(1,\boldsymbol{\omega}).
%\end{align*}
the complete data log-likelihood is
\begin{align}\label{cdll1}
L_{c}(\boldsymbol{\Psi}\vert {\underline{\boldsymbol{y}}}, \underline{\boldsymbol{u}}, \boldsymbol{g})=& \prod_{i=1}^{n}\Bigl[\boldsymbol{\phi}_{p}\bigl(\boldsymbol{y}_i\big \vert\boldsymbol{\mu}+\boldsymbol{\Lambda} \boldsymbol{u}_{i}, [h(g_{i})]^{-1} \Sigma \bigr) \boldsymbol{\phi}_{q}\bigl( \boldsymbol{u}_{i} \big \vert \boldsymbol{0}_{q}, [h(g_{i})]^{-1} \boldsymbol{I}_{q}\bigr)f_{}\bigl(g_{i} \big \vert\boldsymbol{\theta}\bigr)\Bigr]\nonumber\\
\propto& \prod_{i=1}^{n} \bigl[h(g_{i})\bigr]^{\frac{p+q}{2}}\exp\Bigl\{- \frac{h(g_{i})}{2} \Bigl[ \bigl(\boldsymbol{y}_i-\boldsymbol{\mu}-\boldsymbol{\Lambda} \boldsymbol{u}_{i}\bigr)^{\top}\Sigma^{-1}\bigl(\boldsymbol{y}_i-\boldsymbol{\mu}-\boldsymbol{\Lambda} \boldsymbol{u}_{i}\bigr)\nonumber\\
&+\boldsymbol{u}^{\top}_{i}\boldsymbol{I}_{q}\boldsymbol{u}_{i}\Bigr] \Bigr\} f_{}\bigl(g_{i} \big \vert\boldsymbol{\theta}\bigr),
\end{align}
where $\boldsymbol{\Psi}=(\boldsymbol{\mu},\Sigma,\boldsymbol{\Lambda},\boldsymbol{\theta})$ is the model's parameter vector. We write $\boldsymbol{Y} \sim \text{SGX}_{p,q}(\boldsymbol{\mu},\Sigma,\boldsymbol{\Lambda},\boldsymbol{\theta})$ (the term SG in SGX is a short form for skewed Gaussian and X refers to distribution of latent variable $G$), to denote the family of distributions that follow representation (\ref{sngl2}). For example, if in (\ref{sngl2}) $G$ follows a generalized Lindley distribution, then $\text{SGGL}_{p,q}(\boldsymbol{\mu},\Sigma,\boldsymbol{\Lambda},\boldsymbol{\theta})$ refers to the class of skew Gaussian generalized Lindely distribution in which $\boldsymbol{\theta}=(\omega,\beta,\gamma)^{\top}$.
%while $h(x)=x$ and the mixing variable $G$ follows a gamma distribution.
% with PDF 
%\begin{align}
%f(g \vert \boldsymbol{\theta})=\frac {\beta^{\alpha}}{\Gamma(\alpha)} g^{\alpha-1}\exp\{ -\beta g \},
%\end{align}
%where ${\boldsymbol{\theta}}=(\alpha,\beta)^{\top}$.
%where pdf $f_{G}(g \vert\boldsymbol{\theta})$ is given by Table \ref{tab1}.
\begin{thm}\label{skew-Gaussian-models-thm}
Let $\boldsymbol{Y}\sim \text{SGGL}_{p,q}(\boldsymbol{\mu},\Sigma,{\Lambda},\boldsymbol{\theta})$. The PDF of $\boldsymbol{Y}$ is given by
\begin{align*}
f_{}(\boldsymbol{y} \vert \boldsymbol{\Psi})=&\frac{2^{q}\beta^{1-\frac{p}{2}} \omega^{\frac{p}{2}}}{(\beta+\gamma)} \boldsymbol{t}_{p}\Bigl(\sqrt{\frac{\omega}{\beta}}(\boldsymbol{y}-\boldsymbol{\mu})\bigg \vert \boldsymbol{0},{\Omega},2\omega\Bigr)
\boldsymbol{T}_{q}\Bigl(\boldsymbol{m}\sqrt{\frac{p+2\omega}{2\beta+d(\boldsymbol{y})}}\bigg \vert\boldsymbol{0},{\Delta}, p+2\omega\Bigr)\nonumber\\
&+ \frac{2^{q}\gamma\beta^{-\frac{p}{2}} (\omega+1)^{\frac{p}{2}}}{(\beta+\gamma)} \boldsymbol{t}_{p}\Bigl(\sqrt{\frac{\omega+1}{\beta}}(\boldsymbol{y}-\boldsymbol{\mu})\bigg \vert \boldsymbol{0},{\Omega},2\omega+2\Bigr)\nonumber\\
&\times \boldsymbol{T}_{q}\Bigl(\boldsymbol{m}\sqrt{\frac{p+2\omega+2}{2\beta+d(\boldsymbol{y})}}\bigg \vert\boldsymbol{0},{\Delta}, p+2\omega+2\Bigr),
\end{align*}
where $\boldsymbol{\Psi}=(\boldsymbol{\mu}, \Sigma, {\Lambda},\boldsymbol{\theta})$
, $\boldsymbol{\theta}=(\omega,\beta,\gamma)^{\top}$, $d(\boldsymbol{y})=(\boldsymbol{y}-\boldsymbol{\mu})^{\top}{{\Omega}^{-1}}(\boldsymbol{y}-\boldsymbol{\mu})$,
$\boldsymbol{m}={\Lambda}^{\top}{{\Omega}}^{-1}(\boldsymbol{y}-\boldsymbol{\mu})$,
${\Omega}={\Sigma}+{{\Lambda}}{{\Lambda}}^{\top}$, ${\Delta}=\boldsymbol{I}_{q}-{{\Lambda}}^{\top}{\Omega}^{-1}{{\Lambda}}$,  $\boldsymbol{t}_{p}(.\vert\boldsymbol{0},\Omega, 2\omega)$ is the PDF of a $p$-dimensional Student's $t$ with location parameter $\boldsymbol{0}$, dispersion matrix $\Omega$ and $2\omega$ degrees of freedom, and $\boldsymbol{T}_{q}(.\vert\boldsymbol{0},{\Delta}, p+2\omega+2)$ is the CDF of $q$-dimensional Student's $t$ distribution with location parameter $\boldsymbol{0}$, dispersion matrix ${\Delta}$, and $p+2\omega+2$ degrees of freedom.
\end{thm}
{\bf{Proof:}} 
By definition, based on (\ref{hierarchical1}), we have
\begin{align}\label{ypdf}
f_{\boldsymbol{Y}}(\boldsymbol{y}\big \vert{\Psi})
=&\int_{\mathbb{R}^{q+}} {\int_{0}^{\infty}} f_{}(\boldsymbol{y} \vert \boldsymbol{t}, g,{\Psi}) f_{}(\boldsymbol{t}\vert g) f_{}(g \vert \boldsymbol{\theta}) dg d\boldsymbol{t},\nonumber\\
=&\frac{2^{q} \vert {\Sigma} \vert ^{-\frac{1}{2}}}{ (2\pi)^{\frac{p+q}{2}}}\int_{\mathbb{R}^{q+}} {\int_{0}^{\infty}} g^{\frac{p+q}{2}}
\exp\Bigl\{-\frac{g}{2}(\boldsymbol{y}-\boldsymbol{\mu}-{\Lambda}\boldsymbol{t})^{\top}{{\Sigma}}^{-1}(\boldsymbol{y}-\boldsymbol{\mu}-{\Lambda}\boldsymbol{t})\Bigr\}\nonumber\\
&\times\exp\Bigl\{-\frac{g}{2} \boldsymbol{t}^{\top}\boldsymbol{I}_{q} \boldsymbol{t}\Bigr\}f_{}(g \vert \boldsymbol{\theta})dg d\boldsymbol{t}\nonumber
\end{align}
\begin{align}
=&\frac{2^{q}}{ (2\pi)^{\frac{p+q}{2}}\vert{\Sigma}\vert^{\frac{1}{2}}} \int_{\mathbb{R}^{q+}} {\int_{0}^{\infty}}g^{\frac{p+1}{2}}\exp\Bigl\{-\frac{g}{2}\Bigl[d(\boldsymbol{y})+(\boldsymbol{t}-\boldsymbol{m})^{\top} \Delta^{-1}(\boldsymbol{t}-\boldsymbol{m})\Bigr]
\Bigr\}\nonumber\\
&\times f_{}(g \vert \boldsymbol{\theta})dg d\boldsymbol{t}\nonumber\\
=&\frac{2^{q}\beta^{\alpha+1}}{(\beta+\gamma) \Gamma(\alpha)(2\pi)^{\frac{p+q}{2}}\vert{\Sigma}\vert^{\frac{1}{2}}} \int_{\mathbb{R}^{q+}} {\int_{0}^{\infty}}g^{\frac{p+q+2\alpha}{2}-1}\nonumber\\
&\times \exp\Bigl\{-\frac{g}{2}\Bigl[d(\boldsymbol{y})+(\boldsymbol{t}-\boldsymbol{m})^{\top} \Delta^{-1}(\boldsymbol{t}-\boldsymbol{m})+2\beta\Bigr]
\Bigr\}dg d\boldsymbol{t}\nonumber\\
&+\frac{2^{q}\gamma\beta^{\alpha+1}}{\alpha(\beta+\gamma) \Gamma(\alpha)(2\pi)^{\frac{p+q}{2}}|{\Sigma}|^{\frac{1}{2}}} \int_{\mathbb{R}^{q+}} {\int_{0}^{\infty}}g^{\frac{p+q+2\alpha+2}{2}-1}\nonumber\\
&\times \exp\Bigl\{-\frac{g}{2}\Bigl[d(\boldsymbol{y})+(\boldsymbol{t}-\boldsymbol{m})^{\top} \Delta^{-1}(\boldsymbol{t}-\boldsymbol{m})+2\beta\Bigr]
\Bigr\}dg d\boldsymbol{t}\nonumber.
\end{align}
It follows that 
\begin{align}
f_{\boldsymbol{Y}}(\boldsymbol{y}\big \vert {\Psi})=&
% \frac{2^{q}\beta^{\alpha+1}}{(\beta+\gamma) \Gamma(\alpha)(2\pi)^{\frac{p+q}{2}}|{\Sigma}|^{\frac{1}{2}}}
\frac{2^{q}\beta^{\alpha+1} \Gamma\bigl(\frac{p+q+2\alpha}{2}\bigr)|{\Sigma}|^{-\frac{1}{2}}}{(\beta+\gamma)\Gamma(\alpha)(2\pi)^{\frac{p+q}{2}}}  \int_{\mathbb{R}^{q+}}
\frac{d\boldsymbol{t}}{
\Bigl[\frac{2\beta+d(\boldsymbol{y})+(\boldsymbol{t}-\boldsymbol{m})^{\top} \Delta^{-1}(\boldsymbol{t}-\boldsymbol{m})}{2}\Bigr]^{\frac{p+q+2\alpha}{2}}}\nonumber\\
&+\frac{2^{q}\gamma \beta^{\alpha+1} \Gamma\bigl(\frac{p+q+2\alpha+2}{2}\bigr)\vert{\Sigma}\vert^{-\frac{1}{2}}}{\alpha(\beta+\gamma)\Gamma(\alpha)(2\pi)^{\frac{p+q}{2}}}  \int_{\mathbb{R}^{q+}}
\frac{d\boldsymbol{t}}{
\Bigl[\frac{2\beta+d(\boldsymbol{y})+(\boldsymbol{t}-\boldsymbol{m})^{\top} \Delta^{-1}(\boldsymbol{t}-\boldsymbol{m})}{2}\Bigr]^{\frac{p+q+2\alpha+2}{2}}}\nonumber\\
=&I_1+I_2.
\end{align}
Using a change of variable  $\boldsymbol{v}=(\boldsymbol{t}-\boldsymbol{m})\sqrt{p+2\alpha}/\sqrt{2\beta+d(\boldsymbol{y})}$, we can write
\begin{align}
I_1=&\frac{2^{q}\beta^{\alpha+1} \Gamma\bigl(\frac{p+q+2\alpha}{2}\bigr)\vert{\Sigma}\vert^{-\frac{1}{2}}}{(\beta+\gamma)\Gamma(\alpha)(2\pi)^{\frac{p+q}{2}} }  \frac{2^{\frac{p+q+2\alpha}{2}}(p+2\alpha)^{-\frac{q}{2}}}{
\Bigl[2\beta+d(\boldsymbol{y})\Bigr]^{\frac{p+2\alpha}{2}}{}}
{\int_{-\boldsymbol{\infty}}^{\boldsymbol{m}_{0}}} \frac{d\boldsymbol{v}}{\Bigl[1+\frac{\boldsymbol{v}^{\top}\Delta^{-1}\boldsymbol{v}}{p+2\alpha}\Bigr]^{\frac{p+q+2\alpha}{2}}},\nonumber\\
=&\frac{2^{q}\beta^{\alpha+1} \Gamma\bigl(\frac{p+2\alpha}{2}\bigr) \vert{\Sigma}\vert^{-\frac{1}{2}}}{(\beta+\gamma)\Gamma(\alpha)(2\pi)^{\frac{p}{2}}}  \frac{2^{\frac{p+2\alpha}{2}} \vert{\Delta}\vert^{\frac{1}{2}}}{
\Bigl[2\beta+d(\boldsymbol{y})\Bigr]^{\frac{p+2\alpha}{2}}{}}
\boldsymbol{T}_{q}\Bigl(\boldsymbol{m}\sqrt{\frac{p+2\alpha}{2\beta+d(\boldsymbol{y})}}\Big \vert\boldsymbol{0},\Delta ,p+2\alpha\Bigr),\nonumber
\end{align}
where $\boldsymbol{m}_{0}=\boldsymbol{m}\sqrt{p+2\alpha}/\sqrt{2\beta+d(\boldsymbol{y})}$. It follows that
\begin{align}\label{iii}
I_1=&\frac{2^{q}\beta^{\alpha+1} 2^{\frac{p+2\alpha}{2}}(2\alpha)^{\frac{p}{2}}{\vert\Omega}\vert^{\frac{1}{2}}\vert{\Delta}\vert^{\frac{1}{2}}}{(\beta+\gamma)2^{\frac{p}{2}}\vert{\Sigma}\vert^{\frac{1}{2}}} 
\frac{\Gamma\bigl(\frac{p+2\alpha}{2}\bigr)}{ \pi^{\frac{p}{2}} (2\alpha)^{\frac{p}{2}} \Gamma(\alpha){|\Omega}|^{\frac{1}{2}}\Bigl[2\beta+d(\boldsymbol{y})\Bigr]^{\frac{p+2\alpha}{2}}}\nonumber\\
&\times \boldsymbol{T}_{q}\Bigl(\boldsymbol{m}\sqrt{\frac{p+2\alpha}{2\beta+d(\boldsymbol{y})}}\Big \vert\boldsymbol{0},\Delta ,p+2\alpha\Bigr)\nonumber\\
=&\frac{2^{q}\beta^{1-\frac{p}{2}} \alpha^{\frac{p}{2}}}{(\beta+\gamma)} \boldsymbol{t}_{p}\Bigl(\sqrt{\frac{\alpha}{\beta}}(\boldsymbol{y}-\boldsymbol{\mu})\Big \vert \boldsymbol{0},{\Omega},2\alpha\Bigr)
\boldsymbol{T}_{q}\Bigl(\boldsymbol{m}\sqrt{\frac{p+2\alpha}{2\beta+d(\boldsymbol{y})}}\Big \vert\boldsymbol{0},\Delta ,p+2\alpha\Bigr).
\end{align}
It should be noted that for obtaining the term in the RHS of (\ref{iii}), we use this fact that $\vert {\Delta} \vert^{\frac{1}{2}} \Omega \vert ^{\frac{1}{2}}/\vert {\Sigma}\vert^{\frac{1}{2}}=1$. Likewise, we can obtain $I_2$ as
%%%
\begin{align*}
I_2=&\frac{2^{q}\gamma \beta^{\alpha+1} \Gamma\bigl(\frac{p+2\alpha+2}{2}\bigr)}{\alpha(\beta+\gamma)\Gamma(\alpha)(2\pi)^{\frac{p}{2}} \vert{\Sigma}\vert^{\frac{1}{2}}}  \frac{2^{\frac{p+2\alpha+2}{2}} \vert{\Delta}\vert^{\frac{1}{2}}}{
\Bigl[2\beta+d(\boldsymbol{y})\Bigr]^{\frac{p+2\alpha+2}{2}}{}}\nonumber\\
&\times \boldsymbol{T}_{q}\Bigl(\boldsymbol{m}\sqrt{\frac{p+2\alpha+2}{2\beta+d(\boldsymbol{y})}}\Big \vert\boldsymbol{0},\Delta, p+2\alpha\Bigr),\nonumber\\
=&\frac{2^{q} \gamma\beta^{\alpha+1} 2^{\frac{p+2\alpha+2}{2}}(2\alpha+2)^{\frac{p}{2}}{\vert\Omega}\vert^{\frac{1}{2}}\vert{\Delta}\vert^{\frac{1}{2}}}{(\beta+\gamma)2^{\frac{p}{2}}\vert{\Sigma}\vert^{\frac{1}{2}}} \frac{\Gamma\bigl(\frac{p+2\alpha+2}{2}\bigr)\Bigl[2\beta+d(\boldsymbol{y})\Bigr]^{-\frac{p+2\alpha+2}{2}}}{ \pi^{\frac{p}{2}} (2\alpha+2)^{\frac{p}{2}}\alpha \Gamma(\alpha)\vert \Omega \vert^{\frac{1}{2}}}\nonumber\\
&\times \boldsymbol{T}_{q}\Bigl(\boldsymbol{m}\sqrt{\frac{p+2\alpha+2}{2\beta+d(\boldsymbol{y})}}\Big \vert\boldsymbol{0},\Delta, p+2\alpha+2\Bigr)\nonumber\\
=&\frac{2^{q}\gamma\beta^{-\frac{p}{2}} (\alpha+1)^{\frac{p}{2}}}{(\beta+\gamma)} \boldsymbol{t}_{p}\Bigl(\sqrt{\frac{\alpha+1}{\beta}}(\boldsymbol{y}-\boldsymbol{\mu})\Big \vert \boldsymbol{0},{\Omega},2\alpha+2\Bigr)\nonumber\\
&\times \boldsymbol{T}_{q}\Bigl(\boldsymbol{m}\sqrt{\frac{p+2\alpha+2}{2\beta+d(\boldsymbol{y})}}\Big \vert\boldsymbol{0},\Delta, p+2\alpha+2\Bigr).
\end{align*}
The proof is complete.
\begin{cor}
If in (\ref{hierarchical1}) $G\sim {\cal{G}}(\nu/2,\nu/2)$, then $\boldsymbol{Y}\sim \text{SNG}_{p,q}(\nu,\boldsymbol{\mu},\Sigma,{\Lambda})$ follows skew Student's $t$ distribution with PDF given by
\begin{align*}
f_{}(\boldsymbol{y} \vert \boldsymbol{\Psi})=&{2^q} \boldsymbol{t}_{p}\bigl(\boldsymbol{y}\big \vert \boldsymbol{\mu}, {\Omega}, \nu\bigr)
\boldsymbol{T}_{q}\Bigl(\boldsymbol{m}\sqrt{\frac{\nu+p}{\nu+d(\boldsymbol{y})}}\bigg \vert\boldsymbol{0},{\Delta}, p+\nu\Bigr),\nonumber
\end{align*}
where $\boldsymbol{\Psi}=(\nu,\boldsymbol{\mu}, \Sigma, {\Lambda})^{}$. 
\end{cor}
\begin{thm}
Let $\boldsymbol{Y}\sim \text{SNGIG}_{p,q}(\boldsymbol{\mu},\Sigma,{\Lambda},\boldsymbol{\theta})$. The PDF of $\boldsymbol{Y}$ is given by
\begin{align*}
f_{}(\boldsymbol{y} \vert \boldsymbol{\Psi})=&2^q \boldsymbol{h}_{p}\bigl(\boldsymbol{y}\big \vert \boldsymbol{\mu},{\Omega},\tau,\chi,\chi\bigr)
\boldsymbol{H}_{q}\Bigl(\boldsymbol{m}\Bigl[\frac{\chi}{\chi+d(\boldsymbol{y})}\Bigr]^{\frac{1}{4}}\Big \vert\boldsymbol{0},{\Delta}, \tau-\frac{p}{2}, \kappa, \kappa\Bigr),\nonumber
\end{align*}
where $\boldsymbol{\Psi}=(\boldsymbol{\mu}, \Sigma, \boldsymbol{\Lambda},\boldsymbol{\theta})$, $\boldsymbol{\theta}=(\tau,\chi)^{\top}$, $\kappa=\chi\sqrt{1+d(\boldsymbol{y})/\chi}$, $\boldsymbol{h}_{p}(.\big \vert \boldsymbol{\mu},{\Omega},\tau,\chi,\chi )$ and $\boldsymbol{H}_{q}\bigl(\cdot\big \vert\boldsymbol{0},{\Delta},  \tau-p/2, \kappa, \kappa\bigr)$ are the PDF and CDF of a $p$- and $q$-dimensional symmetric hyperbolic distribution, respectively.
\end{thm}
\par
{\bf{Proof:}} See \cite{murray2017hidden}.
\begin{landscape}
\begin{table}
\small
\centering
\caption{Family of heavy-tailed skew models}
\begin{tabular}{l|l|l|p{7.4cm}}
\hline
Characteristics & $\text{SGGL}_{p,q}(\boldsymbol{\mu},\Sigma,{\Lambda},\boldsymbol{\theta})$ &  $\text{SGG}_{p,q}(\boldsymbol{\mu},\Sigma,{\Lambda},\boldsymbol{\theta})$ &  $\text{SGGIG}_{p,q}(\boldsymbol{\mu},\Sigma,{\Lambda},\boldsymbol{\theta})$\\
\hline
family of $G$ &Generalized Lindley (${\cal{GL}}(\omega,\beta,\gamma)$)& ${\cal{G}}(\eta, \zeta)$& generalized inverse Gaussian (${\cal{GIG}}(\tau, \chi,\chi)$)
% $\text{GL}(\omega,\beta,\gamma)$ & $\text{gamma}(\eta, \zeta)$
\\
{parameter}&${\boldsymbol{\theta}}=(\omega,\beta,\gamma)^{\top}$&${\boldsymbol{\theta}}=(\eta, \zeta)^{\top}$&${\boldsymbol{\theta}}=(\tau, \chi)^{\top}$\\
PDF of $G$ &$f_{}(g \vert \boldsymbol{\theta})=\frac{\beta^{\omega+1} g^{\omega-1}(\omega+\gamma g)}{(\beta+\gamma)\Gamma(\omega+1)}\exp\bigl\{-\beta g\bigr\}$&$f_{}(g \vert \boldsymbol{\theta})=\frac {\zeta^{\eta}}{\Gamma(\eta)} g^{\eta-1}\exp\{ -\zeta g \}$ &
$f_{}(g\vert\boldsymbol{\theta}) =\frac {g^{\tau-1}}{2{\cal{K}}_{\tau}({\chi})}\exp\bigl\{-\frac{\chi g}{2}-\frac{\chi}{2g}\bigr\}$ \\
PDF& $f(\boldsymbol{y} \vert \boldsymbol{\Psi})=\frac{2^{q}\beta^{1-\frac{p}{2}} \omega^{\frac{p}{2}}}{(\beta+\gamma)} \boldsymbol{t}_{p}\Bigl(\sqrt{\frac{\omega}{\beta}}(\boldsymbol{y}-\boldsymbol{\mu})\bigg \vert \boldsymbol{0},{\Omega},2\omega\Bigr)$&
$f(\boldsymbol{y} \vert \boldsymbol{\Psi})=2^{q} \bigl(\frac{\eta}{\zeta}\bigr)^{\frac{p}{2}}$&
$f(\boldsymbol{y} \vert \boldsymbol{\Psi})={2^q} \boldsymbol{h}_{p}\Bigl(\boldsymbol{y}\big \vert \boldsymbol{\mu},{\Omega},\tau,\chi,\chi\Bigr)$
\\
&$\times\boldsymbol{T}_{q}\Bigl(\boldsymbol{m}\sqrt{\frac{p+2\omega}{2\beta+d(\boldsymbol{y})}}\bigg \vert\boldsymbol{0},{\Delta}, p+2\omega\Bigr)$
&$\times  \boldsymbol{t}_{p}\Bigl(\sqrt{\frac{\eta}{\zeta}}(\boldsymbol{y}-\boldsymbol{\mu})\bigg \vert \boldsymbol{0},{\Omega},2\eta\Bigr)$&
$\times \boldsymbol{H}_{q}\Bigl(\boldsymbol{m}\Bigl[\frac{\chi}{\chi+d(\boldsymbol{y})}\Bigr]^{\frac{1}{4}}\Big \vert\boldsymbol{0},\Delta,
\tau-\frac{p}{2},\chi\sqrt{1+\frac{d(\boldsymbol{y})}{\chi}},$\\
&$+ \frac{2^{q}\gamma\beta^{-\frac{p}{2}} (\omega+1)^{\frac{p}{2}}}{(\beta+\gamma)} \boldsymbol{t}_{p}\Bigl(\sqrt{\frac{\omega+1}{\beta}}(\boldsymbol{y}-\boldsymbol{\mu})\bigg \vert \boldsymbol{0},{\Omega},2\omega+2\Bigr)$
&$\times\boldsymbol{T}_{q}\Bigl(\boldsymbol{m}\sqrt{\frac{p+2\eta}{2\zeta+d(\boldsymbol{y})}}\bigg \vert\boldsymbol{0},{\Delta}, p+2\eta\Bigr)$&$\chi\sqrt{1+\frac{d(\boldsymbol{y})}{\chi}}\biggr)$
\\
&$\times \boldsymbol{T}_{q}\Bigl(\boldsymbol{m}\sqrt{\frac{p+2\omega+2}{2\beta+d(\boldsymbol{y})}}\bigg \vert\boldsymbol{0},{\Delta} ,p+2\omega+2\Bigr)$&&\\
Representation &$\boldsymbol{Y} \mathop=\limits^d \boldsymbol{\mu}+\frac{{{\Lambda}} \vert \boldsymbol{Z}_{0} \vert}{\sqrt{GL}}+{\Sigma}^{\frac{1}{2}} \frac{\boldsymbol{Z}_1}{\sqrt{GL}}$ & $\boldsymbol{Y} \mathop=\limits^d \boldsymbol{\mu}+\frac{{{\Lambda}} \vert \boldsymbol{Z}_{0} \vert}{\sqrt{G}}+{\Sigma}^{\frac{1}{2}} \frac{\boldsymbol{Z}_1}{\sqrt{G}}$&
$\boldsymbol{Y} \mathop=\limits^d \boldsymbol{\mu}+\frac{{\Lambda}}{\sqrt{GIG}}\big \vert \boldsymbol{Z}_{0}\big \vert+{\Sigma}^{\frac{1}{2}} \frac{\boldsymbol{Z}_1}{\sqrt{GIG}}$\\
%4&{$\text{gamma}(\nu/2.\nu/2)$}&${{\theta}}=\nu$&
%$f_G(g\vert{{\theta}}) = \frac{1}{\Gamma\bigl(\frac{\nu}{2}\bigr)} \bigl( \frac{\nu}{2}\bigr)^{\frac{\nu}{2}}g^{\frac{\nu}{2}-1}\exp\bigl\{ -\frac{\nu}{2}g \bigr\}$&$\text{SGG1}_{p,q}({\theta},\boldsymbol{\mu},\Sigma,\boldsymbol{\Lambda})$
%\\
%5&{$\text{PT}(p/2,\alpha)$}&${{\theta}}=\alpha$&
%$f_{G}(g \vert {\theta})=\frac {\Gamma\bigl(1+\frac{p}{2}\bigr)g^{-\frac{p}{2}}}{\Gamma\bigl(1+\frac{p}{\alpha}\bigr)}f_P(g\vert\alpha)$&$\text{SGPT2}_{p,q}({\theta},\boldsymbol{\mu},\Sigma,\boldsymbol{\Lambda})$
%\\
%6&{$\text{S}\bigl(\alpha/2,1,\bigl(\cos(\pi \alpha/4)\bigr)^{2/\alpha},0\bigr)$}&$\theta=\alpha$&
%$f_{G}(g \vert {\theta})=f_P(g\vert\alpha)$&$\text{SGP2}_{p,q}({\theta},\boldsymbol{\mu},\Sigma,\boldsymbol{\Lambda})$\\
\hline
\end{tabular}
%\item[1] $\rho=\chi\sqrt{1+\frac{d(\boldsymbol{y})}{\chi}}$
\label{skew-Gaussian-models-tab}
\end{table}
\end{landscape}
 \subsubsection{Full conditional $G\vert(\boldsymbol{Y}=\boldsymbol{y}_{i},\cdots)$}
 %${G} \vert (\boldsymbol{Y}=\boldsymbol{y}_{i}, \boldsymbol{U}=\boldsymbol{u}_{i},\boldsymbol{\Psi})$}
Implementing the Gibbs sampling for making Bayesian inference about parameters of the heavy-tailed skew models represented earlier in (\ref{hierarchical1}) involves sampling from the full conditionals of variables $G$ and $\boldsymbol{U}$. Herein, we represent the full conditionals ${G} \vert (\boldsymbol{Y}=\boldsymbol{y}_{i}, \boldsymbol{U}=\boldsymbol{u}_{i},\boldsymbol{\Psi})$ and $\boldsymbol{U}\vert (\boldsymbol{Y}=\boldsymbol{y}_{i}, {G}={g}_{i},\boldsymbol{\Psi})$, and describe how to sample from it under three scenarios shown in Table \ref{skew-Gaussian-models-tab}.
\begin{enumerate}[label=\roman*.]
 \item{\bf{Scenario 1 (mixing random variable is ${\cal{GL}}(\omega,\beta,\gamma)$):}} 
By definition we have
\begin{align}\label{mixfull4}
\pi_{}\bigl(g_i\big \vert \boldsymbol{y}_{i},\boldsymbol{u}_{i},\boldsymbol{\Psi}\bigr)=&\frac{f\bigl(\boldsymbol{y}_{i}, \boldsymbol{u}_{i},g_i\big \vert \boldsymbol{\Psi}\bigr)}{f\bigl(\boldsymbol{y}_{i}, \boldsymbol{u}_{i}\big \vert\boldsymbol{\Psi} \bigr)}\nonumber\\
\propto&
f\bigl(\boldsymbol{y}_{i}, \boldsymbol{u}_{i},g_i\big \vert \boldsymbol{\Psi}\bigr)\nonumber\\
=&f\bigl(\boldsymbol{y}_{i}\big \vert  \boldsymbol{u}_{i},g_i\big \vert \boldsymbol{\Psi}\bigr)f\bigl(\boldsymbol{u}_{i} \big \vert g_i\bigr)f\bigl(g_{i} \big \vert \boldsymbol{\theta}\bigr)\nonumber\\
\propto&  \frac{\beta^{\omega+1} }{(\beta+\gamma)\Gamma(\omega+1)} \Bigl[h(g_{i})\Bigr]^{\frac{p+q}{2}} \Bigl[g_{i}^{\omega -1}(\omega+\gamma g_{i})\Bigr]\exp\Bigl\{-\beta g_{i}\nonumber\\
& - \frac{h(g_{i})}{2} \Bigl[ \bigl(\boldsymbol{y}_i-\boldsymbol{\mu}-\boldsymbol{\Lambda} \boldsymbol{u}_{i}\bigr)^{\top}\Sigma^{-1}\bigl(\boldsymbol{y}_i-\boldsymbol{\mu}-\boldsymbol{\Lambda} \boldsymbol{u}_{i}\bigr)+\boldsymbol{u}^{\top}_{i}\boldsymbol{u}_{i}\Bigr] \Bigr\}.
\end{align}
Setting $h(g_i)=g_i$ in the RHS of (\ref{mixfull4}), it turns out that
\begin{align}\label{mixfull22}
\pi_{}(g_i\vert \boldsymbol{y}_{i},\boldsymbol{u}_{i},\boldsymbol{\Psi})
\propto& g_{i}^{\frac{p+q}{2}+\omega -1}(\omega+\gamma g_{i})\exp\Bigl\{-g_{i} \Bigl[\beta +\frac{\boldsymbol{u}^{\top}_{i}\boldsymbol{u}_{i}}{2} +\bigl(\boldsymbol{y}_i-\boldsymbol{\mu}-{\Lambda} \boldsymbol{u}_{i}\bigr)^{\top} \nonumber\\&\Sigma^{-1}\bigl(\boldsymbol{y}_i-\boldsymbol{\mu}-{\Lambda} \boldsymbol{u}_{i}\bigr)/2\Bigr] \Bigr\}.
\end{align}
Comparing the RHS of (\ref{mixfull22}) with the PDF $f_{}(g \vert \boldsymbol{\theta})$ given in the first column of Table \ref{skew-Gaussian-models-tab}. Then more algebra shows that 
\begin{align}\label{mixfull23}
G \vert \bigl(\boldsymbol{Y}=\boldsymbol{y}_{i},\boldsymbol{U}=\boldsymbol{u}_{i},\boldsymbol{\Psi}\bigr) \sim {\cal{GL}}\bigl(\omega^{*}, \beta^{*}, \gamma^{*}\bigr),
\end{align}
where $\omega^{*} = (p + q)/2 +\omega$, $\beta^{*} =\beta + \bigl(\boldsymbol{y}_i-\boldsymbol{\mu}-{\Lambda}\boldsymbol{u}_{i}\bigr)^{\top}\Sigma^{-1}\bigl(\boldsymbol{y}_i-\boldsymbol{\mu}-\boldsymbol{\Lambda} \boldsymbol{u}_{i}\bigr)/2+\boldsymbol{u}^{\top}_{i}\boldsymbol{u}_{i}/2$, and $\gamma^{*}=\gamma/\omega$.
\item {\bf{Scenario 2 (mixing random variable is ${\cal{G}}(\eta,\zeta)$):}} Following the same argument as in Scenario 1 when $\gamma=0$, $\omega=\eta$, and $\beta=\zeta$, then it turns out from RHS of (\ref{mixfull23}) that 
\begin{align}\label{mixfull24}
G \vert \bigl(\boldsymbol{Y}=\boldsymbol{y}_{i},\boldsymbol{U}=\boldsymbol{u}_{i},\boldsymbol{\Psi}\bigr) \sim {\cal{G}}\bigl(\eta^{*}, \zeta^{*}\bigr),
\end{align}
 where $\eta^{*} =(p + q)/2+\eta$ and $\zeta^{*} =2\zeta + \bigl(\boldsymbol{y}_i-\boldsymbol{\mu}-{\Lambda}\boldsymbol{u}_{i}\bigr)^{\top}\Sigma^{-1}\bigl(\boldsymbol{y}_i-\boldsymbol{\mu}-{\Lambda} \boldsymbol{u}_{i}\bigr)/2+\boldsymbol{u}^{\top}_{i}\boldsymbol{u}_{i}/2$. 
\item {\bf{Scenario 3 (mixing random variable is ${\cal{GIG}}(\tau,\chi,\chi)$):}} If we let  $f_{}\bigl(g_i\big \vert,\boldsymbol{\theta}\bigr)$ in the RHS of (\ref{cdll1}) to be PDF of GIG distribution whose form is given in third column of Table \ref{skew-Gaussian-models-tab}, we have
\begin{align}\label{mixfull11}
\pi_{}\bigl(g_i \big \vert \boldsymbol{y}_{i},\boldsymbol{u}_{i},\boldsymbol{\Psi}\big)
\propto&  
\biggl[\frac {g_{i}^{\tau-1}}{2{\cal{K}}_{\tau}({\chi})}\exp\Bigl\{-\frac { \chi g_{i}}{2}-\frac{\chi}{2g_{i}}\Bigr\}\biggr] \bigl[h(g_{i})\bigr]^{\frac{p+q}{2}}\exp\Bigl\{-\frac{h(g_{i})}{2}\nonumber\\
&\times \Bigl[\bigl(\boldsymbol{y}_i-\boldsymbol{\mu}-\boldsymbol{\Lambda} \boldsymbol{u}_{i}\bigr)^{\top}\Sigma^{-1}\bigl(\boldsymbol{y}_i-\boldsymbol{\mu}-\boldsymbol{\Lambda} \boldsymbol{u}_{i}\bigr)+\boldsymbol{u}^{\top}_{i}\boldsymbol{I}_{q}\boldsymbol{u}_{i}\Bigr] \Bigr\},
\end{align}
If in the RHS of (\ref{mixfull11}) we let $h(g_i)=g_i$, then it turns out that
\begin{align}\label{mixfull12}
\pi_{}\bigl(g_i \big \vert \boldsymbol{y}_{i},\boldsymbol{u}_{i},\boldsymbol{\Psi}\big)
\propto& g_{i}^{\frac{p+q}{2}+\tau-1}\exp\Bigl\{- \epsilon\frac{g_{i}}{2} -\frac{\chi}{2g_{i}}\Bigr\}.
\end{align}
Evidently,
\begin{align*}
{G} \vert \bigl(\boldsymbol{Y}=\boldsymbol{y}_{i},\boldsymbol{U}=\boldsymbol{u}_{i},\boldsymbol{\Psi}\bigr)\sim {\cal{GIG}}\Bigl(\frac{p}{2}+\frac{q}{2}+\tau, \epsilon, \chi\Bigr),
\end{align*}
where $\epsilon=\bigl[ \bigl(\boldsymbol{y}_i-\boldsymbol{\mu}-{\Lambda} \boldsymbol{u}_{i}\bigr)^{\top}\Sigma^{-1}\bigl(\boldsymbol{y}_i-\boldsymbol{\mu}-{\Lambda} \boldsymbol{u}_{i}\bigr)+\boldsymbol{u}^{\top}_{i}\boldsymbol{u}_{i}+\chi\bigr]$.
\end{enumerate}
\subsubsection{Full conditional $\boldsymbol{U}\vert(\boldsymbol{Y}=\boldsymbol{y}_{i},\cdots)$}
 By definition we have
\begin{align*}
\pi_{}\bigl(\boldsymbol{u}_{i}\big \vert \boldsymbol{y}_{i},g_i,\boldsymbol{\Psi}\bigr)=&\frac{f\bigl(\boldsymbol{y}_{i}, \boldsymbol{u}_{i},g_i\big \vert \boldsymbol{\Psi}\bigr)}{f\bigl(\boldsymbol{y}_{i}, {g}_{i}\big \vert\boldsymbol{\Psi} \bigr)}\times \mathbb{I}_{\mathbb{R}^{q+}}(\boldsymbol{u}_{i})\nonumber\\
\propto&
f\bigl(\boldsymbol{y}_{i}, \boldsymbol{u}_{i},g_i\big \vert \boldsymbol{\Psi}\bigr)\times\mathbb{I}_{\mathbb{R}^{q+}}(\boldsymbol{u}_{i})\nonumber\\
=&f\bigl(\boldsymbol{y}_{i}\big \vert  \boldsymbol{u}_{i},g_i\big \vert \boldsymbol{\Psi}\bigr)f\bigl(\boldsymbol{u}_{i} \big \vert g_i\bigr)f\bigl(g_{i} \big \vert \boldsymbol{\theta}\bigr)\times\mathbb{I}_{\mathbb{R}^{q+}}(\boldsymbol{u}_{i})\nonumber\\
\propto&f\bigl(\boldsymbol{y}_{i}\big \vert  \boldsymbol{u}_{i},g_i\big \vert \boldsymbol{\Psi}\bigr)f\bigl(\boldsymbol{u}_{i} \big \vert g_i\bigr)\times\mathbb{I}_{\mathbb{R}^{q+}}(\boldsymbol{u}_{i})\nonumber\\
\propto& -\exp\Bigl\{ \frac{h(g_{i})}{2} \Bigl[ \bigl(\boldsymbol{y}_i-\boldsymbol{\mu}-\boldsymbol{\Lambda} \boldsymbol{u}_{i}\bigr)^{\top}\Sigma^{-1}\bigl(\boldsymbol{y}_i-\boldsymbol{\mu}-\boldsymbol{\Lambda} \boldsymbol{u}_{i}\bigr)\nonumber\\
&+\boldsymbol{u}^{\top}_{i}\boldsymbol{u}_{i}\Bigr] \Bigr\}\times\mathbb{I}_{\mathbb{R}^{q+}}(\boldsymbol{u}_{i}),
\end{align*}
where $\mathbb{I}_{{\cal{S}}}(x)$ is defined in (\ref{indicator-function}). We use the fact that
\begin{align}
&\bigl(\boldsymbol{y}_i-\boldsymbol{\mu}-\boldsymbol{\Lambda} \boldsymbol{u}_{i}\bigr)^{\top}\Sigma^{-1}\bigl(\boldsymbol{y}_i-\boldsymbol{\mu}-\boldsymbol{\Lambda} \boldsymbol{u}_{i}\bigr)+\boldsymbol{u}^{\top}_{i}\boldsymbol{u}_{i}\nonumber\\
&= d\bigl(\boldsymbol{y}_i\bigr)+
\bigl(\boldsymbol{u}_i-\boldsymbol{m}\bigr)^{\top}\Delta^{-1}\bigl(\boldsymbol{u}_i-\boldsymbol{m}\bigr),
\end{align}
where $d\bigl(\boldsymbol{y}_i\bigr)$, $\boldsymbol{m}$, and $\Delta$ are defined in Theorem \ref{skew-Gaussian-models-thm}. Hence,
\begin{align}\label{mixfull7}
\pi_{}\bigl(\boldsymbol{u}_{i}\big \vert \boldsymbol{y}_{i},g_i,\boldsymbol{\Psi}\bigr)
\propto&\exp\Bigl\{ - \frac{h(g_{i})}{2} \bigl(\boldsymbol{u}_i-\boldsymbol{m}\bigr)^{\top}\Delta^{-1}\bigl(\boldsymbol{u}_i-\boldsymbol{m}\bigr) \Bigr\}\times \mathbb{I}_{\mathbb{R}^{q+}}(\boldsymbol{u}_{i}).
\end{align}
This means that
 \begin{align}\label{mixfull8}
\boldsymbol{U} \vert \bigl(\boldsymbol{Y}=\boldsymbol{y}_{i},G={g}_{i},\boldsymbol{\Psi}\bigr) \sim {\cal{TN}}_{q}\Bigl(\boldsymbol{m}, \frac{\Delta}{h(g_{i})}\boldsymbol{0},\boldsymbol{\infty}\Bigr),
\end{align}
where ${\cal{TN}}_{q}(\cdot,\cdot,\cdot,\cdot)$ refers to the family of truncated $q$-dimensional Gaussian distributions introduced in Sub-section \ref{first-moment-truncated-multivariate-Gaussian-distribution}. 

\subsubsection{Bayesian inference for skew Student's $t$ distribution}
As it is seen from second column of Table \ref{skew-Gaussian-models-tab}, the skew Gaussian gamma distribution in which $G\sim{\cal{G}}(\alpha,\beta)$ specializes, for $\alpha=\beta=\nu/2$, to skew Student's $t$ distribution with $\nu$ degrees of freedom for which a Bayesian paradigm has been developed by \citep{parisi2018objective}. Herein, we compute the Bayesian estimators for parameters of $\text{SGG}_{p,q}(\boldsymbol{\mu},\Sigma,\boldsymbol{\Lambda},\boldsymbol{\theta})$ in which $\boldsymbol{\theta}=(\alpha,\beta)^{\top}$. To this end, first, we simulate from full conditionals $g_i\big \vert \bigl(\boldsymbol{\mu}, {\Sigma}, \nu, \boldsymbol{u}_{i}, {\boldsymbol{y}}_{i} \bigr)$ and $\boldsymbol{u}_{i}\big \vert \bigl(\boldsymbol{\mu}, {\Sigma}, \nu,{{g}}_{i}, {\boldsymbol{y}}_{i} \bigr)$ (for $i=1\cdots,n$) in order to construct the complete data $\{(\boldsymbol{y}_{1},\boldsymbol{u}_{1},g_{1}),\cdots,(\boldsymbol{y}_{n},\boldsymbol{u}_{n},g_{n})\}$. It follows from the RHS of (\ref{mixfull24}) that 
\begin{align}\label{mixfull9}
\pi\bigl(g_i \big \vert \boldsymbol{y}_{i},\boldsymbol{u}_{i},\boldsymbol{\Psi}\bigr)= {\cal{G}}(\alpha^{*}, \beta^{*}),
\end{align}
 where $\alpha^{*} =(p + q)/2+\alpha$ and $\beta^{*} =2\beta + \bigl(\boldsymbol{y}_i-\boldsymbol{\mu}-\boldsymbol{\Lambda}\boldsymbol{u}_{i}\bigr)^{\top}\Sigma^{-1}\bigl(\boldsymbol{y}_i-\boldsymbol{\mu}-\boldsymbol{\Lambda} \boldsymbol{u}_{i}\bigr)/2+\boldsymbol{u}^{\top}_{i}\boldsymbol{u}_{i}/2$. Moreover, using $h(g_i)=g_i$, it turns out from (\ref{mixfull8}) that
 \begin{align} \label{mixfull10}
\pi\bigl(\boldsymbol{u}_{i} \Big \vert \boldsymbol{y}_{i},g_{i},\boldsymbol{\Psi}\bigr) ={\cal{TN}}_{q}\Bigl(\boldsymbol{m}_{i}, \frac{{\Delta}}{{g_i}}, \boldsymbol{0},\boldsymbol{\infty}\Bigr),
\end{align}
where $\boldsymbol{m}_{i}={\Lambda}^{\top}\Omega^{-1}(\boldsymbol{y}_{i}-\boldsymbol{\mu})$, ${\Delta}=\boldsymbol{I}_{q}-{{\Lambda}}^{\top}{\Omega}^{-1}{{\Lambda}}$, and ${\Omega}={\Sigma}+{{\Lambda}}{{\Lambda}}^{\top}$.
% follows a truncated normal distribution on $\mathbb{R}^{q+}$ with mean vector $\boldsymbol{m}_{i}=\boldsymbol{\Lambda}^{\top}\Omega^{-1}(\boldsymbol{y}_{i}-\boldsymbol{\mu})$ and covariance matrix  $\boldsymbol{\Delta}=\boldsymbol{I}_{q}-\boldsymbol{\Lambda}^{\top}\Omega^{-1}\boldsymbol{\Lambda}$ 
%%\subsection{Simulating from full conditional $\boldsymbol{\Psi} \vert (\boldsymbol{Y}=\boldsymbol{y}_{i}, \boldsymbol{U}=\boldsymbol{u}_{i}, {G}={g}_{i},\boldsymbol{\Psi}_{(-\boldsymbol{\Theta})})$}
Constructing the complete data, we proceed to generate from the remainder full conditionals 
$\boldsymbol{\mu} \big \vert \bigl({\underline{\boldsymbol{y}}},\underline{\boldsymbol{u}},{\boldsymbol{g}},\boldsymbol{\Psi}_{(-\boldsymbol{\mu})}\bigr)$, 
$\Sigma \big \vert \bigl({\underline{\boldsymbol{y}}},\underline{\boldsymbol{u}},{\boldsymbol{g}},\boldsymbol{\Psi}_{(-{\Sigma})}\bigr)$, 
${\Lambda} \big \vert \bigl({\underline{\boldsymbol{y}}},\underline{\boldsymbol{u}},{\boldsymbol{g}},\boldsymbol{\Psi}_{(-{\Lambda})}\bigr)$,
$\alpha \big \vert \bigl({\underline{\boldsymbol{y}}},\underline{\boldsymbol{u}},{\boldsymbol{g}},\boldsymbol{\Psi}_{(-{\alpha})}\bigr)$, and
$\beta \big \vert \bigl({\underline{\boldsymbol{y}}},\underline{\boldsymbol{u}},{\boldsymbol{g}},\boldsymbol{\Psi}_{(-{\beta})}\bigr)$ in which ${\underline{\boldsymbol{y}}}=(\boldsymbol{y}^{\top}_{1},\cdots,\boldsymbol{y}^{\top}_{n})^{\top}$, $\underline{\boldsymbol{u}}=(\boldsymbol{u}^{\top}_{1},\cdots,\boldsymbol{u}^{\top}_{n})^{\top}$, and ${\boldsymbol{y}}=({g}_{1},\cdots,{g}_{n})^{\top}$. First, we notice that
\begin{align}\label{postpdf}
\pi(\boldsymbol{\Psi} \vert {\underline{\boldsymbol{y}}},\underline{\boldsymbol{u}},{\boldsymbol{g}})&\propto L_{c}(\boldsymbol{\Psi}\vert\underline{\boldsymbol{y}},\underline{\boldsymbol{u}},{\boldsymbol{g}})\pi(\boldsymbol{\theta})\nonumber\\
&= L_{c}(\boldsymbol{\Psi}\vert\underline{\boldsymbol{y}},\underline{\boldsymbol{u}},{\boldsymbol{g}})\pi(\boldsymbol{\mu})\pi(\Sigma)\pi({\Lambda})\pi(\alpha)\pi(\beta),
\end{align}
where, as usual, we assumed that priors are independent and furthermore 
\begin{align*}
\pi(\boldsymbol{\mu})&\sim {\cal{N}}_{p}\bigl({\cal{M}}_{0}, {\cal{S}}_{0}\bigr),\\
\pi(\Sigma)&\sim {\cal{IW}}\bigl({\cal{D}}_{0}, \nu_0\bigr),\\
\pi(\Lambda)&\sim {\cal{N}}_{q}\bigl({\cal{L}}_{0},{\cal{P}}_{0}\bigr),\\
\pi(\alpha)&\sim {\cal{G}}\bigl(a_{0},a_{0}\bigr),\\
\pi(\beta)&\sim {\cal{G}}\bigl(b_{0},b_{0}\bigr).
\end{align*}
As it is seen, for priors of $\alpha$ and $\beta$, we assumed a gamma distribution with common shape and rate parameters. The results below give the full conditionals needed for implementing the Bayesian paradigm.
\begin{itemize}
\item {\bf{full conditional of $\boldsymbol{\mu} \big \vert \bigl({\underline{\boldsymbol{y}}},\underline{\boldsymbol{u}},{\boldsymbol{g}},\boldsymbol{\Psi}_{(-\boldsymbol{\mu})}\bigr)$:}} By considering a conjugate prior, that is $\pi(\boldsymbol{\mu}) \sim {\cal{N}}_{p}({\cal{M}}_{0}, {\cal{S}}_{0})$, more algebra shows that
\begin{align*}
\pi\bigl(\boldsymbol{\mu} \big \vert {\underline{\boldsymbol{y}}},\underline{\boldsymbol{u}},{\boldsymbol{g}},\boldsymbol{\Psi}_{(-\boldsymbol{\mu})}\bigr)={\cal{N}}_{p}\Bigl(\boldsymbol{\mu} \Big \vert {\cal{Q}}_{0}\Bigl[{\cal{S}}^{-1}_{0}{\cal{M}}_{0}+\Sigma^{-1}\sum_{i=1}^{n}g_{i}(\boldsymbol{y}_{i}-{\Lambda}\boldsymbol{u}_{i})^{}\Bigr],{\cal{Q}}_{0}\Bigr),
\end{align*}
where
\begin{align*}
{\cal{Q}}_{0}=\Bigl[{\cal{S}}^{-1}_{0}+\Sigma^{-1}\sum_{i=1}^{n}g_{i}\Bigr]^{-1}.
\end{align*}
\item {\bf{full conditional of 
$\Sigma \big \vert \bigl({\underline{\boldsymbol{y}}},\underline{\boldsymbol{u}},{\boldsymbol{g}},\boldsymbol{\Psi}_{(-{\Sigma})}\bigr)$:}} By considering a conjugate prior, that is ${\Sigma}\sim {\cal{IW}}({\cal{D}}_{0}, \nu_{0})$, we can write
\begin{align*}
\pi\bigl(\Sigma \big \vert {\underline{\boldsymbol{y}}},\underline{\boldsymbol{u}},{\boldsymbol{g}},\boldsymbol{\Psi}_{(-{\Sigma})}\bigr)=  {\cal{IW}}\bigl(\cdot \big \vert {\cal{R}}_{0}, \nu_{0}+n\bigr),
\end{align*}
where ${\cal{R}}_{0}={\cal{D}}_{0} +\sum_{i=1}^{n} g_{i} \bigl(\boldsymbol{y}_i-\boldsymbol{\mu}-{\Lambda} \boldsymbol{u}_{i}\bigr)\bigl(\boldsymbol{y}_i-\boldsymbol{\mu}-{\Lambda} \boldsymbol{u}_{i}\bigr)^{\top}$.
\item {\bf{full conditional of 
${\Lambda}  \big \vert \bigl({\underline{\boldsymbol{y}}},\underline{\boldsymbol{u}},{\boldsymbol{g}},\boldsymbol{\Psi}_{(-{{\Lambda}})}\bigr)$:}} Although a possible choice for prior of ${\Lambda}$ is a matrix normal normal distribution, but we prefer rather to sample from this full conditional in a slightly different method. We perform this by sampling from each column of ${\Lambda}$ separately. Let ${{\Lambda}}_{(c)}$ denote the $c$-th column (for $c=1,\dots, q$) of matrix ${\Lambda}$ and ${\Lambda}_{(-c)}$ is the matrix ${\Lambda}$ when its $c$-th column is removed. Likewise, assume $u_{i(c)}$ denote the $c$-th element of vector $\boldsymbol{u}_{i}$ and $\boldsymbol{u}_{i(-c)}$ is vector $\boldsymbol{u}_{i}$ when its $c$-th element is removed. Suppose $\boldsymbol{\xi}_{i}=\boldsymbol{y}_i-\boldsymbol{\mu}-{{\Lambda}}_{(-c)} \boldsymbol{u}_{i(-c)}$, then it can be seen that    
\begin{align*}
&\bigl(\boldsymbol{y}_i-\boldsymbol{\mu}-{{\Lambda}} \boldsymbol{u}_{i}\bigr)^{\top}\Sigma^{-1}\bigl(\boldsymbol{y}_i-\boldsymbol{\mu}-{{\Lambda}} \boldsymbol{u}_{i}\bigr) \nonumber\\
&=
 \bigl( {\Lambda}_{(c)} {u}_{i(c)}-\boldsymbol{\xi}_{i}\bigr)^{\top}\Sigma^{-1}\bigl({\Lambda}_{(c)} {u}_{i(c)}-\boldsymbol{\xi}_{i}\bigr).
\end{align*}
Replacing above identity in the RHS of (\ref{cdll1}) 
%rather than $\bigl(\boldsymbol{y}_i-\boldsymbol{\mu}-\boldsymbol{\Lambda} \boldsymbol{u}_{i}\bigr)^{\top}\Sigma^{-1}\bigl(\boldsymbol{y}_i-\boldsymbol{\mu}-\boldsymbol{\Lambda} \boldsymbol{u}_{i}\bigr)$
 and simultaneously assuming a multivariate Gaussian conjugate prior for ${\Lambda}_{(c)}$, that is ${{\Lambda}}_{(c)}\sim {\cal{N}}_{p}\bigl({\cal{L}}_{0(c)}, {\cal{P}}_{0(c)}\bigr)$, it turns out that
\begin{align*}
{\Lambda}_{(c)}  \big \vert \bigl({\underline{\boldsymbol{y}}},\underline{\boldsymbol{u}},{\boldsymbol{g}},\Psi_{(-\Lambda_{0(c)})}\bigr) \sim {\cal{N}}_{p}\bigl({\cal{V}},{\cal{W}}\bigr),
\end{align*}
where, for $c=1,\dots, q$, we have 
\begin{align*}
{\cal{V}}&={\cal{W}}\Bigl[{\cal{P}}^{-1}_{0(c)}{\cal{L}}_{0(c)}+
\Sigma^{-1}\sum_{i=1}^{n} g_{i} u_{i(c)}\boldsymbol{\xi}_{i}\Bigr],\\
{\cal{W}}&=\Bigl[{\cal{P}}^{-1}_{0(c)}+\Sigma^{-1}\sum_{i=1}^{n} g_{i}u^{2}_{i(c)}\Bigr]^{-1}.
\end{align*}
After simulating sequence $\{\Lambda_{(1)}, \Lambda_{(2)}, \ldots, \Lambda_{(q)}\}$, the generated skewness matrix ${\Lambda}$ is reconstructed as ${\Lambda}=\bigl[\Lambda_{(1)} \big \vert \Lambda_{(2)}\big \vert \ldots \big \vert \Lambda_{(q)}\bigr]$.
\item {\bf{full conditionals 
${\alpha} \big \vert \bigl({\underline{\boldsymbol{y}}},\underline{\boldsymbol{u}},{\boldsymbol{g}},\boldsymbol{\Psi}_{(-{\alpha})}\bigr)$ and ${\beta} \big \vert \bigl({\underline{\boldsymbol{y}}},\underline{\boldsymbol{u}},{\boldsymbol{g}},\boldsymbol{\Psi}_{(-{\alpha})}\bigr)$:}} We have
% follows that
%from (\ref{cdll1}) 
%The pdf of these full conditionals can be written as 
\begin{align}\label{fullpdftheta1}
\boldsymbol{\theta} \big \vert \bigl({\underline{\boldsymbol{y}}},\underline{\boldsymbol{u}},{\boldsymbol{g}},\boldsymbol{\Psi}_{(-\boldsymbol{\theta})}\bigr)\propto L_{c}(\boldsymbol{\Psi}\vert {\underline{\boldsymbol{y}}}, \underline{\boldsymbol{u}}, \boldsymbol{g}) \pi(\boldsymbol{\theta}),
\end{align}
By assuming independence between marginals of $\pi(\boldsymbol{\theta})$, it follows from RHS of in (\ref{fullpdftheta1}) that
\begin{align}\label{fullpdftheta2}
\pi\bigl(\alpha \big \vert {\underline{\boldsymbol{y}}},\underline{\boldsymbol{u}},{\boldsymbol{g}},\boldsymbol{\Psi}_{(-\alpha)}\bigr)&\propto \prod_{i=1}^{n} f\bigl(g_{i} \big \vert\boldsymbol{\theta}\bigr)\pi(\alpha)\nonumber\\
&= \frac{\beta^{n\alpha}\alpha_{0}^{\alpha_{0}}}{[\Gamma(\alpha)]^n\Gamma(\alpha_{0})}\bigl(
\prod_{i=1}^{n}g_{i}\bigr)^{\alpha-1}\alpha^{\alpha_{0}-1}\exp \bigl\{ -\alpha_{0}\alpha \bigr\}.
\end{align}
It can be easily seen that $\pi\bigl(\alpha \big \vert {\underline{\boldsymbol{y}}},\underline{\boldsymbol{u}},{\boldsymbol{g}},\boldsymbol{\Psi}_{(-\alpha)}\bigr)$ is log-concave for $\alpha_{0} \geq 1$, and hence the ARS approach is suggested for sampling from this full conditional. If one would like to employ the MH approach, one choice for candidate may be ${\cal{G}}(\alpha_{0},\alpha_{0})$. This yields the the acceptance ratio as
\begin{align}
p\bigl(\alpha_{k}, \alpha_{k+1}\bigr)=\min \Biggl\{1,~&\Bigl(\frac{\alpha_{k+1}}{\alpha_{k}}\Bigr)^{\alpha_{0}-1} \Bigl[\frac{\Gamma\bigl(\alpha_{k}\bigr)}{\Gamma\bigl(\alpha_{k+1}\bigr)}\Bigr]^{n}
\beta^{n(\alpha_{k+1}-\alpha_{k})}\nonumber\\
&\times\exp\Bigl\{\alpha_{0}(\alpha_{k}-\alpha_{k+1})\Bigr\}
\bigl(\prod_{i=1}^{n}g_{i}\bigr)^{\alpha_{k+1}-\alpha_{k}}
\Biggr\}.
\end{align}
Similarly, assuming $\pi(\beta) \sim {\cal{G}}\bigl(\beta_{0},\beta_{0}\bigr)$. We have
\begin{align}\label{fullpdfbeta}
\pi\bigl(\beta \big \vert {\underline{\boldsymbol{y}}},\underline{\boldsymbol{u}},{\boldsymbol{g}},\boldsymbol{\Psi}_{(-\beta)}\bigr)&\propto \prod_{i=1}^{n} f\bigl(g_{i} \big \vert\boldsymbol{\theta}\bigr)\pi(\beta)\nonumber\\
&\propto\beta^{n\alpha+\beta_{0}-1}\exp \Bigl\{ -\beta\bigl(\beta_{0}+\sum_{i=1}^{n}g_{i}\bigr) \Bigr\}.
\end{align}
From the RHS of (\ref{fullpdfbeta}), it turns out that 
$\beta \big \vert\bigl({\underline{\boldsymbol{y}}},\underline{\boldsymbol{u}},{\boldsymbol{g}},\boldsymbol{\Psi}_{(-\beta)}\bigr)\sim {\cal{G}}\bigl(\beta_{0}+n\alpha, \beta_{0}+\sum_{i=1}^{n}g_{i}\bigr)$.
\end{itemize}
In what follows, Algorithm \ref{Bayesian inference for SGG} describes how to compute the Bayesian estimator for parameters of ${\text{SGG}}_{p,q}(\boldsymbol{\mu},\Sigma,{\Lambda},\boldsymbol{\theta})$.
\begin{algorithm}
\caption{Bayesian inference for SGG}
    \label{Bayesian inference for SGG}
\begin{algorithmic}[1]
%\Procedure{}{}     %  \Comment{This is a test}
 %   \State System Initialization
 %   \State Read the value 
%    \If{$condition = True$}
%        \State Do this
%        \If{$Condition \geq 1$}
%        \State Do that
%        \ElsIf{$Condition \neq 5$}
%        \State Do another
%        \State Do that as well
%        \Else
%        \State Do otherwise
%        \EndIf
%    \EndIf
%\State Set $t=1$, read $M$ as the {\it{burn-in}} period, and initiate algorithm with 
%\State randomly selected sample $\boldsymbol{x}^{(0)}$
\State Read $N$, $M$, and determine quantities ${\cal{M}}_{0}$, ${\cal{S}}_{0}$, ${\cal{D}}_{0}$, $\nu_0$, ${\cal{L}}_{0}$, ${\cal{P}}_{0}$, $a_0$, and 
\State $\boldsymbol{\theta}_{0}=(a_0,b_0)^{\top}$;
% , and suggest $\boldsymbol{\theta}^{(0)}=\bigl(\theta^{(0)}_{1},\theta^{(0)}_{2},\theta^{(0)}_{3}\bigr)^{\top}$ arbitrarily;
\State Set $t \leftarrow 0$;
\State Set $\boldsymbol{\mu}^{(0)} \leftarrow {\cal{M}}^{0}$, $\Sigma^{(0)}\leftarrow 1/n\sum_{i=1}^{n}\bigl(\boldsymbol{y}_i-\boldsymbol{\mu}^{(0)} \bigr)\bigl(\boldsymbol{y}_i-\boldsymbol{\mu}^{(0)} \bigr)^{\top}$, ${\nu}^{(0)}=2$,
\State and set $\boldsymbol{\Psi}^{(0)}=\bigl(\boldsymbol{\mu}^{(0)}, \Sigma^{(0)}, {\Lambda}^{(0)}, \boldsymbol{\theta}^{(0)}\bigr)$;
    \While{$t \leq N$}  %\Comment{put some comments here}
    \State Set $i=1$;
    \While{$i \leq n$}  %\Comment{put some comments here}
    \State Simulate $g_i$ from ${\cal{G}}(a,b)$ where $a=\alpha^{(t)} +(p+q)/2$
     and
       $b=$
       \State $\Bigl[2\beta^{(t)} + \bigl(\boldsymbol{y}_i-\boldsymbol{\mu}^{(t)} -{\Lambda}^{(t)}\boldsymbol{u}_{i}\bigr)^{\top}\bigl[\Sigma^{(t)} \bigr]^{-1}\bigl(\boldsymbol{y}_i-\boldsymbol{\mu}^{(t)} -{\Lambda}^{(t)}\boldsymbol{u}_{i}\bigr)\Bigr]/2+\boldsymbol{u}^{\top}_{i}\boldsymbol{u}_{i}/2$;
        \State Simulate $\boldsymbol{u}_{i}$ from ${\cal{HN}}_{q}\bigl( \boldsymbol{m}^{(t)}_{i}, \frac{{\Delta}^{(t)}}{{g_i}}\bigr)$ where 
 $\boldsymbol{m}^{(t)}_{i}=\bigl[{\Lambda}^{(t)}\bigr]^{\top}\bigl[\Omega^{(t)}\bigr]^{-1}(\boldsymbol{y}_{i}-$
 \State $\boldsymbol{\mu})$ with ${\Omega}^{(t)}={\Sigma}^{(t)}+{{\Lambda}^{(t)}}\bigl[{{\Lambda}^{(t)}}\bigr]^{\top}$ and ${\Delta}^{(t)}=\boldsymbol{I}_{q}-\bigl[{{\Lambda}^{(t)}}\bigr]^{\top}\bigl[{\Omega^{(t)}}\bigr]^{-1}{{\Lambda}^{(t)}}$;
    \State Set $i \leftarrow i+1$;
        %\Comment{another comment}
        %\State $var3 \leftarrow var4$
             \EndWhile  %\label{roy's loop}
             \State {\bf{end}}
        \State  Generate $\boldsymbol{\mu}^{(t+1)}\sim 
{\cal{N}}_{p}\Bigl({\cal{Q}}_{0}\Bigl[{\cal{S}}^{-1}_{0}{\cal{M}}_{0}+\bigl[\Sigma^{(t)} \bigr]^{-1}\sum_{i=1}^{n}g_{i}(\boldsymbol{y}_{i}-{\Lambda}^{(t)}\boldsymbol{u}_{i})\Bigr],{\cal{Q}}_{0}\Bigr)$
\State where ${\cal{Q}}_{0}=\Bigl[{\cal{S}}^{-1}_{0}+\bigl[\Sigma^{(t)} \bigr]^{-1}\sum_{i=1}^{n}g_{i}\Bigr]^{-1}$;
\State  Generate $\Sigma^{(t+1)}$ from
${\cal{IW}}\bigl({\cal{R}}_{0}, \nu_{0}+n\bigr)$ where ${\cal{R}}_{0}={\cal{D}}_{0}+\sum_{i=1}^{n} g_{i} \bigl(\boldsymbol{y}_{i}-$
\State $\boldsymbol{\mu}^{(t+1)}\bigr)\bigl(\boldsymbol{y}_{i}-\boldsymbol{\mu}^{(t+1)}\bigr)^{\top}$;
        %\Comment{another comment}
        %\State $var3 \leftarrow var4$
    %%%%%%%%%%%%%%%%%%%%%%%%%    
        \State Set $j=1$;
    \While{$j \leq q$}  %\Comment{put some comments here}
\State Set ${\cal{W}}^{-1}=\bigl[{\cal{P}}^{-1}_{0(j)}+\bigl[\Sigma^{-1}\bigr]
^{(t+1)} \sum_{i=1}^{n} g_{i}u^{2}_{i(j)}\bigr]$,
${\cal{V}}={\cal{W}}\bigl[{\cal{P}}^{-1}_{0(j)}{\cal{L}}_{0(j)}+$\\
 \State $\bigl[\Sigma^{(t+1)}\bigr]^{-1} \sum_{i=1}^{n} g_{i} u^{}_{i(j)}{\boldsymbol{\xi}_{i}}
\bigr]$, and $\boldsymbol{\xi}_{i}=\boldsymbol{y}_i-\boldsymbol{\mu}^{(t+1)}-{{\Lambda}}^{(t)}_{(-j)} \boldsymbol{u}_{i(-j)}$;
\State Generate ${{\Lambda}}_{(j)}\sim {\cal{N}}_{p}\bigl({\cal{V}}, {\cal{W}}\bigr)$;  
%, for $c=1,\dots, q$. After simulating sequence 
                 \EndWhile  %\label{roy's loop}
                       \State {\bf{end}}
                       \State Construct ${\Lambda}^{(t+1)}=\bigl[{\Lambda}_{(1)} \big \vert {\Lambda}_{(2)}\big \vert \ldots \big \vert {\Lambda}_{(q)}\bigr]$;
      \State Use the ARS-within-Gibbs sampling technique for simulating 
      $\alpha^{(t+1)}$ while
    \State $\log \pi\bigl(\alpha \big \vert {\underline{\boldsymbol{y}}},\underline{\boldsymbol{u}},{\boldsymbol{g}},\boldsymbol{\Psi}_{(-\alpha)}\bigr) \propto n\alpha \log \beta -n \log \Gamma(\alpha) + (\alpha-1)\sum_{i=1}^{n}\log g_{i}+$
\State $(\alpha_{0}-1)\log \alpha -\alpha_{0}\alpha$
\State Simulate $\beta^{(t+1)}$  from full conditional $\beta \big \vert\bigl({\underline{\boldsymbol{y}}},\underline{\boldsymbol{u}},{\boldsymbol{g}},\boldsymbol{\Psi}_{(-\beta)}\bigr)\sim {\cal{G}}\bigl(\beta_{0}+$
\State $n\alpha^{(t+1)}, \beta_{0}+\sum_{i=1}^{n}g_{i}\bigr)$; 
             \EndWhile  %\label{roy's loop}
           \State {\bf{end}}
  %  \EndWhile  %\label{roy's loop}
            \State Set $\boldsymbol{\Psi}^{(t+1)} \leftarrow \bigl(\boldsymbol{\mu}^{(t+1)}, \Sigma^{(t+1)}, {\Lambda}^{(t+1)},\boldsymbol{\theta}
            ^{(t+1)}\bigr)$ and $t \leftarrow t+1$;
                                   \State {\bf{end}}
   \State  Sequence $\bigl\{\boldsymbol{\Psi}^{(M)},\boldsymbol{\Psi}^{(M+1)},\cdots,\boldsymbol{\Psi}^{(N)}\bigr\}$ is a sample of size $N-M+1$ from
   \State posterior $\pi(\boldsymbol{\Psi} \vert {\underline{\boldsymbol{y}}},\underline{\boldsymbol{u}},{\boldsymbol{g}})$;
  \State The Bayesian estimator of $\boldsymbol{\Psi}$ is given by $\frac{1}{N-M+1}\sum_{t=1}^{N-M+1}\boldsymbol{\Psi}^{(t)}$.
%\EndProcedure
%\State {\bf{end procedure}}
\end{algorithmic}
\end{algorithm}
\begin{example}%\lipsum*[]
Herein, we suppose a sample of $n=500$ observations are drawn from 
$\boldsymbol{Y}\sim {\text{SGG}}_{2,2}(\boldsymbol{\mu}, \Sigma, \boldsymbol{\Lambda}, \boldsymbol{\theta})$ with parameters
\begin{align*}
\boldsymbol{\mu}=
\biggl[\begin{matrix}
2\\ -2
\end{matrix}\biggr],~
\Sigma=
\biggl[\begin{matrix}
0.4&0\\
0& 0.4\\
\end{matrix}\biggr],~
\boldsymbol{\Lambda}=\biggl[\begin{matrix}
-2&2\\
3& 0\\
\end{matrix}\biggr],~
\boldsymbol{\theta}=
\biggl[\begin{matrix}
\alpha\\ \beta
\end{matrix}\biggr]=\biggl[\begin{matrix}
5\\ 5
\end{matrix}\biggr].
\end{align*}
We follow the steps of Algorithm 7 to obtain Bayesian estimator of $\boldsymbol{\Psi}$. 
\end{example}
The pertaining \verb+R+ code for implementing example above is given as follows. As it may seen from \verb+R+ code, we hyperparameters as ${\cal{M}}_{0}=(0,0)^{\top}$, ${\cal{S}}_{0}={\cal{D}}_{0}={\cal{P}}_{0}= \boldsymbol{I}_{p}$, $\nu_{0}= 2$, ${\cal{L}}_{0}=(0,0)^{\top}$, $\beta_{0}=2$, and $\alpha_{0}=2$. The output of the Gibbs sampler is displayed in Figure \ref{fig6}.
\begin{lstlisting}[style=deltaj]
R >	f.a <- function(x, beta, alpha0, y)
+				{
+				 n = length(y)
+				      n*x*log(beta) - n*lgamma(x) + (x-1)*sum(log(y)) + 
+ 					 (alpha0-1)*log(x) - alpha0*x
+				}
R >	fprim.a <- function(x, beta, alpha0, y)
+				{
+				n = length(y) 
+				  n*log(beta) - n*digamma(x) + sum(log(y)) + (alpha0-1)/x - alpha0
+				}
R> rssg <- function(n, alpha, beta, Mu, Sigma, Lambda)
+		{
+			Dim <- length(Mu)
+			Q <- length(Lambda[1,])
+			Y <- matrix(NA, nrow = n, ncol = Dim)
+				for (i in 1:n)
+				{
+					Z <- rgamma(1, shape = alpha, rate = beta)
+					              X <- mvrnorm(n= 1, mu = rep(0, Dim), Sigma = Sigma )
+					u <- abs( rnorm(Q) )
+					Y[i, ] <- Mu + Lambda%*%u/sqrt(Z) + X/sqrt(Z)
+				}
+			 Y
+		}
R> set.seed(20240520)
R> library(MASS); library(tmvtnorm); library(ars)
R> n <- 500; alpha <- 5; beta <- 5; Sigma <- matrix( c(.4,0,0,.4), 2, 2); 
R> Mu <- c(2, -2); Lambda <- matrix( c(-2,3,2,0), 2, 2)
R> Y <- rssg(n, alpha, beta, Mu, Sigma, Lambda ); 
R> g <- rgamma(n, shape = alpha, rate = beta);
R> Q <- dim(Lambda)[2]
R> p <- length(Mu)
R> T <- matrix( abs(rnorm(n*Q, mean = 0, sd = 1)), nrow = n, ncol = Q)
R> N <-10000; M <-5000; Mu0c <-rep(0, p);  beta0 <-2; alpha0 <-2; nu0 <-2; 
R> S0Mu <- diag(p); M0Mu <- rep(0, p); 
R> D0 <- S0c <- P0c <-diag(p);
R> Psi <-matrix(0, nrow = N, ncol = p + p^2 + p*Q + 2 )
R> for(k in 1:N)
+	{
+	Sigmainv <- solve(Sigma)
+	Omega <- Sigma + tcrossprod( Lambda )
+	Omegainv <- solve(Omega)
+	Delta <- diag(Q) - (t(Lambda)%*%Omegainv%*%Lambda)
+			for(i in 1:n)
+			{
+				rate <- mahalanobis(x = Y[i, ], center = Mu + Lambda%*%T[i, ], cov = Sigma )/2 + (T[i,]%*%T[i,])[1]/2 + beta
+				g[i]  <- rgamma(1, shape = alpha + (p + Q)/2, rate = rate )
+				T[i,] <- rtmvnorm(1, mean = c(t(Lambda)%*%Omegainv%*%c( Y[i, ] - Mu )), sigma = Delta/g[i],
+				lower = rep(0, length = Q), upper = rep( Inf, length = Q), algorithm = c("rejection") ) 
+			}
+	Sum_g <- sum(g)
+	M_sum <- rowSums( sapply(1:n, function(i) g[i]*c( Y[i, ] - 
+										Lambda%*%T[i, ]) ) )
+	Q0 <- solve( solve( S0Mu ) + Sigmainv*Sum_g )
+	Mu <- c( mvrnorm(1, mu = Q0%*%( solve( S0Mu )%*%M0Mu + Sigmainv%*%M_sum ), 
+						 				Sigma = Q0) ); 
+	Psi[k, 1:p] <- Mu
+	R <- matrix(0, nrow = p, ncol = p)
+			for(i in 1:n)
+			{
+				R <- R + g[i]*c( Y[i, ] - Mu - Lambda%*%T[i, ])%*%t( c( Y[i, ] - Mu - 
+										Lambda%*%T[i, ])
+			)}
+			Sigma <- solve( rWishart(1, df = n+nu0+1, Sigma = solve(D0 + R) )[,,1] )
+			Psi[k, (p + 1):(p + p^2)] <- as.numeric(Sigma)
+					for(j in 1:Q)
+					{
+						Sumgt <- sum( g*T[, j]^2 )
+						Sc <- solve( solve( S0c ) + Sigmainv*Sumgt )
+							if(Q >=3)
+							{
+								Mc_sum <- rowSums( sapply(1:n,function(i) g[i]*T[i, j]*c( Y[i, ] - 
+																 Mu - 	Lambda[,-j]%*%T[i, -j]) ) )
+								}else{
+								Mc_sum <- rowSums( sapply(1:n, 
+								function(i) g[i]*T[i, j]*c( Y[i, ] - Mu - Lambda[,-j]*T[i, -j]) ) )
+							}
+						Mc <- Sc%*%( solve( S0c )%*%Mu0c + Sigmainv%*%Mc_sum ) 
+						Lambda[, j] <- mvrnorm(n= 1, mu = Mc , Sigma = Sc )
+					}
+	Psi[k, (p + p^2 + 1):(p + p^2 + p*Q)] <- as.numeric(Lambda)
+	alpha <- ars(1, f.a, fprim.a, x=c(0.5, 4, 10, 30), m=4, lb=TRUE, xlb=0, 
+											 ub=FALSE, beta=beta, alpha0=alpha0, y=g )
+	beta <- rgamma( 1, n*alpha + beta0, beta0 + sum(g) )
#   b0 <- 1/(3-(cov(Y)%*%solve(Sigma))[1,1])
+	Psi[k, (p + p^2 + p*Q + 1)] <- alpha
+	Psi[k, (p + p^2 + p*Q + 2)] <- beta
+	}
R > Psi.hat <- apply(Psi[(N-M+1):N,], 2, mean)
R > Mu <- Psi.hat[1:p]
R > Sigma <- Psi.hat[(p+1):(p + p^2)]
R > Lambda <- Psi.hat[(p + p^2 + 1):(p + p^2 + p*Q)]
R > alpha <- Psi.hat[p + p^2 + p*Q + 1]
R > beta <- Psi.hat[p + p^2 + p*Q + 2]
\end{lstlisting}
\begin{figure}
\center
\includegraphics[width=65mm,height=65mm]{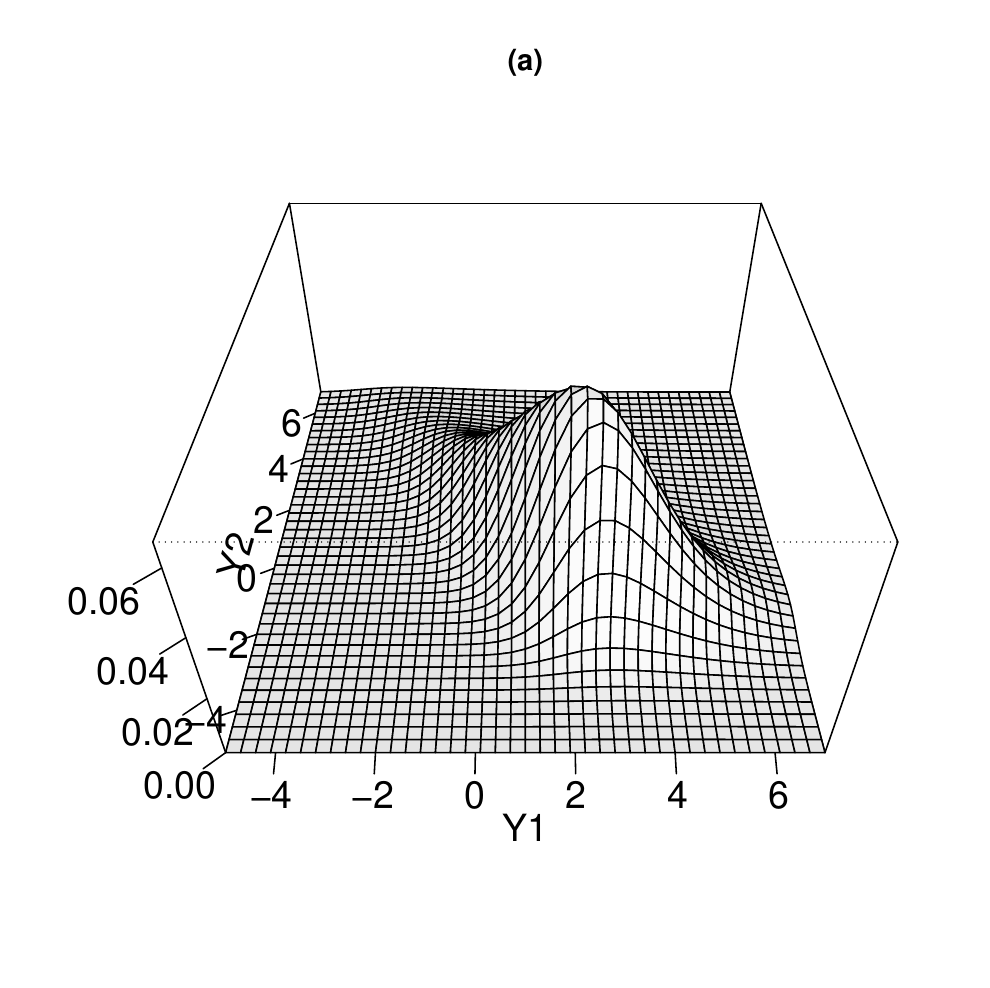}
\includegraphics[width=65mm,height=65mm]{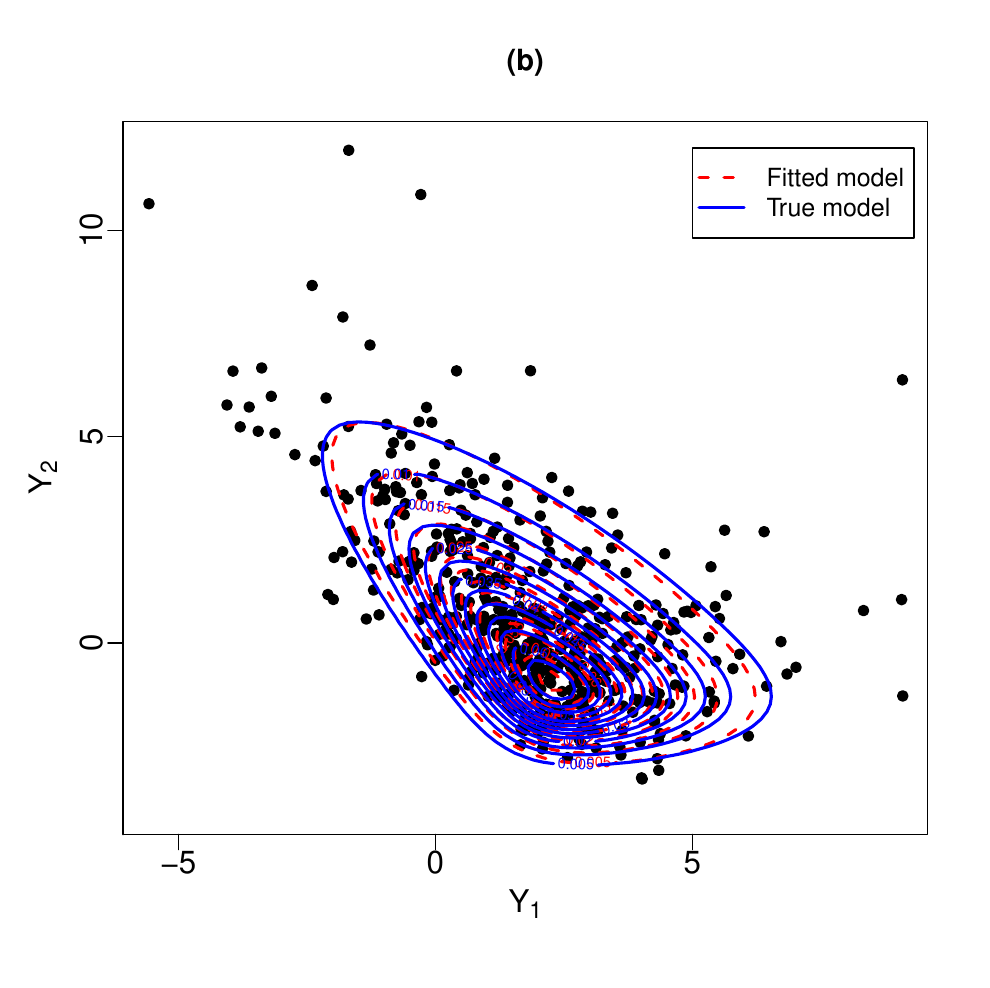}
\caption{The PDF of $\text{SGG}_{2,2}(\boldsymbol{\mu},\Sigma,{\Lambda},\boldsymbol{\theta})$. (a): true model and (b): true and fitted contour plots.}
\label{figst1}
\end{figure}
\begin{figure}
\center
\includegraphics[width=55mm,height=40mm]{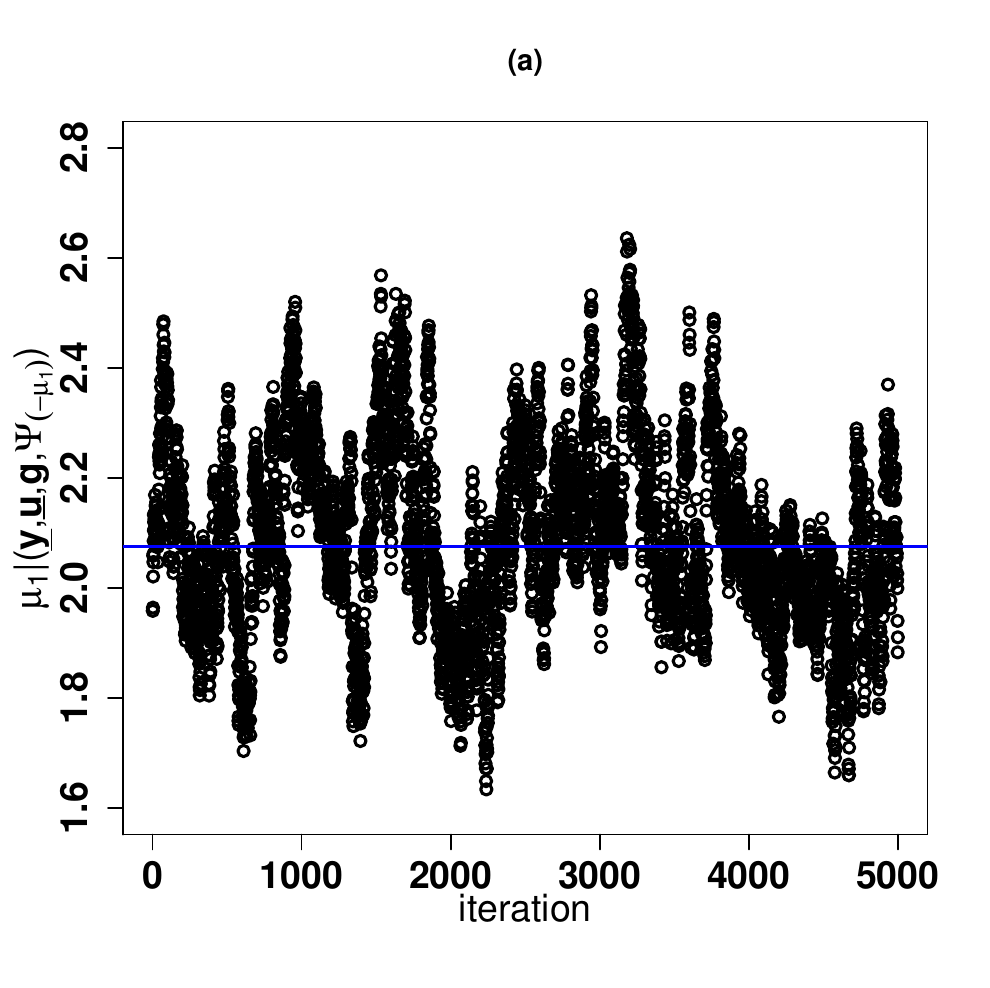}
\includegraphics[width=55mm,height=40mm]{plot-SGG-mu1}\\
\includegraphics[width=55mm,height=40mm]{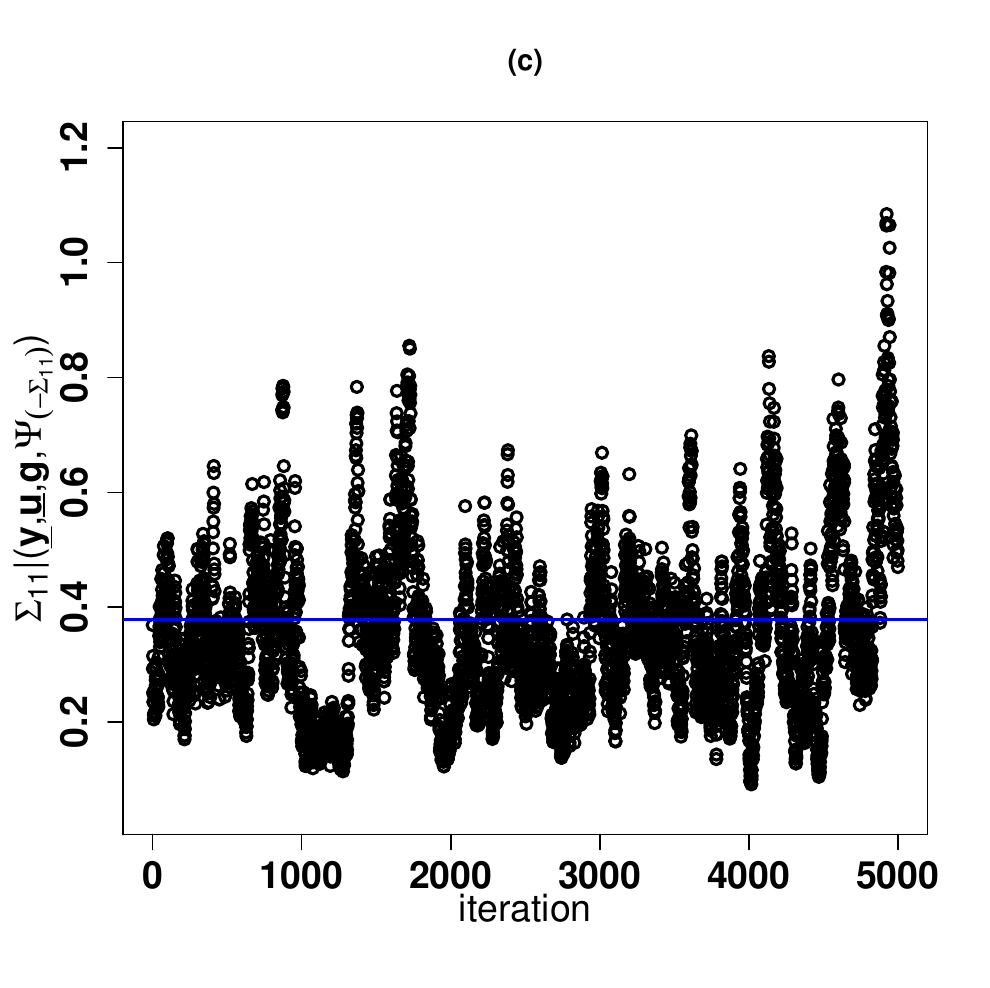}
\includegraphics[width=55mm,height=40mm]{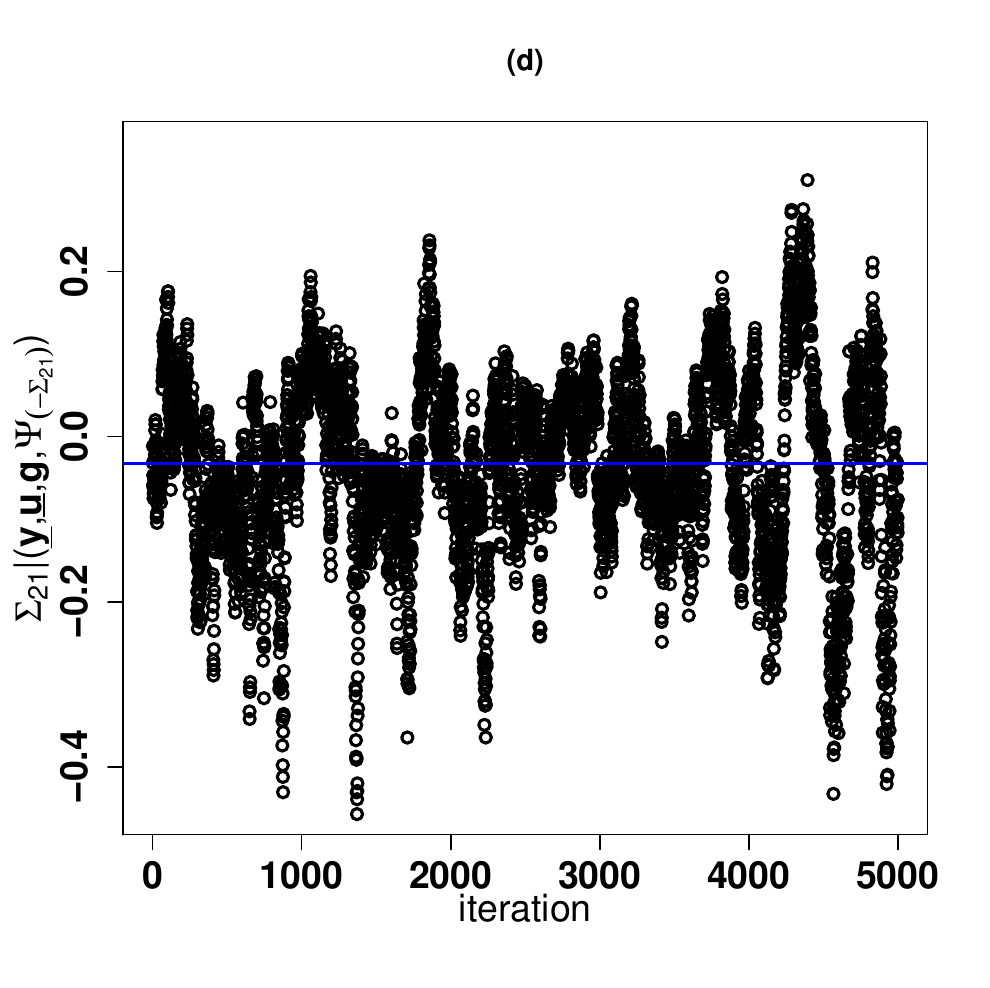}
\caption{Scatterplot of 5000 samples generated from full conditionals involved in Bayesian paradigm of $\text{SGG}_{2,2}(\boldsymbol{\mu},\Sigma,{\Lambda},\boldsymbol{\theta})$.}
\label{figst2}
\end{figure}
\section{Gibbs sampling in Bayesian paradigm with more than two latent variables: Model-based Clustering}
The aim of clustering technique is to divide a non-homogeneous population into $\text{K}$ numbers of homogeneous sub-populations. Obviously, members of a non-homogeneous population have dissimilarities while the clustered or partitioned sub-populations have more similarities. Herein, the term ``similarity'' may refer to some measurable attribute or criterion such as Euclidean distance. If two observations are similar, then they are close together (high degree of similarity) and otherwise, two observations are dissimilar (low degree of similarity).
\begin{figure}[!h]
\center
\includegraphics[width=55mm,height=55mm]{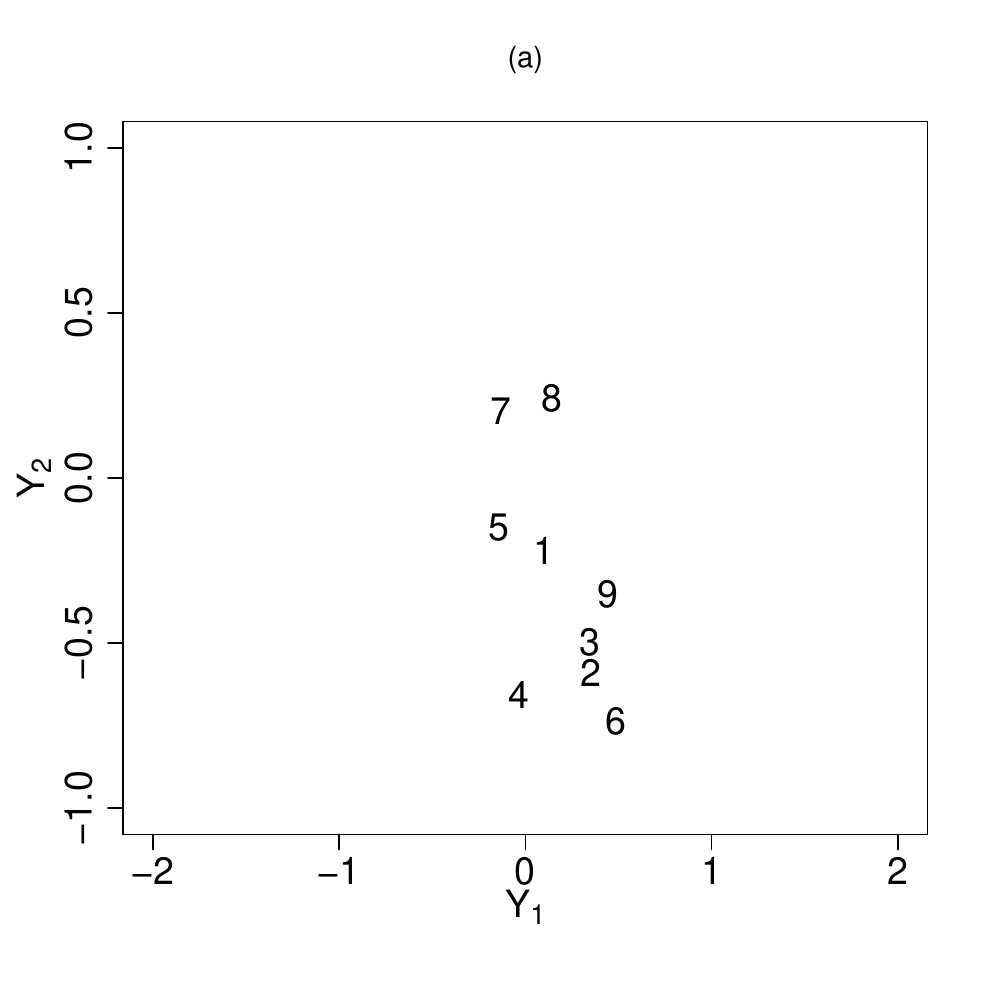}
\includegraphics[width=55mm,height=55mm]{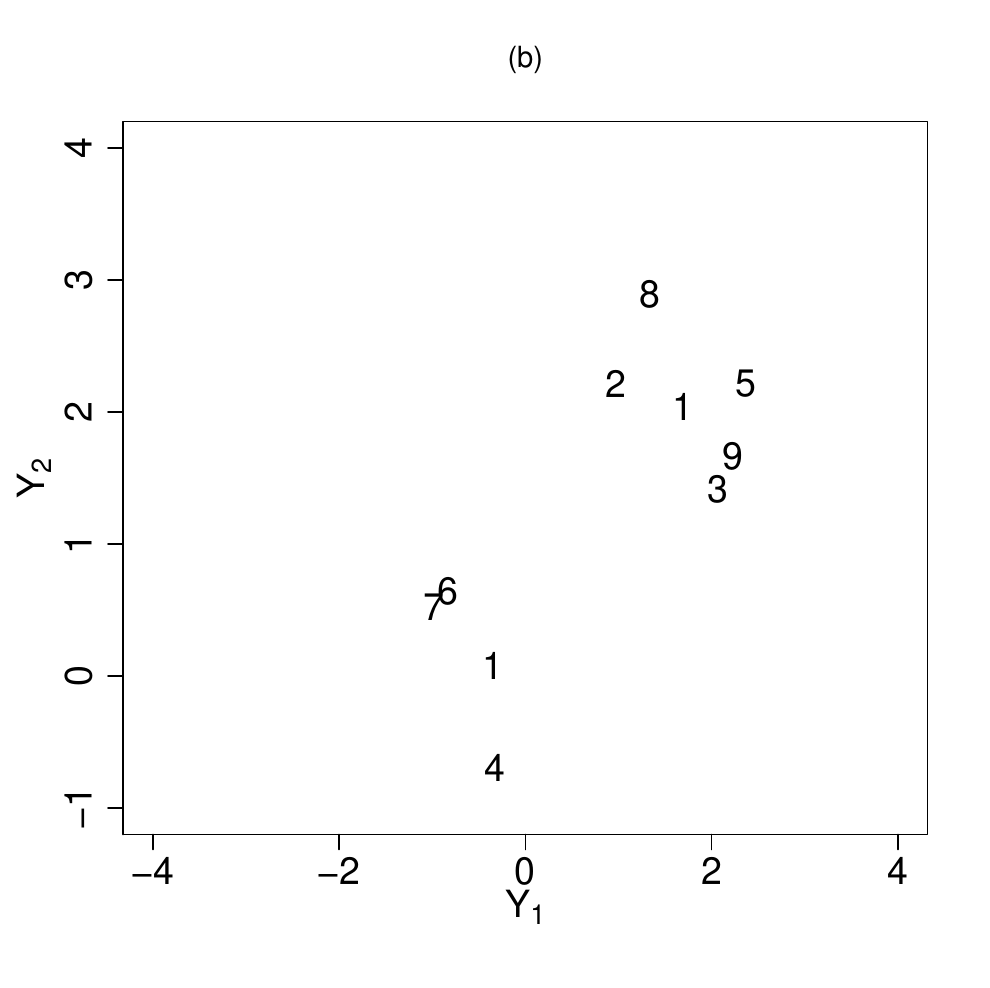}\\
\includegraphics[width=55mm,height=55mm]{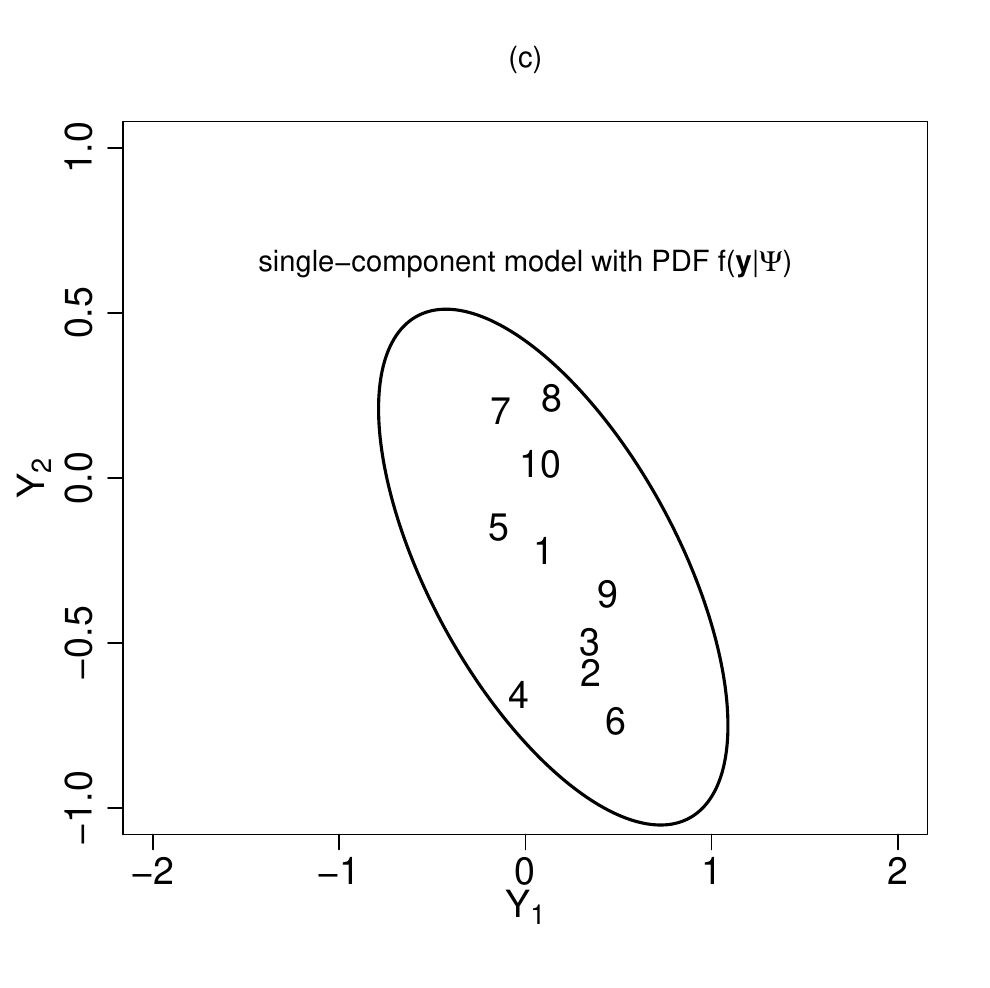}
\includegraphics[width=55mm,height=55mm]{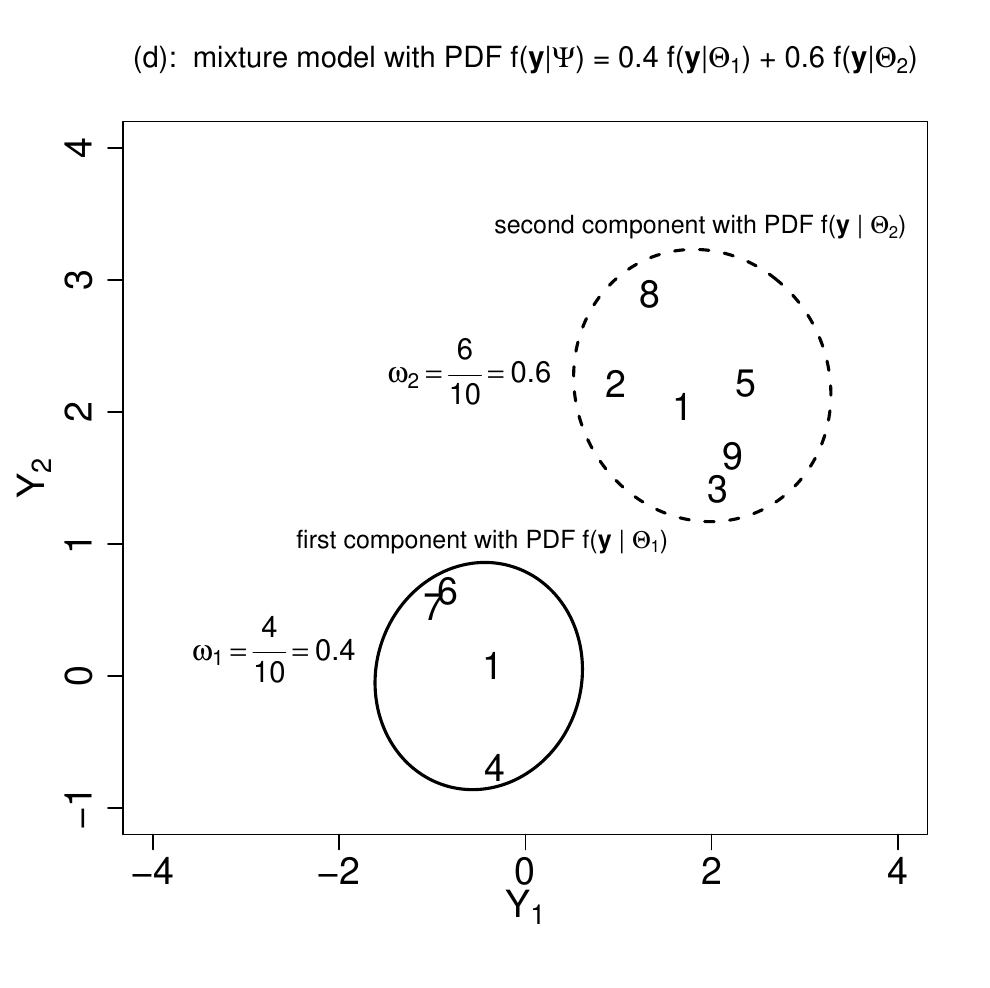}
\caption{(a): Ten data points that can constitute a cluster, (b):
Ten data points that should be clustered into two groups, (c):
Ten data points in part (a) that can be modelled through a single-component (homogeneous) model with PDF $f\bigl(\boldsymbol{y}\bigl \vert \boldsymbol{\Theta}\bigr)$, and (d): Ten data points in part (b) that can be modelled through a two-component (non-homogeneous) mixture model with PDF $0.4 f\bigl(\boldsymbol{y}\bigl \vert \boldsymbol{\Theta}_{1}\bigr)+0.6 f\bigl(\boldsymbol{y}\bigl \vert \boldsymbol{\Theta}_{2}\bigr)$.}
\label{plot-mixture-two-component-schematic}
\end{figure}
\par Figure \ref{plot-mixture-two-component-schematic}(a) displays a set of ten observations, labeled with numbers $1,2,\cdots,10$, that are similar and so can constitute a cluster. On the other hand, consider the situation in which ten observations are non-homogeneous or dissimilar as shown by Figure \ref{plot-mixture-two-component-schematic}(b). Herein, ten  observations with labels $\{7, 6, 1,4\}$ are similar together, but quite dissimilar with observations whose label are $\{2, 8, 1, 3, 9, 5\}$. So, observations in Figure \ref{plot-mixture-two-component-schematic}(b) should be groped into two clusters. We emphasize that the objective of clustering technique is to determine the label of each observation. Furthermore, sometimes the yardstick for measuring the similarity between observations may not be a quantitative tool such as the Euclidean distance. The clustering technique proceeds to divide observations into {\text{K}} groups such that the observations within each group have the maximum degree of similarity while the observations between groups have the maximum degree of dissimilarity. Sometimes the clustering technique is known as semi-supervised or unsupervised machine learning since label of observations are not known and the only available information are observed data \cite{mcnicholas2016mixture}. 
\par The model-based clustering (or non-hierarchical clustering) uses a finite mixture of multivariate statistical distributions of the same family for the purpose of clustering. Hence, the model-based clustering and finite mixture models are often used interchangeably in the literature. As mentioned above, the aim of clustering is to determine the label or the origin of an observation is coming from. It is worth to note that the concepts ``origin'', ``cluster'', ``label'', and ``component'' may be used interchangeably. For more clarifications, Figure \ref{plot-mixture-two-component-schematic}(c) shows a set of ten observations with perfect similarity and, within the model-based clustering framework, we assume that these ten observations follow independently the same multivariate distribution. On the other hand, Figure \ref{plot-mixture-two-component-schematic}(d) shows another set of ten observations with two clusters, and hence within the model-based clustering framework, we prefer to assume these data points are coming from two different multivariate statistical distribution of the same family. In other words, for the first and second clusters we consider two multivariate statistical distributions of the same family whose PDFs are $f\bigl(\boldsymbol{y}\bigl \vert \boldsymbol{\Theta}_{1}\bigr)$ and $f\bigl(\boldsymbol{y}\bigl \vert \boldsymbol{\Theta}_{2}\bigr)$, respectively. The corresponding weights are 0.4 and 0.6 and hence the distribution of ten observations is a mixture of two distributions fitted to two clusters whose PDF is
\begin{align}\label{mixture-PDF2}
g(\boldsymbol{y} \vert \boldsymbol{\Psi})=0.4 f\bigl(\boldsymbol{y}\bigl \vert \boldsymbol{\Theta}_{1}\bigr) + 0.6  f\bigl(\boldsymbol{y}\bigl \vert \boldsymbol{\Theta}_{2}\bigr).
\end{align}
In (\ref{mixture-PDF2}), the vector $\boldsymbol{\omega}=(0.40,0.60)^{\top}$, is known as the weight vector that means 40\% of observations in Figure \ref{plot-mixture-two-component-schematic}(d) are coming from the first component and the remaining 60\% are coming from the second component. Overall, for $p$-dimensional random vector $\boldsymbol{Y}$, the PDF of a $\text{K}$-component mixture model is given by
\begin{align}\label{mixture-PDFk}
g(\boldsymbol{y} \vert \boldsymbol{\Psi})=\sum_{k=1}^{\text{K}} \omega_{k} f\bigl(\boldsymbol{y}\bigl \vert \boldsymbol{\Theta}_{k}\bigr),
\end{align}
where $\boldsymbol{\Psi}=(\boldsymbol{\omega}^{\top},\boldsymbol{\Theta}_1,\cdots, \boldsymbol{\Theta}_\text{K})$ is the parameter space in which $\boldsymbol{\omega}=(\omega_1,\dots,\omega_\text{K})^{\top}$ is vector of mixing parameters such that $\sum_{k=1}^{\text{K}}\omega_{k}=1$. In (\ref{mixture-PDFk}), $f_{\boldsymbol{Y}}(\boldsymbol{y} \vert\boldsymbol{\Theta}_{k})$ is the PDF of $k$-th component and hence the PDF $g(\boldsymbol{y} \vert \boldsymbol{\Psi})$ is proper since 
%for $\boldsymbol{\Theta}_{k}=\bigl(\boldsymbol{\theta}_k, \boldsymbol{\mu}_k, {\Sigma}_k, {\Lambda}_k\bigr)$
\begin{align*}%\label{Gibbs-mixture-PDF2}
\int_{\mathbb{R}^{p}}g(\boldsymbol{y} \vert \boldsymbol{\Psi})d\boldsymbol{y}=\sum_{k=1}^{\text{K}} \omega_{k} \int_{\mathbb{R}^{p}}f\bigl(\boldsymbol{y}\bigl \vert \boldsymbol{\Theta}_{k}\bigr)d\boldsymbol{y}=\sum_{k=1}^{\text{K}} \omega_{k}=1.
\end{align*}
The weight parameter $\boldsymbol{\omega}$ provides the chance of observing sample from components. Sometimes, representation (\ref{mixture-PDFk}) is called the {\it{finite mixture model}} since the quantity ${\text{K}}$ or number of clusters is assumed to be finite. In next two subsections, we proceed to discuss about generating realization from a finite mixture model and how to determine quantity ${\text{K}}$.
\subsection{Generating from Gaussian finite mixture model}
For generating $n$ realizations from a ${\text{K}}$-component mixture model, we may follow one of the two following approaches. In the first method, 
one can generate independently $n_{k}=\lceil n \times \omega_{k} \rceil$ (for $k=1,\cdots, {\text{K}}-1$) realizations from PDF $f\bigl(\boldsymbol{y}\big \vert \boldsymbol{\Theta}_{k}\bigr)$ in which $\lceil n \times \omega_{k}\rceil$ denotes the smallest integer value equal or greater than $n\times \omega_{k}$, and finally $n_{\text{K}}=n-\sum_{k=1}^{{\text{K}}-1} n_{k}$ realizations are drawn independently from PDF $f\bigl(\boldsymbol{y}\big \vert \boldsymbol{\Theta}_{\text{K}}\bigr)$. 
%To this end, herein, we introduce the multinomial distribution. If random vector $\boldsymbol{X}=(X_1,\cdots,X_{\text{K}})^{\top}$ follows a multinomial distribution, then its PDF is given by 
%\begin{align}\label{Multinomial-PDF}
%{\cal{MUL}}\bigl(\boldsymbol{x}\big \vert n, \boldsymbol{\omega}\bigr)=\frac{\Gamma(n+1)}{\prod_{k=1}^{{\text{K}}}\Gamma\bigl(n_{k}+1\bigr)}\omega_{1}^{x_{1}}\omega_{2}^{x_{2}}\cdots \omega_{{\text{K}}}^{x_{{\text{K}}}},
%\end{align}
%where $\boldsymbol{\omega}=\bigl(\omega_1,\cdots,\omega_{\text{K}}\bigr)^{\top}$ is  distribution's parameter vector, $n=\sum_{k=1}^{{\text{K}}}x_{k}$, and $\omega_k>0$ provided that $\sum_{k=1}^{{\text{K}}}\omega_{k}=1$. We write $\boldsymbol{X} \sim {\cal{MUL}}(n,\boldsymbol{\omega})$ to indicate that random vector $\boldsymbol{X}$ follows a multinomial distribution with PDF given by (\ref{Multinomial-PDF}). 
%Moreover, if random vector $\boldsymbol{X}=(X_1,\cdots,X_{\text{K}})^{\top}$ follows a Dirichlet distribution with parameter vector $\boldsymbol{\rho}$, say $\boldsymbol{X}\sim {\cal{DIR}}(\boldsymbol{\rho})$, then the PDF of $\boldsymbol{X}$ is
%\begin{align}\label{Dirichlet-PDF}
%{\cal{DIR}}(\boldsymbol{x}\vert \boldsymbol{\rho})=\frac{\Gamma\bigl(\rho_1 +\cdots+\rho_{{\text{K}}}\bigr)}{\prod_{k=1}^{{\text{K}}}\Gamma\bigl(\rho_{k}\bigr)}x_{1}^{\rho_{1}-1}\times x_{2}^{\rho_{2}-1}\times\cdots \times x_{{\text{K}}}^{\rho_{\text{K}}-1},
%\end{align}
%where $0< x_{k}<1$ and ${\rho}_{k}>0$ (for $k=1,\cdots,{\text{K}}$), provided that $\sum_{k=1}^{{\text{K}}}x_{k}=1$. 
The following pseudo code describes how to draw a sample of size $n$ from a ${\text{K}}$-component Gaussian mixture model under the second method. 
\begin{algorithm}
\caption{Simulating from ${\text{K}}$-component Gaussian mixture model.}
    \label{Simulating from a mixture model-2}
\begin{algorithmic}[1]
\State Read $n$, ${\text{K}}$ (number of components), and $\boldsymbol{\Psi}=(\boldsymbol{\omega},\boldsymbol{\Theta}_1,\cdots, \boldsymbol{\Theta}_\text{K})$;
\State Set $\boldsymbol{Y}$ a vector of length $n$; 
    \State Set $i=1$;
    \While{$i \leq n$}  %\Comment{put some comments here}
\State Sample $\boldsymbol{u}\sim {\cal{MUL}}(1,\boldsymbol{\omega})$;
\State Set $k_{max} =\bigl\{k \vert \boldsymbol{u}[k]=1\bigr\}$ where $\boldsymbol{u}[k]$ is the $k$th element of vector $\boldsymbol{u}$ for 
\State $k=1,\cdots,{\text{K}}$;
\State Sample $\boldsymbol{y}$ from PDF $f\bigl(\cdot\bigl \vert \boldsymbol{\Theta}_{k_{max}}\bigr)$
\State Set $\boldsymbol{Y}[i] \leftarrow \boldsymbol{y}$;
    \State Set $i \leftarrow i+1$;
        %\Comment{another comment}
        %\State $var3 \leftarrow var4$
             \EndWhile  %\label{roy's loop}
             \State {\bf{end}}
        \State  Accept $\boldsymbol{Y}$ as sample of size $n$ from ${\text{K}}$-component mixture model. 
%\EndProcedure
%\State {\bf{end procedure}}
\end{algorithmic}
\end{algorithm}
\begin{example}%\lipsum*[]
Suppose we are interested in generating a sample of $n=300$ observations from 
observations are coming from a three-component bivariate Gaussian mixture model with PDF given by
\begin{align}\label{mixture-three-component-Gaussian-1}
g(\boldsymbol{y} \vert \boldsymbol{\Psi})= \frac{1}{3}{\cal{N}}_{2}\bigl(\boldsymbol{\mu}_{1},\Sigma_{1}\bigr)+\frac{1}{3}{\cal{N}}_{2}\bigl(\boldsymbol{\mu}_{2},\Sigma_{2}\bigr)+\frac{1}{3}{\cal{N}}_{2}\bigl(\boldsymbol{\mu}_{3},\Sigma_{3}\bigr),
\end{align}
where $\boldsymbol{\Psi}=\bigl(\omega_1,\omega_2,\omega_3,\Theta_1,\Theta_2,\Theta_3\bigr)$ is whole parameter vector whose elements are
\begin{align}\label{mixture-three-component-Gaussian-2}
&\Theta_1=\omega_1=\frac{1}{3}, \boldsymbol{\mu}_1=(-3,-3)^{\top},
\Sigma_1=\biggl[\begin{matrix} 1&0.5\\ 0.5& 1 \end{matrix}\biggr], \nonumber\\
&\Theta_2=\omega_2=\frac{1}{3}, \boldsymbol{\mu}_2=(0,0)^{\top},~~~~~
\Sigma_2=\biggl[\begin{matrix} 1&0\\ 0& 1 \end{matrix}\biggr], \nonumber\\
&\Theta_3=\omega_3=\frac{1}{3}, \boldsymbol{\mu}_3=(3,3)^{\top},~~~~~
\Sigma_3=\biggl[\begin{matrix} 1&-0.25\\ -0.25& 1 \end{matrix}\biggr].
\end{align}
Based on Algorithm \ref{Simulating from a mixture model-2}, we use the following \verb+R+ code to generate from three-component bivariate Gaussian mixture model with whole parameter given as above. Figure \ref{plot-mixture-three-component-determining-BIC}(a) displays the corresponding scatterplot.
\end{example}
\begin{lstlisting}[style=deltaj]
R> library(MASS)  # for generating realization from multivariate Gaussian 
R> set.seed(20240713)
R> K <- 3
R> omega <- rep(1/3, K)
R> Mu1 <- rep(-3, p); Mu2 <- rep(0, p); Mu3 <- c(3, -3)
R> Sigma1 <- diag(0.5, p) + matrix(.5, p, p) 
R> Sigma2 <- diag(1, p) 
R> Sigma3 <- diag(1.5, p) + matrix(-0.5, p, p)
R> Sigma2 <-matrix(c(1, 0, 0, 1), 2, 2);
R> Sigma3 <-matrix(c(1,-0.25,-0.25,1), 2, 2);
R> Mu <- list(Mu1, Mu2, Mu3)
R> Sigma <- list(Sigma1, Sigma2, Sigma3)
R> rmixnorm <- function(n, Mu, Sigma, omega)
+	{
+	p <- length( Mu[[1]] )
+	label <- rep(0, n)
+	Y <- matrix(0, nrow = n, ncol = p)
+		for(i in 1:n)
+		{
+		r_MUL <- rmultinom(n = 1, size = 1, omega)
+		k <- apply(r_MUL, 2, which.max)
+		label[i] <- k
+		Y[i, ] <- mvrnorm(n = 1, mu = Mu[[k]], Sigma = Sigma[[k]])
+		}
+	ls <- list( "sample" = Y, "label" = label )
+	return(ls)
+	}
R> Y <- rmixnorm(n, Mu, Sigma, omega)
R> plot( Y$sample, col = Y$label )
\end{lstlisting}
\subsection{Model selection}
\begin{figure}
\center
\includegraphics[width=55mm,height=55mm]{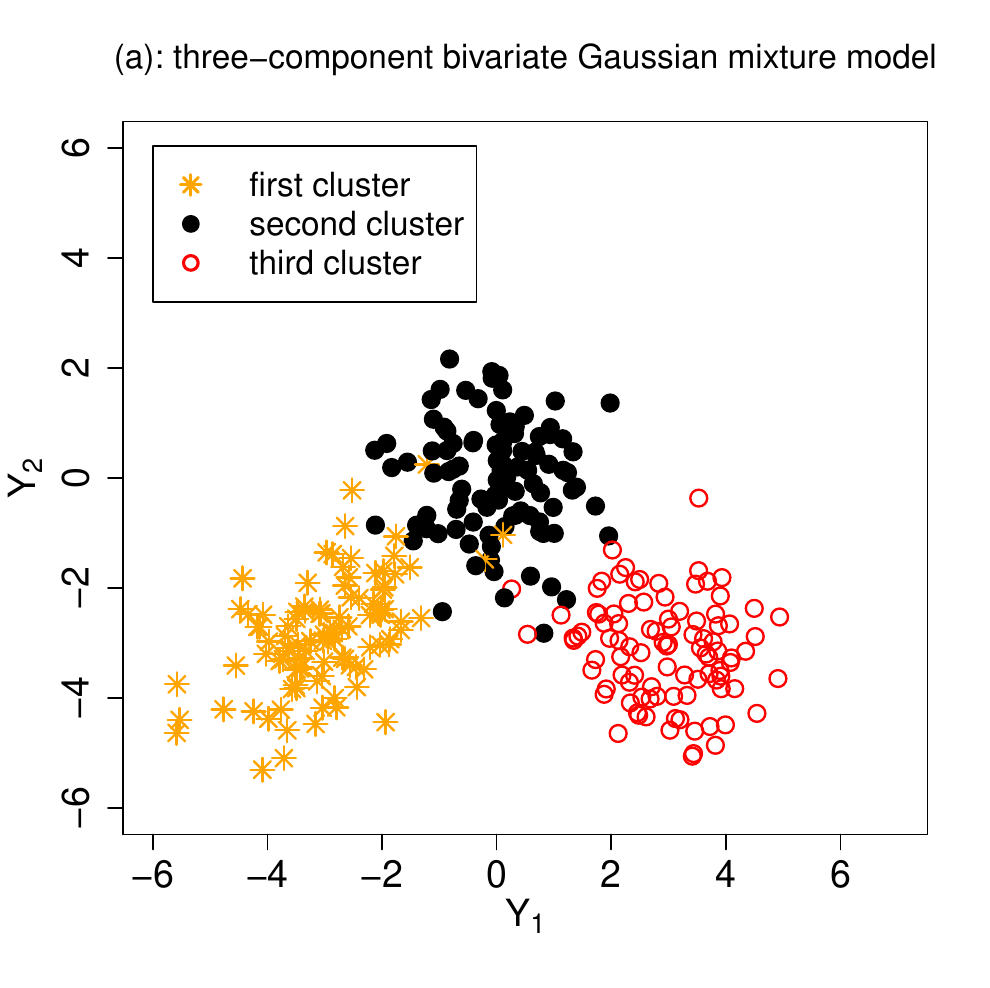}
\includegraphics[width=55mm,height=55mm]{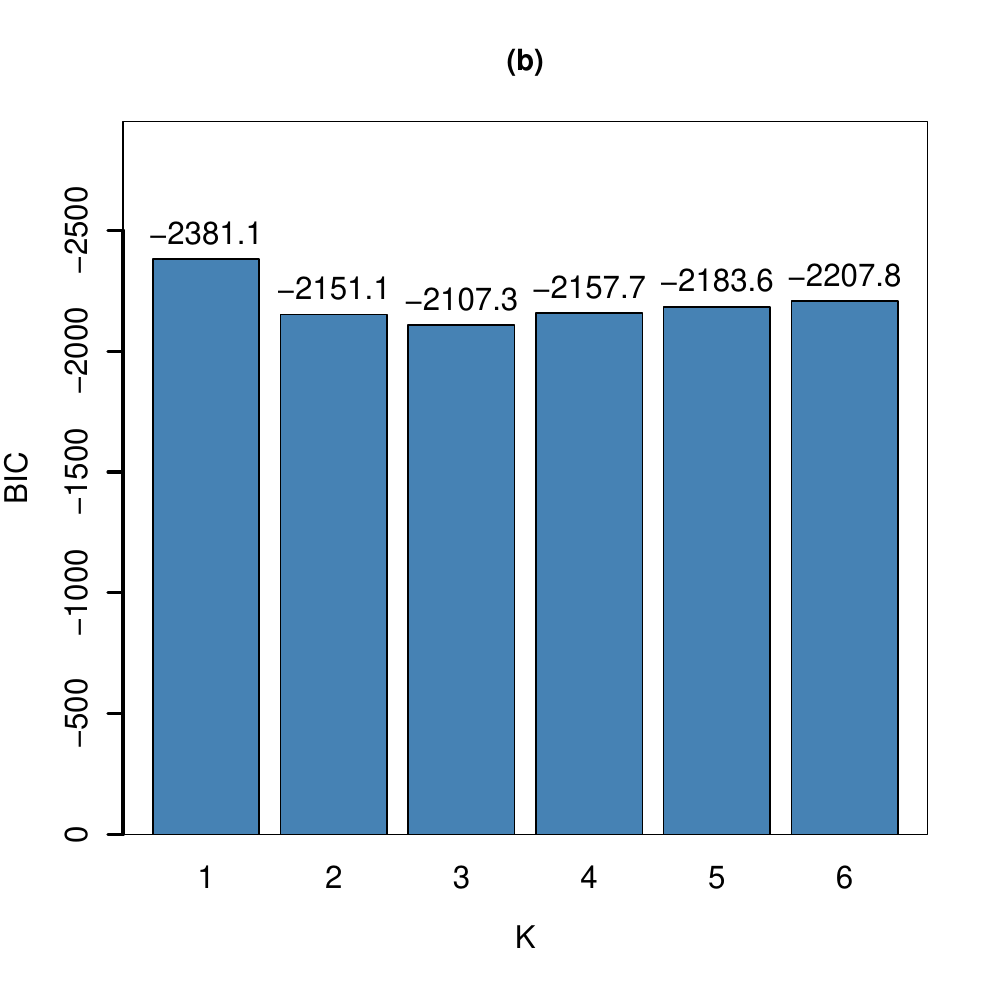}
\caption{(a): Scatterplot of 300 data points drawn from $\boldsymbol{Y}=(Y_1,Y_2)^{\top}$ following a three-component Gaussian mixture model and (b): Computed BIC versus number of clusters. 
%For comparison purposes, the constant 1350 is added to BIC. 
}
\label{plot-mixture-three-component-determining-BIC}
\end{figure}

The true value of ${\text{K}}$ is not known in practice. This quantity can be determined through the Bayesian information criterion (BIC) \citep{schwarz1978estimating}. The process of finding best ${\text{K}}$ is known as the {\it{model selection}} in the literature. Based on $n$ realizations $\boldsymbol{y}_{1},\cdots, \boldsymbol{y}_{n}$ that are assumed to follow independently a distribution with PDF $g(\boldsymbol{y} \vert \boldsymbol{\Psi})$, the BIC is defined as
\begin{align}\label{BIC}
\text{BIC}= 2\sum_{i=1}^{n}\log \sum_{k=1}^{{\text{K}}} \widehat{\omega_{k}} f\bigl(\boldsymbol{y}_{i}\big \vert \widehat{\boldsymbol{\Theta}_k} \bigr)-N \log n,
\end{align}
where $\widehat{{\omega}_k}$ and $\widehat{\boldsymbol{\Theta}_k}$ (for $k=1,\cdots,{{\text{K}}} $) account for the ML estimators of ${{\omega}}_k$ and ${\boldsymbol{\Theta}}_k$, respectively. Furthermore, constant $N$ denotes number of model's free parameters that must be estimated. The BIC is an effective criterion for model selection in Gaussian mixture models, see \cite{fraley2002model,mcnicholas2008parsimonious}, and has been used frequently for other model-based clustering, see\cite{keribin2000consistent,fraley1998many,fraley2002model,
mcnicholas2008parsimonious,mcnicholas2010model}. In practice, for determining ${{\text{K}}}$, we compute BIC defined by (\ref{BIC}) when data points are clustered through other clustering approaches such as k-mean \citep{hartigan1979algorithm} or other approaches of clustering, for $k=1,2,\cdots$. We choose the mixture model for which BIC is the largest. For instance, suppose that we do not know the true value of ${{\text{K}}}$ when 300 observations are coming from a three-component bivariate Gaussian mixture model with true parameters given by (\ref{mixture-three-component-Gaussian-2}). As expected, when ${{\text{K}}}=3$ the BIC gets its maximum value, see Figure \ref{plot-mixture-three-component-determining-BIC}(b). The \verb+R+ function \verb+BIC+ has been developed for determining the true value of ${{\text{K}}}$ for a ${{\text{K}}}$-component Gaussian mixture model.
\begin{lstlisting}[style=deltaj]
R>  dmvnorm <- function(x, mean, sigma)
+  {
+   p <- length(mean)
+   1/sqrt((2*pi)^p*det(sigma))*exp(-mahalanobis(x, mean, sigma)/2)
+  } 
R> BIC <- function(Y, param = NULL, k_max = 6)
+ {
+ dim_y <- dim(Y)
+ n <- dim_y[1]
+ p <- dim_y[2]
+ if( is.null(param) )
+ {
+ 	bic <- rep(0, k_max) 
+ 	for(k in 1:k_max)
+ 	{
+ 	 Mu_hat <- matrix(0, nrow = k, ncol = p)
+ 	 Sigma_hat <- array(0, dim = c(p, p, k) )
+ 	 N <- (k - 1) + k*( p + p*(p + 1)/2 )
+ 	 cluster <- kmeans(Y, k)
+ 	 omega <- cluster$size/n
+ 	 Sum_bic <- matrix(0, nrow = n, ncol = k)
+ 		for(i in 1:k)
+ 		{
+ 		  clust <- Y[cluster$cluster == i, ]
+ 		  Mu_hat[i,]     <- apply(clust, 2, mean)
+ 		  Sigma_hat[,,i] <- cov(clust)
+ 		  Sum_bic[, i]   <- omega[i]*dmvnorm(Y, mean = Mu_hat[i,], 
+																				sigma = Sigma_hat[,,i]) 
+ 		}
+ 	bic[k] <- 2*sum( log( apply(Sum_bic, 1, sum) ) ) - N*log(n)
+ 	}
+ 	out <- data.frame(k = 1:k, bic = bic)
+ 	ls <- list( "model" = out, "K" = which.max(bic) )
+ }else{
+ 	omega <- param[[1]]
+ 	Mu <- param[[2]]
+ 	Sigma <- param[[3]]
+ 	K <- length( Mu )
+ 	Sum_bic <- matrix(0, nrow = n, ncol = K)
+ 	N <- (K - 1) + K*( p + p*(p + 1)/2 )
+ 		 for(k in 1:K)
+ 		 {
+ 		  Sum_bic[, k] <- omega[k]*dmvnorm(Y, mean = Mu[[k]],
+							 												sigma = Sigma[[k]]) 
+ 		 }
+ 	out <- 2*sum( log( apply(Sum_bic, 1, sum) ) ) - N*log(n)
+ 	ls <- list( "bic" = out, "K" = K )
+ }
+ return(ls)
+ }
\end{lstlisting}

\par As noted earlier, determining the label of each observation is the main task of each clustering technique. To this end, we may use the Bayesian (Gibbs sampling) or the expectation-maximization (EM) \citep{dempster1977maximum} algorithm. If the former applied, depending on the model structure, the Gibbs sampling involves one, two, or more latent variables. 
\subsection{Finite mixture of Gaussian distributions}
Evidently the most commonly used statistical distribution for the purpose of model-based clustering is Gaussian mixture model. Suppose $p$-dimensional random vector $\boldsymbol{Y}$ follows a ${\text{K}}$-component Gaussian mixture model. It is clear that
\begin{align*}
\boldsymbol{Y} \big \vert H_{ik}=1&\sim {\cal{N}}_{p}\bigl(\boldsymbol{\mu}_k, \Sigma_k  \bigr),\\
\boldsymbol{H}_i&\sim{\cal{MUL}}\bigl(\boldsymbol{\omega}\big \vert 1\bigr).
\end{align*}
Since $\boldsymbol{H}_i$ is latent variable, hence we encounter with a incomplete data problem. Here $\{\boldsymbol{y}_1,\cdots,\boldsymbol{y}_n\}$ is the sequence of observed data and $\boldsymbol{H}_i=(H_{i1},\cdots,H_{i\text{K}})^{\top}$ is the  missing component's label. Herein ${H}_{ik}=1$ takes on values zero and one. If ${H}_{ik}=1$, then $\boldsymbol{y}_i$ comes from $k$-th component, otherwise $H_{i1}=0$. Hence, the complete data PDF can be represented as 
\begin{align} \label{mixture-three-component-Gaussian-4}
g_c({\underline{\boldsymbol{y}}}\vert \boldsymbol{\Psi})=\prod^{n}_{i=1} g\bigl(\boldsymbol{y}_i, \boldsymbol{H}_i \big \vert \boldsymbol{\Psi}\bigr), 
\end{align} 
where ${\underline{\boldsymbol{y}}}= \{\boldsymbol{y}_1,\cdots, \boldsymbol{y}_{n}\}$ and
\begin{align} \label{mixture-three-component-Gaussian-5}
g\bigl(\boldsymbol{y}_i, \boldsymbol{H}_{i} \big \vert \boldsymbol{\Psi}\bigr)&= 
g\bigl(\boldsymbol{y}_i\big \vert \boldsymbol{H}_{i},\boldsymbol{\Psi}\bigr) \times{\cal{MUL}}\bigl(\boldsymbol{\omega}\big \vert 1\bigr)\nonumber\\
&=\biggl[{\cal{N}}_{p}\Bigl(\boldsymbol{y}_{i}\Big \vert \boldsymbol{\mu}_{k},\Sigma_{k}\Bigr)\biggr]^{H_{ik}}\times\omega_{1}^{H_{i1}}\times \cdots \times \omega_{{\text{K}}}^{H_{i{\text{K}}}}\nonumber\\
&=\prod_{k=1}^{{\text{K}}}\biggl[\omega_{k}{\cal{N}}_{p}\Bigl(\boldsymbol{y}_{i}\Big \vert \boldsymbol{\mu}_{k},\Sigma_{k}\Bigr)\biggr]^{H_{ik}}.
\end{align} 
Replacing the RHS of (\ref{mixture-three-component-Gaussian-5}) into the RHS of (\ref{mixture-three-component-Gaussian-4}), the PDF of complete-data is
\begin{align}\label{mixture-three-component-Gaussian-6}
g_c({\underline{\boldsymbol{y}}}\vert \boldsymbol{\Psi}) = \prod^{n}_{i=1}\prod_{k=1}^{{\text{K}}}\biggl[\omega_{k}{\cal{N}}_{p}\Bigl(\boldsymbol{y}_{i}\Big \vert \boldsymbol{\mu}_{k},\Sigma_{k}\Bigr)\biggr]^{H_{ik}}.
\end{align}
Thus, the complete-data likelihood function is
\begin{align}\label{mixture-three-component-Gaussian-7}
L_{c}(\boldsymbol{\Psi}\vert {\underline{\boldsymbol{y}}},\underline{\boldsymbol{H}}) =&(2\pi)^{-\frac{np}{2}} \times \omega_{k}^{\sum_{i=1}^{n}H_{ik}}\times\cdots\times \omega_{{\text{K}}}^{\sum_{i=1}^{n}H_{i{\text{K}}}}
\times \big \vert\Sigma_{k} \big\vert^{-\frac{\sum_{i=1}^{n}\sum_{k=1}^{{\text{K}}}H_{ik}}{2}}\nonumber\\
&\times \exp\Bigl\{-\frac{1}{2}\sum_{i=1}^{n}\sum_{k=1}^{{\text{K}}}H_{ik}\bigl(\boldsymbol{y}_{i}-\boldsymbol{\mu}_{k}\bigr)^{\top}\Sigma^{-1}_{k}\bigl(\boldsymbol{y}_{i}-\boldsymbol{\mu}_{k}\bigr)\Bigr\},\nonumber\\
\propto& \omega_{1}^{n_{1}}\times\cdots\times \omega_{{\text{K}}}^{n_{\text{K}}}
 \times\big \vert\Sigma_{1} \big\vert^{-\frac{n_{1}}{2}}\times\cdots \times \big \vert\Sigma_{{\text{K}}} \big\vert^{-\frac{n_{{\text{K}}}}{2}}\nonumber\\
&\times \exp\Bigl\{-\frac{1}{2}\sum_{i=1}^{n}\sum_{k=1}^{{\text{K}}}H_{ik}\bigl(\boldsymbol{y}_{i}-\boldsymbol{\mu}_{k}\bigr)^{\top}\Sigma^{-1}_{k}\bigl(\boldsymbol{y}_{i}-\boldsymbol{\mu}_{k}\bigr)\Bigr\},
\end{align}
where $\underline{\boldsymbol{H}}= \{\boldsymbol{H}_1,\cdots, \boldsymbol{H}_{n}\}$ in which $\boldsymbol{H}_i=(H_{i1},\cdots,H_{i\text{K}})^{\top}$ and $n_{k}=\sum_{i=1}^{n}H_{ik}$ provided that $n=\sum_{k=1}^{{\text{K}}}n_{k}$. Given $i$th observed data $\boldsymbol{y}_{i}$, for computing PDF of $\boldsymbol{H}_{i} \vert \boldsymbol{Y}=\boldsymbol{y}_{i}$, it follows from (\ref{mixture-three-component-Gaussian-5}) that
\begin{align*}%\label{mixture-three-component-Gaussian-8}
L_{c}\bigl(\boldsymbol{\Psi} \big \vert {\boldsymbol{y}}_{i}, {\boldsymbol{H}}_{i}\bigr)\propto&g\bigl(\boldsymbol{y}_i\big \vert \boldsymbol{H}_{i},\boldsymbol{\Psi}\bigr) \times{\cal{MUL}}\bigl(\boldsymbol{\omega}\big \vert 1\bigr)\nonumber.
\end{align*}
Obviously, 
\begin{align}\label{mixture-three-component-Gaussian-9}
\boldsymbol{H}_{i} \big \vert \boldsymbol{\Psi}, \boldsymbol{Y}=\boldsymbol{y}_{i}  \sim {\cal{MUL}}\bigl(\boldsymbol{\tau}_{i}\big \vert 1\bigr),
\end{align}
where $\boldsymbol{\tau}_{i}=(\tau_{i1},\cdots,\tau_{i\text{K}})^{\top}$ denotes the $i$th row of $n\times {\text{K}}$ matrix  $\boldsymbol{\tau}$ whose $(i,k)$th element is
\begin{align}\label{mixture-three-component-Gaussian-10}
{\tau}_{i,k}=\frac{\omega_{k} {\cal{N}}_{p}\bigl(\boldsymbol{y}_{i}\big \vert \boldsymbol{\mu}_{k},\Sigma_{k}\bigr)}{\sum_{k=1}^{\text{K}}\omega_{k} {\cal{N}}_{p}\bigl(\boldsymbol{y}_{i}\big \vert \boldsymbol{\mu}_{k},\Sigma_{k}\bigr)}.
\end{align}
%We need to generate form posterior (\ref{mixture-three-component-Gaussian-9}) for $i=1,\cdots,n$, for implementing the Bayesian paradigm. The posterior structure is 
%${\boldsymbol{H}_{i}}$, we consider the priors as follows. 
% \begin{align} 
%\pi\bigl(\boldsymbol{u}_{i} \big \vert \boldsymbol{y}_{i},g_{i},\boldsymbol{\Psi}\bigr) ={\cal{HN}}_{q}\Bigl(\cdot \big \vert \boldsymbol{m}_{i}, \frac{\boldsymbol{\Delta}}{{g_i}}\Bigr),
%\end{align}
%where $\boldsymbol{m}_{i}=\boldsymbol{\Lambda}^{\top}\Omega^{-1}(\boldsymbol{y}_{i}-\boldsymbol{\mu})$ and the short form ${\cal{HN}}_{q}(\cdots \big \vert \boldsymbol{\mu}, \Sigma)$ accounts for the PDF of a half-normal distribution truncated on ${{\mathbb{R}}}^{q+}$ with mean vector $\boldsymbol{\mu}$ and scale matrix $\Sigma$. 
Once we have generated the entire sample (for $i=1,\cdots,n$) from full conditional $\boldsymbol{H}_{i}\big \vert \bigl(\boldsymbol{\Psi}, {\boldsymbol{y}}_{i} \bigr)$, we proceed to generate from full conditionals 
$\boldsymbol{\mu}_{k}\big \vert \bigl({\underline{\boldsymbol{y}}},\underline{\boldsymbol{H}},\boldsymbol{\Psi}_{(-\boldsymbol{\mu}_{k})}\bigr)$ and
$\Sigma_{k} \big \vert \bigl({\underline{\boldsymbol{y}}},\underline{\boldsymbol{H}},\boldsymbol{\Psi}_{(-{\Sigma}_{k})}\bigr)$ (for $k=1,\cdots,{\text{K}}$). To this end, first we have
\begin{align}\label{mixture-three-component-Gaussian-11}
\pi(\boldsymbol{\Psi} \vert {\underline{\boldsymbol{y}}},\underline{\boldsymbol{H}})&\propto L_{c}(\boldsymbol{\Psi}\vert\underline{\boldsymbol{y}},\underline{\boldsymbol{H}})\times\pi\bigl(\boldsymbol{\Psi}_{1}\bigr)\times\cdots \times \pi\bigl(\boldsymbol{\Psi}_{{\text{K}}}\bigr)\nonumber\\
&= L_{c}(\boldsymbol{\Psi}\vert\underline{\boldsymbol{y}},\underline{\boldsymbol{H}})\times\pi\bigl(\boldsymbol{\omega}\bigr)\times \pi\bigl(\boldsymbol{\mu}_{1}\bigr)\times\cdots\times \pi\bigl(\boldsymbol{\mu}_{{\text{K}}}\bigr)\times\pi\bigl(\Sigma_{1}\bigr)\times\cdots \times \pi\bigl(\Sigma_{{\text{K}}}\bigr).
\end{align}
Substituting $L_{c}(\boldsymbol{\Psi}\vert\underline{\boldsymbol{y}},\underline{\boldsymbol{H}})$ given by (\ref{mixture-three-component-Gaussian-7}) into the RHSS of (\ref{mixture-three-component-Gaussian-11}) yields
\begin{align}\label{mixture-three-component-Gaussian-12}
\pi(\boldsymbol{\Psi} \vert {\underline{\boldsymbol{y}}},\underline{\boldsymbol{H}})\propto&
\times\pi\bigl(\boldsymbol{\omega}\bigr)\times \pi\bigl(\boldsymbol{\mu}_{1}\bigr)\times\cdots\times \pi\bigl(\boldsymbol{\mu}_{{\text{K}}}\bigr)\times\pi\bigl(\Sigma_{1}\bigr)\times\cdots \times \pi\bigl(\Sigma_{{\text{K}}}\bigr) \times\nonumber\\
&\omega_{1}^{n_{1}}\times\cdots\times \omega_{{\text{K}}}^{n_{\text{K}}}
 \times\big \vert\Sigma_{1} \big\vert^{-\frac{n_{1}}{2}}\times\cdots \times \big \vert\Sigma_{{\text{K}}} \big\vert^{-\frac{n_{{\text{K}}}}{2}}\nonumber\\
&\times \exp\Bigl\{-\frac{1}{2}\sum_{i=1}^{n}\sum_{k=1}^{{\text{K}}}H_{ik}\bigl(\boldsymbol{y}_{i}-\boldsymbol{\mu}_{k}\bigr)^{\top}\Sigma^{-1}_{k}\bigl(\boldsymbol{y}_{i}-\boldsymbol{\mu}_{k}\bigr)\Bigr\}.
\end{align}
We consider, for $k=1,\cdots,{\text{K}}$, independent conjugate priors as follows.
\begin{align}
\pi(\boldsymbol{\omega})&\sim {\cal{DIR}}\bigl(\boldsymbol{\rho}\bigr),\label{mixture-three-component-Gaussian-13}\\
\pi\bigl(\boldsymbol{\mu}_{k}\bigr)&\sim {\cal{N}}_{p}\bigl({\cal{M}}_{k}, {\cal{S}}_{k}\bigr),\label{mixture-three-component-Gaussian-14}\\
\pi\bigl(\Sigma_{k}\bigr)&\sim {\cal{IW}}\bigl({\cal{D}}_{k}, \nu_k\bigr),\label{mixture-three-component-Gaussian-15}
\end{align}
where ${\cal{DIR}}\bigl(\cdot\bigr)$ is defined in (\ref{Dirichlet-PDF}). In what follows, we give the full conditionals needed for implementing the Gibbs sampling.
\begin{itemize}
\item {\bf{full conditional of $\boldsymbol{\omega} \big \vert \bigl({\underline{\boldsymbol{y}}},\underline{\boldsymbol{H}},\boldsymbol{\Psi}_{(-\boldsymbol{\omega})}\bigr)$:}} Substituting prior $\pi(\boldsymbol{\omega})$ given by (\ref{mixture-three-component-Gaussian-12}) into the RHS of (\ref{mixture-three-component-Gaussian-11}) yields
\begin{align*}
\pi\bigl(\boldsymbol{\omega} \big \vert {\underline{\boldsymbol{y}}},\underline{\boldsymbol{H}}, \boldsymbol{\Psi}_{(-\boldsymbol{\mu})}\bigr) &\propto {\cal{DIR}}\bigl(\boldsymbol{\omega} \big \vert\boldsymbol{\rho}\bigr)\times \omega_{1}^{n_{1}}\times\cdots\times \omega_{{\text{K}}}^{n_{\text{K}}}= {\cal{DIR}}\bigl(\boldsymbol{\omega} \big \vert\boldsymbol{\rho}_{*}\bigr),\nonumber
\end{align*}
where
\begin{align*}
\boldsymbol{\rho}_{*}=\Bigl(\rho_{1}+\sum_{i=1}^{n}H_{i1},\cdots,\rho_{{\text{K}}}+\sum_{i=1}^{n}H_{i{\text{K}}}\Bigr)^{\top}.
\end{align*}
\item {\bf{full conditional of $\boldsymbol{\mu}_{k} \big \vert \bigl({\underline{\boldsymbol{y}}},\underline{\boldsymbol{H}},\boldsymbol{\Psi}_{(-\boldsymbol{\mu}_{k})}\bigr)$:}}  Substitute the prior $\pi\bigl(\boldsymbol{\mu}_{k}\bigr)$ given by (\ref{mixture-three-component-Gaussian-13}) into the RHS of (\ref{mixture-three-component-Gaussian-11}). Then more algebra shows
\begin{align*}
\pi\bigl(\boldsymbol{\mu} \big \vert {\underline{\boldsymbol{y}}},\underline{\boldsymbol{H}}, \boldsymbol{\Psi}_{(-\boldsymbol{\mu}_{k})}\bigr)={\cal{N}}_{p}\Bigl(\boldsymbol{\mu} \Big \vert {\cal{Q}}_{k}\Bigl[{\cal{S}}^{-1}_{k}{\cal{M}}_{k}+\Sigma_{k}^{-1}\sum_{i=1}^{n}H_{ik}\boldsymbol{y}_{i}\Bigr],{\cal{Q}}_{k}\Bigr),
\end{align*}
where
\begin{align*}
{\cal{Q}}_{k}=\Bigl[{\cal{S}}^{-1}_{k}+\Sigma_{k}^{-1}\times \sum_{i=1}^{n}H_{ik}\Bigr]^{-1}.
\end{align*}
%%%%%%%%%%%%%%%%%%%%%
\item {\bf{full conditional of 
${\Sigma}_{k} \big \vert \bigl({\underline{\boldsymbol{y}}},\underline{\boldsymbol{H}},\boldsymbol{\Psi}_{(-\Sigma_{k})}\bigr)$:}} Substitute the prior $\pi\bigl({\Sigma}_{k}\bigr)$ given by (\ref{mixture-three-component-Gaussian-14}) into the RHS of (\ref{mixture-three-component-Gaussian-11}). Then more algebra shows 
\begin{align*}
\pi\bigl({\Sigma} \big \vert {\underline{\boldsymbol{y}}},\underline{\boldsymbol{H}}, \boldsymbol{\Psi}_{(-{\Sigma}_{k})}\bigr) \sim  {\cal{IW}}\bigl({\cal{R}}_{k}, \nu_{k}+\sum_{i=1}^{n} H_{ik}\bigr),
\end{align*}
where ${\cal{R}}_{k}={\cal{D}}_{k} +\sum_{i=1}^{n} H_{ik} \bigl(\boldsymbol{y}_i-\boldsymbol{\mu}_{k}\bigr)\bigl(\boldsymbol{y}_i-\boldsymbol{\mu}_{k}\bigr)^{\top}$.
\end{itemize}
The following pseudo code gives the details for implementing the model-based clustering for finite mixture of Gaussian distributions.
\begin{algorithm} 
\caption{Bayesian inference for finite mixture of Gaussian distributions}
    \label{Bayesian inference for finite mixture of Gaussian distributions}
\begin{algorithmic}[1]
\State Read $N$, $M$, and determine quantities ${\text{K}}$, ${\omega}_{k}$, ${\cal{M}}_{k}$, ${\cal{S}}_{k}$, ${\cal{D}}_{k}$, $\nu_{k}$, for $k=1\cdots,{\text{K}}$ 
\State Set $t = 0$;
\State Determine initial values $\boldsymbol{\omega}^{(0)}$, $\boldsymbol{\mu}^{(0)}$, $\Sigma^{(0)}$, and set $\boldsymbol{\Psi}^{(0)}=\bigl(\boldsymbol{\omega}^{(0)}$,$\boldsymbol{\mu}^{(0)}, \Sigma^{(0)}\bigr)$;
    \While{$t \leq N$}  \Comment{start Gibbs sampling}
    \State Set $i=1$;
    \While{$i \leq n$}  \Comment{start for sampling form missing labels}
     \State Simulate the $i$ row of $n \times {\text{K}}$ matrix $\boldsymbol{H}$ as $           \boldsymbol{H}_{i}\sim{\cal{MUL}}\bigl( \cdot \big \vert \boldsymbol{\tau}_{i}
     \bigr)$ where 
    \State $\boldsymbol{\tau}_{i}=(\tau_{i1},\ldots,\tau_{i\text{K}})^{\top}$ is the $i$th row of  matrix $\boldsymbol{\tau}$ is given by (\ref{mixture-three-component-Gaussian-10});
    \State Set $i = i+1$;
             \EndWhile 
             \State {\bf{end}}
              \State Set, for $k=1,\cdots,{\text{K}}$, $n_{k}=\sum_{i=1}^{n} H_{i,k}$ where $H_{i,k}$ denotes the $(i,k)$th
              \State element of $\boldsymbol{H}$;
\State Simulate $\boldsymbol{\omega}\sim {\cal{DIR}}\bigl(1 \vert \rho_{*}\bigr)$ where $\rho_{*}=(\rho_{1} + n_{1},\cdots,\rho_{{\text{K}}} + n_{{\text{K}}})^{\top}$;
    \State Set $k=1$;
    \While{$k \leq K$}  \Comment{start sampling from full conditionals of  $\boldsymbol{\mu}_k$, and $\Sigma_k$}
        \State  Generate $\boldsymbol{\mu}_{k}^{(t+1)}\sim 
{\cal{N}}_{p}\Bigl({\cal{Q}}_{k}\Bigl[{\cal{S}}^{-1}_{k}{\cal{M}}_{k}+\bigl[\Sigma_{k}^{(t)} \bigr]^{-1}\sum_{i=1}^{n}H_{i,k}\times\boldsymbol{y}_{i}\Bigr],{\cal{Q}}_{k}\Bigr)$
\State where ${\cal{Q}}_{0}=\Bigl[{\cal{S}}^{-1}_{0}+\bigl[\Sigma_{k}^{(t)} \bigr]^{-1}\times n_{k}\Bigr]^{-1}$;
\State  Generate $\Sigma_{k}^{(t+1)}$ from
${\cal{IW}}\bigl( {\cal{R}}_{k}, \nu_{k}+n_k\bigr)$ where ${\cal{R}}_{k}={\cal{D}}_{k}+\sum_{i=1}^{n} H_{i,k} \times$ 
\State $ \bigl(\boldsymbol{y}_{i}-\boldsymbol{\mu}_{k}^{(t+1)}\bigr)\bigl(\boldsymbol{y}_{i}-\boldsymbol{\mu}_{k}^{(t+1)}\bigr)^{\top}$;
    \State Set $k = k+1$;
             \EndWhile 
            \State {\bf{end}}
            \State Set $\boldsymbol{\Psi}^{(t+1)} = \bigl(\boldsymbol{\omega}^{(t+1)}, \boldsymbol{\mu}_{1}^{(t+1)}\cdots,\boldsymbol{\mu}_{{\text{K}}}^{(t+1)}, \Sigma_{1}^{(t+1)}, \cdots,\Sigma_{{\text{K}}}^{(t+1)}\bigr)$ and $t = t+1$;
            \EndWhile
   \State {\bf{end}}
   \State  Sequence $\bigl\{\boldsymbol{\Psi}^{(M+1)},\cdots,\boldsymbol{\Psi}^{(N)}\bigr\}$ is a sample of size $N-M$ from
   \State posterior $\pi\bigl(\boldsymbol{\Psi} \vert {\underline{\boldsymbol{y}}},\underline{\boldsymbol{H}}\bigr)$;
  \State The Bayesian estimator of $\boldsymbol{\Psi}$ is given by $\frac{1}{N-M}\sum_{t=1}^{N-M}\boldsymbol{\Psi}^{(t)}$.
\end{algorithmic}
\end{algorithm} 
\par
\begin{figure}
\center
\includegraphics[width=55mm,height=55mm]{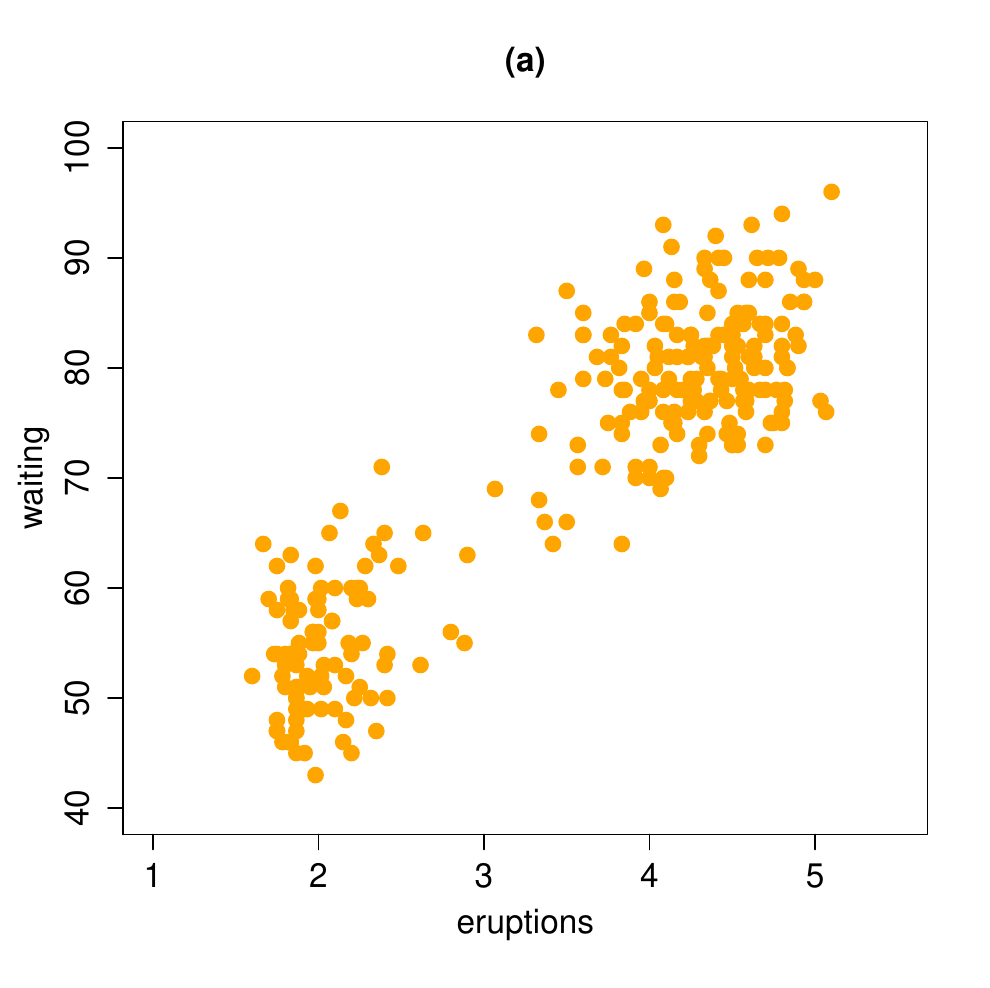}
\includegraphics[width=55mm,height=55mm]{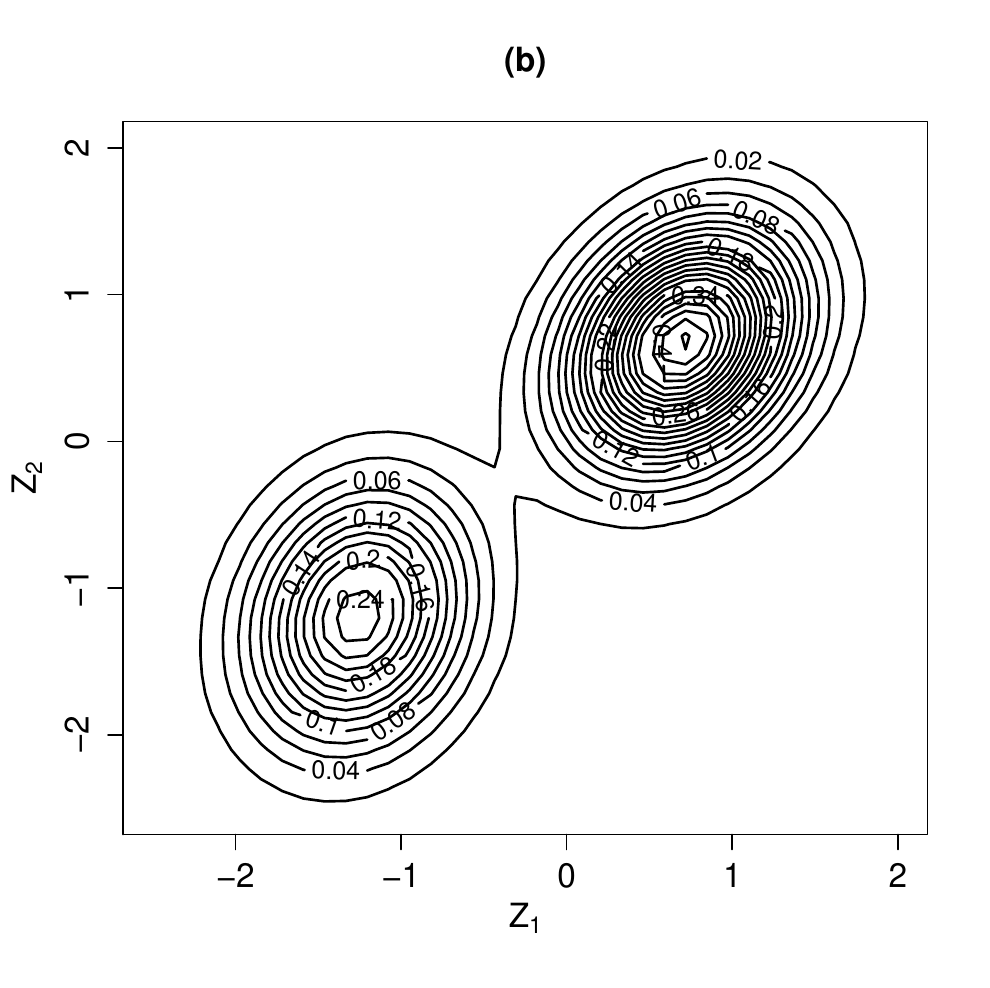}
\includegraphics[width=55mm,height=55mm]{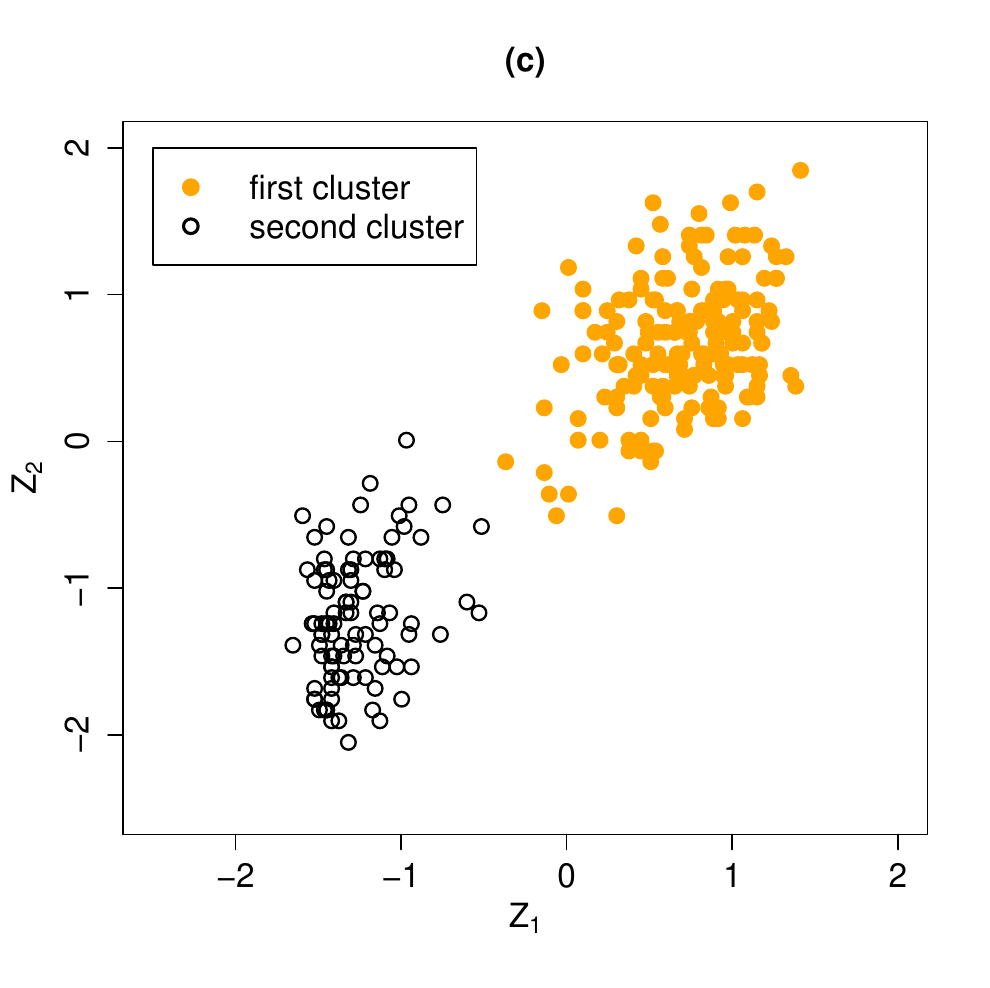}
\caption{(a): Scatterplot of faithful data, (b): contour plot fitted to standardized faithful data, and (c): scatterplot of clustered standardized faithful data. %and fitted contour plot.
}
\label{plot-mixture-two-component-Gaussian-faithful}
\end{figure}

 While Algorithm \ref{Bayesian inference for finite mixture of Gaussian distributions} works reasonably well, its performance is subject to outliers. If clusters have large variances, then the Gibbs sampling may fail to allocate true label to each cluster. This mainly because this fact that Gaussian distribution has light tails and the PDF decays exponentially for observation are far from the center of distribution. For instance, consider a two-component bivariate mixture model with equal weights $\omega_1=\omega_2=0.5$ while the second cluster has large locations and variances. It is easy to check that the elements of second column of matrix $\boldsymbol{\tau}$ in (\ref{mixture-three-component-Gaussian-10}) are very small and hence, $P(H_{i,2}=1) \rightarrow 1$ for $i=1,\cdots,n$. Hence, all observations are allocated to the first cluster. One way to tackle this issue is standardization of observations before clustering \citep{tortora2021model}. The following example gives more detailed information about this case.
\begin{example}%\lipsum*[]
Herein, we are interested in clustering the famous Old Faithful Geyser Data (known as faithful data). This faithful data consists of 272 observations on two variables {\it{duration of the eruption}} in minutes and {\it{waiting time between eruptions}} for the Old Faithful geyser in Yellowstone National Park, Wyoming, USA,  \citep{azzalini1990look}. Figure \ref{plot-mixture-two-component-Gaussian-faithful}(a) displays the scatterplot of this dataset. Our investigation shows that the BIC of -833.7747 suggests that ${\text{K}}=2$ for which the k-mean algorithm pre-clustering results shows that the variances of second variable in both clusters are significantly large as 31.66 and 34.75. Using the \verb+R+ code after standardization given by
\begin{verbatim}
R> data(faithful)
R> Y <- matrix( cbind(faithful$eruptions, faithful$waiting), 
+	      nrow = 272, ncol = 2)
R> Z <- apply( Y, 2, function(x)( x - mean(x) )/sd(x) )
R> bic <- BIC(Z)
R> out_kmeans <- kmeans(Z, bic$K)
R> label <- out_kmeans$cluster
R> sapply(1:K, function(x) var( Z[label == x, ] ))
\end{verbatim}
then, it can be seen that variances of second variable in both clusters are 0.1876  and 0.1874, correspondingly. The R package \verb+mclust+ \citep{mclust} applies the EM algorithm for model-based clustering based on the finite mixture of Gaussian distributions. We notice that the clustering results based on both the Gibbs sampling proposed here and EM algorithm proposed in \citep{mclust} are the same. The most commonly used criterion for measuring the degree of agreement between clustering approaches is the adjusted Rand index (ARI) that varies from 0 (for zero agreement in clustering) to 1 (for perfect agreement in clustering) \citep{hubert1985comparing}. Herein, we use the ARI to compare the Bayesian paradigm and EM algorithm for clustering the faithful data. To this end, among several \verb+R+ packages that can compute the ARI, we call the package \verb+MixGHD+. Fortunately, the agreement between two above approaches in clustering faithful data is perfect, that is ARI=1. Figure \ref{plot-mixture-two-component-Gaussian-faithful}(b) and (c) display the contour and scatterplot fitted to standardized faithful data. The \verb+R+ function \verb+FGM+ given by the following can be used for applying the model-based clustering to the Gaussian finite mixture model.  
\vspace{.5cm}
\begin{lstlisting}[style=deltaj]
R> FMG <- function(Y, K, n_burn, n_sim, param, hyper = NULL, scale = FALSE )
+ {
+ if( scale == TRUE ) Y <- apply( Y, 2, function(x)( x - mean(x) )/sd(x) ) 
+ dim_y <- dim(Y)
+ n <- dim_y[1]
+ p <- dim_y[2]
+ dim_Psi <- K + K*p + K*p^2
+ Psi <- matrix(0, nrow = n_sim, ncol = dim_Psi)
+ 	if( is.null(hyper) )
+ 	{
+ 	rho <- rep(1, K)
+ 	M <- rep( list( rep(0, p) ), K )
+ 	S <- rep( list( diag(1, p) ), K )
+ 	nu <- rep(1, K)
+ 	D0 <- rep( list( diag(10, p) ), K )
+ 	}else{
+ 	rho <- hyper[[1]]
+ 	M  <- hyper[[2]]
+ 	S  <- hyper[[3]]
+ 	D0 <- hyper[[4]]
+ 	nu <- hyper[[5]]
+ 	}
+ 	omega <- param[[1]]
+ 	Mu <- param[[2]]
+ 	Sigma <- param[[3]]
+ 	index_Mu <- (K + 1):(K*p + K)
+ 	index_Sigma <- (K*p + K + 1):(K*(p^2+p+1))
+ 	Psi[1, 1:K ] <- omega
+ 	Psi[1, index_Mu] <- as.numeric( unlist( Mu ) )
+ 	Psi[1, index_Sigma] <- as.numeric( unlist(Sigma) )
+ 	A <- tau <- H <- matrix(0, nrow = n, ncol = K)
+ 			for(j in 2:n_sim)
+ 			{
+ 			for(k in 1:K) A[, k] <- omega[k]*dmvnorm(Y, mean = Mu[[k]], sigma =
+                                       		     															Sigma[[k]])
+ 			tau <- A/apply(A, 1, sum)
+ 			for(i in 1:n) H[i, ] <- rmultinom(n = 1, size = 1, tau[i, ])
+ 			n_k <- apply(H, 2, sum)
+ 			omega <- rDIR( rho + n_k )
+ 			for(k in 1:K){
+ 			index_H <- which(H[, k] == 1)
+ 			Y_k <- Y[index_H, ]
+ 			Q_k <- solve( solve(S[[k]]) + solve(Sigma[[k]])*n_k[k] )
+ 			Sum_Yk <- apply(Y_k, 2, sum)
+ 			Mu_k <- Q_k%*%( solve(S[[k]])%*%M[[k]] + solve(Sigma[[k]])%*%Sum_Yk )
+ 			Mu[[k]] <- mvrnorm(n = 1, mu = Mu_k, Sigma = Q_k )
+ 			R <- matrix(0, nrow = p, ncol = p)
+ 			for(i in index_H){R <- R + c( Y[i, ] - Mu[[k]] )%o%c( Y[i, ] - Mu[[k]] )}
+ 			Sigma[[k]] <- solve( rWishart(1, df = n_k[k] + nu[k], Sigma = 
+ 																	solve(D0[[k]] + R) )[,,1] )
+ 			#print(Sigma[[k]])
+ 			}
+ 			Psi[j, 1:K ] <- omega
+ 			Psi[j, index_Mu] <- as.numeric( unlist( Mu ) )
+ 			Psi[j, index_Sigma] <- as.numeric( unlist(Sigma) )
+ 			}
+ 		omega_hat <- rep(1/K, K)
+ 		Mu_hat    <- rep( list( rep(0, p) ), K )
+ 		Sigma_hat <- rep( list( diag(p) ), K )
+ 		out <- apply(Psi[n_burn:n_sim, ], 2, mean)
+ 		for(k in 1:K)
+   		{
+ 		omega_hat[k]   <- out[k]
+ 		Mu_hat[[k]]    <- out[index_Mu[((k-1)*p + 1):(k*p)]]
+ 		Sigma_hat[[k]] <- matrix( out[index_Sigma[((k-1)*p^2 + 1):(k*p^2)]], 
+																					nrow = p, ncol = p )
+ 		A[, k] <- omega_hat[k]*dmvnorm(Y, mean = Mu_hat[[k]], sigma = 
+													Sigma_hat[[k]])
+ 		}
+ label <- apply(A, 1, which.max)
+ ls <- list( "omega" = omega, "Mu" = Mu_hat, "Sigma" = Sigma_hat,
+													 "label" = label  )
+ return(ls)
+ }
\end{lstlisting}
%{\tiny{
%\begin{verbatim}
%$omega
%[1] 0.6137622 0.3862378
%
%$Mu
%$Mu[[1]]
%[1] 0.7060379 0.6705480
%
%$Mu[[2]]
%[1] -1.259982 -1.197933
%
%
%$Sigma
%$Sigma[[1]]
%          [,1]      [,2]
%[1,] 0.1880074 0.0576521
%[2,] 0.0576521 0.2524663
%
%$Sigma[[2]]
%           [,1]       [,2]
%[1,] 0.16978499 0.03609329
%[2,] 0.03609329 0.30141474
%
%
%$label
%  [1] 1 2 1 2 1 2 1 1 2 1 2 1 1 2 1 2 2 1 2 1 2 2 1 1 1 1 2 1 1 1 1 1 1 1 1 2 2 1 2 1 1 2
% [43] 1 2 1 1 1 2 1 2 1 1 2 1 2 1 1 2 1 1 2 1 2 1 2 1 1 1 2 1 1 2 1 1 2 1 2 1 1 1 1 1 1 2
% [85] 1 1 1 1 2 1 2 1 2 1 2 1 1 1 2 1 2 1 2 1 1 2 1 2 1 1 1 2 1 1 2 1 2 1 2 1 2 1 1 2 1 1
%[127] 2 1 2 1 2 1 2 1 2 1 2 1 2 1 1 2 1 1 1 2 1 2 1 2 1 1 2 1 1 1 1 1 2 1 2 1 2 1 1 1 2 1
%[169] 2 1 2 2 1 1 1 1 1 2 1 1 2 1 1 1 2 1 1 2 1 2 1 2 1 1 1 1 1 1 2 1 2 1 1 2 1 2 1 1 2 1
%[211] 2 1 2 1 1 1 2 1 2 1 2 1 2 1 1 1 1 1 1 1 1 2 1 2 1 2 2 1 1 2 1 2 1 2 1 1 2 1 2 1 2 1
%[253] 1 1 1 1 1 1 2 1 1 1 2 1 2 2 1 1 2 1 2 1
%\end{verbatim}}}
\end{example}
\section{Simulating from truncated Gaussian distribution}\label{Simulating from truncated Gaussian distribution}
A mentioned in Section \ref{trunc-mom}, the truncated distribution is widely used in practice. Generating from truncated multivariate Gaussian distribution may arise in various contexts sch as probit regression \citep{albert1993bayesian}, likelihood estimation for max-stable processes \citep{genton2011likelihood}, fitting mixed effects models with censored data \citep{grun2012modelling}, model-based clustering \citep{maleki2019robust}, regression analysis \citep{rodriguez2004efficient}, logistic regression with covariates subject to measurement error and left censoring \citep{teimouri2024inference}. For more detailed information on application of truncated multivariate Gaussian distribution, we the refer to \cite{botev2017normal} and references therein. Section \ref{trunc-mom} gives the first and second moments of a Gaussian distribution under rectangular constraint. Suppose $p$-dimensional random vector $\boldsymbol{X}=(X_{1},\cdots,X_{p})^{\top}$ follows a truncated Gaussian distribution on region with positive Lebesgue measure $\boldsymbol{R}$ denoted as  
\begin{align}\label{truncated-Gaussian-linear-constraints-1}
{\cal{TN}}_{\boldsymbol{R}}(\boldsymbol{\mu},\Sigma),~~\boldsymbol{R}=\bigl\{ \boldsymbol{x}\in \mathbb{R}^{p} \big \vert \boldsymbol{a} \leq A \boldsymbol{x} \leq \boldsymbol{b}\bigr\},
\end{align}
where $\boldsymbol{a}=(a_{1},\cdots,a_{p})^{\top}$ and $\boldsymbol{b}=(b_{1},\cdots,b_{p})^{\top}$ are vectors of finite or infinite constants. Moreover, matrix $A$ is a $p\times p$ full rank matrix that constructs a set of $p$ independent linear constraints. One can investigate easily, as pointed out by \citep{geweke1991efficient}, that the marginals of $\boldsymbol{X}$ with distribution given by (\ref{truncated-Gaussian-linear-constraints-1}) are not truncated Gaussian, but the full conditionals, that is distribution of $X_{i}$ (for $i=1,\cdots,p$) {\it{conditional}} on all of other members of $\boldsymbol{X}$ are truncated Gaussian. To verify this claim, we shall confine ourselves here to suppose $p=2$ and $A=\boldsymbol{I}_{p}$ in (\ref{truncated-Gaussian-linear-constraints-1}). This means that
\begin{align}\label{truncated-Gaussian-linear-constraints-2}
{\cal{TN}}_{\boldsymbol{R}}(\boldsymbol{\mu},\Sigma),~~\boldsymbol{R}=\bigl\{ \boldsymbol{x}=(x_{1},x_{2})^{\top} \in \mathbb{R}^{2} \big \vert a_{1}\leq x_{1} \leq b_{1}, a_{2}\leq x_{2} \leq b_{2}\bigr\},
\end{align}
where $\boldsymbol{\mu}=(\mu_{1},\mu_{2})^{\top}$ and $\Sigma=\bigl[(\sigma_{11},\sigma_{12})^{\top},(\sigma_{21},\sigma_{22})^{\top}\bigr]$. For computing the marginal distribution of (\ref{truncated-Gaussian-linear-constraints-2}), e.g. $X_{1}$, whose PDF is denoted by ${\cal{TN}}_{R_{1}}(x_{1}\vert \boldsymbol{\mu}, \Sigma)$, we have
\begin{align}\label{truncated-Gaussian-linear-constraints-3}
{\cal{TN}}_{R_{1}}(x_{1}\vert \boldsymbol{\mu}, \Sigma)=&\frac{1}{{\cal{P}}_{G}}\int_{a_{2}}^{b_{2}}
\boldsymbol{\phi}_{2}\bigl(\boldsymbol{x}\vert\boldsymbol{\mu},{\Sigma}\bigr)d x_{2},\nonumber\\
=&\frac{1}{{\cal{P}}_{G}}{\phi}\bigl(x_{1}\big \vert \mu_{1},\sigma_{11}\bigr)
\int_{a_{2}}^{b_{2}}{\phi}\bigl(x_{2}\big \vert {\mu}_{2}^{*}, \sigma_{22}^{*}\bigr)d x_{2},\nonumber\\
=&\frac{1}{{\cal{P}}_{G}}{\phi}\bigl(x_{1}\big \vert \mu_{1},\sigma_{11}\bigr)\Bigl[\Phi\bigl(b_{2}\big \vert {\mu}_{2}^{*}, \sigma_{22}^{*}\bigr)-\Phi\bigl(a_{2}\big \vert {\mu}_{2}^{*}, \sigma_{22}^{*}\bigr)\Bigr],
\end{align}
where $x_{1}\in R_{1}$, 
%in which $R_{i}=\bigl\{x\in \mathbb{R} \big \vert x \in [a_{i}, b_{i}]\bigr \}$ (for $i=1,2$) 
${\mu}_{2}^{*}={\mu}_{2}+\sigma_{12}\sigma_{11}^{-1}(x_{1}- {\mu}_{1})$, $\sigma_{22}^{*}=\sigma_{22}-\sigma_{12}\sigma_{11}^{-1}\sigma_{21}$, and ${\cal{P}}_{G}=\boldsymbol{\Phi}_{2}\bigl(\boldsymbol{b}\big\vert\boldsymbol{\mu}, {\Sigma}\bigr)-\boldsymbol{\Phi}_{2}\bigl(\boldsymbol{a}\big\vert\boldsymbol{\mu}, {\Sigma}\bigr)$. Obviously, since both terms within the square bracket in the RHS of (\ref{truncated-Gaussian-linear-constraints-3}) depend on $x_1$ through ${\mu}_{2}^{*}$, hence ${\cal{TN}}_{R_{1}}(x_{1}\vert \boldsymbol{\mu}, \Sigma)$ is no longer the PDF of a Gaussian distribution unless ${\mu}_{2}^{*}$ is independent of $x_{1}$ that occurs only when $\sigma_{12}=\sigma_{21}=0$. This is while both of the full conditionals $X_{1}\vert X_{2}$ and $X_{2}\vert X_{1}$ correspond to the joint PDF in (\ref{truncated-Gaussian-linear-constraints-2}) are Gaussian. For example, $X_{1}\vert X_{2}\sim {\cal{TN}}_{R_{1}}\bigl({\mu}_{1}^{*}, \sigma_{11}^{*}\bigr)$ where $R_{1}=\bigl\{x_{1} \in \mathbb{R} \big \vert a_1\leq x_1 \leq b_1\bigr\}$, ${\mu}_{1}^{*}={\mu}_{1}+\sigma_{12}\sigma_{22}^{-1}(x_{2}- {\mu}_{2})$,  $\sigma_{11}^{*}=\sigma_{11}-\sigma_{12}\sigma_{22}^{-1}\sigma_{21}$, and $a_2\leq x_2 \leq b_2$. The following proposition generalizes the fact described above for higher dimensions.
\begin{prop}\cite{rodriguez2004efficient}\label{truncated-Gaussian-linear-constraints-4}
Suppose $\boldsymbol{X}\sim{\cal{TN}}_{\boldsymbol{R}}(\boldsymbol{\mu}, \Sigma)$ as defined in (\ref{truncated-Gaussian-linear-constraints-1}). Furthermore $\boldsymbol{Y}=\boldsymbol{c} + B\boldsymbol{X}$ where $\boldsymbol{c}=(c_{1},\cdots,c_{p})^{\top}$ is a vector of finite constants and $B$ is a $p \times p$ full rank matrix. Then,
\begin{enumerate}[label=\roman*.]
\item
$\boldsymbol{Y}\sim {\cal{TN}}_{\boldsymbol{T}}(\boldsymbol{c}+B\boldsymbol{
\mu}, B\Sigma B^{\top}),~~\boldsymbol{T}=\bigl\{ \boldsymbol{y}\in \mathbb{R}^{p} \big \vert \boldsymbol{c} + B A^{-1}\boldsymbol{a}\leq \boldsymbol{y} \leq \boldsymbol{c} + B A^{-1}\boldsymbol{b}\bigr\}$.
\item For a given partition of $\boldsymbol{X}$, $\boldsymbol{\mu}$, and $\Sigma$ given by
\begin{align*}
\boldsymbol{X}=\left[\begin{matrix} 
\boldsymbol{X}_{1}\\
X_{p}
\end{matrix}\right],
\boldsymbol{\mu}=\left[\begin{matrix} 
\boldsymbol{\mu}_{1}\\
\mu_{p}
\end{matrix}\right],
\Sigma=\left[\begin{matrix} 
\Sigma_{11}&\Sigma_{1}\\
\Sigma^{\top}_{1}&\sigma_{pp}\\
\end{matrix}\right],
\end{align*}
we have
\begin{align}\label{x-p}
X_{p}\big \vert \boldsymbol{X}_{1}=\boldsymbol{x}_{1} \sim{\cal{TN}}_{R_{p}}\bigl({\mu}_{p}^{*}, \sigma_{pp}^{*}\bigr),
\end{align} 
where
\begin{align}
{\mu}_{p}^{*}=&{\mu}_{p} +\Sigma^{\top}_{1}\Sigma_{11}^{-1}\bigl(\boldsymbol{x}_{1}- \boldsymbol{\mu}_{1}\bigr),\label{mu-p-star}\\
\sigma_{pp}^{*}=&\sigma_{pp}-\Sigma^{\top}_{1}\Sigma_{11}^{-1}\Sigma_{1},\label{sigma-p-star}\\
R_{p}=&\bigl\{x_{p}\in \mathbb{R} \big \vert \bigl(\boldsymbol{x}_{1},x_{p}\bigr) \in \boldsymbol{R} \bigr\}\nonumber.
\end{align} 
\end{enumerate}
\end{prop}
\par In what follows, we review two algorithms for generating from truncated multivariate Gaussian distribution on a set of linearly independent constraints using the Gibbs sampling. The first algorithm is due to \cite{geweke1991efficient}. For a given covariance matrix $\Sigma$ and a $p \times p$ full rank matrix $A$, suppose $\boldsymbol{Z}\sim {\cal{TN}}_{\boldsymbol{T}}(\boldsymbol{0}, A\Sigma A^{\top})$ follows a truncated Gaussian distribution on region $\boldsymbol{T}=\bigl\{ \boldsymbol{z}\in \mathbb{R}^{p}\big \vert \boldsymbol{\alpha}\leq \boldsymbol{z} \leq \boldsymbol{\beta} \bigr\}$ with positive Lebesgue measure in which 
\begin{align}
\boldsymbol{\alpha}=&\boldsymbol{a}-A\boldsymbol{\mu}\label{c-p-star},\\
\boldsymbol{\beta}=&\boldsymbol{b}-A\boldsymbol{\mu}\label{d-p-star},
\end{align} 
are vectors of finite or infinite constants. It follows from Proposition (\ref{truncated-Gaussian-linear-constraints-4}) (i) that $\boldsymbol{X}=\boldsymbol{\mu}+A^{-1}\boldsymbol{Z}\sim {\cal{TN}}_{\boldsymbol{R}}(\boldsymbol{\mu}, \Sigma)$ where $\boldsymbol{R}=\bigl\{ \boldsymbol{x}\in \mathbb{R}^{p}\big \vert \boldsymbol{a}\leq A\boldsymbol{x} \leq \boldsymbol{b} \bigr\}$ as (\ref{truncated-Gaussian-linear-constraints-1}). 
%\begin{align}\label{truncated-Gaussian-linear-constraints-5}
%\boldsymbol{X}\sim {\cal{TN}}_{\boldsymbol{R}}(\boldsymbol{\mu}, \Sigma),~~\boldsymbol{R}=\bigl\{ \boldsymbol{x}\in \mathbb{R}^{p}\big \vert \boldsymbol{a}\leq A\boldsymbol{x} \leq \boldsymbol{b} \bigr\}.
%\end{align}
Let $\boldsymbol{z}_{-i}$ denote the realization associated with $\boldsymbol{Z}_{-i}=\bigl({Z}_{1},\cdots,{Z}_{i-1},{Z}_{i+1},\cdots,{Z}_{p}\bigr)^{\top}$. It follows from Proposition (\ref{truncated-Gaussian-linear-constraints-4}) (ii) that 
%(\ref{x-p}) that 
\begin{align}\label{truncated-Gaussian-linear-constraints-6}
{Z}_{i}\big \vert \bigl(\boldsymbol{Z}_{-i}= \boldsymbol{z}_{-i}\bigr)\sim {\cal{TN}}_{T_{i}}\bigl({\mu}_{i}^{*},{\sigma}_{ii}^{*}\bigr),~~T_{i}=\bigl\{z_{i} \in \mathbb{R} \big \vert c_{i} \leq z_{i} \leq d_{i}\bigr\},
\end{align}
where elements of vectors $\boldsymbol{c}=\bigl(c_1,\cdots,c_{p}\bigr)^{\top}$ and $\boldsymbol{d}=\bigl(d_1,\cdots,d_{p}\bigr)^{\top}$ are given by (\ref{c-p-star}) and (\ref{d-p-star}), respectively. Furthermore, based on (\ref{mu-p-star}) and (\ref{sigma-p-star}), for $i=1,\cdots,p$, we have
\begin{align}%\label{truncated-Gaussian-linear-constraints-66}
{\mu}_{i}^{*}=&V^{\top}_{1}V_{11}^{-1}\boldsymbol{z}_{-i} \label{mu-p-star-2},\\
{\sigma}_{ii}^{*}=&V_{ii}-V^{\top}_{1}V_{11}^{-1}V_{1},\label{sigma-p-star-2}
\end{align}
in which $V=A\Sigma A^{\top}$. Algorithm \ref{Simulating truncated Gaussian distribution-Geweke's method} describes the method proposed by \cite{geweke1991efficient} for simulating from truncated Gaussian distribution.
\vspace{0.5cm}
\begin{algorithm}
\caption{Simulating truncated Gaussian distribution: Geweke's method} \label{Simulating truncated Gaussian distribution-Geweke's method}
\begin{algorithmic}[1]
\State Read $\boldsymbol{a}$, $\boldsymbol{b}$, $\boldsymbol{\mu}, \Sigma, A$ given by (\ref{truncated-Gaussian-linear-constraints-1}), and $N$ for number of Gibbs sampler runs;
\State Set $t=0$ and suggest the initial vector $\boldsymbol{z}^{(t)}=\bigl(z_{1}^{(t)},\cdots,z_{p}^{(t)}\bigr)^{\top} \in \boldsymbol{T}$ where ${T}_{i}$ (for $i=1,\cdots,p$) is given by (\ref{truncated-Gaussian-linear-constraints-6});
\While{$t < N$}
\State set $i=1$;
\While{$i \leq p$} %\Comment{put some comments here}
\State Draw ${z}_{i}^{(t+1)}$ form full conditional ${Z}_{i}^{(t+1)}\big \vert 
\bigl({Z}_{1}^{(t+1)}={z}_{1}^{(t+1)},\cdots,{Z}_{i-1}^{(t+1)}={z}_{i-1}^{(t+1)},{Z}_{i+1}^{(t)}={z}_{i+1}^{(t)},\cdots,{Z}_{p}^{(t)}={z}_{p}^{(t)}\bigr)$ that follows a truncated Gaussian distribution with PDF given by (\ref{truncated-Gaussian-linear-constraints-6}); 
\State set $i=i+1$;
   \EndWhile
   \State {\bf{end}}
   \State $\boldsymbol{z}^{(t+1)} \leftarrow \boldsymbol{z}^{(t)}$;
      \State Accept $\boldsymbol{x}^{(t+1)}=\boldsymbol{\mu}+A^{-1}\boldsymbol{z}^{(t+1)}$ as a sample of from distribution in (\ref{truncated-Gaussian-linear-constraints-1});
\State set $t=t+1$;
   \EndWhile  %\label{roy's loop}
   \State {\bf{end}}
\end{algorithmic}
\end{algorithm}
\vspace{0.5cm}
The second algorithm is due to \cite{rodriguez2004efficient}. Let $p$-dimensional random vector $\boldsymbol{X}=(X_{1},\cdots,X_{p})^{\top}$ follows a truncated Gaussian distribution on region $\boldsymbol{R}$ denoted as  
\begin{align}\label{truncated-Gaussian-linear-constraints-7}
{\cal{TN}}_{\boldsymbol{R}}(\boldsymbol{\mu},\Sigma),~~\boldsymbol{R}=\bigl\{ \boldsymbol{x}\in \mathbb{R}^{p} \big \vert C \boldsymbol{x} \leq \boldsymbol{b}\bigr\},
\end{align}
where rows of matrix $B$ are not restricted to be linearly independent. If $D$ is a square matrix of full rank for which we have $D\Sigma D^{\top}=\boldsymbol{I}_{p}$, then for transformation $\boldsymbol{Z}=D\boldsymbol{X}$ it follows from Proposition \ref{truncated-Gaussian-linear-constraints-4} (i) that
\begin{align}\label{truncated-Gaussian-linear-constraints-8}
\boldsymbol{Z}\sim {\cal{TN}}_{\boldsymbol{T}}\bigl(D\boldsymbol{\mu},\boldsymbol{I}_{p}\bigr),~~\boldsymbol{T}=\bigl\{ \boldsymbol{z}=D\boldsymbol{x}\in \mathbb{R}^{p} \big \vert E \boldsymbol{z} \leq \boldsymbol{b}\bigr\},
\end{align}
where 
\begin{align}\label{truncated-Gaussian-linear-constraints-9}
E=CD^{-1}.
\end{align}
It is not hard to see that the only choice for $D$ is inverse of the lower-triangular Cholesky decomposition of $\Sigma$. Using the transformation $\boldsymbol{Z}=D\boldsymbol{X}$ avoids computing ${\mu}_{i}^{*}$ and ${\sigma}_{ii}^{*}$ given in (\ref{truncated-Gaussian-linear-constraints-6}) when applying Algorithm \ref{Simulating truncated Gaussian distribution-Geweke's method} proposed by \cite{geweke1991efficient}. Suppose $\boldsymbol{\alpha}=D\boldsymbol{\mu}$, $\boldsymbol{Z}_{-i}=\bigl({Z}_{1},\cdots,{Z}_{i-1},{Z}_{i+1},\cdots,{Z}_{p}\bigr)^{\top}$, and $\boldsymbol{z}_{-i}$ denotes the realization of $\boldsymbol{Z}_{-i}$. Based on Proposition \ref{truncated-Gaussian-linear-constraints-4} (ii), we have
\begin{align}\label{truncated-Gaussian-linear-constraints-10}
Z_{i} \big \vert \bigl(\boldsymbol{Z}_{-i}=\boldsymbol{z}_{-i}\bigr)\sim {\cal{TN}}_{T_{i}}\bigl(\alpha_{i},1\bigr),~~T_{i}=\bigl\{z_{i}\in \mathbb{R} \big \vert E \boldsymbol{z} \leq \boldsymbol{b}\bigr\},
\end{align}
where $\alpha_{i}$ is the $i$th element of vector $\boldsymbol{\alpha}$. Let $E_{-i}$ denote the matrix $E$ given in (\ref{truncated-Gaussian-linear-constraints-9}) when its $i$-th column is removed and $\boldsymbol{e}_{i}$ is the $i$-th column of matrix $E$. Then, the region $T_{i}$ given in (\ref{truncated-Gaussian-linear-constraints-10}) can be represented as 
\begin{align}\label{truncated-Gaussian-linear-constraints-11}
T_{i}=\bigl\{z_{i}\in \mathbb{R} \big \vert  \boldsymbol{e}_{i} z_{i} \leq \boldsymbol{b}- {E}_{-i}\boldsymbol{z}_{-i}\bigr\}.
\end{align}
As pointed out by \cite{rodriguez2004efficient}, since constraints on $\boldsymbol{X}$ construct a convex region in $\mathbb{R}^{p}$, hence the region $T_{i}$ given by (\ref{truncated-Gaussian-linear-constraints-11}) takes the forms $l_{i}\leq z_{i} \leq u_{i}$, $-\infty\leq z_{i} \leq u_{i}$, or $l_{i}\leq u_{i}\leq \infty$ in which $l_{i}$ and $u_{i}$ are, accordingly, the lower and upper bounds of inequalities obtained from (\ref{truncated-Gaussian-linear-constraints-11}). Algorithm \ref{Simulating truncated Gaussian distribution-rodriguez method} describes how to implement the method proposed by \cite{rodriguez2004efficient} for simulating from a truncated Gaussian distribution.
 \vspace{0.5cm}
\begin{algorithm}
\caption{Simulating truncated Gaussian distribution: Rodriguez-Yam et al. method} \label{Simulating truncated Gaussian distribution-rodriguez method}
\begin{algorithmic}[1]
\State Read $\boldsymbol{b}$, $\boldsymbol{\mu}$, $\Sigma$, $C$ given by (\ref{truncated-Gaussian-linear-constraints-7}), and $N$ for number of Gibbs sampler runs;
\State Compute $\boldsymbol{\alpha}=D\boldsymbol{\mu}$ and $E=CD^{-1}$ in which $D$ is the Cholesky decomposition of $\Sigma$; 
\State Set $t=0$ and suggest the initial value $\boldsymbol{z}^{(t)}=\bigl(z_{1}^{(t)},\cdots,z_{p}^{(t)}\bigr)^{\top} \in \boldsymbol{T}$ where $\boldsymbol{T}$ is given by (\ref{truncated-Gaussian-linear-constraints-11});
\While{$t \leq N$}
\State set $i=1$;
\While{$i < p$}  %\Comment{put some comments here}
\State Draw ${z}_{i}^{(t+1)}$ form full conditional ${Z}_{i}^{(t+1)}\big \vert 
\bigl({Z}_{1}^{(t+1)}={z}_{1}^{(t+1)},\cdots,{Z}_{i-1}^{(t+1)}={z}_{i-1}^{(t+1)},{Z}_{i+1}^{(t)}={z}_{i+1}^{(t)},\cdots,{Z}_{p}^{(t)}={z}_{p}^{(t)}\bigr)$ that follows a truncated Gaussian distribution with PDF given by (\ref{truncated-Gaussian-linear-constraints-10}); 
\State set $i=i+1$;
   \EndWhile
   \State {\bf{end}}
   \State $\boldsymbol{Z}^{(t+1)} \leftarrow \boldsymbol{Z}^{(t)}$;
   \State $\boldsymbol{X}^{(t+1)}=D^{-1}\boldsymbol{Z}^{(t+1)} $
\State set $t=t+1$;
   \EndWhile  %\label{roy's loop}
   \State {\bf{end}}
\end{algorithmic}
\end{algorithm}
In what follows, we are willing to review an example given by \citep{rodriguez2004efficient} for simulating from a bivariate truncated Gaussian distribution. It is worthwhile to note that the methods proposed by \cite{geweke1991efficient} and \citep{rodriguez2004efficient} hereafter are denoted as Method 1 and Method 2, respectively.
\begin{example}\label{exam-truncated-Gaussian-linear-constraints-12}%\lipsum[]
Suppose we are interested in simulating from $\boldsymbol{X}=(X_{1},X_{2})^{\top}\sim {\cal{TN}}_{\boldsymbol{R}}(\boldsymbol{\mu},\Sigma)$ where $\boldsymbol{R}=\bigl\{ \boldsymbol{x}\in \mathbb{R}^{2} \big \vert \boldsymbol{x}\geq  \boldsymbol{0}\bigr\}$, $\boldsymbol{\mu}=(\mu_{1},\mu_{2})^{\top}$ and $\Sigma=\bigl[(1,0.8)^{\top},(0.8,1)^{\top}\bigr]$. We discuss the key features for implementing Algorithm \ref{Simulating truncated Gaussian distribution-Geweke's method} and Algorithm \ref{Simulating truncated Gaussian distribution-rodriguez method} as follows.
\begin{itemize}
\item {\bf{Method 1}}: For implementing the first step of Algorithm \ref{Simulating truncated Gaussian distribution-Geweke's method}, we set $A=\bigl[(1,0)^{\top},(0,1)^{\top}\bigr]$, $\boldsymbol{a}=(0,0)^{\top}$, $\boldsymbol{b}=(+\infty,+\infty)^{\top}$, and $N=2000$. The vector of initial values in the second step can be $\boldsymbol{z}^{(0)}=(1,1)^{\top}$ and further, using $\boldsymbol{\mu}=(0,0)^{\top}$, it follows from (\ref{c-p-star}) and (\ref{d-p-star}) that $T_{1}=T_{2}=[0,\infty)$. Based on (\ref{mu-p-star}) and (\ref{sigma-p-star}), the full conditionals in the sixth step of Algorithm \ref{Simulating truncated Gaussian distribution-Geweke's method} are given as follows.
\begin{align}\label{truncated-Gaussian-linear-constraints-13}
&Z_{1}\big \vert \bigl({\boldsymbol{Z}}_{-1}={\boldsymbol{z}}_{-1}\bigr) \sim {\cal{TN}}_{T_{1}}\Bigl(\frac{4}{5}z_{2},\frac{9}{25}\Bigr),\nonumber\\
&Z_{2}\big \vert \bigl({\boldsymbol{Z}}_{-2}={\boldsymbol{z}}_{-2}\bigr) \sim {\cal{TN}}_{T_{2}}\Bigl(\frac{4}{5}z_{1},\frac{9}{25}\Bigr).\nonumber
\end{align}
\vspace{.5cm}
The \verb+R+ function called \verb+rtrunnorm_gibbs(\cdot)+ for implementing Method 1 is given as follows.
\vspace{.5cm}
%\begin{lstlisting}[style=deltaj]
%rtnorm <- function(n, mu, sigma, a, b) 
%{# this function is called in Method 1 and Method 2 
%  u <- runif(n)
%  cdf.a <- pnorm(a, mu, sigma)
%  cdf.b <- pnorm(b, mu, sigma)
%  qnorm( u*(cdf.b - cdf.a) + cdf.a, mean = mu, sd = sigma )  
%}
%\end{lstlisting}
\vspace{.5cm}
\begin{lstlisting}[style=deltaj]
R> rtrunnorm_gibbs <- function(N, Mu, Sigma, a, b, A)
 +{
 + p <- length(Mu)
 + V <- A%*%Sigma%*%t(A) 
 + T1 <- matrix( cbind(a[1:p] - A%*%Mu[1:p], b[1:p] - A%*%Mu[1:p]), nrow = p, ncol = 2)
 + Z <- matrix(0, nrow = N, ncol = p)
 + Z[1, ] <- (a + b)/2 
 + Vinv <- array( 0, dim = c(p-1, p-1, p) )
 + V1   <- matrix(0, nrow = p, ncol = p-1)
 +	for(i in 1:p)
 +	{
 +		Vinv[, , i] <- solve( V[-i, -i] )
 +		V1[i, ] <- as.numeric( V[i, -i]%*%Vinv[, , i] )
 +	}
 +  j <- 1
 +	while(j < N)
 +	{
 +		for(i in p:1)
 +		{
 +			mu.star <- V1[i, ]%*%Z[j, -i]
 +			sigma.star <- sqrt( V[i, i] - V1[i, ]%*%V[i, -i] )
 +			u <- runif(1)
 +			cdf.a <- pnorm(T1[i, 1], mu.star, sigma.star)
 +			cdf.b <- pnorm(T1[i, 2], mu.star, sigma.star)
 +			Z[j, i] <- qnorm( u*(cdf.b - cdf.a) + cdf.a,
 +														mean = mu.star, sd = sigma.star )
 +			Z[j+1, i] <- Z[j, i]
 +		}
 +	j <- j + 1
 +	}
 + Z <- matrix(Mu, nrow = N, ncol = p, byrow = T) + Z%*%solve(A)
 + return(Z)
 +}
\end{lstlisting}
Hence, for simulating truncated multivariate Gaussian using Method 1, we proceed by the following.
\begin{lstlisting}[style=deltaj]
R > N <- 20000 # number of Gibbs sampler iterations
R > X <- Z <- matrix(0, nrow = N, ncol = 2)
R > Mu <- c(0, 0)
R > Sigma <- matrix( c(1, 0.8, 0.8, 1), nrow = 2, ncol = 2)
R > A <- diag(2)
R > a <- rep(0, 2) # truncation lower bounds
R > b <- rep(Inf, 2) # truncation upper bounds
R > rtrunnorm_gibbs(N, Mu, Sigma, a, b, A)
\end{lstlisting}
\item {\bf{Method 2}}: For implementing the first step of Algorithm \ref{Simulating truncated Gaussian distribution-rodriguez method}, we set $\boldsymbol{b}=\boldsymbol{0}$, $C=-\boldsymbol{I}_{2}$, and $N=2000$. For completing the second step, the Cholesky decomposition $D$, of $\Sigma$ and its inverse, are $D=\bigl[(1,4/5)^{\top},(0,3/5)^{\top}\bigr]$ and $D^{-1}=\bigl[(1,-4/3)^{\top},(0,5/3)^{\top}\bigr]$, respectively. Hence, for $\boldsymbol{\mu}=(0,0)^{\top}$, we have $\boldsymbol{\alpha}=D^{-1}\boldsymbol{\mu}=(0,0)^{\top}$ and $E=CD=\bigl[(-1,-4/5)^{\top},(0,-3/5)^{\top}\bigr]$ where ${E}_{-1}=(0,-3/5)^{\top}$ and ${E}_{-2}=(-1,-4/5)^{\top}$. Moreover,
\begin{eqnarray}\label{truncated-Gaussian-linear-constraints-14}
&T_{1}=\biggl\{z_{1} \in \mathbb{R}\bigg \vert \left[\begin{matrix} 
-1\\
-\frac{4}{5}
\end{matrix}\right] 
z_{1} \leq -
\left[\begin{matrix} 
0\\
-\frac{3}{5}
\end{matrix}\right]
z_{2} \bigg\},\nonumber\\
&=\Bigl\{z_{1} \in \mathbb{R}\Big \vert 
z_{1} \geq 0, z_{1} \geq -\frac{3}{4} z_{2} \Big\},~~\nonumber\\
&=\Bigl[\max \Bigl\{0,-\frac{3}{4}z_{2}\Bigr\}, \infty\Bigr),~~~~~~~~~~~~~
\end{eqnarray}
and
\begin{eqnarray}\label{truncated-Gaussian-linear-constraints-15}
&T_{2}=\biggl\{z_{2} \in \mathbb{R}\bigg \vert \left[\begin{matrix} 
0\\
-\frac{3}{5}
\end{matrix}\right] 
z_{2} \leq -
\left[\begin{matrix} 
-1\\
-\frac{4}{5}
\end{matrix}\right]
z_{1} \bigg\},\nonumber\\
&=\Bigl\{z_{2} \in \mathbb{R}\Big \vert 
z_{2} \geq -\frac{4}{3} z_{1} \Big\},~~\nonumber\\
&=\bigl[-\frac{4}{3}z_{1}, \infty\bigr).~~~~~~~~~~~~~~~~~~~
\end{eqnarray}
Using (\ref{truncated-Gaussian-linear-constraints-14}), (\ref{truncated-Gaussian-linear-constraints-15}), and $\boldsymbol{z}^{(0)}=(1,1)^{\top}$ the third step of the Algorithm \ref{Simulating truncated Gaussian distribution-rodriguez method} is complete. Finally, the required full conditionals in the seventh step of this algorithm are given as follows.
\begin{align}%\label{truncated-Gaussian-linear-constraints-16}
&Z_{1}\big \vert \bigl({\boldsymbol{Z}}_{-1}={\boldsymbol{z}}_{-1}\bigr) \sim {\cal{TN}}_{T_{1}}\bigl(\mu_{1},1\bigr),\nonumber\\
&Z_{2}\big \vert \bigl({\boldsymbol{Z}}_{-2}={\boldsymbol{z}}_{-2}\bigr) \sim {\cal{TN}}_{T_{2}}\Bigl(-\frac{4}{3}\mu_{1}+\frac{5}{3}\mu_{2},1\Bigr).\nonumber
\end{align}
The \verb+R+ code for implementing Method 2, when $\boldsymbol{\mu}=(0,0)^{\top}$, is given as follows.
\vspace{.5cm}
\begin{lstlisting}[style=deltaj]
N <- 20000 # number of Gibbs sampler iterations
Z <- matrix(0, nrow = N, ncol = 2)
Mu <- c(0, 0) # mean vector can be changed arbitrarily
Sigma <- matrix( c(1, 0.8, 0.8, 1), nrow = 2, ncol = 2)
C <- -diag(2)
D0 <- t( chol(Sigma) )
D0inv <- solve( D0 )
E <- C%*%D0
alpha <- D0inv%*%Mu
Z[1, ] <- rep(1, 2) # vector of initial values
j <- 1
 while(j < N)
 {
  Z[j + 1, 1] <- rtnorm(1, alpha[1], 1, max(0, -3/4*Z[j, 2]), Inf)
  Z[j + 1, 2] <- rtnorm(1, alpha[2], 1, -4/3*Z[j + 1, 1], Inf)
  j <- j + 1
 }
X <- Z%*%t(D0) # X is vector of realizations following truncated 
                # multivariate Gaussian distribution
\end{lstlisting}
\end{itemize}
\end{example}
%%%%%%%%%%%%%%%%%%%%%%%%%%%%%%%%%%%%%%%%%%%%%%%%%%%%%%%%%%%%%%%%%
%\section*{Appendix}
%\appendix{}
%%%%%%%%%%%%%%%%%%%%%%%%%%%%%%%%%%%%%%%%%%%%%%%%%%%%%%%%%%%%%%%%%
% \bibliographystyle{elsarticle-num} 
 %\bibliographystyle{elsarticle-harv} 
%%%%%%%%%%%%%%%%%%%%%%%%%%%%%%%%%%%%%%%%%%%%%%%%%%%%%%%
% Sample table                                        %
% Source: www1.maths.leeds.ac.uk/latex/TableHelp1.pdf %
%%%%%%%%%%%%%%%%%%%%%%%%%%%%%%%%%%%%%%%%%%%%%%%%%%%%%%%
% \noindent Lorem 
\bibliographystyle{plainnat}
   % basic style, author-year citations
\bibliography{ref}
\end{document}